\documentclass{aastex6}

\usepackage{natbib}
\usepackage{graphicx}

\usepackage{graphics}
\usepackage{amssymb}
\usepackage{amsbsy}
\usepackage{amsfonts}
\usepackage{bm}
\usepackage{mathtools}
\usepackage{empheq}

\usepackage{amsmath}

\usepackage[makeroom]{cancel}
\usepackage{tikz}

\newcommand{\bhat}{\boldsymbol{\hat{b}}}
\newcommand{\bu}{\boldsymbol{u}}
\newcommand{\bV}{\boldsymbol{v}}
\newcommand{\bc}{\boldsymbol{c}}
\newcommand{\bE}{\boldsymbol{E}}
\newcommand{\bb}{\boldsymbol{B}}
\newcommand{\bj}{\boldsymbol{j}}
\newcommand{\bp}{\boldsymbol{p}}
\newcommand{\bq}{\boldsymbol{q}}
\newcommand{\br}{\boldsymbol{r}}

\newcommand{\pr}{\partial}
\newcommand{\nn}{\nonumber}

\newcommand{\bx}{\boldsymbol{x}}
\newcommand{\bX}{\boldsymbol{X}}

\newcommand{\kpar}{k_\parallel}
\newcommand{\kperp}{k_\perp}

\newcommand{\bpar}{\beta_\parallel}
\newcommand{\bee}{\begin{eqnarray}}
\newcommand{\eee}{\end{eqnarray}}

\newcommand{\bxhat}{\hat{b}_x}
\newcommand{\byhat}{\hat{b}_y}
\newcommand{\bzhat}{\hat{b}_z}

\newcommand{\trace}{\textrm{Tr}}

\newcommand{\sign}{\textrm{sign}}

\begin{document}

\title{An introductory guide to fluid models with anisotropic temperatures\\
  Part 1 - CGL description and collisionless fluid hierarchy}

\author{P. Hunana\altaffilmark{1,2}, A. Tenerani\altaffilmark{3},  G. P. Zank\altaffilmark{4,5}, E. Khomenko\altaffilmark{1,2}, M. L. Goldstein\altaffilmark{6}, \\
  G. M. Webb\altaffilmark{4}, P. S. Cally\altaffilmark{7}, M. Collados\altaffilmark{1,2}, M. Velli\altaffilmark{8}, L. Adhikari\altaffilmark{4} }


\altaffiltext{1}{Instituto de Astrof\'isica de Canarias (IAC), La Laguna, Tenerife, 38205, Spain; peter.hunana@gmail.com}
\altaffiltext{2}{Universidad de La Laguna, La Laguna, Tenerife, 38206, Spain}
\altaffiltext{3}{Department of Physics, The University of Texas at Austin, TX 78712, USA}
\altaffiltext{4}{Center for Space Plasma and Aeronomic Research (CSPAR),
  University of Alabama, Huntsville, AL 35805, USA}
\altaffiltext{5}{Department of Space Science, University of Alabama, Huntsville, AL 35899, USA}
\altaffiltext{6}{Space Science Institute, Boulder, CO 80301, USA}
\altaffiltext{7}{School of Mathematics, Monash University, Clayton, Victoria 3800, Australia}
\altaffiltext{8}{Department of Earth, Planetary, and Space Sciences, University of California, Los Angeles, CA 90095, USA}




\begin{abstract}
  We present a detailed guide to advanced collisionless fluid models that incorporate kinetic effects into the fluid framework, and
  that are much closer to the collisionless kinetic description than traditional magnetohydrodynamics. Such fluid models are directly applicable 
  to modeling turbulent evolution of a vast array of astrophysical plasmas, such as the solar corona and the solar wind, the interstellar medium,
  as well as accretion disks and galaxy clusters. The text can be viewed as a detailed guide 
  to Landau fluid models and it is divided into two parts.
  Part 1 is dedicated to fluid models that are obtained by closing the fluid hierarchy with simple (non Landau fluid) closures.
  Part 2 is dedicated to Landau fluid closures. Here in Part 1, we discuss the CGL fluid model in great detail,
  together with fluid models that contain dispersive effects introduced by the Hall term and
  by the finite Larmor radius (FLR) corrections to the pressure tensor. 
  We consider dispersive effects introduced by the non-gyrotropic heat flux vectors. We investigate the parallel and oblique firehose instability,
  and show that the non-gyrotropic heat flux strongly influences the maximum growth rate of these instabilities.     
  Furthermore, we discuss fluid models that contain evolution equations for the gyrotropic heat flux fluctuations and
  that are closed at the 4th-moment level by prescribing a specific form for the distribution function. For the bi-Maxwellian distribution, such a closure
  is known as the ``normal'' closure. We also discuss a fluid closure for the bi-kappa distribution. 
  Finally, by considering one-dimensional Maxwellian fluid closures at higher-order moments,
  we show that such fluid models are always unstable. The last possible non Landau fluid closure is therefore the ``normal'' closure, and beyond the 4th-order moment,
  Landau fluid closures are required.
\end{abstract}

\maketitle
\newpage
\tableofcontents

\newpage
\section{Introduction}
  Fluid models are an extremely important tool in many areas of space physics, astrophysics and laboratory plasmas. 
  Even though many physical systems studied in these fields are almost collisionless, where a proper kinetic description should be
  used, traditional fluid models with isotropic scalar pressures (temperatures), such as the usual magnetohydrodynamic description (MHD)
  and multi-fluid models based on this description, were extremely successful in modeling, interpreting, or at least offering the first insight into many space physics phenomena. 
  For example, it was indeed the simplified fluid approach that allowed \cite{Parker1958c} to predict the existence of the solar
  wind, which was surprisingly several decades later than the breakthrough discoveries in quantum mechanics and relativity.
  In recent years, observational studies re-sparked interest in the temperature anisotropy effects, that cannot be studied with the
  usual MHD fluid descriptions, and we anticipate that the interest will grow even further, once data from the Parker Solar Probe and the future Solar Orbiter
  missions are analyzed.  

  The correct modeling and understanding of collisionless plasmas in a fluid framework, does not concern only systems with plasma temperatures
  far from an isotropic state. It is sometimes forgotten that while the application of fluid theory to strongly collisional systems is intuitively
  obvious, the approach to collisionless (or weakly collisional) systems is not.
  From a linear perspective, the usual MHD description does not converge to the collisionless kinetic
  description even in the low-frequency long-wavelength limit  (even when the kinetic
    distribution function is considered to be an isotropic Maxwellian), since in the absence of collisions the isotropic equation of state is never correct. Additionally,
    Landau damping never completely vanishes (even when electrons are hot). The situation is much more complicated from a non-linear perspective, where on average,
    the effect of Landau damping might be balanced by processes called stochastic plasma echoes \citep{Meyrand2019}.

 As a linear example, the ordering of the phase speeds in MHD is always slow, Alfv\'en, fast,
i.e., $v_s\le v_A \le v_f$. In contrast, in collisionless systems with sufficiently
high plasma beta, the real phase speed of the slow mode can become faster than
the Alfv\'en mode, so that the ordering can become Alfv\'en, slow, fast, i.e., $v_A\le v_s \le v_f$.
Or in another words, the phase speeds of linear eigenmodes that are present in MHD do not
hold in the collisionless (or weakly collisional) regime, and this effect exists even if the temperatures are isotropic.   
From a fluid perspective, the main reason for this discrepancy is that in magnetized collisionless systems,
the pressure \emph{fluctuations} in the directions parallel and perpendicular to the magnetic field lines are not equivalent,
and cannot be described with a single scalar pressure equation. The pressure fluctuations have to be described 
with two different evolution equations for $p_\parallel$ and $p_\perp$, even if the mean pressure values are equal $(p_\parallel^{(0)}=p_\perp^{(0)})$ and no  mean temperature
anisotropy exists.

The simplest collisionless fluid description is the CGL fluid model - named after Chew, Goldberger and Low \cite{Chew1956} - and sometimes also referred to as collisionless MHD.
The CGL dispersion relation is not equivalent to the MHD dispersion relation (even in the case $p_\parallel^{(0)}=p_\perp^{(0)}$), since the pressure fluctuations still remain anisotropic and the 
evolution equations for $p_\parallel$ and $p_\perp$ remain different.
In another words, in collisionless systems the distribution function is free to evolve from
its initial state and to become anisotropic. By ``forcibly'' prescribing only one scalar pressure,
one effectively prescribes a high-collisionality regime, even if collisions
are not prescribed explicitly.  Formally, the MHD equations can indeed be derived
from the collisionless Vlasov equation, i.e. with no explicit collisional operator.
It is therefore often stated, that the MHD description is highly-collisional \emph{implicitly}.
 Moreover, as discussed for example by \cite{Kulsrud1983}, while in the presence of a magnetic field, 
transverse motions can in some circumstances be described by fluid-type equations,
the determination of pressures as well as longitudinal motions a priori requires a kinetic description.

Here we focus on collisionless fluid models.
Nevertheless, weak collisions can be incorporated easily, and calculations just yield additional
terms on the right hand sides of the parallel and perpendicular pressure and heat flux equations. For the simple
 Bhatnagar-Gross-Krook (BGK) collisional operator
 \citep{BGK1954}, see for example \cite{Snyder1997}.
A thorough review of anisotropic fluid models, including the collisional dynamics, was presented by \cite{BarakatSchunk1982}.
A very good discussion about collisionality of various astrophysical plasmas, such as the solar wind, interstellar medium, accretion disks and galaxy clusters,
can be found for example in \cite{Schekochihin2009}. Weakly collisional fluid models also seem to be applicable for modelling the upper solar photosphere and chromosphere,
where a curious situation exists and the proton-proton collisional frequency is roughly equal to the proton cyclotron frequency - see for example Figure 1 in \cite{Khomenko2014}. 
We note that in this 
guide we use the definition of a ``collisionless fluid model'' as being a fluid model that 1) is derived from collisionless Vlasov
equation with zero right hand side; and 2) that has two different pressure equations. Our definition therefore differs
from an alternative view of for example \cite{Zank2014,ZankBook2014}, where models with scattering operator that reflects charged particle scattering by
electromagnetic fluctuations on the right hand side of Vlasov equation are also viewed as collisionless. 

Fluid modeling of collisionless plasmas is an extremely attractive subject, and an enormous amount of theoretical and numerical work was done in this
field in the past. The manuscript presented here has no intention to be a proper review paper in this field. 
In our opinion, no satisfactory easy-to-read introductory
paper exists about collisionless fluid models, and this manuscript attempts to fill such a spot. 
Instead of just stating the major results and discussing what was done in the past and by whom, we make a significant effort to present a
(hopefully consistent) derivation of basic collisionless fluid models. On one hand, the presented calculations might be considered as too detailed in many places.
On the other hand, it is exactly the relatively complicated algebra of collisionless fluid models, that makes the field difficult to enter for new researchers.
The primary goal of this paper is to allow researchers new to the subject to follow the algebra easily.
Instead of spending years, an interested reader should be able to get comfortable with the basics of the subject in a matter of days,
or perhaps a couple of weeks. The text is separated to two parts.
Part 1 is dedicated to fluid models that are obtained by closing the fluid hierarchy with simple (non Landau fluid) closures. Part 2 is dedicated to Landau fluid closures.
\\\\
Here in Part 1, Section \ref{section:PressureTensor},
we start with the detailed derivation of the pressure tensor equation. The pressure tensor $\boldsymbol{p}$ is then decomposed to its gyrotropic 
part $\boldsymbol{p}^g=p_\parallel \bhat\bhat +p_\perp (\boldsymbol{I}-\bhat\bhat)$ (also referred to as $\bp^{CGL}$),
and non-gyrotropic part $\boldsymbol{\Pi}$, the latter usually called the finite Larmor radius
(FLR) corrections to the pressure tensor, or the gyroviscous stress tensor.
In the highly-collisional decomposition $\boldsymbol{p}=p\boldsymbol{I}+\boldsymbol{\Pi}$, see e.g. \cite{Braginskii1958,Braginskii1965},
the quantity $\boldsymbol{\Pi}$ is called the stress tensor. 
The decomposition procedure yields rigorously exact (even though still not closed) evolution equations for $p_\parallel$ and $p_\perp$, see e.g.
\cite{PassotSulem2004b,Goswami2005,Oraevskii1968}. 
Importantly, at the leading order (by neglecting the $\boldsymbol{\Pi}$ and also the non-gyrotropic heat flux contributions $\bq^{\textrm{ng}}$), the equations of
\cite{Chew1956},  hereafter referred to as CGL, are recovered.
We discuss the paper by \cite{Chew1956} and point out that the paper derived the correct form of the pressure equations
with the gyrotropic heat flux contributions included, however, the quantities $q_n,q_s$ used in that paper are related to the usual $q_\parallel,q_\perp$ by relations
$q_n=q_\parallel-3q_\perp$ and $q_s=q_\perp$. By further neglecting the heat flux contributions, the pressure equations can be written in conservative
form. The resulting pressure equations became known as the CGL equations, and
they can be interpreted as the conservation laws for the first and second adiabatic invariants \citep{Kulsrud1983,GurnettBhattacharjee2005}.  

 Furthermore, we discuss the general equations of collisionless multi-fluid models. 
The pressure equations are rigorously exact, even though the system is not closed, since the FLR pressure
tensor and the entire heat flux tensor are not specified; for a quick look see eq. (\ref{eq:Gcontinuity})-(\ref{eq:Gpperp}).
Rewriting the system to a form where the usual CGL conservation laws are on the
left hand side, and all other contributions on the right hand side, nicely represents the complicated plasma heating process that can be encountered, and
that are responsible for the breaking of the adiabatic invariants; see eq. (\ref{eq:VelkaPica1}), (\ref{eq:VelkaPica2}).
We also discuss the conservation of energy. For the case of periodic boundary conditions, the total conservation of energy has a very illuminating form,
that can be found for example in \cite{YangMatthaeus2017_1,YangMatthaeus2017_2}.
That formulation is obtained by considering the \emph{total} ``internal'' energy for each particle species $E^{\textrm{int}}_r=\frac{1}{2}\langle\textrm{Tr}\,\bp_r\rangle$ (where
the brackets represent integration over the entire spatial domain), which only reveals the total plasma heating. 
Here we split the internal energy into its parallel and perpendicular components
$E^{\textrm{int}}_{\parallel r}=\frac{1}{2}\langle p_{\parallel r}\rangle$, $E^{\textrm{int}}_{\perp r}=\langle p_{\perp r}\rangle$,   
and formulate the total conservation of energy with the possibility of anisotropic plasma heating, see eq. (\ref{eq:GpparPer2})-(\ref{eq:EMAGG}).

The CGL description is analyzed in great detail in Section \ref{section:CGL}.
We derive the CGL dispersion relation, and discuss properties of the slow, Alfv\'en and fast modes that are present.
We verify many of the classical results of \cite{Abraham-Shrauner1967}, here written in a slightly more convenient notation
by using the parallel plasma beta $\bpar$, and the temperature anisotropy ratio $a_p=T_\perp/T_\parallel$.
Collisionless plasmas can not reach arbitrarily large values of temperature anisotropy, and the linear CGL eigenmodes can indeed become unstable, with
the associated instabilities referred to as the (parallel and oblique) firehose instability and the mirror instability.
Similarly to MHD, the simple CGL description does not contain any dispersive effects and is technically scale invariant, even though valid
only on the largest scales. The instability thresholds present in the CGL model therefore
can be referred to as ``hard'' thresholds, i.e. thresholds that are obtained in the long-wavelength low-frequency limit. 
The firehose and mirror instabilities are believed to play a crucial role in solar wind dynamics (see e.g. \cite{Hellinger2006,Bale2009}),
and in Figure \ref{fig:CGL-th}, they are plotted in the usual $(\bpar,T_\perp/T_\parallel)$ plane
with logarithmic scales. While the firehose threshold matches the one obtained from kinetic theory, the CGL mirror threshold contains the well-known
factor of 6 error for large $\bpar$ values.  The factor of 6 error can be interpreted as inadequacy of the adiabatic CGL closure in
  the very slow-dynamics context, such as the mirror instability. This is further addressed in Section \ref{sec:StaticClosure}, where we discuss that
  the ``static'' closure, which can be viewed as a generalization of isothermal closure in presence of temperature
  anisotropy and variations of magnetic field strength, reproduces the correct mirror threshold
  \citep{Constantinescu2002,ChustBelmont2006,PassotRubanSulem2006}. 

We discuss the core differences between MHD and CGL, which can be nicely summarized  
with the concept of adiabatic indices $\gamma$, related to the number of degrees of freedom $i$ by $\gamma=(i+2)/i$. While MHD is fully 3D with $\gamma=5/3$,
the CGL can be viewed as composed of 1D and 2D dynamics with $\gamma_\parallel=3$ and $\gamma_\perp=2$. 
We also address the velocity and magnetic field eigenvectors. Similarly to the velocity field eigenvector in MHD (see Figure \ref{fig:MHD}), which shows a
``singular'' behavior for strictly parallel propagation with $V_A=C_s$, the CGL eigenvector (Figure \ref{fig:CGL-u}) shows similar singularity for $\bpar=2/(4-T_\perp/T_\parallel)$. 
We also briefly discuss fluid models with empirical ``free'' polytropic indices $\gamma_\parallel,\gamma_\perp$, studied for example by
\cite{HauSonnerup1993,HauPhan1993,AbrahamShrauner1973}.

In Section \ref{section:HallCGL},
we introduce the simplest dispersive effects by including the Hall term into the induction equation, and study dispersion relations of the Hall-CGL model. 
We focus on the parallel firehose instability,  and show that the instability is indeed associated with the whistler mode,  
see Figures \ref{fig:4} and \ref{fig:5}. We show that negative real frequencies $\omega_r<0$ have to be handled carefully, and that
non-causal analytic solutions (\ref{eq:Hall-IC1})-(\ref{eq:Hall-IC2}) have to be modified to the causal form (\ref{eq:Hall-IC1corr})-(\ref{eq:Hall-IC2corr}).
We briefly discuss the simplest ion-cyclotron resonances, and compare solutions of the Hall-CGL model with solutions
of linear kinetic theory, see Figure \ref{fig:7}.

In Section \ref{section:FLR}, we evaluate the FLR tensor $\boldsymbol{\Pi}$ at several levels of approximation.
The evaluation of the FLR tensor is cumbersome, because the tensor is described by the pressure tensor equation implicitly.     
We first reproduce the fully nonlinear ``inversion'' procedure on how to obtain $\boldsymbol{\Pi}$ from expression $(\bhat\times\boldsymbol{\Pi})+(\bhat\times\boldsymbol{\Pi})^T$ 
that can be found for example in \cite{Hsu1986,PassotSulem2004b,Ramos2005} as a brief note. By applying this inversion procedure
at the pressure tensor equation evaluates the FLR corrections correctly along the magnetic field lines,
but leads to an equation for $\boldsymbol{\Pi}$ that is still implicit. Nevertheless, by evaluating the resulting equation at the leading-order
(technically first order in frequency and wavenumber), leads to an explicit expression for $\boldsymbol{\Pi}$.
We first recover the nonlinear result of \cite{Schekochihin2010}, derived in that paper slightly differently, without using the inversion procedure. 
We point out that the result can be slightly simplified, and by rearranging the expression yields two different useful forms for writing the nonlinear $\boldsymbol{\Pi}$. 
Finally, by using the non-dispersive (MHD) induction equation, we obtain the nonlinear result of \cite{Ramos2005} (see also \cite{MacMahon1965}).  
For further evaluation of nonlinear FLR corrections to higher-orders, an advanced reader is referred to \cite{Ramos2005}.
We continue with the evaluation of the FLR tensor in the linear approximation, i.e. when the magnetic field lines are not too distorted. 
For the 1st-order tensor (here called FLR1), we recover the classical result of \cite{Yajima1966}, which is notably different from the one provided by \cite{Oraevskii1968}.
In the isotropic case, the FLR1 tensor is consistent with the one extracted from the stress tensor of \cite{Braginskii1965},
if the collisional terms are ``ignored''. It is noteworthy that a proper collisionless limit cannot be achieved from the stress tensor of \cite{Braginskii1965}, because
expressions are proportional to $\tau$ (and also $1/\tau$), where $\tau$ is time between two collisions ($\tau \sim 1/\nu$ where $\nu$ is the usual collisional frequency),
so the collisionless limit $\tau\to\infty$ does not work.   

We consider the Hall-CGL-FLR1 fluid model, and we provide dispersion relation for generally oblique propagation, which can be also found in \cite{HunanaZank2017}. 
For higher-order FLR corrections, we only provide analytic dispersion relations for the parallel
propagating whistler and ion-cyclotron modes, as well as the perpendicular fast mode. Nevertheless, we provide linearized, normalized and Fourier transformed equations
written in the x-z plane for all the fluid models. To obtain the dispersion relation for an oblique propagation,
the reader is encouraged to use analytic software such as Maple or Mathematica, or to solve the system numerically.
The 2nd-order FLR corrections (FLR2) are here defined as containing the Hall-term and the time-derivative $\pr\boldsymbol{\Pi}/\pr t$. 
We also consider FLR corrections with the non-gyrotropic heat flux vectors $\boldsymbol{S}^{\parallel}_\perp$, $\boldsymbol{S}^{\perp}_\perp$, that are here
defined as FLR3. The precision of various FLR corrections can be compared nicely by considering the perpendicular fast mode in the long-wavelength limit. 
The comparison is especially meaningful, when written in the notation of \cite{DelSarto2017}, see eq. (\ref{eq:DelSarto1})-(\ref{eq:Smolyakov}).
We proceed by showing that the FLR3 corrections (with the second-order non-gyrotropic heat flux vectors)
indeed recover the fully kinetic dispersion relation for the fast mode in the long-wavelength (low-frequency) limit,
a result reported by \cite{Smolyakov1985}.   
Instead of expanding the pressure tensor equation, one can derive very precise linear FLR corrections from linear kinetic theory, which is not addressed here, 
and the reader is referred to papers by \cite{PassotSulem2007,SulemPassot2015} and references therein.

In Section \ref{section:Firehose}, we investigate the parallel and oblique firehose instability.
The FLR and Hall dispersive effects are crucial for the stabilization of the firehose instability at small scales, 
and a comprehensive discussion can be found in \cite{HunanaZank2017}. 
That paper was essentially extracted from this guide (with many figures that we do not republish here), and an interested reader who wants to focus
on the firehose instability can find further information there.
Nevertheless, here we briefly investigate improvements that can be made by considering the FLR2 and FLR3 corrections,
see Figures \ref{fig:firehose-bump}-\ref{fig:both_firehose}. Importantly, we show that the non-gyrotropic heat flux vectors in the FLR3 model,
partially reproduce the large ``bump'' in the imaginary phase speed (growth rate normalized to the wavenumber), when the plasma is close to the
long wavelength limit ``hard'' firehose threshold, see Figure \ref{fig:firehose-bump}. 
The firehose instability in a fluid formalism was also investigated by
\cite{WangHau2003,WangHau2010,Schekochihin2010,Rosin2011} and references therein. 

In Section \ref{section:HeatFlux}, we derive the heat flux tensor equation. The heat flux tensor is then decomposed into its gyrotropic and non-gyrotropic parts,
$\bq=\bq^{\textrm{g}}+\bq^{\textrm{ng}}$. The procedure yields evolution equations for the gyrotropic
parallel and perpendicular heat fluxes, $q_\parallel$ and $q_\perp$, that contain the tensor of the 4th-order moment $\boldsymbol{r}$. 
It is emphasized that, if one wants to keep the non-gyrotropic $\boldsymbol{\Pi}$ contributions in the scalar heat flux equations, one needs to keep the non-gyrotropic
contributions of the 4th-order moment $\boldsymbol{r}^{\textrm{ng}}$ as well, since there are several possible cancellations even at the linear level. 
The non-gyrotropic heat flux $\bq^{\textrm{ng}}$ can be further decomposed to the non-gyrotropic heat flux vectors $\boldsymbol{S}^{\parallel}_\perp$,
$\boldsymbol{S}^{\perp}_\perp$ and tensor $\boldsymbol{\sigma}$. 
The detailed algebra of the non-gyrotropic heat flux vectors $\boldsymbol{S}^{\parallel}_\perp$, $\boldsymbol{S}^{\perp}_\perp$, i.e. how to express them
through lower-order moments, is presented  in the Appendix \ref{sec:NONGheat}.
The first-order expressions are obtained at the nonlinear level and the second-order expressions at the linear level.
We do not address how to decompose the tensor $\boldsymbol{\sigma}$ through lower-order moments. Such calculations
require a complicated ``inversion'' procedure for a 3rd-rank tensor $(\bhat\times\boldsymbol{\sigma})^S$, and an advanced reader is referred to \cite{Ramos2005}.

In Section \ref{section:4th}, we consider the 4th-order moment $\br$ which is a tensor of 4th-rank, $r_{ijkl}$.
The moment is again decomposed to its gyrotropic and non-gyrotropic parts, $\br=\br^{\textrm{g}}+\br^{\textrm{ng}}$ and the gyrotropic part  
has three scalar components, $r_{\parallel\parallel}$, $r_{\parallel\perp}$ and $r_{\perp\perp}$. 
We show that for a bi-Maxwellian distribution function, the gyrotropic components can indeed be evaluated as $r_{\parallel\parallel}=3p_\parallel^2/\rho$,
$r_{\parallel\perp}=p_\parallel p_\perp/\rho$, and $r_{\perp\perp}=2 p_\perp^2/\rho$. This constitutes a ``normal'' closure, a name suggested by \cite{ChustBelmont2006}.
By using a similar procedure to the one provided by \cite{Grad1949} for dilute gases, it is possible to express the non-gyrotropic
bi-Maxwellian $\boldsymbol{r}^{\textrm{ng}}$ through a combination of $\boldsymbol{p}^g$ and $\boldsymbol{\Pi}$, see e.g. \cite{Oraevskii1968}.

In Section \ref{section:CGL2}, a dispersion relation of a fluid model closed with the bi-Maxwellian ``normal'' closure is provided for generally oblique propagation,
and we call this model 2nd-order CGL (CGL2). We specifically focus on the mirror instability, since this simple fluid model (without any Landau damping)
corrects the erroneous 1/6 factor in the ``hard'' mirror threshold found in the basic CGL description, a result also reported by \cite{DzhalilovKuznetsov2011}.
The mirror instability is not addressed to a higher level of sophistication in this guide. However, we note
that capturing the mirror growth rate (when the threshold is crossed) sufficiently well requires Landau fluid models \citep{Snyder1997}
and the stabilization at small scales requires FLR corrections \citep{PassotSulem2007,SulemPassot2015}. 
We also provide the dispersion relation of the Hall-CGL2 fluid model.  Finally, the CGL2 model can be simplified by considering slow-dynamics regime and
constructing generalized isothermal closure that is called the ``static'' closure, yielding the simplest fluid model that captures the correct mirror threshold
 \citep{Constantinescu2002,ChustBelmont2006,PassotRubanSulem2006}.

In Section \ref{section:BiKappa}, we consider the bi-Kappa distribution function. We show that the closure at the 4th-order moment is constructed by
$r_{\parallel\parallel}=\alpha_\kappa 3p_\parallel^2/\rho$,
$r_{\parallel\perp}=\alpha_\kappa p_\parallel p_\perp/\rho$, and $r_{\perp\perp}=\alpha_\kappa 2 p_\perp^2/\rho$, where the coefficient
$\alpha_\kappa=(\kappa-3/2)/(\kappa-5/2)$, and the closure is valid for $\kappa>5/2$. We call this closure and the associated fluid model
as ``BiKappa'', and we discuss its dispersion relations. Even though the linear modes are generally different in this fluid model than in the CGL2 fluid model,
we show that the ``hard'' thresholds for the parallel and oblique firehose instability, and for the highly-oblique mirror instability, are not affected by 
and are independent of the $\kappa$ value. The Hall-BiKappa fluid dispersion relation is
also provided. We also provide the first-order non-gyrotropic heat flux vectors $\boldsymbol{S}^{\parallel}_\perp$, $\boldsymbol{S}^{\perp}_\perp$.
We do not calculate the $\boldsymbol{r}^{\textrm{ng}}$ for the bi-Kappa distribution, and therefore we do not provide the second order non-gyrotropic heat flux vectors.

In Section \ref{section:Landau},
we discuss the core differences between the usual fluid models and kinetic theory. Namely, we discuss why the usual fluid models do not contain collisionless
damping mechanisms, such as the Landau damping, regardless of the order to which the fluid hierarchy is developed.
The effect of Landau damping is present in the collisionless Vlasov equation, and the crucial difference between the usual fluid hierarchy and kinetic calculations just
lies in the technique how the Vlasov equation is integrated over the velocity space. 
We introduce preliminary ideas how the Landau fluid closures will be constructed. For example, for closures performed at the 4th-order moment
instead of the ``normal'' closure, one needs to consider perturbations around this state,
and prescribe $r_{\parallel\parallel}=3p_\parallel^2/\rho +\widetilde{r}_{\parallel\parallel} $,
$r_{\parallel\perp}=p_\parallel p_\perp/\rho+\widetilde{r}_{\parallel\perp}$, and $r_{\perp\perp}=2 p_\perp^2/\rho+\widetilde{r}_{\perp\perp}$.
The deviations $\widetilde{r}_{\parallel\parallel}$, $\widetilde{r}_{\parallel\perp}$, $\widetilde{r}_{\perp\perp}$ will be calculated from linear kinetic theory in Part 2,
by performing Landau fluid closures.

In Section \ref{section:nth}, we derive evolution equation for a general n-th order fluid moment (a tensor with $3^n$ components). 
We then consider fluid models in the simplified 1D geometry that can be viewed as an electrostatic case (or as a propagation along the mean magnetic field),
that are closed at a general n-th order level by a Maxwellian fluid closure.
A dispersion relation is obtained, which for $n>4$ always yields some solutions that are unstable. It is therefore concluded that the last non-Landau fluid closure
is the ``normal'' closure and that for $n>4$, Landau fluid closures are required. This surprising result, first reported in \cite{HunanaPRL2018}, serves as motivation
for Part 2, which is a detailed guide to Landau fluid closures.

\newpage
\section{Pressure tensor equation} \label{section:PressureTensor}
Collisionless plasmas are described by the Vlasov equation, which in CGS units reads 
\begin{equation} \label{eq:Vlasov}
\frac{\pr f_r}{\pr t}+ \bV\cdot\nabla f_r +\frac{q_r}{m_r}(\bE+\frac{1}{c}\bV\times\bb)\cdot\nabla_v f_r =0,
\end{equation}
and which describes how a distribution function $f_r(\bx,\bV,t)$ evolves in time. The $r$ is the index of species and $r=p$ for
protons, $r=e$ for electrons, etc. The $q_r$ is the particle charge, $m_r$ the particle mass, $c$ the speed of light,
$\bE$ the electric field vector and $\bb$ the magnetic field vector. The species index $r$ can sometimes be confusing in lengthy tensor calculations
with multiple indices and for clarity we will often drop it and reintroduce it when required.
To derive the fluid equations, we need to integrate (perform an averaging) at each spatial point over the ``kinetic'' velocity $\bV$. 
It is important to realize that the distribution function just describes the probability to find a particle with velocity $\bV$ at the
position $\bx,t$ and that the ``kinetic'' velocity $\bV$ entering the distribution function $f(\bx,\bV,t)$
is a completely independent variable from $\bx,t$, i.e.
\begin{equation}
  \frac{\pr v_i}{ \pr t}=0; \qquad \frac{\pr v_i}{\pr x_j}=0.
\end{equation}  
Also, the magnetic and electric fields $\bb(\bx,t),\bE(\bx,t)$ are macroscopic quantities that
do not depend on $\bV$ and can be moved outside of velocity integrals over $d^3 v$, or in another words $\pr B_i /\pr v_j =0$ and $\pr E_i / \pr v_j =0$. 
The definitions of the fluid moments are
\begin{eqnarray}
  n &=& \int f d^3 v; \label{eq:defDens}\\
  n\bu &=& \int \bV f d^3 v; \label{eq:defVel}\\
  \bp &=& m \int (\bV-\bu)(\bV-\bu) f d^3 v; \label{eq:defPT}\\
  \boldsymbol{q} &=& m \int (\bV-\bu)(\bV-\bu)(\bV-\bu) f d^3 v; \label{eq:defHFT}\\
  \boldsymbol{r} &=& m \int (\bV-\bu)(\bV-\bu)(\bV-\bu)(\bV-\bu)f d^3 v, \label{eq:def4th}
\end{eqnarray}  
where we have omitted the tensor product notation that is sometimes written down explicitly
as $\bu\bu=\bu\otimes\bu$ and in the index notation $(\bu\bu)_{ij}=u_iu_j$. If two vectors or tensors are next to each other
without an operator between them, a tensor product is always assumed. The number density $n$ is a scalar,
the fluid velocity $\bu$ is a vector, the pressure tensor $\bp$ is a tensor of second rank ($3\times3$ matrix),
the heat flux tensor $\boldsymbol{q}$ is a tensor of third rank (with $3\times 3\times 3$ components), the tensor 
$\boldsymbol{r}$ is of fourth rank (with $3\times 3\times 3\times 3$ components), etc. Directly from the definitions, it is
obvious that all fluid tensors must be symmetric in all of their indices, i.e. $p_{ij}=p_{ji}$, $q_{ijk}=q_{ikj}=...$
The fluctuating velocity is defined as
\begin{equation}
  \bc=\bV-\bu,
\end{equation}
and should not be confused with the speed of light $c$.

The second important concept that is used in deriving the fluid hierarchy, is the use of the usual Gauss-Ostrogradsky (divergence) theorem.
The divergence theorem is used in velocity space and, written in a form that is typically encountered when calculating the fluid hierarchy, it reads
\begin{equation} \label{eq:GaussOstrogradsky}
\int_V \nabla_v \cdot (f\boldsymbol{A}) d^3 v = \int_S f \boldsymbol{A} \cdot d\boldsymbol{S}.
\end{equation}
The $\boldsymbol{A}$ is a general n-th order tensor, $f$ is a distribution function, the left hand side is a 3D integral calculated over the entire velocity volume $V$
and the right hand side is a surface integral calculated over a boundary of that volume $d\boldsymbol{S}=\boldsymbol{\hat{n}} dS$, where $\boldsymbol{\hat{n}}$ is
a unit normal vector to the local surface area pointing outwards. When such an integral is encountered in the fluid hierarchy, the result is always assumed to
be zero. The volume integrals are from $v=-\infty$ to $v=\infty$ in each velocity component and the integration on the right hand side of
(\ref{eq:GaussOstrogradsky}) is therefore performed over the velocity surface area at infinity. The necessary (but technically not sufficient) condition for the integral to
be negligible is
\begin{equation}
  \lim_{v\to \infty} f(v)=0.
\end{equation}  
Since the area $dS \sim v^2 dv$, the sufficient condition for the integral to vanish, can be estimated more precisely as
\begin{equation} \label{eq:limit}
\lim_{v\to \infty} f(v) A(v) v^2 =0. 
\end{equation}
When calculating the fluid hierarchy, the encountered expressions are always $A(v)\sim v^n$, where $n$ is a positive integer.   
Then, for example, for a Maxwellian distribution $f(v)\sim e^{-v^2}$ the limit (\ref{eq:limit}) is always zero for all $n$.
Even for a slower converging distribution $f(v)\sim e^{-|v|}$, the limit is
always zero. For distribution functions proportional to a power law, for example $f(v)\sim (v^2)^{-(\kappa+1)}$ as for a kappa-distribution, 
the situation is more restrictive, with some minimum required values of $\kappa$. When we will calculate
the n-th order fluid moment (Section \ref{section:nth}), we will see that two surface integrals are encountered, one with $A(v)\sim v^n$ and one with $A(v)\sim v^{n+1}$.
Therefore, for a kappa-distribution, the strict condition for $\lim_{v\to\infty} (v^2)^{-(\kappa+1)} v^{n+1} v^2 =0$ yields the requirement $\kappa>(n+1)/2$.
For example, many fluid models discussed in this guide are closed at the 4th-order moment $\boldsymbol{r}$, so $n=4$, which implies the requirement $\kappa>5/2$.
At first sight, the required limit (\ref{eq:limit}) can be considered only a technical mathematical detail, since physically,
one can argue that particles with enormously large energies will not be measured/observed, and some physical mechanism that is responsible
for a cut-off of the distribution function at finite energies, can be usually assumed. Or in another words, even observational studies that fit the data
with a kappa-distribution do not assume that the fit is valid all the way up to $v\to\infty$, and usually a cut-off is implicitly assumed.
However, once we calculate the 4th-order moments for a kappa distribution, we will see that for example $r_{\parallel\parallel}=\alpha_\kappa 3p_\parallel^2/\rho$,
where $\alpha_\kappa=(\kappa-3/2)/(\kappa-5/2)$, so the restriction $\kappa>5/2$ returns, and has to be applied regardless.    
When the limit (\ref{eq:limit}) is satisfied, the neglect of the surface integrals (\ref{eq:GaussOstrogradsky}) is therefore based on solid
theoretical principles, and it is not an approximate or an ad-hoc choice, the surface integrals are really zero. 
Nevertheless, for complete clarity in the upcoming fluid hierarchy calculations,
we will differentiate between expressions that are zero exactly, and between the surface integrals (\ref{eq:GaussOstrogradsky})
that are zero asymptotically, by using $=0$ and $\to 0$.  

We start directly with the derivation of the pressure tensor equation, since a detailed derivation of the density and momentum equations can be found in many books.
To derive the pressure tensor equation, it is possible to multiply the Vlasov equation by $m c_i c_j$ or $m v_i v_j$
(another possibility is $m c_i v_j$). Here we will use the first choice which is slightly easier to present in detail,
because the second choice requires the use of density and momentum equations to cancel several terms.
It is useful to derive the following identities
\begin{eqnarray}
 \int \bc fd^3v &=& \int (\bV-\bu)f d^3v = \int \bV fd^3v -\bu\int fd^3v = n\bu-\bu n =0;\\
  m \int \bc\bV f d^3v &=& m \int (\bV-\bu)(\bV-\bu+\bu) f d^3v =
  m \int (\bV-\bu)(\bV-\bu) f d^3 v + m \bu \underbrace{\int (\bV-\bu)f d^3v}_{=0} = \bp;\\
  m \int \bc\bc\bV f d^3v &=& m\int \bc\bc(\bV-\bu+\bu) fd^3v = \boldsymbol{q}+\bp\bu.
\end{eqnarray}
We write down the derivatives with respect to time $\pr/\pr t$ and velocity $\pr/\pr v_i$ explicitly, but we abbreviate
the derivative with respect to spatial coordinates as $\pr/\pr x_i\equiv \pr_i$.
We will need
\begin{eqnarray}
  \frac{\pr}{\pr t} c_i &=& \frac{\pr}{\pr t}(v_i-u_i) = -\frac{\pr}{\pr t} u_i;\\
  \frac{\pr}{\pr t} (c_i c_j) &=& -c_i\frac{\pr u_j}{\pr t} -c_j\frac{\pr u_i}{\pr t};\\
  \pr_k (c_i c_j) &=& -c_i \pr_k u_j -c_j\pr_k u_i;\\
  \frac{\pr}{\pr v_k} c_i &=& \frac{\pr}{\pr v_k}(v_i-u_i) = \frac{\pr v_i}{\pr v_k} = \delta_{ik};\\
  \frac{\pr}{\pr v_k} (c_i c_j) &=& c_i\delta_{jk} + c_j \delta_{ik};\\
  \frac{\pr}{\pr v_k} [c_i c_j (\bV\times\bb)_k] &=& \frac{\pr}{\pr v_k} (c_i c_j)(\bV\times\bb)_k
  + c_i c_j \frac{\pr}{\pr v_k} ( \epsilon_{klm}v_l B_m) = (\delta_{ik}c_j+\delta_{jk}c_i)(\bV\times\bb)_k
  + c_i c_j \underbrace{\epsilon_{klm} \delta_{lk}}_{=0} B_m \nn \\
  &=& c_j (\bV\times\bb)_i +c_i (\bV\times\bb)_j,
  \end{eqnarray}
where in the last identity we used the result $\epsilon_{klm} \delta_{lk} =0$ since the Levi-Civita tensor $\epsilon_{ijk}$ is antisymmetric and
the Kronecker $\delta_{ij}$ is a symmetric tensor.
Integrating the first term of the Vlasov equation yields
\begin{eqnarray}
\bigcirc{ }\!\!\!\!\mbox{\small 1} &=& m\int c_i c_j \frac{\pr f}{\pr t}d^3 v= \frac{\pr }{\pr t}\Big( \underbrace{m\int c_i c_j f d^3 v}_{=p_{ij}} \Big)
- m \int f \frac{\pr}{\pr t}(c_i c_j) d^3v = \frac{\pr }{\pr t} p_{ij} + m \frac{\pr u_j}{\pr t}\underbrace{\int fc_i d^3v}_{=0}
+m \frac{\pr u_i}{\pr t}\underbrace{\int fc_j d^3v}_{=0} \nn \\
&=& \frac{\pr }{\pr t} p_{ij}.
\end{eqnarray}
The second term of the Vlasov equation yields
\begin{eqnarray}
\bigcirc{ }\!\!\!\!\mbox{\small 2} &=& 
  m\int c_i c_j \underbrace{\bV\cdot\nabla}_{=v_k\pr_k}f d^3v = \pr_k \Big(\underbrace{m\int c_i c_j v_k f d^3v}_{=q_{ijk}+p_{ij}u_k}\Big)
  - m\int v_k f \pr_k (c_i c_j) d^3v
  = \pr_k (q_{ijk} + p_{ij} u_k ) + (\pr_k u_j)\underbrace{m\int c_i v_k f d^3v}_{=p_{ik}} \nn\\
&&  + (\pr_k u_i) \underbrace{m \int c_j v_k f d^3v}_{=p_{jk}}
  = \pr_k (\underbrace{q_{ijk}}_{=q_{kij}} + p_{ij} u_k ) +p_{ik}\pr_k u_j + p_{jk}\pr_k u_i
  =  \pr_k (q_{kij} + u_k p_{ij}) +p_{ik}\pr_k u_j + p_{jk}\pr_k u_i\nn\\
  &=& \big[ \nabla\cdot(\boldsymbol{q}+\bu\bp) +\bp\cdot\nabla \bu + (\bp\cdot\nabla \bu)^{\textrm{T}} \big]_{ij}.
\end{eqnarray}
To go back and forth between the index notation and the vector notation in a fully consistent matter, it is important
to establish some conventions that were not required in a simple fluid models. One important convention that is typically used
is that the divergence of a tensor is meant to be through its first component, i.e. for a general tensor $X_{ijk...n}$, the
divergence $\nabla\cdot\boldsymbol{X}$ means $\pr_i X_{ijk...n}$. This convention is the reason why in the above expression
we used that $q_{ijk}=q_{kij}$ and also the ordering of $\bu$ and $\bp$ inside $\nabla\cdot(\bu\bp)$ reflects that in the index
notation we have $\pr_k(u_k p_{ij})$.\\
The third term of the Vlasov equation calculates
\begin{eqnarray}
&&\bigcirc{ }\!\!\!\!\mbox{\small 3}=
  q\int c_i c_j \bE\cdot\nabla_v f d^3v = q\int c_i c_j E_k \frac{\pr}{\pr v_k} f d^3 v =
  q E_k \underbrace{\int \frac{\pr}{\pr v_k} \big( c_i c_j f\big) d^3v}_{\rightarrow 0} - q E_k \int \frac{\pr}{\pr v_k} \big( c_i c_j \big) f d^3v \nn\\
&&  = -q E_k \int \frac{\pr}{\pr v_k}(c_i c_j) f d^3v = -qE_k\int (\delta_{ik}c_j+\delta_{jk}c_i) f d^3v =
  -qE_i\underbrace{\int c_j f d^3v}_{=0} - qE_j\underbrace{\int c_i f d^3v}_{=0} = 0.
\end{eqnarray}  
Finally, the fourth term of the Vlasov equation calculates
\begin{eqnarray}
&& \bigcirc{ }\!\!\!\!\mbox{\small 4}=
 \frac{q}{c}\int c_i c_j(\bV\times\bb)\cdot \nabla_v f d^3v = \frac{q}{c}\int c_i c_j(\bV\times\bb)_k \frac{\pr f}{\pr v_k} d^3v
  =\frac{q}{c} \underbrace{\int\frac{\pr}{\pr v_k}\big[ c_i c_j (\bV\times\bb)_k f\big] d^3v}_{\rightarrow 0} \nn\\
&&  -\frac{q}{c} \int\frac{\pr}{\pr v_k}\big[ c_i c_j (\bV\times\bb)_k \big] f d^3v
= -\frac{q}{c}\int \big[ c_j (\bV\times\bb)_i + c_i (\bV\times\bb)_j\big] f d^3v =
-\frac{q}{c}\int c_j \epsilon_{ikl} v_k B_l f d^3v \nn \\
&& - \frac{q}{c}\int c_i \epsilon_{jkl}v_k B_l f d^3v 
 = -\frac{q}{c}\epsilon_{ikl} B_l \underbrace{\int c_j  v_k f d^3v}_{=\frac{1}{m}p_{jk}}
  - \frac{q}{c}\epsilon_{jkl} B_l \underbrace{\int c_i v_k f d^3v}_{=\frac{1}{m}p_{ik}} =
  -\frac{q}{mc}\big[ \epsilon_{ikl}B_l p_{jk} +\epsilon_{jkl} B_l p_{ik}\big] \nn \\
&& = \frac{q}{mc}\big[ \epsilon_{ilk}B_l p_{kj} +\epsilon_{jlk} B_l p_{ki}\big] 
  = \frac{q}{mc}\big[ (\bb\times\bp)_{ij} +(\bb\times\bp)_{ji}\big] = \frac{q}{mc}\big[ \bb\times\bp +(\bb\times\bp)^{\textrm{T}}\big]_{ij}. \label{eq:KulsrudTypo}
\end{eqnarray}
In the expression above, one encounters a vector product of a vector with a matrix, $\bb\times\bp$. This operator might appear unusual
at first, but it is just a generalization of a vector product of two vectors, it is defined as $(\bb\times\bp)_{ij} = \epsilon_{ikl} B_k p_{lj}$ and the
result is a matrix. Similarly, a vector product between a vector and a tensor of 3rd rank is defined as
$(\bb\times\bq)_{ijk} = \epsilon_{ilm} B_l q_{mjk}$ and for a tensor of
n-th rank $(\bb\times\boldsymbol{X})_{i...jk} = \epsilon_{ilm} B_l X_{m...jk}$ (i.e., the second index in $\epsilon$ is the vector $\bb$ index
and the third index in $\epsilon$ is the first index of tensor $\boldsymbol{X}$). However, a definition of a vector product where a matrix is applied on a vector
reads $(\bp\times\bb)_{ij}=\epsilon_{jkl} p_{ik} \bb_l$, and in general $\bp\times\bb=-(\bb\times\bp^T)^T$, but since the $\bp$ is here symmetric,
$\bp\times\bb=-(\bb\times\bp)^T$. The vector product (cross product) is addressed further  in the Appendix \ref{sec:AppendixB}.
The term (\ref{eq:KulsrudTypo}) is sometimes rewritten as
\begin{equation}
  \bb\times\bp +(\bb\times\bp)^{\textrm{T}} = \bb\times\bp - \bp\times\bb. \label{eq:KulsrudCorrect}
\end{equation}
Combining all the results together
$\bigcirc{ }\!\!\!\!\mbox{\small 1}+\bigcirc{ }\!\!\!\!\mbox{\small 2}+\bigcirc{ }\!\!\!\!\mbox{\small 3} + \bigcirc{ }\!\!\!\!\mbox{\small 4}=0$,
yields the entire pressure tensor equation as
\begin{equation} \label{eq:Ptensor}
\frac{\pr \bp}{\pr t} + \nabla\cdot(\bu\bp+\boldsymbol{q})+\bp\cdot\nabla \bu + (\bp\cdot\nabla \bu)^{\textrm{T}}
+\frac{q}{m c}\Big[\bb\times\bp + (\bb\times\bp)^{\textrm{T}} \Big] =0.
\end{equation}
One notices that, for example, the pressure tensor equation of \cite{Kulsrud1983}, eq. 57, contains a minus sign typo in the last term,
which should be written as (\ref{eq:KulsrudCorrect}). That it is indeed a typo and not a definition of $\bp\times\bb$ is evident from his subsequent eq. 59.
It is useful to introduce a symmetric operator that acts on a matrix $\boldsymbol{A}$ according to
\begin{equation} \label{eq:Sym2D}
  \boldsymbol{A}^S \equiv \boldsymbol{A}+\boldsymbol{A}^{\textrm{T}}; \qquad A_{ij}^S \equiv A_{ij}+A_{ji}, 
\end{equation}
which yields a more compact form of the pressure tensor equation
\begin{equation} \label{eq:PtensorF}
\frac{\pr \bp}{\pr t} + \nabla\cdot(\bu\bp+\boldsymbol{q})+ \Big[\bp\cdot\nabla\bu +\frac{q}{m c}\bb\times\bp\Big]^S =0.
\end{equation}

\subsection{Pressure tensor decomposition}
By introducing a unit vector in the direction of the local magnetic field $\bhat=\bb/|\bb|$
and the gyrofrequency $\Omega=q B_0 / (m c)$, the last term in the pressure tensor equation (\ref{eq:Ptensor})
can be rewritten as
\begin{equation}
\Omega\frac{|\bb|}{B_0}\Big[\bhat\times\bp + (\bhat\times\bp)^{\textrm{T}} \Big]. \label{eq:term}
\end{equation}
At very low frequencies $\omega\ll \Omega$ (and at very long spatial scales) this term will dominate and must be equal to
zero. The simplest possibility that makes this term to be equal to zero is  
$\bp^{\textrm{iso}} = p \boldsymbol{I}$, where $p$ is the scalar pressure and $\boldsymbol{I}$ is the unit matrix. In index
notation $p_{ij}^{\textrm{iso}}=p\delta_{ij}$, and we can verify that
\begin{eqnarray}
  && (\bhat\times\bp^{\textrm{iso}})_{ij} = \epsilon_{ikl}\hat{b}_k p_{lj}^{\textrm{iso}} = \epsilon_{ikl}\hat{b}_k p \delta_{lj} =
  p \epsilon_{ikj}\hat{b}_k = -p \epsilon_{ijk}\hat{b}_k;\\
  &&  \big[ \bhat\times\bp^{\textrm{iso}} + (\bhat\times\bp^{\textrm{iso}})^{\textrm{T}} \big]_{ij} = -p \epsilon_{ijk}\hat{b}_k -p \epsilon_{jik}\hat{b}_k
  = -p \hat{b}_k (\epsilon_{ijk}-\epsilon_{ijk}) =0.
\end{eqnarray}  
There is however a much more general solution that makes the term (\ref{eq:term}) equal to zero, which is 
\begin{equation} \label{eq:PD}
\bp^\textrm{g} = p_{\parallel}\bhat\bhat + p_{\perp}(\boldsymbol{I}-\bhat\bhat),
\end{equation}
where the ``g'' stands for gyrotropic. For quick analytic calculations it is sometimes easier to use
$\bp^\textrm{g} = (p_\parallel - p_\perp)\bhat\bhat +p_\perp\boldsymbol{I}$
since we have to deal with $\bhat\bhat$ only once instead of twice. We need to verify that the term (\ref{eq:term}) indeed
disappears,
\begin{eqnarray}
 && (\bhat\times\bp^\textrm{g})_{ij} = \epsilon_{ikl}\hat{b}_k p_{lj}^{\textrm{g}} = \epsilon_{ikl}\hat{b}_k
  \Big( (p_\parallel-p_\perp)\hat{b}_l \hat{b}_j +p_\perp \delta_{lj} \Big) =
  (p_\parallel-p_\perp)\hat{b}_j\underbrace{\epsilon_{ikl}\hat{b}_k\hat{b}_l}_{=0} +p_\perp \epsilon_{ikj}\hat{b}_k =
  -p_\perp \epsilon_{ijk}\hat{b}_k;\\
=>\quad  &&  \big[ \bhat\times\bp^\textrm{g} + (\bhat\times\bp^\textrm{g})^{\textrm{T}} \big]_{ij} = -p_\perp \hat{b}_k (\epsilon_{ijk}+\epsilon_{jik})=0.
\end{eqnarray}  
It is easy to see that the parallel and perpendicular pressures can be extracted from the gyrotropic pressure tensor matrix
by performing double contractions according to 
\begin{equation} \label{eq:Pdecomp}
p_{\parallel} = \bp^\textrm{g} : \bhat\bhat, \quad \textrm{and} \quad p_{\perp} = \bp^\textrm{g} :(\boldsymbol{I}-\bhat\bhat)/2,
\end{equation}
The double contraction (the double dot product, represented by the colon) is a very useful operator that is frequently encountered
in higher order fluid models. For two matrices
it is defined as $\boldsymbol{A}:\boldsymbol{B}=A_{ij}B_{ij}$ and yields a scalar.
Sometimes, the double contraction is defined as $\boldsymbol{A}:\boldsymbol{B}=A_{ij}B_{ji}$,
which for symmetric matrices is equivalent to the previous definition.
As a reminder, the usual matrix product of two matrices is $(\boldsymbol{A}\cdot \boldsymbol{B})_{ij}=A_{ik}B_{kj}$ and yields a matrix.
It is useful to write down the following identities that involve the double contraction,
\begin{eqnarray}
&&  \bhat\bhat:\bhat\bhat =1; \label{eq:DCid1}\\
  && \boldsymbol{I}:\bhat\bhat =1; \label{eq:DCid2}\\
&&  (\boldsymbol{I}-\bhat\bhat):\bhat\bhat =0; \label{eq:DCid3}\\
  && \boldsymbol{I}:\boldsymbol{I} =3; \label{eq:DCid4}\\
&&  (\boldsymbol{I}-\bhat\bhat):(\boldsymbol{I}-\bhat\bhat)/2= 1. \label{eq:DCid5}
\end{eqnarray}
Also interestingly, the trace of a matrix $\boldsymbol{A}$ can be expressed through the double contraction with $\boldsymbol{I}$,
i.e. $\boldsymbol{A}:\boldsymbol{I}=A_{ij}\delta_{ij}=A_{ii}=\trace(\boldsymbol{A})$.
The pressure decomposition (\ref{eq:Pdecomp}) can now be verified easily.   
Even more revealing is to apply the decomposition (\ref{eq:Pdecomp}) directly at
the definition of the \emph{entire} pressure tensor (\ref{eq:defPT}), which yields 
\begin{eqnarray}
  \bp:\bhat\bhat &=& m \int \bc\bc f d^3v :\bhat\bhat = 
   m\int (\bV\cdot\bhat-\bu\cdot\bhat)^2 fd^3v = m\int (v_\parallel-u_\parallel)^2 f d^3v  \equiv p_\parallel; \label{eq:Pbue1}\\
  \bp:(\boldsymbol{I}-\bhat\bhat)/2 &=& \frac{m}{2} \int |\bV_\perp-\bu_\perp|^2 f d^3v \equiv p_\perp, \label{eq:Pbue2}
\end{eqnarray}  
where we have used the fact that the magnitude of the 3D velocity vector can be decomposed as
$|\bV-\bu|^2=|\bV_\perp-\bu_\perp|^2+(v_\parallel-u_\parallel)^2$, or equivalently $|\bc|^2=|\bc_\perp|^2+c_\parallel^2$.

A distribution function is called isotropic when the direction of the velocity vector $\bV$ does not matter, and the distribution function
depends only on magnitude $|\bV|$, as for example the Maxwellian distribution $e^{-|\bV|^2}$.
A distribution function is called \emph{gyrotropic} when the direction of the perpendicular velocity
vector $\bV_\perp$ does not matter, and the function depends only on $|\bV_\perp|$, as for example the bi-Maxwellian distribution $e^{-\alpha_\parallel v_\parallel^2} e^{-\alpha_\perp |\bV_\perp|^2}$.
In another words, the gyrotropic distribution function is isotropic only in its transverse velocity components.
The same vocabulary is used for the fluid moments, and a fluid moment is called gyrotropic when it involves integrals over $|\bc_\perp|^{2i}$, $i=0,1,2,3\ldots$.
For the pressure tensor, there are only two possibilities, represented by the parallel and perpendicular pressure (\ref{eq:Pbue1}), (\ref{eq:Pbue2}).

The part of the pressure tensor extracted by the decomposition
(\ref{eq:Pdecomp}) is therefore called the gyrotropic pressure $\bp^\textrm{g}$, sometimes also called the CGL pressure $\bp^{CGL}$.
The gyrotropic approximation is sufficient at very long spatial scales, where the Larmor radius (=the gyroradius) of
particles gyrating around the magnetic field is small, and the non-gyrotropic contributions become negligible.
However, at sufficiently small spatial scales comparable to the gyroradius, the non-gyrotropic pressure contributions become
significant and we represent these through a tensor $\boldsymbol{\Pi}$, that is called the Finite Larmor Radius (FLR) corrections
to the (gyrotropic) pressure tensor, or sometimes non-gyrotropic or gyroviscous stress tensor.
The entire pressure tensor is thus decomposed according to
\begin{eqnarray} \label{eq:Pfull}
\bp = p_{\parallel}\bhat\bhat + p_{\perp}(\boldsymbol{I}-\bhat\bhat) + \boldsymbol{\Pi},
\end{eqnarray}
and for clarity we write down the decomposition explicitly in matrix notation
\begin{eqnarray}
  \boldsymbol{p}=
  p_\parallel \left( \begin{array}{ccc}
    \bxhat\bxhat, & \bxhat\byhat, & \bxhat\bzhat \\
    \byhat\bxhat, & \byhat\byhat, & \byhat\bzhat \\
    \bzhat\bxhat, & \bzhat\byhat, & \bzhat\bzhat
  \end{array} \right)
+ p_\perp \left(
\begin{array}{ccc}
1-\bxhat\bxhat, & -\bxhat\byhat, & -\bxhat\bzhat \\
-\byhat\bxhat, & 1-\byhat\byhat, & -\byhat\bzhat \\
-\bzhat\bxhat, & -\bzhat\byhat, & 1-\bzhat\bzhat
\end{array}
\right) 
+ \left(
\begin{array}{ccc}
\Pi_{xx}, & \Pi_{xy}, & \Pi_{xz} \\
\Pi_{yx}, & \Pi_{yy}, & \Pi_{yz} \\
\Pi_{zx}, & \Pi_{zy}, & \Pi_{zz}
\end{array} \right).
\end{eqnarray}
In the momentum equation, the pressure tensor enters through its divergence, which is useful to break down to its components. Since
$\nabla\cdot(\bhat\bhat)=\bhat(\nabla\cdot\bhat)+(\bhat\cdot\nabla)\bhat$, and $\nabla\cdot(p_\perp\boldsymbol{I})=\nabla p_\perp$,
the divergence of the pressure tensor calculates
\begin{eqnarray} \label{eq:divPcalc}
  \nabla\cdot\bp &=& \nabla\cdot(p_\perp \boldsymbol{I})+ \nabla\cdot\big[(p_\parallel-p_\perp)\bhat\bhat\big] +\nabla\cdot\boldsymbol{\Pi} \nn\\
  &=& \nabla p_\perp +(p_\parallel-p_\perp)\big[  \bhat(\nabla\cdot\bhat)+(\bhat\cdot\nabla)\bhat \big] +\bhat (\bhat\cdot\nabla) (p_\parallel-p_\perp)
  +\nabla\cdot\boldsymbol{\Pi}. 
\end{eqnarray}
 Importantly, we have seen in (\ref{eq:Pbue1}), (\ref{eq:Pbue2}) that the scalar parallel and perpendicular pressures are extracted
by applying the double contractions at the entire pressure tensor (and not necessarily only at the gyrotropic part).  The decomposition (\ref{eq:Pfull}) can be rewritten as
\begin{eqnarray} \label{eq:PfullDec}
  \bp = \big(\bp : \bhat\bhat\big) \bhat\bhat + \big( \bp: (\boldsymbol{I}-\bhat\bhat)/2 \big) (\boldsymbol{I}-\bhat\bhat)
  + \boldsymbol{\Pi}.
\end{eqnarray}
By applying the double contraction $:\bhat\bhat$ to (\ref{eq:PfullDec}) yields the first requirement for the FLR tensor 
\begin{equation} \label{eq:FLRid1}
\boldsymbol{\Pi}:\bhat\bhat =0.
\end{equation}  
Similarly, by applying $:(\boldsymbol{I}-\bhat\bhat)/2$ to (\ref{eq:PfullDec}) yields the second requirement for the FLR tensor,
\begin{equation} \label{eq:FLRid2}
\boldsymbol{\Pi}:(\boldsymbol{I}-\bhat\bhat)=0.
\end{equation}
By further using the first requirement, the second requirement can be rewritten as
\begin{equation}
\boldsymbol{\Pi}:\boldsymbol{I}=\textrm{Tr}(\boldsymbol{\Pi}) = 0. 
\end{equation}  
By using the pressure decomposition (\ref{eq:Pfull}) in the pressure tensor equation (\ref{eq:Ptensor}), and using the fact that the gyrotropic part of
pressure satisfies $\bhat\times\bp^g+(\bhat\times\bp^g)^T=0$, the pressure tensor equation can be rewritten as 
\begin{equation} \label{eq:PTfinal}
\frac{\pr \bp}{\pr t} + \nabla\cdot(\bu\bp+\boldsymbol{q})+\bp\cdot\nabla \bu + (\bp\cdot\nabla \bu)^{\textrm{T}}
+\Omega\frac{|\bb|}{B_0}\Big[\bhat\times\boldsymbol{\Pi} + (\bhat\times\boldsymbol{\Pi})^{\textrm{T}} \Big] =0.
\end{equation}
We can now derive the time dependent equations for parallel and perpendicular pressures by applying the usual double
contractions on this pressure tensor equation.
\subsection{Parallel pressure equation}
We calculate the double contraction with $\bhat\bhat$ term by term.
We will need the following identities
\begin{eqnarray}
  \frac{\pr}{\pr t}(\bhat\bhat) :\bhat\bhat = 0; \qquad  \pr_k(\bhat\bhat):\bhat\bhat = 0.
\end{eqnarray}  
The first term calculates
\begin{eqnarray}
  \frac{\pr\bp}{\pr t}:\bhat\bhat &=& \frac{\pr}{\pr t}\Big( p_\parallel \hat{b}_i\hat{b}_j
  +p_\perp (\delta_{ij}-\hat{b}_i\hat{b}_j) + \Pi_{ij}\Big) \hat{b}_i\hat{b}_j =
  \frac{\pr p_\parallel}{\pr t}\underbrace{|\bhat|^2|\bhat|^2}_{=1}
  + p_\parallel \underbrace{\frac{\pr}{\pr t}(\hat{b}_i\hat{b}_j)\hat{b}_i\hat{b}_j}_{=0}
  +\frac{\pr p_\perp}{\pr t} \underbrace{(\delta_{ij}-\hat{b}_i\hat{b}_j)\hat{b}_i\hat{b}_j}_{=0} \nn\\
  && + p_\perp \underbrace{\hat{b}_i\hat{b}_j\frac{\pr}{\pr t}(\delta_{ij}-\hat{b}_i\hat{b}_j)}_{=0}
  +\frac{\pr}{\pr t}(\underbrace{\Pi_{ij}\hat{b}_i\hat{b}_j}_{=0})
  - \Pi_{ij}\frac{\pr}{\pr t}(\hat{b}_i\hat{b}_j) = \frac{\pr p_\parallel}{\pr t} - \boldsymbol{\Pi}:\frac{\pr}{\pr t}(\bhat\bhat).
\end{eqnarray}
The second term calculates
\begin{eqnarray}
 \nabla\cdot(\bu\bp):\bhat\bhat &=& \pr_k (u_k p_{ij}) \hat{b}_i\hat{b}_j = \pr_k \big( u_k \underbrace{p_{ij} \hat{b}_i\hat{b}_j}_{=p_\parallel}\big)
 -u_k p_{ij}\pr_k(\hat{b}_i\hat{b}_j) = \pr_k(u_k p_\parallel) - u_k p_\parallel \underbrace{\hat{b}_i\hat{b}_j \pr_k (\hat{b}_i\hat{b}_j)}_{=0}\nn\\
 && -p_\perp u_k \underbrace{(\delta_{ij}-\hat{b}_i\hat{b}_j)\pr_k(\hat{b}_i\hat{b}_j)}_{=0} -\Pi_{ij} u_k\pr_k(\hat{b}_i\hat{b}_j)
 = \nabla\cdot(p_\parallel\bu)-\boldsymbol{\Pi}: \Big( \bu\cdot\nabla(\bhat\bhat)\Big);\\
 (\nabla\cdot\bq): \bhat\bhat &=& (\nabla\cdot\bq)_{ij} \hat{b}_i\hat{b}_j = \bhat\cdot(\nabla\cdot\bq)\cdot\bhat.
\end{eqnarray}  
The third term calculates
\begin{eqnarray}
  (\bp\cdot\nabla\bu) :\bhat\bhat &=& (\bp\cdot\nabla\bu)_{ij} \hat{b}_i\hat{b}_j = (p_{ik} \pr_k u_j) \hat{b}_i\hat{b}_j
  = (\pr_k u_j) p_{ik}\hat{b}_i\hat{b}_j = (\pr_k u_j) \Big[p_\parallel \hat{b}_i\hat{b}_k
    +p_\perp (\delta_{ik}-\hat{b}_i\hat{b}_k)  + \Pi_{ik}\Big] \hat{b}_i\hat{b}_j \nn\\
  &=&  (\pr_k u_j) \Big[ p_\parallel \hat{b}_k\hat{b}_j + p_\perp \underbrace{(\hat{b}_k\hat{b}_j - \hat{b}_k\hat{b}_j)}_{=0}
    + \Pi_{ik} \hat{b}_i\hat{b}_j\Big] = p_\parallel \hat{b}_k (\pr_k u_j)\hat{b}_j + \Pi_{ik} (\pr_k u_j) \hat{b}_i\hat{b}_j \nn\\
  &=& p_\parallel \bhat\cdot\nabla\bu\cdot\bhat +(\boldsymbol{\Pi}\cdot\nabla\bu ):\bhat\bhat.
\end{eqnarray}
The fourth term calculates similarly
\begin{eqnarray}
  (\bp\cdot\nabla\bu)^{\textrm{T}} :\bhat\bhat &=& (\bp\cdot\nabla\bu)_{ji} \hat{b}_i\hat{b}_j =
  (p_{jk} \pr_k u_i) \hat{b}_i\hat{b}_j
  = (\pr_k u_i) p_{jk}\hat{b}_i\hat{b}_j = (\pr_k u_i) \Big[p_\parallel \hat{b}_j\hat{b}_k
    +p_\perp (\delta_{jk}-\hat{b}_j\hat{b}_k)  + \Pi_{jk}\Big] \hat{b}_i\hat{b}_j \nn\\
  &=&  (\pr_k u_i) \Big[ p_\parallel \hat{b}_k\hat{b}_i + p_\perp \underbrace{(\hat{b}_i\hat{b}_k - \hat{b}_k\hat{b}_i)}_{=0}
    + \Pi_{jk} \hat{b}_i\hat{b}_j\Big] = p_\parallel \hat{b}_k (\pr_k u_i)\hat{b}_i + \Pi_{jk} (\pr_k u_i) \hat{b}_i\hat{b}_j \nn\\
  &=& p_\parallel \bhat\cdot\nabla\bu\cdot\bhat +(\boldsymbol{\Pi}\cdot\nabla\bu )_{ji} \hat{b}_i\hat{b}_j
   = p_\parallel \bhat\cdot\nabla\bu\cdot\bhat +(\boldsymbol{\Pi}\cdot\nabla\bu)^{\textrm{T}} :\bhat\bhat,
\end{eqnarray}
and the final fifth term becomes
\begin{eqnarray}
  \Big[\bhat\times\boldsymbol{\Pi} + (\bhat\times\boldsymbol{\Pi})^{\textrm{T}} \Big] :\bhat\bhat
  &=& \Big[(\bhat\times\boldsymbol{\Pi})_{ij} + (\bhat\times\boldsymbol{\Pi})_{ji} \Big] \hat{b}_i\hat{b}_j
  = \Big[ \epsilon_{ikl}\hat{b}_k \Pi_{lj} + \epsilon_{jkl}\hat{b}_k \Pi_{li} \Big] \hat{b}_i\hat{b}_j \nn\\
  &=& \hat{b}_j \Pi_{lj} \underbrace{\epsilon_{ikl} \hat{b}_i\hat{b}_k}_{=0}
  + \hat{b}_i \Pi_{li} \underbrace{\epsilon_{jkl} \hat{b}_j\hat{b}_k}_{=0} =0.
\end{eqnarray}  
The entire equation for the parallel pressure therefore reads
\begin{eqnarray}
  \frac{\pr p_\parallel}{\pr t} + \nabla\cdot(p_\parallel \bu) +2p_\parallel \bhat\cdot\nabla\bu\cdot\bhat
  + \bhat\cdot(\nabla\cdot\bq)\cdot\bhat
    - \boldsymbol{\Pi} :\Big[\Big( \frac{\pr}{\pr t} +\bu\cdot\nabla \Big)\big(\bhat\bhat\big) \Big]
   +\Big[ \boldsymbol{\Pi}\cdot\nabla\bu + (\boldsymbol{\Pi}\cdot\nabla\bu)^{\textrm{T}}\Big] :\bhat\bhat = 0.
\end{eqnarray}
Using the symmetric operator (\ref{eq:Sym2D}) and the definition for convective derivative $\frac{d}{dt}\equiv \frac{\pr}{\pr t} +\bu\cdot\nabla$,
the equation is written in a simple form as
\begin{eqnarray} \label{eq:PparFLR}
  \frac{\pr p_\parallel}{\pr t} + \nabla\cdot(p_\parallel \bu) +2p_\parallel \bhat\cdot\nabla\bu\cdot\bhat
  + \bhat\cdot(\nabla\cdot\bq)\cdot\bhat
    - \boldsymbol{\Pi} : \frac{d}{dt}\big(\bhat\bhat\big) + (\boldsymbol{\Pi}\cdot\nabla\bu)^S :\bhat\bhat = 0,
\end{eqnarray}
which corresponds to eq. 26 of \cite{PassotSulem2004b}, eq. 9 of \cite{Goswami2005} (see also \cite{PassotSulem2007,PSH2012,Oraevskii1968}).
\subsection{Perpendicular pressure equation}
It is possible to either apply the double contraction $:(\boldsymbol{I}-\bhat\bhat)/2$ on the pressure tensor equation, and directly obtain the
equation for $\pr p_\perp/\pr t$, or to apply the Trace (the double contraction $:\boldsymbol{I}$), obtain the equation
for $\pr p_\parallel/\pr t + 2\pr p_\perp/\pr t$, and subtract the previously obtained equation for $\pr p_\parallel/\pr t$.
Both approaches are equivalent, since $p_\perp = (\textrm{Tr}(\bp) -p_\parallel)/2$.
In another words, directly calculating the trace of the pressure tensor yields
\begin{equation}
  \trace \bp = \delta_{ij} p_{ij} = p_{ii} = p_\parallel \underbrace{\hat{b}_i\hat{b}_i}_{=1}
  + p_\perp \underbrace{(\delta_{ii} -\hat{b}_i\hat{b}_i)}_{=2}
  + \underbrace{\Pi_{ii}}_{=0} = p_\parallel + 2p_\perp, 
\end{equation}
where we have used that $\trace\boldsymbol{\Pi}=\Pi_{ii}=0$.
Applying the trace operator at the pressure tensor equation (\ref{eq:PTfinal}) term by term yields for
the first term 
\begin{eqnarray}
  \trace \frac{\pr}{\pr t}\bp = \delta_{ij} \frac{\pr}{\pr t} p_{ij} = \frac{\pr}{\pr t} p_{ii} =
  \frac{\pr}{\pr t} (p_\parallel + 2p_\perp);
\end{eqnarray}  
for the second term
\begin{eqnarray}
  \trace  \big( \nabla\cdot(\bu\bp)\big) &=& \delta_{ij} \pr_k (u_k p_{ij}) = \pr_k (u_k p_{ii}) =
  u_k \pr_k p_{ii} + p_{ii}\pr_k u_k = \bu\cdot\nabla (p_\parallel + 2p_\perp) + (p_\parallel + 2p_\perp)\nabla\cdot\bu;\\
 \trace (\nabla\cdot\bq) &=& \textrm{unchanged};  
\end{eqnarray}  
the third term
\begin{eqnarray}
  \trace \big( \bp\cdot\nabla\bu\big) &=& \delta_{ij} (\bp\cdot\nabla\bu)_{ij} =
  p_{ik}\pr_k u_i = \Big( (p_\parallel-p_\perp)\hat{b}_i\hat{b}_k + p_\perp \delta_{ik} + \Pi_{ik} \Big) \pr_k u_i =
  (p_\parallel-p_\perp)\hat{b}_k (\pr_k u_i) \hat{b}_i + p_\perp \pr_i u_i\nn\\
  &+& \Pi_{ik} \pr_k u_i = (p_\parallel-p_\perp)\bhat\cdot\nabla\bu\cdot\bhat + p_\perp\nabla\cdot\bu
  + \trace(\boldsymbol{\Pi}\cdot\nabla\bu);
\end{eqnarray}
the fourth term is equivalent to the third term
\begin{eqnarray}
  \trace \big( \bp\cdot\nabla\bu\big)^{\textrm{T}} = (\bp\cdot\nabla\bu)_{ii} =
  (p_\parallel-p_\perp)\bhat\cdot\nabla\bu\cdot\bhat + p_\perp\nabla\cdot\bu
  + \trace(\boldsymbol{\Pi}\cdot\nabla\bu)^{\textrm{T}};
\end{eqnarray}
even though we kept the last expression in transpose form so that we can use the symmetric operator after
we combine all the terms. Finally the fifth term
\begin{eqnarray}
  \trace \big( \bhat \times\boldsymbol{\Pi} \big) &=&\delta_{ij} \epsilon_{ikl} \hat{b}_k \Pi_{lj}
  =\hat{b}_k \underbrace{\epsilon_{ikl} \Pi_{li}}_{=0} =0;\\
  \trace \big( \bhat \times\boldsymbol{\Pi} \big)^\textrm{T} &=& \trace \big( \bhat \times\boldsymbol{\Pi} \big) =0.
\end{eqnarray}  
Combining all the terms, using the parallel pressure equation (\ref{eq:PparFLR}) and dividing by 2 yields the time dependent
equation for the perpendicular pressure that reads
\begin{eqnarray}
  \frac{\pr p_\perp}{\pr t} &+& \nabla\cdot(p_\perp\bu) + p_\perp\nabla\cdot\bu -p_\perp \bhat\cdot\nabla\bu\cdot\bhat
  +\frac{1}{2}\Big[\trace\nabla\cdot\bq -\bhat\cdot(\nabla\cdot\bq)\cdot\bhat\Big] \nn\\
  &+&\frac{1}{2}\Big[ \trace(\boldsymbol{\Pi}\cdot\nabla\bu)^S
  + \boldsymbol{\Pi}:\frac{d}{d t}(\bhat\bhat) -(\boldsymbol{\Pi}\cdot\nabla\bu)^S:\bhat\bhat \Big] =0.
\end{eqnarray}
The result corresponds to for example eq. 25 of \cite{PassotSulem2004b}, eq. 8 of \cite{Goswami2005}.
It is important to note that the parallel and perpendicular pressure equations are exact equations. These equations
are valid regardless of what kind of higher order closure will be adopted later.
It is useful to rewrite the term $\boldsymbol{\Pi}:\frac{d}{d t}(\bhat\bhat)$ slightly so that we can directly use the induction equation
and eliminate the time derivative. The term can be rewritten as
\begin{eqnarray}
  \boldsymbol{\Pi}:\frac{d}{d t}(\bhat\bhat) =  \Pi_{ij}\frac{d}{dt}(\hat{b}_i\hat{b}_j)
  = \Pi_{ij} \Big(\hat{b}_i \frac{d}{dt}\hat{b}_j + \hat{b}_j\frac{d}{dt}\hat{b}_i\Big)
  = \hat{b}_i \Pi_{ij} \frac{d}{dt}\hat{b}_j + \hat{b}_j \underbrace{\Pi_{ij}}_{\Pi_{ji}} \frac{d}{dt}\hat{b}_i
  = 2\bhat\cdot\boldsymbol{\Pi} \cdot \frac{d\bhat}{dt}.
\end{eqnarray}
We will need the following identity for the time derivative of the unit vector (see later in the text),
\begin{eqnarray}
  \frac{d}{d t}\bhat = \frac{1}{|\bb|}\Big[  \frac{d\bb}{d t} -\bhat \Big( \bhat\cdot \frac{d \bb}{d t} \Big) \Big],
\end{eqnarray}
which, when used in the above expression together with $\bhat\cdot\boldsymbol{\Pi}\cdot\bhat =0$, allows us to write
\begin{eqnarray}
  \boldsymbol{\Pi}:\frac{d}{d t}(\bhat\bhat) = \frac{2}{|\bb|}\bhat\cdot\boldsymbol{\Pi} \cdot \frac{d\bb}{d t},
\end{eqnarray}
which is an expression of general validity since no specific form of the induction equation was assumed yet. 
In the next section we will calculate the
decomposition of the heat flux tensor $\bq$ and we will show that if only the gyrotropic heat flux components
$q_\parallel,q_\perp$ are considered, the heat flux terms entering the pressure equations read
\begin{eqnarray}
&&  \bhat\cdot(\nabla\cdot\bq^\textrm{g})\cdot\bhat = \nabla\cdot(q_\parallel\bhat) - 2q_\perp \nabla\cdot\bhat; \label{eq:HFcontr1} \\
&& \frac{1}{2}\Big[\trace\nabla\cdot\bq^\textrm{g} -\bhat\cdot(\nabla\cdot\bq^\textrm{g})\cdot\bhat\Big] = \nabla\cdot(q_\perp\bhat) + q_\perp \nabla\cdot\bhat. \label{eq:HFcontr2} 
\end{eqnarray}  
Therefore, by splitting the heat flux into gyrotropic and non-gyrotropic parts, $\bq=\bq^{\textrm{g}}+\bq^{\textrm{ng}}$, the pressure equations read
\begin{eqnarray}
  \frac{\pr p_\parallel}{\pr t} &+& \nabla\cdot(p_\parallel \bu) +2p_\parallel \bhat\cdot\nabla\bu\cdot\bhat
  +\nabla\cdot(q_\parallel\bhat) - 2q_\perp \nabla\cdot\bhat +\bhat\cdot(\nabla\cdot\bq^{\textrm{ng}})\cdot\bhat \nn\\
    &-& \frac{2}{|\bb|}\bhat\cdot\boldsymbol{\Pi} \cdot \frac{d\bb}{d t} + (\boldsymbol{\Pi}\cdot\nabla\bu)^S :\bhat\bhat = 0; \label{eq:PparEnd}\\
  \frac{\pr p_\perp}{\pr t} &+& \nabla\cdot(p_\perp\bu) + p_\perp\nabla\cdot\bu -p_\perp \bhat\cdot\nabla\bu\cdot\bhat
  +\nabla\cdot(q_\perp\bhat) + q_\perp \nabla\cdot\bhat + \frac{1}{2}\Big[\trace\nabla\cdot\bq^{\textrm{ng}} -\bhat\cdot(\nabla\cdot\bq^{\textrm{ng}})\cdot\bhat\Big]\nn\\
  &+&\frac{1}{2}\Big[ \trace(\boldsymbol{\Pi}\cdot\nabla\bu)^S
  + \frac{2}{|\bb|}\bhat\cdot\boldsymbol{\Pi} \cdot \frac{d\bb}{d t} -(\boldsymbol{\Pi}\cdot\nabla\bu)^S:\bhat\bhat \Big] =0. \label{eq:PperpEnd}
\end{eqnarray}
Terms with the non-gyrotropic heat flux $\bq^{\textrm{ng}}$ can be further split to non-gyrotropic heat flux vectors $\boldsymbol{S}^{\parallel}_\perp$, $\boldsymbol{S}^{\perp}_\perp$
and heat flux tensor $\boldsymbol{\sigma}$, which is addressed  in the Appendix \ref{sec:NONGheat}, see equations (\ref{eq:QngPressure1}), (\ref{eq:QngPressure2}).
Terms containing the FLR pressure tensor $\boldsymbol{\Pi}$ are usually called the FLR stress forces, and these forces can generate complicated
plasma heating processes. The FLR stress forces can generate both parallel and perpendicular plasma
heating that can be determined only from fully nonlinear numerical simulations, since all the terms disappear at the linear level.
This process is also sometimes called ``stochastic heating''. The importance of stochastic heating was shown by \cite{Laveder2013},
who performed 1.5 dimensional numerical simulations with the sophisticated FLR Landau fluid model of \cite{PassotSulem2007,PSH2012}. 
\subsection{On the paper by Chew, Goldberger, Low 1956}
The very influential paper by \cite{Chew1956} that is typically cited for the CGL equations (with zero heat fluxes)
actually also derived the pressure equations when the gyrotropic heat fluxes are considered. It is important
to emphasize that even though the scalar pressures $p_n$, $p_s$ in the notation of that paper are equal to the usual
pressures $p_\parallel$, $p_\perp$, this is not the case for the heat fluxes. The gyrotropic heat flux tensor in that paper
is decomposed into components $q_n$, $q_s$ according to (Chew et al. 1956, eq. 34)
\begin{equation}
q_{ijk}^\textrm{g} = q_n \hat{b}_i \hat{b}_j \hat{b}_k + q_s \big(\delta_{ij}\hat{b}_k+\delta_{jk}\hat{b}_i+\delta_{ik}\hat{b}_j\big).
\end{equation}
In contrast, the usual gyrotropic heat flux decomposition to $q_\parallel$, $q_\perp$ reads (see later in the text)
\begin{equation}
q_{ijk}^\textrm{g} = q_\parallel \hat{b}_i \hat{b}_j \hat{b}_k + q_\perp\big( \delta_{ij}\hat{b}_k +\delta_{jk}\hat{b}_i+\delta_{ik}\hat{b}_j
- 3\hat{b}_i\hat{b}_j\hat{b}_k \big),
\end{equation}
i.e. the last term is not present in their decomposition. However, this does not mean that their pressure equations
(\cite{Chew1956}, eq. 31, 32) are incorrect. It only implies that their equations are written in terms of
\begin{equation} \label{eq:CGLheat}
  q_n=q_\parallel-3q_\perp; \qquad q_s=q_\perp.
\end{equation}
By neglecting the FLR corrections $\boldsymbol{\Pi}$ and $\bq^{\textrm{ng}}$, the parallel and perpendicular pressure
equations (\ref{eq:PparEnd}), (\ref{eq:PperpEnd}) simplify to
\begin{eqnarray}
  \frac{d p_\parallel}{d t} &+& p_\parallel \nabla\cdot \bu +2p_\parallel \bhat\cdot\nabla\bu\cdot\bhat
  +\nabla\cdot(q_\parallel\bhat) - 2q_\perp \nabla\cdot\bhat = 0; \label{eq:PparSimple}\\
  \frac{d p_\perp}{d t} &+&  2 p_\perp\nabla\cdot\bu -p_\perp \bhat\cdot\nabla\bu\cdot\bhat
  +\nabla\cdot(q_\perp\bhat) + q_\perp \nabla\cdot\bhat =0. \label{eq:PperpSimple}
\end{eqnarray}
By using (\ref{eq:CGLheat}), the heat flux contributions in the parallel pressure equation above can be rewritten as
\begin{eqnarray}
\nabla\cdot(q_\parallel\bhat) - 2q_\perp \nabla\cdot\bhat = \nabla\cdot \big[(q_n+q_s)\bhat\big] +2\bhat\cdot\nabla q_s,
\end{eqnarray}  
which agrees with the parallel pressure equation obtained by \cite{Chew1956}, eq. 31. The pressure equations of
Chew et al. 1956 are therefore correct, but it has to be remembered that $q_n$ is not the usual parallel heat flux
$q_\parallel=m\int (v_\parallel-u_\parallel)^3 f d^3v$.\footnote{Sometimes the parallel heat flux is defined with the factor of 1/2 as 
  $q_\parallel=\frac{m}{2}\int (v_\parallel-u_\parallel)^3 f d^3v$, but this definition should be avoided.} Actually, the authors do not even call the quantities $q_n, q_s$ heat flux
components, but components of a ``pressure-transport'' tensor.

To conclude this subsection, we will go slightly ahead and use equations
(\ref{eq:PparCool}), (\ref{eq:PperpCool}) (without the zero right hand sides) that are the usual CGL pressure equations and 
valid only at long spatial scales when the Hall term in the induction equation is neglected. Under this assumption,
the pressure equations (\ref{eq:PparSimple}), (\ref{eq:PperpSimple}) are rewritten as
\begin{eqnarray}
  \frac{d}{dt}\left( \frac{p_{\parallel} |\bb|^2}{\rho^3}\right) &=& -\frac{|\bb|^2}{\rho^3}
  \left[ \nabla\cdot(q_\parallel\bhat) - 2q_\perp \nabla\cdot\bhat \right]; \label{eq:PparSuper}\\
  \frac{d}{d t} \left(\frac{p_{\perp}}{\rho|\bb|}\right) &=&
  - \frac{1}{\rho|\bb|}\left[ \nabla\cdot(q_\perp\bhat) + q_\perp \nabla\cdot\bhat\right]. \label{eq:PperpSuper}
\end{eqnarray}
It is often forgotten that the paper by Chew, Goldberger, Low 1956 derived the correct pressure equations with
gyrotropic heat flux contributions. The authors however did not write the equations in the form (\ref{eq:PparSuper}),
(\ref{eq:PperpSuper}), and left them in the form (\ref{eq:PparSimple}), (\ref{eq:PperpSimple}), using the ``conservative''
form only after prescribing the closure $q_\parallel=q_\perp=0$ (their eq. 35, 36): 
\begin{eqnarray}
  \frac{d}{dt}\left( \frac{p_{\parallel} |\bb|^2}{\rho^3}\right) &=& 0; \label{eq:cgl1}\\
  \frac{d}{d t} \left(\frac{p_{\perp}}{\rho|\bb|}\right) &=& 0. \label{eq:cgl2}
\end{eqnarray}
Over time, equations (\ref{eq:cgl1}), (\ref{eq:cgl2}) became known as the CGL description.  Equations (\ref{eq:cgl1}), (\ref{eq:cgl2})
  are often replaced by the equations of state  $p_{\parallel} |\bb|^2/\rho^3=\textrm{const.}$ and $p_{\perp}/(\rho|\bb|)=\textrm{const.}$,
  see e.g. \cite{Mjolhus2009}.  Additionally, similarly to the definition of (appropriately normalized) entropy in MHD, $s=\ln(p/\rho^\gamma)$,
    it is useful to define parallel and perpendicular entropy in the CGL model, according to
\begin{equation}
    s_\parallel=\frac{1}{3}\ln\Big(\frac{p_{\parallel} |\bb|^2}{\rho^3}\Big); \qquad
    s_\perp=\frac{2}{3}\ln\Big(\frac{p_{\perp}}{\rho|\bb|}\Big).
\end{equation}
The total entropy, $s_\parallel+s_\perp=\ln(p_\parallel^{1/3}p_\perp^{2/3}\rho^{-5/3})$, is equivalent to the MHD entropy when $p_\parallel=p_\perp$,
see for example eq. 9-11 in \cite{Abraham-Shrauner1967}.   
\subsection{Physical meaning of CGL equations} \label{sec:PhysMeaningCGL}
The exact derivation of the CGL equation (\ref{eq:cgl1}), (\ref{eq:cgl2}) will be completed in the next section. Here we briefly
discuss the physical meaning of these equations. 
The CGL equations (\ref{eq:cgl2}), (\ref{eq:cgl1}) can be interpreted as the conservation of the 1st and 2nd adiabatic invariants,
as is nicely discussed for example by \cite{Kulsrud1983} and by \cite{GurnettBhattacharjee2005}.
Considering many particles with velocity components $v_\parallel$ and $v_\perp$ with respect to the mean magnetic field, the parallel and perpendicular pressures
can be estimated as 
\begin{equation} \label{eq:pEst}
p_\parallel \sim \langle v_\parallel^2 \rangle \rho;\qquad p_\perp \sim \langle v_\perp^2 \rangle \rho,
\end{equation}  
where the brackets represent an average over all particles. To complete the estimate, we need expressions for $\langle v_\parallel^2 \rangle$
and $\langle v_\perp^2 \rangle$, which come from the conservation of the adiabatic invariants.
The 1st adiabatic invariant is the conservation of magnetic moment $\mu$ of a particle that is gyrating periodically around a mean magnetic field.
The 2nd adiabatic invariant, sometimes also called the longitudinal invariant, is associated with a particle bouncing periodically between two
magnetic mirror points. The conserved quantity is the integral over the parallel momentum of a particle $J\equiv \oint_a^b mv_\parallel dl$, where the magnetic
mirrors are located at ``a'' and ``b'' and the integral is performed along the magnetic field line. The distance between the two magnetic mirrors at ``a'' and ``b''
can be labeled as L. The conservation of the 1st and 2nd adiabatic invariants is therefore
\begin{equation}
\mu \equiv \frac{mv_\perp^2}{2B} = \textrm{const.}; \qquad  J \equiv mv_\parallel L = \textrm{const.}
\end{equation}  
Since we are considering only non-relativistic particles, the particle mass ``m'' is also a constant.
The conservation of magnetic moment $\mu$ therefore implies $v_\perp^2\sim B$, and using this in (\ref{eq:pEst}) yields
$p_\perp\sim B \rho$, recovering (\ref{eq:cgl2}). The CGL equation (\ref{eq:cgl2}) therefore corresponds to conservation of magnetic moment, i.e., the 1st adiabatic invariant.
The conservation of the second adiabatic invariant J is more tricky, since we need to somehow estimate the non-intuitive length L between the two magnetic mirrors,
and no magnetic mirrors are explicitly assumed, since we are just dealing with somewhat random magnetic field lines. The length L can be nevertheless estimated from two
fundamental physical principles. Consider a magnetic flux tube (a deformed cylinder) with cross sectional area A and length L. The conservation of the total magnetic
flux through the area A implies $AB = \textrm{const.}$, which yields an estimate for the area $A\sim 1/B$. The second principle is the conservation of the
total mass of particles that are \emph{completely trapped} in that flux tube and that can not escape, $mnAL=\textrm{const.}$, yielding 
$L\sim 1/(\rho A)$. The use of $A\sim 1/B$ yields the final estimate for the non-intuitive length $L\sim B/\rho$. The conservation of the 2nd adiabatic invariant
$J$ therefore implies $v_\parallel\sim 1/L \sim \rho/B$. Using this result in (\ref{eq:pEst}) implies $p_\parallel \sim \rho^3/B^2$, recovering (\ref{eq:cgl1}).
The CGL equation (\ref{eq:cgl1}) corresponds to conservation of the second adiabatic invariant J, and the equation is valid for particles that are completely trapped
and therefore cannot carry a heat flux. As we discuss later, the consideration of the Hall-term, the heat flux, and the FLR stress forces (the stochastic heating),
leads to the breaking of these two adiabatic invariants. Nevertheless, exact conservation laws can still be derived, 
and the very simple CGL equations (\ref{eq:cgl1}), (\ref{eq:cgl2}) have to be modified with non-zero right hand sides.

In Part 2 of our guide (see section 4.1), we
show that the conservation of magnetic moment is very useful for understanding the form of the perturbed distribution function $f^{(1)}$ in the gyrotropic limit.
By performing calculations in the laboratory reference frame, we show that to obtain the correct $f^{(1)}$ in the gyrotropic limit actually requires the
usual complicated kinetic integration around the unperturbed orbit (see Appendix of Part 2) and only then the gyrotropic limit can be imposed. In contrast,
performing calculations in the guiding-center reference frame allows one to prescribe the conservation of magnetic moment from the beginning, and obtain the
correct $f^{(1)}$ in the gyrotropic limit perhaps more intuitively. 
Conservation of adiabatic invariants is important in many areas of space physics and astrophysics. For example, conservation of adiabatic invariants is used to
construct models that describe particles trapped inside of magnetic islands and that are being accelerated during magnetic island contraction
and merging \citep{Drake2013,ZankRecon2014}.

\subsection{Exact equations of anisotropic multi-fluid models}
 Before we discuss solutions of specific fluid models, it is beneficial to summarize the most general equations
of multi-fluid models that were derived with no simplifications at all. By reintroducing the species index r,  
the equations that were directly obtained by integrating the collisionless Vlasov equation read
\begin{eqnarray}
&& \frac{\pr \rho_r}{\pr t} + \nabla\cdot (\rho_r \bu_r )=0; \label{eq:Gcontinuity}\\
&& \frac{\pr \bu_r}{\pr t} +\bu_r \cdot\nabla \bu_r +\frac{1}{\rho_r}\nabla\cdot \bp_r 
  -\frac{q_r}{m_r} \left( \bE +\frac{1}{c}\bu_r\times\bb \right)=0; \label{eq:Gmomentum}\\
&&  \frac{\pr p_{\parallel r}}{\pr t} + \nabla\cdot(p_{\parallel r} \bu_r) +2p_{\parallel r} \bhat\cdot\nabla\bu_r\cdot\bhat
  + \bhat\cdot(\nabla\cdot\bq_r)\cdot\bhat
    - \frac{2}{|\bb|}\bhat\cdot\boldsymbol{\Pi}_r \cdot \frac{d_r\bb}{d t} + (\boldsymbol{\Pi}_r\cdot\nabla\bu_r)^S :\bhat\bhat = 0; \label{eq:Gppar}\\
&&  \frac{\pr p_{\perp r}}{\pr t} + \nabla\cdot(p_{\perp r}\bu_r) + p_{\perp r}\nabla\cdot\bu_r -p_{\perp r} \bhat\cdot\nabla\bu_r\cdot\bhat
  +\frac{1}{2}\Big[\trace\nabla\cdot\bq_r -\bhat\cdot(\nabla\cdot\bq_r)\cdot\bhat\Big] \nn\\
&&\qquad  +\frac{1}{2}\Big[ \trace(\boldsymbol{\Pi}_r\cdot\nabla\bu_r)^S
  + \frac{2}{|\bb|}\bhat\cdot\boldsymbol{\Pi}_r \cdot \frac{d_r\bb}{d t} -(\boldsymbol{\Pi}_r\cdot\nabla\bu_r)^S:\bhat\bhat \Big] =0,\label{eq:Gpperp}
\end{eqnarray}
 where the convective derivative $d_r/dt =\pr/\pr t+\bu_r\cdot\nabla$.
In the above equations, the pressure tensor for each species was just decomposed to a gyrotropic contributions $p_{\parallel r}$, $p_{\perp r}$
and the rest of the pressure tensor (the FLR pressure tensor $\boldsymbol{\Pi}_r$), according to
\begin{equation}
\bp_r  =p_{\parallel r}\bhat\bhat + p_{\perp r}(\boldsymbol{I}-\bhat\bhat) + \boldsymbol{\Pi}_r,
\end{equation}
which is a rigorous decomposition not introducing any simplifications. The heat flux $\boldsymbol{q}_r$ contributions are here left at the most general level
without any simplifications either. The equations are accompanied by the Maxwell's equations,
that for now we write down here with no simplifications to emphasize that no Maxwell's equations were used in deriving the above system, 
\begin{eqnarray}
  && \nabla\cdot\bE=4\pi\rho_c = 4\pi \sum_r q_r n_r; \qquad \nabla\cdot\bb=0;\\
  && \frac{\pr \bb}{\pr t} = -c\nabla\times\bE; \qquad
  \bj = \sum_r q_r n_r \bu_r = \frac{c}{4\pi} \nabla\times\bb - \frac{1}{4\pi}\frac{\pr\bE}{\pr t}. \label{eq:GMaxwell}
\end{eqnarray}
Equations (\ref{eq:Gcontinuity})-(\ref{eq:GMaxwell}) represent exact (even though not closed) kinetic Vlasov-Maxwell system formulated in fluid variables,
and the description is valid for any general distribution function and for all considered spatial and temporal scales.
All advanced collisionless multi-fluid models have to be based in one way or another
on these equations. One can concentrate on the proton dynamics or one can concentrate on the electron dynamics and simplify these equations
accordingly for these spatial scales. Naturally, one can consider complicated multi-fluid descriptions composed of many particle species.  

For a new reader who just jumped straight to this section, the symmetric operator ``S'' acts on a matrix $A_{ij}$ according to $A_{ij}^S=A_{ij}+A_{ji}$.
If one does not like the symmetric operator in the expressions above, the symmetric operator is actually not necessary, since
\begin{eqnarray}
 \big( \boldsymbol{\Pi}\cdot\nabla\bu  \big)_{ij} &=& \Pi_{ik} \pr_k u_j;\nn\\
 \big( \boldsymbol{\Pi}\cdot\nabla\bu  \big)_{ij}^S &=& \Pi_{ik} \pr_k u_j + \Pi_{jk} \pr_k u_i;\\
 \big( \boldsymbol{\Pi}\cdot\nabla\bu  \big)^S :\bhat\bhat &=& 2  \hat{b}_i \Pi_{ik} (\pr_k u_j)\hat{b}_j
 = 2 \big( \boldsymbol{\Pi}\cdot\nabla\bu  \big):\bhat\bhat = 2 \bhat \cdot \big( \boldsymbol{\Pi}\cdot\nabla\bu  \big)\cdot \bhat;\\
 \textrm{Tr} \big( \boldsymbol{\Pi}\cdot\nabla\bu  \big)^S &=& 2 \Pi_{ik} \pr_k u_i = 2\boldsymbol{\Pi}:\nabla \bu = 2 (\boldsymbol{\Pi}\cdot\nabla)\cdot \bu, 
\end{eqnarray}
where we also got rid of the double contractions. Note that $\Pi_{ik}=\Pi_{ki}$ and also $p_{ik}=p_{ki}$.  


 Equations (\ref{eq:Gcontinuity})-(\ref{eq:Gpperp}) contain two very important catches - the entire heat flux tensor $\boldsymbol{q}_r$ and the nongyrotropic pressure
tensor $\boldsymbol{\Pi}_r$ are not specified yet. In this moment, the entire class of collisionless fluid models separates to many possible classes
and sub-classes, depending on how precisely one wants to evaluate the heat flux tensor $\boldsymbol{q}_r$ and
the FLR pressure tensor $\boldsymbol{\Pi}_r$. There are several complications how to correctly model these two quantities, but before we discuss this, we want to
point out that by using the simple general identities (\ref{eq:PperpGen}), (\ref{eq:PparGen}) (that are nothing more than calculating derivatives and using the
density equation for each species separately), the above pressure equations can be rewritten to the following form
\begin{eqnarray}
  \frac{d_r}{dt}\left( \frac{p_{\parallel r} |\bb|^2}{\rho_r^3}\right) &=& \frac{|\bb|^2}{\rho_r^3}\bigg\{ 2p_{\parallel r}\nabla\cdot\bu_r - 2p_{\parallel r}\bhat\cdot\nabla\bu_r\cdot\bhat
   +2\frac{p_{\parallel r}}{|\bb|}\bhat\cdot\frac{d_r\bb}{dt} -\bhat\cdot (\nabla\cdot\bq_r)\cdot\bhat \nn \\
    && \qquad\quad +\frac{2}{|\bb|}\bhat\cdot\boldsymbol{\Pi}_r\cdot\frac{d_r\bb}{dt} - (\boldsymbol{\Pi}_r\cdot\nabla\bu_r)^S:\bhat\bhat \bigg\};\label{eq:VelkaPica1}\\
  \frac{d_r}{d t} \left(\frac{p_{\perp r}}{\rho_r |\bb|}\right) &=&  \frac{1}{\rho_r|\bb|}\bigg\{ -p_{\perp r}\nabla\cdot\bu_r + p_{\perp r}\bhat\cdot\nabla\bu_r\cdot\bhat
    -\frac{p_{\perp r}}{|\bb|}\bhat\cdot\frac{d_r\bb}{dt} -\frac{1}{2}\Big[\trace\nabla\cdot\bq_r -\bhat\cdot(\nabla\cdot\bq_r)\cdot\bhat\Big] \nn\\
&&\qquad  -\frac{1}{2}\Big[ \trace(\boldsymbol{\Pi}_r\cdot\nabla\bu_r)^S
  + \frac{2}{|\bb|}\bhat\cdot\boldsymbol{\Pi}_r \cdot \frac{d_r\bb}{d t} -(\boldsymbol{\Pi}_r\cdot\nabla\bu_r)^S:\bhat\bhat \Big] \bigg\}.\label{eq:VelkaPica2}
\end{eqnarray}
Again, no simplifications were introduced and no Maxwell's equations were used yet. The equations are exact. The equations represent the complicated plasma heating processes
that are responsible for anisotropic plasma heating, and that are in general very difficult to classify.
The left hand sides of (\ref{eq:VelkaPica1})-(\ref{eq:VelkaPica2}) are the familiar CGL equations that represent conservations of the first and second adiabatic
invariants, and all the expressions on the right hand sides break these adiabatic invariants. On the right
hand side, we have the $\pr\bb/\pr t$ that couples various species together through the Maxwell’s equations, and
that also introduces the Hall-term responsible for the simplest dispersive effects.
The heat flux tensor $\bq_r$ (a tensor of 3rd rank), is decomposed to its gyrotropic and non-gyrotropic (FLR) part
$\bq_r=\bq_r^{\textrm{g}}+\bq_r^{\textrm{ng}}$. 
At large scales, the gyrotropic heat flux can be viewed as a gateway for the simplest forms of collisionless damping
mechanisms, known as Landau damping, that can be further separated to a ``pure'' electrostatic Landau damping,
and its magnetic analogue, the transit-time damping - see Part 2 of the text, subsection
``Coulomb force \& mirror force (Landau damping \& transit time damping)''. Nevertheless, fluid models that contain the gyrotropic heat flux fluctuations, but that
do not contain any Landau damping can be constructed as well (see CGL2 model later in the text). 
The FLR pressure tensor $\boldsymbol{\Pi}_r$ introduces further dispersive effects and therefore also modifies the
collisionless damping rates, as well as the heat flux $\bq_r$ modifies the FLR pressure corrections $\boldsymbol{\Pi}_r$. The
quantities $\bq_r$ and $\boldsymbol{\Pi}_r$ are therefore generally coupled. Importantly, the FLR pressure tensor $\boldsymbol{\Pi}_r$
introduces complicated turbulent plasma heating processes, referred to as stochastic heating. Since
these FLR stress forces completely disappear at the linear level in the above pressure equations, they
can be explored only with fully nonlinear numerical simulations ($\boldsymbol{\Pi}_r$
still enters at the linear level in the momentum equation and modifies the dispersion relations).

From a fluid perspective, there are several major difficulties how to correctly model the quantities $\boldsymbol{\Pi}_r$ and $\bq_r$.
The heat flux tensor $\boldsymbol{q}_r$ is described by an infinite hierarchy of higher-order fluid moments, and one needs to find an appropriate way, how to
close the system. The FLR pressure tensor $\boldsymbol{\Pi}_r$ is described by equations that
are implicit, and the FLR corrections have to be typically
expanded on temporal and spatial scales in order to obtain tractable expressions that can be used for numerical simulations. 
This expansion will naturally restrict the area of validity of such fluid models to those scales considered.
For example, for the proton-electron plasma where the spatial and temporal scales are largely separated because of the mass ratio $m_p/m_e=1836$, 
simplest first order FLR corrections expanded around the proton scales $k \rho_i, \omega/\Omega_p$ will long lose their validity at the electron scales.    
The most complicated Landau fluid models that use linear kinetic theory to evaluate $\boldsymbol{\Pi}$ and that contain the Bessel functions \cite{PassotSulem2007,PSH2012,SulemPassot2015},
have technically no restriction for wavenumbers $k_\perp \rho_i$, but the first order frequency restriction is there nevertheless.

\subsection{Conservation of energy}
The equations  (\ref{eq:VelkaPica1})-(\ref{eq:VelkaPica2}), that correctly split the entire plasma heating to parallel and perpendicular heating, are indeed quite
complicated. To gain further insight, let's briefly consider a case in which we are not interested in parallel and perpendicular heating
(whose net effect can be possibly zero),
and we are only interested in the heating of the entire system. Such formulations are sometimes used to interpret plasma heating in fully kinetic simulations 
\cite{YangMatthaeus2017_1,YangMatthaeus2017_2}. By summing together the pressure equations (\ref{eq:Gppar}), (\ref{eq:Gpperp}),
we are interested in the evolution equation $\pr (p_\parallel+2p_\perp)/\pr t$, or in another words we are interested in $\pr \textrm{Tr} \bp/\pr t$.
Summing the equations together (actually the trace of the entire pressure tensor equation was calculated few pages back, that is how the
$p_\perp$ equation was derived) yields
\begin{eqnarray}\label{eq:Eint_1}
  \frac{\pr}{\pr t} (p_{\parallel r}+2p_{\perp r}) +\nabla\cdot \Big( (p_{\parallel r}+2p_{\perp r})\bu_r \Big) +\nabla\cdot(\textrm{Tr} \boldsymbol{q}_r)
  + 2\Big( (\underbrace{\boldsymbol{p}_r^g+\boldsymbol{\Pi}_r}_{=\bp_r})\cdot\nabla\Big)\cdot \bu_r = 0.
\end{eqnarray} 
Note that $\textrm{Tr} \nabla\cdot\boldsymbol{q}_r = \nabla\cdot(\textrm{Tr} \boldsymbol{q}_r)$, and also $\textrm{Tr}\boldsymbol{\Pi}_r=0$. By integrating over the
entire spatial volume of the plasma, one can define the internal energy (or thermal energy) for each particle species as 
\begin{equation} 
  E^{\textrm{int}}_r = \frac{1}{2} \int \textrm{Tr}\, \boldsymbol{p}_r \; d^3 x =  \frac{1}{2} \int (p_{\parallel r}+2p_{\perp r}) \; d^3 x
  \equiv \frac{1}{2}\langle p_{\parallel r}+2p_{\perp r} \rangle.
\end{equation}
By considering a special case of periodic boundary conditions, i.e. typical numerical simulations of turbulence in a periodic box, one can
use the Gauss-Ostrogradsky theorem (\ref{eq:GaussOstrogradsky}), now in a spatial domain, and for periodic boundary conditions all divergence operators
in (\ref{eq:Eint_1}) vanish, yielding
\begin{equation} \label{eq:Eint_2}
\frac{\pr}{\pr t} E^{\textrm{int}}_r =  - \langle \Big( (\boldsymbol{p}_r^g+\boldsymbol{\Pi}_r)\cdot\nabla\Big)\cdot \bu_r \rangle.
\end{equation}
Equations (\ref{eq:Eint_1}), (\ref{eq:Eint_2}) are equivalent to equations 4 and 7 of \cite{YangMatthaeus2017_1}.
Here the equations are just written in such a way that one can explicitly see the pressure
components $p_\parallel, p_\perp, \boldsymbol{\Pi}$, but the equations are equivalent.
Following \cite{YangMatthaeus2017_1}, one can also easily derive the evolution equations for the ``kinetic'' energy and the electromagnetic energy
\begin{equation}
 E^{\textrm{kin}}_r = \frac{1}{2} \langle \rho_r |\bu_r|^2 \rangle; \qquad  E^{\textrm{mag}} = \frac{1}{8\pi} \langle \Big( |\bb|^2+|\bE|^2  \Big) \rangle.
\end{equation}
For the kinetic energy, by first applying $\bu_r\cdot$ at the momentum equation (\ref{eq:Gmomentum})
\begin{equation}
  \bu_r\cdot \Big( \rho_r \frac{\pr \bu_r}{\pr t} + \rho_r\bu_r\cdot\nabla\bu_r \Big) = -\bu_r\cdot (\nabla\cdot\boldsymbol{p}_r)
  + \underbrace{q_r n_r \bu_r}_{=\boldsymbol{j}_r} \cdot \bE,
\end{equation}
which further implies
\begin{eqnarray}
  \frac{1}{2}\frac{\pr}{\pr t} \big(\rho|\bu|^2\big) &=& \frac{1}{2}\Big( \frac{\pr \rho}{\pr t}|\bu|^2 +\rho \frac{\pr |\bu|^2}{\pr t} \Big)
  = -\frac{1}{2}\nabla\cdot (\rho\bu) |\bu|^2 +\rho\bu\cdot \frac{\pr \bu}{\pr t} = -\frac{1}{2}\nabla\cdot \big(\rho\bu|\bu|^2)
  + \bu\cdot \Big( \rho \frac{\pr \bu}{\pr t} + \rho\bu\cdot\nabla\bu \Big)\nn \\
  &=& -\frac{1}{2}\nabla\cdot \big(\rho_r\bu_r |\bu_r|^2)
  -\nabla\cdot(\boldsymbol{p}_r \bu_r) + (\boldsymbol{p}_r\cdot\nabla)\cdot\bu_r + \boldsymbol{j}_r \cdot \bE,
\end{eqnarray}
and for the electromagnetic energy
\begin{eqnarray}
  \frac{1}{8\pi}\frac{\pr}{\pr t} \Big( |\bb|^2+|\bE|^2  \Big) &=& \frac{1}{4\pi}\frac{\pr}{\pr t} \Big( \bb\cdot \frac{\pr\bb}{\pr t}+\bE\cdot \frac{\pr\bE}{\pr t}  \Big)
  = -\frac{c}{4\pi}\Big( \bb\cdot(\nabla\times\bE)-\bE\cdot(\nabla\times\bb) \Big) -\boldsymbol{j}\cdot\bE\nn\\
  &=& -\frac{c}{4\pi}\nabla\cdot(\bE\times\bb)-\boldsymbol{j}\cdot\bE.
\end{eqnarray}
Therefore, by integrating over the entire volume and assuming periodic boundary conditions, the total conservation of energy can be expressed as \citep{YangMatthaeus2017_1}
\begin{eqnarray} \label{eq:Ekin_2}
  \frac{\pr}{\pr t} E^{\textrm{int}}_r &=&  - \langle (\boldsymbol{p}_r\cdot\nabla)\cdot \bu_r \rangle;\\
  \frac{\pr}{\pr t} E^{\textrm{kin}}_r &=& +\langle (\boldsymbol{p}_r\cdot\nabla)\cdot \bu_r \rangle + \langle \boldsymbol{j}_r \cdot \bE \rangle;\\
  \frac{\pr}{\pr t} E^{\textrm{mag}}  &=& - \langle \boldsymbol{j}\cdot\bE \rangle, \label{eq:Ekin_2x}
\end{eqnarray}
which beautifully clarifies how the energy can be transfered. Nevertheless, we point out that it is exactly the splitting into the parallel and perpendicular heating that is
so complicated, since even with periodic boundary conditions, only few terms in the pressure equations (\ref{eq:Gppar}), (\ref{eq:Gpperp}) vanish.
Considering anisotropic heating, and splitting the internal (thermal) energy into parallel and perpendicular parts
\begin{equation}
  E^{\textrm{int}}_r = E^{\textrm{int}}_{\parallel r} + E^{\textrm{int}}_{\perp r};\qquad
  E^{\textrm{int}}_{\parallel r} = \frac{1}{2}\langle p_{\parallel r} \rangle; \qquad E^{\textrm{int}}_{\perp r} = \langle p_{\perp r} \rangle,
\end{equation}
the conservation of total energy can be expressed as
\begin{eqnarray}
  \frac{\pr}{\pr t} E^{\textrm{int}}_{\parallel r} &=&  - \langle p_{\parallel r} \bhat \cdot \nabla\bu_r\cdot\bhat \rangle
  - \frac{1}{2}\langle \bhat\cdot (\nabla\cdot\bq_r)\cdot \bhat \rangle
    + \langle \frac{\bhat}{|\bb|}\cdot\boldsymbol{\Pi}_r \cdot \frac{d_r\bb}{d t}\rangle - \langle \bhat\cdot (\boldsymbol{\Pi}_r\cdot\nabla\bu_r) \cdot\bhat\rangle; \label{eq:GpparPer2}\\
  \frac{\pr}{\pr t} E^{\textrm{int}}_{\perp r} &=& - \langle p_{\perp r}\nabla\cdot\bu_r\rangle +\langle p_{\perp r} \bhat\cdot \nabla\bu_r\cdot\bhat \rangle
  +\frac{1}{2} \langle \bhat\cdot(\nabla\cdot\bq_r)\cdot\bhat \rangle 
  - \langle \frac{\bhat}{|\bb|}\cdot\boldsymbol{\Pi}_r \cdot \frac{d_r\bb}{d t}\rangle +\langle \bhat\cdot(\boldsymbol{\Pi}_r\cdot\nabla\bu_r)\cdot\bhat \rangle \nn\\
  && \quad -\langle (\boldsymbol{\Pi}_r\cdot\nabla)\cdot\bu_r \rangle;\label{eq:GpperpPer2}\\
  \frac{\pr}{\pr t} E^{\textrm{kin}}_r &=& +\langle (\boldsymbol{p}_r\cdot\nabla)\cdot \bu_r \rangle + \langle \boldsymbol{j}_r \cdot \bE \rangle;\\
  \frac{\pr}{\pr t} E^{\textrm{mag}}  &=& - \langle \boldsymbol{j}\cdot\bE \rangle, \label{eq:EMAGG}
\end{eqnarray}
where
\begin{equation}
 \langle (\boldsymbol{p}_r\cdot\nabla)\cdot\bu_r \rangle = \langle p_{\parallel r}\bhat\cdot \nabla\bu_r \cdot\bhat\rangle - \langle p_{\perp r}\bhat\cdot \nabla\bu_r \cdot\bhat\rangle
  +\langle p_{\perp r} \nabla\cdot\bu_r\rangle + \langle (\boldsymbol{\Pi}_r\cdot\nabla)\cdot\bu_r\rangle.
\end{equation}
The anisotropic plasma heating is obviously a very complicated process.

\clearpage
\section{CGL description} \label{section:CGL}
The CGL model is the simplest possible fluid model that incorporates anisotropic temperatures. In contrast to MHD, CGL contains two pressure equations.
For very good MHD reviews, from the perspective of turbulence in the solar wind, and the use of MHD in modeling heliospheric and astrophysical plasmas, we recommend
the reviews
by \cite{Goldstein1995,TuMarsch1995,Zank1999,Zhou2004,BrunoCarbone2013}. 
The CGL model is obtained by using the parallel and perpendicular pressure equations (\ref{eq:PparEnd}), (\ref{eq:PperpEnd}) 
and prescribing zero heat flux $q_\parallel=q_\perp=0$, $\bq^{\textrm{ng}}=0$, and by neglecting the FLR pressure tensor $\boldsymbol{\Pi}=0$. 
The pressure equations greatly simplify to the form
\begin{eqnarray}
&&  \frac{\pr p_\parallel}{\pr t} + \nabla\cdot(p_\parallel \bu) +2p_\parallel \bhat\cdot\nabla\bu\cdot\bhat= 0; \label{eq:PparCGL}\\
&&  \frac{\pr p_\perp}{\pr t} + \nabla\cdot(p_\perp\bu) + p_\perp\nabla\cdot\bu -p_\perp \bhat\cdot\nabla\bu\cdot\bhat=0. \label{eq:PperpCGL}
\end{eqnarray}
By using the notation for the convective derivative $d/dt=\pr/\pr t + \bu\cdot\nabla$, the pressure equations can be rewritten as
\begin{eqnarray}
&&  \frac{d p_\parallel}{d t} + p_\parallel \nabla\cdot \bu +2p_\parallel \bhat\cdot\nabla\bu\cdot\bhat= 0; \label{eq:PparCGL2}\\
&&  \frac{d p_\perp}{d t} + 2 p_\perp\nabla\cdot\bu -p_\perp \bhat\cdot\nabla\bu\cdot\bhat=0. \label{eq:PperpCGL2}
\end{eqnarray}
Note that the CGL pressure equations are fully nonlinear and that the nonlinearities are actually of 4th order, i.e. CGL pressure
equations are more nonlinear than the usual MHD pressure equation $dp/dt+\gamma p\nabla\cdot\bu =0$,
where the nonlinearity is only of 2nd order. At first sight, expressions $\bhat\cdot\nabla\bu\cdot\bhat$ might appear to be 
a little unusual when written without any brackets. From MHD, one is familiar with nonlinear expressions of the type $\bu\cdot\nabla\bu$,
which are also usually written without any brackets and can be interpreted in two equivalent forms, either as $(\bu\cdot\nabla)\bu$
(a scalar $\bu\cdot\nabla$ applied on a vector $\bu$; or $\bu\cdot(\nabla\bu)$ (a vector $\bu$ multiplied by a matrix $\nabla\bu$).
In a similar fashion, the expression $\bhat\cdot\nabla\bu\cdot\bhat$ is meant to be interpreted as $(\bhat\cdot\nabla\bu)\cdot\bhat$ or
$\bhat\cdot(\nabla\bu)\cdot\bhat$; but importantly, it is not meant to be interpreted as $\bhat\cdot\nabla(\bu\cdot\bhat)$, because the derivative
is meant to act only on $\bu$ in this case. For direct numerical simulations, the 4th order nonlinearity explicitly reads
\begin{eqnarray}
  \bhat\cdot\nabla\bu\cdot\bhat &=& \hat{b}_i (\nabla\bu)_{ij} \hat{b}_j = \hat{b}_i \hat{b}_j \pr_i u_j =
  \hat{b}_x (\hat{b}_x \pr_x u_x + \hat{b}_y\pr_x u_y + \hat{b}_z\pr_x u_z) \nn\\
  &+& \hat{b}_y (\hat{b}_x \pr_y u_x + \hat{b}_y\pr_y u_y + \hat{b}_z\pr_y u_z)
  +\hat{b}_z (\hat{b}_x \pr_z u_x + \hat{b}_y\pr_z u_y + \hat{b}_z\pr_z u_z),
\end{eqnarray}
and is of course a scalar. 
So far, we have not made any assumptions about particle species, and have only integrated the Vlasov equation and derived the density, momentum and
scalar pressure equations, which can be done for each species separately. 
By re-introducing the species index $r$, where $r=p$ for protons, $r=e$ for electrons etc., the n-fluid CGL-type equations become
\begin{eqnarray}
&& \frac{\pr \rho_r}{\pr t} + \nabla\cdot (\rho_r \bu_r )=0; \label{eq:continuity}\\
&& \frac{\pr \bu_r}{\pr t} +\bu_r \cdot\nabla \bu_r +\frac{1}{\rho_r}\nabla\cdot \bp_r 
-\frac{q_r}{m_r} \left( \bE +\frac{1}{c}\bu_r\times\bb \right)=0; \label{eq:momentum}\\
&& \frac{\pr p_{\parallel r}}{\pr t} + \nabla\cdot(\bu_r p_{\parallel r}) 
+2p_{\parallel r} \bhat\cdot\nabla \bu_r \cdot \bhat=0; \label{eq:pressure1}\\
&& \frac{\pr p_{\perp r}}{\pr t} + \nabla\cdot(\bu_r p_{\perp r}) + p_{\perp r}\nabla\cdot \bu_r 
-p_{\perp r} \bhat\cdot \nabla \bu_r \cdot \bhat=0. \label{eq:pressure2}
\end{eqnarray}
The above CGL pressure equations are valid regardless of the form of the induction equation $\pr\bb /\pr t$ and are therefore
very general and valid for a wide range of CGL-type n-fluid models. 
It is advisable to keep them in this form when considering direct numerical simulations or solving linear dispersion relations.
The above equations are accompanied by Maxwell's equations. Specifically, 1) Gauss's law $\nabla\cdot\bE=4\pi\rho_c$ where 
the total charge density $\rho_c=\sum_r q_r n_r$, 2) $\nabla\cdot\bb=0$, 3) Faraday's induction equation
$\nabla\times\bE=-\frac{1}{c}\frac{\pr\bb}{\pr t}$ and 4) Amp\`ere's law   
$\nabla \times \bb =\frac{4\pi}{c}\bj +\frac{1}{c}\frac{\pr \bE}{\pr t}$, where the total current $\bj=\sum_r q_r n_r \bu_r$.
In many areas of space physics and astrophysics it is often very useful to eliminate very high frequency effects occurring 
at frequencies higher than the plasma frequency. This is achieved by prescribing local charge neutrality (so called quasi-neutrality
since the plasma is still ionized) by $\rho_c=0$ and by neglecting the second term in the Amp\`ere's law
$\frac{1}{c}\frac{\pr \bE}{\pr t}$ known as Maxwell's displacement current (or Maxwell's correction to Amp\`ere's law). 
This eliminates high frequency effects such as for example plasma oscillations known as Langmuir waves. This assumption is
also equivalent to assuming that the Alfv\'en speed is much less than the speed of light $V_A/c\ll 1$.
 It is important to emphasize that the Gauss's law $\nabla\cdot\bE=4\pi\rho_c$ is \emph{replaced} by
  the quasi-neutrality condition $\rho_c=0$ and no requirement is imposed on $\nabla\cdot\bE$ (other than it is small).
  For further discussion concerning the charge neutrality and $\nabla\cdot\bE$,
see for example \cite{Braginskii1965} page 264, \cite{Webb2007} eq. 7, and \cite{PassotSulem2007} eq. A1. 
The Maxwell's equations in this approximation simplify to
\begin{eqnarray}
&&  \nabla\cdot\bb=0;\\
&&  \frac{\pr\bb}{\pr t} = -c \nabla\times\bE;\\
&&  \bj = \sum_r q_r n_r \bu_r = \frac{c}{4\pi} \nabla \times \bb \label{eq:MaxJ},
\end{eqnarray}  
and this form of Maxwell's equations is considered in most of the fluid models that we discuss, regardless of future complexities
arising from FLR corrections, heat flux, Landau damping and so on.
 The exception is section in Part 2 ``Electron Landau damping of the Langmuir mode'', where the displacement current has to be retained.
Here we further focus only on a plasma consisting of protons and electrons, with charges $q_p=e$ and $q_e=-e$, but models with
more species can of course be considered. Charge neutrality implies that the proton and electron number densities are equal,
i.e. $n_p=n_e=n$. By using the definition of current $\bj=en\bu_p-en\bu_e$, the electron velocities are related to the proton velocities by
\begin{equation}
\bu_e = \bu_p-\frac{1}{en}\bj,
\end{equation}
which after using (\ref{eq:MaxJ}) yields
\begin{equation} \label{eq:Ve}
\bu_e=\bu_p-\frac{1}{en}\frac{c}{4\pi}\nabla\times\bb.
\end{equation}
By using the momentum equation (\ref{eq:momentum}) for electrons, the electric field can be expressed as
\begin{equation} \label{eq:E}
  \bE = -\frac{1}{c}\bu_e\times\bb -\frac{1}{en}\nabla\cdot\bp_e -\frac{m_e}{e}\big( \frac{\pr\bu_e}{\pr t} + \bu_e\cdot\nabla\bu_e \big).
\end{equation}
The last term represents the effects of electron inertia and is for example responsible for the effect that at small spatial scales
the frequency $\omega$ of the parallel whistler mode converges to the electron cyclotron frequency
$\Omega_e$ instead of increasing to infinity. When considering only relatively low
frequencies, it is very useful to neglect the electron inertia term.\footnote{ The magnitude of terms on the r.h.s. of (\ref{eq:E}) can be easily compared
  to the advection term by estimating
  \begin{equation}
    \frac{|\frac{1}{en}\nabla\cdot\bp_e|}{|\frac{1}{c}\bu_e\times\bb|}\sim \frac{\rho_e}{l}\frac{v_{\textrm{th}e}}{u_e}; \qquad
    \frac{|\frac{m_e}{e}\frac{d\bu_e}{dt}|}{|\frac{1}{c}\bu_e\times\bb|}\sim \frac{1}{\Omega_e \tau},
  \end{equation}
where $l,\tau$ are characteristic length-scale and time-scale, $\rho_e$ the electron gyroradius and $v_{\textrm{th}e}$ the electron thermal velocity.}
This yields somewhat simpler analytic expressions, but most
importantly, it provides a great benefit for direct numerical simulations since the time step is not restricted by the requirement
to fully resolve the electron motion. This of course restricts the validity of a model to
 frequencies that are sufficiently smaller than the electron cyclotron frequency.
By neglecting the electron inertia and using (\ref{eq:Ve}), the electric field can be expressed as 
\begin{equation} \label{eq:Efield}
\bE = -\frac{1}{c}\bu_p\times\bb +\frac{1}{4\pi en}(\nabla\times\bb)\times\bb-\frac{1}{en}\nabla\cdot\bp_e.
\end{equation}
Using this expression in the proton momentum equation and in the induction equation eliminates the electric field from the
entire system. The system of equations therefore reads 
\begin{eqnarray}
&& \frac{\pr \rho_p}{\pr t} + \nabla\cdot (\rho_p \bu_p )=0;\\
&& \frac{\pr \bu_p}{\pr t} +\bu_p \cdot\nabla \bu_p +\frac{1}{\rho_p}\nabla\cdot (\bp_p + \bp_e )
-\frac{1}{4\pi\rho_p} (\nabla\times\bb)\times\bb=0;\\
&&  \frac{\pr \bb}{\pr t} = \nabla\times(\bu_p\times\bb)-\frac{c}{4\pi e}\nabla\times\Big[ \frac{1}{n}(\nabla\times\bb)\times\bb\Big]
  +\frac{c}{e}\nabla\times\Big[ \frac{1}{n}\nabla\cdot\bp_e\Big], \label{eq:IndB}
\end{eqnarray}
and is accompanied by the parallel and perpendicular pressure equations (\ref{eq:pressure1}), (\ref{eq:pressure2}) for the proton and
electron species. In the induction equation (\ref{eq:IndB}), the first term is the familiar term from the MHD induction equation and considering only this
first term yields the classical CGL model (where in addition the electrons are prescribed to be cold to eliminate $\boldsymbol{p}_e$ from the momentum
equation). The second term in the induction equation represents the Hall-term, and fluid models
containing this term are generally referred to as Hall-CGL fluid models. The third term represents the electron pressure contributions and in CGL plasmas
the term is in general nonzero, since
\begin{equation}
\nabla\cdot\bp_e = (p_{\parallel e}-p_{\perp e})\big( \bhat\cdot\nabla\bhat + \bhat\nabla\cdot\bhat\big) +\bhat\bhat\cdot\nabla(p_{\parallel e}-p_{\perp e}) + \nabla p_{\perp e}.
\end{equation}
In contrast, by prescribing the electrons to be isotropic and isothermal (a typical assumption used in hybrid simulations)
with pressure $\boldsymbol{p}_e=p_e\boldsymbol{I}$, $p_e=n T_e^{(0)}$, so that $\nabla\cdot\bp_e=\nabla p_e = T_e^{(0)}\nabla n$,
eliminates the pressure term since $\nabla\times [ \frac{1}{n}\nabla n ]= \nabla\times [ \nabla ln(n) ]=0$.
By prescribing the electrons to be isotropic but not isothermal, with pressure $\boldsymbol{p}_e=p_e\boldsymbol{I}$, the term initially does not
disappear and it is equal to
\begin{equation} \label{eq:battery}
\frac{c}{e}\nabla\times\Big[ \frac{1}{n}\nabla\cdot\bp_e\Big] = -\frac{c}{en^2} (\nabla n)\times(\nabla p_e).
\end{equation}  
The term only disappears after the equation of state is prescribed, for example for $p_e\sim n^\gamma$. The term (\ref{eq:battery}) is sometimes
called the ``battery'' term \citep{Biermann1950,Kulsrud1997,Khomenko2017}, since it is argued that for any deviations from an
ideal equation of state (caused for example by shocks), the term can produce magnetic field fluctuations even if there is no magnetic field initially.
 
\subsection{Normalized equations and definitions}
As with MHD and Hall-MHD, it is often useful to work with normalized equations. The density, velocity, magnetic field and pressure
are normalized with respect to $\rho_0,u_0,B_0,p_\parallel^{(0)}$.
The length is normalized with respect to (for now) arbitrary $x_0$ and the time with respect to $t_0=x_0/u_0$. The normalized quantities are then
\begin{eqnarray}
&&  \widetilde{\rho}=\frac{\rho}{\rho_0}; \qquad \widetilde{\bu}=\frac{\bu}{u_0}; \qquad \widetilde{\bb}=\frac{\bb}{B_0};
\qquad  \widetilde{p}_\parallel=\frac{p_\parallel}{p_\parallel^{(0)}}; \qquad \widetilde{p}_\perp=\frac{p_\perp}{p_\parallel^{(0)}};\label{eq:NormDef}\\
  && \widetilde{\bx} = \frac{\bx}{x_0}; \qquad \widetilde{\nabla}=\nabla x_0; \qquad \widetilde{t}=t\frac{u_0}{x_0};
  \qquad \frac{\pr}{\pr \widetilde{t}} = \frac{\pr}{\pr t} \frac{x_0}{u_0}.  
\end{eqnarray}
Obviously $\widetilde{n}=\widetilde{\rho}$. After dropping the tilde, the normalization yields the same density and pressure equations, whereas
the normalized momentum and induction equation become 
\begin{eqnarray}
&& \frac{\pr \bu_p}{\pr t} +\bu_p \cdot\nabla \bu_p +\frac{p_\parallel^{(0)}}{\rho_0 u_0^2} \frac{1}{\rho_p}\nabla\cdot (\bp_p+\bp_e)
-\frac{V_A^2}{u_0^2} \frac{1}{\rho_p}(\nabla\times\bb)\times\bb=0;\\
&&  \frac{\pr \bb}{\pr t} = \nabla\times(\bu\times\bb) -\frac{V_A^2}{\Omega_p u_0 x_0} \nabla\times\Big[ \frac{1}{n}(\nabla\times\bb)\times\bb\Big]
  +\frac{p_\parallel^{(0)}}{\Omega_p \rho_0 u_0 x_0}\nabla\times\Big[ \frac{1}{n}\nabla\cdot\bp_e\Big],
\end{eqnarray}
where the Alfv\'en speed $V_A=B_0/\sqrt{4\pi\rho_0}$. 
The momentum equation implies that it is beneficial to choose the (so far unspecified) normalizing velocity $u_0$ to be the Alfv\'en speed $u_0=V_A$. 
If the 2nd and 3rd terms are neglected in the induction equation (i.e. when the usual MHD induction equation is used), the CGL system is independent of length-scale $x_0$,
similarly to the MHD system. If the Hall term is considered, it is beneficial to choose (so far unspecified) normalizing
length-scale $x_0$ to be the ion inertial length 
\begin{equation}
  d_i=\frac{V_A}{\Omega_p}. 
\end{equation}
Therefore by choosing $x_0=d_i$, the normalizations that we use in all fluid models in this paper read
\begin{eqnarray} \label{eq:NormDef2}
\widetilde{\bu}=\frac{\bu}{V_A}; \qquad \widetilde{\bx} = \frac{\bx}{d_i}; \qquad \widetilde{t}=t\frac{V_A}{d_i}=t\Omega_p. 
\end{eqnarray}
The normalization also has the advantage that, when transferred to Fourier space, one obtains the dispersion relations
for normalized wavenumber $\widetilde{k}$ and frequency $\widetilde{\omega}$ as
\begin{equation} \label{eq:NormKO}
\widetilde{k} = \frac{k V_A}{\Omega_p},\qquad \widetilde{\omega} = \frac{\omega}{\Omega_p}.
\end{equation}
It is useful to define the parallel and perpendicular thermal speeds
\begin{equation} \label{eq:vthDef}
v_{\textrm{th}\parallel} = \sqrt{\frac{2 T_{\parallel}^{(0)}}{m_p}}; \qquad v_{\textrm{th}\perp} = \sqrt{\frac{2 T_{\perp}^{(0)}}{m_p}},
\end{equation}
and the definition of parallel and perpendicular temperatures are $T_{\parallel}=p_{\parallel}/n$ and $T_{\perp}=p_{\perp}/n$.
We use the usual convention with the Boltzmann constant $k_B=1$.\footnote{Without this
  convention the temperature is defined as $T_{\parallel\perp}=p_{\parallel\perp}/(nk_B)$ and the thermal speeds
  $v_{\textrm{th} \parallel\perp} = \sqrt{2 k_B T_{\parallel\perp}^{(0)}/m_p}$, so the plasma $\beta_{\parallel}$, $\beta_{\perp}$ expressions are the same as above.}
The parallel and perpendicular plasma beta are defined according to 
\begin{equation}
  \bpar = \frac{v_{\textrm{th}\parallel}^2}{V_A^2} = \frac{2p_\parallel^{(0)}}{\rho_0 V_A^2}=\frac{p_\parallel^{(0)}}{B_0^2/(8\pi)};
  \qquad \beta_\perp = \frac{v_{\textrm{th}\perp}^2}{V_A^2} = \frac{2p_\perp^{(0)}}{\rho_0 V_A^2}=\frac{p_\perp^{(0)}}{B_0^2/(8\pi)} =\bpar \frac{T_\perp^{(0)}}{T_\parallel^{(0)}},
\end{equation}  
and later we will often use the abbreviation $a_p$ for the proton temperature anisotropy ratio 
\begin{equation} \label{eq:TempRatio}
a_p = \frac{T_\perp^{(0)}}{T_\parallel^{(0)}}=\frac{p_\perp^{(0)}}{p_\parallel^{(0)}}.
\end{equation}
The normalized momentum and induction equations therefore read (tilde are dropped)
\begin{eqnarray}
&& \frac{\pr \bu_p}{\pr t} +\bu_p \cdot\nabla \bu_p +\frac{\bpar}{2} \frac{1}{\rho_p}\nabla\cdot (\bp_p+\bp_e)
  -\frac{1}{\rho_p}(\nabla\times\bb)\times\bb=0;\\
&&  \frac{\pr \bb}{\pr t} = \nabla\times(\bu\times\bb) - \nabla\times\Big[ \frac{1}{n}(\nabla\times\bb)\times\bb\Big]
  +\frac{\bpar}{2}\nabla\times\Big[ \frac{1}{n}\nabla\cdot\bp_e\Big].
\end{eqnarray}
Another useful quantity that we will use later in the text is the proton Larmor radius (the gyroradius) $\rho_i$, defined according to
\begin{equation}
\rho_i = \frac{v_{\textrm{th}\perp}}{\Omega_p}. 
\end{equation}
We note that normalizations do not have to be done with respect to mean values, and normalizations with respect
to fluctuating (turbulent) quantities are used for example in the Nearly Incompressible (NI) models of \cite{ZankNI2017,ZM1993}.
\subsection{Classical CGL model with cold electrons}
Here we want to consider the simplest possible CGL model for the proton species. We assume the electrons to be massless and cold ($\bp_e=0$), and
we also neglect the Hall-term in the induction equation. Since only  proton species are present, for simplicity we can drop the proton index ``p''.
Let's work in physical units for a moment. The classical CGL model with cold electrons therefore reads (written in physical units)
\begin{eqnarray}
&& \frac{\pr \rho}{\pr t} + \nabla\cdot (\rho \bu )=0; \label{eq:CGL_first}\\
&& \frac{\pr \bu}{\pr t} +\bu \cdot\nabla \bu +\frac{1}{\rho}\nabla\cdot \bp
-\frac{1}{4\pi\rho} (\nabla\times\bb)\times\bb=0;\\
&& \frac{\pr p_\parallel}{\pr t} + \nabla\cdot(p_\parallel \bu) +2p_\parallel \bhat\cdot\nabla\bu\cdot\bhat= 0; \label{eq:PparCGL_cl}\\
&&  \frac{\pr p_\perp}{\pr t} + \nabla\cdot(p_\perp\bu) + p_\perp\nabla\cdot\bu -p_\perp \bhat\cdot\nabla\bu\cdot\bhat=0; \label{eq:PperpCGL_cl}\\
&&  \frac{\pr \bb}{\pr t} = \nabla\times(\bu\times\bb). \label{eq:CGL_last}
\end{eqnarray}
Similarly to the usual MHD model, the CGL equations are scale-invariant and do not contain any information about the physical length scale.
A length scale is introduced by considering the Hall-term in the induction equation, and we discuss the Hall-CGL model
in the next section. Also, the absence of the Hall-term allows the CGL pressure equations to be rewritten into ``conservative''
form as
\begin{equation}
\frac{d}{dt}\left( \frac{p_{\parallel} |\bb|^2}{\rho^3}\right)=0; \qquad
\frac{d}{dt}\left(\frac{p_{\perp}}{\rho|\bb|}\right) =0, \label{eq:CGLconserv}
\end{equation}
where $d/dt$ is the convective derivative. This formulation is the most often cited form of the CGL pressure equations
(often the $|\bb|$ is abbreviated only as $B$, which can lead to confusion on how to correctly calculate the derivatives). We want to show the
equivalence of (\ref{eq:CGLconserv}) and (\ref{eq:PparCGL_cl}), (\ref{eq:PperpCGL_cl}).
We need the following identity,
\begin{eqnarray}
  \frac{\pr}{\pr t}|\bb| &=& \frac{\pr}{\pr t} (B_x^2+B_y^2+B_z^2)^{1/2}
  = \frac{1}{|\bb|} \bb\cdot \frac{\pr \bb}{\pr t} = \bhat\cdot \frac{\pr \bb}{\pr t},
\end{eqnarray}
and the result is of course a scalar. Similarly for the spatial and convective derivative
\begin{equation}
  \pr_i |\bb| = \bhat \cdot ( \pr_i \bb );\qquad
  \frac{d}{d t}|\bb| = \Big(\frac{\pr}{\pr t}+u_i \pr_i\Big) |\bb| =\bhat \cdot \frac{d \bb}{d t} \label{eq:dtb}.
\end{equation}
Later we will need
\begin{equation}
\frac{d}{dt} |\bb|^2 = \frac{d}{dt} (B_x^2+B_y^2+B_z^2) = 2\bb\cdot\frac{d\bb}{dt}.
\end{equation}
Now we can directly calculate
\begin{eqnarray}
  \frac{d}{d t} \left(\frac{p_{\perp}}{\rho|\bb|}\right) &=& \frac{1}{\rho|\bb|}\frac{d p_\perp}{dt}
  - \frac{p_\perp}{|\bb| \rho^2} \frac{d\rho}{dt} - \frac{p_\perp}{\rho|\bb|^2} \frac{d |\bb|}{dt}\nn\\
  &=& \frac{1}{\rho|\bb|}\left( \frac{d p_\perp}{dt} - \frac{p_\perp}{\rho} \frac{d\rho}{dt} - \frac{p_\perp}{|\bb|} \frac{d |\bb|}{dt} \right),
\end{eqnarray}
and, by using the density equation $d\rho/dt=-\rho\nabla\cdot\bu$ and the identity (\ref{eq:dtb}), we obtain
\begin{eqnarray}
  \frac{d}{d t} \left(\frac{p_{\perp}}{\rho|\bb|}\right) = \frac{1}{\rho|\bb|}\left( \frac{d p_\perp}{dt} + p_\perp\nabla\cdot\bu
  - \frac{p_\perp}{|\bb|} \bhat \cdot \frac{d \bb}{d t} \right). \label{eq:PperpGen}
\end{eqnarray}
The above equation is completely general since we did not use the induction equation so far and we will use this equation
in the next section when we consider the Hall-CGL model. In a similar way one derives a completely general identity
\begin{eqnarray}
  \frac{d}{dt}\left( \frac{p_{\parallel} |\bb|^2}{\rho^3}\right) &=& \frac{|\bb|^2}{\rho^3} \left(
  \frac{dp_\parallel}{dt} + \frac{p_\parallel}{|\bb|^2}\frac{d|\bb|^2}{dt} - 3\frac{p_\parallel}{\rho} \frac{d\rho}{dt}\right)\nn\\
  &=&  \frac{|\bb|^2}{\rho^3} \left( \frac{dp_\parallel}{dt} + 3p_\parallel \nabla\cdot\bu +\frac{2p_\parallel}{|\bb|}\bhat\cdot\frac{d\bb}{dt}\right).
  \label{eq:PparGen}
\end{eqnarray}
Now we use the induction equation
\begin{equation}
  \frac{\pr\bb}{\pr t} = \nabla\times(\bu\times\bb) = \bu\underbrace{\nabla\cdot\bb}_{=0} -\bb\nabla\cdot\bu +\bb\cdot\nabla\bu -\bu\cdot\nabla\bb,
\end{equation}
that allows us to calculate
\begin{eqnarray}
  \frac{d\bb}{d t} &=& \frac{\pr\bb}{\pr t} + \bu\cdot\nabla\bb = -\bb\nabla\cdot\bu +\bb\cdot\nabla\bu;\\
  \frac{\bhat}{|\bb|} \cdot \frac{d\bb}{d t} &=& -\nabla\cdot\bu + \bhat\cdot \nabla\bu\cdot\bhat,
\end{eqnarray}
and which, when used in (\ref{eq:PparGen})-(\ref{eq:PperpGen}) yields the final result
\begin{eqnarray}
  \frac{d}{dt}\left( \frac{p_{\parallel} |\bb|^2}{\rho^3}\right) &=& \frac{|\bb|^2}{\rho^3}
  \left( \frac{dp_\parallel}{dt} + p_\parallel \nabla\cdot\bu + 2p_\parallel \bhat\cdot \nabla\bu\cdot\bhat \right)=0; \label{eq:PparCool}\\
  \frac{d}{d t} \left(\frac{p_{\perp}}{\rho|\bb|}\right) &=&  \frac{1}{\rho|\bb|}\left( \frac{d p_\perp}{dt} + 2p_\perp\nabla\cdot\bu
  -p_\perp \bhat\cdot \nabla\bu\cdot\bhat \right)=0. \label{eq:PperpCool}
\end{eqnarray}
Equations (\ref{eq:PparCool})-(\ref{eq:PperpCool}) verify that the CGL pressure equations (\ref{eq:CGLconserv}) written in the ``conservative'' form and the directly derived
CGL equations (\ref{eq:PparCGL_cl})-(\ref{eq:PperpCGL_cl}) are equivalent. We will see that the right hand side of the usual CGL equations (\ref{eq:CGLconserv})
is nonzero if the Hall-term is considered, and that it is further modified by the heat flux contributions and by the FLR stress forces.
%
\subsection{Linearized CGL equations}
\noindent
To obtain the dispersion relations, the equations need to be linearized and transformed to Fourier space.  
Equations (\ref{eq:CGL_first})-(\ref{eq:CGL_last}) are linearized with respect to mean values
$\rho_0; \bu^{(0)} \equiv \langle \bu\rangle =0; p_{\parallel\perp}^{(0)}; \bb_0=(0,0,B_0); \bhat_0=(0,0,1)$ where the magnetic field $B_0$ is assumed to be in the z-direction.
We assume that the mean value of the velocity is zero (the zero mean velocity value $\bu^{(0)}=0$ should not be confused with normalizing velocity
as in (\ref{eq:NormDef}), which is later chosen to be $V_A$).
The terms $\bhat\cdot\nabla\bu\cdot\bhat$ are linearized as $\bhat\cdot\nabla\bu\cdot\bhat \overset{\textrm{\tiny lin}}{=}\pr_z u_z$ and the linearized system reads
\begin{eqnarray}
&& \frac{\pr \rho}{\pr t} + \rho_0 \nabla\cdot \bu =0; \\
&& \frac{\pr \bu}{\pr t} +\frac{1}{\rho_0}\nabla\cdot \bp
-\frac{1}{4\pi\rho_0} (\nabla\times\bb)\times \bb_0=0;\\
&& \frac{\pr p_\parallel}{\pr t} + p_\parallel^{(0)}\nabla\cdot \bu +2p_\parallel^{(0)} \pr_z u_z= 0;\\
&&  \frac{\pr p_\perp}{\pr t}  + 2p_\perp^{(0)}\nabla\cdot\bu -p_\perp^{(0)} \pr_z u_z=0;\\
&&  \frac{\pr \bb}{\pr t} = \nabla\times(\bu\times\bb_0).
\end{eqnarray}
The divergence of the pressure tensor is
\begin{eqnarray}
  (\nabla\cdot\bp)_i = \pr_j p_{ji} = \hat{b}_i\hat{b}_j \pr_j p_\parallel + (\delta_{ij}-\hat{b}_i\hat{b}_j)\pr_j p_\perp
  +(p_\parallel-p_\perp)(\hat{b}_i \pr_j\hat{b}_j+\hat{b}_j\pr_j\hat{b}_i),
\end{eqnarray}  
and the first step of linearization yields, by components,
\begin{eqnarray}
  (\nabla\cdot\bp)_x &\overset{\textrm{\tiny lin}}{=}& \pr_x p_\perp + (p_\parallel^{(0)}-p_\perp^{(0)})\pr_z \hat{b}_x;\\
  (\nabla\cdot\bp)_y &\overset{\textrm{\tiny lin}}{=}& \pr_y p_\perp + (p_\parallel^{(0)}-p_\perp^{(0)})\pr_z \hat{b}_y;\\
  (\nabla\cdot\bp)_z &\overset{\textrm{\tiny lin}}{=}& \pr_z p_\parallel + (p_\parallel^{(0)}-p_\perp^{(0)}) (\nabla\cdot\bhat +\pr_z \hat{b}_z),
\end{eqnarray}  
which can be conveniently written in matrix notation as
\begin{eqnarray}
 \nabla\cdot\bp \overset{\textrm{\tiny lin}}{=}
 \nabla\cdot\left( \begin{array}{ccc}
    p_\perp & 0 & 0 \\
    0 & p_\perp & 0 \\
    0 & 0 & p_\parallel
 \end{array} \right) +
 (p_\parallel^{(0)}-p_\perp^{(0)})\nabla\cdot
 \left( \begin{array}{ccc}
    0 & 0 & \hat{b}_x \\
    0 & 0 & \hat{b}_y \\
    \hat{b}_x & \hat{b}_y & 2 \hat{b}_z
 \end{array} \right)\Big|_{lin}.
\end{eqnarray}
Notice that since $\pr_i |\bb| = \bhat\cdot(\pr_i\bb)$, the derivatives of unit vectors are calculated according to
\begin{eqnarray}
  \pr_i \bhat = \pr_i \frac{\bb}{|\bb|} = \frac{1}{|\bb|}\Big[ \pr_i \bb - \frac{\bb}{|\bb|}\pr_i|\bb| \Big]
  =\frac{1}{|\bb|}\left[ \pr_i \bb - \bhat \Big(\bhat\cdot(\pr_i\bb) \Big)\right], 
\end{eqnarray}  
and in the linear approximation one has
\begin{eqnarray}
  \pr_i \bhat \overset{\textrm{\tiny lin}}{=} \frac{1}{B_0} \Big[ \pr_i\bb - \bhat_0 \pr_i B_z \Big],
\end{eqnarray}  
which by components reads
\begin{eqnarray}
  \pr_i \hat{b}_x \overset{\textrm{\tiny lin}}{=} \frac{1}{B_0}\pr_i B_x; \qquad
  \pr_i \hat{b}_y \overset{\textrm{\tiny lin}}{=} \frac{1}{B_0}\pr_i B_y; \qquad
  \pr_i \hat{b}_z \overset{\textrm{\tiny lin}}{=} \frac{1}{B_0}(\pr_i B_z - \pr_z B_z),
\end{eqnarray}  
where, importantly, $\pr_z \hat{b}_z \overset{\textrm{\tiny lin}}{=}0$. The divergence of the pressure tensor in the liner approximation
is therefore expressed as
\begin{eqnarray}
 \nabla\cdot\bp \overset{\textrm{\tiny lin}}{=}
 \nabla\cdot\left( \begin{array}{ccc}
    p_\perp & 0 & 0 \\
    0 & p_\perp & 0 \\
    0 & 0 & p_\parallel
 \end{array} \right) +
 \frac{1}{B_0}(p_\parallel^{(0)}-p_\perp^{(0)})\nabla\cdot
 \left( \begin{array}{ccc}
    0 & 0 & B_x \\
    0 & 0 & B_y \\
    B_x & B_y & 0
 \end{array} \right),
 \end{eqnarray}
where the second term contributes if the temperature anisotropy is present, i.e. when $p_\parallel^{(0)}\neq p_\perp^{(0)}$.
The vectors $(\nabla\times\bb)\times\bb_0$ and $\nabla\times(\bu\times\bb_0)$ are straightforward to calculate and are
\begin{eqnarray}
(\nabla\times\bb)\times\bb_0 = B_0 \left( \begin{array}{c}
    -\pr_x B_z +\pr_z B_x\\ -\pr_y B_z+\pr_z B_y \\ 0 \end{array} \right); \qquad
 \nabla\times(\bu\times\bb_0) =  B_0 \left( \begin{array}{c}
   \pr_z u_x \\ \pr_z u_y \\ -\pr_x u_x - \pr_y u_y \end{array} \right). \label{eq:LinParts}
\end{eqnarray}
The entire set of linearized CGL equations reads
\begin{eqnarray}
&& \frac{\pr \rho}{\pr t} + \rho_0 \nabla\cdot \bu =0; \label{eq:CGL-rhoAng}\\
&& \frac{\pr u_x}{\pr t} +\frac{1}{\rho_0}\pr_x p_\perp +\frac{1}{B_0\rho_0}(p_\parallel^{(0)}-p_\perp^{(0)})\pr_z B_x
  -\frac{B_0}{4\pi\rho_0} (-\pr_x B_z+\pr_z B_x)=0;\\
&& \frac{\pr u_y}{\pr t} +\frac{1}{\rho_0}\pr_y p_\perp +\frac{1}{B_0\rho_0}(p_\parallel^{(0)}-p_\perp^{(0)})\pr_z B_y
  -\frac{B_0}{4\pi\rho_0} (-\pr_y B_z+\pr_z B_y)=0;\\
  && \frac{\pr u_z}{\pr t} +\frac{1}{\rho_0}\pr_z p_\parallel +\frac{1}{B_0\rho_0}(p_\parallel^{(0)}-p_\perp^{(0)})(\pr_x B_x+\pr_y B_y)=0;\\
  && \frac{\pr B_x}{\pr t} = B_0 \pr_z u_x;\qquad \frac{\pr B_y}{\pr t} = B_0 \pr_z u_y;\qquad \frac{\pr B_z}{\pr t} = -B_0 \pr_x u_x-B_0 \pr_y u_y;\label{eq:CGL-Bxz}\\
&& \frac{\pr p_\parallel}{\pr t} + p_\parallel^{(0)}(\pr_x u_x+\pr_y u_y) +3p_\parallel^{(0)} \pr_z u_z= 0; \label{eq:PparLin}\\
&&  \frac{\pr p_\perp}{\pr t}  + 2p_\perp^{(0)}(\pr_x u_x+\pr_y u_y) +p_\perp^{(0)} \pr_z u_z=0. \label{eq:PperpLin}
\end{eqnarray}
Additionally, the 3 components of the induction equation are not independent, but
satisfy the $\nabla\cdot\bb=0$ constraint.
In order to find the linear dispersion relations, without any loss of generality, we consider wave propagation in the x-z plane and assume
that there is no variation with respect to the y-coordinate (i.e. one assumes that all expressions containing $\pr_y$ are zero;
alternatively, one can consider propagation in the y-z plane, where all expressions containing $\pr_x$ would be zero). 
 By exploring the above system, it is noteworthy that the density
equation does not play any role in determining the dispersion relation, since no other equation contains the variable $\rho$.
\subsection{Alfv\'en mode}  
On considering (\ref{eq:CGL-rhoAng})-(\ref{eq:PperpLin})  written in the x-z plane with $\pr_y=0$, it is apparent that the equations
for $u_y$ and $B_y$ decouple from the entire system and by applying $\pr/\pr t$ at the equation for $u_y$ yields
\begin{eqnarray}
\frac{\pr^2 u_y}{\pr t^2} - \frac{1}{\rho_0}\Big(  \frac{B_0^2}{4\pi}-p_\parallel^{(0)}+p_\perp^{(0)} \Big) \pr_z^2 u_y=0,
\end{eqnarray}
which is a wave equation describing the propagation of (generally oblique) Alfv\'en waves in the CGL description.
The same wave equation can be obtained for the component $B_y$.
For isotropic mean pressures $p_\parallel^{(0)}=p_\perp^{(0)}$ the equation is equivalent to the MHD description with
the usual Alfv\'en speed $V_A=B_0/(4\pi\rho_0)^{1/2}$. Since
\begin{eqnarray} \label{eq:footKB}
  \frac{\beta_\parallel}{2} = \frac{p_\parallel^{(0)}}{\rho_0 V_A^2}; \qquad a_p \frac{\beta_\parallel}{2} = \frac{p_\perp^{(0)}}{\rho_0 V_A^2},
\end{eqnarray}  
where $a_p$ is the temperature ratio (\ref{eq:TempRatio}), the Alfv\'en wave equation can be rewritten as
\begin{eqnarray} \label{eq:AlfvenPica2}
\frac{\pr^2 u_y}{\pr t^2} - V_A^2\Big[ 1 +\frac{\beta_\parallel}{2}(a_p -1) \Big] \pr_z^2 u_y=0.
\end{eqnarray}
Considering a wave propagating obliquely in the x-z plane with the wavevector $\boldsymbol{k}=(k_\perp,0,k_\parallel)$ at an angle $\theta$ with respect to the z-direction
(the $B_0$ direction), the parallel and perpendicular wavenumbers are defined as
\begin{equation} \label{eq:ks}
k_\parallel = k\cos\theta; \qquad k_\perp = k\sin\theta,
\end{equation}  
and the transformation to Fourier space is performed according to  (see Appendix \ref{sec:AppendixA})
\begin{equation}
\frac{\pr}{\pr t}\rightarrow -i\omega; \qquad \pr_z\rightarrow ik_\parallel; \qquad \pr_x\rightarrow ik_\perp.  
\end{equation}
The wave equation (\ref{eq:AlfvenPica2}) is then
\begin{equation}
\omega ^2 u_y - V_A^2\Big[ 1 +\frac{\beta_\parallel}{2}(a_p -1) \Big] k^2\cos^2\theta u_y=0,
\end{equation}  
which yields the dispersion relation for the Alfv\'en waves
\begin{eqnarray} \label{eq:CGLAlfven}
\omega = \pm k\cos\theta V_A \sqrt{1+\frac{1}{2}\bpar(a_p-1)} = \pm k_\parallel V_A \sqrt{1+\frac{1}{2}\bpar(a_p-1)}.
\end{eqnarray}
It is important to emphasize that similarly to MHD, the Alfv\'en mode propagates in all the oblique directions,
with frequency $\omega$ that is proportional to the projection of the wavenumber $\boldsymbol{k}$ to the direction of $B_0$, i.e. $k_\parallel=k\cos\theta$. 
If the expression under the square root in (\ref{eq:CGLAlfven}) becomes negative, i.e. if $1+\bpar(a_p-1)/2<0$, the frequency $\omega$
is imaginary and the Alfv\'en mode becomes unstable. This instability, that affects the obliquely propagating Alfv\'en mode
is known as the \emph{oblique firehose instability}.
The necessary (but not sufficient) condition for the instability to develop is
$a_p=T_\perp^{(0)}/T_\parallel^{(0)}<1$. The firehose instability criterion can be rewritten in many forms, the most common being
\begin{equation} \label{eq:firehose}
1-\frac{\bpar}{2}+\frac{1}{2}\bpar a_p <0; \qquad  \frac{T_\perp^{(0)}}{T_\parallel^{(0)}} + \frac{2}{\bpar}-1 <0; \qquad
\bpar -\beta_\perp >2; \qquad p_\parallel^{(0)}-p_\perp^{(0)} > \frac{B_0^2}{4\pi}.
\end{equation}
The first expression directly implies that if $\bpar \le 2$, no firehose instability can exist. The firehose instability therefore exists
only for a relatively high plasma beta $\bpar>2$ and only if the parallel temperature is higher than the perpendicular temperature.
Importantly, the firehose instability criterium in the CGL model is equivalent to the one found from linear kinetic theory in the long-wavelength limit 
 (for example, \cite{Rosenbluth1956,Chandrasekhar1958,Parker1958-firehose,VedenovSagdeev1958}).
Alfv\'en wave propagation only affects components $u_y$ and $B_y$ which are decoupled from the rest of the system,
and yields the other (eigenvector) components as $u_x, u_z, B_x, B_z$ being zero. The $u_y$ component does not enter the density equation and
Alfv\'en mode fluctuations therefore do not produce any density fluctuations or pressure fluctuations.
The Alfv\'en mode is therefore incompressible. It is also useful to define the magnetic compressibility $\chi_b(k_\parallel,k_\perp)$ in Fourier space, 
that measures the relative ratio of magnetic energy in the parallel component versus the total magnetic energy
\begin{equation} \label{eq:MagComp}
\chi_b(k_\parallel,k_\perp) = \frac{|B_z|^2}{|B_x|^2+|B_y|^2+|B_z|^2}.
\end{equation}
Since the Alfv\'en mode produces only $B_y$ fluctuations, with $B_x=B_z=0$, the magnetic compressibility for this mode
is
\begin{equation}
  \chi_b(k_\parallel,k_\perp) =0,
\end{equation}
for all the propagation directions. Sometimes the magnetic compressibility is defined as a ratio of parallel and
perpendicular energies $|B_z|^2/(|B_x|^2+|B_y|^2)$, this quantity however
has a slight disadvantage in that the values can become very large (and actually equal to infinity), which makes plotting of this quantity
problematic. In contrast, the magnetic compressibility as defined in (\ref{eq:MagComp}), has a nicely bounded range of values between 0 and 1.
A similar quantity is defined for the ratio of the energy in velocity fluctuations
\begin{equation}
\chi_u(k_\parallel,k_\perp) = \frac{|u_z|^2}{|u_x|^2+|u_y|^2+|u_z|^2},
\end{equation}
and the Alfv\'en mode has $\chi_u=0$ for all propagation directions, since only the perpendicular velocity component $u_y$ is nonzero. 
\subsection{Slow and fast modes}
By applying $\pr/\pr t$ at the equations for $\pr u_x/\pr t$ and $\pr u_z/\pr t$ and using the evolution equations for
$B_x,B_z,p_\parallel,p_\perp$ one obtains
\begin{eqnarray}
&&  \frac{\pr^2 u_x}{\pr t^2} - \frac{p_\perp^{(0)}}{\rho_0}\Big( 2\pr_x^2u_x +\pr_x\pr_z u_z\Big) +
  \Big( \frac{p_\parallel^{(0)}}{\rho_0} - \frac{p_\perp^{(0)}}{\rho_0} -V_A^2 \Big) \pr_z^2 u_x -V_A^2 \pr_x^2 u_x = 0;\\
&& \frac{\pr^2 u_z}{\pr t^2} - \frac{3p_\parallel^{(0)}}{\rho_0}\pr_z^2 u_z - \frac{p_\perp^{(0)}}{\rho_0}\pr_x \pr_z u_x =0.
\end{eqnarray}  
The easiest way how to solve this coupled system is to transform to Fourier space, which yields
\begin{eqnarray}
\left( \begin{array}{cc}
  \omega^2-\big( \frac{2p_\perp^{(0)}}{\rho_0}+V_A^2\big)k_\perp^2+\big(\frac{p_\parallel^{(0)}}{\rho_0}-\frac{p_\perp^{(0)}}{\rho_0}-V_A^2\big)k_\parallel^2;
    & \qquad - \frac{p_\perp^{(0)}}{\rho_0}k_\perp k_\parallel\\
    - \frac{p_\perp^{(0)}}{\rho_0}k_\perp k_\parallel; & \qquad \omega^2-\frac{3p_\parallel^{(0)}}{\rho_0}k_\parallel^2
\end{array} \right)
\left( \begin{array}{c} u_x \\ u_z \end{array} \right) = \left( \begin{array}{c} 0 \\ 0 \end{array} \right). \label{eq:SFmatrix}
\end{eqnarray}  
The system has nontrivial solutions only if the determinant of the matrix is zero, yielding the CGL dispersion relation
of 4th order in frequency $\omega$ that contains the slow and fast modes
\begin{eqnarray}
&&  \omega^4 - \omega^2 \Big[ \Big(\frac{2p_\perp^{(0)}}{\rho_0}+V_A^2\Big)k_\perp^2 +
    \Big(\frac{2 p_\parallel^{(0)}}{\rho_0}+\frac{p_\perp^{(0)}}{\rho_0}+V_A^2\Big)k_\parallel^2 \Big] \nn\\
  && +3k_\parallel^2 \frac{p_\parallel^{(0)}}{\rho_0} \Big[ \Big( \frac{2p_\perp^{(0)}}{\rho_0}-\frac{{p_\perp^{(0)}}^2}{3 p_\parallel^{(0)} \rho_0} +V_A^2\Big) k_\perp^2
     -  \Big(\frac{p_\parallel^{(0)}}{\rho_0}-\frac{p_\perp^{(0)}}{\rho_0}-V_A^2\Big) k_\parallel^2 \Big] =0.
\end{eqnarray}
One can proceed in several ways  with the above dispersion relation.

In the specific cases of strictly parallel ($k_\perp=0$)
and strictly perpendicular ($k_\parallel=0$) propagation directions it is actually easier to work directly with the system (\ref{eq:SFmatrix}),
because the off-diagonal components in the matrix are zero and the dynamics in the $u_x$ and $u_z$ components decouples. For strictly parallel propagation
one solution is in the $u_x$ component, with the dispersion relation identical to the Alfv\'en wave (\ref{eq:CGLAlfven}).
Since the magnetic and velocity fluctuations are only in the $u_x$, $b_x$ components,
the compressibility $\chi_b=0$ and $\chi_u=0$ for this mode. The second solution for parallel propagation is in the $u_z$ component, and it is called the
sound mode, or the ion-acoustic mode, with dispersion relation $\omega=\pm k_\parallel \sqrt{3p_\parallel^{(0)}/\rho_0}=\pm k_\parallel V_A \sqrt{3\bpar/2}$.
The parallel ion-acoustic mode has $\chi_u=1$ and $\chi_b=0$. 
Considering strictly perpendicular propagation, the waves in the $u_z$ component vanish with the solution $\omega=0$. The waves in the
$u_x$ component are fast magnetosonic waves with dispersion relation $\omega=\pm k_\perp \sqrt{V_A^2+2p_\perp^{(0)}/\rho_0} = \pm k_\perp V_A \sqrt{1+a_p \bpar}$, and 
$\chi_u=0$ and $\chi_b=1$.
It is also possible to define the parallel and perpendicular sound speeds $C_\parallel=\sqrt{3p_\parallel^{(0)}/\rho_0}$;
$C_\perp=\sqrt{2p_\perp^{(0)}/\rho_0}$, which yields the parallel acoustic mode dispersion $\omega=\pm k_\parallel C_\parallel$ and the perpendicular fast mode
dispersion $\omega=\pm k_\perp \sqrt{V_A^2+C_\perp^2}$.
\footnote{Sometimes the parallel and perpendicular sounds speeds are defined as $C_\parallel=\sqrt{p_\parallel^{(0)}/\rho_0}$;
$C_\perp=\sqrt{p_\perp^{(0)}/\rho_0}$, which yields the parallel acoustic mode dispersion $\omega=\pm k_\parallel \sqrt{3}C_\parallel$ and the perpendicular fast mode
dispersion $\omega=\pm k_\perp \sqrt{V_A^2+2C_\perp^2}$ }

In the general case of oblique propagation, we choose to use the normalized frequencies and wavenumbers (\ref{eq:NormKO}),
and rewrite the dispersion relation as
\begin{eqnarray} 
  && \widetilde{\omega}^4-A_2\widetilde{\omega}^2+A_0 = 0; \label{eq:SF}\\
&&   A_2 = \widetilde{k}_\perp^2 \left(1+a_p\bpar\right) + \widetilde{k}_\parallel^2 \left(1+\bpar+\frac{1}{2}a_p\bpar\right); \nonumber \\
&& A_0 = \frac{3}{2}\widetilde{k}_\parallel^2 \bpar \left[ \widetilde{k}_\parallel^2 \left(1-\frac{1}{2}\bpar +\frac{1}{2}a_p\bpar\right) + 
\widetilde{k}_\perp^2 \left(1+a_p\bpar-\frac{1}{6}a_p^2\bpar\right)  \right].\nonumber
\end{eqnarray}
It is sometimes useful to introduce the parallel Alfv\'en phase speed $v_{A\parallel}$, and its normalization $\widetilde{v}_{A\parallel} \equiv v_{A\parallel}/V_A$,
so that
\begin{equation}
v_{A\parallel} = V_A \sqrt{1+\frac{\bpar}{2}(a_p-1)}; \qquad \widetilde{v}_{A\parallel}^2 = 1+\frac{\bpar}{2}(a_p-1), 
\end{equation}
since $\widetilde{v}_{A\parallel}^2$ can be used to write the CGL dispersion relation in a shorter form, as done later in (\ref{eq:SFbigmatrix}).
The solution of the polynomial (\ref{eq:SF}) is simply $\widetilde{\omega}^2 = (A_2\pm \sqrt{A_2^2-4A_0})/2$, and
it is obvious that the coefficient $A_2$ is always
positive, since both the plasma beta and the temperature anisotropy ratio must be always positive.
It is also possible to show that the discriminant under the square root $A_2^2-4A_0 \ge 0$ (proof shown few lines below).
The solution with the minus sign is called
the slow mode and the solution with the plus sign is called the fast mode. For the slow mode, $\widetilde{\omega}^2$ becomes negative
(and $\widetilde{\omega}$ complex) when $A_0 <0$, implying that the slow mode becomes unstable when
\begin{equation} \label{eq:SFunstable}
\left[ \kpar^2 \left(1-\frac{1}{2}\bpar +\frac{1}{2}a_p\bpar\right) + 
\kperp^2 \left(1+a_p\bpar-\frac{1}{6}a_p^2\bpar\right)  \right] < 0,
\end{equation}
where we suppress the tildes on the wavenumbers since the above condition is valid for both $\widetilde{k}$ and $k$ forms of wavenumbers.
The above condition implies that there are two competing instabilities. The highly parallel propagation with $k_{\parallel}\gg k_\perp$ yields
instability threshold equivalent to (\ref{eq:firehose}) and the associated instability is called the
\emph{parallel firehose instability}. The highly oblique propagation with $k_\perp\gg k_\parallel$ yields the \emph{mirror instability}
and the threshold in the CGL description reads
\footnote{To obtain the mirror instability threshold, it is not possible to consider strictly perpendicular propagation with $k_\parallel=0$
 because the coefficient $A_0$ would be exactly zero. The mirror threshold is revealed only in the highly oblique limit $k_\perp\gg k_\parallel$.}   
\begin{equation} \label{eq:CGLmirror}
1+a_p\bpar-\frac{1}{6}a_p^2\bpar <0; \qquad \frac{1}{6}\frac{T_\perp^{(0)}}{T_\parallel^{(0)}}-1-\frac{1}{\beta_\perp} >0; \qquad
\frac{1}{6}\frac{\beta_\perp^2}{\bpar}>1+\beta_\perp.
\end{equation}
It is well documented that the mirror instability threshold in the CGL description is not equivalent to the one found in linear kinetic theory. In particular,
the CGL mirror threshold contains a factor of 6 error \citep{Abraham-Shrauner1967,Tajiri1967,Kulsrud1983,FerriereAndre2002},
since the correct threshold obtained from kinetic theory reads 
\begin{equation} \label{eq:KineticMirror}
1+a_p\bpar-a_p^2\bpar<0; \qquad \frac{T_\perp^{(0)}}{T_\parallel^{(0)}}-1-\frac{1}{\beta_\perp} >0; \qquad \frac{\beta_\perp^2}{\bpar}>1+\beta_\perp.
\end{equation}
The threshold (\ref{eq:CGLmirror}) implies that the necessary condition for the mirror instability to exist in the CGL description is
$a_p=T_\perp^{(0)}/T_\parallel^{(0)}>6$, whereas in kinetic theory the necessary condition is $a_p>1$.
 A very good discussion about the physical mechanism of the mirror instability and the role of resonant particles can be found in \cite{Southwood1993,Kivelson1996}.
  The correct mirror instability threshold is obtained when using the normal closure discussed later on in the text or, more simply,
  its quasi-static limit \citep{Constantinescu2002,ChustBelmont2006,PassotRubanSulem2006}. In the opposite case where ions are cold,
  a similar electron temperature anisotropy instability is obtained which is called the field-swelling instability, see e.g. \cite{BasuCoppi1984}. 

Returning to the oblique dispersion relation, we show that indeed $A_2^2-4A_0 \ge 0$. By using (\ref{eq:ks}) the wavenumber $k$ can be pulled out
of the $A_0,A_2$ coefficients and the dispersion relation can be rewritten in terms of a phase speed as
\begin{eqnarray}
&&  (\widetilde{\omega}/\widetilde{k})^4-A_2 (\widetilde{\omega}/\widetilde{k})^2+A_0 = 0; \label{eq:sfGen}\\
&&  A_2 = 1+a_p\bpar\big(1-\frac{1}{2}\cos^2\theta) + \bpar\cos^2\theta; \\
&&  A_0 = \frac{3}{2}\bpar\cos^2\theta \Big[ 1+a_p\bpar\big(1-\frac{1}{2}\cos^2\theta) -\frac{1}{2}\bpar\cos^2\theta -\frac{1}{6}a_p^2\bpar\sin^2\theta \Big],
\end{eqnarray}
and (after staring for a sufficiently long time) it can be shown that 
\begin{equation} \label{eq:miracle}
A_2^2-4A_0 = \Big[ 1+a_p\bpar\big(1-\frac{1}{2}\cos^2\theta) -2\bpar\cos^2\theta \Big]^2 +a_p^2 \bpar^2 \sin^2\theta\cos^2\theta,
\end{equation}  
which is obviously always non-negative. 
The dispersion relation for the slow and fast waves in the CGL description can be written in the compact form
\begin{eqnarray} 
  \left(\frac{\widetilde{\omega}}{\widetilde{k}}\right)_{sf}^2 =
  \left( \frac{\omega}{kV_A}\right)_{sf}^2 &=& \frac{1}{2}\Big[1+a_p\bpar\big(1-\frac{1}{2}\cos^2\theta) + \bpar\cos^2\theta \Big] \nn\\
  && \pm \frac{1}{2}\sqrt{ \Big[ 1+a_p\bpar\big(1-\frac{1}{2}\cos^2\theta) -2\bpar\cos^2\theta \Big]^2 +a_p^2 \bpar^2 \sin^2\theta\cos^2\theta },
  \label{eq:CGLSF}
\end{eqnarray}
which is equivalent to dispersion relation of \cite{Abraham-Shrauner1967}, eq. 30, here written in the convenient units of
$\bpar$ and $a_p=T_\perp^{(0)}/T_\parallel^{(0)}$ (for a direct comparison with units used in that paper,
the relations are $\bpar=2S_\parallel$ and $a_p\bpar=2S_\perp$). The magnetic compressibility of the slow and fast modes can be calculated
easily directly from the induction equations (\ref{eq:CGL-Bxz}). The slow and fast modes generate only components $B_x$ and $B_z$ with
the component $B_y=0$. The transformation of (\ref{eq:CGL-Bxz}) to Fourier space yields $-\omega B_x = B_0 k_\parallel u_x$, and $-\omega B_z = -B_0 k_\perp u_x$,
and by applying the magnitude operator yields the magnetic compressibility
\begin{equation}
\chi_b (k_\parallel,k_\perp) = \frac{k_\perp^2}{k_\parallel^2+k_\perp^2} = \sin^2\theta.
\end{equation}
For quasi-parallel propagation angles the magnetic compressibility $\chi_b$ is therefore not a good indicator for mode recognition, since
all three modes have $\chi_b$ close to zero. The magnetic compressibility is however an excellent tool  for oblique propagation angles, 
because the Alfv\'en mode has $\chi_b=0$ regardless of the propagation direction, whereas for the slow and fast mode $\chi_b$ increases
with $\theta$ towards $\chi_b=1$. We will leave discussion about the CGL velocity eigenvector (the ratio $\chi_u$) to a later subsection since the discussion
is a bit technical, and we instead focus now on an interesting effect that can be directly obtained by considering the CGL dispersion relations.
\begin{figure*}
$$\includegraphics[width=0.6\linewidth]{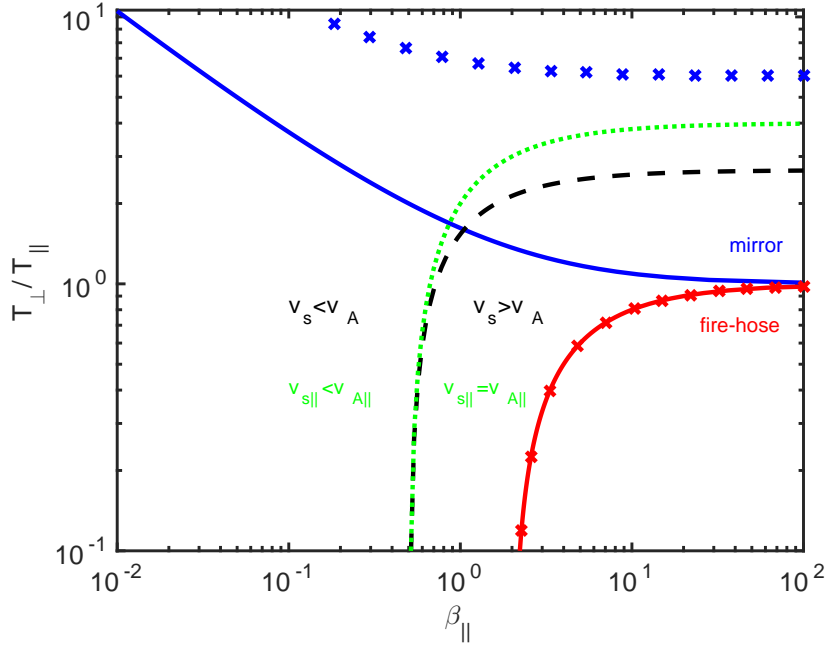}$$
  \caption{Analytic ``hard'' thresholds of the mirror instability (blue) and the firehose instability (red) at long wavelengths.
  Solid lines are solutions of the linear kinetic theory (bi-Maxwellian) and 
  crosses are solutions of the linear CGL model. The firehose instability is described correctly (both the oblique firehose and the parallel firehose),
  whereas the mirror instability contains an asymptotic factor of 6 error for large $\bpar$ values. The black dashed line is the CGL threshold (\ref{eq:SFthresh2}), where 
  for all oblique directions the slow mode speed $v_s$ matches the Alfv\'en mode speed $v_A$, and which separates two regions where $v_s<v_A$ and $v_s>v_A$.
  The green dotted curve is a special case for parallel propagation (\ref{eq:3point}) where $v_{s\parallel}=v_{A\parallel}$. For parallel propagation the slow mode
  speed cannot exceed the Alfv\'en mode speed, but can only match it, and the green dotted curve separates two regions
  where $v_{s\parallel}<v_{A\parallel}$ and $v_{s\parallel}=v_{A\parallel}$.  For a similar plot in linear scale, see Figure 1 of \cite{Abraham-Shrauner1967}.} \label{fig:CGL-th}
\end{figure*}
\subsection{Slow mode can become faster than Alfv\'en mode} \label{sec:SAreversal}
There is a very interesting phenomenon present in the CGL description that is also observed in the kinetic description but that is
absent in the MHD description. For a sufficiently high plasma beta, the slow mode phase speed can be larger than the Alfv\'en mode phase speed.
To obtain the critical plasma beta when this happens, one can write the dispersion relation for the Alfv\'en wave as
$(\widetilde{\omega}/\widetilde{k})_{A}^2 = (1-\bpar/2+a_p\bpar/2 ) \cos^2\theta \equiv A_A$ and  
using this expression with the dispersion relation (\ref{eq:sfGen}), the critical condition when the slow mode speed
matches the Alfv\'en speed is $(A_2-\sqrt{A_2^2-4A_0})/2 = A_A$, and since $A_2^2-4A_0\ge 0$, this further
yields $A_A^2-A_2A_A+A_0= 0$.
It takes quite a few algebraic operations to simplify the result, the easiest way is to collect terms with different powers of $\bpar$ and then simplify.
A very valuable approach available now is to use  mathematical software such as Maple or Mathematica for guidance,
and for example the Maple command simplify(expr,trig) immediately simplifies the result, which can then be of course
verified with pencil and paper. The result is
\begin{eqnarray} \label{eq:SFthresh}
A_A^2-A_2A_A+A_0 = -\sin^2\theta\cos^2\theta \Big[ 1+\bpar\Big(\frac{3}{2}a_p-2\Big) +\bpar^2a_p\Big(\frac{3}{4}a_p-2\Big) \Big] = 0.
\end{eqnarray}
The expression is naturally satisfied for cases of strictly parallel and strictly perpendicular propagation.
For $\theta=\pi/2$, both the slow mode and the Alfv\'en mode have zero frequency. For $\theta=0$,
the slow mode speed can be only smaller or equal to the speed of the Alfv\'en mode (never higher), depending on another condition that will be
easier to clarify later when specifically considering the parallel propagation limit.
Now, on which side of the threshold (\ref{eq:SFthresh}) does the speed of the slow mode become faster than the speed of the Alfv\'en mode? 
Working with inequalities is a bit trickier. By assuming that both modes have real frequencies (and are not firehose or mirror unstable),
the condition is derived to be
\begin{eqnarray} \label{eq:SFthresh3}
-\sin^2\theta\cos^2\theta \Big[ 1+\bpar\Big(\frac{3}{2}a_p-2\Big) +\bpar^2a_p\Big(\frac{3}{4}a_p-2\Big) \Big] \ge 0,
\end{eqnarray}
which agrees with the result obtained by Abraham-Shrauner 1967, eq. 42.
The expression (\ref{eq:SFthresh3}) reveals that there is a condition that can be satisfied for all directions of propagation and that reads
\begin{eqnarray} \label{eq:SFthresh2}
 1+\bpar\Big(\frac{3}{2}a_p-2\Big) +\bpar^2a_p\Big(\frac{3}{4}a_p-2\Big) \le 0.
\end{eqnarray}
This is a quadratic equation in $\bpar$ which can be solved easily for a given $a_p$ (or solved for $a_p$ for a given $\bpar$).
To visualize this equation, the threshold is plotted in Figure \ref{fig:CGL-th} as a black dashed line. 
A particularly useful result is obtained for isotropic temperatures ($a_p=1$). In this case the inequality (\ref{eq:SFthresh2})
becomes $1-\bpar/2 -\bpar^2 5/4 \le 0$ and the critical plasma beta at which the slow mode speed becomes larger
than the Alfv\'en mode speed is
\begin{equation}
\bpar^{\textrm{crit}} \ge \frac{\sqrt{21}-1}{5} \approx 0.72. 
\end{equation}
Thus, this effect exists in the CGL description even when the temperatures are isotropic. 
\subsection{CGL parallel propagation}
When numerically solving the CGL dispersion relations for finite (nonzero) propagation angles,
the slow and fast mode expressions always get the correct ordering with $\omega_s\le \omega_f$.
It is sometimes confusing about how the dispersion relations hold in the limit of parallel propagation $\theta\rightarrow 0$
which we will address here. In this limit, the dispersion relations (\ref{eq:CGLSF}) for the slow and fast modes can be directly evaluated as
\begin{eqnarray} 
  \left(\frac{\widetilde{\omega}}{\widetilde{k}}\right)_{sf}^2 &=& \frac{1}{2}\Big[1+\frac{1}{2}a_p\bpar + \bpar \Big]
  \pm \frac{1}{2}\sqrt{ \Big[ 1+\frac{1}{2}a_p\bpar -2\bpar\Big]^2 }.
\end{eqnarray}
Now, because $\sqrt{a^2}=|a|$, for further calculation one needs to
determine the sign of the expression  under the square root. For $1+\frac{1}{2}a_p\bpar -2\bpar >0$, i.e. when $\bpar<2/(4-a_p)$,
the above equations yield
\begin{eqnarray} \label{eq:CGLparD1}
  (\widetilde{\omega}/\widetilde{k})_{s}^2 = \frac{3}{2}\bpar;
  \qquad (\widetilde{\omega}/\widetilde{k})_{f}^2 = 1+\frac{\bpar}{2}(a_p-1)=(\widetilde{\omega}/\widetilde{k})_{A}^2,
\end{eqnarray}
so the slow mode is the sound/acoustic mode and the fast mode coincides with the Alfv\'en mode. However, for $1+\frac{1}{2}a_p\bpar -2\bpar <0$,
i.e. when $\bpar>2/(4-a_p)$, the dispersion relations yield
\begin{eqnarray} \label{eq:CGLparD2}
  (\widetilde{\omega}/\widetilde{k})_{s}^2 = 1+\frac{\bpar}{2}(a_p-1)=(\widetilde{\omega}/\widetilde{k})_{A}^2; \qquad
  (\widetilde{\omega}/\widetilde{k})_{f}^2 = \frac{3}{2}\bpar,
\end{eqnarray}
so the slow mode coincides with the Alfv\'en mode and the fast mode is the sound/acoustic mode. In this region, the slow mode phase speed will always stay
equal to the Alfv\'en mode phase speed, regardless of further increases of $\bpar$. In the special case when
$1+\frac{1}{2}a_p\bpar -2\bpar=0$, which is equivalent to
\begin{equation}
 \qquad \bpar=\frac{2}{4-a_p}; \qquad a_p=4-\frac{2}{\bpar}, \label{eq:3point}
\end{equation}
all three modes propagate with the same phase speed
$(\widetilde{\omega}/\widetilde{k})_{Asf}^2 = \frac{3}{2}\bpar$. The criterion (\ref{eq:3point}) can be satisfied
only for $\bpar\ge 1/2$ and in the limit $\bpar\rightarrow \infty$ the asymptote is $a_p=4$, so the possible range of $a_p$ is $[0,4]$.
This curve is plotted in Figure \ref{fig:CGL-th} as a green dotted line. The curve separates two regions. In the first region $\bpar<2/(4-a_p)$ and $v_{s\parallel}<v_{A\parallel}$.
In the second region $\bpar>2/(4-a_p)$ and $v_{s\parallel}=v_{A\parallel}$. The interesting result to remember is that
for isotropic temperatures ($a_p=1$) and parallel propagation, the critical plasma beta at which the slow mode speed matches the Alfv\'en mode speed is
\begin{equation}
\bpar^{\textrm{crit}}(0) \ge \frac{2}{3}.
\end{equation}  
\subsection{CGL with $T_\perp=T_\parallel$ is not equivalent to MHD}
It is important to emphasize that even for isotropic temperatures $a_p=1$, the CGL dispersion (\ref{eq:CGLSF}) is not equivalent to the MHD dispersion
for the slow and fast waves. 
The MHD dispersion is usually written in the form (see (\ref{MHDappendix})  in the Appendix \ref{sec:AppendixC})
\begin{eqnarray} \label{eq:MHD1}
\left(\frac{\omega}{k}\right)^2 = \frac{1}{2}(V_A^2 + C_s^2) \pm \frac{1}{2}\sqrt{(V_A^2 + C_s^2)^2 - 4 V_A^2 C_s^2\cos^2\theta}, 
\end{eqnarray}
or alternatively
\begin{eqnarray} \label{eq:MHD2}
\left(\frac{\omega}{k}\right)^2 = \frac{1}{2}(V_A^2 + C_s^2) \pm \frac{1}{2}\sqrt{(V_A^2 - C_s^2)^2 + 4 V_A^2 C_s^2\sin^2\theta}, 
\end{eqnarray}
which nicely shows that the expression under the square root is always greater or equal to zero. 
The MHD sound speed is defined as $C_s^2=\gamma p^{(0)}/\rho_0$, where $\gamma=5/3$. Rewritten with parameters we use here $C_s^2/V_A^2=\gamma\bpar/2 = 5\bpar/6$, the
MHD dispersion for the slow and fast waves reads
\begin{eqnarray} 
\left(\frac{\omega}{kV_A}\right)_{MHD}^2 = \frac{1}{2}(1 + \frac{\gamma}{2}\bpar) \pm \frac{1}{2}\sqrt{(1 + \frac{\gamma}{2}\bpar)^2 - 2\gamma\bpar\cos^2\theta}, 
\end{eqnarray}
or alternatively
\begin{eqnarray}
\left(\frac{\omega}{kV_A}\right)_{MHD}^2 = \frac{1}{2}(1 + \frac{\gamma}{2}\bpar) \pm \frac{1}{2}\sqrt{(1 - \frac{\gamma}{2}\bpar)^2 + 2\gamma\bpar\sin^2\theta}. 
\end{eqnarray}
In the CGL description with isotropic temperatures $(a_p=1)$ the dispersion relation (\ref{eq:CGLSF}) for slow and fast waves reads  
\begin{eqnarray} 
  \left( \frac{\omega}{kV_A}\right)^2_{CGL} &=& \frac{1}{2}\Big[1+\bpar\big(1+\frac{1}{2}\cos^2\theta) \Big] 
  \pm \frac{1}{2}\sqrt{ \Big[ 1+\bpar\big(1-\frac{5}{2}\cos^2\theta) \Big]^2 +\bpar^2 \sin^2\theta\cos^2\theta },
 \end{eqnarray}
or alternatively
\begin{eqnarray} 
  \left( \frac{\omega}{kV_A}\right)^2_{CGL} &=& \frac{1}{2}\Big[1+\bpar\big(1+\frac{1}{2}\cos^2\theta) \Big] 
  \pm \frac{1}{2}\sqrt{ \Big[ 1+\bpar\big(1+\frac{1}{2}\cos^2\theta)\Big]^2 -6\bpar\cos^2\theta\Big(1+\frac{5}{6}\bpar\sin^2\theta\Big)}.
\end{eqnarray}
We wrote down all major possibilities on purpose, since we wanted to clearly show that regardless of the choice of $\bpar$ and $\gamma$,
the MHD dispersion relation can not match the CGL dispersion relation simultaneously for all propagation directions $\theta$. One could even play with an abstract idea to
make the MHD $\gamma$ dependent on angle $\theta$, and possibly $\bpar$. Even then, there is no obvious way to modify the MHD dispersion relation so that
it behaves like the CGL model.  Additionally, the CGL description leads to the development of pressure anisotropy even when the initial pressure is isotropic.

\subsection{MHD velocity eigenvector}
\begin{figure*}[t]
  $$\includegraphics[width=0.48\linewidth]{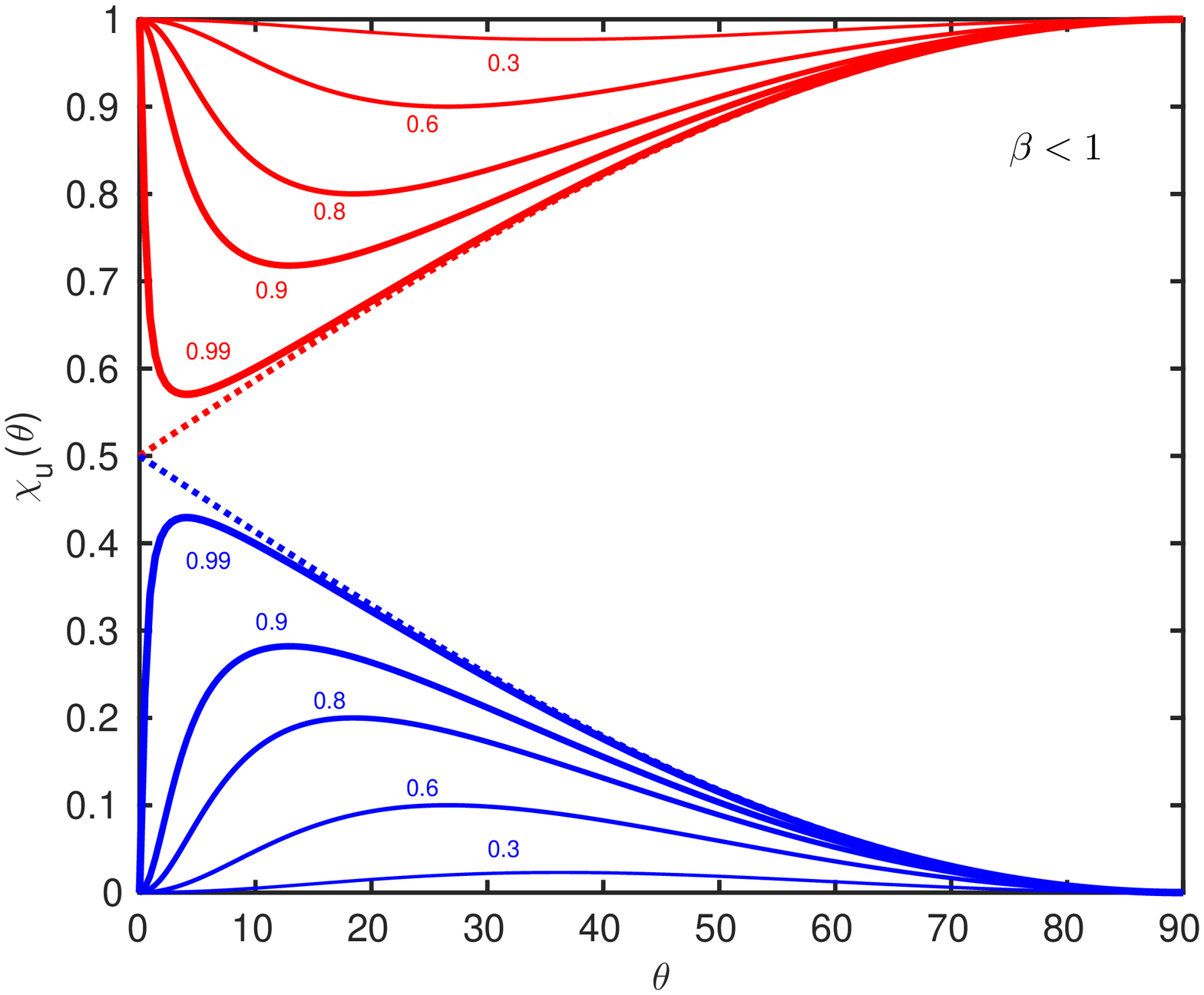}
  \hspace{0.02\textwidth}\includegraphics[width=0.48\linewidth]{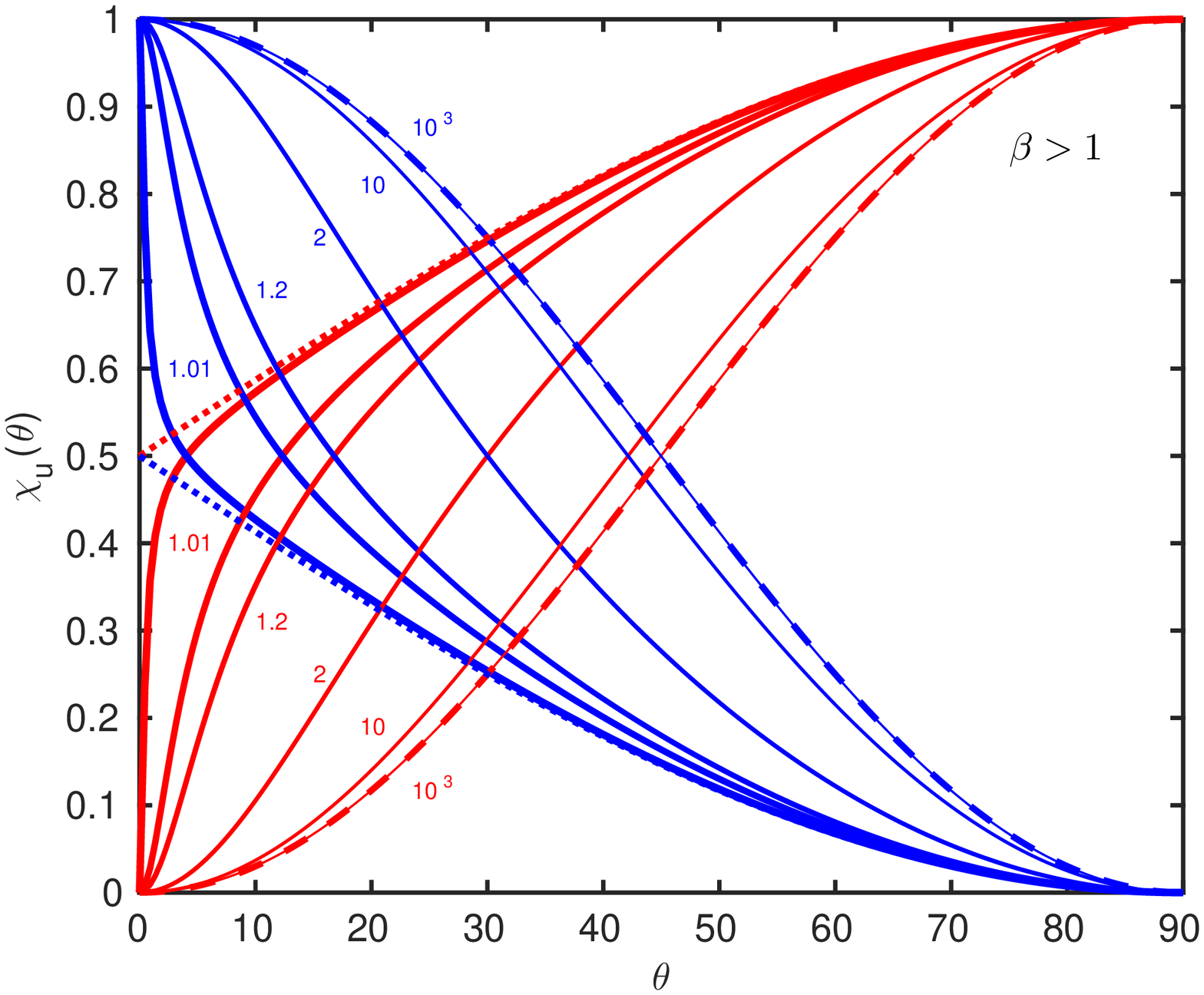}$$
\caption{MHD velocity ratio $\chi_u(\theta)=|u_\parallel|^2/|u|^2$ for the slow mode (red lines) and the fast mode (blue lines) for different ratios
of $C_s^2/V_A^2=\beta_{MHD}$. The left figure is for $\beta_{MHD}<1$, and the thickness of the lines increases with $\beta_{MHD}$ when approaching the critical
value of $\beta_{MHD}=1$. The plotted lines have $\beta_{MHD}=0.3; 0.6; 0.8; 0.9; 0.99$. The right figure
is for $\beta_{MHD}>1$ and the thickness of the lines decreases with $\beta_{MHD}$ when going away from the critical value of $\beta_{MHD}=1$.
The plotted lines have $\beta_{MHD}=1.01; 1.1; 1.2; 2.0; 10; 10^3$. In both figures, the dotted lines are the critical case $\beta_{MHD}=1$
with solutions (\ref{eq:MHDspecial}). In the right figure the dashed lines represent the limit $\beta_{MHD}\gg 1$ with solutions
(\ref{eq:MHDlarge}) and the dashed lines nicely overlap with a thin solid line solutions obtained for $\beta_{MHD}=10^3$.
The Alfv\'en mode has $\chi_u(\theta)=0$ regardless of plasma beta and is not plotted.} \label{fig:MHD}
\end{figure*}

Before we proceed with the discussion of the CGL velocity eigenvector (which can appear to be a bit tricky),
for the sake of clarity it is useful to consider the simpler MHD case first.  
By following the steps outlined above, it is easy to show that again the Alfv\'en mode dynamics in the $u_y$ decouples from the dynamics in the other directions and the
velocity eigenvector matrix for the slow and fast waves in MHD is
\begin{eqnarray}
\left( \begin{array}{cc}
\omega^2-(V_A^2 k^2 +C_s^2k_\perp^2) & \qquad - C_s^2 k_\parallel k_\perp\\
    - C_s^2 k_\parallel k_\perp & \qquad \omega^2 - C_s^2 k_\parallel^2
\end{array} \right)
\left( \begin{array}{c} u_x \\ u_z \end{array} \right) = \left( \begin{array}{c} 0 \\ 0 \end{array} \right),
\end{eqnarray}
or alternatively
\begin{eqnarray}
\left( \begin{array}{cc}
(\omega/k)^2-(V_A^2 +C_s^2\sin^2\theta) & \qquad - C_s^2 \sin\theta\cos\theta\\
    - C_s^2 \sin\theta\cos\theta & \qquad (\omega/k)^2 - C_s^2 \cos^2\theta
\end{array} \right)
\left( \begin{array}{c} u_x \\ u_z \end{array} \right) = \left( \begin{array}{c} 0 \\ 0 \end{array} \right). \label{eq:MHDuMatrix}
\end{eqnarray}
This yields the MHD dispersion relations for the slow and fast mode in the form of (\ref{eq:MHD1}) or alternatively (\ref{eq:MHD2}).
It possible to either express the ratio $u_x/u_z$ or $u_z/u_x$. We want to calculate the velocity ratio $\chi_u$ and we choose the first possibility
that allows us to express that for the slow and fast modes (since $u_y=0$)
\begin{eqnarray}
  \chi_u (k_\parallel,k_\perp) &=& \frac{|u_z|^2}{|u_x|^2+|u_z|^2} = \frac{1}{|\frac{u_x}{u_z}|^2+1}.
\end{eqnarray}
The ratio $u_x/u_z$ can be found for example from the second row of the matrix as
\begin{equation} \label{eq:uxuzMHD}
\frac{u_x}{u_z} = \frac{(\omega/k)^2 - C_s^2 \cos^2\theta}{C_s^2\sin\theta\cos\theta},
\end{equation}
or equivalently from the first row of the matrix as
\begin{equation} \label{eq:uxuzMHD2}
\frac{u_x}{u_z} = \frac{C_s^2\sin\theta\cos\theta}{(\omega/k)^2-(V_A^2 +C_s^2\sin^2\theta)}.
\end{equation}  
The possible range of the ratio $u_x/u_z$ is from $-\infty$ to $+\infty$ and the possible range for $\chi_u$ is from $0$ to $1$.  
Solutions can be easily divided into two categories - depending on the behavior for parallel propagation - and are plotted in Figure \ref{fig:MHD}.
For the sake of clarity, we only consider propagation with positive wavenumbers $k_\parallel$ and $k_\perp$, so the possible range
of angle $\theta$ is from $0$ to $\pi/2$ and $\sin\theta\ge 0$, $\cos\theta\ge 0$.
Consider first the case $C_s<V_A$. For parallel propagation, the dispersion relation (\ref{eq:MHD2}) directly yields that
for the fast mode $(\omega/k)^2_f |_{\theta=0} = V_A^2$ and for the slow mode $(\omega/k)^2_s|_{\theta=0} = C_s^2$.
Using the expression (\ref{eq:uxuzMHD}) then yields that for the fast mode $(u_x/u_z)_f |_{\theta=0} =+\infty$
and for the slow mode $(u_x/u_z)_s|_{\theta=0}= 0$. This implies that for $C_s<V_A$, all the $\chi_u(\theta)$ curves
start at
\begin{equation}
C_s<V_A: \qquad  \chi_u^f(\theta=0)=0; \qquad \chi_u^s(\theta=0)=1,
\end{equation}
regardless of the actual value of $C_s/V_A<1$.
The second case $C_s>V_A$ is straightforward, since for parallel propagation everything is just turned around;
the fast mode dispersion relation is $(\omega/k)^2_f |_{\theta=0} = C_s^2$ and the slow mode dispersion relation is $(\omega/k)^2_s|_{\theta=0} = V_A^2$. Implying that
for $C_s>V_A$, all the $\chi_u(\theta)$ curves start at
\begin{equation}
C_s>V_A:\qquad  \chi_u^f(\theta=0)=1; \qquad \chi_u^s(\theta=0)=0,
\end{equation}  
regardless of the actual value of $C_s/V_A>1$. The perpendicular propagation limit is even easier since the split into two categories is not
required and for both cases the dispersion (\ref{eq:MHD1}) yields  $(\omega/k)^2_f |_{\theta=\pi/2} = V_A^2+C_s^2$ and $(\omega/k)^2_s |_{\theta=\pi/2} =0$,
implying that for the fast mode $(u_x/u_z)_f|_{\theta=\pi/2}=+\infty$ and that for the slow mode $(u_x/u_z)_s|_{\theta=\pi/2}=0$. All the $\chi_u(\theta)$ 
curves therefore ``end'' at
\begin{equation}
  \chi_u^f(\theta=\pi/2)=0; \qquad \chi_u^s(\theta=\pi/2)=1,
\end{equation}
whether the ratio $C_s/V_A<1$ or $C_s/V_A>1$. The behavior of the $\chi_u(\theta)$ curves is plotted in Figure \ref{fig:MHD}.

In MHD, the ratio $C_s^2/V_A^2$ is sometimes
used as the definition of plasma beta, which we will call here $\beta_{MHD}\equiv C_s^2/V_A^2$. As can be seen in Figure \ref{fig:MHD}, for $\beta_{MHD}$
approaching 1, the curves display a ``singular'' behavior, and it is difficult to guess what will happen when $\beta_{MHD}$
reaches the value of 1 exactly. This special case is addressed in the next subsection. There are also two special cases,
one for $C_s\ll V_A$ or $\beta_{MHD}\ll 1$ and one for $C_s\gg V_A$ or $\beta_{MHD}\gg 1$.
Both limits $C_s\ll V_A$ and $C_s\gg V_A$ are easily obtained by rewriting the dispersion relation (\ref{eq:MHD1}) to a form
\begin{equation}
 \left(\frac{\omega}{k}\right)^2 =  \frac{1}{2}(V_A^2+C_s^2)\pm \frac{1}{2}(V_A^2+C_s^2) \sqrt{1-\frac{4V_A^2C_s^2}{(V_A^2+C_s^2)^2}\cos^2\theta}. 
\end{equation}
In both limits the second term under the square root is small and the expression can be expanded as $\sqrt{1-\epsilon}=1-\epsilon/2$,
yielding the slow and fast mode dispersion relations
\begin{equation}
  C_s\ll V_A \; \textrm{or}\; C_s\gg V_A: \qquad \left(\frac{\omega}{k}\right)_f^2 = V_A^2+C_s^2 - \frac{V_A^2 C_s^2}{V_A^2+C_s^2}\cos^2\theta;
  \qquad \left(\frac{\omega}{k}\right)_s^2 = \frac{V_A^2 C_s^2}{V_A^2+C_s^2}\cos^2\theta,
\end{equation}  
which further yields dispersion relations (\ref{eq:MHDsmallBeta}), (\ref{eq:MHDlargeBeta}).
Note that the limit $C_s\ll V_A$ can be easily obtained directly from the velocity matrix (\ref{eq:MHDuMatrix}) by neglecting
the off-diagonal terms that decouples the two modes. In contrast, the limit $C_s\gg V_A$ is not that obvious and one needs to 
calculate the determinant. 

In the limit $\beta_{MHD}\ll 1$, the MHD dispersion relation simplify to
\begin{equation}
\beta_{MHD}\ll 1:\qquad (\omega/k)_f^2 = V_A^2+C_s^2\sin^2\theta; \qquad (\omega/k)_s^2=C_s^2\cos^2\theta. \label{eq:MHDsmallBeta}
\end{equation}  
The expression (\ref{eq:uxuzMHD}) yields for the fast mode $(u_x/u_z)_f(\theta)=\infty$ and the expression (\ref{eq:uxuzMHD2}) yields
for the slow mode $(u_x/u_z)_s(\theta)=0$, implying
\begin{equation}
\beta_{MHD}\ll 1:\qquad \chi_u^f(\theta)=0;\qquad \chi_u^s(\theta)=1.
\end{equation}  
In the opposite limit $\beta_{MHD}\gg 1$, the MHD dispersions simplify to
\begin{equation}
\beta_{MHD}\gg 1: \qquad (\omega/k)_f^2 = C_s^2+V_A^2\sin^2\theta; \qquad (\omega/k)_s^2=V_A^2\cos^2\theta. \label{eq:MHDlargeBeta}
\end{equation}
In this limit, the expression (\ref{eq:uxuzMHD}) yields $(u_x/u_z)_f(\theta)=(1+V_A^2/C_s^2)\tan\theta$, that describes the behavior of the fast mode
for large beta values and the complete limit is $(u_x/u_z)_f(\theta) = \tan\theta$. For the slow mode the expression (\ref{eq:uxuzMHD}) yields
$(u_x/u_z)_s(\theta)=-(1-V_A^2/C_s^2)\cot\theta$ and the complete limit is $(u_x/u_z)_s(\theta)=-\cot\theta$. The $\chi_u(\theta)$ solutions for large
beta values therefore converge to
\begin{equation} \label{eq:MHDlarge}
\beta_{MHD}\gg 1: \qquad \chi_u^f(\theta)=\cos^2\theta; \qquad \chi_u^s(\theta) = \sin^2\theta,
\end{equation}  
and these two curves are plotted in Figure \ref{fig:MHD} as dashed lines.
\subsubsection{MHD special case $V_A=C_s$}
Consider the very peculiar case of parallel propagation $\theta=0$ with $V_A=C_s$. The velocity eigenvector matrix (\ref{eq:MHDuMatrix}), when evaluated exactly
at these values reads
\begin{eqnarray}
\left( \begin{array}{cc}
0 & \qquad 0\\
    0 & \qquad 0
\end{array} \right)
\left( \begin{array}{c} u_x \\ u_z \end{array} \right) = \left( \begin{array}{c} 0 \\ 0 \end{array} \right). \label{eq:MHDu0}
\end{eqnarray}
It is not possible to determine the ratio $u_x/u_z$ and the eigenvector must be found by taking the limit $\theta\rightarrow 0$.
For $V_A=C_s$ the dispersion relation for the slow and fast modes (\ref{eq:MHD2}) simplifies to $(\omega/k)^2=V_A^2\pm V_A^2|\sin\theta|$.
As before, for the sake of simplicity we consider only positive wavenumbers, so the propagation angle $\theta$ is in the range $[0,\pi/2]$ and
$\sin\theta\ge 0$, $\cos\theta\ge 0$. Evaluating the velocity matrix (\ref{eq:MHDuMatrix}) at $V_A=C_s$ only,
it is possible to pull the $\sin\theta$ out of the system so that
\begin{eqnarray}
\left( \begin{array}{cc}
\pm 1-\sin\theta & \qquad -\cos\theta\\
    -\cos\theta & \qquad \pm 1 +\sin\theta
\end{array} \right) V_A^2\sin\theta
\left( \begin{array}{c} u_x \\ u_z \end{array} \right) = \left( \begin{array}{c} 0 \\ 0 \end{array} \right),
\end{eqnarray}
which nicely demonstrates that for $\theta\rightarrow 0$ the system approaches (\ref{eq:MHDu0}). 
Nevertheless, we can now calculate the velocity ratio (for example from the second row) as
\begin{equation} \label{eq:MHDu}
\frac{u_x}{u_z} = \frac{\pm 1+\sin\theta}{\cos\theta},
\end{equation}
which can be evaluated at $\theta=0$. For the fast mode $(u_x/u_z)^f(0)=1$ and for the slow mode $(u_x/u_z)^s(0)=-1$.
This implies that for the peculiar case of $V_A=C_s$, the velocity ratio $\chi_u^{sf}(0)=1/2$ for both the fast and slow modes.
The expression (\ref{eq:MHDu}) can of course be directly obtained by using the result $(\omega/k)^2=V_A^2\pm V_A^2\sin\theta$ in the expression
(\ref{eq:uxuzMHD}) and by canceling the $\sin{\theta}$ in the nominator and denominator.  
The expression (\ref{eq:MHDu}) is valid for any angle $\theta$. However, for the slow mode it becomes singular at $\theta=\pi/2$,
and has to be rearranged by multiplying and dividing with $1+\sin\theta$
(or equivalently, use the first row of the matrix). To have useful expression for the entire range of $\theta$,
we therefore split the expressions for the slow and fast modes as
\begin{equation}
(u_x/u_z)^s = -\frac{\cos\theta}{1+\sin\theta}; \qquad (u_x/u_z)^f = \frac{1+\sin\theta}{\cos\theta},
\end{equation}
which yields in this peculiar case of $V_A=C_s$, that the ratio $\chi_u$ for the slow and fast modes can be expressed as
\begin{equation} \label{eq:MHDspecial}
V_A=C_s: \qquad \chi_u^s (\theta)= \frac{1}{2}(1+\sin\theta); \qquad \chi_u^f (\theta)= \frac{1}{2}(1-\sin\theta).
\end{equation}
These solutions are plotted in Figure \ref{fig:MHD} as dotted lines.
\subsection{CGL velocity eigenvector}
\begin{figure*}
$$\includegraphics[width=0.48\linewidth]{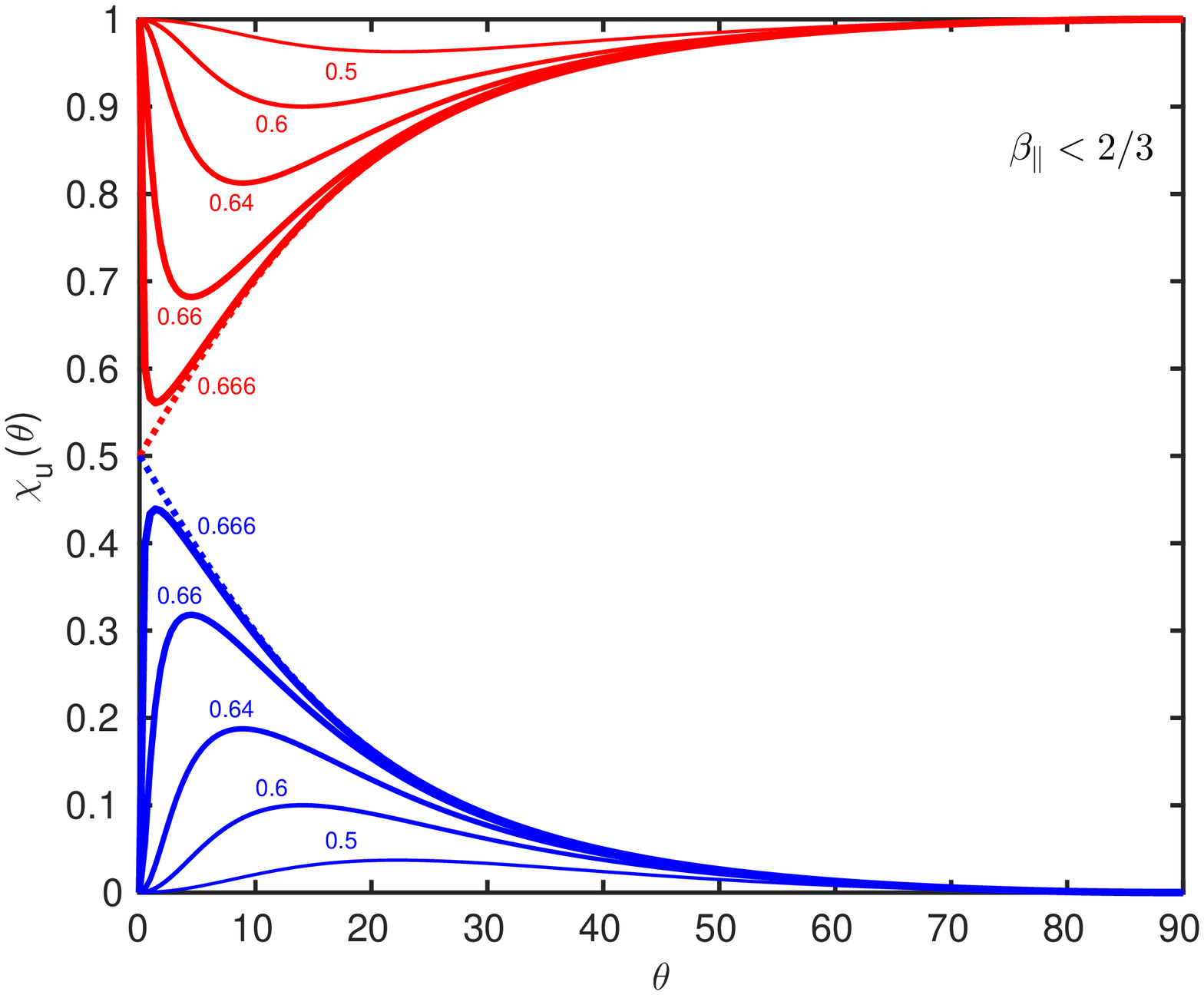}
\hspace{0.02\textwidth}\includegraphics[width=0.48\linewidth]{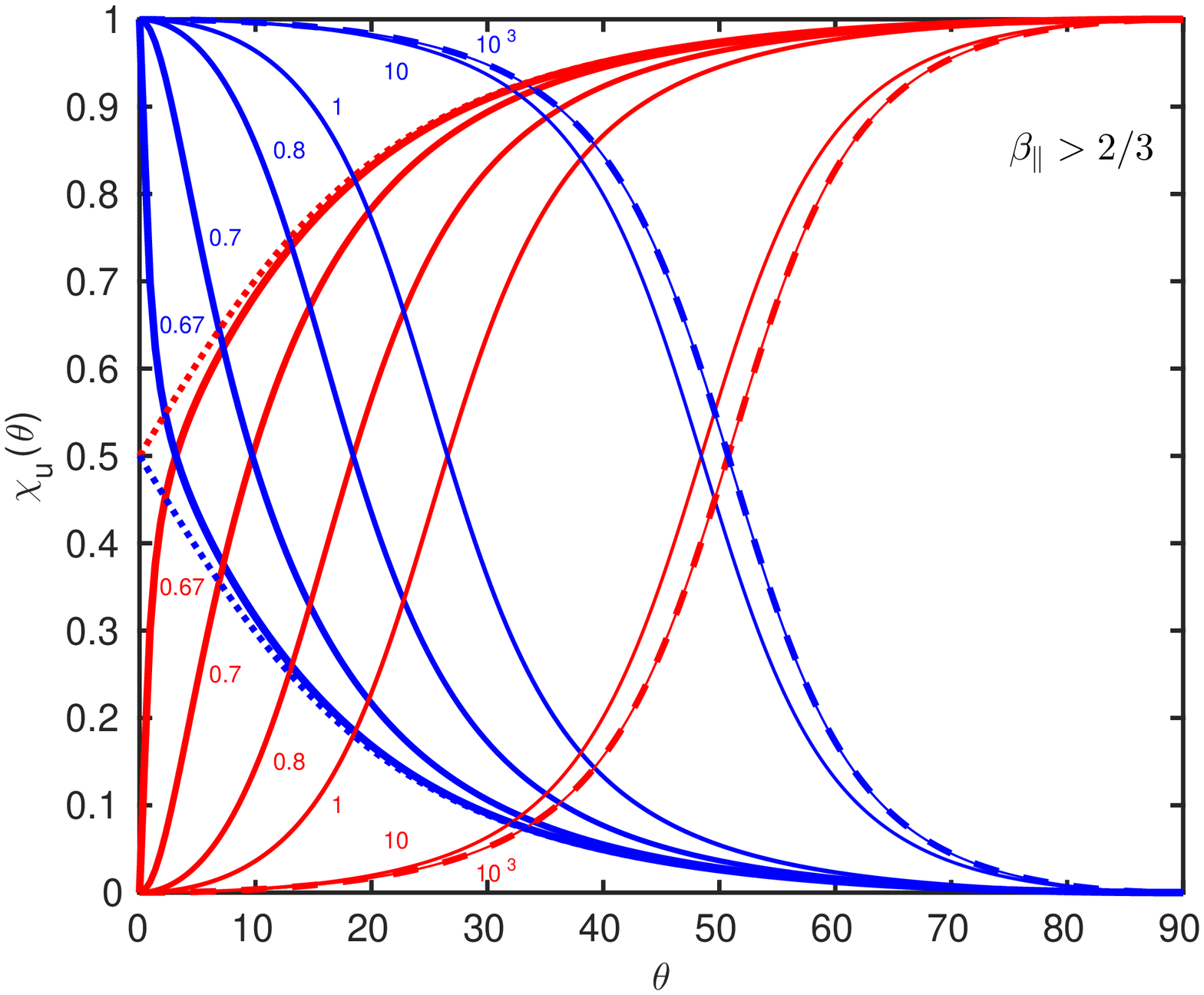}$$
\caption{CGL velocity ratio $\chi_u(\theta)=|u_\parallel|^2/|u|^2$ for the slow mode (red lines) and the fast mode (blue lines) for different values
  of $\bpar$, with isotropic temperature $T_\perp=T_\parallel$ (or the temperature anisotropy ratio $a_p=1$).
  The left figure is for $\bpar<2/3$, and the thickness of the lines increases with $\bpar$ when approaching the critical
value of $\bpar=2/3$. The plotted lines have $\bpar=0.5; 0.6; 0.64; 0.66; 0.666$. 
The right figure is for $\bpar>2/3$ and the thickness of the lines decreases with $\bpar$ when going away from the critical value of $\bpar=2/3$.
The plotted lines have $\bpar=0.67; 0.7; 0.8; 1.0; 10, 10^3$. In both figures, the dotted lines are the critical case $\bpar=2/3$
with solutions (\ref{eq:CGLspecialU}). In the right figure the dashed lines represent the limit $\bpar\gg 1$ with solutions
(\ref{eq:Finaly1}), (\ref{eq:Finaly2}) and the dashed lines nicely overlap with a thin solid line solutions obtained for $\bpar=10^3$.
The Alfv\'en mode has $\chi_u(\theta)=0$ regardless of plasma beta and is not plotted.} \label{fig:CGL-u}
\end{figure*}

To find the ratio for velocity components $\chi_u$ in the CGL description,
it is useful for a moment to work with abbreviated symbols,
and we write the matrix (\ref{eq:SFmatrix}) for the $u_x$ and $u_z$ components as
\begin{eqnarray}
\left( \begin{array}{cc}
  (\widetilde{\omega}/\widetilde{k})^2-a;
    & \qquad - b\\
    - b & \qquad (\widetilde{\omega}/\widetilde{k})^2-c
\end{array} \right)
\left( \begin{array}{c} u_x \\ u_z \end{array} \right) = \left( \begin{array}{c} 0 \\ 0 \end{array} \right), \label{eq:CGLmat}
\end{eqnarray}
where the coefficients are
\begin{eqnarray}
  a=1+a_p\bpar(1-\frac{1}{2}\cos^2\theta)-\frac{\bpar}{2}\cos^2\theta;\qquad
  b=\frac{\bpar}{2}a_p\sin\theta\cos\theta; \qquad
  c=\frac{3}{2}\bpar\cos^2\theta.
\end{eqnarray}  
The determinant must be zero which implies $(\widetilde{\omega}/\widetilde{k})^4-(\widetilde{\omega}/\widetilde{k})^2(a+c)+ac-b^2=0$, and has a solution
\begin{equation} \label{eq:miracle2}
(\widetilde{\omega}/\widetilde{k})^2 = \frac{1}{2}(a+c)\pm\frac{1}{2}\sqrt{(a+c)^2-4ac+4b^2} = \frac{1}{2}(a+c)\pm\frac{1}{2}\sqrt{(a-c)^2+4b^2}. 
\end{equation} 
The last step was actually used in obtaining (\ref{eq:miracle}) which is otherwise quite difficult to see.
Using this result in the above matrix, the eigenvector matrix becomes
\begin{eqnarray}
\left( \begin{array}{cc}
- \frac{1}{2}(a-c)\pm\frac{1}{2}\sqrt{(a-c)^2+4b^2} ;
    & - b\\
    - b & \frac{1}{2}(a-c)\pm\frac{1}{2}\sqrt{(a-c)^2+4b^2}
\end{array} \right)
\left( \begin{array}{c} u_x \\ u_z \end{array} \right) = \left( \begin{array}{c} 0 \\ 0 \end{array} \right), \label{eq:UUmatrix}
\end{eqnarray}
where
\begin{eqnarray}
a-c=1+a_p\bpar(1-\frac{1}{2}\cos^2\theta)-2\bpar\cos^2\theta. \label{eq:acrit}
\end{eqnarray}
The ratio of $u_x/u_z$ can be expressed (for example from the second row) as
\begin{eqnarray}
  \frac{u_x}{u_z} &=& \frac{1}{b} \Big[(\widetilde{\omega}/\widetilde{k})^2 -c\Big]=\frac{1}{2b}\Big[a-c\pm\sqrt{(a-c)^2+4b^2}\Big]\nn\\ 
  &=& \frac{1}{\bpar a_p\sin\theta\cos\theta}
  \Big[1+a_p\bpar(1-\frac{1}{2}\cos^2\theta)-2\bpar\cos^2\theta \nn\\
 &&   \pm \sqrt{\Big(1+a_p\bpar(1-\frac{1}{2}\cos^2\theta)-2\bpar\cos^2\theta\Big)^2  +a_p^2\bpar^2\sin^2\theta\cos^2\theta} \,\,    \Big], \label{eq:umf1} 
\end{eqnarray}
or alternatively (from the first row) as
\begin{eqnarray}
  \frac{u_x}{u_z} &=& \frac{b}{(\widetilde{\omega}/\widetilde{k})^2-a} = \frac{2b}{-(a-c)\pm\sqrt{(a-c)^2+4b^2}}\nn\\
  &=&   \bpar a_p\sin\theta\cos\theta
  \Big[-1-a_p\bpar(1-\frac{1}{2}\cos^2\theta)+2\bpar\cos^2\theta \nn\\
 &&   \pm \sqrt{\Big(1+a_p\bpar(1-\frac{1}{2}\cos^2\theta)-2\bpar\cos^2\theta\Big)^2  +a_p^2\bpar^2\sin^2\theta\cos^2\theta} \,\,    \Big]^{-1}. \label{eq:umf2}
\end{eqnarray}  
The behavior of these solutions is plotted in Figure \ref{fig:CGL-u}, where we plot solutions for isotropic temperatures $a_p=1$.
Similarly to MHD, two classes of solutions can be characterized based upon the behavior
for parallel propagation $\theta=0$, depending if the quantity $a-c$ (\ref{eq:acrit}) evaluated at $\theta=0$ is positive or negative.
The quantity $(a-c)|_{\theta=0}=1+\frac{1}{2}a_p\bpar-2\bpar$ was discussed in the section for parallel propagation and this quantity
represents the green dotted curve in Figure \ref{fig:CGL-th} that separates the system into two regions.  
For $(a-c)|_{\theta=0}>0$, i.e. in the region $\bpar<2/(4-a_p)$, for the slow mode (use eq. \ref{eq:umf2}) the ratio $(u_x/u_z)^s(0)=0$ and
for the fast mode (use eq. \ref{eq:umf1}) the ratio $(u_x/u_z)^f(0)=+\infty$. In this region, all the $\chi_u(\theta)$ curves start at
\begin{equation}
\bpar<2/(4-a_p): \qquad \chi_u^s(0)=1;  \qquad \chi_u^f(0)=0.
\end{equation}  
For $(a-c)|_{\theta=0}<0$, i.e. in the region $\bpar>2/(4-a_p)$, for the slow mode (use eq. \ref{eq:umf1}) the ratio
$(u_x/u_z)^s(0)=+\infty$ and for the fast mode (use eq. \ref{eq:umf2}) the ratio $(u_x/u_z)^f(0)=0$.
In this region, all the $\chi_u(\theta)$ curves start at
\begin{equation}
\bpar>2/(4-a_p): \qquad \chi_u^s(0)=0; \qquad \chi_u^f(0)=1.
\end{equation}  
For perpendicular propagation $\theta=\pi/2$ the quantity $(a-c)|_{\theta=\pi/2}=1+a_p\bpar$ is always positive and the separation into two regions is
not necessary. For the slow mode $(u_x/u_z)^s(\pi/2)=0$ and for the fast mode $(u_x/u_z)^f(\pi/2)=+\infty$. All the $\chi_u(\theta)$ curves, 
regardless of the value of $\bpar$ therefore end at
\begin{equation}
\qquad \chi_u^s(\pi/2)=1; \qquad \chi_u^f(\pi/2)=0.
\end{equation}  
This summarizes the general behavior of the solutions. However, as can be seen from Figure \ref{fig:CGL-u},
there is a special case when the parameters approach $\bpar=2/(4-a_p)$, i.e. when $(a-c)|_{\theta=0}=0$, that will be discussed in the next subsection.
One can again consider the two limiting cases when $\bpar$ is small or large. The $\bpar\ll 1$ case yields dispersion relations
\footnote{One can directly work with the dispersion relation (\ref{eq:miracle2}) and for $\bpar\ll 1$ and $a_p$ bounded and less than an extremely large values,
  the parameter $(a-c)$ is positive and $(a-c)\sim 1$, on the other hand the parameters $b\ll 1$ and $c\ll 1$,
  so an expansion of the square root is possible as
\begin{equation}
  \frac{1}{2}(a+c)\pm\frac{1}{2}(a-c)\sqrt{1+\frac{4b^2}{(a-c)^2}} \simeq \frac{1}{2}(a+c)\pm\frac{1}{2}(a-c)\Big(1+\frac{2b^2}{(a-c)^2}\Big)
  \simeq \frac{1}{2}(a+c)\pm\frac{1}{2} \Big[ (a-c) +\frac{2b^2}{a}\Big].
\end{equation}
The fast mode $(\widetilde{\omega}/\widetilde{k})^2_f \simeq a+\frac{b^2}{a} \simeq a$, and the slow mode
$(\widetilde{\omega}/\widetilde{k})^2_s \simeq c-\frac{b^2}{a} \simeq c$, which yields low $\bpar$ dispersion relations.
With ``less confidence'' the same result can be directly obtained from the velocity matrix (\ref{eq:CGLmat}) by neglecting the off-diagonal terms.}
\begin{eqnarray}
  \bpar\ll 1: \qquad (\widetilde{\omega}/\widetilde{k})^2_f &=& 1+a_p\bpar ( 1-\frac{1}{2}\cos^2\theta)-\frac{\bpar}{2}\cos^2\theta;\\
  (\widetilde{\omega}/\widetilde{k})^2_s &=& \frac{3}{2}\bpar\cos^2\theta. \label{eq:CGLlowBeta}
\end{eqnarray}
Using expression (\ref{eq:umf2}) for the slow mode yields $(u_x/u_z)^s(\theta)=0$ and expression (\ref{eq:umf1}) for the fast mode yields
$(u_x/u_z)^f(\theta)=\infty$, implying that in low beta limit the $\chi_u(\theta)$ solutions are
\begin{equation}
\bpar \ll 1:\qquad \chi_u^s(\theta)=1; \qquad \chi_u^f(\theta)=0,
\end{equation}
and the constant solutions approaching this limit are visible in the plot of Figure \ref{fig:CGL-u}. The opposite $\bpar\gg 1$ limit in the CGL description is not particularly
revealing since there is no obvious way to simplify the expansion, even for isotropic temperatures. For $a_p=1$, the $\bpar\gg 1$ expansion
yields 
\begin{eqnarray} 
  \bpar \gg 1:\qquad && (\widetilde{\omega}/\widetilde{k})_{sf}^2 =\frac{1}{2}\Big( 1+\bpar(1+\frac{1}{2}\cos^2\theta)\Big)
  \pm \frac{1}{2}\Big[ \bpar \sqrt{A} +\frac{1-\frac{5}{2}\cos^2\theta}{\sqrt{A}} \Big], \label{eq:blaaa}\\
  && A=\big( 1-\frac{5}{2}\cos^2\theta\big)^2+\sin^2\theta\cos^2\theta, 
\end{eqnarray}
where we neglected a term proportional to $1/\bpar$.
The expression results in the correct parallel limit $(\widetilde{\omega}/\widetilde{k})_f^2(0)=\frac{3}{2}\bpar$,
$(\widetilde{\omega}/\widetilde{k})_s^2(0)=1$ and the correct perpendicular limit $(\widetilde{\omega}/\widetilde{k})_f^2(\pi/2)=1+\bpar$,
$(\widetilde{\omega}/\widetilde{k})_s^2(\pi/2)=0$. If the perpendicular fast mode limit is sufficient to be
$(\widetilde{\omega}/\widetilde{k})_f^2(\pi/2)=\bpar$ (which is acceptable since $\bpar\gg 1$), only the terms 
proportional to $\bpar$ can be retained and the fast mode dispersion relation can be simplified to
\begin{equation}
  \bpar \gg 1:\qquad (\widetilde{\omega}/\widetilde{k})_{f}^2 =
  \frac{1}{2}\bpar\Big[ 1+\frac{1}{2}\cos^2\theta+\sqrt{\big( 1-\frac{5}{2}\cos^2\theta\big)^2+\sin^2\theta\cos^2\theta}  \Big].
\end{equation}
The expression (\ref{eq:umf1}) yields the velocity ratio of the fast mode as
\begin{equation} \label{eq:HbetaCGLvel}
  \bpar \gg 1: \qquad \left(\frac{u_x}{u_z}\right)^f =
  \frac{1-\frac{5}{2}\cos^2\theta+\sqrt{\big(1-\frac{5}{2}\cos^2\theta   \big)^2+\sin^2\theta\cos^2\theta}}{\sin\theta\cos\theta},
\end{equation}
that is independent of $\bpar$. 
For the slow mode wave the dispersion relation cannot be further simplified because if we collect only terms proportional to $\bpar$,
the resulting expression will be nonzero everywhere except exactly at $\theta=0$ (and also $\theta=\pi/2$).
For $\theta=0$, the entire contribution of $(\widetilde{\omega}/\widetilde{k})_{s}^2(0) =1$ comes
from terms in (\ref{eq:blaaa}) that are not proportional to $\bpar$. Nevertheless, the analytic velocity ratio (\ref{eq:HbetaCGLvel}) 
for the fast mode is sufficiently simple and to obtain analytic $\chi_u^s(\theta)$ for the slow mode, it is actually possible to cheat a little by realizing
that the slow and fast mode curves are always symmetric and that $\chi_u^s(\theta)+\chi_u^f(\theta)=1$. Our final large beta results therefore are
\begin{eqnarray}
  \chi_u^f(\theta) &=&
  \bigg[ 1+\bigg(\frac{1-\frac{5}{2}\cos^2\theta+\sqrt{\big(1-\frac{5}{2}\cos^2\theta\big)^2+\sin^2\theta\cos^2\theta}}{\sin\theta\cos\theta}\bigg)^2 \bigg]^{-1};
  \label{eq:Finaly1}\\
  \chi_u^s(\theta) &=& 1-\chi_u^f(\theta). \label{eq:Finaly2}
\end{eqnarray}
Solutions are plotted in Figure \ref{fig:CGL-u} as dashed lines and fit the solutions of the full dispersion relations with $\bpar=10^3$ very nicely.
\subsubsection{CGL special case $\bpar = 2/(4-T_\perp/T_\parallel)$}
In this peculiar case when all 3 modes propagate in the parallel direction with the same phase speed, one again faces complications 
when calculating the velocity eigenvector, as with the MHD peculiar case when $V_A=C_s$. This CGL special case
can be encountered only for $\bpar\ge 1/2$ because the temperature anisotropy must be $a_p\ge 0$.
We start with the velocity eigenvector matrix (\ref{eq:UUmatrix}) and for the special case considered $1+\frac{1}{2}a_p\bpar -2\bpar =0$,
the expression (\ref{eq:acrit}) simplifies to $a-c=(4\bpar-1)\sin^2\theta$.
One can already see that for strictly parallel propagation the entire matrix will be zero
and it will not be possible to evaluate the eigenvector. Similarly to the MHD case, the $\sin\theta$ can be pulled out of the matrix
or alternatively one can use (\ref{eq:umf1}), (\ref{eq:umf2}) and calculate the limit. For the special case considered here, the parameter
$b=\frac{1}{2}(4\bpar-2)\sin\theta\cos\theta$. Considering again for simplicity only positive wavenumbers,
the expression (\ref{eq:umf1}) yields in this particular case
\begin{equation} \label{eq:CGLspecialU}
\frac{u_x}{u_z} = \frac{4\bpar-1}{4\bpar-2}\tan\theta\pm \sqrt{ \left(\frac{4\bpar-1}{4\bpar-2}\right)^2 \tan^2\theta+1}.
\end{equation}  
The expression can be evaluated at $\theta=0$, which yields for the slow mode $(u_x/u_z)^s(0)=-1$ and for the fast mode
$(u_x/u_z)^f(0)=+1$, implying that both modes start at
\begin{equation}
\bpar=2/(4-a_p): \qquad \chi_u^{s}(0)=\frac{1}{2}; \qquad \chi_u^{f}(0)=\frac{1}{2}.
\end{equation}  
For the perpendicular propagation the discussion is unchanged from the previous general case and yields $\chi_u^s(\pi/2)=1$ and $\chi_u^f(\pi/2)=0$.
\subsection{Empirical models with polytropic indices $\gamma_\parallel$, $\gamma_\perp$} \label{sec:EmpirModels}
 The differences between MHD and CGL can be further clarified with polytropic indices. The linearized MHD pressure equation reads
\begin{equation}
  \frac{\pr p}{\pr t} + \gamma p^{(0)} (\pr_x u_x +\pr_y u_y + \pr_z u_z) =0. \label{eq:Pmhd0}
\end{equation}
One can introduce the concept of polytropic indices $\gamma_\parallel$ and $\gamma_\perp$ to the CGL pressure equations
(\ref{eq:PparLin}), (\ref{eq:PperpLin}) and write
\begin{eqnarray}
&& \frac{\pr p_\parallel}{\pr t} + p_\parallel^{(0)}(\pr_x u_x+\pr_y u_y) +\gamma_\parallel p_\parallel^{(0)} \pr_z u_z= 0;\label{eq:Pcgl0}\\
&&  \frac{\pr p_\perp}{\pr t}  + \gamma_\perp p_\perp^{(0)}(\pr_x u_x+\pr_y u_y) +p_\perp^{(0)} \pr_z u_z=0, \label{eq:Pcgl1}
\end{eqnarray}  
where for the CGL model $\gamma_\parallel=3$ and $\gamma_\perp=2$.
\footnote{One can then define the parallel and perpendicular sound speeds
as $C_\parallel=\sqrt{\gamma_\parallel p_{\parallel}^{(0)}/\rho_0}$ and $C_\perp=\sqrt{\gamma_\perp p_{\perp}^{(0)}/\rho_0}$, since in this model
the parallel acoustic mode will have a dispersion $\omega=\pm C_\parallel k_\parallel$ and the perpendicular fast mode $\omega=\pm k_\perp \sqrt{V_A^2+C_\perp^2}$,
as can be easily verified.}
The reasoning for such a construction is to enable a smooth transition between two extreme cases - the
CGL model on one side, and the isothermal model $\gamma_\parallel=1$ and $\gamma_\perp=1$ on the other side. 
The important observation is that in the parallel pressure equation the $\gamma_\parallel$ acts only on the parallel velocity component $u_z$,
and in the perpendicular pressure equation the $\gamma_\perp$ acts only on the perpendicular velocity components $u_x,u_y$.
In MHD, the $\gamma$ acts on all three components $u_x,u_y,u_z$. The values of polytropic indices are directly related to the number of degrees of
freedom $i$, through the well known relation $\gamma=(i+2)/i$. In the MHD description the number of degrees of freedom is $i=3$ and so $\gamma=5/3$.
In contrast, the CGL description can be interpreted as being composed of a (strongly coupled) mixture of one-dimensional and
two-dimensional dynamics, where for the parallel direction $i=1$ and $\gamma_\parallel=3$ and for the perpendicular direction $i=2$ and $\gamma_\perp=2$.
The subtle appearing, but nevertheless fundamental difference between (\ref{eq:Pmhd0}) and (\ref{eq:Pcgl0}), (\ref{eq:Pcgl1}) is the core of
many differences between the MHD and CGL descriptions, such as for example the effect of the slow mode becoming faster than the Alfv\'en mode.
The two systems remain different even if the mean CGL pressures (or temperatures) are prescribed to be isotropic
$p_\parallel^{(0)}=p_\perp^{(0)}$ and equal to the mean MHD pressure $p^{(0)}$. This means that the MHD does not match the CGL description even if
the distribution function is isotropic, as for example Maxwellian.

We briefly consider models with polytropic indices $\gamma_\parallel,\gamma_\perp$ that are based on the linearized pressure
equations (\ref{eq:Pcgl0}), (\ref{eq:Pcgl1}). If these pressure equations are ``un-linearized'', i.e. if one performs a ``reverse engineering''
procedure to obtain a nonlinear equations (a procedure that is possible to do because of the known guidance by the CGL model),
the nonlinear equations read
\begin{eqnarray}
&& \frac{d p_\parallel}{d t} + p_\parallel \nabla\cdot \bu +(\gamma_\parallel-1) p_\parallel \bhat\cdot\nabla\bu\cdot\bhat= 0;\\
&&  \frac{d p_\perp}{d t} + \gamma_\perp p_\perp\nabla\cdot\bu -(\gamma_\perp-1) p_\perp \bhat\cdot\nabla\bu\cdot\bhat=0,
\end{eqnarray}  
which can be rewritten as
\begin{eqnarray}
  \frac{d}{dt}\left( \frac{p_\parallel |\bb|^{\gamma_\parallel-1}}{\rho^{\gamma_\parallel}}\right) =0; \qquad
  \frac{d}{dt}\left( \frac{p_\perp}{\rho |\bb|^{\gamma_\perp-1}}\right) =0. \label{eq:gammamodel}
\end{eqnarray}
Empirical fluid models of this type were studied for example by \cite{HauSonnerup1993} and \cite{HauPhan1993} with the motivation that
 spacecraft observations in the Earth's magnetosheath often show behavior that cannot be explained by the CGL values
of $\gamma_\parallel=3,\gamma_\perp=2$ or by the isothermal values of $\gamma_\parallel=1, \gamma_\perp=1$, but can be fitted by the
values that lie somewhere in between (and actually quite close to the isothermal case).
For a direct comparison with the CGL model, the equations (\ref{eq:gammamodel}) can be rewritten to a form
\begin{eqnarray}
&&  \frac{d}{dt}\left( \frac{p_\parallel |\bb|^2}{\rho^3}\right)
  =-(\gamma_\parallel-3)\left(\frac{p_\parallel |\bb|^2}{\rho^3}\right)\frac{\rho}{|\bb|}\frac{d}{dt}\left(\frac{|\bb|}{\rho}\right); \\
  &&  \frac{d}{dt}\left( \frac{p_\perp}{\rho |\bb|}\right)
  =(\gamma_\perp-2)\left(\frac{p_\perp}{\rho|\bb|}\right)\frac{1}{|\bb|}\frac{d}{dt}|\bb|,
\end{eqnarray}  
or to an interesting alternative form 
\begin{eqnarray}
&&  \frac{d}{dt}\ln \left( \frac{p_\parallel |\bb|^2}{\rho^3}\right) = -(\gamma_\parallel-3)\frac{d}{dt}\ln \left(\frac{|\bb|}{\rho}\right);\\
&&  \frac{d}{dt}\ln \left( \frac{p_\perp}{\rho |\bb|}\right) = (\gamma_\perp-2) \frac{d}{dt} \ln |\bb|.
\end{eqnarray}
Obviously, for CGL values of $\gamma_\parallel=3,\gamma_\perp=2$ the right hand sides disappear and one obtains the classical CGL expressions.
The right hand sides of the above expressions can be compared with equations of Chew et al. 1956 (\ref{eq:PparSuper}), (\ref{eq:PperpSuper})
which contain the gyrotropic heat flux contributions. Therefore, the deviations from the CGL values of $\gamma_\parallel=3,\gamma_\perp=2$
can be considered to offer a very simple empirical model (perhaps too simple) that shows how the heat flux contributions modify the CGL dynamics.    
The dispersion relations for this fluid model can be found in \cite{HauPhan1993} and also in \cite{AbrahamShrauner1973}, who studied even
more general empirical model with four polytropic indices (essentially, the $\gamma_\parallel$, $\gamma_\perp$
acting on $|\bb|$ and $\rho$ in (\ref{eq:gammamodel}) can be further split to two independent parameters).
 Interestingly, complex polytropic indices have been suggested to model Landau damping \citep{BelmontMazelle1992}.  
By considering the  linearized CGL equations (\ref{eq:CGL-rhoAng})-(\ref{eq:CGL-Bxz}) and linearized polytropic
pressure equations (\ref{eq:Pcgl0})-(\ref{eq:Pcgl1}) written in the x-z plane yields the dispersion relations for this model.
The Alfv\'en wave is still decoupled from the rest of the system and the
Alfv\'en wave dispersion relation is unaffected by the different pressure equations, therefore also
yielding the same oblique firehose instability threshold. What is modified is the dispersion relation of the slow and fast waves. The velocity equations
read
\begin{eqnarray}
&&  \frac{\pr^2 u_x}{\pr t^2} - \frac{p_\perp^{(0)}}{\rho_0}\Big( \gamma_\perp \pr_x^2u_x +\pr_x\pr_z u_z\Big) +
  \Big( \frac{p_\parallel^{(0)}}{\rho_0} - \frac{p_\perp^{(0)}}{\rho_0} -V_A^2 \Big) \pr_z^2 u_x -V_A^2 \pr_x^2 u_x = 0;\\
&& \frac{\pr^2 u_z}{\pr t^2} - \gamma_\parallel \frac{p_\parallel^{(0)}}{\rho_0}\pr_z^2 u_z - \frac{p_\perp^{(0)}}{\rho_0}\pr_x \pr_z u_x =0,
\end{eqnarray}
and the transformation to Fourier space yields the velocity matrix
\begin{eqnarray}
\left( \begin{array}{cc}
  \omega^2-\big( \gamma_\perp \frac{p_\perp^{(0)}}{\rho_0}+V_A^2\big)k_\perp^2+\big(\frac{p_\parallel^{(0)}}{\rho_0}-\frac{p_\perp^{(0)}}{\rho_0}-V_A^2\big)k_\parallel^2;
    & \qquad - \frac{p_\perp^{(0)}}{\rho_0}k_\perp k_\parallel\\
    - \frac{p_\perp^{(0)}}{\rho_0}k_\perp k_\parallel; & \qquad \omega^2-\gamma_\parallel\frac{p_\parallel^{(0)}}{\rho_0}k_\parallel^2
\end{array} \right)
\left( \begin{array}{c} u_x \\ u_z \end{array} \right) = \left( \begin{array}{c} 0 \\ 0 \end{array} \right),
\end{eqnarray}
or alternatively
\begin{eqnarray}
\left( \begin{array}{cc}
  \widetilde{\omega}^2-\big( \gamma_\perp a_p \frac{\bpar}{2} +1 \big)\widetilde{k}_\perp^2
  -\big(1+\frac{\bpar}{2}(a_p-1)\big)\widetilde{k}_\parallel^2;
    & \qquad - \frac{1}{2}a_p\bpar \widetilde{k}_\perp \widetilde{k}_\parallel\\
    - \frac{1}{2}a_p\bpar \widetilde{k}_\perp \widetilde{k}_\parallel; & \qquad \widetilde{\omega}^2-\gamma_\parallel\frac{\bpar}{2} \widetilde{k}_\parallel^2
\end{array} \right)
\left( \begin{array}{c} u_x \\ u_z \end{array} \right) = \left( \begin{array}{c} 0 \\ 0 \end{array} \right).
\end{eqnarray}
Calculating the determinant yields a dispersion relation for the slow and fast modes in the form
\begin{eqnarray} 
  && \widetilde{\omega}^4-A_2\widetilde{\omega}^2+A_0 = 0; \label{eq:GenPolypPica}\\
  &&   A_2 = \widetilde{k}_\perp^2 \left(1+\frac{\gamma_\perp}{2} a_p\bpar\right) + \widetilde{k}_\parallel^2 \left(1+\frac{1}{2}\bpar(\gamma_\parallel-1)
  +\frac{1}{2}a_p\bpar\right); \nonumber \\
&& A_0 = \frac{\gamma_\parallel}{2}\widetilde{k}_\parallel^2 \bpar \left[ \widetilde{k}_\parallel^2 \left(1-\frac{1}{2}\bpar +\frac{1}{2}a_p\bpar\right) + 
\widetilde{k}_\perp^2 \left(1+\frac{\gamma_\perp}{2} a_p\bpar-\frac{1}{2\gamma_\parallel} a_p^2\bpar\right)  \right].\nonumber
\end{eqnarray}
Solutions for parallel propagation $(k_\perp=0)$ are the usual CGL Alfv\'en mode and the ion-acoustic mode
  $\widetilde{\omega}=\pm \sqrt{\frac{\gamma_\parallel}{2}\bpar}\widetilde{k}_\parallel$, and for perpendicular propagation $(k_\parallel=0)$ the
fast mode $\widetilde{\omega}=\pm \sqrt{1+\frac{\gamma_\perp}{2}a_p\bpar}\widetilde{k}_\perp$.
Similarly to the CGL description it can be shown that $A_2^2-4A_0\ge 0$ because of the trick (\ref{eq:miracle2}), implying that
the slow mode becomes unstable when $A_0<0$, i.e. when
\begin{equation}
\left[ k_\parallel^2 \left(1-\frac{1}{2}\bpar +\frac{1}{2}a_p\bpar\right) + 
k_\perp^2 \left(1+\frac{\gamma_\perp}{2} a_p\bpar-\frac{1}{2\gamma_\parallel} a_p^2\bpar\right)  \right] <0.
\end{equation}  
The two competing instabilities are again the parallel firehose instability and the highly oblique mirror instability.
The parallel firehose instability has the same threshold as obtained previously. However, the mirror instability threshold reads
\begin{equation} \label{eq:gammaMirror}
1+\frac{\gamma_\perp}{2} a_p\bpar-\frac{1}{2\gamma_\parallel} a_p^2\bpar <0,
\end{equation}
which is consistent with the result of \cite{HauSonnerup1993}, their eq. 14, written in the form
\footnote{Note that the equation in the cited paper has opposite sign, which is not a typo, authors just call the mirror threshold
  a condition when a mode is stable - opposite to a more usual condition when a mode becomes unstable. Similarly, authors write the
condition for the firehose stable region as $\bpar \le 2+\beta_\perp$.}
\begin{equation}
\gamma_\parallel\bpar < \frac{\beta_\perp^2}{2+\gamma_\perp\beta_\perp}.
\end{equation}
For $\gamma_\parallel=3, \gamma_\perp=2$, the condition (\ref{eq:gammaMirror}) is naturally equivalent to the CGL mirror
threshold. However, the condition offers a temptingly simple result to keep the perpendicular polytropic index unchanged $\gamma_\perp=2$
and modify the parallel polytropic index to $\gamma_\parallel=1/2$. In this way, the mirror threshold matches the correct
kinetic threshold (\ref{eq:KineticMirror}). Nevertheless, this is just an interesting empirical concept, it fixes the highly oblique
mirror threshold but at the same time moves the slow and fast mode dispersion relations in other directions, for example the parallel sound speed
in such a model is $C_\parallel^2=\frac{1}{2}p_\parallel^{(0)}/\rho_0$, which appears to be unrealistically low.
Such a 'tuning' of models with free parameters - only in one particular direction  and only at the level of dispersion relations -
can be sometimes useful for interpretation of observational data, however, such models will usually lead to unphysical situations. 
We will not consider models with polytropic indices further in this text.
Instead, to obtain a better match with kinetic description, it is advisable to use the correct CGL values of $\gamma_\parallel=3, \gamma_\perp=2$ and
focus on fluid models that derive the correct heat flux contributions $q_\parallel$ and $q_\perp$.   
\subsection{CGL and radially expanding solar wind}
Similarly to MHD, the CGL fluid model can be very useful for understanding evolutions of temperatures in the radially expanding solar wind.
A good discussion can be found for example in \cite{Matteini2007,Matteini2013} and references therein.
Let's first consider the MHD case. Combining the pressure equation $dp/dt+\gamma\nabla\cdot\bu=0$ with the density equation $d\rho/dt+\rho\nabla\cdot\bu=0$ yields
\begin{equation}
\textrm{MHD:}\qquad \quad\frac{d}{dt}\left(\frac{p}{\rho^\gamma}\right) =0; \quad => \quad \frac{d}{dt}\left(\frac{T}{\rho^{\gamma-1}}\right) =0,
\end{equation}  
which in a steady-state (or co-moving with the convective derivative) implies $T \sim \rho^{\gamma-1}$. 
By considering simple radially expanding solar wind, one prescribes that the density $\rho$ evolves with the heliocentric distance $r$ as
$\rho\sim 1/r^2$, and by using $\gamma=5/3$, the ideal MHD model yields (in the absence of any dissipation and associated turbulent heating)
\begin{equation}
T \sim r^{-4/3}.
\end{equation}
In contrast, the CGL equations (\ref{eq:CGLconserv}) are rewritten as
\begin{equation}
\textrm{CGL:}\qquad \quad \frac{d}{dt}\left( \frac{T_{\parallel} |\bb|^2}{\rho^2}\right)=0; \qquad
\frac{d}{dt}\left(\frac{T_{\perp}}{|\bb|}\right) =0,
\end{equation}
which in a steady-state implies
\begin{equation}
T_\parallel\sim \frac{\rho^2}{|\bb|^2}; \qquad T_\perp \sim |\bb|; \qquad \frac{T_\perp}{T_\parallel}=\frac{|\bb|^3}{\rho^2}. 
\end{equation}  
For the magnetic field it is possible to consider profiles of roughly $|\bb|\sim 1/r^2$, or $|\bb|\sim 1/r$
(one can be more precise and consider the Parker spiral profile). The first magnetic field profile yields 
\begin{equation} \label{eq:Wind1}
|\bb|\sim 1/r^2;\quad=>\quad \qquad T_\parallel \sim \textrm{const.}; \qquad T_\perp \sim 1/r^2; \qquad \frac{T_\perp}{T_\parallel}\sim 1/r^2.
\end{equation}
The second magnetic field profile yields
\begin{equation} \label{eq:Wind2}
|\bb|\sim 1/r;\quad=>\quad \qquad T_\parallel \sim 1/r^2; \qquad T_\perp \sim 1/r; \qquad \frac{T_\perp}{T_\parallel}\sim r.
\end{equation}
There is also a curious case when the magnetic field profile is prescribed to be $|\bb|\sim r^{-4/3}$, i.e. a profile between
two cases of $1/r$ and $1/r^2$, that yields
\begin{equation} \label{eq:TempCrit}
|\bb|\sim r^{-4/3};\quad=>\quad \qquad T_\parallel \sim r^{-4/3}; \qquad T_\perp \sim r^{-4/3}; \qquad \frac{T_\perp}{T_\parallel}\sim \textrm{const.},
\end{equation}
and the temperature anisotropy stays constant. The estimations are of course valid only until a firehose threshold or a mirror threshold is reached.  
For example, evolution according to (\ref{eq:Wind1}) will lead the system to a firehose threshold, and evolution according to (\ref{eq:Wind2}) will lead the system
to a mirror threshold. In general, if we fix the density profile to $\rho\sim 1/r^2$, any magnetic field profile steeper than $r^{-4/3}$ will evolve
the system towards a firehose threshold, and a magnetic field profile shallower than $r^{-4/3}$ will evolve the system towards a mirror threshold.
The estimations do not reveal presence of the firehose instability or the mirror instability in the CGL model, since the solutions are written in a steady-state,
and concern only evolution of mean temperatures (i.e. the solutions should be written with $T_\parallel^{(0)},T_\perp^{(0)}$).
To find an instability, one of course needs to consider evolution equations for fluctuating variables and analyze associated dispersion relations, since
it is the fluctuations that become unstable. Further discussion can be found in \cite{HunanaZank2017}.

From a linear perspective, the CGL model has to be accompanied by the Hall term and FLR corrections, so that the instabilities
are stabilized at small spatial scales. However, considering nonlinear evolution with finite amplitudes of fluctuations, even the non-dispersive CGL model is stabilized.
For example, considering parallel propagation \cite{Tenerani2017} has shown that the firehose instability
criterion reads (see eq. 9 in that paper)
\begin{equation} \label{eq:TeneraniFirehose}
V_A^2 + \frac{1}{\rho_0} \frac{p_\perp^{(0)}-p_\parallel^{(0)}}{1+|\frac{\bb_\perp}{B_0}|^2}<0; \qquad => \qquad 1+\frac{\bpar}{2}\frac{a_p-1}{1+|\frac{\bb_\perp}{B_0}|^2}<0, 
\end{equation}  
i.e., for infinitely small amplitudes of the magnetic field, the usual firehose criterion is obtained. However, when the CGL system reaches firehose unstable regime
and the amplitude of fluctuations starts to grow, the firehose criterion is modified according to (\ref{eq:TeneraniFirehose}). 
Therefore, the amplitude of fluctuations can not grow without bounds, and beyond some critical amplitude, the system becomes again stabilized.  
The condition (\ref{eq:TeneraniFirehose}) nicely demonstrates how the CGL system is stabilized during nonlinear evolution, once the firehose instability is reached. 
For further discussion, see \cite{Tenerani2017,Tenerani2018}. 

To finish this subsection, the heuristically generalized CGL model with polytropic indices $\gamma_\parallel$, $\gamma_\perp$ yields that in the
steady state the mean temperatures evolve according to
\begin{equation}
  T_\parallel = \left( \frac{\rho}{|\bb|} \right)^{\gamma_\parallel-1}; \qquad T_\perp = |\bb|^{\gamma_\perp-1};
  \qquad \frac{T_\perp}{T_\parallel}=\frac{|\bb|^{\gamma_\parallel+\gamma_\perp -2}}{\rho^{\gamma_\parallel-1}},
\end{equation}
and such heuristic models can yield a wide array of possible temperature profiles.
\subsection{CGL model - summary}
 The usual ``combining'' of evolution equations (by performing $\pr/\pr t$ and substituting one equation to another) is
very useful to gain insight into the linear eigenmodes of a considered system.
However, for more advanced fluid models, such a procedure can become very analytically involved,
 and it is much easier to calculate the determinant of the entire system with analytic software, such as Maple or Mathematica. Therefore, for more advanced
  fluid models we often do not write down the final dispersion relation for arbitrary propagation angle (which might be uneconomically large to write down),
  and we only provide normalized, linearized in the x-z plane and Fourier transformed equations. For easy comparison with more advanced fluid models,
  it is beneficial to write the final CGL equations in the same form. 
\emph{It is useful to work in normalized units, and we drop writing the tilde}. The normalized, linearized in the x-z plane, 
and Fourier transformed CGL equations read 
\begin{eqnarray}
&& -\omega \rho + k_\perp u_x + k_\parallel u_z =0;\\
&& -\omega u_x +\frac{\bpar}{2}k_\perp p_\perp -v_{A\parallel}^2 k_\parallel B_x  +k_\perp B_z =0;\\
&& -\omega u_y - v_{A\parallel}^2 k_\parallel B_y=0;\\
  && -\omega u_z +\frac{\bpar}{2}k_\parallel p_\parallel +\frac{\bpar}{2}(1-a_p)k_\perp B_x=0;\\
  && -\omega B_x -k_\parallel u_x =0;\qquad  -\omega B_y - k_\parallel u_y =0;\qquad  -\omega B_z + k_\perp u_x =0;\\
&& -\omega p_\parallel + k_\perp u_x +3 k_\parallel u_z= 0;\\
&&  -\omega p_\perp  + 2a_p k_\perp u_x +a_p k_\parallel u_z=0,
\end{eqnarray}
where we have introduced the normalized parallel Alfv\'en speed
$v_{A\parallel}^2 \equiv 1+\frac{\bpar}{2}(a_p-1)$ (tilde on $v_{A\parallel}$ is dropped too). 
The determinant can be easily calculated and factorized in Maple, yielding the CGL dispersion relation in normalized units
(tilde are dropped)
\begin{eqnarray}
  &&  \Big(\omega^2-k_\parallel^2 v_{A\parallel}^2\Big)\Big(\omega^4-A_2 \omega^2+A_0\Big) = 0; \label{eq:SFbigmatrix}\\
 &&   A_2 = k_\perp^2 \left(1+a_p\bpar\right) + k_\parallel^2 \left(v_{A\parallel}^2+\frac{3}{2}\bpar\right); \nonumber \\
 && A_0 = \frac{3}{2}k_\parallel^2 \bpar \left[ k_\parallel^2 v_{A\parallel}^2 + 
  k_\perp^2 \left(1+a_p\bpar-\frac{1}{6}a_p^2\bpar\right)  \right], \nn
\end{eqnarray}
which of course agrees with the previously obtained  
CGL solutions for the Alfv\'en mode (\ref{eq:CGLAlfven}), and the slow \& fast modes (\ref{eq:SF}).

\newpage
\section{Hall-CGL model} \label{section:HallCGL}
Here we consider a slightly generalized CGL model when the Hall-term is included in the induction equation.
 Concerning the isotropic Hall-MHD model, a good comparison with solutions of kinetic thoery can be found for example in \cite{Howes2009}.
Before we proceed with the discussion
of dispersion relations, we want to point out how the Hall-term and the electron pressure contributions and the electron inertia contributions
modify the usual ``CGL conservation equations'' for protons. The general electric field was already written down in (\ref{eq:E}).
For a moment it is beneficial to consider this most general form including the
electron inertia and the electron pressure contributions and separate the electric field into two parts
\begin{eqnarray}
  \bE &=& -\frac{1}{c}\bu_p\times\bb + \bE_H\\
  \bE_H &=& \frac{1}{4\pi en}(\nabla\times\bb)\times\bb-\frac{1}{en}\nabla\cdot\bp_e -\frac{m_e}{e}\big( \frac{\pr\bu_e}{\pr t} + \bu_e\cdot\nabla\bu_e \big).
  \label{eq:Ehalll}
\end{eqnarray}
The notation $\bE_H$ is perhaps slightly misleading since only the first term in (\ref{eq:Ehalll}) is usually called the Hall electric field.
Note that the first term in $\bE_H$ still has to be used in the proton
momentum equation even for the simplest CGL (or MHD) models. The induction equation then reads 
\begin{equation}
\frac{\pr \bb}{\pr t} = \nabla\times(\bu_p\times\bb)-c\nabla\times \bE_H,
\end{equation}
which is then rewritten with the use of convective derivative as
\begin{eqnarray}
  \frac{d\bb}{dt} &=& -\bb\nabla\cdot\bu_p +\bb\cdot\nabla\bu_p - c\nabla\times\bE_H;\\
  \frac{\bhat}{|\bb|} \cdot \frac{d\bb}{d t} &=& -\nabla\cdot\bu_p + \bhat\cdot \nabla\bu_p\cdot\bhat
  - \frac{c\bhat}{|\bb|}\cdot (\nabla\times \bE_H).
\end{eqnarray} 
The last expression is then used in the completely general equations (\ref{eq:PperpGen}), (\ref{eq:PparGen}),
which yields
\begin{eqnarray}
  \frac{d}{dt}\left( \frac{p_{\parallel p} |\bb|^2}{\rho_p^3}\right) &=& \frac{|\bb|^2}{\rho_p^3}
  \left( \frac{dp_{\parallel p}}{dt} + p_{\parallel p}\nabla\cdot\bu_p + 2p_{\parallel p} \bhat\cdot \nabla\bu_p\cdot\bhat -2p_{\parallel p}\frac{c\bhat}{|\bb|}\cdot (\nabla\times \bE_H)\right);\\
  \frac{d}{d t} \left(\frac{p_{\perp p}}{\rho_p|\bb|}\right) &=&  \frac{1}{\rho_p|\bb|}\left( \frac{d p_{\perp p}}{dt} + 2p_{\perp p}\nabla\cdot\bu_p
  -p_{\perp p} \bhat\cdot \nabla\bu_p\cdot\bhat +p_{\perp p} \frac{c\bhat}{|\bb|}\cdot (\nabla\times \bE_H) \right).
\end{eqnarray}
By canceling pressure equations (\ref{eq:PparCGL2}), (\ref{eq:PperpCGL2}), that are also the pressure equations of the Hall-CGL model, one obtains that for a general
Hall-CGL model the electric field $\bE_H$ modifies the conservation laws according to
\begin{eqnarray}
  \frac{d}{dt}\left( \frac{p_{\parallel p} |\bb|^2}{\rho_p^3}\right) &=& \frac{p_{\parallel p} |\bb|^2}{\rho_p^3} \bigg[-2 \frac{c\bhat}{|\bb|}\cdot (\nabla\times \bE_H)\bigg];\\
  \frac{d}{d t} \left(\frac{p_{\perp p}}{\rho_p |\bb|}\right) &=&  \frac{p_{\perp p}}{\rho_p|\bb|}\bigg[\frac{c\bhat}{|\bb|}\cdot (\nabla\times \bE_H)\bigg],
\end{eqnarray}
implying the Hall term breaks the 1st and 2nd adiabatic invariants. 
\subsection{Hall-CGL model with cold electrons}
Here we want to study the simplest Hall-CGL model for the proton species. We neglect the electron inertia and we make the electrons cold with $\bp_e=0$.
The full model is described by the usual CGL equations (\ref{eq:CGL_first})-(\ref{eq:PperpCGL_cl}), but the induction equation now reads
\begin{equation}
\frac{\pr \bb}{\pr t} = \nabla\times(\bu_p\times\bb)-\frac{c}{4\pi e}\nabla\times \Big(\frac{1}{n}(\nabla\times\bb)\times\bb \Big).
\end{equation}
The induction equation is linearized as
\begin{equation}
\frac{\pr \bb}{\pr t} \overset{\textrm{\tiny lin}}{=} \nabla\times(\bu_p\times\bb_0)-\frac{c}{4\pi en_0}\nabla\times \Big((\nabla\times\bb)\times\bb_0 \Big).
\end{equation}
By using (\ref{eq:LinParts}), direct calculation of the second term yields
\begin{eqnarray}
\nabla\times\Big((\nabla\times\bb)\times\bb_0 \Big)=  B_0 \left( \begin{array}{c}
   \pr_y\pr_z B_z-\pr_z^2B_y \\ -\pr_x\pr_z B_z+\pr_z^2 B_x \\ \pr_x\pr_z B_y -\pr_y\pr_z B_x \end{array} \right),
\end{eqnarray}
and by using $c/(4\pi e n_0)=V_A^2/(\Omega_p B_0)$, the Hall-term contributions are
\begin{eqnarray}
c\nabla\times\bE_H  \overset{\textrm{\tiny lin}}{=} \frac{V_A^2}{\Omega_p} \left( \begin{array}{c}
   \pr_y\pr_z B_z-\pr_z^2B_y \\ -\pr_x\pr_z B_z+\pr_z^2 B_x \\ \pr_x\pr_z B_y -\pr_y\pr_z B_x \end{array} \right).
\end{eqnarray}  
The entire linearized induction equation reads
\begin{eqnarray}
\frac{\pr \bb}{\pr t} \overset{\textrm{\tiny lin}}{=}  B_0 \left( \begin{array}{c}
  \pr_z u_x \\ \pr_z u_y \\ -\pr_x u_x - \pr_y u_y \end{array} \right) -\frac{V_A^2}{\Omega_p}
  \left( \begin{array}{c} \pr_y\pr_z B_z-\pr_z^2B_y \\ -\pr_x\pr_z B_z+\pr_z^2 B_x \\ \pr_x\pr_z B_y -\pr_y\pr_z B_x \end{array} \right),
\end{eqnarray}  
which when written in the x-z plane (all $\pr_y=0$) yields
\begin{eqnarray}
  \frac{\pr B_x}{\pr t} &=& B_0 \pr_z u_x + \frac{V_A^2}{\Omega_p} \pr_z^2 B_y;\\
  \frac{\pr B_y}{\pr t} &=& B_0 \pr_z u_y + \frac{V_A^2}{\Omega_p} (\pr_x\pr_z B_z - \pr_z^2 B_x);\\
  \frac{\pr B_z}{\pr t} &=& -B_0 \pr_x u_x - \frac{V_A^2}{\Omega_p} \pr_x\pr_z B_y.
\end{eqnarray}  
These equations replace the $\pr B/\pr t$ equations in the linearized CGL  system written in the x-z plane. 
And the trouble is immediately apparent. In contrast to the CGL system, we are no longer able to separate the Alfv\'en mode which was before
exclusively in the $B_y$, $u_y$ components. Now all the components are coupled and all three modes, the Alfv\'en mode, the slow mode and the fast mode are coupled. 
This is not surprising, since the same situation is in the Hall-MHD. In this case, we have no other choice and we have to calculate the dispersion relation for all
three modes. The dispersion relation can be solved analytically only for special cases
and in general, the dispersion relation has to be solved numerically. Let's quickly consider the two special
cases of parallel and perpendicular propagation that can be solved analytically.
\subsection{Parallel propagation} \label{sec:HallCGL_parallel}
For propagation parallel to the mean magnetic field, all $\pr_x=0$ and the familiar sound mode /ion acoustic mode decouples in the $u_z$, $p_\parallel$ components as
\begin{equation}
\frac{\pr^2 u_z}{\pr t^2} -\frac{3p_\parallel^{(0)}}{\rho_0}\pr_z^2 u_z =0,
\end{equation}
with the familiar dispersion relation $\omega^2=C_\parallel^2 k_\parallel^2$. The dispersion relation of this mode does not depend on the Hall-term.
However, the other two modes are now coupled through
\begin{eqnarray}
&&  \frac{\pr u_x}{\pr t} -\frac{1}{B_0}\Big( V_A^2 -\frac{p_\parallel^{(0)}}{\rho_0}+\frac{p_\perp^{(0)}}{\rho_0} \Big) \pr_z B_x =0;\label{eq:HallColdUx}\\
&&  \frac{\pr u_y}{\pr t} -\frac{1}{B_0}\Big( V_A^2 -\frac{p_\parallel^{(0)}}{\rho_0}+\frac{p_\perp^{(0)}}{\rho_0} \Big) \pr_z B_y =0;\label{eq:HallColdUy}\\
&&  \frac{\pr B_x}{\pr t} = B_0 \pr_z u_x + \frac{V_A^2}{\Omega_p} \pr_z^2 B_y; \label{eq:HallColdBx}\\
&&  \frac{\pr B_y}{\pr t} = B_0 \pr_z u_y - \frac{V_A^2}{\Omega_p} \pr_z^2 B_x. \label{eq:HallColdBy}
\end{eqnarray}
If the last terms in the B-field equations are neglected, i.e. if the Hall-term is neglected, the two modes decouple and one naturally recovers the CGL
parallel propagating Alfv\'en modes, both propagating with the phase speed
\begin{equation}
v_{A\parallel}^2=V_A^2 -\frac{p_\parallel^{(0)}}{\rho_0}+\frac{p_\perp^{(0)}}{\rho_0} = V_A^2 \Big( 1+\frac{\bpar}{2}(a_p-1)\Big).
\end{equation}
Applying $\pr_t$ at the B-field equations yields
\begin{eqnarray}
&&  \frac{\pr^2 B_x}{\pr t^2} = v_{A\parallel}^2 \pr_z^2 B_x + \frac{V_A^2}{\Omega_p} \pr_z^2 \frac{\pr B_y}{\pr t};\\
&&  \frac{\pr^2 B_y}{\pr t^2} = v_{A\parallel}^2 \pr_z^2 B_y - \frac{V_A^2}{\Omega_p} \pr_z^2 \frac{\pr B_x}{\pr t},
\end{eqnarray}
which when transformed to Fourier space reads
\begin{eqnarray}
\left( \begin{array}{cc}
  \omega^2-v_{A\parallel}^2 k_\parallel^2; &\qquad +i\frac{V_A^2}{\Omega_p}k_\parallel^2 \omega\\
    -i\frac{V_A^2}{\Omega_p}k_\parallel^2 \omega; & \qquad \omega^2-v_{A\parallel}^2 k_\parallel^2
\end{array} \right)
\left( \begin{array}{c} B_x \\ B_y \end{array} \right) = \left( \begin{array}{c} 0 \\ 0 \end{array} \right), \label{eq:Hall_matrix}
\end{eqnarray}
and the zero determinant requirement implies
\begin{equation} \label{eq:ICWdisper}
\omega^4 - \omega^2 k_\parallel^2\Big( 2 v_{A\parallel}^2 +k_\parallel^2 \frac{V_A^4}{\Omega_p^2}\Big) +k_\parallel^4 v_{A\parallel}^4 =0.
\end{equation}
The solutions of the dispersion relation (quadratic polynomial in $\omega^2$) is simply
\begin{eqnarray}
  \omega_W^2 = k_\parallel^2 V_A^2 \bigg[ 1+\frac{\bpar}{2}(a_p-1) +\frac{k_\parallel^2 V_A^2}{2\Omega_p^2}+ \frac{|k_\parallel| V_A}{\Omega_p}\sqrt{1+\frac{\bpar}{2}(a_p-1)
      +\frac{k_\parallel^2 V_A^2}{4\Omega_p^2}  } \;\; \bigg]; \label{eq:Wpar1} \\
  \omega_{IC}^2 = k_\parallel^2 V_A^2 \bigg[ 1+\frac{\bpar}{2}(a_p-1) +\frac{k_\parallel^2 V_A^2}{2\Omega_p^2}- \frac{|k_\parallel| V_A}{\Omega_p}\sqrt{1+\frac{\bpar}{2}(a_p-1)
      +\frac{k_\parallel^2 V_A^2}{4\Omega_p^2}  } \;\; \bigg], \label{eq:ICpar1}
\end{eqnarray}  
where importantly, the $\sqrt{\kpar^2}=|\kpar|$ was used. 
The solution (\ref{eq:Wpar1}) is the \emph{whistler} mode, and the solution (\ref{eq:ICpar1}) is the \emph{ion-cyclotron} mode. 
For isotropic temperatures ($a_p=1$), the result is equal to the Hall-MHD dispersion relation.
At long wavelengths ($k_\parallel\rightarrow 0$) the solutions become non-dispersive and connect to the usual CGL Alfv\'en modes.
For parallel propagation, the Hall-term is therefore responsible for splitting the two modes that otherwise would propagate at the same speed, 
while it does not influence the ion-acoustic mode. Or in another words, using the vocabulary of slow, Alfv\'en, fast,
depending on plasma $\bpar$ and temperature anisotropy $a_p$, the Hall-term either splits the parallel propagating Alfv\'en and fast mode,
or the slow and Alfv\'en mode. 
At very short spatial scales $(k_\parallel \gg 1)$, the limits are
\footnote{Note that the limit for the ion-cyclotron mode has to be calculated with a trick $a-b = (a^2-b^2)/(a+b)$ since both $a,b$ go to $\infty$.}
\begin{equation}
k_\parallel\rightarrow\pm \infty: \qquad  \omega_{IC}^2 = \Omega_p^2 \Big(1+\frac{\bpar}{2}(a_p-1)\Big)^2; \qquad \omega_W^2 = k_\parallel^4 \frac{V_A^4}{\Omega_p^2}. \label{eq:Hallres}
\end{equation}
For isotropic temperatures $a_p=1$ the frequency of the ion-cyclotron mode (that should be actually called the proton-cyclotron mode) converges towards the proton cyclotron
frequency $\pm \Omega_p$. For anisotropic temperatures, the frequency of the mode starts to be $\bpar$ dependent and will converge to frequencies that can be higher or lower
than the $\Omega_p$. The frequency of the whistler mode is not bounded and technically
goes to infinity, which is a consequence of neglecting the electron inertia. Importantly, both modes are stable for high wavenumbers. 
The solutions (\ref{eq:Wpar1}), (\ref{eq:ICpar1}) are analytically correct, but similarly to the Hall-MHD, the solutions can be rewritten to a more useful form, 
by realizing that the dispersion equation (\ref{eq:ICWdisper}) can be decomposed as
\footnote{One can double check the result by working with quantities $B^\pm = iB_x \pm B_y$, yielding
 $\Big( \omega^2\pm\frac{k_\parallel^2 V_A^2}{\Omega_p}\omega -k_\parallel^2 v_{A\parallel}^2\Big)B^\mp =0$.}
\begin{equation}
  \Big( \omega^2+\frac{k_\parallel^2 V_A^2}{\Omega_p}\omega -k_\parallel^2 v_{A\parallel}^2\Big)
  \Big( \omega^2-\frac{k_\parallel^2 V_A^2}{\Omega_p}\omega -k_\parallel^2 v_{A\parallel}^2\Big) = 0. \label{eq:Hall-parF}
\end{equation}
The first bracket yields two solutions
\begin{eqnarray}
&&  \omega_{IC,1} = -\frac{k_\parallel^2 V_A^2}{2\Omega_p} + |k_\parallel| V_A \sqrt{1+\frac{\bpar}{2}(a_p-1) +\frac{k_\parallel^2V_A^2}{4\Omega_p^2} }; \qquad - \textrm{Green line};\label{eq:Hall-IC1}\\
  &&  \omega_{W,1} = -\frac{k_\parallel^2 V_A^2}{2\Omega_p} - |k_\parallel| V_A \sqrt{1+\frac{\bpar}{2}(a_p-1) +\frac{k_\parallel^2V_A^2}{4\Omega_p^2} }; \qquad - \textrm{dashed Magenta line},
  \label{eq:Hall-W1}
\end{eqnarray}
and the second bracket yields two solutions
\begin{eqnarray}
  &&  \omega_{W,2} = +\frac{k_\parallel^2 V_A^2}{2\Omega_p} + |k_\parallel| V_A \sqrt{1+\frac{\bpar}{2}(a_p-1) +\frac{k_\parallel^2V_A^2}{4\Omega_p^2} }; \qquad - \textrm{Blue line};\label{eq:Hall-W2}\\
  &&  \omega_{IC,2} = +\frac{k_\parallel^2 V_A^2}{2\Omega_p} - |k_\parallel| V_A \sqrt{1+\frac{\bpar}{2}(a_p-1) +\frac{k_\parallel^2V_A^2}{4\Omega_p^2} }; \qquad - \textrm{dashed Cyan line}.\label{eq:Hall-IC2}
\end{eqnarray}
 It is straightforward to check that $(\omega-\omega_{W,1})(\omega-\omega_{W,2})=0$ yields the whistler mode solution (\ref{eq:Wpar1}), and
$(\omega-\omega_{IC,1})(\omega-\omega_{IC,2})=0$ yields the ion-cyclotron mode solution (\ref{eq:ICpar1}). Often, solutions (\ref{eq:Hall-IC1})-(\ref{eq:Hall-IC2}) are
written in an alternative form where the absolute value on $k_\parallel$ is removed (which obviously can be done), and the solutions are just slightly re-arranged.   
Additionally, for more complicated fluid models, we want to write solution in one line instead of four lines. In this example, we just write a shortcut that two solutions are 
\begin{equation} \label{eq:ICW_pica}
\omega = \pm\frac{k_\parallel^2 V_A^2}{2\Omega_p} + k_\parallel V_A \sqrt{1+\frac{\bpar}{2}(a_p-1) +\frac{k_\parallel^2V_A^2}{4\Omega_p^2} },
\end{equation}
with another two solutions obtained by substituting $\omega$ with $-\omega$.

\subsubsection{Firehose instability} \label{sec:FInstability}
\begin{figure*}
$$\includegraphics[width=0.48\linewidth]{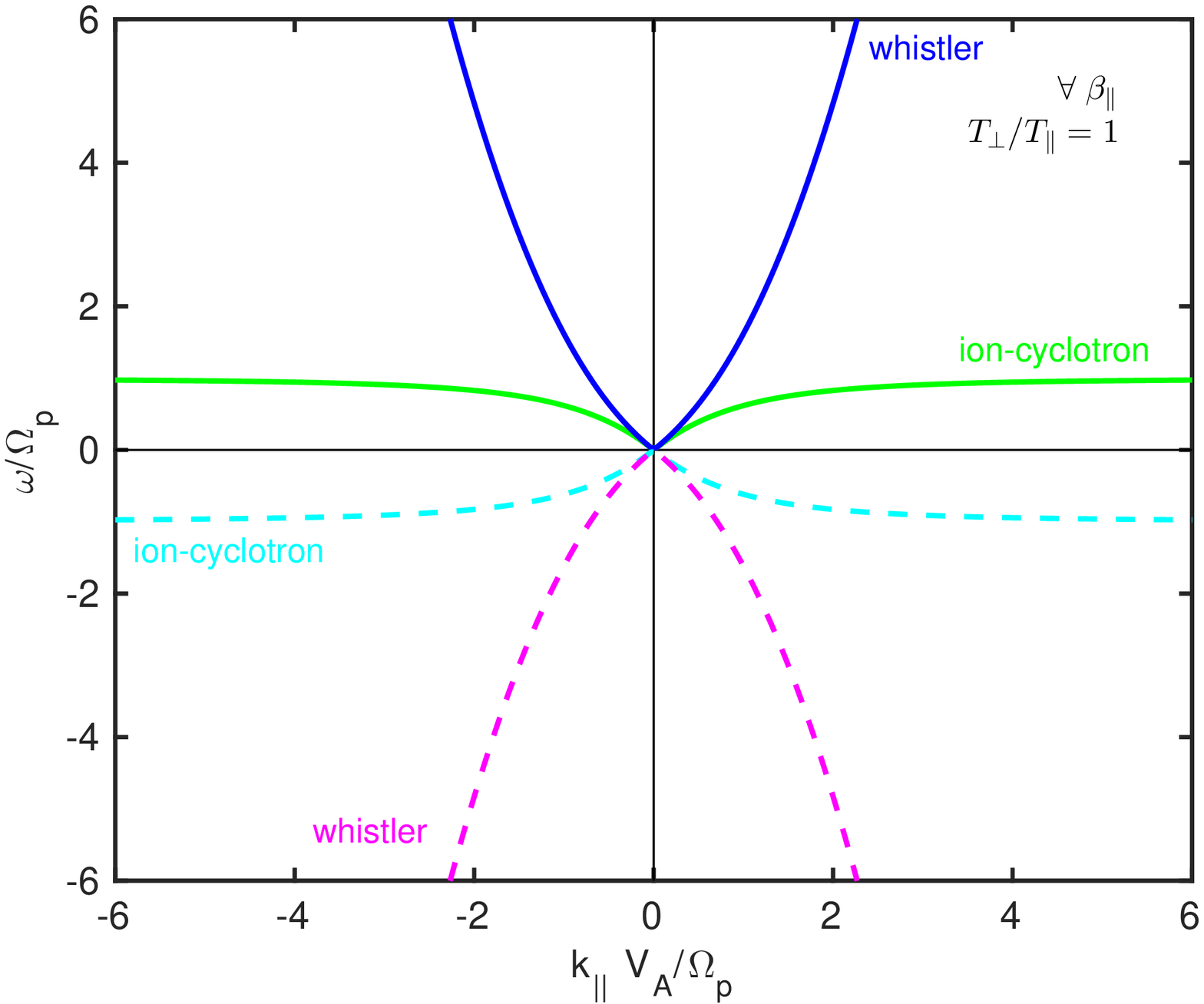}\hspace{0.03\textwidth}\includegraphics[width=0.48\linewidth]{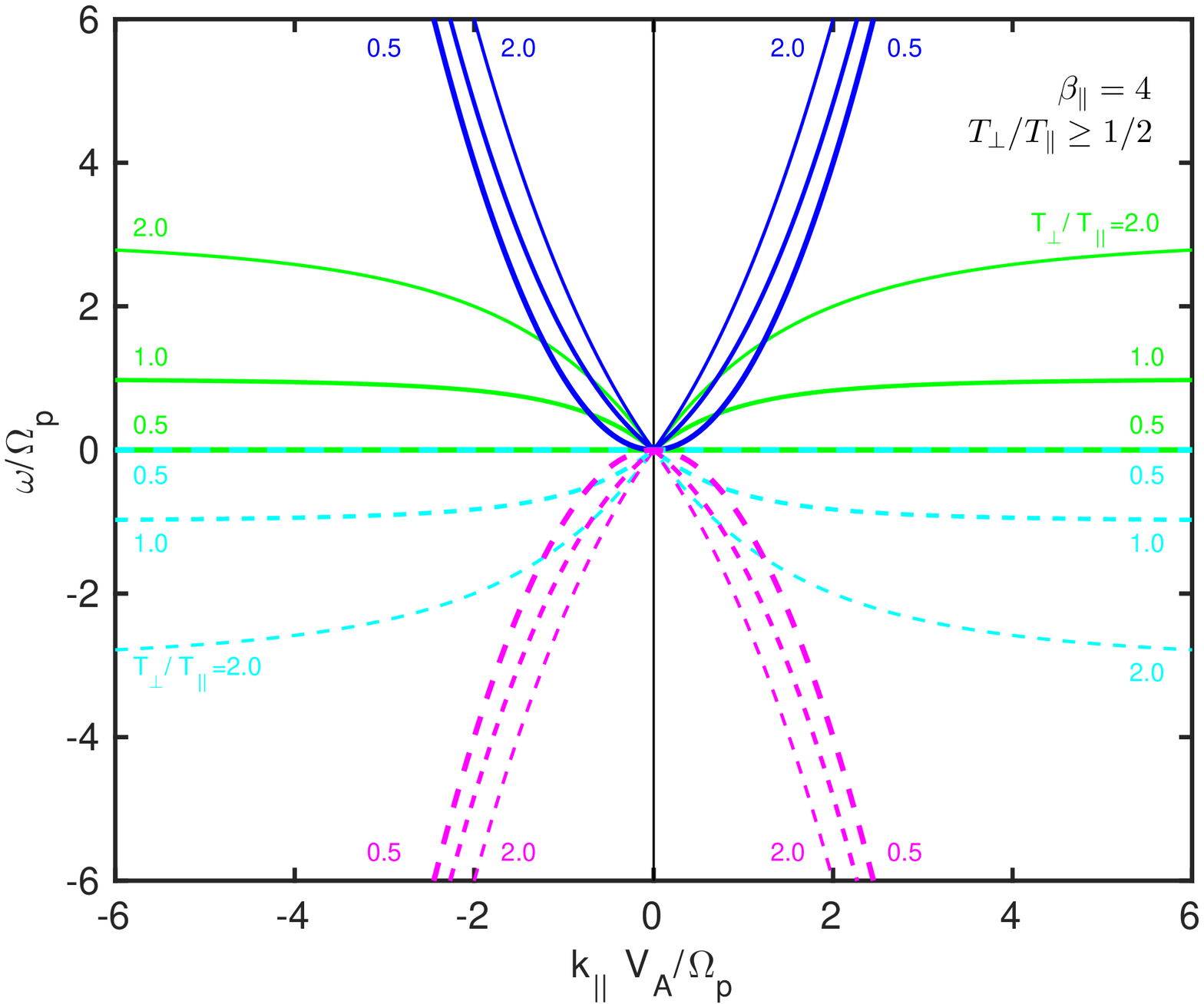}$$
  \caption{Dispersion relations for the Hall-CGL model (with cold electrons), for parallel propagation $\theta=0$. The color coding of lines is according to
    (\ref{eq:Hall-IC1})-(\ref{eq:Hall-IC2}).
    Left panel: isotropic temperatures $a_p=T_\perp/T_\parallel=1.0$. The dispersion relations do not depend on the value of $\bpar$, which is a consequence of neglecting
  the FLR pressure corrections, etc. The figure is actually equivalent to the Hall-MHD model. Right panel: $\bpar$ is fixed with the value $\bpar=4.0$
  and the temperature anisotropy is varied as $a_p=2.0; 1.0; 0.5$. The thickness of the curves increases with decreasing $a_p$ as we approach the 
  firehose threshold at $a_p=0.5$. All the curves have purely real frequency $\omega$.} \label{fig:4}
\end{figure*}

\begin{figure*}
  $$\includegraphics[width=0.48\linewidth]{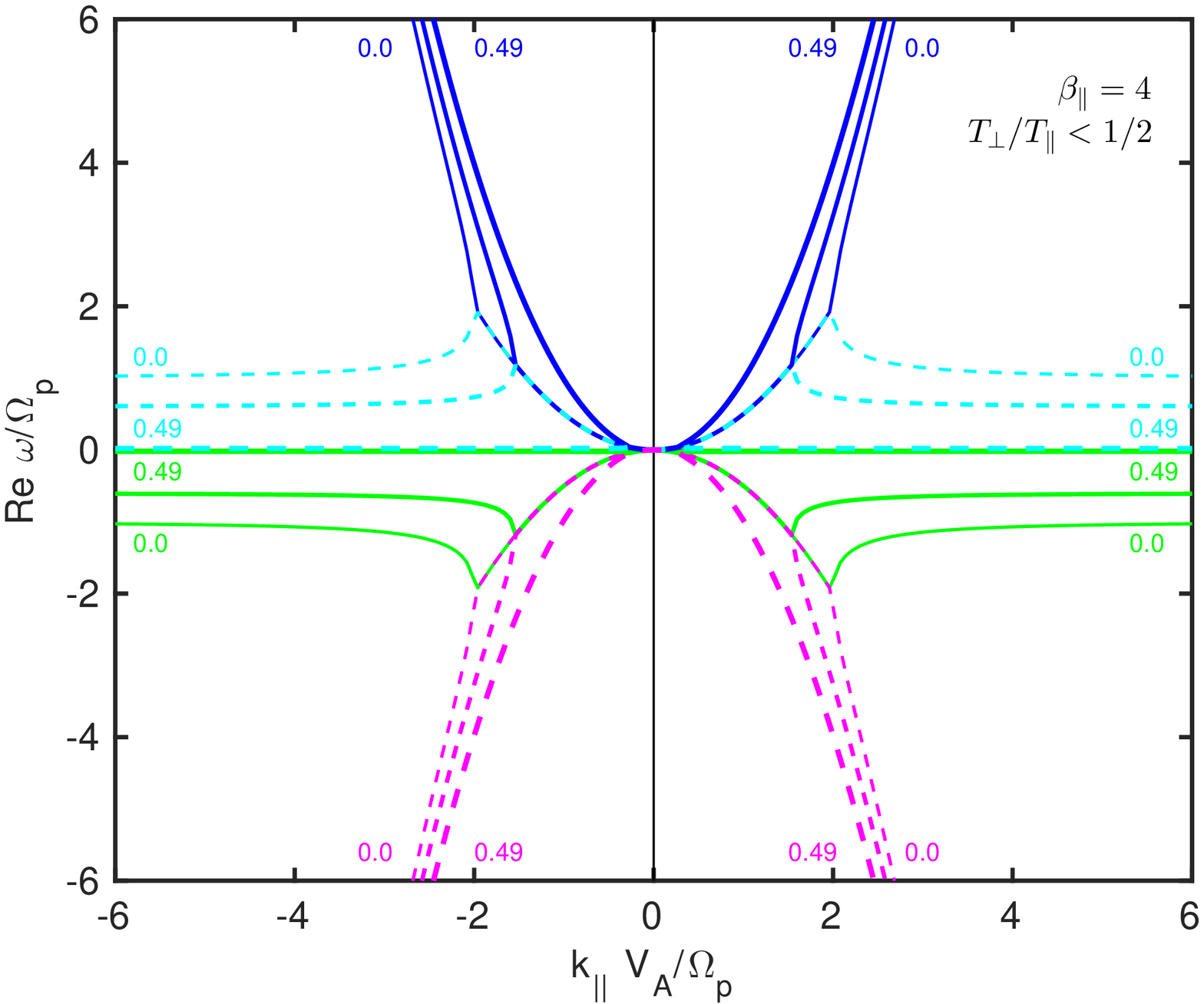}\hspace{0.03\textwidth}\includegraphics[width=0.48\linewidth]{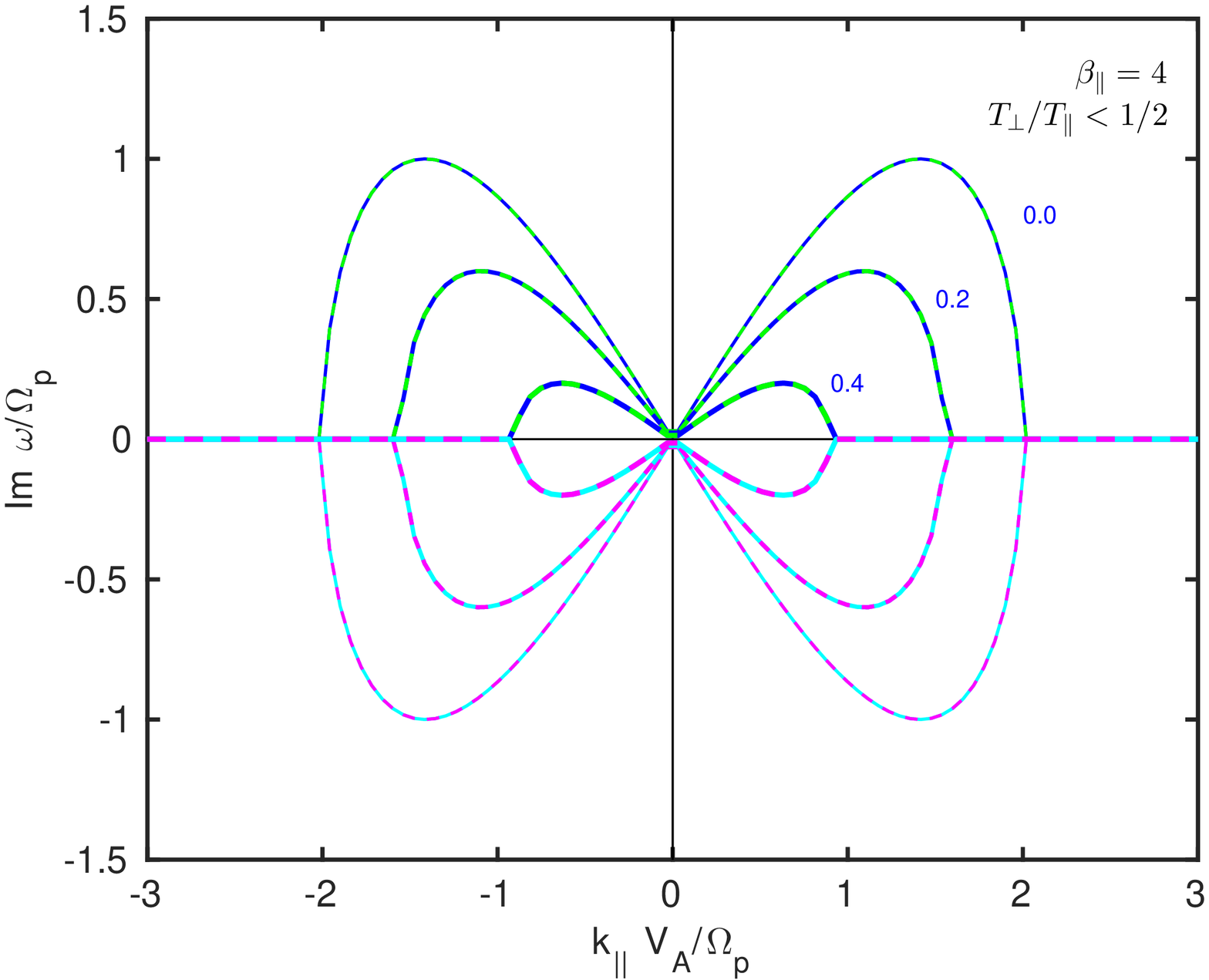}$$
  \caption{Dispersion relations for the Hall-CGL model (with cold electrons), for parallel propagation $\theta=0$, with fixed $\bpar=4$, after crossing the firehose threshold
    $a_p=T_\perp/T_\parallel<0.5$. The color coding of lines is according to
    (\ref{eq:Hall-IC1})-(\ref{eq:Hall-IC2}). The thickness of the lines decreases as going away from the threshold $a_p=0.5$.
    Left panel: real frequency and the curves have $a_p=0.49; 0.2; 0.0$. Right panel: imaginary frequency and the curves have
    $a_p=0.4; 0.2; 0.0$. The mode with $a_p=0.49$ shown in the left panel has the imaginary part of the frequency which is very close to zero
    for the entire range of $k_\parallel$. For this reason
    this mode is not plotted in the right panel, where instead a mode with $a_p=0.4$ is plotted. A mode that has $\textrm{Im}\, \omega>0$ is unstable and growing in time.
    A mode that has $\textrm{Im}\, \omega<0$ is stable and damped.  According to the figure, whistler modes with $\omega_r>0$ are unstable,
    and ion-cyclotron modes with $\omega_r<0$ are unstable. It is assumed that $\sqrt{-1}=+i$. The solutions for $\omega_r<0$ are
      here non-causal, and the right panel should be re-plotted with causal solutions (\ref{eq:Hall-IC1corr})-(\ref{eq:Hall-IC2corr}),
      so that whistler modes are always unstable, and ion-cyclotron modes stable. } \label{fig:5}
\end{figure*}

Now we can discuss the firehose instability. If the quantity $1+\frac{\bpar}{2}(a_p-1)>0$, then all the solutions (\ref{eq:Hall-IC1})-(\ref{eq:Hall-IC2})
have frequencies that are purely real for the entire range of wavenumbers.
 For isotropic temperatures $a_p=1$, the dispersion relations do not depend on the value of $\bpar$, which is a consequence of neglecting the FLR pressure corrections as will be discussed later,
and all solutions are equivalent to the Hall-MHD dispersion relations.
 Solutions (\ref{eq:Hall-IC1})-(\ref{eq:Hall-IC2}) with $a_p=1$ are plotted in Figure \ref{fig:4} left panel, where we plotted both positive and negative
wavenumbers and positive and negative frequencies to clearly understand how the modes are connected to each other. The magenta and cyan lines are plotted as dashed lines,
  so that the growth rates in the firehose-unstable regime (plotted in Figure \ref{fig:5}) are clearly visible.
For anisotropic temperatures $a_p\neq 1$, the solutions are naturally $\bpar$ dependent. To show solutions in a fire-hose unstable regime, one has to choose $\bpar>2$ since the
instability does not exist for lower values of $\bpar$. In Figure \ref{fig:4}, right panel, we have chosen a value of $\bpar=4$ and varied the temperature
anisotropy from $a_p=2$ down to a critical value of $a_p=0.5$. In this regime, all the modes have purely real frequencies. 

Importantly, if the temperature anisotropy parameter $a_p$ is further decreased, the frequencies of all the modes will become complex numbers (with a real and imaginary part)
in the region where
\begin{equation} \label{eq:HallCGL-firehose}
1+\frac{\bpar}{2}(a_p-1) +\frac{k_\parallel^2V_A^2}{4\Omega_p^2} <0,
\end{equation}
which can be viewed as a modified Hall-CGL firehose instability threshold that is now length-scale dependent. At long spatial scales ($k_\parallel\rightarrow 0$),
the above criterium is naturally equivalent to the CGL firehose criterium. 
The term proportional to $k_\parallel^2$ in (\ref{eq:HallCGL-firehose}) is always positive and as $k_\parallel$
increases, after some critical wavenumber the frequency of modes will become purely real. 
The Hall-term is therefore responsible for the stabilization of the firehose instability at sufficiently high wavenumbers (small spatial scales).
The solutions in this firehose unstable regime are plotted in Figure \ref{fig:5}, where again $\bpar=4$ and the temperature anisotropy is varied from $a_p=0.49$ to the
lowest possible value of $a_p=0.0$. The left panel shows real frequencies and the right panel shows imaginary frequencies.   

It is of interest to determine, which mode becomes unstable. According to our Fourier transform (see Appendix \ref{sec:AppendixA}),
a mode with positive imaginary frequency is unstable and growing, and a mode with negative imaginary frequency is stable and damped.
If the firehose threshold (\ref{eq:HallCGL-firehose}) is satisfied, for some range of $\kpar$ the second terms in solutions
(\ref{eq:Hall-IC1})-(\ref{eq:Hall-IC2}) become purely imaginary
and the first terms stay purely real. Consider one of the four possible quadrants in the Figure \ref{fig:5} left, for example the quadrant with $k_\parallel>0; \omega_r>0$.
Up to some critical $k_\parallel$ where the Hall-CGL firehose criterium is satisfied, both the  blue mode and the cyan mode propagate with the same real frequency 
$\omega_r=+k_\parallel^2 V_A^2/(2\Omega_p)$. The differences are in the imaginary frequency,  see Figure \ref{fig:5} right,
where in this quadrant the blue mode is unstable and the cyan mode is stable (and damped). After a critical $\kpar$ (for example for $a_p=0$ it is $\kpar V_A/\Omega_p=2$),
both modes become purely real and evolve with different real frequencies, here the blue mode is obviously the whistler mode and the cyan mode is the ion-cyclotron mode. 
The same situation is in the quadrant $k_\parallel<0; \omega_r>0$. We can therefore conclude
that for positive real frequencies, $\omega_r>0$, the whistler mode is unstable and the ion-cyclotron mode is stable and damped.

Now we could say that considering both positive and negative wavenumbers and frequencies is redundant, and
  that we are finished. However, from a perspective of this guide, we consider useful to show that one might easily come into contradictions,
  if negative frequencies are not handled with sufficient care. Importantly, according to Figure \ref{fig:5}, for negative real frequencies $\omega_r<0$,
  the whistler mode is stable and the ion-cyclotron mode is unstable.
  This is indeed contradictory. Keeping the $k_\parallel>0$, a change from positive to negative $\omega_r$ represents a change
  of direction of propagation along $B_0$. One can argue, that in a homogeneous system (e.g. excluding propagation in a stratified fluid),
  a change of direction of propagation should not yield a change of stability. 
  Additionally, kinetic theory and simulations show, that it is the whistler mode that is firehose unstable
  for both $\omega_r>0$ and $\omega_r<0$, and the ion-cyclotron mode is stable.
  It is important to emphasize that from a perspective of solutions (\ref{eq:Hall-IC1})-(\ref{eq:Hall-IC2}),
  at the range of wavenumbers where the firehose instability is present, the whistler and ion-cyclotron modes propagate (in a given quadrant)
  with the same real frequency. Moreover, both modes have the same polarization of electric and magnetic field (see Section \ref{sec:polarization}).
  Both modes are completely degenerate, and not distinguishable. Thus, how the modes are continued for higher wavenumbers
  once the firehose instability disappears, is not obvious (since $\sqrt{-1}$ can be possibly both $+i$ or $-i$).
  One should consider all the possible solutions and verify, which solutions are physically plausible.
  Therefore, after the firehose threshold is crossed, the analytic solutions (\ref{eq:Hall-IC1})-(\ref{eq:Hall-IC2}) are not correctly separated.
  Obviously, more ``physical information'' is needed to distinguish between the ion-cyclotron and whistler modes in the
  firehose unstable regime, and to determine which solutions are causal and which are non-causal.

\subsubsection{Generalization to causal (correct) solutions}
The mind boggling puzzle, why the Hall-CGL solutions (\ref{eq:Hall-IC1})-(\ref{eq:Hall-IC2}) are in contradiction to kinetic theory
  and to common sense (a discrepancy found by one of the Referees of this text), was 
  beautifully solved by Paul Cally. The easiest way to obtain correct solutions is to introduce a very small ``causal'' dissipation into the system,
  and remove it later. Alternatively, it should be possible to obtain the same result by using
  Laplace-Fourier transforms on temporal-spatial scales and considering an initial value problem.

We introduce causality, by adding a tiny amount of dissipation on the right hand sides of momentum equations
(\ref{eq:HallColdUx})-(\ref{eq:HallColdUy}), in the following form
\begin{eqnarray}
&&  \frac{\pr u_x}{\pr t} -v_{A\parallel}^2 \pr_z \frac{B_x}{B_0} = + \epsilon \frac{2V_A^2}{\Omega_p} \pr_z^2 u_x;\\
&&  \frac{\pr u_y}{\pr t} -v_{A\parallel}^2 \pr_z \frac{B_y}{B_0} = + \epsilon \frac{2V_A^2}{\Omega_p} \pr_z^2 u_y,
\end{eqnarray}  
where $\epsilon$ is a dimensionless real small positive number, for example $\epsilon=10^{-8}$. The dissipation will be later completely removed
by the limit $\lim_{\epsilon\to 0^+}$. The $V_A^2/\Omega_p$ is there so that
the parameter $\epsilon$ is indeed dimensionless (since here we usually normalize with respect to $V_A$ and $\Omega_p$, but other normalizations
can be of course considered). The factor of 2 is there for pure convenience, so that the final dispersion relations
(\ref{eq:Hall-IC1corr})-(\ref{eq:Hall-IC2corr}) are of the utmost beauty, and that we don't have to later redefine $\epsilon$. 
The momentum equations are accompanied by the induction equations (\ref{eq:HallColdBx})-(\ref{eq:HallColdBy}). The reader is encouraged to verify the calculations
by switching to normalized variables $\widetilde{\kpar}=\kpar V_A/\Omega_p$, $\widetilde{\omega}=\omega/\Omega_p$,
$\widetilde{v}_{A\parallel}^2=1+\frac{\bpar}{2}(a_p-1)$, we continue in physical units. 

By adding the dissipation, the dispersion relation (\ref{eq:Hall-parF}) is modified to the following form
\begin{eqnarray}
&  \Big[ \omega^2+\frac{k_\parallel^2 V_A^2}{\Omega_p} \omega(1+2i\epsilon) -k_\parallel^2 v_{A\parallel}^2 +2i\epsilon \frac{\kpar^4 V_A^4}{\Omega_p^2}\Big] (iB_x-B_y) = 0;\\
&  \Big[ \omega^2-\frac{k_\parallel^2 V_A^2}{\Omega_p} \omega(1-2i\epsilon) -k_\parallel^2 v_{A\parallel}^2 -2i\epsilon \frac{\kpar^4 V_A^4}{\Omega_p^2}\Big] (iB_x+B_y) = 0.
\end{eqnarray}
It is useful to define the Hall-CGL firehose threshold 
\begin{equation}  
  \Delta\equiv 1+\frac{\bpar}{2}(a_p-1) +\frac{k_\parallel^2V_A^2}{4\Omega_p^2},
\end{equation}
and similar generalizations can be done for models with FLR corrections. Then, since $\epsilon$ is small,  
it is easy to show that the non-causal solutions (\ref{eq:Hall-IC1})-(\ref{eq:Hall-IC2}) are ``corrected'' by the tiny dissipation to the following causal form
\begin{eqnarray}
  \omega_{IC,1} &=& -\frac{k_\parallel^2 V_A^2}{2\Omega_p} +  |k_\parallel| V_A  \lim_{\epsilon\to 0^+}\sqrt{\Delta -i\epsilon}; \qquad - \textrm{Green line};\label{eq:Hall-IC1corr}\\
  \omega_{W,1} &=&  -\frac{k_\parallel^2 V_A^2}{2\Omega_p} -  |k_\parallel| V_A  \lim_{\epsilon\to 0^+}\sqrt{\Delta -i\epsilon}; \qquad - \textrm{dashed Magenta line};
  \label{eq:Hall-W1corr}\\
  \omega_{W,2} &=&  +\frac{k_\parallel^2 V_A^2}{2\Omega_p} +  |k_\parallel| V_A  \lim_{\epsilon\to 0^+}\sqrt{\Delta +i\epsilon}; \qquad - \textrm{Blue line};\label{eq:Hall-W2corr}\\
  \omega_{IC,2} &=& +\frac{k_\parallel^2 V_A^2}{2\Omega_p} -  |k_\parallel| V_A  \lim_{\epsilon\to 0^+}\sqrt{\Delta +i\epsilon}; \qquad - \textrm{dashed Cyan line}.\label{eq:Hall-IC2corr}
\end{eqnarray}
The reader is encouraged to plot these solutions, for example with $\epsilon=10^{-8}$, and verify that the right panel of Figure \ref{fig:5}, is indeed
corrected. I.e., that the whistler is unstable for both $\omega_r>0$ and $\omega_r<0$. We want to further address the limits.
For clarity, let's first consider the following simplest case   
  \begin{equation} \label{eq:square}
  \lim_{\epsilon\to 0^+} \sqrt{-1\pm i\epsilon} = \frac{\epsilon}{2}\pm i,
  \end{equation}
where one can further suppress the small real $\epsilon/2$ on the right hand side. The limit (\ref{eq:square}) provides an answer, when the
  $\sqrt{-1}$ is equal to $+i$ or $-i$.  The result (\ref{eq:square}) can be easily generalized, and limits in expressions
  (\ref{eq:Hall-IC1corr})-(\ref{eq:Hall-IC2corr}) are calculated according to  
  \begin{equation}
  \lim_{\epsilon\to 0^+} \sqrt{\Delta\pm i\epsilon} = \begin{dcases}
    \sqrt{\Delta}, & \quad \text{$\Delta>0$};\\
    0, & \quad \text{$\Delta=0$}; \\
    \pm i\sqrt{-\Delta}, & \quad \text{$\Delta<0$}. 
  \end{dcases}
  \end{equation}
  Obviously, the whistler solutions (\ref{eq:Hall-W1corr}), (\ref{eq:Hall-W2corr}) are unstable, and the ion-cyclotron solutions
  (\ref{eq:Hall-IC1corr}), (\ref{eq:Hall-IC2corr}) are damped. Only now we can conclude with confidence,
  that the parallel firehose instability as an instability of the whistler mode.

  We note that if instead of the momentum equations, we added the dissipation (diffusivity) to the induction equations,
  the solutions (\ref{eq:Hall-IC1corr})-(\ref{eq:Hall-IC2corr}) are recovered with $\epsilon\to-\epsilon$.
  I.e., by adding a dissipation to the induction equations, the whistler mode would be always damped and the ion-cyclotron mode unstable.
  The result makes sense with respect to kinetic theory, since the ion-cyclotron resonances enter through the velocity moment (by integration over the
  perturbation of the distribution function $f^{(1)}$), or more precisely through the current $\boldsymbol{j}=\sum_r q_r n_r \bu_r$, see Part 2.

\subsubsection{Polarization} \label{sec:polarization}
The polarization of the magnetic field can be easily checked, actually without plugging any dispersion relations into the matrix (\ref{eq:Hall_matrix}),
since the matrix can be represented simply as
\begin{eqnarray}
\left( \begin{array}{cc}
  a &\qquad +ib\\
    -ib & \qquad a
\end{array} \right)
\left( \begin{array}{c} B_x \\ B_y \end{array} \right) = \left( \begin{array}{c} 0 \\ 0 \end{array} \right), \label{eq:Hall_matrix2}
\end{eqnarray}
where $a= \omega^2-v_{A\parallel}^2 k_\parallel^2$ and $b=V_A^2 k_\parallel^2 \omega/\Omega_p$. It is useful to define the polarization as the angle between
the complex $B_x$ and $B_y$ components, which is calculated as $\textrm{Arg}(B_y/B_x)$. 
If the result is smaller than zero, the wave is left polarized. If the
result is larger than zero, the wave is right polarized. The determinant of the matrix (\ref{eq:Hall_matrix2}) is $a^2-b^2=0$, so $a=\pm b$.
The case $a=-b$ (which is directly equal to the first bracket in (\ref{eq:Hall-parF})  with solutions (\ref{eq:Hall-IC1}), (\ref{eq:Hall-W1})) yields that $B_y=-iB_x$, so that
\begin{equation} \label{eq:Bpol_pic1}
\frac{B_y}{B_x} = -i = e^{-i\frac{\pi}{2}}; \quad => \quad \textrm{Arg}\frac{B_y}{B_x} = -\frac{\pi}{2} <0 \qquad (\textrm{left polarized}). 
\end{equation}
The second case $a=+b$ (which is equal to the second bracket  in (\ref{eq:Hall-parF})  with solutions (\ref{eq:Hall-W2}), (\ref{eq:Hall-IC2})) yields that $B_y=+iB_x$, so that
\begin{equation} \label{eq:Bpol_pic2}
\frac{B_y}{B_x} = +i = e^{+i\frac{\pi}{2}}; \quad => \quad \textrm{Arg}\frac{B_y}{B_x} = +\frac{\pi}{2} >0 \qquad (\textrm{right polarized}). 
\end{equation}
Similarly, one can calculate the polarization of electric field $\boldsymbol{E}$, and (when not specified that the $\boldsymbol{B}$ field is considered),
the words left/right polarized are usually meant for the $\boldsymbol{E}$ field.
The ratio $E_y/E_x$ can be easily calculated, for example from equations (\ref{eq:Efield_Rit}), (\ref{eq:Efield_Rit2}) (written there in the normalized form),
which when multiplied by $\omega$ and use of momentum equations, yields for the special case of parallel propagation considered here
\begin{equation}
\frac{E_y}{E_x}=\frac{-\widetilde{v}_{A\parallel}^2+i\widetilde{\omega}\frac{B_y}{B_x}}{\widetilde{v}_{A\parallel}^2 \frac{B_y}{B_x} +i\widetilde{\omega}}.   
\end{equation}
When plotting many solutions, it is useful to additionally divide the angle by $\pi$,
and define the $\boldsymbol{B}$ and $\boldsymbol{E}$ field polarizations as
\begin{equation} \label{eq:PolDef}
\mathcal{P}_B = \frac{1}{\pi}\textrm{Arg}\frac{B_y}{B_x}; \qquad \mathcal{P}_E = \frac{1}{\pi}\textrm{Arg}\frac{E_y}{E_x}.
\end{equation}
Electric field polarization defined this way was used for example by \cite{Hunana2013} to investigate properties of (highly oblique) kinetic Alfv\'en waves in various
fluid models and kinetic description; and also by \cite{CamporealeBurgess2017} who compared various linear modes in hybrid, gyrokinetic and fully kinetic descriptions.
\footnote{Similar definition was used by \cite{Sahraoui2012}, who however plot a bit confusing $\textrm{Arg}(E_y)/\pi$, since in the WHAMP code the electric field is normalized
  so that $\textrm{Arg}(E_x)=0$.}
When $\mathcal{P}_E<0$ the wave is called \emph{left polarized} and when  $\mathcal{P}_E>0$, the wave is called \emph{right polarized}. 
For the first group (\ref{eq:Bpol_pic1}) this yields $E_y/E_x=-i$ and $\mathcal{P}_E=-1/2<0$ (left polarized wave), and for the second group (\ref{eq:Bpol_pic2})
$E_y/E_x=+i$ and $\mathcal{P}_E=+1/2>0$ (right polarized wave). 

 Finally, it is of interest to note another possible definition of polarization, that is defined with respect to positive real frequency, i.e.
  the electric field polarization (\ref{eq:PolDef}) is redefined to  
\begin{equation} \label{eq:PolDef2}
\mathcal{P}_E = \frac{\sign(\omega_r)}{\pi}\textrm{Arg}\frac{E_y}{E_x}.
\end{equation}
A similar definition of polarization with $\sign(\omega_r)$ is used for example in the book by \cite{Gary1993}, page 91.
With such a definition, \emph{the whistler mode is always right polarized},
both in the firehose-stable and firehose-unstable regimes, since for the whistler mode (\ref{eq:Hall-W2}, blue line) the real frequency remains $\omega_r>0$, and for the
whistler mode (\ref{eq:Hall-W1}, magenta line) $\omega_r<0$. The situation is more confusing for the ion-cyclotron mode, where the firehose-stable and
firehose-unstable regimes have to be considered separately. 1) In the firehose-stable regime, both ion-cyclotron modes are left polarized,
since for the mode (\ref{eq:Hall-IC1}, green line) $\omega_r>0$, and for the mode (\ref{eq:Hall-IC2}, cyan line) $\omega_r<0$. 2) In contrast, in the firehose-unstable regime,   
both ion-cyclotron modes are right polarized, since for the mode (\ref{eq:Hall-IC1}, green line) $\omega_r<0$, and for the mode (\ref{eq:Hall-IC2}, cyan line) $\omega_r>0$. 
  
\subsubsection{Simplest ion-cyclotron resonances}
When talking about ``resonances'', one often expects to see expressions with a denominator that becomes zero. 
 Generally speaking, the simplest ``singular'' expressions representing resonances just arise from a technique used in deriving the dispersion relations.
If one eliminates the electric field $\bE$ from the beginning and expresses the fluid model through a matrix multiplied by a vector
$(\rho,u_{x},u_{y},u_{z},b_x,b_y,b_z,p_{\parallel}, p_{\perp})$, leads to dispersion relations that are not singular explicitly. 
In contrast, expressing the model through a matrix multiplied by a vector $(E_x, E_y, E_z)$, yields equivalent dispersion relations where the
resonances are shown explicitly. The dispersion  equation (\ref{eq:Hall-parF}) can be easily rewritten as
\begin{equation}
  \frac{k_\parallel^2}{\omega^2} = -\frac{\Omega_p/V_A^2}{\omega- \Omega_p\big(1+\frac{\bpar}{2}(a_p-1)) }; \qquad
  \frac{k_\parallel^2}{\omega^2} = +\frac{\Omega_p/V_A^2}{\omega+ \Omega_p\big(1+\frac{\bpar}{2}(a_p-1)) },
\end{equation}
which, in the cold plasma limit ($\bpar=0$) or in this case also for isotropic temperatures $(a_p=1)$, simplifies to
\begin{equation}
  \frac{k_\parallel^2}{\omega^2} = -\frac{\Omega_p/V_A^2}{\omega- \Omega_p}; \qquad
  \frac{k_\parallel^2}{\omega^2} = +\frac{\Omega_p/V_A^2}{\omega+ \Omega_p}, \label{eq:ICresPic}
\end{equation}
making it more evident that we have the simplest ion-cyclotron resonances here too. Indeed, for $k\rightarrow\infty$ the frequency
$\omega\rightarrow \Omega_p$ (first case) and $\omega\rightarrow -\Omega_p$ (second case).
For a plasma with finite $\bpar$ and nonisotropic temperatures, the $k\rightarrow\infty$ limit yields more general ion-cyclotron resonances (\ref{eq:Hallres}).

\begin{figure*}
$$\includegraphics[width=0.48\linewidth]{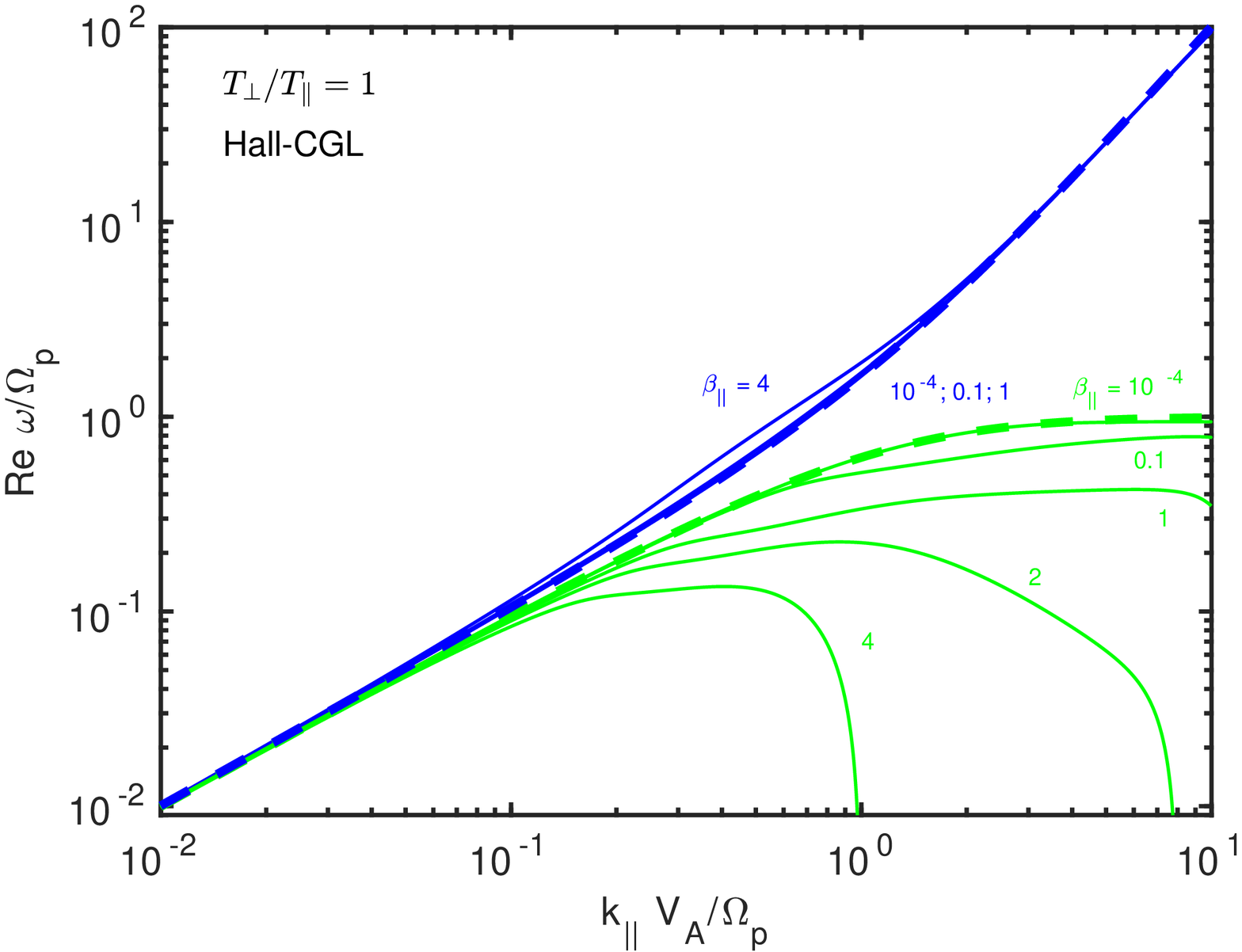}\hspace{0.03\textwidth}\includegraphics[width=0.48\linewidth]{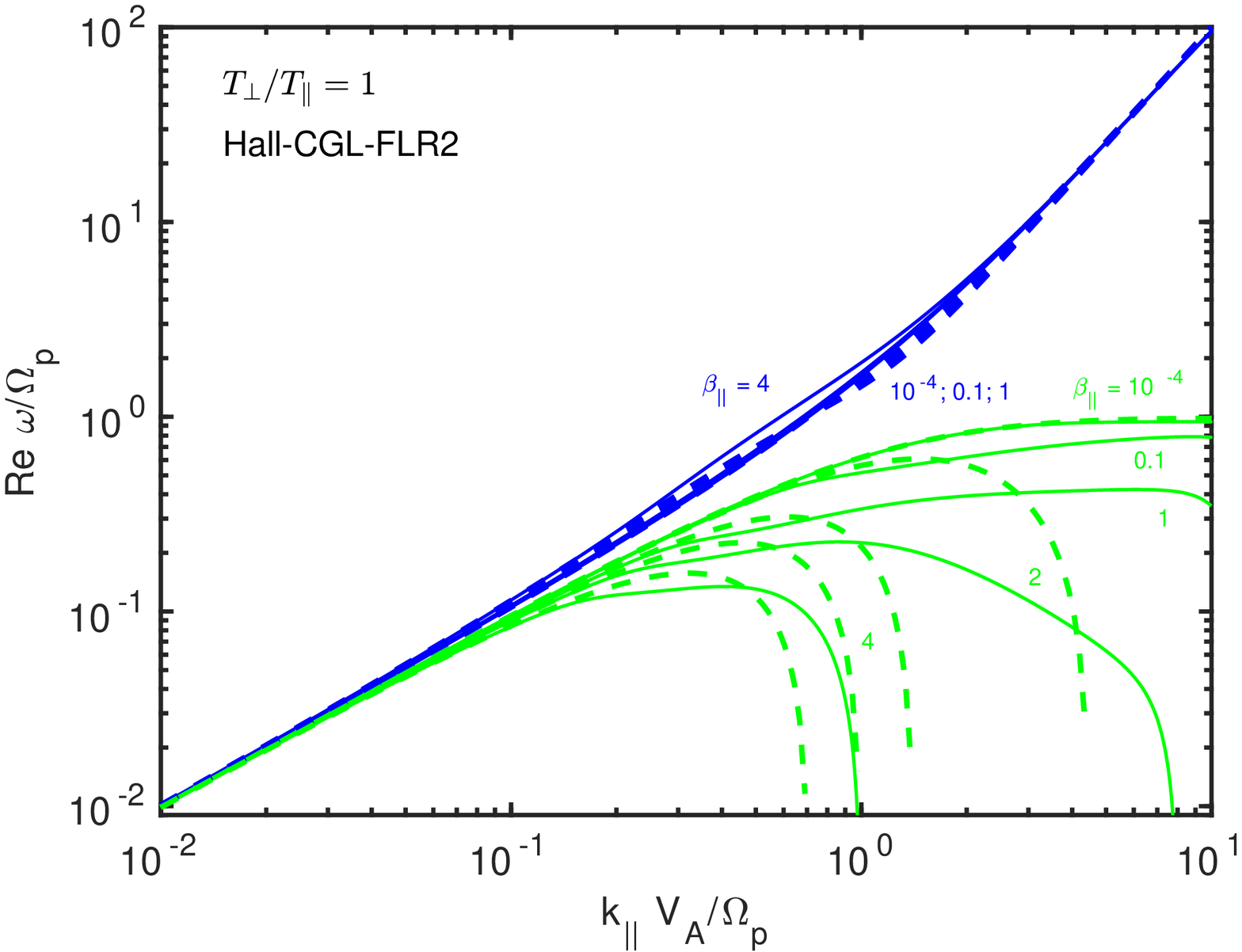}$$
  \caption{Left panel: Comparison of solutions of full kinetic theory obtained with the WHAMP code (solid lines) and the simple dispersion of the Hall-CGL model (dashed lines)
    for the whistler mode (blue) and the ion-cyclotron mode (green). The proton temperatures are isotropic $T_\perp/T_\parallel=1$, electrons are cold
    and the angle of propagation $\theta=0$. Real frequency is plotted. The $\bpar$ is varied as $\bpar=10^{-4}; 0.1; 1; 2; 4$. The kinetic solutions show
    that while the frequency of whistler mode is not much $\bpar$ dependent (only the case $\bpar=4$ is slightly different), the frequency of the ion-cyclotron mode
    is strongly $\bpar$ dependent since the mode experiences strong damping.
    The Hall-CGL model has only the simplest ion-cyclotron resonance, and the Hall-CGL
    ion-cyclotron mode matches the kinetic ion-cyclotron mode only for the lowest $\bpar=10^{-4}$. Right panel: Solutions of the Hall-CGL-FLR2 model are plotted. Note
  that the real frequency of the ion-cyclotron mode is crudely reproduced.} \label{fig:7}
\end{figure*}

To better understand what effects are present and what effects are absent in the Hall-CGL model, we plot fully kinetic solutions obtained with the WHAMP code
in Figure \ref{fig:7}. The proton temperature is prescribed to be isotropic $T_\perp/T_\parallel=1$ and the electrons are prescribed to be cold with $T_e/T_p=0$ (in the
WHAMP code, the value is chosen to be $10^{-8}$). In the WHAMP code it is
necessary to prescribe the ratio of the parallel proton thermal speed to the speed of light, and we choose $v_{\textrm{th}\parallel}/c = 10^{-4}$.
The solid lines are kinetic solutions with different $\bpar=10^{-4}; 0.1; 1; 2; 4$. The blue curve is the whistler mode and the green curve is the ion-cyclotron
mode. Solutions of the Hall-CGL model are shown in the left panel of Figure \ref{fig:7} and, for isotropic temperatures the solutions
are completely $\bpar$ independent and are represented with two dashed curves.
It is shown that the kinetic
solutions for the whistler mode are almost $\bpar$ independent and only the mode with $\bpar=4$ deviates from the other whistler solutions, and only between $k d_i=0.1-2.0$.
For comparison, in the right panel of Figure \ref{fig:7} we plot solutions of the Hall-CGL-FLR2 model (the model is discussed later in the text). 

\subsection{Hall-CGL dispersion relation for arbitrary propagation angle}
To make everything as clear as possible, let's write down one more time the normalized equations
we are solving, including the linearized and Fourier transformed equations in the x-z plane. The nonlinear Hall-CGL equations
with cold electrons are (dropping the tilde)
\begin{eqnarray}
&& \frac{\pr \rho_p}{\pr t} + \nabla\cdot (\rho \bu_p )=0; \\
&& \frac{\pr \bu_p}{\pr t} +\bu_p \cdot\nabla \bu_p +\frac{\bpar}{2} \frac{1}{\rho_p}\nabla\cdot\bp_p
  -\frac{1}{\rho_p}(\nabla\times\bb)\times\bb=0;\\
&&  \frac{\pr \bb}{\pr t} - \nabla\times(\bu_p\times\bb) + \nabla\times\Big[ \frac{1}{\rho_p}(\nabla\times\bb)\times\bb\Big]=0;\\
&& \frac{\pr p_{\parallel p}}{\pr t} + \nabla\cdot(p_{\parallel p}\bu_p) +2p_{\parallel p}\bhat\cdot\nabla\bu_p\cdot\bhat= 0;\\
&&  \frac{\pr p_{\perp p}}{\pr t} + \nabla\cdot(p_{\perp p}\bu_p) + p_{\perp p}\nabla\cdot\bu_p -p_{\perp p} \bhat\cdot\nabla\bu_p\cdot\bhat=0.
\end{eqnarray}
Linearizing in normalized units is even easier, because (temporarily introducing back the tilde for clarity) the
$\widetilde{p}_{\parallel p}^{(0)}=1$; $\widetilde{p}_{\perp p}^{(0)}=a_p$; $\widetilde{\rho}_0=1$; $\widetilde{B}_0=1$.
Linearizing these equations, writing them in the x-z plane, and Fourier transforming yields (dropping also the proton index 'p')
\begin{eqnarray}
&& -\omega \rho + k_\perp u_x + k_\parallel u_z =0;\\
&& -\omega u_x +\frac{\bpar}{2}k_\perp p_\perp -v_{A\parallel}^2 k_\parallel B_x  +k_\perp B_z =0;\\
&& -\omega u_y - v_{A\parallel}^2 k_\parallel B_y =0;\\
  && -\omega u_z +\frac{\bpar}{2}k_\parallel p_\parallel +\frac{\bpar}{2}(1-a_p)k_\perp B_x=0;\\
  && -\omega B_x -k_\parallel u_x -i k_\parallel^2 B_y=0; \label{eq:Indd11}\\
  && -\omega B_y - k_\parallel u_y -ik_\parallel k_\perp B_z +ik_\parallel^2 B_x=0;\\
  && -\omega B_z + k_\perp u_x+ik_\parallel k_\perp B_y=0; \label{eq:Indd33}\\
&& -\omega p_\parallel + k_\perp u_x +3 k_\parallel u_z= 0;\\
&&  -\omega p_\perp  + 2a_p k_\perp u_x +a_p k_\parallel u_z=0,
\end{eqnarray}
where the normalized $v_{A\parallel}^2=1+\frac{\bpar}{2}(a_p-1)$.
Alternatively, it is sometimes illuminating to work with the general induction equation where the electric field is used
\begin{eqnarray}
&&  -\omega B_x -k_\parallel E_y =0; \label{eq:Indd1}\\
&&  -\omega B_y -k_\perp E_z + k_\parallel E_x =0;\\
&&  -\omega B_z +k_\perp E_y = 0, \label{eq:Indd3}
\end{eqnarray}  
and in this case with cold electrons the electric field components are
\begin{eqnarray}
  E_x &=& -u_y -i k_\perp B_z +i k_\parallel B_x; \label{eq:Efield_Rit}\\
  E_y &=& u_x + ik_\parallel B_y; \label{eq:Efield_Rit2}\\
  E_z &=& 0.
\end{eqnarray}
The induction equation (\ref{eq:Indd1})-(\ref{eq:Indd3}) with the above electric field components is of course equivalent to the induction equation
(\ref{eq:Indd11})-(\ref{eq:Indd33}). All equations are normalized, the tilde are dropped, and the electric field $\widetilde{E}=E/E_0$, where $E_0=V_A B_0/c$.
The Hall-CGL dispersion relation reads
\begin{eqnarray}
&&  \omega^6-A_4\omega^4+A_2\omega^2-A_0=0;\\
  &&  A_4= k_\parallel^2\Big( \frac{3}{2}\bpar+2v_{A\parallel}^2\Big) +k_\perp^2\Big(1+a_p\bpar\Big) + k^2 k_\parallel^2; \nn \\
  && A_2 = k_\parallel^2 \Big[ k_\parallel^2 v_{A\parallel}^2(v_{A\parallel}^2+3\bpar)
    +k_\perp^2\Big( v_{A\parallel}^2(1+a_p\bpar) +\frac{3}{2}\bpar(1+a_p\bpar-\frac{1}{6}a_p^2\bpar) \Big) \Big] + k^2 k_\parallel^2\bpar (\frac{3}{2}k_\parallel^2+a_p k_\perp^2); \nn \\
  && A_0 = \frac{3}{2}\bpar k_\parallel^4 v_{A\parallel}^2 \Big[k_\parallel^2 v_{A\parallel}^2 +k_\perp^2(1+a_p\bpar-\frac{1}{6}a_p^2\bpar) \Big] + k^2 k_\parallel^4 k_\perp^2 \frac{5}{4}a_p \bpar^2. \nn
\end{eqnarray}
where the wavenumber $k^2=k_\parallel^2+k_\perp^2$. Perhaps the nicest form is to keep the CGL contributions on the l.h.s. (so the terms can be factorized to form the
CGL dispersion relation) and put the Hall term contributions on the r.h.s., which yields the Hall-CGL dispersion relation 
\begin{eqnarray} \label{eq:HallCGL-dispR}
  &&  \Big(\omega^2-k_\parallel^2 v_{A\parallel}^2\Big)\Big(\omega^4-A_2{\omega}^2+A_0\Big) =
  {k}^2 {k}_\parallel^2 \Big[ {\omega}^4-\bpar(\frac{3}{2}{k}_\parallel^2+a_p {k}_\perp^2)\omega^2
    + {k}_\parallel^2 {k}_\perp^2 \frac{5}{4}a_p\bpar^2\Big];\\
 &&   A_2 = {k}_\perp^2 \left(1+a_p\bpar\right) + {k}_\parallel^2 \left(v_{A\parallel}^2+\frac{3}{2}\bpar\right); \nonumber \\
 && A_0 = \frac{3}{2}{k}_\parallel^2 \bpar \left[ {k}_\parallel^2 v_{A\parallel}^2 + 
 {k}_\perp^2 \left(1+a_p\bpar-\frac{1}{6}a_p^2\bpar\right)  \right].\nn
\end{eqnarray}
Solutions for the parallel propagation ($k_\perp=0$) are of course the whistler and the ion-cyclotron mode
\begin{equation} \label{eq:ICWpica2}
\omega = \pm \frac{\kpar^2}{2} +\kpar\sqrt{v_{A\parallel}^2+\frac{\kpar^2}{4}},
\end{equation}
with other two solutions obtained by exchanging $\omega$ with $-\omega$. By bringing back the tilde to the normalized solution (\ref{eq:ICWpica2}), 
the result (\ref{eq:ICW_pica}) in physical units is easily recovered. The other parallel solution is the ion-acoustic mode $\omega=\pm \sqrt{3\bpar/2}\kpar$.
For perpendicular propagation ($\kpar=0$), the fast mode solution is $\omega=\pm k_\perp \sqrt{1+a_p\bpar}$, which is equivalent to CGL since the
Hall-term disappears. 

We investigated solutions of the Hall-CGL model in much greater detail in our paper ``On the Parallel and Oblique Firehose Instability in Fluid Models''
\citep{HunanaZank2017}, a paper which was essentially pulled out of this guide. We do not want to repeat the discussion with many associated figures here,
and a reader who is further interested in the Hall-CGL firehose instability can find much more information in that paper.

\newpage
\section{FLR corrections to the pressure tensor} \label{section:FLR}
The finite Larmor radius (FLR) corrections to the pressure tensor can be derived in several levels of approximation.
The pressure tensor equation (\ref{eq:PTfinal}) can be rewritten as
\begin{equation} \label{eq:FLR_full}
  \bhat\times\boldsymbol{\Pi} + (\bhat\times\boldsymbol{\Pi})^{\textrm{T}} = - \frac{1}{\Omega}\frac{B_0}{|\bb|} \Big[
\frac{\pr \bp}{\pr t} + \nabla\cdot(\bu\bp+\boldsymbol{q})+\bp\cdot\nabla \bu + (\bp\cdot\nabla \bu)^{\textrm{T}}
    \Big].
\end{equation}
This equation is exact and since $\bp=\bp^\textrm{g}+\boldsymbol{\Pi}$, the FLR tensor $\boldsymbol{\Pi}$ is described implicitly.
\subsection{Fully nonlinear FLR corrections}
If preservation of all nonlinearities is desired, this implicit equation can be further rewritten by the following procedure that
can be found for example in \cite{Hsu1986,PassotSulem2004b,Ramos2005} as a brief note.
By applying $\bhat\times$ to the left hand side, the first term calculates
\begin{eqnarray}
  [\bhat\times(\bhat\times\boldsymbol{\Pi})]_{ij} &=& \epsilon_{ikl}\hat{b}_k(\bhat\times\boldsymbol{\Pi})_{lj} =
  \epsilon_{ikl}\hat{b}_k \epsilon_{lrs}\hat{b}_r\Pi_{sj} = (\delta_{ir}\delta_{ks} - \delta_{is}\delta_{kr}) \hat{b}_k \hat{b}_r \Pi_{sj}
  = \hat{b}_k \hat{b}_i \Pi_{kj} - \underbrace{\hat{b}_k \hat{b}_k}_{=1} \Pi_{ij} \nn\\
  &=& \hat{b}_i \hat{b}_k \Pi_{kj} -\Pi_{ij},
\end{eqnarray}
where we have used that  $\epsilon_{ikl}\epsilon_{lrs} = \delta_{ir}\delta_{ks} - \delta_{is}\delta_{kr}$. The second term on the left
hand side of (\ref{eq:FLR_full}) calculates quite differently and we will need an identity
\begin{eqnarray}
  \epsilon_{ikl}\epsilon_{jrs} = \delta_{ij}( \delta_{kr}\delta_{ls}-\delta_{ks}\delta_{lr})
  -\delta_{ir}(\delta_{kj}\delta_{ls}-\delta_{ks}\delta_{lj}) +\delta_{is}(\delta_{kj}\delta_{lr}-\delta_{kr}\delta_{lj}).
\end{eqnarray}
Let's also remind ourselves that in the index notation $\Pi_{ss}=\trace\boldsymbol{\Pi}=0$ and
$\hat{b}_k\hat{b}_l\Pi_{kl}=\boldsymbol{\Pi}:\bhat\bhat=0$. The second term calculates
\begin{eqnarray}
  [\bhat\times(\bhat\times\boldsymbol{\Pi})^{\textrm{T}}]_{ij} &=& \epsilon_{ikl}\hat{b}_k(\bhat\times\boldsymbol{\Pi})^{\textrm{T}}_{lj} =
  \epsilon_{ikl}\hat{b}_k(\bhat\times\boldsymbol{\Pi})_{jl} = \epsilon_{ikl} \hat{b}_k \epsilon_{jrs} \hat{b}_r \Pi_{sl}
  = \epsilon_{ikl}\epsilon_{jrs} \hat{b}_k\hat{b}_r \Pi_{sl} \nn\\
  &=& \delta_{ij}\big(\hat{b}_k\hat{b}_k \underbrace{\Pi_{ss}}_{=0} -\underbrace{\hat{b}_k\hat{b}_l \Pi_{kl}}_{=0}\big)
  -\big(\hat{b}_j\hat{b}_i \underbrace{\Pi_{ss}}_{=0} - \hat{b}_k\hat{b}_i \Pi_{kj} \big)
  +\big( \hat{b}_j\hat{b}_l \Pi_{il} - \hat{b}_k\hat{b}_k \Pi_{ij} \big) \nn\\
  &=&  \hat{b}_i\hat{b}_k \Pi_{kj} +\hat{b}_j \Pi_{il} \hat{b}_l - \Pi_{ij}. 
\end{eqnarray}
The entire left hand side therefore reads
\begin{eqnarray}
  \big[\bhat\times(\bhat\times\boldsymbol{\Pi}) + \bhat\times(\bhat\times\boldsymbol{\Pi})^{\textrm{T}}\big]_{ij}
  = -2 \Pi_{ij} +2 \hat{b}_i\hat{b}_l \Pi_{lj} + \hat{b}_j\Pi_{il} \hat{b}_l.
\end{eqnarray}
By performing a usual matrix (single) dot product of this result with matrix $\bhat\bhat$ yields
\begin{eqnarray}
&&  \bigg[ \big[\bhat\times(\bhat\times\boldsymbol{\Pi}) + \bhat\times(\bhat\times\boldsymbol{\Pi})^{\textrm{T}} \big]\cdot (\bhat\bhat) \bigg]_{ij}=
  \big[\bhat\times(\bhat\times\boldsymbol{\Pi}) + \bhat\times(\bhat\times\boldsymbol{\Pi})^{\textrm{T}} \big]_{ik} (\bhat\bhat)_{kj} \nn\\
&&    = \big[ -2 \Pi_{ik} +2 \hat{b}_i\hat{b}_l \Pi_{lk} + \hat{b}_k\Pi_{il} \hat{b}_l  \big] \hat{b}_k \hat{b}_j
 = -2\Pi_{ik} \hat{b}_k \hat{b}_j +2\hat{b}_i\hat{b}_j \underbrace{\Pi_{lk}\hat{b}_l\hat{b}_k}_{=0}
 +\Pi_{il}\hat{b}_l\hat{b}_j \underbrace{\hat{b}_k\hat{b}_k}_{=1} = -2\Pi_{ik} \hat{b}_k \hat{b}_j + \Pi_{il}\hat{b}_l\hat{b}_j \nn\\
 && =  -2\Pi_{ik} \hat{b}_k \hat{b}_j + \Pi_{ik}\hat{b}_k\hat{b}_j = -\Pi_{ik} \hat{b}_k \hat{b}_j.
\end{eqnarray}
This further yields that
\begin{eqnarray}
 \bigg[ \big[\bhat\times(\bhat\times\boldsymbol{\Pi}) + \bhat\times(\bhat\times\boldsymbol{\Pi})^{\textrm{T}} \big]\cdot
  \big(\boldsymbol{I}+3\bhat\bhat\big) \bigg]_{ij} &=& -2\Pi_{ij} +2\hat{b}_i\hat{b}_k\Pi_{kj} + \hat{b}_j\Pi_{ik}\hat{b}_k -3\Pi_{ik} \hat{b}_k \hat{b}_j \nn\\
  &=&-2\Pi_{ij} +2\hat{b}_i\hat{b}_k\Pi_{kj} -2\Pi_{ik} \hat{b}_k \hat{b}_j,
\end{eqnarray}
and adding this result together with its transpose implies
\begin{eqnarray}
\bigg[ \big[\bhat\times(\bhat\times\boldsymbol{\Pi}) + \bhat\times(\bhat\times\boldsymbol{\Pi})^{\textrm{T}} \big]\cdot
  \big(\boldsymbol{I}+3\bhat\bhat\big) \bigg]^S_{ij} &=& -2\Pi_{ij} +2\hat{b}_i\hat{b}_k\Pi_{kj} -2\Pi_{ik} \hat{b}_k \hat{b}_j
-2\Pi_{ji} +2\hat{b}_j\hat{b}_k\Pi_{ki} -2\Pi_{jk} \hat{b}_k \hat{b}_i\nn\\
&=& - 4\Pi_{ij}, 
\end{eqnarray}
where we have used that the $\Pi_{ij}=\Pi_{ji}$. One therefore derives that the FLR tensor can be extracted from the left hand side
of equation (\ref{eq:FLR_full}) by performing an ``inversion'' procedure
\begin{eqnarray} \label{eq:PIextract}
\boldsymbol{\Pi} = -\frac{1}{4}\Big[ \big[\bhat\times(\bhat\times\boldsymbol{\Pi}) + \bhat\times(\bhat\times\boldsymbol{\Pi})^{\textrm{T}} \big]\cdot
  \big(\boldsymbol{I}+3\bhat\bhat\big) \Big]^S.
\end{eqnarray}
We can introduce matrix $\boldsymbol{\kappa}$ that will represent the right hand side of equation (\ref{eq:FLR_full}) as
\begin{eqnarray}
&&  \bhat\times\boldsymbol{\Pi} + (\bhat\times\boldsymbol{\Pi})^{\textrm{T}} = -\boldsymbol{\kappa}; \label{eq:kappa_defX}\\
&&  \boldsymbol{\kappa} = \frac{1}{\Omega}\frac{B_0}{|\bb|} \Big[
\frac{\pr \bp}{\pr t} + \nabla\cdot(\bu\bp +\boldsymbol{q})+\bp\cdot\nabla \bu + (\bp\cdot\nabla \bu)^{\textrm{T}}
    \Big]. \label{eq:kappa_def}
\end{eqnarray}
By performing operations (\ref{eq:PIextract}) at this equation yields the FLR tensor in the following form
\begin{eqnarray} \label{eq:FLRpica}\boxed{
\boldsymbol{\Pi} = \frac{1}{4}\Big[ \big(\bhat\times\boldsymbol{\kappa}\big)\cdot \big(\boldsymbol{I}+3\bhat\bhat\big) \Big]^S.}
\end{eqnarray}
The expression (\ref{eq:FLRpica}) is completely general since no simplifications were introduced, the expression is exact. However, the
equation is still \emph{implicit}, since on the right hand side the tensor $\boldsymbol{\kappa}$ contains the full pressure tensor
$\bp=\bp^\textrm{g}+\boldsymbol{\Pi}$. Nevertheless, the equation is extremely useful, once the expansion of the r.h.s. is performed, as we will do below.
 Note that the brackets are not unique, since 
\begin{equation}
  \big(\bhat\times\boldsymbol{\kappa}\big)\cdot \big(\boldsymbol{I}+3\bhat\bhat\big) =  \bhat\times \big[ \boldsymbol{\kappa} \cdot \big(\boldsymbol{I}+3\bhat\bhat\big)\big]
  = \bhat\times \boldsymbol{\kappa} \cdot \big(\boldsymbol{I}+3\bhat\bhat\big),
\end{equation}
where the last choice leaves the brackets unspecified (since both options are allowed), similarly to $\bhat\cdot\nabla\bu$.
Alternatively, one can get rid off the operator $\cdot\boldsymbol{I}$ (which acting on a matrix yields the same matrix), and split the result into two parts
\begin{eqnarray} \label{eq:FLRpicaRamos}
\boldsymbol{\Pi} = \frac{1}{4}\big(\bhat\times\boldsymbol{\kappa}\big)^S+ \frac{3}{4}\big(\bhat\times\boldsymbol{\kappa}\cdot\bhat\bhat \big)^S.
\end{eqnarray}
Another possible form can be obtained by subtracting the evolution equations for the gyrotropic pressure components $\pr p_\parallel/\pr t$ and $\pr p_\perp/\pr t$. 
The scalar pressure
equations were derived by applying $:\bhat\bhat$ and $:(\boldsymbol{I}-\bhat\bhat)/2$ on the pressure tensor equation.
The subtraction can be formally represented by introducing ``overbar'' projection operator $\bar{\boldsymbol{a}}$ that
projects any $(3\times 3)$ matrix $\boldsymbol{a}$ to 
\begin{eqnarray}
  \bar{\boldsymbol{a}} = \boldsymbol{a} - [\boldsymbol{a}:\bhat\bhat] \bhat\bhat -
          [\boldsymbol{a}:(\boldsymbol{I}-\bhat\bhat)/2] (\boldsymbol{I}-\bhat\bhat). \label{eq:ugly_projection}
\end{eqnarray}  
The definition is of course motivated by the pressure decomposition (\ref{eq:PfullDec}). By applying this operator on (\ref{eq:kappa_defX}), the l.h.s. remains
the same, yielding $(\bhat\times\boldsymbol{\Pi})^S=-\bar{\boldsymbol{\kappa}}$.
The solution (\ref{eq:FLRpica}) for the FLR tensor therefore rewrites
\begin{eqnarray} \label{eq:FLRpica2}
\boldsymbol{\Pi} = \frac{1}{4}\Big[ \big(\bhat\times\bar{\boldsymbol{\kappa}}\big)\cdot \big(\boldsymbol{I}+3\bhat\bhat\big) \Big]^S.
\end{eqnarray}
The subtraction of scalar pressure equations is motivated by the observation, that when working in the linear approximation directly with
equation (\ref{eq:kappa_defX}), i.e. without performing the inversion procedure, the scalar pressure equation have to be subtracted at the end. However,
by performing the inversion procedure, the scalar pressure equations are subtracted ``automatically'' during calculations, and it is actually easier
to work directly with (\ref{eq:FLRpica}) instead of (\ref{eq:FLRpica2}). 

Therefore, let's work with (\ref{eq:FLRpica}). The leading-order nonlinear FLR corrections are obtained by neglecting
the non-gyrotropic contributions $\boldsymbol{\Pi}$ and $\boldsymbol{q}^{\textrm{ng}}$ on the right hand side of (\ref{eq:FLRpica}), i.e. by prescribing 
\begin{eqnarray} \label{eq:FLRpicaSkoro}
  \boldsymbol{\Pi} &=& \frac{1}{4}\Big[ \big(\bhat\times \boldsymbol{\kappa}^{(1)} \big)\cdot \big(\boldsymbol{I}+3\bhat\bhat\big) \Big]^S;\\
  \boldsymbol{\kappa}^{(1)} &=& \frac{1}{\Omega}\frac{B_0}{|\bb|} \Big[
   \frac{\pr \bp^\textrm{g}}{\pr t} + \nabla\cdot(\bu\bp^\textrm{g}+\boldsymbol{q}^\textrm{g})+\bp^\textrm{g} \cdot\nabla \bu + (\bp^\textrm{g} \cdot\nabla \bu)^{\textrm{T}}
    \Big].
\end{eqnarray}
 Therefore, the only non-gyrotropic quantities that were retained on the r.h.s. of (\ref{eq:FLRpica}), are the perpendicular components of the velocity $\bu$.
The slightly unappealing factors $B_0/|\bb|$ in all the expressions can be removed, by redefining the cyclotron frequency with $|\bb|$ instead of $B_0$,
so that $\Omega_{\scriptscriptstyle|\bb|}=e|\bb|/(mc)$. Let's calculate (\ref{eq:FLRpicaSkoro}) step by step. For the brevity of calculations, it is useful to introduce
convective derivative $d/dt=\pr/\pr t+\bu\cdot\nabla$ and work with
\begin{equation} \label{eq:kappa_rit}
\boldsymbol{\kappa}^{(1)} = \frac{1}{\Omega_{\scriptscriptstyle|\bb|}} \Big[
   \frac{d \bp^\textrm{g}}{d t} + \bp^{\textrm{g}} \nabla\cdot\bu +\nabla\cdot\boldsymbol{q}^\textrm{g}+\bp^\textrm{g} \cdot\nabla \bu + (\bp^\textrm{g} \cdot\nabla \bu)^{\textrm{T}}
    \Big].
\end{equation}
We want to present complete derivation with the least amount of algebra involved, and we first want to obtain the form  of \cite{Schekochihin2010}.
Let's first write each term in $\boldsymbol{\kappa}^{(1)}$ in the index notation.
By using
\begin{eqnarray}
p_{ij}^g &=& (p_\parallel-p_\perp)\hat{b}_i\hat{b}_j+p_\perp\delta_{ij};\nn\\
q_{ijk}^\textrm{g} &=& (q_\parallel-3q_\perp) \hat{b}_i \hat{b}_j \hat{b}_k + q_\perp\big( \delta_{ij}\hat{b}_k +\delta_{jk}\hat{b}_i+\delta_{ik}\hat{b}_j\big),\nn
\end{eqnarray}
in the following form
\begin{eqnarray}
  \Big(\frac{d \bp^{\textrm{g}}}{dt}\Big)_{ij} &=& \hat{b}_i \hat{b}_j \frac{d}{dt}(p_\parallel-p_\perp) +(p_\parallel-p_\perp)\frac{d}{dt}(\hat{b}_i \hat{b}_j) +\delta_{ij} \frac{dp_\perp}{dt};\\
  \big( \bp^{\textrm{g}}\nabla\cdot\bu\big)_{ij} &=& \Big((p_\parallel-p_\perp) \hat{b}_i \hat{b}_j + \delta_{ij} p_\perp \Big)\nabla\cdot\bu;\\
  \big( \bp^{\textrm{g}}\cdot\nabla\bu\big)_{ij} &=& (p_\parallel-p_\perp) \hat{b}_i \bhat\cdot\nabla u_j +p_\perp \pr_i u_j;\\
  \big( \bp^{\textrm{g}}\cdot\nabla\bu\big)^T_{ij} &=& (p_\parallel-p_\perp) \hat{b}_j \bhat\cdot\nabla u_i +p_\perp \pr_j u_i;\\
  \big( \nabla\cdot\bq^{\textrm{g}}\big)_{ij} &=&  \hat{b}_i \hat{b}_j \nabla\cdot\big( (q_\parallel-3q_\perp)\bhat \big) +(q_\parallel-3q_\perp)\bhat\cdot\nabla ( \hat{b}_i \hat{b}_j)
  +\delta_{ij} \nabla\cdot(q_\perp\bhat) + \pr_i(q_\perp \hat{b}_j) +\pr_j(q_\perp \hat{b}_i)
\end{eqnarray}
and grouping similar terms in $\boldsymbol{\kappa}^{(1)}$ together yields
\begin{eqnarray}
  \kappa^{(1)}_{ij} &=& \frac{1}{\Omega_{\scriptscriptstyle|\bb|}} \Big\{ p_\perp (\pr_i u_j + \pr_j u_i) + \pr_i (q_\perp \hat{b}_j) + \pr_j(q_\perp \hat{b}_i)\nn\\
  && + (p_\parallel-p_\perp) \Big[ \frac{d}{dt}(\hat{b}_i \hat{b}_j) +\hat{b}_i \bhat\cdot\nabla u_j+\hat{b}_j \bhat\cdot\nabla u_i \Big]
  +(q_\parallel-3q_\perp)\bhat\cdot\nabla(\hat{b}_i \hat{b}_j)\nn\\
  && + \hat{b}_i \hat{b}_j \Big[ \frac{d}{dt}(p_\parallel-p_\perp) + (p_\parallel-p_\perp)\nabla\cdot\bu +\nabla\cdot\big( (q_\parallel-3q_\perp)\bhat\big) \Big]
  +\delta_{ij} \Big[ \frac{dp_\perp}{dt} +p_\perp\nabla\cdot\bu +\nabla\cdot(q_\perp\bhat)\Big] \Big\}. \label{eq:FLRPicaPica}
\end{eqnarray}
The first line in the above expression is left unchanged, since it is the final result, defined below as a matrix
$W_{ij}= p_\perp (\pr_i u_j+\pr_j u_i) + \pr_j (q_\perp \hat{b}_i) +\pr_i (q_\perp \hat{b}_j)$, that can be written as
$\boldsymbol{W}=p_\perp \nabla\bu +\nabla(q_\perp\bhat) + \big( p_\perp \nabla\bu +\nabla(q_\perp\bhat)  \big)^T = \big( p_\perp \nabla\bu +\nabla(q_\perp\bhat)  \big)^S$.
However, calculating the rest of the expression according to (\ref{eq:FLRpicaSkoro}) simplifies several terms, and for example all terms in
the third line of (\ref{eq:FLRPicaPica}) vanish.
The terms in the 3rd line containing $\hat{b}_i\hat{b}_j$ vanish after applying $\bhat\times$, since $(\bhat\times (\bhat\bhat))_{ij}=\epsilon_{ikl} \hat{b}_k \hat{b}_{l} \hat{b}_j=0$.
The terms in the 3rd line containing $\delta_{ij}$ vanish after applying the symmetric operator since
\begin{eqnarray}
  \Big[ (\bhat\times\boldsymbol{I})\cdot(\boldsymbol{I}+3\bhat\bhat)\Big]_{ij}^S &=&  \Big[ (\bhat\times\boldsymbol{I})_{im} (\boldsymbol{I}+3\bhat\bhat)_{mj}\Big]^S
  = \Big[ \epsilon_{ikl} \hat{b}_k \delta_{lm} (\delta_{mj}+3\hat{b}_m\hat{b}_j) \Big]^S =
  \Big[ \epsilon_{ikj} \hat{b}_k + 3\underbrace{\epsilon_{ikl} \hat{b}_k\hat{b}_l}_{=0} \hat{b}_j \Big]^S \nn\\
  &=& \epsilon_{ikj} \hat{b}_k + \epsilon_{jki} \hat{b}_k = \epsilon_{ikj} \hat{b}_k - \epsilon_{ikj} \hat{b}_k =0.
\end{eqnarray}
We therefore need to concentrate only on the 2nd line of (\ref{eq:FLRPicaPica}), and slightly rewrite that one. For example, the first term in the 2nd line
of (\ref{eq:FLRPicaPica}), the term $\frac{d}{dt}(\hat{b}_i\hat{b}_j)$, calculates
\begin{eqnarray}
  \Big[ \Big(\bhat\times \frac{d}{d t}(\bhat\bhat) \Big) \cdot(\boldsymbol{I}+3\bhat\bhat)\Big]_{ij}^S &=&
  \Big[ \Big(\bhat\times \frac{d}{d t}(\bhat\bhat) \Big)_{im} (\boldsymbol{I}+3\bhat\bhat)_{mj}\Big]^S
  = \Big[ \epsilon_{ikl} \hat{b}_k \frac{d}{d t}(\hat{b}_l\hat{b}_m) (\delta_{mj}+3\hat{b}_m \hat{b}_j )\Big]^S \nn\\
  &=&  \Big[ \epsilon_{ikl} \hat{b}_k \Big( \cancel{\hat{b}_l \frac{d \hat{b}_m }{d t}} + \hat{b}_m \frac{d \hat{b}_l }{d t} \Big) (\delta_{mj}+3\hat{b}_m \hat{b}_j )\Big]^S
  = \Big[ \epsilon_{ikl} \hat{b}_k \frac{d \hat{b}_l }{d t} (\hat{b}_{j}+3 \hat{b}_j )\Big]^S \nn\\
  &=&  4 \Big[ \Big( \bhat\times  \frac{d \bhat }{d t}\Big)_i \hat{b}_j  \Big]^S
  = 4\Big[ \Big( \bhat\times  \frac{d \bhat }{d t}\Big) \bhat  \Big]^S_{ij}. \label{eq:2ndline1}
\end{eqnarray}
The second term in the 2nd line of (\ref{eq:FLRPicaPica}), the term $\hat{b}_i \bhat\cdot\nabla u_j$, vanishes after applying $\bhat\times$, since
\begin{eqnarray}
  \Big[ \bhat\times \big( \bhat(\bhat\cdot\nabla\bu)   \big)  \Big]_{ij} &=& \epsilon_{ikl} \hat{b}_k \big( \bhat(\bhat\cdot\nabla\bu) \big)_{lj}
  =\underbrace{\epsilon_{ikl} \hat{b}_k  \hat{b}_l}_{=0} (\bhat\cdot\nabla \bu)_j =0.
\end{eqnarray}  
The third term in the 2nd line of (\ref{eq:FLRPicaPica}), the term $\hat{b}_j \bhat\cdot\nabla u_i$ however does not vanish.
Note that $b_j\bhat\cdot\nabla u_i = \big( (\bhat\cdot\nabla\bu) \bhat \big)_{ij}$ and the third term calculates 
\begin{eqnarray}
  \Big[ \Big(\bhat\times \big( (\bhat\cdot\nabla\bu)\bhat   \big) \Big) \cdot(\boldsymbol{I}+3\bhat\bhat)\Big]_{ij}^S &=&
  \Big[ \Big(\bhat\times \big( (\bhat\cdot\nabla\bu)\bhat   \big) \Big)_{im} (\boldsymbol{I}+3\bhat\bhat)_{mj} \Big]^S =
  \Big[ \epsilon_{ikl} \hat{b}_k  \big( (\bhat\cdot\nabla \bu)\bhat \big)_{lm}  (\delta_{mj}+3\hat{b}_m \hat{b}_j) \Big]^S \nn\\
  &=&  \Big[ \epsilon_{ikl} \hat{b}_k   (\bhat\cdot\nabla u_l) \hat{b}_m   (\delta_{mj}+3\hat{b}_m \hat{b}_j) \Big]^S
  = 4 \Big[ \epsilon_{ikl} \hat{b}_k   (\bhat\cdot\nabla u_l) \hat{b}_j \Big]^S \nn \\
  &=& 4 \Big[ \Big(\bhat\times (\bhat\cdot\nabla \bu)\Big)_i \hat{b}_j  \Big]^S  =4 \Big[ \Big(\bhat\times (\bhat\cdot\nabla \bu)\Big)\bhat  \Big]_{ij}^S.
\end{eqnarray}
The fourth term in the 2nd line of (\ref{eq:FLRPicaPica}), calculates in the same fashion than the second and third term
since $\bhat\cdot\nabla(\hat{b}_i\hat{b}_j) = \hat{b}_i (\bhat\cdot\nabla \hat{b}_j) + (\bhat\cdot\nabla \hat{b}_i) \hat{b}_j$, yielding 
\begin{eqnarray}
  \Big[ \Big(\bhat\times \big( \bhat\cdot\nabla(\bhat\bhat)   \big) \Big) \cdot(\boldsymbol{I}+3\bhat\bhat)\Big]_{ij}^S
  &=& 4 \Big[ \Big(\bhat\times (\bhat\cdot\nabla \bhat)\Big) \bhat \Big]_{ij}^S.
\end{eqnarray}
Collecting all the results together yields the FLR tensor  
\begin{eqnarray}
  \boldsymbol{\Pi} &=& \frac{1}{4\Omega_{\scriptscriptstyle |\bb|}}\Big[ \big(\bhat\times \boldsymbol{W} \big)\cdot \big(\boldsymbol{I}+3\bhat\bhat\big) \Big]^S
  + \frac{1}{\Omega_{\scriptscriptstyle |\bb|}} \Big[ (\bhat \times\boldsymbol{w})\bhat \Big]^S; \label{eq:FLR_Scheko}\\
  \boldsymbol{W} &=&\big[ p_\perp \nabla\bu +\nabla(q_\perp\bhat)  \big]^S; \label{eq:WmatrixDef}\\
  \boldsymbol{w} &=& (p_\parallel-p_\perp)\Big( \frac{d\bhat}{dt}+\bhat\cdot\nabla\bu \Big) +(q_\parallel-3q_\perp)\bhat\cdot\nabla\bhat,\label{eq:wvectorDef}
\end{eqnarray}
where $\boldsymbol{W}$ is a matrix and $\boldsymbol{w}$ is a vector.
The result is equivalent to equations (5)-(8) of \cite{Schekochihin2010}, derived in the Appendix of that paper in a somewhat simpler way.
In \cite{Schekochihin2010}, instead of our notation for matrix $\boldsymbol{W}$ and vector $\boldsymbol{w}$,
the symbols $\boldsymbol{S}$ and $\boldsymbol{\sigma}$ are used, where
$\boldsymbol{S}=\boldsymbol{W}$ and $\boldsymbol{\sigma}=-\boldsymbol{w}$. We wanted to avoid the use of $\boldsymbol{S}$ and $\boldsymbol{\sigma}$, since
these symbols are typically used when decomposing the heat flux tensor. For a reader who just jumped to this result, the symmetric operator is defined
as $A_{ij}^S=A_{ij}+A_{ji}$. 
For clarity, for any vector $\boldsymbol{w}$ and any matrix $\boldsymbol{W}$, the following identities hold
\begin{eqnarray}
  [(\bhat\times\boldsymbol{w})\bhat]^S &=& (\bhat\times\boldsymbol{w})\bhat+\bhat(\bhat\times\boldsymbol{w});\\
  \Big[ \bhat\times \boldsymbol{W} \cdot \big(\boldsymbol{I}+3\bhat\bhat\big) \Big]^S
  &=&\bhat\times \boldsymbol{W} \cdot \big(\boldsymbol{I}+3\bhat\bhat\big) -\big(\boldsymbol{I}+3\bhat\bhat\big)\cdot  \boldsymbol{W}^T \times\bhat.
\end{eqnarray}
The last identity follows from an identity for general matrix $\boldsymbol{A}$, that reads $(\bhat\times\boldsymbol{A})^T=-\boldsymbol{A}^T\times\bhat$,
and for a matrix $\boldsymbol{A}=\boldsymbol{B}\cdot\boldsymbol{C}$ the $\boldsymbol{A}^T=\boldsymbol{C}^T\cdot\boldsymbol{B}^T$, implying
$(\bhat\times\boldsymbol{B}\cdot\boldsymbol{C})^T=-\boldsymbol{C}^T\cdot\boldsymbol{B}^T\times\bhat$.

The result (\ref{eq:FLR_Scheko})-(\ref{eq:wvectorDef}) can be slightly simplified. The matrix $\boldsymbol{W}$ in the index notation reads
\begin{equation}
W_{ij} = p_\perp (\pr_i u_j +\pr_j u_i) +q_\perp(\pr_i \hat{b}_j + \pr_j \hat{b}_i) + (\pr_iq_\perp)\hat{b}_j +(\pr_jq_\perp)\hat{b}_i,
\end{equation}  
and by applying $\bhat\times$ eliminates only the last term
\begin{eqnarray}
  \big( \bhat\times \boldsymbol{W})_{ij} = \epsilon_{ikl}\hat{b}_k W_{lj}
  = \epsilon_{ikl}\hat{b}_k  \Big[  p_\perp (\pr_l u_j +\pr_j u_l) +q_\perp(\pr_l \hat{b}_j + \pr_j \hat{b}_l) + (\pr_lq_\perp)\hat{b}_j +\cancel{(\pr_jq_\perp)\hat{b}_l} \Big],
\end{eqnarray}
implying that the matrix $\boldsymbol{W}$ could be possibly redefined to
\begin{equation} \label{eq:SchekoX}
\boldsymbol{W} = \big[ p_\perp \nabla\bu +q_\perp \nabla \bhat \big]^S + (\nabla q_\perp) \bhat.
\end{equation}  
However, now the matrix is not symmetric.
Note that the brackets in the expression (\ref{eq:FLR_Scheko}) are not unique, and one can for example pull the $\bhat\times$ out, so that the expression reads
\begin{eqnarray}
  \boldsymbol{\Pi} = \frac{1}{\Omega_{\scriptscriptstyle |\bb|}} \Big\{ \bhat\times \Big[ \frac{1}{4} \boldsymbol{W} \cdot \big(\boldsymbol{I}+3\bhat\bhat\big) 
  +  \boldsymbol{w}\bhat \Big] \Big\}^S. \label{eq:FLR_SchekoF}
\end{eqnarray}
Let's multiply the last term in (\ref{eq:SchekoX}) by $\cdot(\boldsymbol{I}+3\bhat\bhat)$, it calculates
\begin{equation}
\Big((\nabla q_\perp) \bhat\Big) \cdot(\boldsymbol{I}+3\bhat\bhat) = 4 (\nabla q_\perp) \bhat.
\end{equation}
Obviously, it is beneficial to move this term to the vector $\boldsymbol{w}$. Therefore, \emph{redefining} $\boldsymbol{W}$ and $\boldsymbol{w}$, the solution reads 
\begin{empheq}[box=\fbox]{align}
  \boldsymbol{\Pi} &= \frac{1}{4\Omega_{\scriptscriptstyle |\bb|}}\Big[ \big(\bhat\times \boldsymbol{W} \big)\cdot \big(\boldsymbol{I}+3\bhat\bhat\big) \Big]^S
  + \frac{1}{\Omega_{\scriptscriptstyle |\bb|}} \Big[ (\bhat \times \boldsymbol{w})\bhat \Big]^S; \label{eq:FLR_Nice}\\
  \boldsymbol{W} &=\big[ p_\perp \nabla\bu +q_\perp\nabla\bhat  \big]^S; \label{eq:WUS}\\
  \boldsymbol{w} &= (p_\parallel-p_\perp)\Big( \frac{d\bhat}{dt}+\bhat\cdot\nabla\bu \Big) +\nabla q_\perp +(q_\parallel-3q_\perp)\bhat\cdot\nabla\bhat.\label{eq:FLR_Nice2}
\end{empheq}
Now the matrix $\boldsymbol{W}$ is again symmetric, and $\boldsymbol{w}$ is just a vector. This solution with matrix (\ref{eq:WUS}) is slightly nicer
than the solution with matrix (\ref{eq:WmatrixDef}).
Alternatively, one can write the solution in its full form
\begin{eqnarray}
  \boldsymbol{\Pi} = \frac{1}{\Omega_{\scriptscriptstyle |\bb|}} \bigg\{ \bhat\times \bigg[ \frac{1}{4}\big[ p_\perp \nabla\bu +q_\perp\nabla \bhat  \big]^S
    \cdot \big(\boldsymbol{I}+3\bhat\bhat\big) 
  + (p_\parallel-p_\perp)\Big( \frac{d\bhat}{dt}+\bhat\cdot\nabla\bu \Big)\bhat +(\nabla q_\perp)\bhat +(q_\parallel-3q_\perp)(\bhat\cdot\nabla\bhat)\bhat \bigg] \bigg\}^S.\nn\\
\end{eqnarray}

The nonlinear solution for the FLR tensor $\boldsymbol{\Pi}$ can be further re-arranged.
For example, to obtain the form reported by \cite{Ramos2005}, one needs to separate a part that is obtained
by performing $\boldsymbol{\Pi}\cdot\bhat$. The solution (\ref{eq:FLR_Nice}) in the index notation reads
\begin{eqnarray}
  \Pi_{ij} = && \frac{1}{\Omega_{\scriptscriptstyle |\bb|}} \Big\{ \epsilon_{ikl}\hat{b}_k \Big[ \frac{1}{4}W_{lm}\big( \delta_{mj} +3\hat{b}_m\hat{b}_j\big) +w_l\hat{b}_j \Big] \nn\\
  && \qquad +  \epsilon_{jkl}\hat{b}_k \Big[ \frac{1}{4}W_{lm}\big( \delta_{mi} +3\hat{b}_m\hat{b}_i\big) +w_l\hat{b}_i \Big] \Big\},
\end{eqnarray}
and direct calculation yields
\begin{eqnarray}
  \Pi_{ij}\hat{b}_j &=& \frac{1}{\Omega_{\scriptscriptstyle |\bb|}} \Big\{ \epsilon_{ikl}\hat{b}_k
  \Big[ \frac{1}{4}W_{lm}\big( \delta_{mj} +3\hat{b}_m\hat{b}_j\big) +w_l\hat{b}_j \Big] \hat{b}_j \Big\}
  = \frac{1}{\Omega_{\scriptscriptstyle |\bb|}} \epsilon_{ikl}\hat{b}_k \Big( W_{lm}\hat{b}_m +w_l\Big);\\  
  \boldsymbol{\Pi}\cdot\bhat &=& \frac{1}{\Omega_{\scriptscriptstyle |\bb|}} \bhat \times\Big( \boldsymbol{W}\cdot\bhat +\boldsymbol{w}\Big)
  \equiv \frac{1}{\Omega_{\scriptscriptstyle |\bb|}} \bhat\times\boldsymbol{h}, \label{eq:FLR_bhat}
\end{eqnarray}
where it was useful to define new vector $\boldsymbol{h} \equiv \boldsymbol{W}\cdot\bhat+\boldsymbol{w}$, by adding the quantity
\begin{equation}
\boldsymbol{W}\cdot\bhat = p_\perp \Big( (\nabla\bu)\cdot\bhat +\bhat\cdot\nabla \bu \Big) +q_\perp \bhat\cdot\nabla\bhat
\end{equation}
to the vector $\boldsymbol{w}$.
By separating the $\boldsymbol{\Pi}\cdot\bhat$ part, the solution (\ref{eq:FLR_Nice}) for $\boldsymbol{\Pi}$ is therefore re-arranged to the following form
\begin{empheq}[box=\fbox]{align}
  \boldsymbol{\Pi} &= \frac{1}{4\Omega_{\scriptscriptstyle |\bb|}}
  \Big[ (\bhat\times \boldsymbol{W})\cdot(\boldsymbol{I}-\bhat\bhat) \Big]^S
  +\frac{1}{\Omega_{\scriptscriptstyle |\bb|}} \Big[ (\bhat\times \boldsymbol{h}) \bhat \Big]^S ;\label{eq:FLR_Ramos}\\
  \boldsymbol{W} &=\big[ p_\perp \nabla\bu +q_\perp\nabla\bhat  \big]^S; \label{eq:WUS2}\\
  \boldsymbol{h} &= p_\parallel \Big( \frac{d\bhat}{dt}+\bhat\cdot\nabla\bu\Big) -p_\perp \Big( \frac{d\bhat}{dt}-(\nabla\bu)\cdot\bhat\Big)
  +\nabla q_\perp +(q_\parallel-2q_\perp)\bhat\cdot\nabla\bhat. \label{eq:h_Ramos}
\end{empheq}
The solution (\ref{eq:FLR_Ramos}) has a nice advantage that it is decomposed with respect to the magnetic field
lines. I.e. projecting the $\boldsymbol{\Pi}$ along the field lines by performing $\boldsymbol{\Pi}\cdot\bhat$ cancels the first term, 
and directly yields vector $\bhat\times\boldsymbol{h}/\Omega_{\scriptscriptstyle |\bb|}$, that could be defined as vector $\vec{\boldsymbol{\Pi}}_z$. 
\subsubsection{Employing non-dispersive induction equation}
Considering the long-wavelength low-frequency limit, 
the solution (\ref{eq:FLR_Nice2}) for vector $\boldsymbol{w}$ and the solution (\ref{eq:h_Ramos}) for vector $\boldsymbol{h}$ can be further simplified,
by using the usual non-dispersive CGL/MHD induction equation,
that can be written in the following form (see later in the text)
\begin{equation}
  \frac{d\bhat}{dt} =\bhat\cdot\nabla\bu -\bhat \Big[ \bhat\cdot(\nabla\bu)\cdot\bhat\Big].
\end{equation}
By further applying $\bhat\times$ at the induction equation, the last term on the r.h.s. disappears
\begin{equation}
  \bhat\times \frac{d\bhat}{dt} =\bhat\times (\bhat\cdot\nabla\bu),
\end{equation}
and vectors (\ref{eq:FLR_Nice2}), (\ref{eq:h_Ramos}) simplify to 
\begin{eqnarray}
  \boldsymbol{w} &=& 2 (p_\parallel-p_\perp)\bhat\cdot\nabla\bu +\nabla q_\perp +(q_\parallel-3q_\perp)\bhat\cdot\nabla\bhat; \\
  \boldsymbol{h} &=& 2p_\parallel \bhat\cdot\nabla\bu -p_\perp \Big( \bhat\cdot\nabla\bu-(\nabla\bu)\cdot\bhat\Big)
  +\nabla q_\perp +(q_\parallel-2q_\perp)\bhat\cdot\nabla\bhat. \label{eq:McMahonH}
\end{eqnarray}  
The result (\ref{eq:McMahonH}) can be further rewritten with vorticity $\boldsymbol{\omega}=\nabla\times\bu$, by using identity
$(\nabla\bu)\cdot\bhat-\bhat\cdot\nabla\bu=\bhat\times(\nabla\times\bu)=\bhat\times\boldsymbol{\omega}$, finally yielding 
\begin{equation} \label{eq:FLR_Ramos2}
\boldsymbol{h} = 2p_\parallel \bhat\cdot\nabla\bu+ p_\perp \bhat\times\boldsymbol{\omega} +\nabla q_\perp +(q_\parallel-2q_\perp)\bhat\cdot\nabla\bhat.
\end{equation}
Results (\ref{eq:FLR_Ramos}), (\ref{eq:WUS2}), (\ref{eq:FLR_Ramos2})  are equivalent to eq. 49-51 of \cite{Ramos2005}
 (see also eq. 11-13 of \cite{MacMahon1965}).
Further evaluation of nonlinear FLR corrections to higher precision is out of scope of this simple guide, and the interested reader is referred to \cite{Ramos2005}. 
We will continue in a much simpler linear approximation.

\subsection{Simplest FLR corrections (FLR1)}
We could continue by evaluating the obtained nonlinear results in the linear approximation. However, the nonlinear FLR calculations were quite complicated
and it is very useful to clarify, how to derive the simplest FLR corrections directly in the linear approximation.  
Let's start again, and work with the general equation (\ref{eq:FLR_full}).
To derive the simplest finite Larmor radius corrections to the pressure tensor, one has to cancel the $\boldsymbol{\Pi}$ contributions on the
right hand side of (\ref{eq:FLR_full}) and also neglect the heat flux contributions, which yields
\begin{equation} \label{eq:FLR_LS}
  \bhat\times\boldsymbol{\Pi} + (\bhat\times\boldsymbol{\Pi})^{\textrm{T}} = - \frac{1}{\Omega}\frac{B_0}{|\bb|} \Big[
\frac{\pr \bp^\textrm{g}}{\pr t} + \bu\cdot\nabla\bp^\textrm{g} + \bp^\textrm{g}\nabla\cdot\bu +\bp^\textrm{g}\cdot\nabla \bu + (\bp^\textrm{g}\cdot\nabla \bu)^{\textrm{T}}
    \Big].
\end{equation}
Considering complete linearization (so that we can easily go to Fourier space), this is equivalent to performing expansion to the first order
in frequency $\omega/\Omega$ and to the first order in wavenumber $r_L k$. We do not want to perform complete linearization just yet. However, to move further,
we introduce simplification by not evaluating the FLR corrections along the magnetic field lines, but along the ambient
magnetic field $B_0$ that is prescribed to be in the z-direction, and therefore $\bhat=(0,0,1)\equiv\bhat_0$. This simplification typically used
in numerical simulations is appropriate if the magnetic field lines are not too distorted with respect to the mean magnetic field.
Here we will distinguish between complete linearization and evaluation along $\bhat_0$.  However, it is important to note that the procedure 
of evaluating some terms with $\bhat_0$, while other nonlinearities are kept, is of course not a fully consistent procedure.
With the approach presented here, the only fully consistent
procedure that simplifies the general equation (\ref{eq:FLR_full}), is complete linearization. Additionally, as discussed by \cite{Passot2014}, 
neglecting magnetic field line distortion may in some instances lead to spurious instabilities.

Let's evaluate (\ref{eq:FLR_LS}) along $\bhat_0$. On the left hand side we calculate each component in the matrix separately.
Since $(\bhat\times\boldsymbol{\Pi})_{ij}=\epsilon_{ikl}\hat{b}_k \Pi_{lj}$, it is obvious that non-zero contributions are obtained only if
$\hat{b}_k=\hat{b}_z=1$, so the index $k=z$, and $(\bhat\times\boldsymbol{\Pi})_{ij}=\epsilon_{izl}\Pi_{lj}$. Now, for example the $'xx'$
component calculates $(\bhat\times\boldsymbol{\Pi})_{xx}=\epsilon_{xzl}\Pi_{lx}$ and the only nonzero contribution is obtained for index $l=y$,
which yields $(\bhat\times\boldsymbol{\Pi})_{xx}=\epsilon_{xzy}\Pi_{yx}=-\epsilon_{xyz}\Pi_{yx}=-\Pi_{yx}=-\Pi_{xy}$, where in the last step we used
that the FLR tensor is symmetric.
The entire left hand side $[\bhat\times\boldsymbol{\Pi} +(\bhat\times\boldsymbol{\Pi})^{\textrm{T}}]_{ij} = \epsilon_{izl}\Pi_{lj} + \epsilon_{jzl}\Pi_{li}$
and it is straightforward to calculate each component, which yields
\begin{eqnarray}
  \bhat\times\boldsymbol{\Pi} + (\bhat\times\boldsymbol{\Pi})^{\textrm{T}} \Big|_{\bhat=\bhat_0} =
 \left( \begin{array}{ccc}
    -2\Pi_{xy}, & \Pi_{xx}-\Pi_{yy}, & -\Pi_{yz} \\
    \Pi_{xx}-\Pi_{yy}, & +2\Pi_{xy}, & \Pi_{xz} \\
    -\Pi_{yz}, & \Pi_{xz}, & 0
 \end{array} \right).
\end{eqnarray}
The gyrotropic pressure tensor is of course
\begin{eqnarray}
  \boldsymbol{p}^\textrm{g} \Big|_{\bhat=\bhat_0} =
 \left( \begin{array}{ccc}
    p_\perp & 0 & 0 \\
    0 & p_\perp & 0 \\
    0 & 0 & p_\parallel
  \end{array} \right). \label{eq:PgEval}
\end{eqnarray}  
The first term on the r.h.s. of (\ref{eq:FLR_LS}) calculates
\begin{eqnarray}
  \frac{\pr}{\pr t} p^\textrm{g}_{ij} = \hat{b}_i\hat{b}_j \frac{\pr}{\pr t}p_\parallel + (\delta_{ij}-\hat{b}_i\hat{b}_j) \frac{\pr}{\pr t}p_\perp
  + (p_\parallel-p_\perp)\big( \hat{b}_i \frac{\pr}{\pr t}\hat{b}_j + \hat{b}_j \frac{\pr}{\pr t}\hat{b}_i\big),
\end{eqnarray}
and using the matrix notation
\begin{eqnarray}
 \Big(\frac{\pr}{\pr t}\boldsymbol{p}^\textrm{g} \Big)\Big|_{\bhat=\bhat_0} =
 \frac{\pr}{\pr t} \left( \begin{array}{ccc}
    p_\perp & 0 & 0 \\
    0 & p_\perp & 0 \\
    0 & 0 & p_\parallel
 \end{array} \right) +
 (p_\parallel-p_\perp)
 \left( \begin{array}{ccc}
    0 & 0 & \frac{\pr \hat{b}_x}{\pr t} \\
    0 & 0 & \frac{\pr \hat{b}_y}{\pr t} \\
    \frac{\pr \hat{b}_x}{\pr t} & \frac{\pr \hat{b}_y}{\pr t} & 2\frac{\pr \hat{b}_z}{\pr t}
 \end{array} \right)\Big|_{\bhat=\bhat_0}. \label{eq:IDKnow1}
\end{eqnarray}
The second term in (\ref{eq:IDKnow1}) contributes because of the induction equation $\pr \bb /\pr t= - c \nabla\times\bE$. For now, we will not use
the induction equation and keep our FLR calculations in the above form. Let's continue with the second term on the r.h.s of (\ref{eq:FLR_LS}).
When completely linearized this term does not contribute, but for now we will keep it and combine it with the first term into a convective derivative
$\pr\bp^\textrm{g}/\pr t + \bu\cdot\nabla\bp^\textrm{g} = d\bp^\textrm{g}/dt$, so that $\pr/\pr t$ in (\ref{eq:IDKnow1}) is replaced by $d/dt$. 
The remaining terms on the r.h.s. of (\ref{eq:FLR_LS}) are straightforward to evaluate since no derivatives of $\bhat$ are encountered,  and the third term
$\bp^\textrm{g}\nabla\cdot\bu\big|_{\bhat=\bhat_0}$ is equal to matrix (\ref{eq:PgEval}) multiplied by $\nabla\cdot\bu$.
The last two terms of (\ref{eq:FLR_LS}) are together evaluated as
\begin{eqnarray}
\bp^\textrm{g}\cdot\nabla\bu + (\bp^\textrm{g}\cdot\nabla\bu)^{\textrm{T}} \Big|_{\bhat=\bhat_0}  = \left( \begin{array}{ccc}
    2p_\perp \pr_x u_x; & p_\perp (\pr_x u_y + \pr_y u_x); & p_\perp \pr_x u_z + p_\parallel \pr_z u_x\\
    p_\perp (\pr_y u_x+\pr_x u_y); & 2 p_\perp \pr_y u_y; & p_\perp \pr_y u_z+p_\parallel\pr_z u_y \\
    p_\parallel \pr_z u_x + p_\perp\pr_x u_z; & p_\parallel \pr_z u_y+p_\perp\pr_y u_z; & 2p_\parallel \pr_z u_z
  \end{array} \right). \label{eq:FLR_term4}
\end{eqnarray}
Now we need to subtract the scalar CGL pressure equations obtained previously. Since
$\bhat\cdot\nabla\bu\cdot\bhat|_{\bhat=\bhat_0} =\pr_z u_z$, the scalar pressure equations read 
\begin{eqnarray}
  \frac{d p_\parallel}{d t} + p_\parallel \nabla\cdot \bu +2p_\parallel \pr_z u_z= 0;\qquad \frac{d p_\perp}{d t} + 2p_\perp \nabla\cdot\bu -p_\perp \pr_z u_z=0.
\end{eqnarray}
By canceling terms coming from these two equations (affecting only the diagonal components),
the entire equation (\ref{eq:FLR_LS}) rewrites 
\begin{eqnarray}
 \left( \begin{array}{ccc}
    -2\Pi_{xy}, & \Pi_{xx}-\Pi_{yy}, & -\Pi_{yz} \\
    \Pi_{xx}-\Pi_{yy}, & +2\Pi_{xy}, & \Pi_{xz} \\
    -\Pi_{yz}, & \Pi_{xz}, & 0
 \end{array} \right) &=& -\frac{1}{\Omega}
(p_\parallel-p_\perp)
 \left( \begin{array}{ccc}
    0 & 0 & \frac{d \hat{b}_x}{d t} \\
    0 & 0 & \frac{d \hat{b}_y}{d t} \\
    \frac{d \hat{b}_x}{d t} & \frac{d \hat{b}_y}{d t} & \frac{d \hat{b}_z}{d t}
 \end{array} \right)\Big|_{\bhat=\bhat_0} \nn\\
&& -\frac{1}{\Omega}
    \left( \begin{array}{ccc}
    p_\perp (\pr_x u_x-\pr_y u_y); & p_\perp (\pr_x u_y + \pr_y u_x); & p_\perp \pr_x u_z + p_\parallel \pr_z u_x\\
    p_\perp (\pr_y u_x+\pr_x u_y); & p_\perp (\pr_y u_y-\pr_x u_x); & p_\perp \pr_y u_z+p_\parallel\pr_z u_y \\
    p_\parallel \pr_z u_x + p_\perp\pr_x u_z; & p_\parallel \pr_z u_y+p_\perp\pr_y u_z; & 0
  \end{array} \right). \label{eq:Syst1}
\end{eqnarray}
The identity $\boldsymbol{\Pi}:\bhat\bhat=0$ evaluated for $\bhat=\bhat_0$ implies that $\Pi_{zz}=0$ and
the identity $\trace\boldsymbol{\Pi}=0$ further implies that $\Pi_{xx}=-\Pi_{yy}$. The components of the
pressure tensor therefore read
\begin{eqnarray}
&& \Pi_{xx} = -\Pi_{yy} = -\frac{p_\perp}{2\Omega} (\pr_x u_y + \pr_y u_x); \qquad \Pi_{xy} = \frac{p_\perp}{2\Omega} (\pr_x u_x-\pr_y u_y);\qquad  \Pi_{zz} =0;  \nn\\
  && \Pi_{xz} = -\frac{1}{\Omega}\Big( (p_\parallel-p_\perp)\frac{d\hat{b}_y}{dt} + p_\parallel \pr_z u_y + p_\perp \pr_y u_z \Big);\qquad
   \Pi_{yz} = \frac{1}{\Omega}\Big( (p_\parallel-p_\perp)\frac{d\hat{b}_x}{dt} + p_\parallel \pr_z u_x + p_\perp \pr_x u_z \Big).
\end{eqnarray}  
Finally, the time derivative of the unit vector  
\begin{eqnarray}
  \frac{d}{d t}\bhat &=& \frac{d}{d t} \frac{\bb}{|\bb|} = \frac{1}{|\bb|} \frac{d\bb}{d t}
  -\frac{\bb}{|\bb|^2} \frac{d|\bb|}{d t} = \frac{1}{|\bb|}\Big[  \frac{d\bb}{d t} -\bhat \frac{d|\bb|}{d t} \Big]
  = \frac{1}{|\bb|}\Big[  \frac{d\bb}{d t} -\bhat \Big( \bhat\cdot \frac{d \bb}{d t} \Big) \Big],
\end{eqnarray}
which evaluated for $\bhat=\bhat_0=(0,0,1)$, $|\bb|=B_0$ yields
\begin{eqnarray}
&&  \frac{d \hat{b}_x }{d t} \Big|_{\bhat=\bhat_0}= \frac{1}{B_0} \frac{d B_x}{d t}\Big|_{\bhat=\bhat_0}; \qquad 
 \frac{d \hat{b}_y }{d t} \Big|_{\bhat=\bhat_0}= \frac{1}{B_0} \frac{d B_y}{d t}\Big|_{\bhat=\bhat_0}; \qquad
 \frac{d \hat{b}_z }{d t} \Big|_{\bhat=\bhat_0}= \frac{1}{B_0} \big(\frac{d B_z}{d t} - \frac{d B_z}{d t}\big) =0.
\end{eqnarray}
The general induction equation reads
\begin{eqnarray}
  \frac{d\bb}{dt} = -\bb\nabla\cdot\bu + \bb\cdot\nabla\bu - c\nabla\times\bE_H,
\end{eqnarray}
which evaluates in the $\bhat=\bhat_0$ approximation as
\begin{eqnarray}
  \frac{dB_x}{dt}\Big|_{\bhat=\bhat_0} &=& B_0 \pr_z u_x -c(\nabla\times\bE_H)_x;\qquad
  \frac{dB_y}{dt}\Big|_{\bhat=\bhat_0} = B_0 \pr_z u_y -c(\nabla\times\bE_H)_y; \nn\\
  \frac{dB_z}{dt}\Big|_{\bhat=\bhat_0} &=& -B_0\nabla\cdot\bu +B_0 \pr_z u_z -c(\nabla\times\bE_H)_z.
\end{eqnarray}
These expressions are used in the FLR tensor components $\Pi_{xz}$ and $\Pi_{yz}$, and the FLR tensor reads
\begin{eqnarray}
&& \Pi_{xx} = -\Pi_{yy} = -\frac{p_\perp}{2\Omega} (\pr_x u_y + \pr_y u_x);\qquad
  \Pi_{xy} = \frac{p_\perp}{2\Omega} (\pr_x u_x-\pr_y u_y);\qquad  \Pi_{zz} =0; \nn\\
  && \Pi_{xz} = -\frac{1}{\Omega}\Big[ (p_\parallel-p_\perp)\Big(\pr_z u_y-\frac{c}{B_0}(\nabla\times\bE_H)_y \Big) + p_\parallel \pr_z u_y + p_\perp \pr_y u_z \Big];\nn\\
  && \Pi_{yz} =  \frac{1}{\Omega}\Big[ (p_\parallel-p_\perp)\Big(\pr_z u_x-\frac{c}{B_0}(\nabla\times\bE_H)_x  \Big) + p_\parallel \pr_z u_x + p_\perp \pr_x u_z \Big].\label{eq:FLR1Hall}
\end{eqnarray}
However, since the FLR
tensor was derived with a precision to only first order in wavenumber k, to make everything consistent,
the Hall contributions in the induction equation are usually (but not always) neglected. We therefore have
$d\hat{b}_x/dt = \pr_z u_x$ and $d\hat{b}_y/dt= \pr_z u_y$ which yields the leading-order FLR pressure tensor evaluated
along the magnetic field $\bhat=(0,0,1)$ in the form
\begin{eqnarray}
&& \Pi_{xx} = -\Pi_{yy} = -\frac{p_\perp}{2\Omega} (\pr_x u_y + \pr_y u_x);\qquad 
  \Pi_{xy} = \frac{p_\perp}{2\Omega} (\pr_x u_x-\pr_y u_y); \qquad \Pi_{zz} =0;\nn\\
  && \Pi_{xz} = -\frac{1}{\Omega}\Big( (2p_\parallel-p_\perp)\pr_z u_y + p_\perp \pr_y u_z \Big);\qquad
  \Pi_{yz} = \frac{1}{\Omega}\Big( (2p_\parallel-p_\perp)\pr_z u_x + p_\perp \pr_x u_z \Big),
\end{eqnarray}
which is equivalent to equation 5 of \cite{Yajima1966}. Perhaps surprisingly, the result is also equivalent to the highly collisional
eq. (2.20)-(2.24) of \cite{Braginskii1965},
after one prescribes $p_\parallel=p_\perp$ and after terms containing the collisional time $\tau$ are ``ignored''
($\tau \sim 1/\nu$ where $\nu$ is the usual collisional frequency). 
A proper collisionless limit $\tau\to\infty$ can not be obtained from the expressions of \cite{Braginskii1965}, since the results
are proportional to $\tau$ (and also $1/\tau$). 

Furthermore, evaluation along $\bhat=(0,0,1)$ basically means that the magnetic field terms of the system were linearized and other parts of the system
were not. The expansion procedure preserved some nonlinearities,
but other nonlinearities that are of the same order were implicitly neglected.  
Therefore, the only fully consistent procedure is to evaluate the FLR tensor in the linear approximation, yielding
\begin{eqnarray}
&& \Pi_{xx}^{(1)} = -\Pi_{yy}^{(1)} = -\frac{p_\perp^{(0)}}{2\Omega} (\pr_x u_y + \pr_y u_x);\qquad
  \Pi_{xy}^{(1)} = \frac{p_\perp^{(0)}}{2\Omega} (\pr_x u_x-\pr_y u_y);\qquad \Pi_{zz}^{(1)} =0;  \nn\\
  && \Pi_{xz}^{(1)} = -\frac{1}{\Omega}\Big( (2p_\parallel^{(0)}-p_\perp^{(0)})\pr_z u_y + p_\perp^{(0)} \pr_y u_z \Big);\qquad
  \Pi_{yz}^{(1)} = \frac{1}{\Omega}\Big( (2p_\parallel^{(0)}-p_\perp^{(0)})\pr_z u_x + p_\perp^{(0)} \pr_x u_z \Big). \label{eq:FLR1s}
\end{eqnarray}
Notably, the tensor is very different than eq. (3.10)-(3.13) of \cite{Oraevskii1968}. 
 The FLR corrections to the pressure tensor (these ones or more complicated ones) were used in a number of direct numerical simulations, see for example
\cite{Borgogno2009,Hunana2011,Kobayashi2017,PerronePassot2018}, or for the isotropic case see for example \cite{Ghosh2015}, and references therein.

\subsection{Hall-CGL-FLR1 fluid model}
The normalized, linearized, Fourier transformed system of equations written in x-z plane reads (dropping tilde everywhere)
\begin{eqnarray}
&& -\omega \rho + k_\perp u_x + k_\parallel u_z =0;\\
&& -\omega u_x +\frac{\bpar}{2}k_\perp p_\perp -v_{A\parallel}^2 k_\parallel B_x  +k_\perp B_z +\frac{\bpar}{2}\big( k_\perp \Pi_{xx} +k_\parallel \Pi_{xz} \big)=0;\\
&& -\omega u_y - v_{A\parallel}^2 k_\parallel B_y +\frac{\bpar}{2}\big(k_\perp \Pi_{xy}+k_\parallel \Pi_{yz}  \big)=0;\\
&& -\omega u_z +\frac{\bpar}{2}k_\parallel p_\parallel +\frac{\bpar}{2}(1-a_p)k_\perp B_x + \frac{\bpar}{2}k_\perp \Pi_{xz}=0;\\
&& -\omega B_x -k_\parallel u_x -i k_\parallel^2 B_y=0;\\
&& -\omega B_y - k_\parallel u_y -ik_\parallel k_\perp B_z +ik_\parallel^2 B_x=0;\\
&& -\omega B_z + k_\perp u_x+ik_\parallel k_\perp B_y=0;\\
&& -\omega p_\parallel + k_\perp u_x +3 k_\parallel u_z= 0;\\
&&  -\omega p_\perp  + 2a_p k_\perp u_x +a_p k_\parallel u_z=0,
\end{eqnarray}
where $v_{A\parallel}^2=1+\frac{\bpar}{2}(a_p-1)$, and the first-order FLR tensor $\boldsymbol{\Pi}=\boldsymbol{\Pi}^{(1)}$ is 
\begin{eqnarray}
  && \Pi_{xx}^{(1)} = -\frac{a_p}{2} ik_\perp u_y;\qquad  
  \Pi_{xy}^{(1)} = \frac{a_p}{2} ik_\perp u_x; \qquad \Pi_{zz}^{(1)} =0;  \nn\\
  && \Pi_{xz}^{(1)} = -(2-a_p)ik_\parallel u_y;\qquad  \Pi_{yz}^{(1)} =  (2-a_p)ik_\parallel u_x + a_pik_\perp u_z.\label{eq:FLR1sss}
\end{eqnarray}
The dispersion relation for the Hall-CGL-FLR1 fluid model can be written as
\begin{eqnarray}
  &&  \Big(\omega^2-k_\parallel^2 v_{A\parallel}^2\Big)\Big(\omega^4-A_2\omega^2+A_0\Big) =
  k^2 k_\parallel^2 \Big[ \omega^4-\bpar(\frac{3}{2}k_\parallel^2+a_p k_\perp^2)\omega^2
    + k_\parallel^2 k_\perp^2 \frac{5}{4}a_p\bpar^2\Big] + \mathcal{P}^{\textrm{FLR}}; \label{eq:HallCGLdisp}\\
 &&   A_2 = k_\perp^2 \left(1+a_p\bpar\right) + k_\parallel^2 \left(v_{A\parallel}^2+\frac{3}{2}\bpar\right); \nonumber \\
 && A_0 = \frac{3}{2}k_\parallel^2 \bpar \left[ k_\parallel^2 v_{A\parallel}^2  + 
     k_\perp^2 \left(1+a_p\bpar-\frac{1}{6}a_p^2\bpar\right)  \right], \label{eq:CGLparameters}
\end{eqnarray}
where the polynomial on the right hand side
\begin{eqnarray}
  \mathcal{P}^{\textrm{FLR}} &=& A_4' \omega^4 - A_2'\omega^2 +A_0';\\
  A_4' &=& \bpar^2\Big[ k_\parallel^4 \Big(1-\frac{a_p}{2}\Big)^2 +k_\perp^4\frac{a_p^2}{16}+k_\parallel^2 k_\perp^2 a_p \Big(1-\frac{a_p}{2}\Big) \Big];\\
  A_2' &=& \bpar k_\parallel^2 \Bigg\{ k_\parallel^6 \bpar \Big(1-\frac{a_p}{2}\Big)^2
  +k_\parallel^4 \Big[ \frac{3}{2}\bpar^2 \Big(1-\frac{a_p}{2} \Big)^2 +k_\perp^2\bpar\Big( 1-\frac{a_p^2}{4}\Big) +v_{A\parallel}^2(a_p-2) \Big]  \nn\\
  && +k_\parallel^2 k_\perp^2\Big[ -1+\frac{a_p}{2}+\bpar\frac{a_p}{4}\Big( 1-a_p+\bpar(1-\frac{a_p}{2}) \Big)
    -v_{A\parallel}^2\Big( 1-\bpar\frac{a_p}{2}(1-\frac{a_p}{2}) \Big) +k_\perp^2 a_p\bpar (1-\frac{7}{16}a_p) \Big]   \nn \\
  && +k_\perp^4 \frac{a_p}{4}\Big[  -1+2\bpar-v_{A\parallel}^2-a_p\bpar\Big( 1+\bpar(a_p-\frac{15}{8})-\frac{1}{4}k_\perp^2\Big)  \Big] \Bigg\};\\
  A_0' &=& \bpar^2 k_\parallel^4\Bigg\{ k_\parallel^6\frac{3}{2}\bpar\Big(1-\frac{a_p}{2}\Big)^2
  +k_\parallel^4 \Big(1-\frac{a_p}{2}\Big)\Big[ -3v_{A\parallel}^2+k_\perp^2\bpar\frac{3}{2}\Big(1-\frac{a_p}{2}+\frac{a_p^2}{6}\Big) \Big] \nn\\
  && +k_\parallel^2 k_\perp^2 \Big[ k_\perp^2\bpar a_p^2 (\frac{21}{32}-\frac{5}{16}a_p) -v_{A\parallel}^2\frac{3}{2}(1-\frac{a_p}{2}+\frac{1}{6}a_p^2 )
    -\Big(1-\frac{a_p}{2}\Big)\frac{3}{2}\Big( 1+\frac{1}{6}a_p\bpar-\frac{1}{6}a_p^2\bpar\Big) \Big] \nn\\
  && +k_\perp^4 a_p \Big[ k_\perp^2 a_p\bpar(\frac{13}{32}-\frac{3}{16}a_p) +\frac{v_{A\parallel}^2}{8}(1-2a_p) -\frac{1}{8}+\frac{3}{16}a_p\bpar(1-a_p)\Big] \Bigg\}.
\end{eqnarray}
The solution for strictly parallel propagation is
\begin{equation}
  \omega = \pm \frac{k_\parallel^2}{2}\Big(1+\bpar(1-\frac{a_p}{2})\Big) + k_\parallel \sqrt{1+\frac{\bpar}{2}(a_p-1)
    +\frac{k_\parallel^2}{4} \Big(1-\bpar(1-\frac{a_p}{2}) \Big)^2 },\\
\end{equation}
For strictly perpendicular propagation the solution is
\begin{equation} \label{eq:FLR1-Halll}
\omega = \pm k_\perp \sqrt{1+a_p\bpar +\frac{k_\perp^2\bpar^2 a_p^2}{16}},
\end{equation}
and the result is consistent with eq. 17 of \cite{Yajima1966},
after neglecting the electron inertia in that paper (by $\omega_0\rightarrow\infty$). The strictly perpendicular propagation is an
excellent example to point out the deficiencies of the Hall-CGL-FLR1 fluid model, or actually the CGL-FLR1 model, since the Hall contributions
are zero for $k_\parallel=0$. The result (\ref{eq:FLR1-Halll}) is also equal to eq. 34 of \cite{DelSarto2017}, who discuss in detail that compared to
solutions of a truncated kinetic Vlasov system, and also of a fluid model with more precise FLR corrections,
the solution (\ref{eq:FLR1-Halll}) actually has an opposite sign in front of the FLR term $\sim k_\perp^2$.
For further refinement with FLR2, see eq. (\ref{eq:FLR2-22}) and the subsequent discussion.
\subsection{CGL-FLR1 fluid model (no Hall term)}
If the Hall term is neglected, the general dispersion relation of the CGL-FLR1 fluid model reads
\begin{equation}
\Big(\omega^2-k_\parallel^2 v_{A\parallel}^2\Big)\Big(\omega^4-A_2\omega^2+A_0\Big) = \mathcal{P}^{\textrm{FLR-only}},
\end{equation}
where the CGL parameters $A_2,A_0$ are (\ref{eq:CGLparameters}) and the FLR1 polynomial on the right hand side has the form
\begin{eqnarray}
  \mathcal{P}^{\textrm{FLR-only}} &=& \omega^2 \bpar^2 (A_2'' \omega^2 -A_0'');\\
  A_2'' &=& k_\parallel^4 \Big(1-\frac{a_p}{2}\Big)^2 +k_\perp^4\frac{a_p^2}{16}+k_\parallel^2 k_\perp^2 a_p \Big(1-\frac{a_p}{2}\Big);\\
  A_0'' &=& k_\parallel^2 \Bigg\{ k_\parallel^4 \frac{3}{2}\bpar \Big(1-\frac{a_p}{2}\Big)^2
  +k_\parallel^2 k_\perp^2 \frac{a_p}{2}\Big(v_{A\parallel}^2+\frac{\bpar}{2}\Big)\Big(1-\frac{a_p}{2}\Big)
  +k_\perp^4 \frac{a_p}{2}\Big[ 1-\frac{a_p}{2}+a_p\bpar \Big(\frac{15}{16}-\frac{a_p}{2}\Big) \Big] \Bigg\},
\end{eqnarray}
where $v_{A\parallel}^2=1+\frac{\bpar}{2}(a_p-1)$. For strictly parallel propagation, the solution for the whistler and ion-cyclotron waves reads
\begin{equation}
\omega = \pm k_\parallel^2 \frac{\bpar}{2}\Big(1-\frac{a_p}{2}\Big) +k_\parallel \sqrt{v_{A\parallel}^2+\frac{k_\parallel^2\bpar^2}{4}\Big(1-\frac{a_p}{2}\Big)^2}.
\end{equation}
For strictly perpendicular propagation the solution is equivalent to (\ref{eq:FLR1-Halll}) since the Hall contributions vanish.

\subsection{2nd-order FLR corrections (FLR2)}
If a more precise FLR corrections are desired, it is relatively easy to increase the precision to the second order in
frequency $\omega/\Omega$ while keeping the precision in wavenumber $k r_L$ at first order. The pressure tensor equation
(\ref{eq:FLR_full}) is simplified to
\begin{equation}
  \bhat\times\boldsymbol{\Pi} + (\bhat\times\boldsymbol{\Pi})^{\textrm{T}} = - \frac{1}{\Omega}\frac{B_0}{|\bb|} \Big[
    \frac{\pr}{\pr t}(\bp^\textrm{g}+\boldsymbol{\Pi}) + \nabla\cdot(\bu\bp^\textrm{g} +\boldsymbol{q}^\textrm{g})
    + \bp^\textrm{g}\cdot\nabla\bu + (\bp^\textrm{g}\cdot\nabla \bu)^{\textrm{T}}
    \Big]. \label{eq:FLR2nd}
\end{equation}
For clarity, let's first consider the case with the heat flux $\boldsymbol{q}=0$. Compared to the first-order FLR corrections in the
previous section, we just have one more matrix present, $\frac{\pr}{\pr t}\boldsymbol{\Pi}$. As before, the FLR tensor must be symmetric and
also satisfy conditions $\Pi_{zz}=0$ and $\Pi_{yy}=-\Pi_{xx}$. Following the previous derivation, the system (\ref{eq:Syst1}) now includes also the
time derivative of the FLR tensor on the right hand side. Let's write down the system directly in the linear approximation. The system reads 
\begin{eqnarray}
 \left( \begin{array}{ccc}
    -2\Pi_{xy}, & 2\Pi_{xx}, & -\Pi_{yz} \\
    2\Pi_{xx}, & 2\Pi_{xy}, & \Pi_{xz} \\
    -\Pi_{yz}, & \Pi_{xz}, & 0
 \end{array} \right) &=& -\frac{1}{\Omega}\frac{\pr}{\pr t}
 \left( \begin{array}{ccc}
    \Pi_{xx} & \Pi_{xy} & \Pi_{xz} \\
    \Pi_{xy} & -\Pi_{xx} & \Pi_{yz} \\
    \Pi_{xz} & \Pi_{yz} & 0
 \end{array} \right)
 -\frac{1}{\Omega}
(p_\parallel^{(0)}-p_\perp^{(0)})
 \left( \begin{array}{ccc}
    0 & 0 & \frac{\pr \hat{b}_x}{\pr t} \\
    0 & 0 & \frac{\pr \hat{b}_y}{\pr t} \\
    \frac{\pr \hat{b}_x}{\pr t} & \frac{\pr \hat{b}_y}{\pr t} & 0
 \end{array} \right) \nn\\
&& -\frac{1}{\Omega}
    \left( \begin{array}{ccc}
    p_\perp^{(0)} (\pr_x u_x-\pr_y u_y); & p_\perp^{(0)} (\pr_x u_y + \pr_y u_x); & p_\perp^{(0)} \pr_x u_z + p_\parallel^{(0)} \pr_z u_x\\
    p_\perp^{(0)} (\pr_y u_x+\pr_x u_y); & p_\perp^{(0)} (\pr_y u_y-\pr_x u_x); & p_\perp^{(0)} \pr_y u_z+p_\parallel^{(0)}\pr_z u_y \\
    p_\parallel^{(0)} \pr_z u_x + p_\perp^{(0)}\pr_x u_z; & p_\parallel^{(0)} \pr_z u_y+p_\perp^{(0)}\pr_y u_z; & 0
  \end{array} \right).
\end{eqnarray}
There are only four FLR components that have to be considered and which can be rewritten separately as
\begin{eqnarray}
&&  \frac{\pr}{\pr t} \Pi_{xx} -2\Omega \Pi_{xy} +  p_\perp^{(0)} (\pr_x u_x-\pr_y u_y) =0;\\
&&  \frac{\pr}{\pr t} \Pi_{xy} +2\Omega \Pi_{xx} +  p_\perp^{(0)} (\pr_x u_y + \pr_y u_x)=0;\\
&&  \frac{\pr}{\pr t} \Pi_{xz} -\Omega \Pi_{yz} + p_\perp^{(0)} \pr_x u_z + p_\parallel^{(0)} \pr_z u_x +(p_\parallel^{(0)}-p_\perp^{(0)}) \frac{\pr \hat{b}_x}{\pr t}=0;\\
&&  \frac{\pr}{\pr t} \Pi_{yz} +\Omega \Pi_{xz} + p_\perp^{(0)} \pr_y u_z+p_\parallel^{(0)}\pr_z u_y +(p_\parallel^{(0)}-p_\perp^{(0)}) \frac{\pr \hat{b}_y}{\pr t}=0. 
\end{eqnarray}
The FLR components could be considered as independent fluid quantities described by the time-dependent equations as written above.
However, from a linear perspective this introduces additional linear modes, and from direct numerical simulations perspective it is preferred not
to introduce additional four time-dependent equations  (nevertheless, numerical simulations with the full pressure tensor equation are sometimes performed, see
e.g. \cite{Wang2015,DelSarto2016,Ng2017}). The above equations are therefore typically simplified by expanding the FLR tensor to the
first and second-order as $\boldsymbol{\Pi}=\boldsymbol{\Pi}^{(1)}+\boldsymbol{\Pi}^{(2)}$ and the time derivative of the second-order tensor is
neglected, i.e. $\frac{\pr}{\pr t}\boldsymbol{\Pi}^{(2)}=0$ and the system reads
\begin{eqnarray}
&&  \frac{\pr}{\pr t} \Pi_{xx}^{(1)} -2\Omega (\Pi_{xy}^{(1)}+\Pi_{xy}^{(2)}) +  p_\perp^{(0)} (\pr_x u_x-\pr_y u_y) =0;\\
&&  \frac{\pr}{\pr t} \Pi_{xy}^{(1)} +2\Omega (\Pi_{xx}^{(1)}+\Pi_{xx}^{(2)})+  p_\perp^{(0)} (\pr_x u_y + \pr_y u_x)=0;\\
  &&  \frac{\pr}{\pr t} \Pi_{xz}^{(1)} -\Omega (\Pi_{yz}^{(1)}+\Pi_{yz}^{(2)}) + p_\perp^{(0)} \pr_x u_z + p_\parallel^{(0)} \pr_z u_x
  +(p_\parallel^{(0)}-p_\perp^{(0)}) \Big( \pr_z u_x -\frac{c}{B_0}(\nabla\times\bE_H)_x \Big)=0;\\
  &&  \frac{\pr}{\pr t} \Pi_{yz}^{(1)} +\Omega (\Pi_{xz}^{(1)}+ \Pi_{xz}^{(2)}) + p_\perp^{(0)} \pr_y u_z+p_\parallel^{(0)}\pr_z u_y
  +(p_\parallel^{(0)}-p_\perp^{(0)}) \Big( \pr_z u_y -\frac{c}{B_0}(\nabla\times\bE_H)_y \Big)=0, 
\end{eqnarray}
where we also used the linearized induction equation in the last two equations.
Now there are two possibilities to handle the Hall-term contributions coming from the induction equation. It is possible to either define
$\boldsymbol{\Pi}^{(1)}$ to be equal to set (\ref{eq:FLR1s}) and move the Hall-term to $\boldsymbol{\Pi}^{(2)}$, or it is possible to keep the Hall term in
$\boldsymbol{\Pi}^{(1)}$ that is equal to linearized set (\ref{eq:FLR1Hall}). The first choice, i.e. if $\boldsymbol{\Pi}^{(1)}$ is set (\ref{eq:FLR1s}) yields
the second-order components
\begin{eqnarray}
&& \Pi_{xx}^{(2)} = -\frac{1}{2\Omega} \frac{\pr}{\pr t} \Pi_{xy}^{(1)}; \qquad 
 \Pi_{xy}^{(2)} = \frac{1}{2\Omega} \frac{\pr}{\pr t} \Pi_{xx}^{(1)}; \nn \\
&& \Pi_{xz}^{(2)} = -\frac{1}{\Omega} \frac{\pr}{\pr t} \Pi_{yz}^{(1)} + \frac{1}{\Omega}(p_\parallel^{(0)}-p_\perp^{(0)}) \frac{c}{B_0}(\nabla\times\bE_H)_y; \nn\\
&& \Pi_{yz}^{(2)} = \frac{1}{\Omega} \frac{\pr}{\pr t} \Pi_{xz}^{(1)} - \frac{1}{\Omega}(p_\parallel^{(0)}-p_\perp^{(0)}) \frac{c}{B_0}(\nabla\times\bE_H)_x. \label{eq:FLR2_o4}
\end{eqnarray}
The second choice, i.e. when $\boldsymbol{\Pi}^{(1)}$ is equal to linearized set (\ref{eq:FLR1Hall}) yields
\begin{eqnarray}
 \Pi_{xx}^{(2)} = -\frac{1}{2\Omega} \frac{\pr}{\pr t} \Pi_{xy}^{(1)};\quad 
 \Pi_{xy}^{(2)} = \frac{1}{2\Omega} \frac{\pr}{\pr t} \Pi_{xx}^{(1)};\quad
 \Pi_{xz}^{(2)} = -\frac{1}{\Omega} \frac{\pr}{\pr t} \Pi_{yz}^{(1)};\quad
 \Pi_{yz}^{(2)} = \frac{1}{\Omega} \frac{\pr}{\pr t} \Pi_{xz}^{(1)}.
\end{eqnarray}
The difference between the two approaches therefore is that the first approach neglects the time derivative of the Hall-term. It is noted that
the second approach is not necessarily better or more accurate than the first approach. For example for parallel propagation, the frequency of the
whistler mode seems to be better described by the first approach and the solutions are closer to kinetic theory.
\subsection{Hall-CGL-FLR2 fluid model}
The second order FLR tensor (\ref{eq:FLR2_o4}) is normalized and written in Fourier space according to
\begin{eqnarray}
  && \Pi_{xx}^{(2)} = +\frac{i}{2} \omega \Pi_{xy}^{(1)};\qquad 
   \Pi_{xy}^{(2)} = -\frac{i}{2} \omega \Pi_{xx}^{(1)};\qquad \Pi_{zz}^{(2)} =0; \nn\\
  && \Pi_{xz}^{(2)} = +i\omega \Pi_{yz}^{(1)}+(1-a_p)(\nabla\times \boldsymbol{E}_H)_y;\qquad
   \Pi_{yz}^{(2)} = -i\omega\Pi_{xz}^{(1)}-(1-a_p)(\nabla\times \boldsymbol{E}_H)_x, \label{eq:FLR2sss}
\end{eqnarray}
where for cold (and massless) electrons in the x-z plane
\begin{eqnarray}
  (\nabla\times \boldsymbol{E}_H)_x = +k_\parallel^2 B_y; \qquad (\nabla\times \boldsymbol{E}_H)_y = k_\parallel k_\perp B_z-k_\parallel^2 B_x; \qquad
  (\nabla\times \boldsymbol{E}_H)_z = -k_\parallel k_\perp B_y. \label{eq:FLR2-Hall}
\end{eqnarray}  
We purposely wrote the equations with $E_H$ so that the generalization to a more elaborate electric field will be straightforward.
Note that the Hall contributions (\ref{eq:FLR2-Hall}) in the FLR tensor completely vanish for isotropic temperatures $a_p=1$,
regardless of the form of $E_H$, i.e. regardless if the electrons are cold or not. 

To write down the solution for parallel propagation $(k_\perp=0)$ it is useful for brevity to introduce the following quantity
\begin{equation}
  v_b\equiv\bpar\Big(1-\frac{a_p}{2} \Big),
\end{equation}
and the parallel propagating whistler and ion-cyclotron modes factorize as
\begin{eqnarray} \label{eq:ICresFLR2}
  \omega^2 \Big(1+v_b k_\parallel^2\Big) \mp \omega k_\parallel^2 \Big(1+v_b(1+k_\parallel^2) \Big)
  -k_\parallel^2 \Big(v_{A\parallel}^2-\frac{\bpar}{2}k_\parallel^2 \Big)=0,
\end{eqnarray}  
and the two solutions are
\begin{eqnarray} \label{eq:FLR2-IC}
  \omega = \frac{1}{1+v_b k_\parallel^2} \bigg[ \pm \frac{k_\parallel^2}{2}\Big( 1+v_b(1+k_\parallel^2) \Big)
    +k_\parallel \sqrt{ \Big(1+\frac{\bpar}{2}(a_p-1)-\frac{\bpar}{2}k_\parallel^2\Big)(1+v_b k_\parallel^2)+
    \frac{k_\parallel^2}{4}\Big( 1+v_b(1+k_\parallel^2)  \Big)^2 } \bigg],
\end{eqnarray}  
with two further solutions corresponding to $-\omega$ on the left hand side. It is perhaps useful to rewrite expression (\ref{eq:ICresFLR2}) to the
``ion-cyclotron'' resonance form
\begin{equation}
\frac{k_\parallel^2}{\omega^2} = \frac{1+v_b k_\parallel^2}{\pm \omega \Big(1+v_b(1+k_\parallel^2)\Big) + \Big(v_{A\parallel}^2-\frac{\bpar}{2}k_\parallel^2 \Big)},
\end{equation}
that clearly shows that the ion-cyclotron resonances (in the parallel direction) become more and more complicated.
For strictly perpendicular propagation $(k_\parallel=0)$, where notably the Hall contributions (\ref{eq:FLR2-Hall}) vanish, the solution of the model is
\begin{eqnarray} \label{eq:FLR2-22}
\omega = \pm \frac{k_\perp}{1+k_\perp^2\frac{a_p}{8}\bpar} \sqrt{1+a_p\bpar +k_\perp^2 \frac{a_p}{8}\bpar \Big(1+\frac{3}{2}a_p\bpar \Big) }.
\end{eqnarray}  
It is useful to compare this result with the solution obtained for FLR1 corrections, eq. (\ref{eq:FLR1-Halll}). For direct comparison, expanding the
solution (\ref{eq:FLR2-22}) for small $|k_\perp|\ll 1$ yields
\begin{equation} \label{eq:FLR2-222}
\omega = \pm k_\perp \sqrt{1+a_p \bpar - k_\perp^2\frac{a_p}{8}\bpar \Big(1+\frac{a_p}{2}\bpar\Big)  }.
\end{equation}
Importantly, in comparison to FLR1 solution (\ref{eq:FLR1-Halll}), the FLR2 solution (\ref{eq:FLR2-222}) has an opposite sign in front of the
FLR term $\sim k_\perp^2$. The equation (\ref{eq:FLR2-222}) is equivalent to eq. 32 of \cite{DelSarto2017}, obtained by solving the pressure tensor
equation without the expansion to $\boldsymbol{\Pi}^{(1)}$ and $\boldsymbol{\Pi}^{(2)}$, and by expanding for small wavenumber afterwards.
The difference between the FLR1 solution (\ref{eq:FLR1-Halll}) and the FLR2 solution (\ref{eq:FLR2-222}) is perhaps even more clear, when written in
physical units for $\omega$ and $k$, and by using the gyroradius $\rho_i=v_{\textrm{th} \perp}/\Omega_p$, in the following form
\begin{eqnarray}
  \textrm{FLR1}: && \qquad \omega^2 = k_\perp^2 \Big[ V_A^2 + v_{\textrm{th}\perp}^2 \Big(1+\frac{k_\perp^2\rho_i^2}{16} \Big)\Big]; \label{eq:DelSarto1}\\
  \textrm{FLR2}: && \qquad \omega^2 = k_\perp^2 \Big[ V_A^2 \Big(1-\frac{k_\perp^2\rho_i^2}{8}\Big) + v_{\textrm{th}\perp}^2 \Big(1-\frac{k_\perp^2\rho_i^2}{16}\Big)\Big]; \label{eq:DelSarto2}\\
  \textrm{Kinetic}: && \qquad \omega^2 = k_\perp^2 \Big[ V_A^2 \Big(1-\frac{k_\perp^2\rho_i^2}{8}\Big) + v_{\textrm{th}\perp}^2 \Big(1-\frac{5}{16} k_\perp^2\rho_i^2 \Big)\Big].
  \label{eq:Smolyakov}
\end{eqnarray}
Equations (\ref{eq:DelSarto1}), (\ref{eq:DelSarto2}), (\ref{eq:Smolyakov}) are eq. 34, 32, 56 of \cite{DelSarto2017}.
As discussed in that paper, the FLR1 model does not only fail to capture the correction to the
Alv\'en velocity, it also introduces correction to the thermal velocity with incorrect sign.   
The kinetic result can be obtained for example from eq. 2.10 of \cite{Smolyakov1985},
when written for cold electrons $(\beta_e=0)$, and rewritten
to our notation (thermal speed in that paper does not have a factor of 2, so $(\rho_i^2)^{(MS)}=\rho_i^2/2$). 

Coming back to the generally oblique propagation, the second approach to write the FLR2 tensor, i.e. when the time derivative of the Hall term is not
neglected and kept in the first-order tensor $\boldsymbol{\Pi}^{(1)}$ with components
\begin{eqnarray}
  && \Pi_{xx}^{(1)} = -\frac{a_p}{2} ik_\perp u_y;\qquad 
   \Pi_{xy}^{(1)} = \frac{a_p}{2} ik_\perp u_x; \qquad  \Pi_{zz}^{(1)}=0; \nn\\
  && \Pi_{xz}^{(1)} = -(2-a_p)ik_\parallel u_y+(1-a_p)(\nabla\times E_H)_y;\nn\\
  && \Pi_{yz}^{(1)} =  (2-a_p)ik_\parallel u_x + a_pik_\perp u_z-(1-a_p)(\nabla\times E_H)_x,
\end{eqnarray}
yields the second-order tensor
\begin{eqnarray}
  \Pi_{xx}^{(2)} = +\frac{i}{2} \omega \Pi_{xy}^{(1)}; \quad \Pi_{xy}^{(2)} = -\frac{i}{2} \omega \Pi_{xx}^{(1)};\quad
  \Pi_{xz}^{(2)} = +i\omega \Pi_{yz}^{(1)}; \quad \Pi_{yz}^{(2)} = -i\omega\Pi_{xz}^{(1)}; \quad \Pi_{zz}^{(2)} =0.
\end{eqnarray}
The resulting dispersion relation for strictly parallel propagation is slightly different
\begin{eqnarray}
  \omega^2 \Big(1+v_b k_\parallel^2\Big) \mp \omega k_\parallel^2 \Big(1+v_b+\frac{\bpar}{2}k_\parallel^2 \Big)
  -k_\parallel^2 \Big(v_{A\parallel}^2-\frac{\bpar}{2}k_\parallel^2 \Big)=0,
\end{eqnarray} 
with explicit solutions
\begin{eqnarray}
  \omega = \frac{1}{1+v_b k_\parallel^2} \bigg[ \pm \frac{k_\parallel^2}{2}\Big( 1+v_b+\frac{\bpar}{2} k_\parallel^2 \Big)
    + k_\parallel \sqrt{ \Big(1+\frac{\bpar}{2}(a_p-1)-\frac{\bpar}{2} k_\parallel^2\Big)(1+v_b k_\parallel^2)+
    \frac{k_\parallel^2}{4}\Big( 1+v_b+\frac{\bpar}{2} k_\parallel^2  \Big)^2 } \bigg].
\end{eqnarray}
For isotropic temperatures $(a_p=1)$ the result is equivalent to (\ref{eq:FLR2-IC}) since the Hall term contributions for the FLR tensor disappear
for all propagation directions. Also, for perpendicular propagation, the dispersion relation is naturally equal to (\ref{eq:FLR2-22}). 

\subsection{FLR corrections with gyrotropic heat flux}
The equation (\ref{eq:FLR2nd}) contains the divergence of the gyrotropic heat flux, $\nabla\cdot\bq^\textrm{g}$, which in the index notation calculates
\begin{eqnarray}
 (\nabla\cdot\bq^\textrm{g})_{ij} &=& \pr_k q_{kij}^\textrm{g} = \pr_k q_{ijk}^\textrm{g}
  = \pr_k \Big[ q_\parallel \hat{b}_i\hat{b}_j\hat{b}_k +
    q_\perp\Big( \delta_{ij}\hat{b}_k +\delta_{jk}\hat{b}_i+\delta_{ki}\hat{b}_j - 3\hat{b}_i\hat{b}_j\hat{b}_k \Big) \Big] \nn\\
  &=& \pr_k \Big[ (q_\parallel-3q_\perp) \hat{b}_i\hat{b}_j\hat{b}_k +
    q_\perp\Big( \delta_{ij}\hat{b}_k +\delta_{jk}\hat{b}_i+\delta_{ki}\hat{b}_j \Big) \Big] \nn\\
  &=& \hat{b}_i\hat{b}_j\hat{b}_k \pr_k (q_\parallel-3q_\perp) + (q_\parallel-3q_\perp) \Big( \hat{b}_j\hat{b}_k\pr_k\hat{b}_i + \hat{b}_i\hat{b}_k\pr_k\hat{b}_j
  + \hat{b}_i\hat{b}_j\pr_k\hat{b}_k \Big) \nn \\
  && \quad + \Big( \delta_{ij}\hat{b}_k\pr_k +\hat{b}_i\pr_j +\hat{b}_j\pr_i\Big)q_\perp + q_\perp\Big(\delta_{ij}\pr_k \hat{b}_k +\pr_j \hat{b}_i +\pr_i \hat{b}_j\Big). \label{eq:bqgnabla}
\end{eqnarray}
Evaluation along magnetic field $\bhat=\bhat_0=(0,0,1)$ yields $\hat{b}_k\pr_k\rightarrow \pr_z$ and 
the matrix is evaluated as
\begin{eqnarray}
  (\nabla\cdot\bq^\textrm{g})_{xx} &=&  \pr_z q_\perp +q_\perp (\nabla\cdot\bhat + 2\pr_x \hat{b}_x);\\
  (\nabla\cdot\bq^\textrm{g})_{xy} &=& q_\perp (\pr_y \hat{b}_x + \pr_x \hat{b}_y ) = (\nabla\cdot\bq^\textrm{g})_{yx};\\
  (\nabla\cdot\bq^\textrm{g})_{xz} &=& (q_\parallel-2q_\perp)\pr_z\hat{b}_x + \pr_x q_\perp +q_\perp \pr_x \hat{b}_z = (\nabla\cdot\bq^\textrm{g})_{zx};\\
  (\nabla\cdot\bq^\textrm{g})_{yz} &=& (q_\parallel-2q_\perp)\pr_z\hat{b}_y +\pr_y q_\perp +q_\perp \pr_y \hat{b}_z = (\nabla\cdot\bq^\textrm{g})_{zy};\\
  (\nabla\cdot\bq^\textrm{g})_{yy} &=& \pr_z q_\perp +q_\perp ( \nabla\cdot\bhat + 2\pr_y \hat{b}_y );\\
  (\nabla\cdot\bq^\textrm{g})_{zz} &=& \pr_z q_\parallel + (q_\parallel-2q_\perp) \nabla\cdot\bhat,
\end{eqnarray}
where in the last 'zz' component we also used $\pr_z\hat{b}_z=0$, but for brevity we left the divergence
$\nabla\cdot\bhat=\pr_x\hat{b}_x+\pr_y\hat{b}_y$ intact in all expressions. Note that $(\nabla\cdot\bq^\textrm{g})_{yy}\neq -(\nabla\cdot\bq^\textrm{g})_{xx}$, which is not a
problem as will be described below.
 The pressure tensor equation is here approximated as
\begin{equation}
\frac{\pr (\bp^\textrm{g}+\boldsymbol{\Pi})}{\pr t} + \nabla\cdot(\bu\bp^\textrm{g}+\boldsymbol{q}^\textrm{g})+\bp^\textrm{g}\cdot\nabla \bu + (\bp^\textrm{g}\cdot\nabla \bu)^{\textrm{T}}
+\Omega\Big[\bhat\times\boldsymbol{\Pi} + (\bhat\times\boldsymbol{\Pi})^{\textrm{T}} \Big] =0,
\end{equation}
and evaluated with respect to $\bhat=\bhat_0=(0,0,1)$. All components were already evaluated and the calculation is straightforward, however,
we need to subtract from diagonal components the scalar pressure equations that contain the gyrotropic heat flux contributions and
that evaluated along $\bhat_0$ read
\begin{eqnarray}
&&  \frac{d p_\parallel}{d t} + p_\parallel \nabla\cdot \bu +2p_\parallel \pr_z u_z +\pr_z q_\parallel +(q_\parallel-2q_\perp)\nabla\cdot\bhat= 0;\\
&&  \frac{d p_\perp}{d t} + 2p_\perp \nabla\cdot\bu -p_\perp \pr_z u_z+\pr_z q_\perp +2q_\perp \nabla\cdot\bhat=0.
\end{eqnarray}
The requirement again is that $\Pi_{zz}=0$ and $\Pi_{yy}=-\Pi_{xx}$. Direct calculation yields
\begin{eqnarray}
&&  \frac{\pr}{\pr t} \Pi_{xx} -2\Omega \Pi_{xy} +  p_\perp (\pr_x u_x-\pr_y u_y) +q_\perp (\pr_x\hat{b}_x-\pr_y\hat{b}_y) =0;\\
&&  \frac{\pr}{\pr t} \Pi_{xy} +2\Omega \Pi_{xx} +  p_\perp (\pr_x u_y + \pr_y u_x)+q_\perp( \pr_x\hat{b}_y+\pr_y\hat{b}_x) =0;\\
  &&  \frac{\pr}{\pr t} \Pi_{xz} -\Omega \Pi_{yz} + p_\perp\pr_x u_z + p_\parallel\pr_z u_x +(p_\parallel-p_\perp) \frac{d \hat{b}_x}{d t}
  +(q_\parallel-2q_\perp)\pr_z\hat{b}_x +q_\perp\pr_x\hat{b}_z+\pr_x q_\perp =0; \label{eq:FLR-qparR}\\
  &&  \frac{\pr}{\pr t} \Pi_{yz} +\Omega \Pi_{xz} + p_\perp\pr_y u_z+p_\parallel\pr_z u_y +(p_\parallel-p_\perp) \frac{d \hat{b}_y}{d t}
  +(q_\parallel-2q_\perp)\pr_z\hat{b}_y +q_\perp\pr_y\hat{b}_z + \pr_y q_\perp=0.\label{eq:FLR-qperpR} 
\end{eqnarray}
Note that after subtraction of the perpendicular pressure equation, the $'xx'$ and $'yy'$ components are again anti-symmetric,
i.e. $(\nabla\cdot\bq^\textrm{g})_{yy}-\pr_z q_\perp - 2q_\perp \nabla\cdot\bhat = - q_\perp (\pr_x \hat{b}_x-\pr_y \hat{b}_y)$, and the system
indeed satisfies that $\Pi_{yy}=-\Pi_{xx}$. Also, considering the $'zz'$ component, after subtraction of the parallel pressure equation
all the terms cancel, i.e. $(\nabla\cdot\bq^\textrm{g})_{zz}- \pr_z q_\parallel -(q_\parallel-2q_\perp)\nabla\cdot\bhat =0$
and the system indeed satisfies $\Pi_{zz}=0$. 

Finally, expressing the above system in the linear approximation, with the assumption that mean heat flux values are zero,
i.e. $q_\parallel^{(0)}=0$, $q_\perp^{(0)}=0$, the only gyrotropic heat flux contributions that remain are $\pr_x q_\perp$ and $\pr_y q_\perp$,
 in equations (\ref{eq:FLR-qparR}), (\ref{eq:FLR-qperpR}). 
The FLR tensor is again separated to $\boldsymbol{\Pi}^{(1)}+\boldsymbol{\Pi}^{(2)}$ and similarly to the Hall-term contributions,
it is again a matter of choice if the heat flux contributions are pushed to $\boldsymbol{\Pi}^{(2)}$ or kept in $\boldsymbol{\Pi}^{(1)}$.
We prefer the first choice, i.e. when $\boldsymbol{\Pi}^{(1)}$ is equivalent to (\ref{eq:FLR1s}), which yields the second order tensor
\begin{eqnarray}
&& \Pi_{xx}^{(2)} = -\frac{1}{2\Omega} \frac{\pr}{\pr t} \Pi_{xy}^{(1)};\qquad
 \Pi_{xy}^{(2)} = \frac{1}{2\Omega} \frac{\pr}{\pr t} \Pi_{xx}^{(1)};\nn\\
  && \Pi_{xz}^{(2)} = -\frac{1}{\Omega} \frac{\pr}{\pr t} \Pi_{yz}^{(1)} + \frac{1}{\Omega}(p_\parallel^{(0)}-p_\perp^{(0)}) \frac{c}{B_0}(\nabla\times\bE_H)_y
  -\frac{1}{\Omega}\pr_y q_\perp;\nn\\
  && \Pi_{yz}^{(2)} = \frac{1}{\Omega} \frac{\pr}{\pr t} \Pi_{xz}^{(1)} - \frac{1}{\Omega}(p_\parallel^{(0)}-p_\perp^{(0)}) \frac{c}{B_0}(\nabla\times\bE_H)_x
  +\frac{1}{\Omega}\pr_x q_\perp. \label{eq:FLR-GyroX}
\end{eqnarray}
Note that we could have derived the contributions $\pr_x q_\perp$ and $\pr_y q_\perp$ in a much quicker way, if the expression
(\ref{eq:bqgnabla}) was linearized from the beginning. However, we wanted to demonstrate that if nonlinear FLR corrections are considered, the heat flux
nonlinearities can significantly complicate the dynamics, even if simplified evaluation along $\bhat_0$ is performed.
Right now, we can not use these FLR corrections to obtain a dispersion relation, since we have no means to determine a closure
for the gyrotropic heat flux $q_\perp$. We would have to consider CGL2 fluid model or more complicated Landau fluid models,
that contains evolution equations for $q_\parallel$ and $q_\perp$. Additionally, the perpendicular propagating fast mode,
as well as the parallel propagating ion-cyclotron and whistler modes, are not influenced by the gyrotropic heat flux.  
Instead, as the last step, we will consider contributions of the non-gyrotropic heat flux. 
\subsection{FLR corrections with non-gyrotropic heat flux vectors (FLR3)} \label{sec:FLR3x}
It is possible to further increase precision of the FLR corrections by considering non-gyrotropic heat flux contributions.
Detailed calculations with the non-gyrotropic heat flux vectors $\boldsymbol{S}^\parallel_\perp$ and $\boldsymbol{S}^\perp_\perp$ are presented in the Appendix \ref{sec:NONGheat}.
At the linear level, these non-gyrotropic heat flux vectors contribute to the linearized pressure equations
\begin{eqnarray}
  &&  \frac{\pr p_\parallel}{\pr t} +p_\parallel^{(0)}\nabla\cdot\bu +2p_\parallel^{(0)} \pr_z u_z +\pr_z q_\parallel
  + \pr_x \big( S^{\parallel}_\perp \big)_x + \pr_y \big( S^{\parallel}_\perp \big)_y =0;\\
  && \frac{\pr p_\perp}{\pr t} +2p_\perp^{(0)}\nabla\cdot\bu -p_\perp^{(0)} \pr_z u_z +\pr_z q_\perp +\pr_x \big( S^{\perp}_\perp \big)_x + \pr_y \big( S^{\perp}_\perp \big)_y=0,
\end{eqnarray} 
and also to the linearized equations for the FLR tensor
\begin{eqnarray}
&&  \frac{\pr}{\pr t} \Pi_{xx} -2\Omega \Pi_{xy} +  p_\perp^{(0)} (\pr_x u_x-\pr_y u_y)  +\frac{1}{2}\big[ \pr_x (S^\perp_\perp)_x- \pr_y (S^\perp_\perp)_y \big]=0;\\
&&  \frac{\pr}{\pr t} \Pi_{xy} +2\Omega \Pi_{xx} +  p_\perp^{(0)} (\pr_x u_y + \pr_y u_x) +\frac{1}{2} \big[ \pr_y (S^\perp_\perp)_x +\pr_x (S^\perp_\perp)_y\big]=0;\\
  &&  \frac{\pr}{\pr t} \Pi_{xz} -\Omega \Pi_{yz} + p_\perp^{(0)}\pr_x u_z + p_\parallel^{(0)}\pr_z u_x +(p_\parallel^{(0)}-p_\perp^{(0)}) \frac{\pr \hat{b}_x}{\pr t}
  +\pr_x q_\perp +\pr_z (S^\parallel_\perp)_x  =0;\\
  &&  \frac{\pr}{\pr t} \Pi_{yz} +\Omega \Pi_{xz} + p_\perp^{(0)}\pr_y u_z+p_\parallel^{(0)}\pr_z u_y +(p_\parallel^{(0)}-p_\perp^{(0)}) \frac{\pr \hat{b}_y}{\pr t}
   + \pr_y q_\perp+\pr_z (S^\parallel_\perp)_y =0. 
\end{eqnarray}
These equations are equivalent to equations 18-21 of \cite{Goswami2005} (by noting that the vector $\boldsymbol{S}^\perp = q_\perp\bhat +\boldsymbol{S}^\perp_\perp$,
so  at the linear level $S^\perp_z=q_\perp$). The non-gyrotropic heat flux vectors are separated to the first and second order
\begin{eqnarray}
  \boldsymbol{S}^\parallel_\perp=\boldsymbol{S}^{\parallel (1)}_\perp + \boldsymbol{S}^{\parallel (2)}_\perp; \qquad
  \boldsymbol{S}^\perp_\perp=\boldsymbol{S}^{\perp (1)}_\perp + \boldsymbol{S}^{\perp (2)}_\perp.
\end{eqnarray}
By using the nonlinear results for the first-order vectors derived in the Appendix \ref{sec:NONGheat}, eq. (\ref{eq:SP2015typo}), (\ref{eq:SperpPerpF}),
yields that at the linear level
\begin{eqnarray}
  \boldsymbol{S}^{\parallel (1)}_\perp &=&  \frac{1}{\Omega} \bhat_0\times \Big[ p_\perp^{(0)} \nabla\Big(\frac{p_\parallel}{\rho}\Big)
  +2\frac{p_\parallel^{(0)}}{\rho_0}(p_\parallel^{(0)}-p_\perp^{(0)}) \pr_z \bhat \Big]; \label{eq:SparFLR}\\
  \boldsymbol{S}_\perp^{\perp(1)}  &=& \frac{1}{\Omega} \bhat_0\times \Big[
    2p_\perp^{(0)} \nabla \Big( \frac{p_\perp}{\rho}\Big) \Big], \label{eq:SperpFLR}
\end{eqnarray}
where the gradients are meant to be further linearized. Nevertheless, the above form is useful to point out that the non-gyrotropic heat fluxes
are proportional to the gradients of the temperature. For clarity, partially linearized expressions are also written down by components in the Appendix \ref{sec:NONGheat},
see equations
(\ref{eq:SparPerpPLy}), (\ref{eq:SparPerpPLx}) and (\ref{eq:SperpPerpPLy}), (\ref{eq:SperpPerpPLx}). 
It is important to emphasize that the expressions (\ref{eq:SparFLR}), (\ref{eq:SperpFLR})
were derived for a bi-Maxwellian distribution function. As discussed later in the text, this is achieved by prescribing closures for the gyrotropic 4th-order moment
in the form $r_{\parallel\parallel} = \frac{3p_\parallel^2}{\rho} +\widetilde{r}_{\parallel\parallel}$, $ r_{\parallel\perp} = \frac{p_\parallel p_\perp}{\rho}+\widetilde{r}_{\parallel\perp}$, 
and $r_{\perp\perp} = \frac{2 p_\perp^2}{\rho}+\widetilde{r}_{\perp\perp}$. In (\ref{eq:SparFLR}), (\ref{eq:SperpFLR}), we additionally neglected the perturbations $\widetilde{r}$
(since right now we do not want to consider models where these perturbations are evaluated from linear kinetic theory).
\emph{Therefore, the FLR corrections with the non-gyrotropic heat flux, that we call here FLR3, are distribution function specific.}
Nevertheless, as discussed later in the text, very similar closures can be also obtained for the bi-Kappa distribution function, just with additional coefficients
$\alpha_\kappa=(\kappa-3/2)/(\kappa-5/2)$, and expressions similar to (\ref{eq:SparFLR}), (\ref{eq:SperpFLR}) can be derived.

The second-order vectors are also derived in the Appendix \ref{sec:NONGheat},
this time directly in the linear approximation, eq. (\ref{eq:Spar_2nd}), (\ref{eq:Sperp_2nd}) and the vectors read 
\begin{eqnarray}
  (\boldsymbol{S}^{\parallel}_{\perp})^{(2)}_k &=& \frac{1}{\Omega} \bhat_0 \times \Big[ \frac{\pr}{\pr t} (\boldsymbol{S}^{\parallel}_{\perp})^{(1)}_k
  + 2\frac{p_\parallel^{(0)}}{\rho_0} \pr_z \Pi_{kz}^{(1)} \Big]; \label{eq:Spar_2nd2}\\
 (\boldsymbol{S}^{\perp}_{\perp})^{(2)}_k &=& \frac{1}{\Omega} \bhat_0 \times \Big[ \frac{\pr}{\pr t} (\boldsymbol{S}^{\perp}_{\perp})^{(1)}_k
+ \frac{p_\perp^{(0)}}{\rho_0}\Big( \pr_x \Pi_{xk}^{(1)} +\pr_y \Pi_{yk}^{(1)}  \Big) \Big]. \label{eq:Sperp_2nd2}
\end{eqnarray}
As a reminder, $\Pi_{ij}=\Pi_{ji}$.
For clarity, we wrote the expressions above explicitly in the index notation. The non-gyrotropic heat flux vectors are perpendicular to $\bhat_0=(0,0,1)$, and
the non-zero components are for index $k=x,y$. The expressions can be written in a more elegant vector form, 
by defining vector $\vec{\boldsymbol{\Pi}}_z\equiv (\Pi_{xz},\Pi_{yz},\Pi_{zz}=0)$, vector $\nabla_\perp = (\pr_x, \pr_y, 0)$, and matrix
\begin{eqnarray}
\boldsymbol{\Pi}_\perp \equiv \left( \begin{array}{ccc}
    \Pi_{xx}, & \Pi_{xy}, & 0 \\
    \Pi_{yx}, & \Pi_{yy}, & 0 \\
    0, & 0, & 0
 \end{array} \right),
\end{eqnarray}
so that $\nabla_\perp\cdot\boldsymbol{\Pi}_\perp = (\pr_x \Pi_{xx}+\pr_y \Pi_{yx}, \pr_x \Pi_{xy}+\pr_y \Pi_{yy})$, yielding
\begin{eqnarray}
  \boldsymbol{S}^{\parallel (2)}_{\perp} &=& \frac{1}{\Omega} \bhat_0 \times \Big[ \frac{\pr}{\pr t} \boldsymbol{S}^{\parallel (1)}_{\perp}
  + 2\frac{p_\parallel^{(0)}}{\rho_0} \pr_z\vec{\boldsymbol{\Pi}}_z^{(1)} \Big]; \label{eq:Spar_2nd3}\\
 \boldsymbol{S}^{\perp (2)}_\perp  &=& \frac{1}{\Omega} \bhat_0 \times \Big[ \frac{\pr}{\pr t} \boldsymbol{S}^{\perp (1)}_{\perp}
+ \frac{p_\perp^{(0)}}{\rho_0} \nabla_\perp\cdot\boldsymbol{\Pi}_\perp^{(1)} \Big], \label{eq:Sperp_2nd3}
\end{eqnarray}
which are equations 53, 54 of \cite{PSH2012}. 

Normalizing the equations (dropping the tilde), Fourier transforming, and writing them in the x-z plane (with $\pr_y=0$) yields
\begin{eqnarray}
&&  -i\omega \Pi_{xx} -2 \Pi_{xy} +  a_p ik_\perp u_x +\frac{1}{2} ik_\perp (S^\perp_\perp)_x=0;\\
&&  -i\omega \Pi_{xy} +2 \Pi_{xx} +  a_p ik_\perp u_y  +\frac{1}{2} ik_\perp (S^\perp_\perp)_y =0;\\
  &&  -i\omega \Pi_{xz} - \Pi_{yz} + a_p ik_\perp u_z + ik_\parallel u_x +(1-a_p) \Big( ik_\parallel u_x -(\nabla\times\bE_H)_x \Big)
  +ik_\perp q_\perp +ik_\parallel (S^\parallel_\perp)_x  =0;\label{eq:PixzRit}\\
  &&  -i\omega \Pi_{yz} + \Pi_{xz} + ik_\parallel u_y +(1-a_p) \Big( ik_\parallel u_y -(\nabla\times\bE_H)_y \Big) 
   +ik_\parallel (S^\parallel_\perp)_y =0, \label{eq:PiyzRit}
\end{eqnarray}
where the terms with Hall electric field components are specified in (\ref{eq:FLR2-Hall}). 
 At this moment we do not consider higher-order fluid models with gyrotropic heat fluxes such as CGL2 or Landau fluids (see later in the text)
  and to have a closed model, here we prescribe $q_\perp=0$, $q_\parallel=0$.
It is  useful to briefly explore what fluid models are obtained, if we decide to keep only the first-order vectors
$\boldsymbol{S}^{\parallel (1)}_\perp$, $\boldsymbol{S}^{\perp (1)}_\perp$, and ignore the second-order corrections $\boldsymbol{S}^{\parallel (2)}_\perp$, $\boldsymbol{S}^{\perp (2)}_\perp$. By considering only the $\boldsymbol{S}^{\parallel (1)}_\perp$, $\boldsymbol{S}^{\perp (1)}_\perp$ contributions, we can either put them into
to $\boldsymbol{\Pi}^{(1)}$ or $\boldsymbol{\Pi}^{(2)}$.
 However, it can be shown that both options yield dispersion relation for the perpendicular fast mode (long-wavelength, in physical units)
\begin{equation} \label{eq:PerpPIC2}
\omega^2 = k_\perp^2 \Big[ V_A^2 \Big(1-\frac{k_\perp^2\rho_i^2}{8}\Big) + v_{\textrm{th}\perp}^2 \Big(1+\frac{k_\perp^2\rho_i^2}{16}\Big)\Big].
\end{equation}
Rather amusingly, in comparison to the FLR2 result (\ref{eq:DelSarto2}), the wrong sign in the correction of the thermal speed is back!

\subsection{Hall-CGL-FLR3 fluid model} \label{sec:Hall-CGL-FLR3}
Obviously, if matching with kinetic theory for the fast perpendicular mode is of upmost importance (for small wavenumbers),
we have no other choice and we have to keep the second-order
non-gyrotropic heat flux contributions $\boldsymbol{S}^{\parallel (2)}_\perp$, $\boldsymbol{S}^{\perp (2)}_\perp$. We call this FLR pressure tensor as FLR3,
and as stated previously, the FLR3 pressure tensor is distribution function specific. 
This model is very important to us and especially for the following discussion of the firehose instability. In order to be completely clear on which equations
we are solving and also so that our results can be easily reproduced, let's state the entire model in all of its beauty.
The linearized, normalized, Fourier transformed equations written in the x-z plane read
\begin{eqnarray}
&& -\omega \rho + k_\perp u_x + k_\parallel u_z =0;\nn\\
&& -\omega u_x +\frac{\bpar}{2}k_\perp p_\perp -v_{A\parallel}^2 k_\parallel B_x  +k_\perp B_z +\frac{\bpar}{2}\big( k_\perp \Pi_{xx} +k_\parallel \Pi_{xz} \big)=0;\nn\\
&& -\omega u_y - v_{A\parallel}^2 k_\parallel B_y +\frac{\bpar}{2}\big(k_\perp \Pi_{xy}+k_\parallel \Pi_{yz}  \big)=0;\nn\\
&& -\omega u_z +\frac{\bpar}{2}k_\parallel p_\parallel +\frac{\bpar}{2}(1-a_p)k_\perp B_x + \frac{\bpar}{2}k_\perp \Pi_{xz}=0;\nn\\
&& -\omega B_x -k_\parallel u_x -i k_\parallel^2 B_y=0;\nn\\
&& -\omega B_y - k_\parallel u_y -ik_\parallel k_\perp B_z +ik_\parallel^2 B_x=0;\nn\\
&& -\omega B_z + k_\perp u_x+ik_\parallel k_\perp B_y=0;\nn\\
&& -\omega p_\parallel + k_\perp u_x +3 k_\parallel u_z+\cancel{k_\parallel q_\parallel} +k_\perp (S^\parallel_\perp)_x= 0;\nn\\
&&  -\omega p_\perp  + 2a_p k_\perp u_x +a_p k_\parallel u_z +\cancel{k_\parallel q_\perp} +k_\perp (S^\perp_\perp)_x=0, \label{eq:HallFLR3-P}
\end{eqnarray}
where each component of the (FLR) pressure tensor $\boldsymbol{\Pi}$, and each component of the (FLR) heat flux vectors
$\boldsymbol{S}^{\parallel}_\perp$, $\boldsymbol{S}^{\perp}_\perp$, is separated to
\begin{equation}
\boldsymbol{\Pi}=\boldsymbol{\Pi}^{(1)}+\boldsymbol{\Pi}^{(2)}; \qquad  \boldsymbol{S}^\parallel_\perp=\boldsymbol{S}^{\parallel (1)}_\perp + \boldsymbol{S}^{\parallel (2)}_\perp; \qquad
  \boldsymbol{S}^\perp_\perp=\boldsymbol{S}^{\perp (1)}_\perp + \boldsymbol{S}^{\perp (2)}_\perp. \nn
\end{equation}
The components of the $\boldsymbol{\Pi}$ tensor are given by
\begin{eqnarray}
\begin{tabular}{ l  l }
  $\Pi_{xx}^{(1)} = -\frac{a_p}{2} ik_\perp u_y$; &
  $\quad \Pi_{xx}^{(2)} = +i\frac{\omega}{2} \Pi_{xy}^{(1)} -i\frac{k_\perp}{4}(S^\perp_\perp)^{(1)+(2)}_y$;\\
  $\Pi_{xy}^{(1)} = \frac{a_p}{2} ik_\perp u_x$; &
  $\quad\Pi_{xy}^{(2)} = -i\frac{\omega}{2} \Pi_{xx}^{(1)}+ i\frac{k_\perp}{4}(S^\perp_\perp)^{(2)}_x$;\\
  $\Pi_{xz}^{(1)} = -(2-a_p)ik_\parallel u_y$; &
  $\quad \Pi_{xz}^{(2)} = +i\omega \Pi_{yz}^{(1)} + (1-a_p)(\nabla\times\bE_H)_y -ik_\parallel (S^\parallel_\perp)^{(1)+(2)}_y$;\\
  $\Pi_{yz}^{(1)} =  (2-a_p)ik_\parallel u_x + a_pik_\perp u_z$; &
  $\quad\Pi_{yz}^{(2)} = -i\omega \Pi_{xz}^{(1)} - (1-a_p)(\nabla\times\bE_H)_x+\cancel{ik_\perp q_\perp}+ik_\parallel (S^\parallel_\perp)^{(1)+(2)}_x$,
\end{tabular}
\end{eqnarray}
where the Hall electric field contributions for cold massless electrons read
\begin{eqnarray} \label{eq:Hall_FLR3}
  (\nabla\times \boldsymbol{E}_H)_x = +k_\parallel^2 B_y; \qquad (\nabla\times \boldsymbol{E}_H)_y = k_\parallel k_\perp B_z-k_\parallel^2 B_x,
\end{eqnarray}
and the components of the non-gyrotropic (FLR) heat fluxes are given by
\begin{eqnarray}
\begin{tabular}{l l}
$(S^\parallel_\perp)^{(1)}_x = -\bpar (1-a_p) i k_\parallel B_y$; & \qquad $(S^\parallel_\perp)^{(2)}_x = +i\omega (S^\parallel_\perp)^{(1)}_y-\bpar i k_\parallel \Pi_{yz}^{(1)}$; \\
  $(S^\parallel_\perp)^{(1)}_y = a_p\frac{\bpar}{2}i k_\perp (p_\parallel-\rho) + \bpar (1-a_p) ik_\parallel B_x$; & \qquad
  $(S^\parallel_\perp)^{(2)}_y = -i\omega (S^\parallel_\perp)^{(1)}_x+\bpar i k_\parallel \Pi_{xz}^{(1)}$; \\
$(S_\perp^\perp)_x^{(1)} = 0$; & \qquad $(S^\perp_\perp)^{(2)}_x = +i\omega (S^\perp_\perp)^{(1)}_y -a_p \frac{\bpar}{2} ik_\perp \Pi_{xy}^{(1)}$; \\
  $(S_\perp^\perp)_y^{(1)} = \bpar a_p i k_\perp (p_\perp -a_p\rho)$; &\qquad $(S^\perp_\perp)^{(2)}_y = -\cancel{i\omega (S^\perp_\perp)^{(1)}_x} +a_p\frac{\bpar}{2} i k_\perp \Pi_{xx}^{(1)}$.
  \label{eq:SsuperCool1}
\end{tabular}  
\end{eqnarray}  
Importantly, the $\boldsymbol{\Pi}^{(2)}$ expressions contain both the 1st and 2nd order contributions from the heat flux vectors
$\boldsymbol{S}^{\parallel}_\perp$, $\boldsymbol{S}^{\perp}_\perp$.
To explicitly see where the closure for the gyrotropic heat fluxes $q_\parallel=0,q_\perp=0$ was performed, we scratched the terms involving these quantities.
Additionally, we also scratched contributions from $(S_\perp^\perp)_x^{(1)}$, which are here zero at the linear level. 

Let's check the solution of the Hall-CGL-FLR3 fluid model for the perpendicular propagation ($k_\parallel=0$), which in this case reads 
\begin{equation}
  \omega^2 = \frac{k_\perp^2}{(1+k_\perp^2\frac{a_p}{8}\bpar)(1+a_p\bpar k_\perp^2)} \Big[ 1+a_p\bpar
    +k_\perp^2 a_p \bpar \Big( 1+\frac{13}{16}a_p\bpar +\frac{9}{128}a_p^2\bpar^2 k_\perp^2 \Big)   \Big].
\end{equation}
As a ``sanity check'', since the solution stays always positive for all the wavenumbers and we do not have any instability, the model appears to be good.
For small wavenumbers, the expansion yields
\begin{equation}
\omega^2 = k_\perp^2 \Big[ 1+a_p\bpar -k_\perp^2 a_p\bpar \big(\frac{1}{8}+\frac{5}{16}a_p\bpar \Big) +\cdots \Big],
\end{equation}
and in physical units
\begin{equation} \label{eq:PerpPIC3}
\textrm{FLR3:} \qquad \omega^2 = k_\perp^2 \Big[ V_A^2 \Big(1-\frac{k_\perp^2\rho_i^2}{8}\Big) + v_{\textrm{th}\perp}^2 \Big(1-\frac{5}{16}k_\perp^2\rho_i^2\Big)\Big].
\end{equation}
Therefore, the Hall-CGL-FLR3 fluid model finally matches the analytic result from kinetic theory!  
We note that if the $\boldsymbol{\Pi}$ contributions are neglected in the second-order heat flux expressions
(\ref{eq:SsuperCool1}), instead of the correct $-5/16$ FLR correction to the thermal speed, one obtains $-7/16$. The necessity to keep the
second-order heat flux contributions to recover the kinetic result for the perpendicular fast mode was discovered by \cite{Smolyakov1985}. 

Checking the solution for the parallel propagating ($k_\perp=0$) ion-cyclotron and whistler modes yields the following dispersion relation
\begin{eqnarray}  \label{eq:ICresFLR3_4}
 && \omega^2 \Big(1+v_b k_\parallel^2\Big) \mp \omega k_\parallel^2 \Big[1+v_b(1+k_\parallel^2) +k_\parallel^2 \bpar^2 (\frac{3}{2}-a_p)\Big]\nn\\
 &&\qquad\qquad -k_\parallel^2 \Big[v_{A\parallel}^2-\frac{\bpar}{2}k_\parallel^2 \big(1+\bpar(1-a_p)\big) +\frac{\bpar^2}{2}(a_p-2)k_\parallel^4\Big]=0,
\end{eqnarray}
where the quantities $v_b=\bpar(1-\frac{a_p}{2})$ and $v_{A\parallel}^2=1+\frac{\bpar}{2}(a_p-1)$. 
We will use this dispersion relation to study the firehose instability. For a general oblique propagation direction, the dispersion relation is
obviously way too large to write down, and we recommend to use analytic software such as Maple or Mathematica. 
\subsubsection*{Moving the Hall-term to $\boldsymbol{\Pi}^{(1)}$}
For completeness, just in case we want to investigate later the influence of the Hall-term, moving it to $\boldsymbol{\Pi}^{(1)}$ yields
the following dispersion relation for the parallel propagation
\begin{eqnarray}  \label{eq:ICresFLR3_5}
 && \omega^2 \Big(1+v_b k_\parallel^2\Big) \mp \omega k_\parallel^2 \Big[1+v_b+k_\parallel^2\frac{\bpar}{2} +k_\parallel^2 \bpar^2 (\frac{3}{2}-a_p)\Big]\nn\\
 &&\qquad\qquad -k_\parallel^2 \Big[v_{A\parallel}^2-\frac{\bpar}{2}k_\parallel^2 \big(1+\bpar(1-a_p)\big) -\frac{\bpar^2}{2}k_\parallel^4\Big]=0.
\end{eqnarray}



\newpage
\section{Parallel and oblique firehose instability} \label{section:Firehose}
\subsection{Parallel propagation}
For clarity, it is useful to summarize all 3 major models that describe the parallel propagating ion-cyclotron and whistler modes. 
By prescribing $k_\perp=0$ in the equations of the Hall-CGL-FLR3 fluid model, the model greatly simplifies. The parallel magnetic field $B_z=0$, the ion-acoustic mode decouples,
and both the first and second order contributions to $\Pi_{xx}=0$, $\Pi_{xy}=0$, $\boldsymbol{S}^\perp_\perp=0$. 
The normalized, Fourier transformed equations written in the x-z plane read
\begin{eqnarray}
&& -\omega u_x -v_{A\parallel}^2 k_\parallel B_x  +\frac{\bpar}{2} k_\parallel \Pi_{xz}=0;\nn\\
&& -\omega u_y - v_{A\parallel}^2 k_\parallel B_y +\frac{\bpar}{2} k_\parallel \Pi_{yz}=0;\nn\\
&& -\omega B_x -k_\parallel u_x -i k_\parallel^2 B_y=0;\nn\\
&& -\omega B_y - k_\parallel u_y +ik_\parallel^2 B_x=0, \label{eq:ParallelFirehoseKurva}
\end{eqnarray}
where the components of the non-gyrotropic (FLR) pressure tensor $\boldsymbol{\Pi}$ are given by
\begin{eqnarray}
\begin{tabular}{ l  l }
  $\Pi_{xz}^{(1)} = -(2-a_p)ik_\parallel u_y$; &
  $\quad \Pi_{xz}^{(2)} = +i\omega \Pi_{yz}^{(1)} - (1-a_p)k_\parallel^2 B_x  -ik_\parallel (S^\parallel_\perp)^{(1)+(2)}_y$;\\
  $\Pi_{yz}^{(1)} = + (2-a_p)ik_\parallel u_x $; &
  $\quad\Pi_{yz}^{(2)} = -i\omega \Pi_{xz}^{(1)} - (1-a_p)k_\parallel^2 B_y +ik_\parallel (S^\parallel_\perp)^{(1)+(2)}_x$,
\end{tabular}
\end{eqnarray}
and the components of the non-gyrotropic (FLR) heat flux $\boldsymbol{S}^\parallel_\perp$ read
\begin{eqnarray}
\begin{tabular}{l l}
$(S^\parallel_\perp)^{(1)}_x = -\bpar (1-a_p) i k_\parallel B_y$; & \qquad $(S^\parallel_\perp)^{(2)}_x = +i\omega (S^\parallel_\perp)^{(1)}_y-\bpar i k_\parallel \Pi_{yz}^{(1)}$; \\
  $(S^\parallel_\perp)^{(1)}_y = + \bpar (1-a_p) ik_\parallel B_x$; & \qquad
  $(S^\parallel_\perp)^{(2)}_y = -i\omega (S^\parallel_\perp)^{(1)}_x+\bpar i k_\parallel \Pi_{xz}^{(1)}$.
\end{tabular}  
\end{eqnarray}
When the entire $\boldsymbol{\Pi}$ is neglected, yields the Hall-CGL solution
\begin{equation} \label{eq:firehose-HallCGL}
\omega =\pm \frac{\kpar^2}{2}+\kpar \sqrt{v_{A\parallel}^2+\frac{\kpar^2}{4} }.
\end{equation}  
Neglecting the $\boldsymbol{\Pi}^{(2)}$ contributions yields the Hall-CGL-FLR1 solution
\begin{equation} \label{eq:firehose-FLR1}
  \omega = \pm \frac{k_\parallel^2}{2}\Big(1+v_b\Big) + k_\parallel \sqrt{v_{A\parallel}^2  +\frac{k_\parallel^2}{4} \Big(1-v_b\Big)^2 },
\end{equation}
and neglecting the heat flux $\boldsymbol{S}^\parallel_\perp$ yields the Hall-CGL-FLR2 solution 
\begin{eqnarray}
  \omega = \frac{1}{1+v_b k_\parallel^2} \bigg[ \pm \frac{k_\parallel^2}{2}\Big( 1+v_b(1+k_\parallel^2) \Big)
    +k_\parallel \sqrt{ \Big(v_{A\parallel}^2-\frac{\bpar}{2}k_\parallel^2\Big)(1+v_b k_\parallel^2)+
    \frac{k_\parallel^2}{4}\Big( 1+v_b(1+k_\parallel^2)  \Big)^2 } \bigg]. \label{eq:firehose-FLR2}
\end{eqnarray}
Finally, the full model yields Hall-CGL-FLR3 solution
\begin{eqnarray}
  \omega &=& \frac{1}{1+v_b k_\parallel^2} \bigg\{ \pm \frac{k_\parallel^2}{2}\Big[ 1+v_b(1+k_\parallel^2) +\kpar^2\bpar^2(\frac{3}{2}-a_p)\Big]\nn\\
   && +k_\parallel \sqrt{ \Big[v_{A\parallel}^2-\frac{\bpar}{2}k_\parallel^2\Big( 1+\bpar(1-a_p)\Big) +\frac{\bpar^2}{2}(a_p-2)k_\parallel^4 \Big](1+v_b k_\parallel^2)+
    \frac{k_\parallel^2}{4}\Big[ 1+v_b(1+k_\parallel^2)+\kpar^2\bpar^2(\frac{3}{2}-a_p)  \Big]^2 }\; \bigg\}. \nn\\ \label{eq:firehose-FLR3}
\end{eqnarray}
The quantities $v_b$ and $v_{A\parallel}^2$ used in the solutions are defined as
\begin{equation}
v_b=\bpar(1-\frac{a_p}{2}); \qquad v_{A\parallel}^2=1+\frac{\bpar}{2}(a_p-1).\nn
\end{equation}  
Importantly, the solutions are written here for $\omega$ (and not $\omega^2$), and all the models of course yield 4 solutions, the other two solutions
are obtained by substituting $\omega$ with $-\omega$. For a reader who just jumped to this section, and is confused on how the solutions were split,
see section \ref{sec:HallCGL_parallel} where the Hall-CGL model is discussed with final equation (\ref{eq:ICW_pica}). The solutions can be written in various
forms, and one possibility is to keep $|\kpar|$ in front of the square roots.

\begin{figure*}[!htpb]
$$\includegraphics[width=0.48\linewidth]{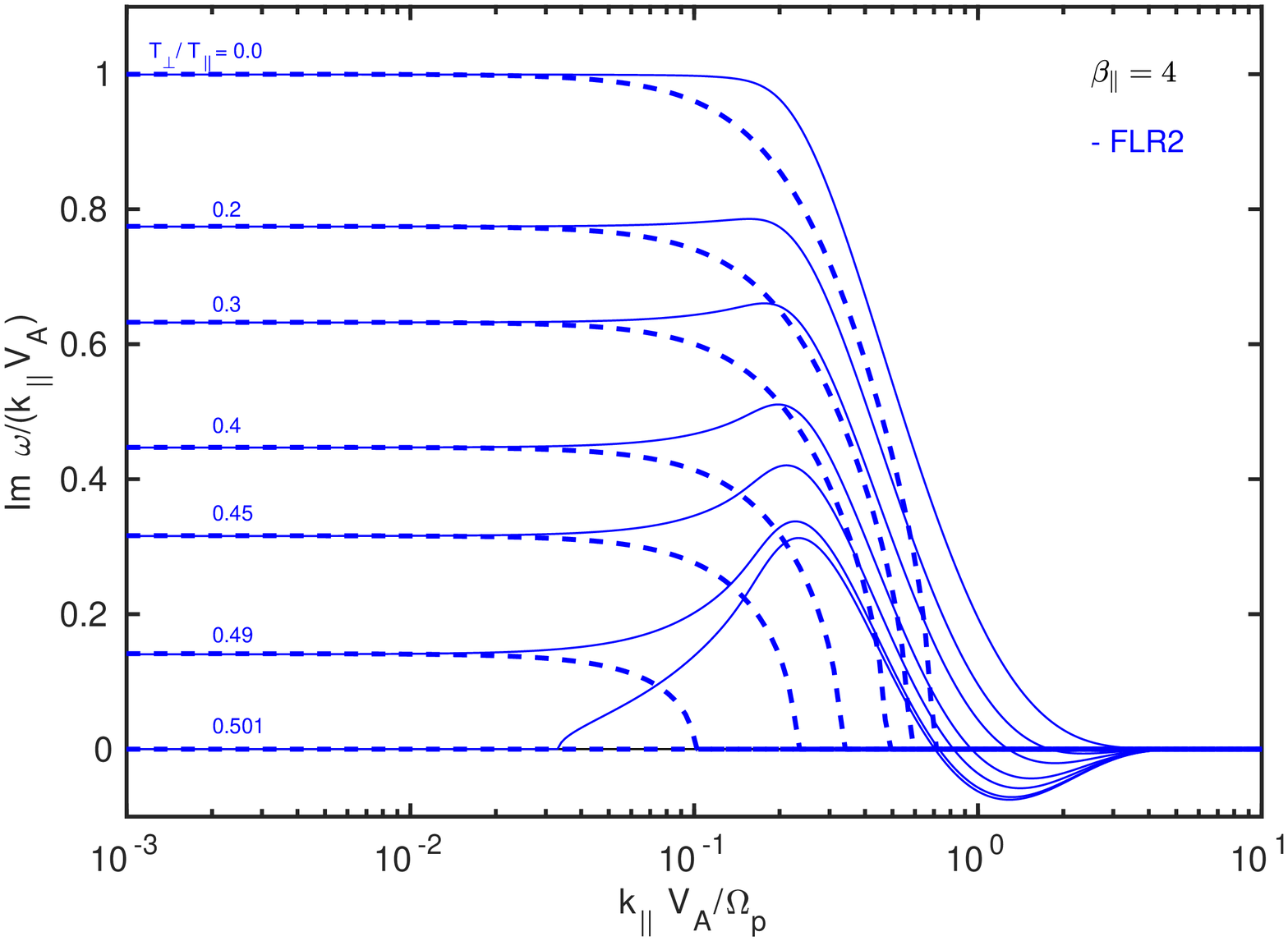}\hspace{0.03\textwidth}\includegraphics[width=0.48\linewidth]{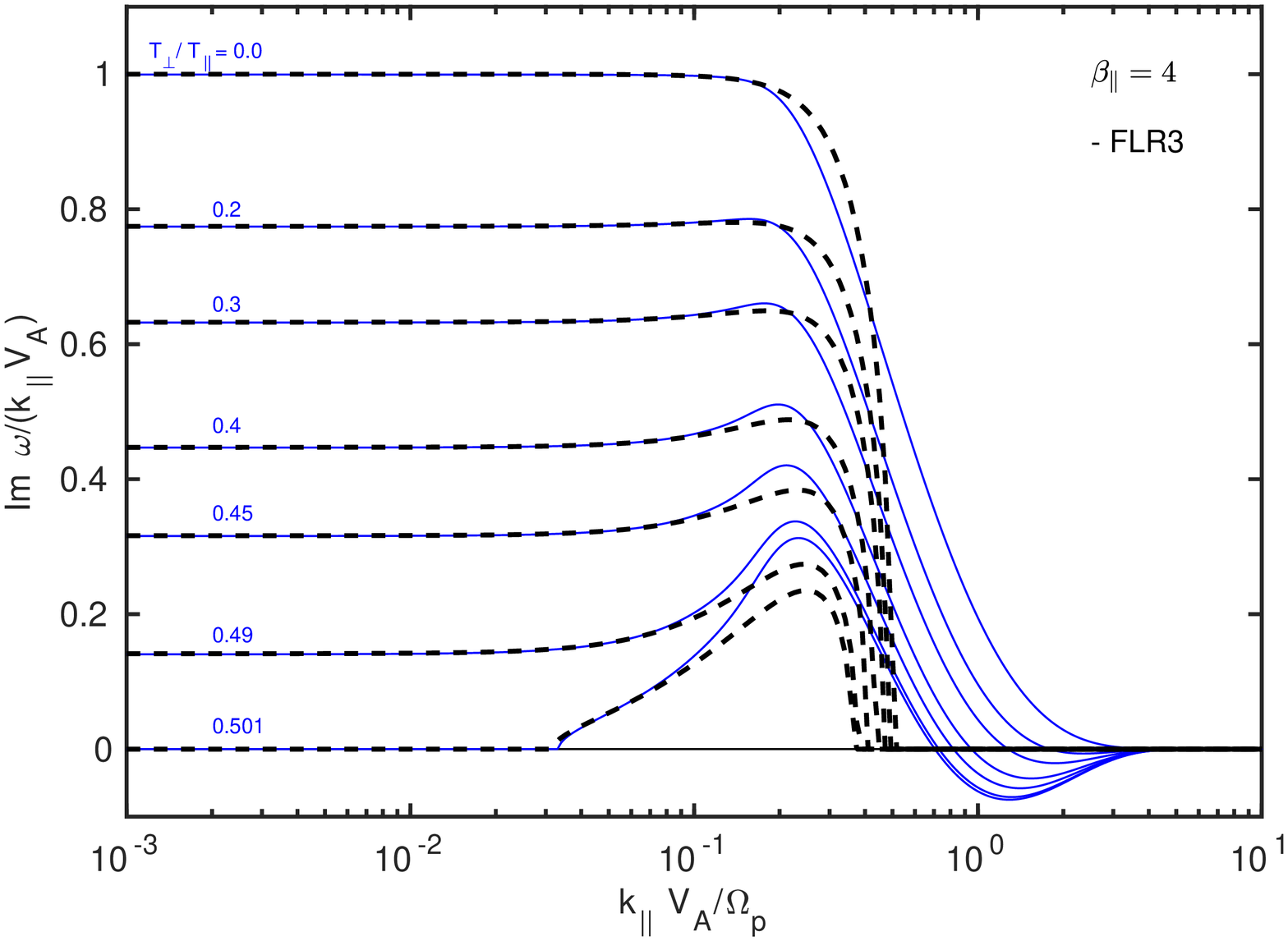}$$
  \caption{Imaginary phase speed (growth rate normalized to the wavenumber) of the parallel firehose instability. The $\bpar=4$,
    and temperature anisotropy is varied so that the whistler mode is in the firehose unstable regime. Solid lines (blue) are kinetic solutions, obtained with
    the WHAMP code. Dashed lines are fluid solutions. Left figure: FLR2 solutions (blue), Right figure: FLR3 solutions (black). It is shown that in contrast to
    the FLR2 model, the FLR3 model reproduces the large ``bump'' when close to the long-wavelength ``hard'' firehose threshold $T_\perp/T_\parallel=0.5$.} \label{fig:firehose-bump}
\end{figure*}

The (parallel and oblique) firehose instability was studied in detail by \cite{HunanaZank2017},
who focused on the Hall-CGL and Hall-CGL-FLR1 models. One of the conclusions reached in that paper was that the main reason for the relatively large
discrepancy between usual fluid models and kinetic description is the appearance of a huge ``bump'' in the imaginary phase speed,
when close to the firehose threshold. The situation is demonstrated in Figure \ref{fig:firehose-bump}, where kinetic solutions obtained by the WHAMP code (blue solid lines)
are compared to solutions of the Hall-CGL-FLR2 model (left figure, blue dashed lines) and the Hall-CGL-FLR3 fluid model (right figure, black dashed lines). 
The plasma $\bpar=4$, so the long-wavelength firehose threshold, that we call ``hard threshold'', is at $a_p=0.5$. The temperature anisotropy is varied from
$a_p=0$ to $a_p=0.501$. For solutions with the Hall-CGL and Hall-CGL-FLR1 models
see Figure 3 and 4 of \cite{HunanaZank2017}. In the WHAMP code the value of $a_p=0$ cannot be used, and $a_p=10^{-4}$ was chosen instead.
Also, since here we concentrate on proton dynamics, the influence of electrons in the WHAMP code was eliminated by making the electrons cold (with $T_e/T_p=10^{-8}$).

\begin{figure*}[!htpb]
$$\includegraphics[width=0.48\linewidth]{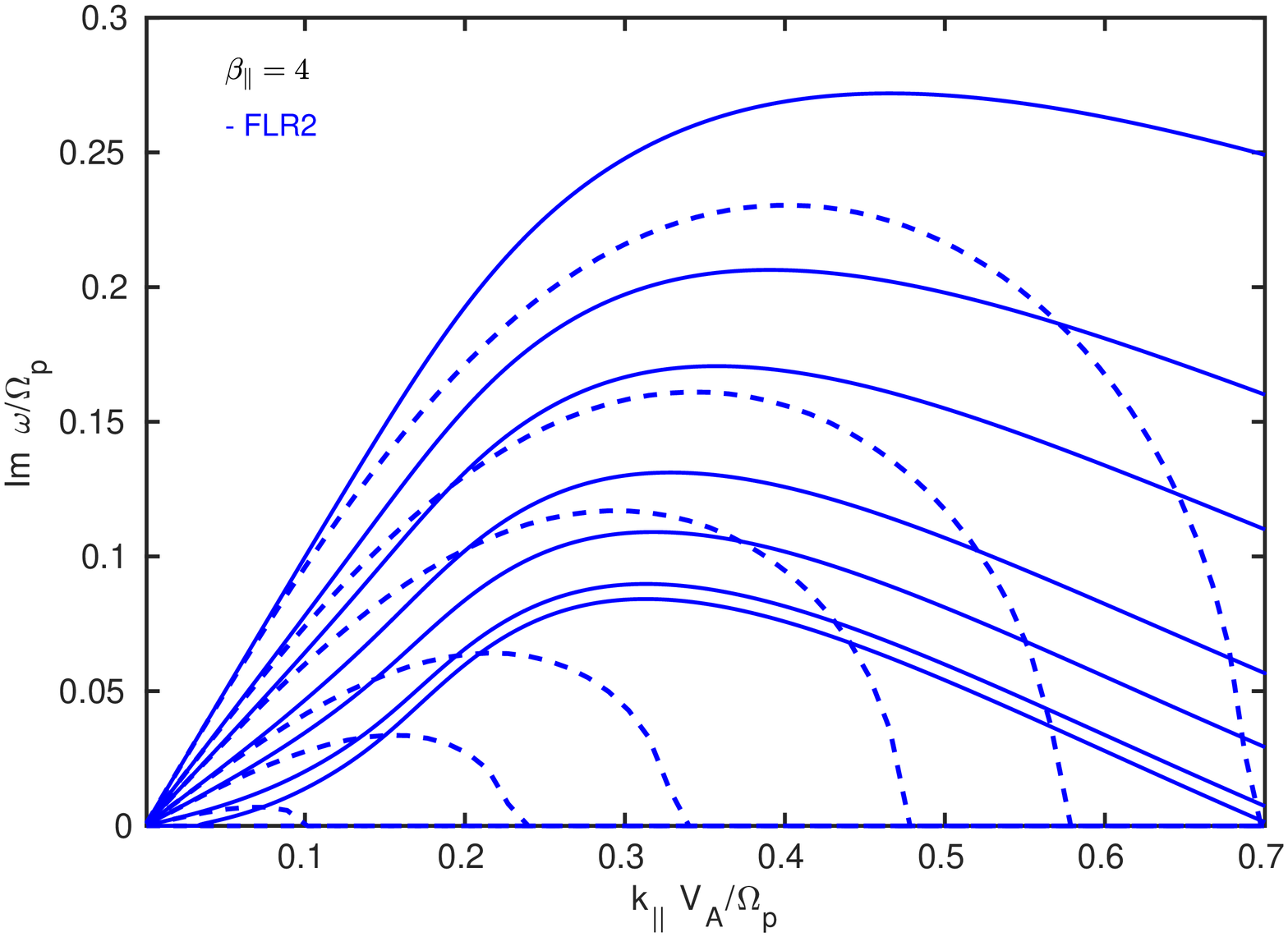}\hspace{0.03\textwidth}\includegraphics[width=0.48\linewidth]{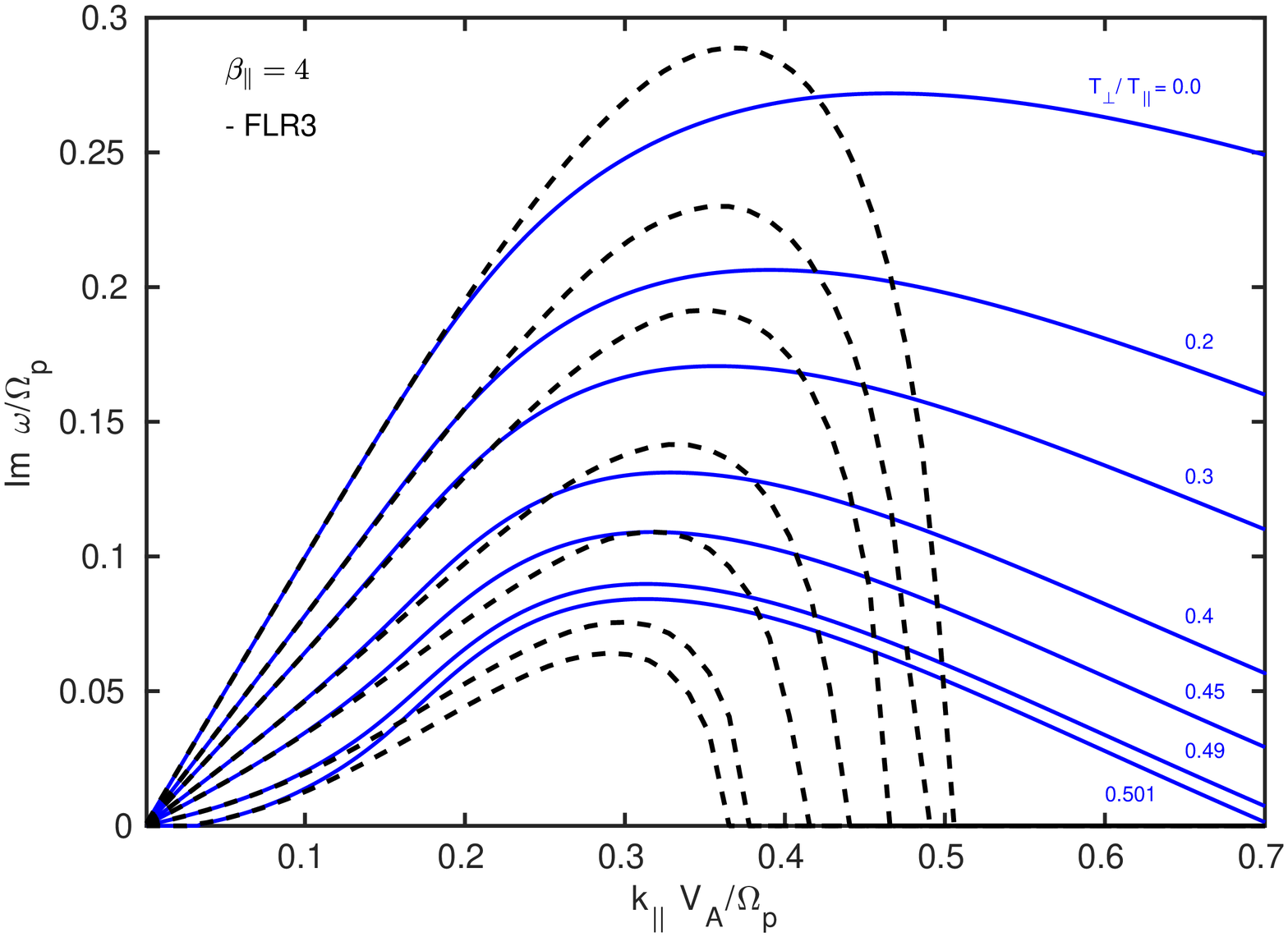}$$
  \caption{Same parameters as in Figure \ref{fig:firehose-bump}, but the growth rate is plotted, and linear scale is used for the x-axis.
    Notice the excellent precision of the FLR3 model for small wavenumbers up to $kd_i=0.1-0.2$. In comparison to kinetic theory, the
    fluid solutions are stabilized much more ``rapidly''.
    Nevertheless, the value of the maximum growth rate, and the wavenumber
    where the maximum growth rate is achieved, is surprisingly close to kinetic theory. This is an excellent result for a fluid model, which
  does not contain collisionless ion-cyclotron damping.}
  \label{fig:firehose-bump2}
\end{figure*}

\begin{figure*}[!htpb]
$$\includegraphics[width=0.48\linewidth]{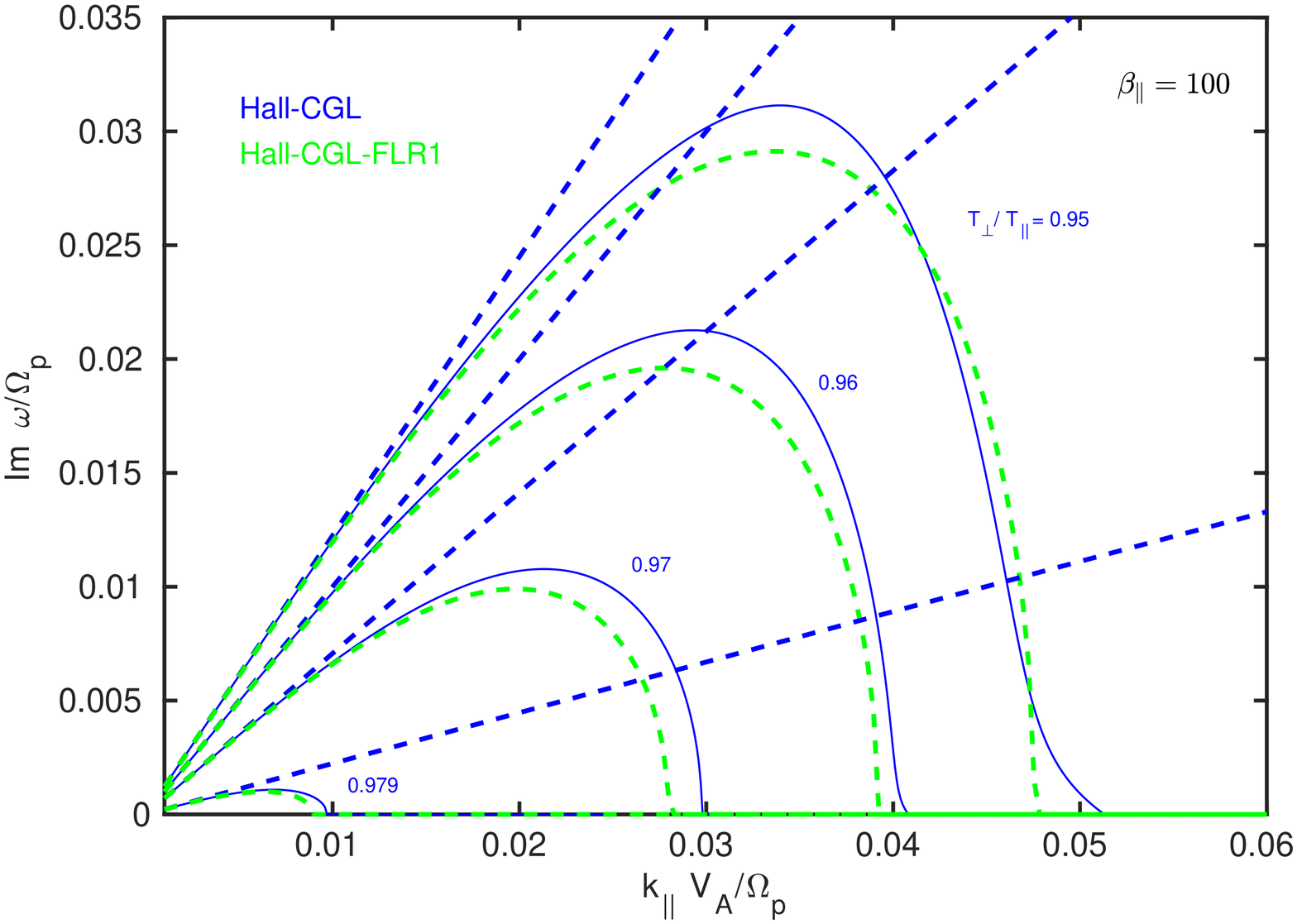}\hspace{0.03\textwidth}\includegraphics[width=0.48\linewidth]{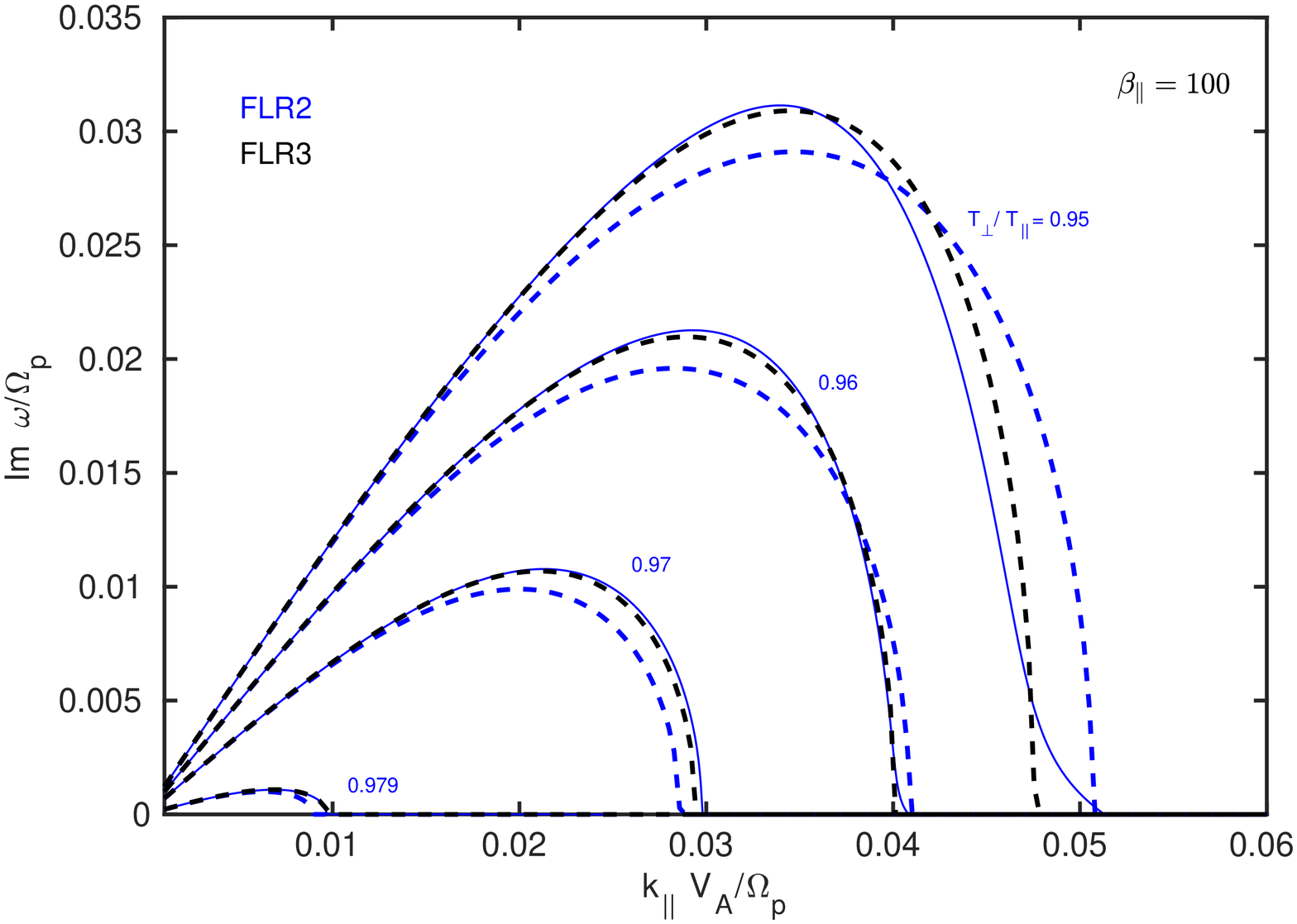}$$
  \caption{Growth rate of the parallel firehose instability for $\bpar=100$. Kinetic solutions are solid blue lines. The temperature
    anisotropy is varied as $a_p=0.95; 0.96; 0.97; 0.979$ (the hard firehose threshold is at $a_p=0.98$). Four different fluid models are plotted, and all fluid solutions
    have dashed lines. Left figure: Hall-CGL (blue), Hall-CGL-FLR1 (green). Right figure: Hall-CGL-FLR2 (blue), Hall-CGL-FLR3 (black).}
  \label{fig:firehose-beta100}
\end{figure*}

Very surprisingly, Figure \ref{fig:firehose-bump} shows that the large ``bump'' is reproduced by the FLR3 model, and the precision is quite good! 
The situation is further analyzed in Figure \ref{fig:firehose-bump2}, where instead of the imaginary
phase speed, the growth rate is plotted.  
The same conclusion is obtained. For the temperature anisotropy $a_p=0.501$, all simpler fluid models (Hall-CGL, FLR1, FLR2) are fully stable for all range of
wavenumbers. In contrast, the FLR3 model still develops a strong firehose instability.
\emph{Therefore, it is the non-gyrotropic heat flux $\boldsymbol{S}^\parallel_\perp$ of the FLR3 model that is responsible for the ``bump'' (for the case of strictly parallel propagation).}
The appearance of the ``bump'' can be understood analytically, by evaluating the fluid dispersion relations 
exactly at the ``hard'' firehose threshold $a_p=1-\frac{2}{\bpar}$, where the quantities $v_{A\parallel}^2=0$ and $v_b=1+\frac{\bpar}{2}$.
At the hard firehose threshold, the expression under the square root of the FLR2 solution (\ref{eq:firehose-FLR2})
can be factorized as
\begin{equation}
\kpar^2\Big[ 1-\frac{\bpar}{4}+\frac{\kpar^2}{2}\big(1+\frac{\bpar}{2}\big) \Big]^2 +\kpar^2\frac{\bpar}{2},
\end{equation}
implying that the FLR2 model is always stable for all values of $\kpar$ and $\bpar$. Such a factorization can not be achieved for the FLR3 model,
and the solution (\ref{eq:firehose-FLR3}) still becomes unstable at some range of wavenumbers, where    
\begin{equation}
  \Big[v_{A\parallel}^2-\frac{\bpar}{2}k_\parallel^2\Big( 1+\bpar(1-a_p)\Big) +\frac{\bpar^2}{2}(a_p-2)k_\parallel^4 \Big](1+v_b k_\parallel^2)+
    \frac{k_\parallel^2}{4}\Big[ 1+v_b(1+k_\parallel^2)+\kpar^2\bpar^2(\frac{3}{2}-a_p)  \Big]^2<0, \label{eq:firehose-FLR333}
\end{equation}
which is the (strictly parallel) firehose instability criterium of the FLR3 model. We note that the model can become unstable also for the temperature anisotropy $a_p>1$.
This can be perhaps considered as some remnant of the ion-cyclotron anisotropy instability, but we did not study the situation further since the instability should not
be reproduced correctly. The parallel firehose instability
for high plasma beta value $\bpar=100$ is shown in Figure \ref{fig:firehose-beta100}. The left Figure is from \cite{HunanaZank2017}, and
it shows solutions of the Hall-CGL model (blue dashed lines) and of the Hall-CGL-FLR1 model (green dashed lines). The kinetic solutions are blue solid lines.
It is shown that the FLR corrections are crucial for the correct stabilization of the firehose instability. The right Figure \ref{fig:firehose-beta100} shows refinement
with FLR2 (blue dashed lines) and FLR3 tensors (black dashed lines). Obviously, for high plasma beta values, the maximum growth rate of the (strictly) parallel firehose
instability is captured very precisely by the Hall-CGL-FLR3 model. 
The maximum growth rate (for parallel propagation) can be easily found analytically only for the Hall-CGL and Hall-CGL-FLR1 models.
For models with the FLR2 and FLR3 tensors it is easier to find the maximum growth rate numerically. For example, let's consider the Hall-CGL model. Assuming
firehose unstable regime, the frequency of the Hall-CGL solution (\ref{eq:firehose-HallCGL}) can be split into $\omega=\omega_r+i\omega_i$, where
the imaginary part $\omega_i=\kpar\sqrt{-v_{A\parallel}^2-\kpar^2/4}$. Then by calculating $\pr\omega_i/\pr\kpar=0$, yields the wavenumber $k_{\parallel \textrm{max}}$ and
the maximum growth rate $\gamma_{\textrm{max}}$ in the following form
\begin{equation}
k_{\parallel \textrm{max}}^2 = -2v_{A\parallel}^2; \qquad \gamma_{\textrm{max}} = -v_{A\parallel}^2 = -\Big( 1+\frac{\bpar}{2}(a_p-1)\Big).
\end{equation}
Similarly, the Hall-CGL-FLR1 model yields
\begin{equation}
  k_{\parallel \textrm{max}}^2 = -\frac{2v_{A\parallel}^2}{(1-v_b)^2}; \qquad \gamma_{\textrm{max}} = \frac{v_{A\parallel}^2}{v_{A\parallel}^2-\frac{\bpar}{2}}
  =  \frac{1+\frac{\bpar}{2}(a_p-1)}{1+\frac{\bpar}{2}(a_p-2)}.
\end{equation}
\subsection{Oblique propagation}
\begin{figure*}[!htpb]
  $$\includegraphics[width=0.48\linewidth]{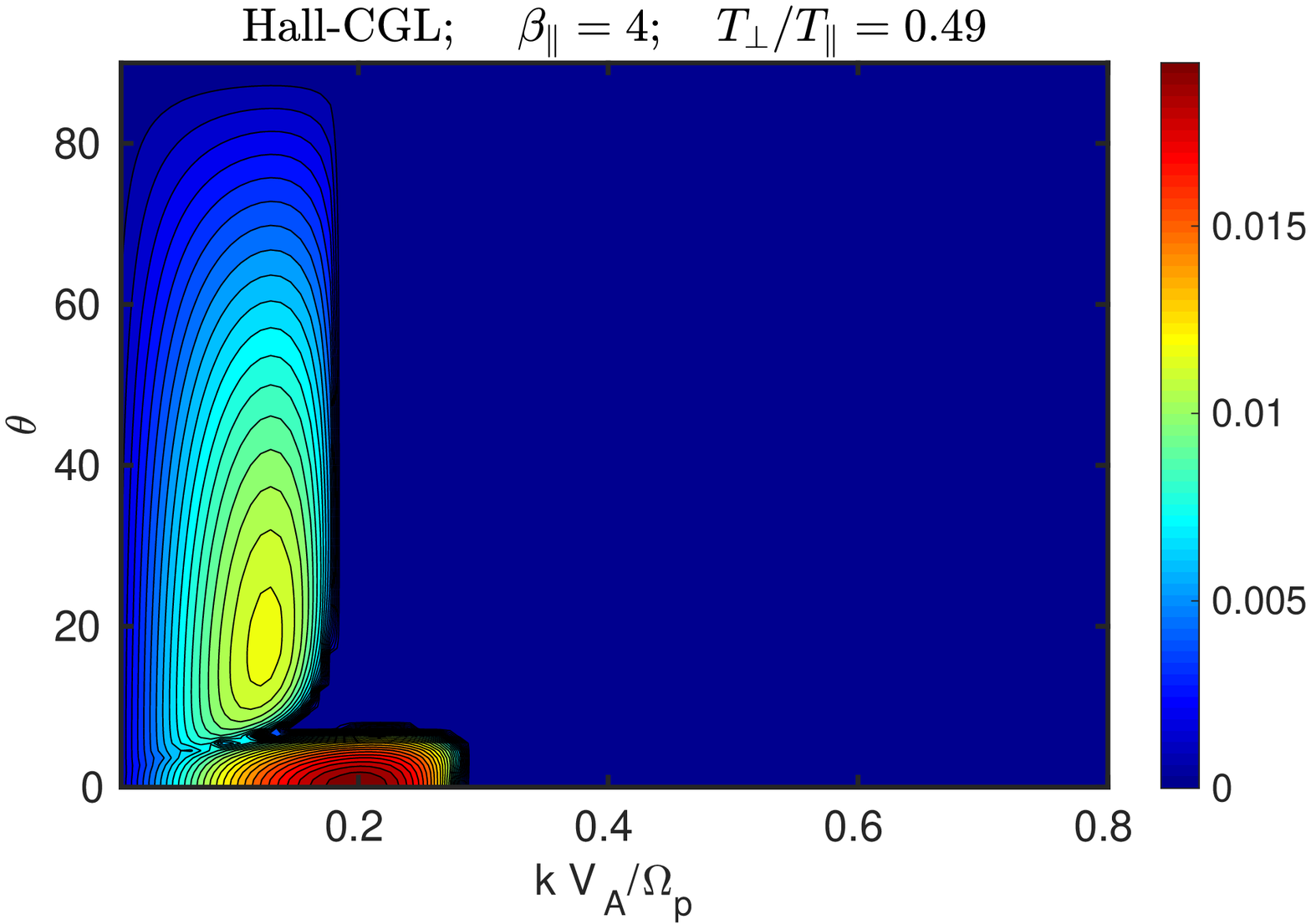}\hspace{0.03\textwidth}\includegraphics[width=0.48\linewidth]{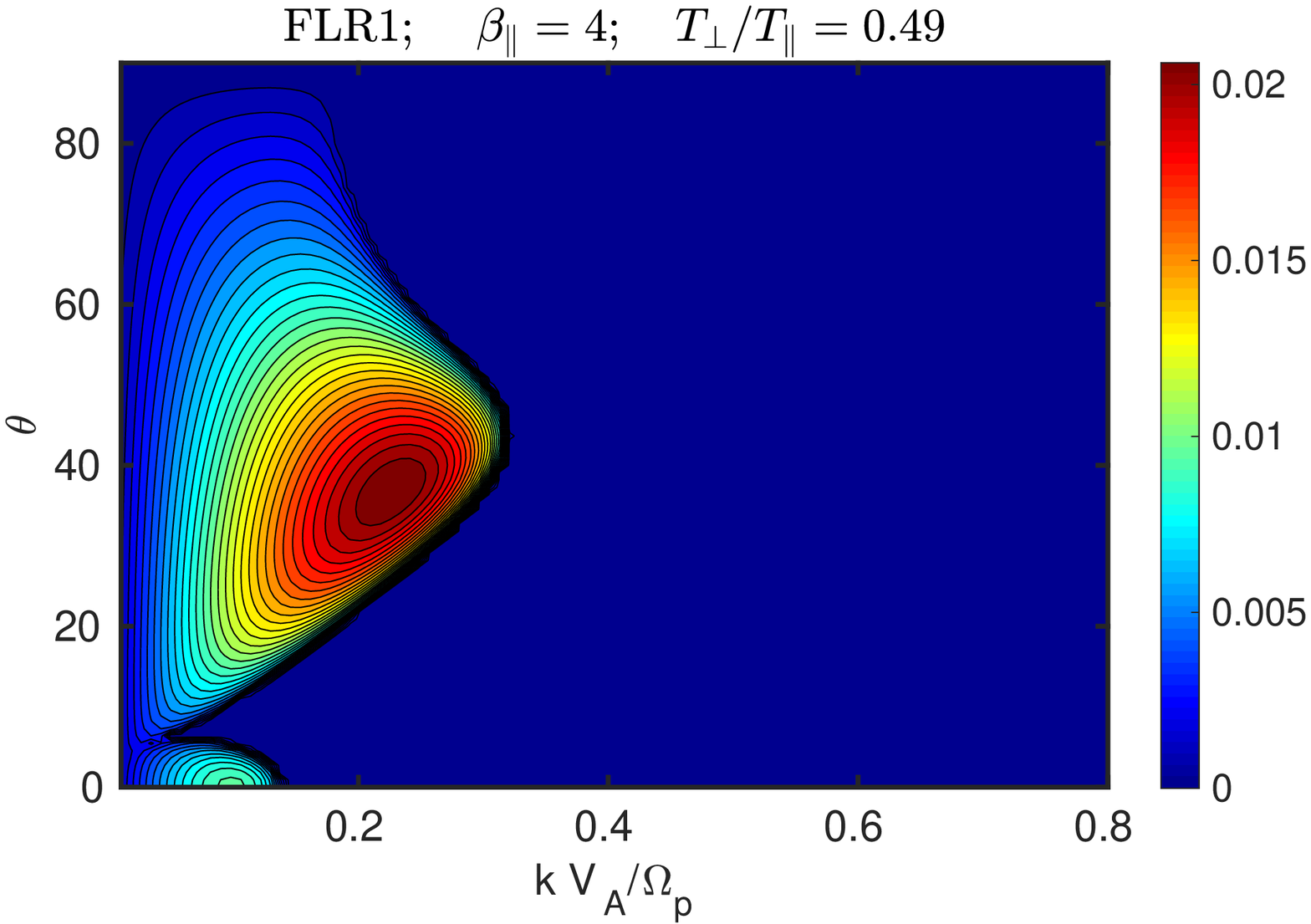}$$
  $$\includegraphics[width=0.48\linewidth]{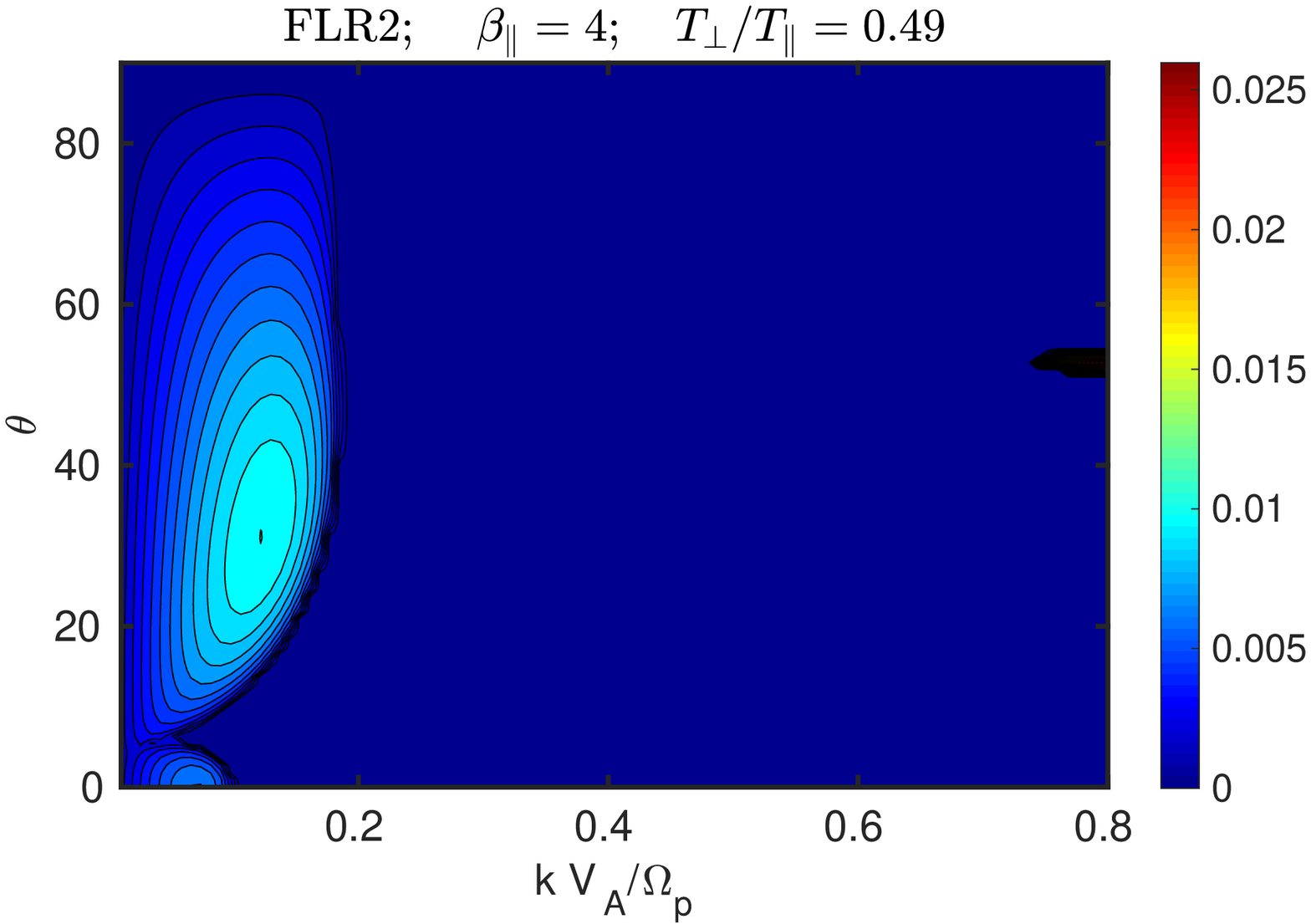}\hspace{0.03\textwidth}\includegraphics[width=0.48\linewidth]{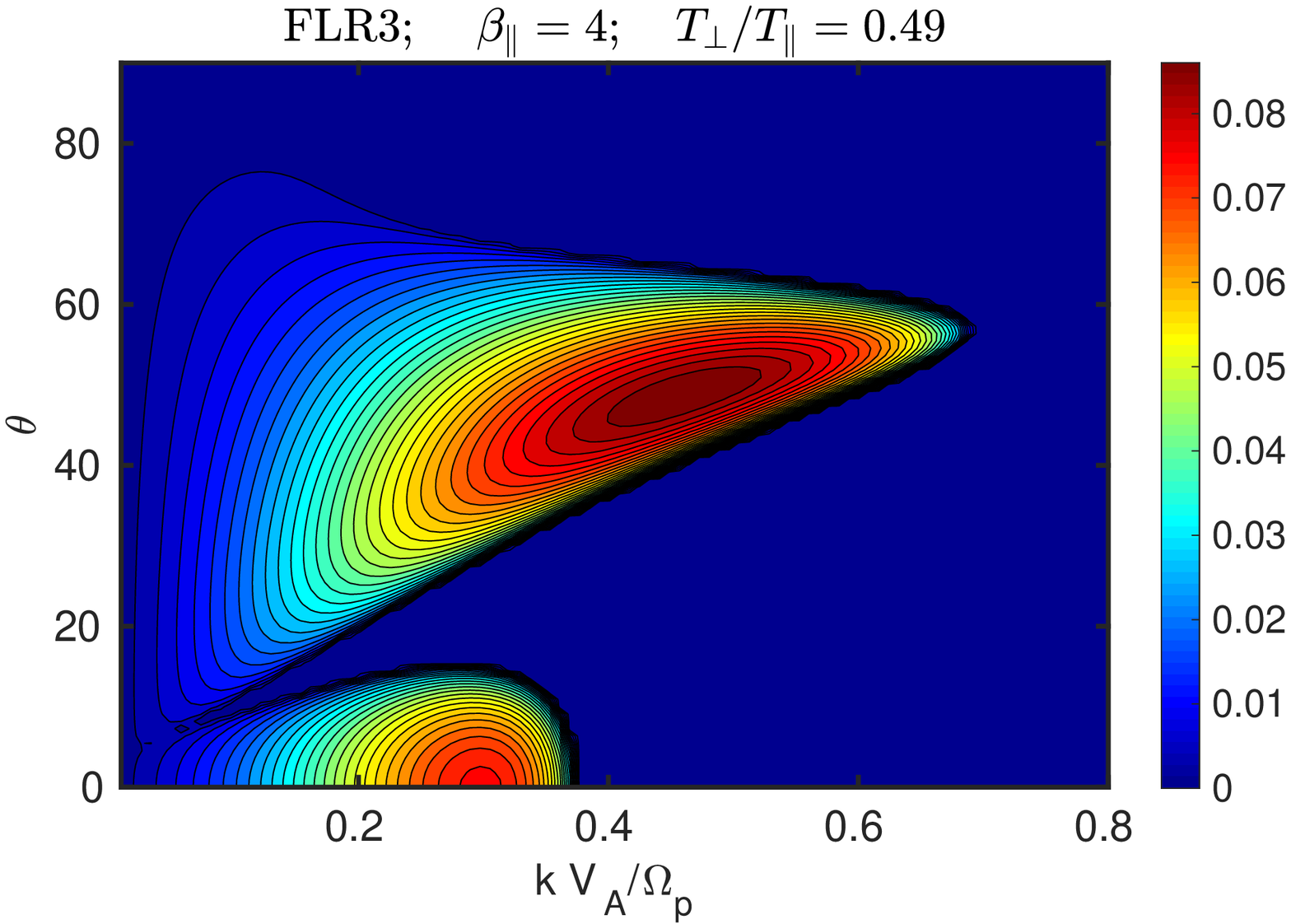}$$
  $$\includegraphics[width=0.48\linewidth]{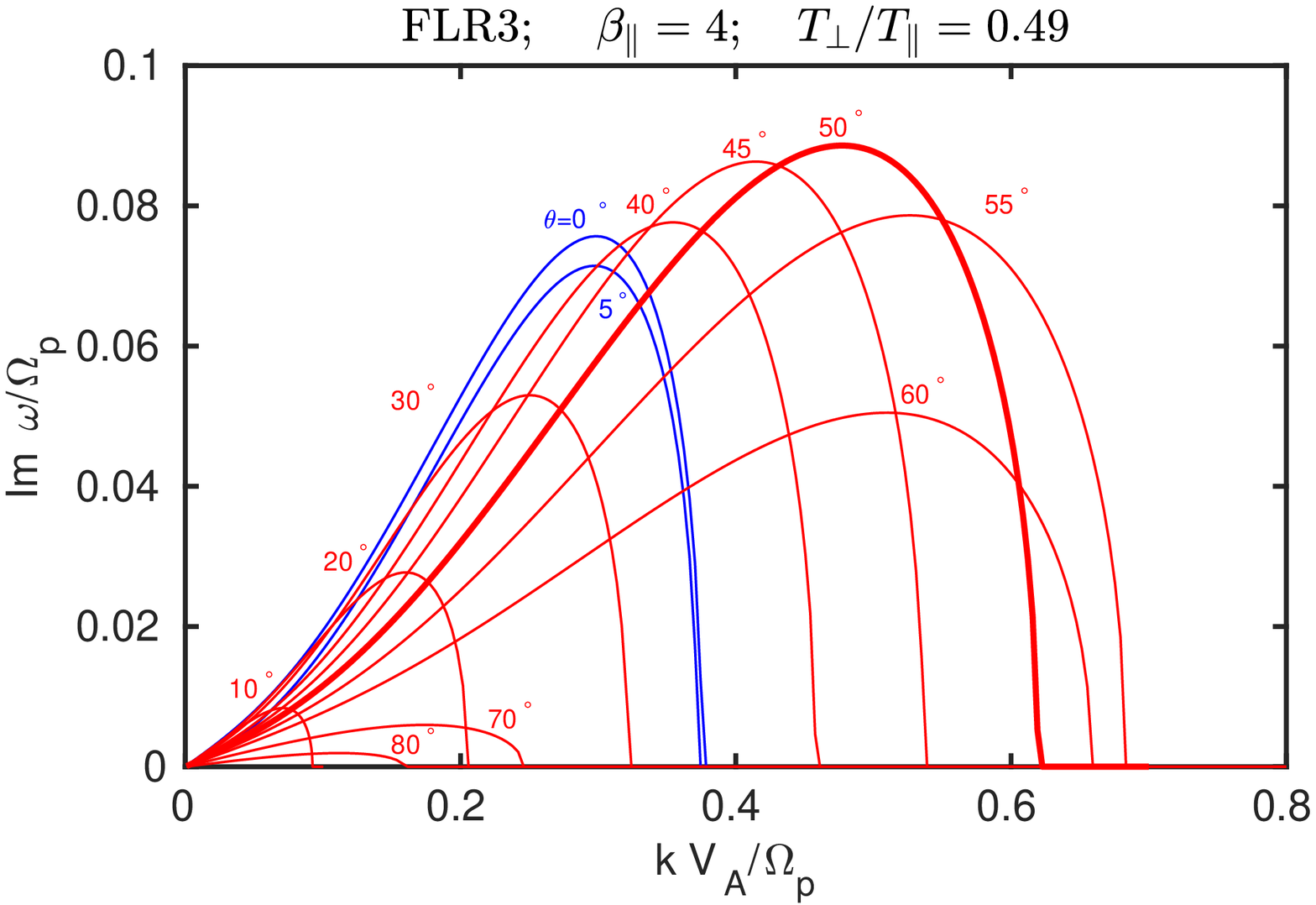}\hspace{0.03\textwidth}\includegraphics[width=0.48\linewidth]{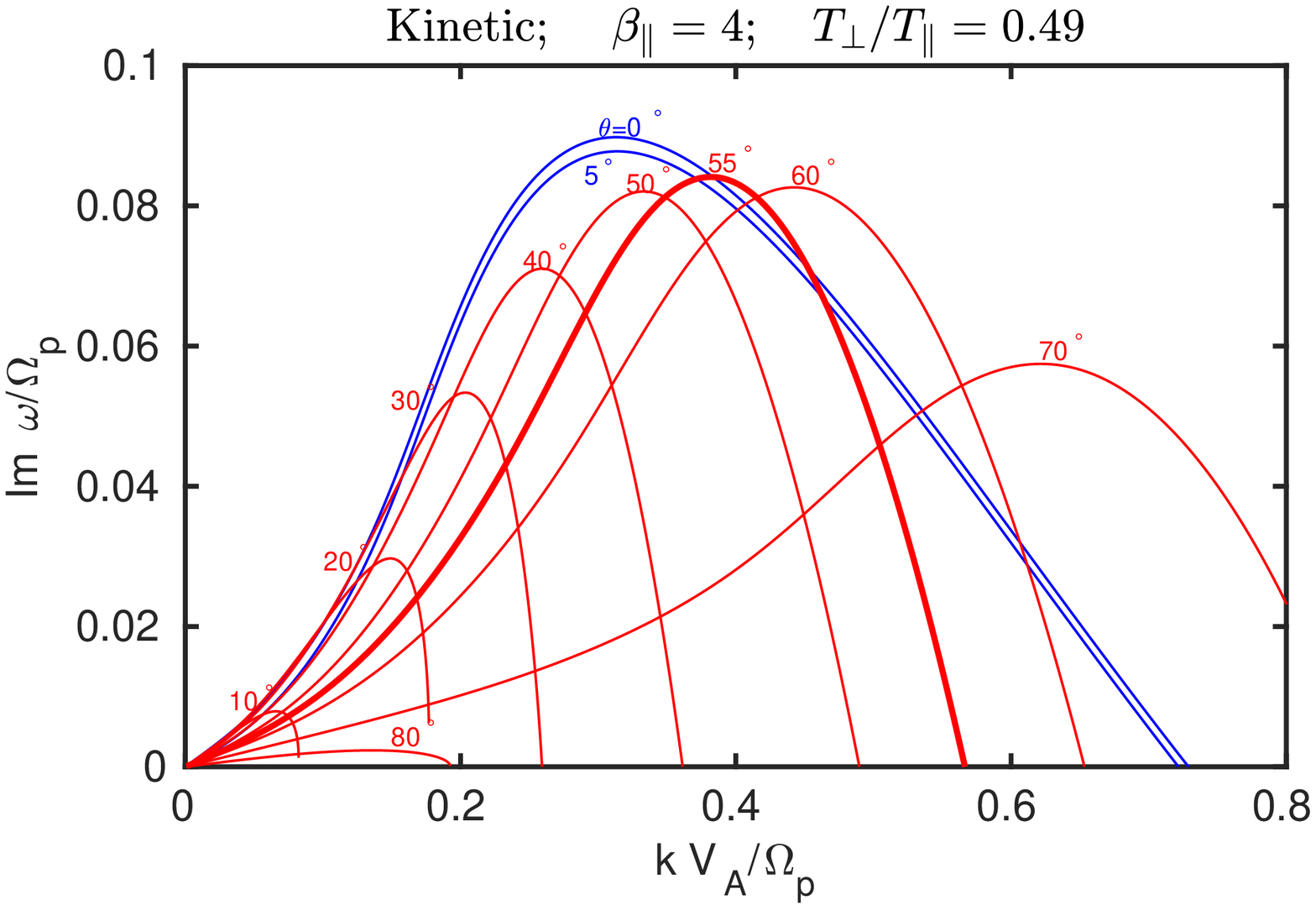}$$
  \caption{The growth rate plotted in the $(k,\theta)$ plane, with fixed $\bpar=4$ and $a_p=0.49$ (the hard firehose
    threshold is at $a_p=0.5$), showing the parallel and oblique firehose instability.
    Four different fluid models are plotted. Top left: Hall-CGL; top right: Hall-CGL-FLR1; middle left: Hall-CGL-FLR2,
    middle right: Hall-CGL-FLR3. We do not provide
    contour plot for kinetic theory. Nevertheless, bottom figures show solutions for several propagation angles. Bottom left: Hall-CGL-FLR3 model, bottom right: kinetic.}
  \label{fig:OF-beta4-ap049}
\end{figure*}

Here we briefly investigate the parallel and oblique firehose instability for oblique propagation directions. Contour plots of the growth rate in $(k,\theta)$ plane are shown
in Figure \ref{fig:OF-beta4-ap049}. The plasma $\bpar=4$ and the temperature anisotropy $a_p=0.49$, so all the fluid models are in the firehose unstable regime.
We compare four different fluid models, Hall-CGL (top left), and enhancement with FLR1 (top right), FLR2 (middle left) and FLR3 (middle right).
The FLR2 solution shows an additional instability at higher wavenumbers that we did not study further.
Note the large differences between the four solutions. Importantly, the FLR3 solution shows large enhancement of the growth rate. 
We use the WHAMP code for kinetic calculations, and we do not provide contour plot for the kinetic theory. Instead, we plot the growth rate for several propagation angles,
so that the comparison with the contour plots can be made easily. The bottom left figure is the FLR3 model, and the bottom right figure is the kinetic result.
It is noted that similarly to other fluid models with higher-order FLR corrections, the FLR3 model can develop secondary instabilities
at scales below the proton gyroscale.
\subsection{Hellinger's contours for Hall-CGL-FLR3 model} \label{HellingerFigures}
\begin{figure*}[!htpb]
  $$\includegraphics[width=0.49\linewidth]{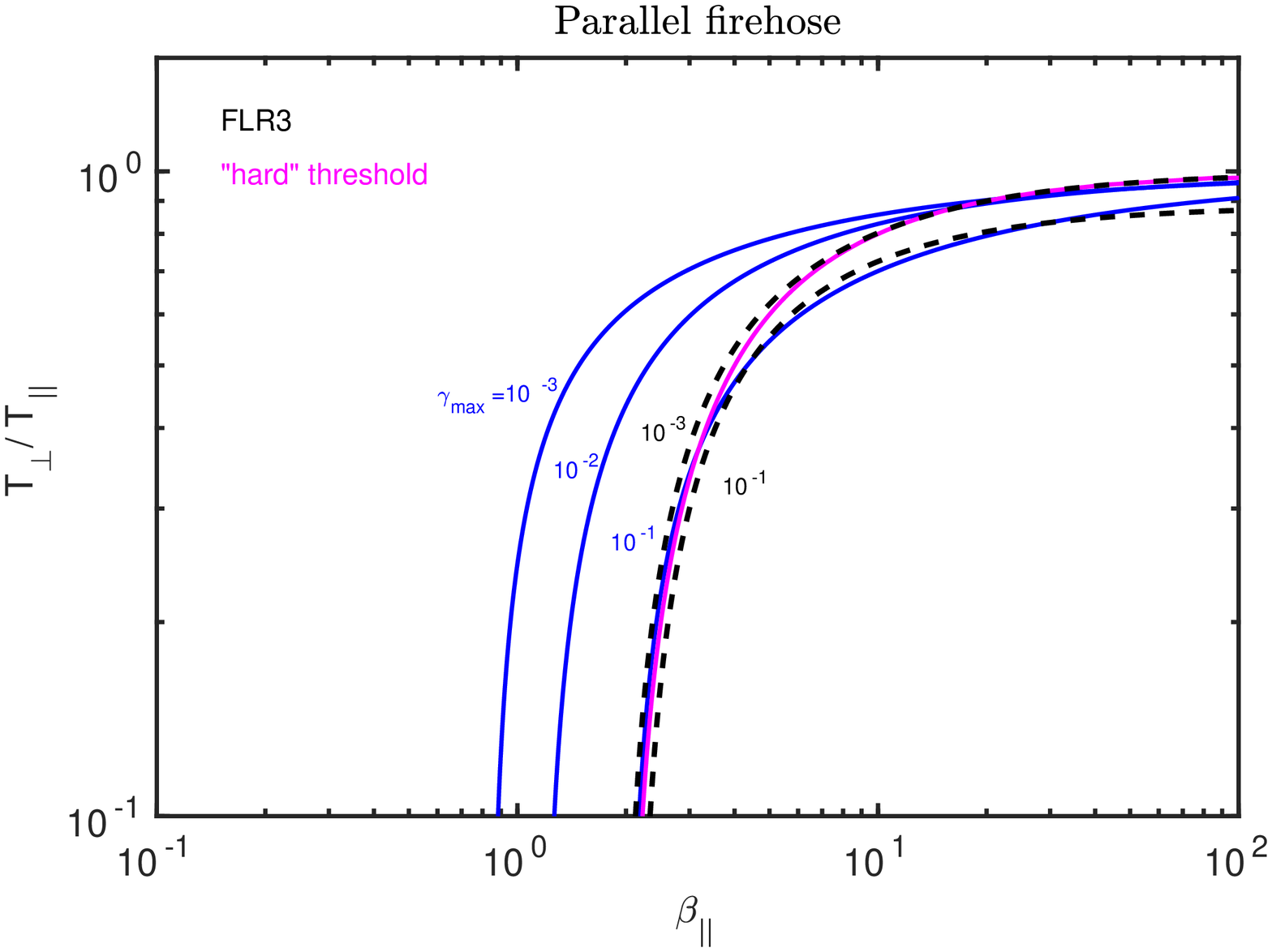}\hspace{0.03\textwidth}\includegraphics[width=0.49\linewidth]{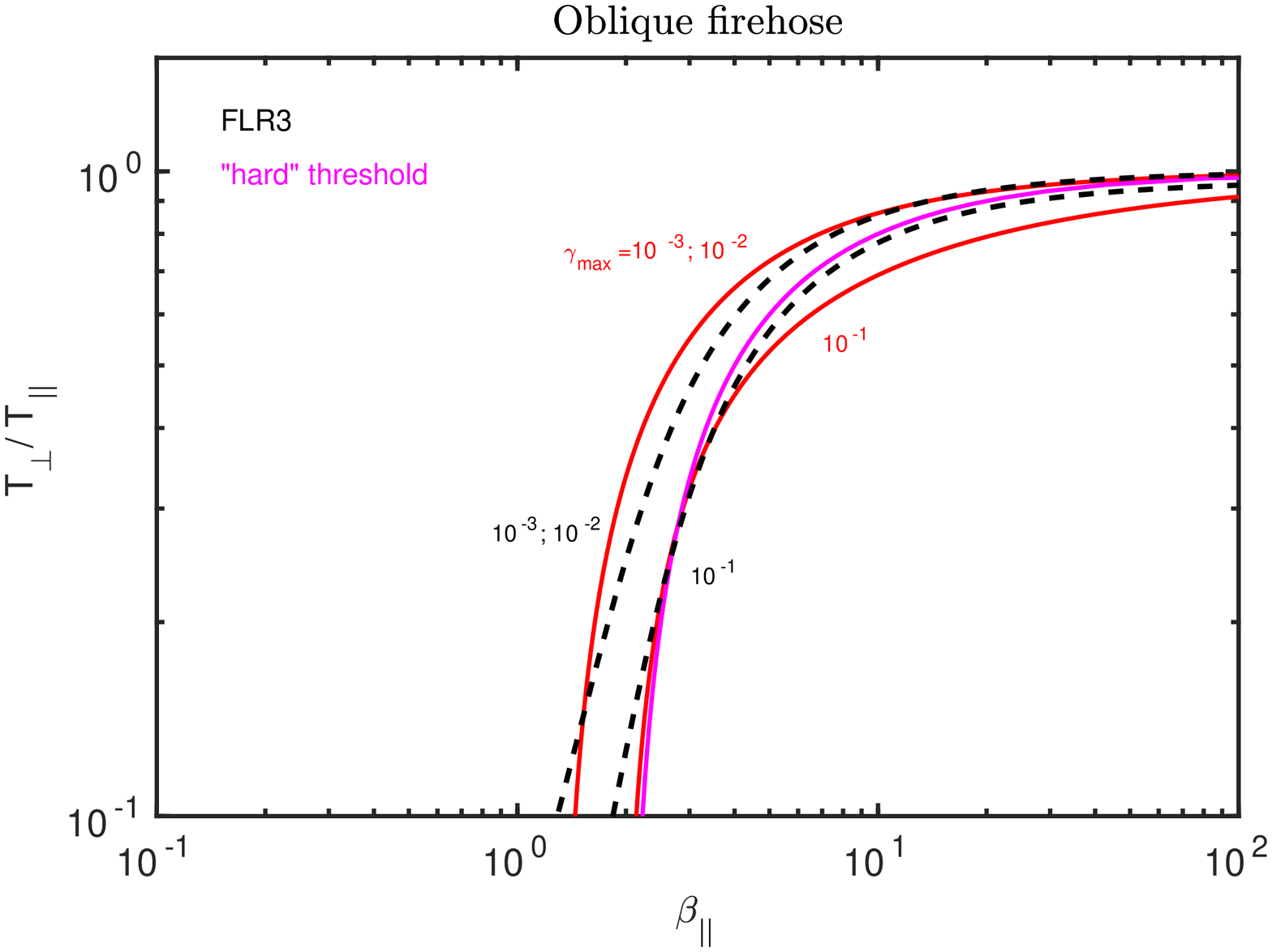}$$
  $$\includegraphics[width=0.48\linewidth]{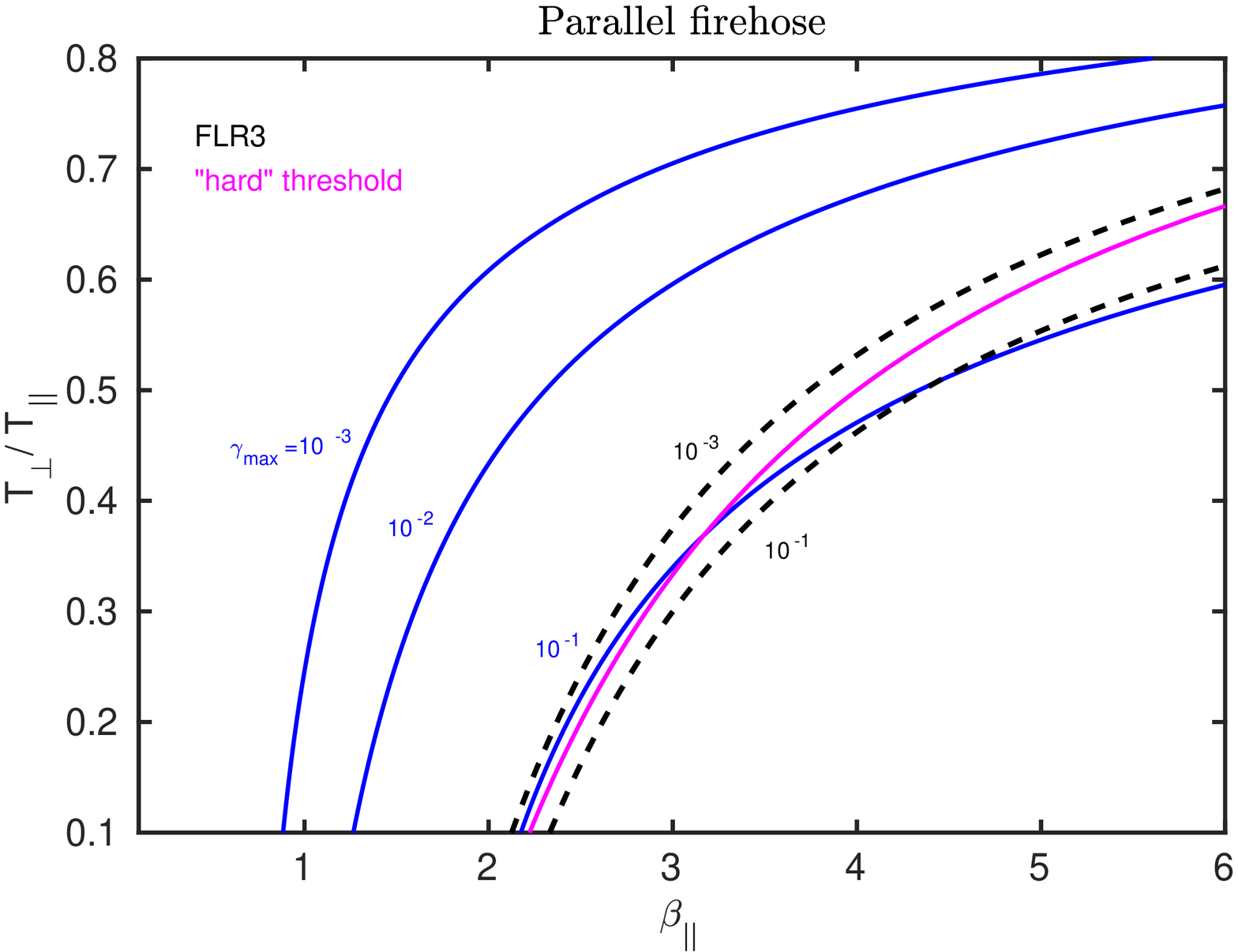}\hspace{0.03\textwidth}\includegraphics[width=0.48\linewidth]{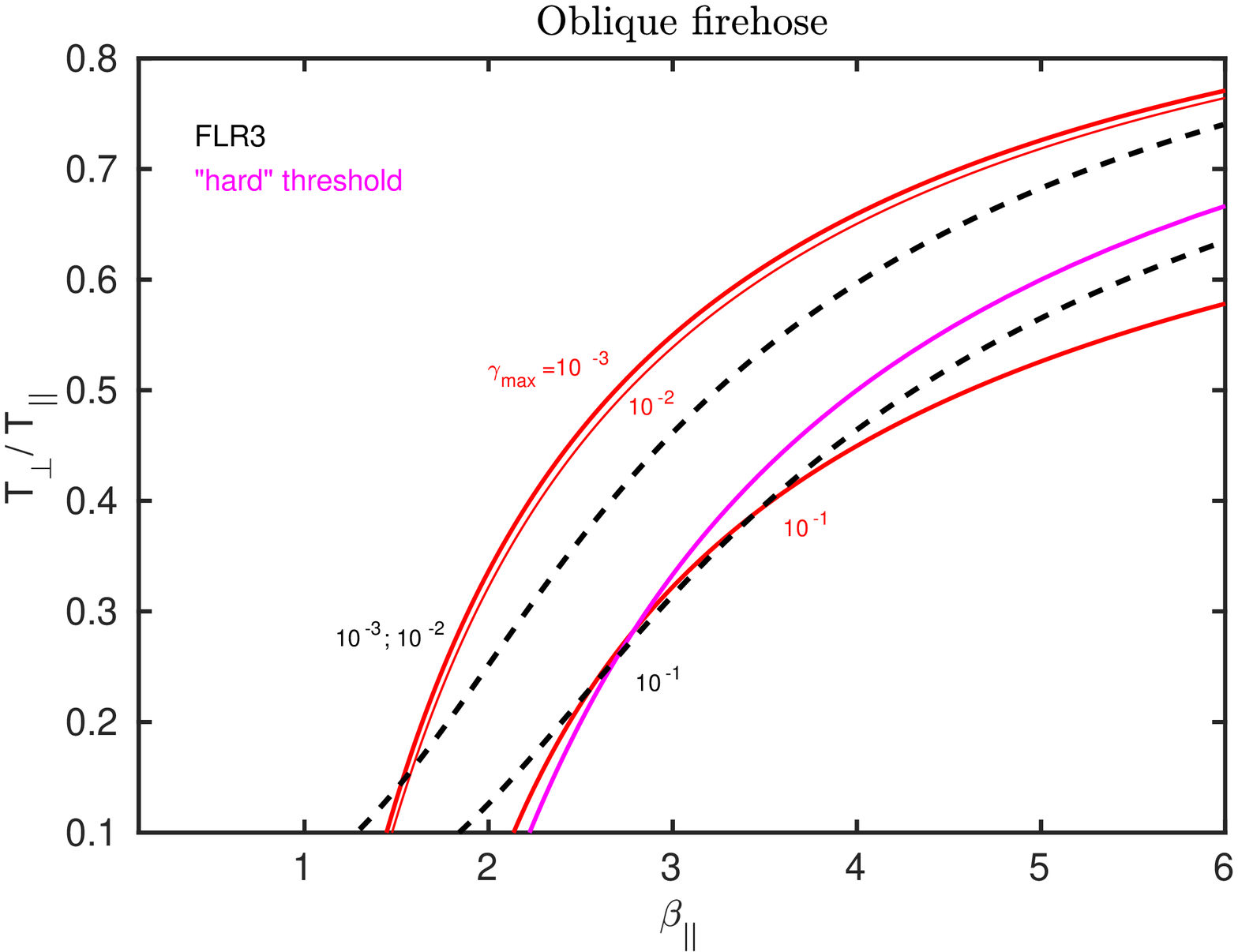}$$
  \caption{Top panels: Solutions for the parallel (left) and oblique (right) firehose instability in the $(\bpar,a_p)$ plane for a prescribed
    maximum growth rate $\gamma_{\textrm{max}}=10^{-3}; 10^{-2}; 10^{-1}$.
    Solid blue and red lines are kinetic solutions from \cite{Hellinger2006}.
    Black dashed lines are solutions of the Hall-CGL-FLR3 model. The magenta line is the ``hard'' (long-wavelength limit) firehose threshold.
  Bottom panels: Same as top panels, but with linear scales for both axis, and only showing results for $\bpar<6$.}
  \label{fig:both_firehose}
\end{figure*}

\begin{figure*}[!htpb]
  $$\includegraphics[width=0.42\linewidth]{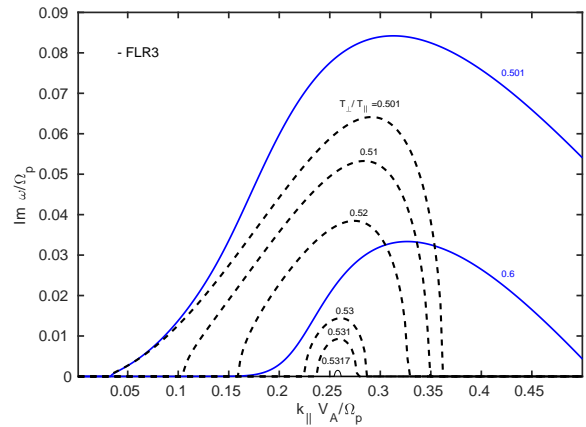}$$
  \caption{Parallel firehose instability, $\bpar=4$. It is shown that the Hall-CGL-FLR3 model (black dashed lines)
    indeed develops firehose instability even for values of $a_p>0.5$, i.e. when there is no instability in the long-wavelength limit.
    Nevertheless, by increasing the value of $a_p$ beyond $0.5$, the growth rate in the FLR3 model quickly falls off, and the instability disappears for $a_p>0.5317$.
    In contrast, kinetic theory (blue solid lines) develops strong firehose instability even at $a_p=0.6$.
  } \label{fig:veryclose}
\end{figure*}

Here we prescribe a fixed value for the maximum growth rate, $\gamma_{\textrm{max}}=10^{-3}; 10^{-2}; 10^{-1}$, and plot solutions in the $(\bpar,a_p)$ plane, 
which is shown in Figure \ref{fig:both_firehose}. The left panels show the parallel firehose instability, and the right panels show the oblique firehose instability.
The top panels are plotted with the usual logarithmic scales, and the bottom panels with linear scales. Additionally, only solutions with $\bpar<6$ are shown at the bottom panels.
Solutions for the Hall-CGL-FLR3 fluid model are black dashed lines, and kinetic solutions are solid blue lines (left panels) and solid red lines (right panels).
The kinetic solutions were provided to us by P. Hellinger (private communication) and are identical to solutions shown in Figure 1 of \cite{Hellinger2006}.
It is shown that the $\gamma_{\textrm{max}}=10^{-3}; 10^{-2}$ contours clearly lie below the hard firehose threshold, i.e.,
the firehose instability in the FLR3 fluid model develops at some range of wavenumbers, even if the model is stable in the long-wavelength limit.
Importantly, the kinetic contours of \cite{Hellinger2006} were not calculated for cold electrons, but for isotropic electrons with $\beta_e=1$.
This does not matter for the parallel propagation, since isotropic electron pressure does not influence the dispersion relations for the parallel propagating
whistler and ion-cyclotron modes. Only the effect of electron inertia will enter, however, the effect should be negligible at the scales considered here.   
Unfortunately, the solution for the parallel firehose instability can not be further improved by currently developed fluid models.     
The oblique case is different, since even isotropic electron pressure will influence the dynamics. The contours for the oblique firehose instability should be therefore
recalculated by using a proper proton-electron two fluid model. The oblique case could be further improved,
by considering FLR contributions from the non-gyrotropic heat flux tensor $\boldsymbol{\sigma}$ that were neglected here (contributions for parallel propagation are zero).
For the oblique case, one can also use higher-order fluid models with the gyrotropic heat flux fluctuations $q_\parallel, q_\perp$ (such as the CGL2 model discussed later),
or even Landau fluid models. The main reason for the relatively large discrepancies found in the contours for the parallel firehose instability  
is demonstrated in Figure \ref{fig:veryclose}.

\clearpage
\newpage
\section{Heat flux tensor equation} \label{section:HeatFlux}
Multiply the Vlasov equation by $mc_ic_jc_k$ and integrate over the velocity space. Naturally, the other possibilities
are to multiply by $mc_ic_jv_k$, $mc_iv_jv_k$ or $mv_iv_jv_k$, but none of them is more revealing and we will use the first choice.
It is convenient to define the symmetric operator $^S$ that acts on a tensor of third rank according to
\begin{equation}
  A^S_{ijk}=A_{ijk}+A_{jki}+A_{kij},
\end{equation}
i.e. the operator represents all possible cyclic permutations.
For the first term we will need the following identities
\begin{eqnarray}
  \frac{\pr}{\pr t}(c_ic_jc_k) &=& -\frac{\pr u_i}{\pr t}c_jc_k - c_i\frac{\pr u_j}{\pr t}c_k - c_ic_j\frac{\pr u_k}{\pr t}; \\
  m\int f \frac{\pr}{\pr t}(c_ic_jc_k) d^3v &=& -\frac{\pr u_i}{\pr t}\underbrace{m\int c_j c_k fd^3v}_{p_{jk}}-\frac{\pr u_j}{\pr t}
  \underbrace{m\int c_ic_kfd^3v}_{p_{ik}}
  -\frac{\pr u_k}{\pr t}\underbrace{m\int c_ic_jfd^3v}_{p_{ij}} \nn\\
  &=& - \frac{\pr u_i}{\pr t}p_{jk} - \frac{\pr u_j}{\pr t}p_{ik}- \frac{\pr u_k}{\pr t}p_{ij},
\end{eqnarray}  
and the entire first term of the integrated Vlasov equation calculates
\begin{eqnarray}
\bigcirc{ }\!\!\!\!\mbox{\small 1} &=&  m\int c_ic_jc_k\frac{\pr f}{\pr t}d^3v = \frac{\pr}{\pr t}\big( \underbrace{m\int c_ic_jc_k f d^3v}_{q_{ijk}} \big)
- m\int f\frac{\pr}{\pr t}(c_ic_jc_k) d^3v \nn\\
&=& \frac{\pr}{\pr t}q_{ijk} + \frac{\pr u_i}{\pr t}p_{jk} + \frac{\pr u_j}{\pr t}p_{ik} + \frac{\pr u_k}{\pr t}p_{ij}
= \frac{\pr}{\pr t}q_{ijk} + \Big[ \frac{\pr \bu}{\pr t}\bp \Big]_{ijk}^S.
\end{eqnarray}
For the second term we will need identities
\begin{eqnarray}
  m\int \bc\bc\bc\bV f d^3v &=& \br + \bq\bu;\\
  m\int c_ic_jc_k v_l f d^3v &=& m\int c_ic_jc_k(v_l-u_l+u_l)fd^3v = \underbrace{m\int c_ic_jc_kc_l fd^3v}_{r_{ijkl}} +
  u_l \underbrace{m\int c_ic_jc_k f d^3v}_{q_{ijk}} \nn\\
  &=& r_{ijkl} + q_{ijk} u_l;\\
   \pr_l (c_ic_jc_k) &=& -c_jc_k\pr_l u_i-c_ic_k\pr_l u_j-c_ic_j\pr_l u_k;\\
  m\int f v_l \pr_l(c_ic_jc_k) d^3v &=& -(\pr_lu_i)\underbrace{m\int c_jc_k v_l f d^3v}_{q_{jkl}+p_{jk}u_l}
  - (\pr_l u_j)\underbrace{m\int c_ic_k v_l f d^3v}_{q_{ikl}+p_{ik}u_l}
  -(\pr_l u_k) \underbrace{m \int c_i c_j v_l f d^3v}_{q_{ijl}+p_{ij}u_l} \nn\\
  &=& -(q_{jkl}+p_{jk}u_l)\pr_l u_i - (q_{ikl}+p_{ik}u_l)\pr_l u_j - (q_{ijl}+p_{ij}u_l)\pr_l u_k,
\end{eqnarray}
and the second term calculates
\begin{eqnarray}
  \bigcirc{ }\!\!\!\!\mbox{\small 2} &=& m\int c_ic_jc_kv_l\pr_lf d^3v =
  \pr_l \big( m\int c_ic_jc_k v_l f d^3v\big) -m\int f v_l \pr_l(c_ic_jc_k) d^3v \nn \\
  &=& \pr_l(r_{lijk} +u_l q_{ijk})
  + (q_{jkl}+p_{jk}u_l)\pr_l u_i + (q_{ikl}+p_{ik}u_l)\pr_l u_j + (q_{ijl}+p_{ij}u_l)\pr_l u_k \nn\\
  &=& \big[\nabla\cdot(\br + \bu\bq)\big]_{ijk} + \big[(\bq+\bp\bu)\cdot\nabla\bu\big]_{jki} + \big[(\bq+\bp\bu)\cdot\nabla\bu\big]_{ikj}
  + \big[(\bq+\bp\bu)\cdot\nabla\bu\big]_{ijk} \nn\\
  &=&  \big[\nabla\cdot(\br + \bu\bq)\big]_{ijk} + \big[(\bq+\bp\bu)\cdot\nabla\bu\big]_{ijk}^S.
\end{eqnarray}
Note again that the divergence of a tensor operates through its first component
$\pr_l(r_{lijk} +u_l q_{ijk}) = [\nabla\cdot(\br+\bu\bq)]_{ijk}$. Similarly, if the operator acts on a tensor from the right hand side,
the most natural way is to define that it operates through the last component, i.e. $[\bq\cdot\nabla]_{ij} = q_{ijl}\pr_l$. 
In the expressions above $[\bq\cdot\nabla\bu]_{ijk}=q_{ijl}\pr_l u_k$ and
$[\bp\bu\cdot\nabla\bu]_{ijk}=(\bp\bu)_{ijl}\pr_l u_k = p_{ij}u_l\pr_l u_k$. The last term can also be rewritten
as $p_{ij}(\bu\cdot\nabla\bu)_k$, where the expression $(\bu\cdot\nabla\bu)_k = u_l\pr_l u_k$ is familiar from MHD.
For the third term we will need identities
\begin{eqnarray}
 \frac{\pr c_i}{\pr v_l} &=& \delta_{il};\\ 
 \frac{\pr}{\pr v_l}(c_i c_j c_k) &=& \delta_{il}c_jc_k +c_i\delta_{jl}c_k + c_ic_j\delta_{kl},
\end{eqnarray}
and the entire third term calculates
\begin{eqnarray}
  \bigcirc{ }\!\!\!\!\mbox{\small 3} &=& q\int c_ic_jc_k E_l \frac{\pr f}{\pr v_l} d^3k
  = q E_l \underbrace{\int \frac{\pr}{\pr v_l}(c_ic_jc_k f) d^3v}_{\rightarrow 0} -q E_l\int \frac{\pr}{\pr v_l}(c_ic_jc_k) f d^3v  \nn\\
  &=& -qE_l \int(\delta_{il}c_jc_k+\delta_{jl}c_ic_k+\delta_{kl}c_ic_j) f d^3v
  = -\frac{q}{m}\big( E_i p_{jk}+E_j p_{ik} + E_k p_{ij} \big) \nn\\
  &=& -\frac{q}{m}\big[ \bE\bp\big]_{ijk}^S.
\end{eqnarray}
For the fourth term we will need identities
\begin{eqnarray}
\frac{\pr}{\pr v_l}(\bV\times\bb)_l &=& 0;\\
  \frac{\pr}{\pr v_l} [c_ic_jc_k(\bV\times\bb)_l ] &=& 
   c_j c_k (\bV\times\bb)_i + c_i c_k (\bV\times\bb)_j + c_i c_j (\bV\times\bb)_k;\\
  \int c_j (\bV\times\bb)_i f d^3v &=&  -\frac{1}{m}(\bb\times\bp)_{ij};\\
  \int c_j c_k (\bV\times\bb)_i fd^3v &=& -\frac{1}{m}\big[\bb\times\big(\bq+\bu\bp\big) \big]_{ijk},
\end{eqnarray}  
and the entire fourth term calculates
\begin{eqnarray}
\bigcirc{ }\!\!\!\!\mbox{\small 4} &=& \frac{q}{c}\int c_i c_j c_k (\bV\times\bb)_l \frac{\pr f}{\pr v_l} d^3 v=
\frac{q}{c}\underbrace{\int \frac{\pr}{\pr v_l}\big[ c_ic_jc_k (\bV\times\bb)_l f\big] d^3v}_{\rightarrow 0}
-\frac{q}{c}\int f \frac{\pr}{\pr v_l}\big[ c_ic_jc_k (\bV\times\bb)_l \big] d^3v \nn\\
&=& -\frac{q}{c}\Big\{ \int f c_j c_k (\bV\times\bb)_i d^3v + \int f c_k c_i (\bV\times\bb)_j d^3v
+\int f c_i c_j (\bV\times\bb)_k d^3v \Big\} \nn\\
&=& \frac{q}{mc} \Big\{ \big[\bb\times\big(\bq+\bu\bp\big)\big]_{ijk} + \big[\bb\times\big(\bq+\bu\bp\big)\big]_{jki}
  + \big[\bb\times\big(\bq+\bu\bp\big)\big]_{kij}\Big\} \equiv \frac{q}{mc} \big[ \bb\times\big(\bq+\bu\bp\big)\big]^S_{ijk}.
\end{eqnarray}
Combining all the results together
$\bigcirc{ }\!\!\!\!\mbox{\small 1}+\bigcirc{ }\!\!\!\!\mbox{\small 2}+\bigcirc{ }\!\!\!\!\mbox{\small 3} + \bigcirc{ }\!\!\!\!\mbox{\small 4}=0$,
the entire heat flux tensor equation obtained by direct integration of the Vlasov equation reads
\begin{eqnarray} \label{eq:Qdirect}
  \frac{\pr\bq}{\pr t} +\nabla\cdot(\br + \bu\bq) + \Big[ \frac{\pr \bu}{\pr t}\bp  
  + (\bq+\bp\bu)\cdot\nabla\bu-\frac{q}{m} \bE\bp +\frac{q}{mc} \bb\times\big(\bq+\bu\bp\big)\Big]^S =0.
\end{eqnarray}  
Now we will need to use 3 different momentum equations that will cancel various terms. We will also need identity
\begin{equation} \label{eq:MomId1}
\big[(\bu\times\bb)\bp\big]_{ijk} =  -\big[\bb\times(\bu\bp)\big]_{ijk}.
\end{equation}
 We need to multiply the momentum equation for ${\pr u_i}/{\pr t}$ by $p_{jk}$, equation for ${\pr u_j}/{\pr t}$ by $p_{ki}$,
and equation for ${\pr u_k}/{\pr t}$ by $p_{ij}$. 
All the three momentum equations that we want to subtract from the heat flux equation can be written together as
\begin{eqnarray}
&& \Big[ \Big(  \frac{\pr\bu}{\pr t} + \bu\cdot\nabla\bu + \frac{1}{mn} \nabla\cdot\bp - \frac{q}{m}\bE
    - \frac{q}{mc}\bu\times\bb \Big)\bp \Big]^S_{ijk}=0,
\end{eqnarray}
and because of identity (\ref{eq:MomId1}), this is equivalent to
\begin{eqnarray} \label{eq:MomId2}
&& \Big[ \Big( \frac{\pr\bu}{\pr t} + \bu\cdot\nabla\bu + \frac{1}{mn} \nabla\cdot\bp - \frac{q}{m}\bE\Big)\bp
  + \frac{q}{mc}\big[\bb\times(\bu\bp)\big] \Big]^S_{ijk}= 0.
\end{eqnarray}
Note that because of the symmetric operator, it does not matter if the tensor $\bp$ is applied on the momentum
equation from the left or right, since all the expressions are symmetric in this regard, i.e. for example
$\big[(\nabla\cdot\bp)\bp\big]^S=\big[\bp(\nabla\cdot\bp)\big]^S$.
By subtracting (\ref{eq:MomId2}) from (\ref{eq:Qdirect}), the final heat flux tensor equation reads 
\begin{eqnarray} 
  \frac{\pr}{\pr t}\bq +\nabla\cdot\big(\br+\bu\bq\big)
  + \Big[\bq\cdot\nabla\bu + \frac{q}{mc}\bb\times\bq - \frac{1}{\rho}\bp(\nabla\cdot\bp)\Big]^S = 0. \label{eq:HFgenF}
\end{eqnarray}  
By defining the cyclotron frequency vector $\boldsymbol{\Omega}=\frac{q\bb}{mc}$, this equation
identifies with equation (A5) in \cite{ChustBelmont2006}. By using scalar cyclotron frequency defined with respect to $|\bb|$ as
$\Omega=\frac{q|\bb|}{mc}$, the heat flux tensor equation is written explicitly in the index notation
\begin{eqnarray}
&&  \frac{\pr}{\pr t}q_{ijk} + \pr_l \big( r_{lijk} +u_l q_{ijk}\big)
  + q_{ijl}\pr_l u_k + q_{jkl}\pr_l u_i + q_{kil}\pr_l u_j
  +\Omega \hat{b}_l \big(\epsilon_{ilm} q_{mjk} + \epsilon_{jlm} q_{mki} +\epsilon_{klm} q_{mij} \big)   \nn\\
&&  -\frac{1}{mn}\big( p_{ij}\pr_l p_{lk} + p_{jk}\pr_l p_{li} + p_{ki}\pr_l p_{lj}\big) =0,
\end{eqnarray}
which is equivalent to equation (4) in \cite{Goswami2005} with the term
\begin{eqnarray}
  (\bhat\times\bq)_{ijk}^S &=&  \epsilon_{ilm}\hat{b}_l q_{mjk} + \epsilon_{jlm}\hat{b}_l q_{mki} +\epsilon_{klm}\hat{b}_l q_{mij} \nn\\
  &=& -\hat{b}_l \big(\epsilon_{iml} q_{jkm} + \epsilon_{jml} q_{ikm} +\epsilon_{kml} q_{ijm} \big).
\end{eqnarray}  

\subsection{Heat flux tensor decomposition}
Considering only the gyrotropic part of the heat flux tensor $(3\times 3\times 3)$ cube,
the heat flux is decomposed to the scalar parallel and perpendicular
heat flux components $q_\parallel, q_\perp$ according to
\begin{equation} \label{eq:Qdec1}
\bq^\textrm{g} = q_\parallel \bhat\bhat\bhat + q_\perp \big[ (\boldsymbol{I}-\bhat\bhat)\bhat \big]^S,
\end{equation}  
and which in the index notation reads  
\begin{equation} \label{eq:Qdec2}
  q_{ijk}^\textrm{g} = q_\parallel \hat{b}_i\hat{b}_j\hat{b}_k +
  q_\perp\Big( \delta_{ij}\hat{b}_k +\delta_{jk}\hat{b}_i+\delta_{ki}\hat{b}_j - 3\hat{b}_i\hat{b}_j\hat{b}_k \Big).  
\end{equation}
This part of the pressure tensor represents only the gyrotropic part, and the full heat flux decomposition can be written as
$\bq = \bq^{\textrm{g}}+\bq^{\textrm{ng}}$. In the heat flux tensor equation, the term $\bb\times\bq$
is proportional to the cyclotron frequency $\Omega=\frac{qB_0}{mc}$ and the equation rewrites
\begin{eqnarray} \label{eq:HF1}
  \frac{\pr}{\pr t}\bq +\nabla\cdot\big(\br+\bu\bq\big)
  + \Big[\bq\cdot\nabla\bu + \Omega\frac{|\bb|}{B_0}\bhat\times\bq - \frac{1}{\rho}\bp(\nabla\cdot\bp)\Big]^S = 0. 
\end{eqnarray}
Situation is now similar to the previously studied pressure tensor. At long spatial scales (low frequencies $\omega$),
this term will dominate and the gyrotropic contribution must be equal to zero
\begin{equation}
\big[ \bhat\times\bq^\textrm{g} \big ]^S=0.
\end{equation}
At first look, it is not that obvious that the decomposition (\ref{eq:Qdec1}) satisfy this equation. It is however
possible to verify that indeed
\begin{eqnarray}
 (\bhat\times\bq^\textrm{g})_{ijk} &=& \epsilon_{irs}\hat{b}_r q_{sjk}^\textrm{g} = \epsilon_{irs} \hat{b}_r \Big[ q_\parallel \hat{b}_s\hat{b}_j\hat{b}_k +
    q_\perp\Big( \delta_{sj}\hat{b}_k +\delta_{jk}\hat{b}_s+\delta_{ks}\hat{b}_j - 3\hat{b}_s\hat{b}_j\hat{b}_k \Big) \Big] =
  q_\parallel \hat{b}_j \hat{b}_k \underbrace{\epsilon_{irs} \hat{b}_r \hat{b}_s}_{=0} \nn\\
  &+&q_\perp \Big(
  \epsilon_{irj} \hat{b}_r \hat{b}_k +\delta_{jk} \underbrace{\epsilon_{irs}  \hat{b}_r \hat{b}_s}_{=0}
  +\epsilon_{irk} \hat{b}_r \hat{b}_j -3\hat{b}_j \hat{b}_k\underbrace{\epsilon_{irs} \hat{b}_r \hat{b}_s}_{=0} \Big)
  = q_\perp \hat{b}_r (\epsilon_{irj} \hat{b}_k + \epsilon_{irk} \hat{b}_j);\\
  (\bhat\times\bq^\textrm{g})_{jki} &=& q_\perp \hat{b}_r (\epsilon_{jrk} \hat{b}_i + \epsilon_{jri} \hat{b}_k);\\
  (\bhat\times\bq^\textrm{g})_{kij} &=& q_\perp \hat{b}_r (\epsilon_{kri} \hat{b}_j + \epsilon_{krj} \hat{b}_i).
\end{eqnarray}  
Putting all terms together
\begin{eqnarray}
  \big[ \bhat\times\bq^\textrm{g} \big ]^S_{ijk} &=& \big[ \bhat\times\bq^\textrm{g} \big ]_{ijk} + \big[ \bhat\times\bq^\textrm{g} \big ]_{jki} + \big[ \bhat\times\bq^\textrm{g} \big ]_{kij} =
  q_\perp \hat{b}_r \big[ \hat{b}_i \underbrace{(\epsilon_{jrk}+\epsilon_{krj})}_{=0} + \hat{b}_j \underbrace{(\epsilon_{irk}+\epsilon_{kri})}_{=0}
    +\hat{b}_k \underbrace{(\epsilon_{irj}+\epsilon_{jri})}_{=0} \big] \nn\\
  &=& 0, \label{eq:GgyroZero}
\end{eqnarray}  
and we see that all three parts of the symmetric operator are required to make this term equal to zero. Therefore, the exact heat flux tensor equation reads
\begin{eqnarray} \label{eq:Qngtensor}
  \frac{\pr}{\pr t}\bq +\nabla\cdot\big(\br+\bu\bq\big)
  + \Big[\bq\cdot\nabla\bu + \Omega\frac{|\bb|}{B_0}\bhat\times\bq^\textrm{ng} - \frac{1}{\rho}\bp(\nabla\cdot\bp)\Big]^S = 0. 
\end{eqnarray}
Since this should be an introductory text, we do not want to be bothered right now with the complicated algebra of the non-gyrotropic heat flux $\bq^{\textrm{ng}}$.
For the clarity of the presented material, here we separate the non-gyrotropic heat flux to a separate term $\boldsymbol{Q}^{\textrm{ng}}$, and write the heat flux
tensor equation in the following form
\begin{eqnarray} \label{eq:HFsimple}
  \frac{\pr}{\pr t}\bq^\textrm{g} +\nabla\cdot\big(\br+\bu\bq^\textrm{g}\big)
  + \Big[\bq^\textrm{g} \cdot\nabla\bu - \frac{1}{\rho}\bp(\nabla\cdot\bp)\Big]^S +\boldsymbol{Q}^{\textrm{ng}}= 0, 
\end{eqnarray}
where
\begin{equation} \label{eq:QngTrick}
\boldsymbol{Q}^{\textrm{ng}} =   \frac{\pr}{\pr t}\bq^\textrm{ng} +\nabla\cdot\big(\bu\bq^\textrm{ng}\big)
  + \Big[\bq^\textrm{ng} \cdot\nabla\bu + \Omega\frac{|\bb|}{B_0}\bhat\times\bq^\textrm{ng} \Big]^S.
\end{equation}  
We will address the non-gyrotropic heat flux contributions in the Appendix \ref{sec:NONGheat}.

In a similar fashion to the pressure decomposition (\ref{eq:Pdecomp}), we are going to frequently apply double contractions with $\bhat\bhat$ and
$(\boldsymbol{I}-\bhat\bhat)/2$. Applying these operators on the heat flux tensor $q_{ijk}$, must yield quantities that are vectors. 
It is therefore logical to define parallel and perpendicular heat flux \emph{vectors}
\begin{equation} \label{eq:HFvectorss}
  \boldsymbol{S}^\parallel \equiv \bq:\bhat\bhat; \qquad \boldsymbol{S}^\perp \equiv \bq:(\boldsymbol{I}-\bhat\bhat)/2.
\end{equation}
The scalar parallel and perpendicular heat flux components $q_\parallel, q_\perp$
(which are the only parts that are gyrotropic) are obtained by further projecting these heat flux vectors along the magnetic field lines,
i.e. by performing $\cdot\bhat$, so that 
\begin{eqnarray} \label{eq:Qdecomp}
q_\parallel = (\bq:\bhat\bhat)\cdot\bhat; \qquad q_\perp = (\bq:(\boldsymbol{I}-\bhat\bhat)/2)\cdot\bhat.
\end{eqnarray}
Briefly considering only the gyrotropic heat flux (\ref{eq:Qdec2}) on the right hand side, it is easy to verify that the parallel decomposition indeed works
\begin{eqnarray}
  (\bq^\textrm{g}:\bhat\bhat)_k &=& q_{kij}^\textrm{g} \hat{b}_i\hat{b}_j = q_{ijk}^\textrm{g} \hat{b}_i\hat{b}_j =
  \Big[ q_\parallel \hat{b}_i\hat{b}_j\hat{b}_k +
    q_\perp\Big( \delta_{ij}\hat{b}_k +\delta_{jk}\hat{b}_i+\delta_{ki}\hat{b}_j - 3\hat{b}_i\hat{b}_j\hat{b}_k \Big)  \Big]\hat{b}_i\hat{b}_j \nn\\
  &=& q_\parallel \hat{b}_k + q_\perp (\hat{b}_k+\hat{b}_k+\hat{b}_k-3\hat{b}_k ) = q_\parallel \hat{b}_k; \\
  \bq^\textrm{g}:\bhat\bhat &=& q_\parallel \bhat;\\
  (\bq^\textrm{g}:\bhat\bhat)\cdot\bhat &=& q_\parallel.
\end{eqnarray}
For the perpendicular decomposition, it is useful to specifically calculate components
\begin{equation}
  q_{iik}^\textrm{g} = q_\parallel\hat{b}_k +q_\perp\Big( \underbrace{\delta_{ii}}_{=3} \hat{b}_k +\hat{b}_k +\hat{b}_k -3\hat{b}_k \Big)
  = q_\parallel\hat{b}_k + 2 q_\perp \hat{b}_k,
\end{equation}
so that
\begin{eqnarray}
&&  Tr \bq^\textrm{g} = \bq^\textrm{g}:\boldsymbol{I} = q_\parallel\bhat + 2q_\perp \bhat;\\
&&  \bq^\textrm{g}:(\boldsymbol{I}-\bhat\bhat)/2 = (\bq^\textrm{g}:\boldsymbol{I} - \bq^\textrm{g}:\bhat\bhat)/2 = (q_\parallel\bhat + 2q_\perp \bhat - q_\parallel\bhat)/2 = q_\perp \bhat;\\
&&  (\bq^\textrm{g}:(\boldsymbol{I}-\bhat\bhat)/2)\cdot\bhat  =q_\perp,
\end{eqnarray}  
which verifies that the perpendicular decomposition (\ref{eq:Qdecomp}) is satisfied for $\bq^\textrm{g}$. Now we apply the decomposition (\ref{eq:Qdecomp}) directly at the
definition of the entire heat flux tensor (\ref{eq:defHFT}), which yields
\begin{eqnarray}
  (\bq:\bhat\bhat)_k &=& (m\int\bc\bc\bc fd^3v :\bhat\bhat)_k = m\int c_ic_jc_k f d^3v \hat{b}_i \hat{b}_j =
  m\int (v_i-u_u)\hat{b}_i (v_j-u_j)\hat{b}_j (v_k-u_k)f d^3v \nn\\
  &=& m\int (v_\parallel-u_\parallel)^2 (v_k-u_k)f d^3v \equiv (\boldsymbol{S}^\parallel)_k; \\
  (\bq:\bhat\bhat)\cdot\bhat &=& m\int (v_\parallel-u_\parallel)^3 f d^3v \equiv q_\parallel;
\end{eqnarray}
and the perpendicular decomposition
\begin{eqnarray}
  (\bq:\boldsymbol{I}/2)_k &=& \frac{m}{2}\int c_i c_j c_k f d^3v \delta_{ij} = \frac{m}{2} \int |\bc|^2 c_k f d^3v
  =  \frac{m}{2} \int |\bV-\bu|^2 (v_k-u_k) f d^3v;\\
  (\bq:(\boldsymbol{I}-\bhat\bhat)/2)_k &=& \frac{m}{2}\int \Big( |\bV-\bu|^2 -(v_\parallel-u_\parallel)^2 \Big) (v_k-u_k) f d^3v \nn\\
  &=&  \frac{m}{2}\int |\bV_\perp-\bu_\perp|^2 (v_k-u_k) f d^3v \equiv (\boldsymbol{S}^\perp)_k;\\
  (\bq:(\boldsymbol{I}-\bhat\bhat)/2)\cdot\bhat &=& \frac{m}{2}\int |\bV_\perp-\bu_\perp|^2 (v_\parallel-u_\parallel) f d^3v \equiv q_\perp.
\end{eqnarray}
The decomposition (\ref{eq:Qdecomp}) obviously works for the entire heat flux $\bq$, as well as for
the gyrotropic part $\bq^\textrm{g}$. Similarly to the pressure decomposition, this further yields required properties that the non-gyrotropic heat flux
$\bq^{\textrm{ng}}$ must satisfy. By using (\ref{eq:Qdecomp}), the entire heat flux decomposition reads
\begin{eqnarray}
  \bq &=& q_\parallel \bhat\bhat\bhat + q_\perp \big[ (\boldsymbol{I}-\bhat\bhat)\bhat \big]^S + \bq^{\textrm{ng}};\\
  \bq &=& \big[(\bq:\bhat\bhat)\cdot\bhat\big]\bhat\bhat\bhat + \big[(\bq:(\boldsymbol{I}-\bhat\bhat)/2)\cdot\bhat \big]
  \big[ (\boldsymbol{I}-\bhat\bhat)\bhat \big]^S + \bq^{\textrm{ng}}.
\end{eqnarray}  
Obviously, it would be useful to introduce a triple-contraction operator and instead of $(\bq:\bhat\bhat)\cdot\bhat$ to
write something like  $\bq \,\, \vdots \,\, \bhat\bhat\bhat$, but we do not want to introduce new notations. Also, an alternative and
perhaps prettier expression is to move the $\cdot\bhat$ to the left hand side, as $(\bq:\bhat\bhat)\cdot\bhat = \bhat\cdot \bq:\bhat\bhat$,   
but we will keep the first choice. By applying $:\bhat\bhat$ and $\cdot\bhat$ at the above equation yields
\begin{equation}
(\bq:\bhat\bhat)\cdot\bhat = (\bq:\bhat\bhat)\cdot\bhat + (\bq^{\textrm{ng}}:\bhat\bhat)\cdot\bhat,
\end{equation}
implying the first requirement for the non-gyrotropic heat flux
\begin{equation}
(\bq^{\textrm{ng}}:\bhat\bhat)\cdot\bhat = 0.
\end{equation}  
Similarly, the second requirement is obtained by applying $:(\boldsymbol{I}-\bhat\bhat)$ and $\cdot\bhat$, yielding
\begin{equation}
(\bq^{\textrm{ng}}:(\boldsymbol{I}-\bhat\bhat))\cdot\bhat =0.
\end{equation}  
By using the first requirement, the second requirement simplifies to
\begin{equation}
(\bq^{\textrm{ng}}:\boldsymbol{I})\cdot\bhat = \trace \bq^{\textrm{ng}} \cdot\bhat =0,
\end{equation}
where obviously the trace and $\cdot\bhat$ operators commute. The two requirements in the index notation read
\begin{equation}
q_{ijk}^{\textrm{ng}} \hat{b}_i \hat{b}_j \hat{b}_k =0; \qquad q_{iik}^{\textrm{ng}} \hat{b}_k =0.
\end{equation}
Instead of decomposing $\bq = \bq^{\textrm{g}}+\bq^{\textrm{ng}}$, an alternative and very useful decomposition of the entire heat flux tensor reads
\begin{equation}
\bq = \boldsymbol{S}+\boldsymbol{\sigma},
\end{equation}  
with the requirement $\boldsymbol{\sigma}:\bhat\bhat=0$ and $\boldsymbol{\sigma}:(\boldsymbol{I}-\bhat\bhat)=0$ (or equivalently $\boldsymbol{\sigma}:\boldsymbol{I}=0$).
The heat flux vectors (\ref{eq:HFvectorss}) therefore satisfy $\boldsymbol{S}^\parallel =\boldsymbol{S}:\bhat\bhat$ and
$\boldsymbol{S}^\perp = \boldsymbol{S}:(\boldsymbol{I}-\bhat\bhat)/2$. The heat flux vectors contain both gyrotropic and non-gyrotropic contributions.
Since the gyrotropic contributions are obtained by projecting these vectors along the magnetic field lines $q_\parallel=\boldsymbol{S}^\parallel\cdot\bhat$,
$q_\perp = \boldsymbol{S}^\perp \cdot\bhat$, it is useful to introduce the following decomposition
\begin{equation}
\boldsymbol{S}^\parallel \equiv q_\parallel\bhat + \boldsymbol{S}^\parallel_\perp; \qquad
\boldsymbol{S}^\perp \equiv q_\perp \bhat + \boldsymbol{S}_\perp^\perp.
\end{equation}
The vectors $\boldsymbol{S}^\parallel_\perp$, $\boldsymbol{S}^\perp_\perp$ are referred to as the \emph{non-gyrotropic heat flux vectors},
and their algebra is addressed in the Appendix \ref{sec:NONGheat}. Here we only state that the $\bq^{\textrm{ng}}$ can be decomposed to vectors
$\boldsymbol{S}^\parallel_\perp$, $\boldsymbol{S}^\perp_\perp$ and tensor $\boldsymbol{\sigma}$ according to
\begin{equation}
\bq^{\textrm{ng}} = \Big[ \boldsymbol{S}^\parallel_\perp \bhat\bhat  \Big]^S + \frac{1}{2}\Big[ \boldsymbol{S}^\perp_\perp (\boldsymbol{I}-\bhat\bhat)\Big]^S +\boldsymbol{\sigma}.
\end{equation}
The entire heat flux tensor $\boldsymbol{\sigma}$ is of course non-gyrotropic.

Now we need to verify the heat flux contributions (\ref{eq:HFcontr1}), (\ref{eq:HFcontr2}) that we used in the pressure equations
(\ref{eq:PparEnd}), (\ref{eq:PperpEnd}). 
To calculate the heat flux contributions to the pressure equations, we will need
\begin{eqnarray}
  \trace(\nabla\cdot\bq^\textrm{g}) &=& \delta_{ij} (\nabla\cdot\bq^\textrm{g})_{ij} = \delta_{ij} \pr_k q_{kij}^\textrm{g} =  \pr_k q_{iik}^\textrm{g}
  = \pr_k \big( q_\parallel\hat{b}_k + 2 q_\perp \hat{b}_k \big) = \nabla\cdot(q_\parallel\bhat) +2\nabla\cdot(q_\perp\bhat);\\
  q_{ijk}^\textrm{g}\hat{b}_i\hat{b}_j &=& q_\parallel \hat{b}_k;\\
 q_{ijk}^\textrm{g}\pr_k(\hat{b}_i\hat{b}_j) &=& \Big[q_\parallel \hat{b}_i\hat{b}_j\hat{b}_k +
  q_\perp\Big( \delta_{ij}\hat{b}_k +\delta_{jk}\hat{b}_i+\delta_{ki}\hat{b}_j - 3\hat{b}_i\hat{b}_j\hat{b}_k \Big) \Big]\pr_k(\hat{b}_i\hat{b}_j)
 =  q_\parallel\hat{b}_k \underbrace{\hat{b}_i\hat{b}_j \pr_k(\hat{b}_i\hat{b}_j)}_{=0}
 + q_\perp \hat{b}_k \underbrace{\pr_k(\hat{b}_i\hat{b}_i)}_{=0} \nn\\
 &+& q_\perp \hat{b}_i\pr_j(\hat{b}_i\hat{b}_j)  + q_\perp \hat{b}_j\pr_i(\hat{b}_i\hat{b}_j)
 -3q_\perp \hat{b}_k \underbrace{\hat{b}_i\hat{b}_j \pr_k(\hat{b}_i\hat{b}_j)}_{=0} = q_\perp \pr_j\hat{b}_j
 + q_\perp \hat{b}_j\underbrace{\hat{b}_i\pr_j\hat{b}_i}_{=0} + q_\perp \pr_i\hat{b}_i
 + q_\perp \hat{b}_i\underbrace{\hat{b}_j\pr_i\hat{b}_j}_{=0} \nn\\
 &=& 2q_\perp \nabla\cdot\bhat,
\end{eqnarray}
and the contributions are
\begin{eqnarray}
 \bhat\cdot(\nabla\cdot\bq^\textrm{g})\cdot\bhat &=& \hat{b}_i (\pr_k q_{kij}^\textrm{g} )\hat{b}_j = \pr_k (q_{ijk}^\textrm{g}\hat{b}_i\hat{b}_j )
  - q_{ijk}^\textrm{g}\pr_k( \hat{b}_i\hat{b}_j ) = \pr_k(q_\parallel \hat{b}_k) - 2q_\perp \nabla\cdot\bhat \nn\\
  &=& \nabla\cdot(q_\parallel\bhat) - 2q_\perp \nabla\cdot\bhat;\\
\frac{1}{2}\Big[\trace\nabla\cdot\bq^\textrm{g} -\bhat\cdot(\nabla\cdot\bq^\textrm{g})\cdot\bhat\Big] &=& \nabla\cdot(q_\perp\bhat) + q_\perp \nabla\cdot\bhat.  
\end{eqnarray}
The left hand side of the above equation can be also written naturally as $(\nabla\cdot\bq^\textrm{g}):(\boldsymbol{I}-\bhat\bhat)/2$,
which is consistent with an alternative derivation of the perpendicular pressure equation, where instead of doing Trace and subtracting the parallel pressure equation,
one can directly perform $:(\boldsymbol{I}-\bhat\bhat)/2$.
\subsection{Parallel heat flux equation}
Now we apply the decomposition (\ref{eq:Qdecomp}) on the heat flux tensor equation (\ref{eq:HFsimple}) step by step.
We consider only gyrotropic heat flux components.
Starting with the equation for the parallel heat flux, the first term calculates
\begin{eqnarray}
  (\frac{\pr \bq^\textrm{g}}{\pr t}:\bhat\bhat)\cdot\bhat &=&
  \frac{\pr}{\pr t}\Big[ q_\parallel \hat{b}_i\hat{b}_j\hat{b}_k +
    q_\perp\Big( \delta_{ij}\hat{b}_k +\delta_{jk}\hat{b}_i+\delta_{ki}\hat{b}_j - 3\hat{b}_i\hat{b}_j\hat{b}_k \Big) \Big] \hat{b}_i \hat{b}_j \hat{b}_k
  = \frac{\pr q_\parallel}{\pr t} \underbrace{\hat{b}_i\hat{b}_j\hat{b}_k \hat{b}_i\hat{b}_j\hat{b}_k}_{=1} \nn\\
  &+& q_\parallel \underbrace{\hat{b}_i\hat{b}_j\hat{b}_k\frac{\pr}{\pr t}(\hat{b}_i\hat{b}_j\hat{b}_k)}_{=0}
  +\frac{\pr q_\perp}{\pr t} (\underbrace{1+1+1-3}_{=0}) + q_\perp \Big[ \underbrace{\hat{b}_k\frac{\pr}{\pr t}\hat{b}_k}_{=0}
    + \underbrace{\hat{b}_i\frac{\pr}{\pr t}\hat{b}_i}_{=0} +\underbrace{\hat{b}_j\frac{\pr}{\pr t}\hat{b}_j}_{=0}
    - 3 \underbrace{\hat{b}_i\hat{b}_j\hat{b}_k\frac{\pr}{\pr t}(\hat{b}_i\hat{b}_j\hat{b}_k)}_{=0} \Big] \nn\\
    &=& \frac{\pr q_\parallel}{\pr t};
\end{eqnarray}
the second term calculates
\begin{eqnarray}
  (\nabla\cdot(\bu\bq^\textrm{g}):\bhat\bhat)\cdot\bhat &=& \pr_l (u_l q_{ijk}^\textrm{g}) \hat{b}_i\hat{b}_j\hat{b}_k =
  (\pr_l u_l) \underbrace{q_{ijk}^\textrm{g} \hat{b}_i\hat{b}_j\hat{b}_k}_{=q_\parallel} + u_l \underbrace{(\pr_l q_{ijk}^\textrm{g}) \hat{b}_i\hat{b}_j\hat{b}_k}_{=\pr_l q_\parallel} \nn\\
  &=& q_\parallel \nabla\cdot\bu + \bu\cdot \nabla q_\parallel = \nabla\cdot(q_\parallel \bu); \\
  ((\nabla\cdot\boldsymbol{r}):\bhat\bhat)\cdot\bhat &=& (\pr_l r_{lijk}) \hat{b}_i\hat{b}_j\hat{b}_k = \textrm{unchanged}.
\end{eqnarray}
Note that in all expressions here, the operator $\cdot\bhat$ can be naturally moved to the left as $\bhat\cdot$. The third term calculates
\begin{eqnarray}
 ((\bq^\textrm{g}\cdot\nabla\bu):\bhat\bhat)\cdot\bhat &=& (\bq^\textrm{g}\cdot\nabla\bu)_{ijk} \hat{b}_i\hat{b}_j\hat{b}_k =
  q_{ijl}^\textrm{g} (\pr_l u_k)\hat{b}_i\hat{b}_j\hat{b}_k
  = (\pr_l u_k) \Big[ q_\parallel \hat{b}_i\hat{b}_j\hat{b}_l +
    q_\perp\Big( \delta_{ij}\hat{b}_l +\delta_{jl}\hat{b}_i+\delta_{li}\hat{b}_j - 3\hat{b}_i\hat{b}_j\hat{b}_l \Big) \Big]\hat{b}_i\hat{b}_j\hat{b}_k;\nn\\
  &=& (\pr_l u_k) \Big[ q_\parallel \hat{b}_l\hat{b}_k +q_\perp \underbrace{\Big(\hat{b}_l\hat{b}_k+ \hat{b}_l\hat{b}_k+
      \hat{b}_l\hat{b}_k -3 \hat{b}_l\hat{b}_k \Big)}_{=0} \Big] = q_\parallel \hat{b}_l (\pr_l u_k) \hat{b}_k = q_\parallel \bhat\cdot\nabla\bu\cdot\bhat;\\
  \big[\bq^\textrm{g}\cdot\nabla\bu\big]^S_{ijk} \hat{b}_i\hat{b}_j\hat{b}_k &=& \big[ q_{ijl}^\textrm{g}\pr_l u_k+q_{jkl}^\textrm{g}\pr_l u_i+q_{kil}^\textrm{g}\pr_l u_j
    \big]\hat{b}_i\hat{b}_j\hat{b}_k = \underbrace{q_{ijl}^\textrm{g}\hat{b}_i\hat{b}_j}_{=q_\parallel \hat{b}_l} (\pr_l u_k) \hat{b}_k
  +\underbrace{q_{jkl}^\textrm{g}\hat{b}_j\hat{b}_k}_{=q_\parallel \hat{b}_l} (\pr_l u_i)\hat{b}_i
  +\underbrace{q_{kil}^\textrm{g}\hat{b}_k\hat{b}_i}_{=q_\parallel \hat{b}_l} (\pr_l u_j)\hat{b}_j \nn\\
  &=&  3 q_\parallel \bhat\cdot\nabla\bu\cdot\bhat.
\end{eqnarray}
For the final fourth term we will need
\begin{eqnarray}
(\nabla\cdot\bp)\cdot\bhat &=& \hat{b}_k \pr_l p_{lk} =  \hat{b}_k \pr_l \Big[ (p_\parallel-p_\perp)\hat{b}_l\hat{b}_k + p_\perp\delta_{lk} +\Pi_{lk}\Big] \nn\\
  &=& \hat{b}_k \Big[ \hat{b}_l\hat{b}_k \pr_l(p_\parallel-p_\perp) + (p_\parallel-p_\perp)\pr_l (\hat{b}_l\hat{b}_k)
    + \pr_k p_\perp + \pr_l \Pi_{lk} \Big] \nn\\
  &=& \Big[ \hat{b}_l \pr_l(p_\parallel-p_\perp)
    + (p_\parallel-p_\perp)\underbrace{\hat{b}_k \pr_l (\hat{b}_l\hat{b}_k)}_{=\nabla\cdot\bhat}
    + \hat{b}_k\pr_k p_\perp + (\pr_l \Pi_{lk})\hat{b}_k \Big] \nn\\
  &=& \bhat\cdot\nabla p_\parallel + (p_\parallel-p_\perp)\nabla\cdot\bhat + (\nabla\cdot\boldsymbol{\Pi})\cdot\bhat, \label{eq:BkPrlPlk}
\end{eqnarray}  
and the fourth term calculates
\begin{eqnarray}
  \big[\bp (\nabla\cdot\bp)\big]_{ijk}\hat{b}_i\hat{b}_j\hat{b}_k &=& p_{ij} (\pr_l p_{lk}) \hat{b}_i\hat{b}_j\hat{b}_k =
  \underbrace{p_{ij} \hat{b}_i\hat{b}_j}_{=p_\parallel} \hat{b}_k \pr_l p_{lk} = p_\parallel \hat{b}_k \pr_l p_{lk} \nn\\
  &=& p_\parallel \bhat\cdot\nabla p_\parallel + p_\parallel (p_\parallel-p_\perp)\nabla\cdot\bhat + p_\parallel (\nabla\cdot\boldsymbol{\Pi})\cdot\bhat;\\
  \big[\bp (\nabla\cdot\bp)\big]^S_{ijk}\hat{b}_i\hat{b}_j\hat{b}_k &=& \Big[ p_{ij}\pr_l p_{lk} + p_{jk}\pr_l p_{li} + p_{ki}\pr_l p_{lj}
    \Big] \hat{b}_i\hat{b}_j\hat{b}_k \nn \\
  &=& \underbrace{p_{ij} \hat{b}_i\hat{b}_j}_{=p_\parallel} \hat{b}_k \pr_l p_{lk} +
  \underbrace{p_{jk}\hat{b}_j \hat{b}_k}_{=p_\parallel} \hat{b}_i\pr_l p_{li} + \underbrace{p_{ki}\hat{b}_k\hat{b}_i}_{=p_\parallel} \hat{b}_j \pr_l p_{lj}
  = 3 p_\parallel \hat{b}_k \pr_l p_{lk};\\
 -\frac{1}{\rho} \big[\bp (\nabla\cdot\bp)\big]^S_{ijk}\hat{b}_i\hat{b}_j\hat{b}_k &=& -3\frac{p_\parallel}{\rho} \bhat\cdot\nabla p_\parallel
  -3\frac{p_\parallel}{\rho} (p_\parallel-p_\perp)\nabla\cdot\bhat - 3\frac{p_\parallel}{\rho} (\nabla\cdot\boldsymbol{\Pi})\cdot\bhat;
\end{eqnarray}
Combining all the terms yields equation for the scalar parallel heat flux
\begin{eqnarray} 
&&  \frac{\pr q_\parallel}{\pr t}+\nabla\cdot(q_\parallel \bu) + \bhat\cdot(\nabla\cdot\boldsymbol{r}):\bhat\bhat
  + 3 q_\parallel \bhat\cdot\nabla\bu\cdot\bhat -3\frac{p_\parallel}{\rho} \bhat\cdot\nabla p_\parallel
  -3\frac{p_\parallel}{\rho} (p_\parallel-p_\perp)\nabla\cdot\bhat \nn\\
&&  - 3\frac{p_\parallel}{\rho} (\nabla\cdot\boldsymbol{\Pi})\cdot\bhat +Q^{\textrm{ng}}_\parallel =0, \label{eq:QparGen}
\end{eqnarray}
where $Q^{\textrm{ng}}_\parallel\equiv (\boldsymbol{Q}^{\textrm{ng}}:\bhat\bhat)\cdot\bhat$.
The heat flux equations contain the fourth-order moment $r_{ijkl}$, which we will consider in the next section and at this stage it is unspecified.
\subsection{Perpendicular heat flux equation}
We will apply the Trace operator on (\ref{eq:HFsimple}) and also perform $\cdot\bhat$ step by step. In the index notation,
we are basically multiplying the entire equation by $\delta_{ij}\hat{b}_k$. The first term calculates
\begin{eqnarray}
  \trace\frac{\pr}{\pr t}\bq^\textrm{g} &=& \frac{\pr}{\pr t} \trace \bq^\textrm{g} = \frac{\pr}{\pr t}(q_\parallel\bhat+2q_\perp\bhat)
  = \bhat\frac{\pr}{\pr t}(q_\parallel+2q_\perp) + (q_\parallel+2q_\perp)\frac{\pr}{\pr t}\bhat;\\
  (\trace\frac{\pr}{\pr t}\bq^\textrm{g})\cdot\bhat &=& \underbrace{\bhat\cdot\bhat}_{=1} \frac{\pr}{\pr t}(q_\parallel+2q_\perp)
  + (q_\parallel+2q_\perp)\underbrace{\frac{\pr\bhat}{\pr t}\cdot\bhat}_{=0} = \frac{\pr}{\pr t}(q_\parallel+2q_\perp).
\end{eqnarray}  
The second term $\nabla\cdot(\br+\bu\bq^\textrm{g})$ calculates
\begin{eqnarray}
  \big[\trace \nabla\cdot(\bu\bq^\textrm{g})\big]_k &=& \delta_{ij} \pr_l (u_l q_{ijk}^\textrm{g}) = \pr_l (u_l q_{iik}^\textrm{g});\\
  \big[\trace \nabla\cdot(\bu\bq^\textrm{g})\big] \cdot\bhat &=& 
  \pr_l (u_l q_{iik}^\textrm{g}) \hat{b}_k = \pr_l (u_l q_{iik}^\textrm{g} \hat{b}_k )
  - u_l \underbrace{q_{iik}^\textrm{g}\pr_l\hat{b}_k}_{=0} = \pr_l [u_l (q_\parallel + 2q_\perp)] \nn\\
  &=& (q_\parallel + 2q_\perp) \nabla\cdot\bu + \bu\cdot\nabla(q_\parallel + 2q_\perp);\\
  \big[ \trace \nabla\cdot\br \big]_k &=& \delta_{ij} \pr_l r_{lijk} = \pr_l r_{liik};\\
  \big[ \trace \nabla\cdot\br \big]\cdot\bhat &=& (\pr_l r_{liik})\hat{b}_k=\textrm{unchanged}.
\end{eqnarray}  
The components $r_{liik}=r_{iikl}$ are nothing else but the trace of the moment $\br$
\begin{eqnarray}
\big[ \trace \br \big]_{kl} = \delta_{ij} r_{ijkl} = r_{iikl}.
\end{eqnarray}  

The third term in (\ref{eq:HFsimple}) calculates
\begin{eqnarray}
  \big[ \bq^\textrm{g}\cdot\nabla \bu \big]^S_{ijk} &=& q_{ijl}^\textrm{g} \pr_l u_k + q_{jkl}^\textrm{g}\pr_l u_i + q_{kil}^\textrm{g}\pr_l u_j;\\
  \trace \big[ \bq^\textrm{g}\cdot\nabla \bu \big]^S_{ijk} &=& q_{iil}^\textrm{g} \pr_l u_k + q_{ikl}^\textrm{g}\pr_l u_i + \underbrace{q_{kil}^\textrm{g}}_{=q_{ikl}^\textrm{g}}\pr_l u_i
  = q_{iil}^\textrm{g} \pr_l u_k + 2q_{ikl}^\textrm{g}\pr_l u_i\\
  \trace \big[ \bq^\textrm{g}\cdot\nabla \bu \big]^S_{ijk} \hat{b}_k &=& q_{iil}^\textrm{g} (\pr_l u_k) \hat{b}_k + 2q_{ikl}^\textrm{g}\hat{b}_k \pr_l u_i
  = (q_\parallel+2q_\perp)\hat{b}_l (\pr_l u_k) \hat{b}_k + 2\big[ q_\parallel \hat{b}_i\hat{b}_l +q_\perp (\delta_{il}-\hat{b}_i\hat{b}_l) \big] \pr_l u_i\nn\\
  &=& (q_\parallel+2q_\perp)\hat{b}_l (\pr_l u_k) \hat{b}_k + 2(q_\parallel-q_\perp) \hat{b}_l (\pr_l u_i) \hat{b}_i +2q_\perp \pr_i u_i \nn\\
  &=& 3 q_\parallel \bhat\cdot\nabla\bu\cdot\bhat + 2q_\perp \nabla\cdot\bu,
\end{eqnarray}
where we have used that $q_{iil}^\textrm{g}=(q_\parallel+2q_\perp)\hat{b}_l$ and
$q_{ikl}^\textrm{g}\hat{b}_k = q_\parallel \hat{b}_i\hat{b}_l +q_\perp (\delta_{il}-\hat{b}_i\hat{b}_l)$. 
The fourth term in (\ref{eq:HFsimple}) calculates
\begin{eqnarray}
  \big[\bp (\nabla\cdot\bp)\big]_{ijk}^S &=& p_{ij}\pr_l p_{lk} + p_{jk}\pr_l p_{li} + p_{ki}\pr_l p_{lj};\\
  \trace \big[\bp (\nabla\cdot\bp)\big]_{ijk}^S &=& p_{ii}\pr_l p_{lk} + p_{ik}\pr_l p_{li} + \underbrace{p_{ki}}_{=p_{ik}}\pr_l p_{li}
  = p_{ii}\pr_l p_{lk} + 2p_{ik}\pr_l p_{li};\\
  \trace \big[\bp (\nabla\cdot\bp)\big]_{ijk}^S \hat{b}_k &=& p_{ii}\hat{b}_k\pr_l p_{lk} + 2p_{ik}\hat{b}_k \pr_l p_{li}, \label{eq:tired1}
\end{eqnarray}
and we have to calculate each term separately. Since $p_{ii}=p_\parallel+2p_\perp$
and by using already calculated equation (\ref{eq:BkPrlPlk}) for $\hat{b}_k\pr_l p_{lk}$ and identities
$p_{ik}\hat{b}_k = p_\parallel\hat{b}_i+\Pi_{ik}\hat{b}_k$ and $\Pi_{ik}\hat{b}_i\hat{b}_k=0$ one obtains
\begin{eqnarray}
  p_{ii}\hat{b}_k\pr_l p_{lk} &=& (p_\parallel+2p_\perp)
  \Big[\bhat\cdot\nabla p_\parallel + (p_\parallel-p_\perp)\nabla\cdot\bhat + (\nabla\cdot\boldsymbol{\Pi})\cdot\bhat\Big];\\
  p_{ik}\hat{b}_k \pr_l p_{li} &=& (p_\parallel\hat{b}_i+\Pi_{ik}\hat{b}_k)
  \Big[ \pr_i p_\perp +(p_\parallel-p_\perp)(\hat{b}_i\pr_l \hat{b}_l + \hat{b}_l\pr_l\hat{b}_i)
    +\hat{b}_i\hat{b}_l\pr_l(p_\parallel-p_\perp) +\pr_l\Pi_{li}\Big] \nn\\
  &=& p_\parallel \Big[\bhat\cdot\nabla p_\parallel + (p_\parallel-p_\perp)\nabla\cdot\bhat + (\nabla\cdot\boldsymbol{\Pi})\cdot\bhat\Big]
  + \Pi_{ik}\hat{b}_k \Big[ \pr_i p_\perp +(p_\parallel-p_\perp)\hat{b}_l\pr_l\hat{b}_i +\pr_l\Pi_{li} \Big] \nn\\
  &=& p_\parallel \Big[\bhat\cdot\nabla p_\parallel + (p_\parallel-p_\perp)\nabla\cdot\bhat + (\nabla\cdot\boldsymbol{\Pi})\cdot\bhat\Big]
  + \Big[ \nabla p_\perp + (p_\parallel-p_\perp)\bhat\cdot\nabla\bhat +\nabla\cdot\boldsymbol{\Pi}\Big] \cdot\boldsymbol{\Pi}\cdot\bhat;
\end{eqnarray}  
and the final result of (\ref{eq:tired1}) is
\begin{eqnarray}
  \trace \big[\bp (\nabla\cdot\bp)\big]_{ijk}^S \hat{b}_k &=&
   (3p_\parallel+2p_\perp) \Big[\bhat\cdot\nabla p_\parallel + (p_\parallel-p_\perp)\nabla\cdot\bhat + (\nabla\cdot\boldsymbol{\Pi})\cdot\bhat\Big]\nn\\
  && + 2\Big[ \nabla p_\perp + (p_\parallel-p_\perp)\bhat\cdot\nabla\bhat +\nabla\cdot\boldsymbol{\Pi}\Big] \cdot\boldsymbol{\Pi}\cdot\bhat.
\end{eqnarray}
Collecting all the terms, one obtains
\begin{eqnarray}
&&  \frac{\pr}{\pr t}(q_\parallel+2q_\perp) + (q_\parallel + 2q_\perp) \nabla\cdot\bu + \bu\cdot\nabla(q_\parallel + 2q_\perp)
   + \big( \trace \nabla\cdot\br \big)\cdot\bhat + 3 q_\parallel \bhat\cdot\nabla\bu\cdot\bhat + 2q_\perp \nabla\cdot\bu \nn\\
&&  - \frac{1}{\rho} (3p_\parallel+2p_\perp) \Big[\bhat\cdot\nabla p_\parallel + (p_\parallel-p_\perp)\nabla\cdot\bhat + (\nabla\cdot\boldsymbol{\Pi})\cdot\bhat\Big]
   -\frac{2}{\rho}\Big[ \nabla p_\perp + (p_\parallel-p_\perp)\bhat\cdot\nabla\bhat +\nabla\cdot\boldsymbol{\Pi}\Big] \cdot\boldsymbol{\Pi}\cdot\bhat \nn\\
&&   +(Q^{\textrm{ng}}_\parallel+2Q^{\textrm{ng}}_\perp)=0,
\end{eqnarray}  
and subtracting the parallel heat flux equation (\ref{eq:QparGen}) and dividing by two yields the perpendicular heat flux equation
\begin{eqnarray}
&&  \frac{\pr}{\pr t}q_\perp + \bu\cdot\nabla q_\perp + 2q_\perp \nabla\cdot\bu 
  + \frac{1}{2}\Big[ \big( \trace \nabla\cdot\br \big)\cdot\bhat - \bhat\cdot(\nabla\cdot\boldsymbol{r}):\bhat\bhat \Big] \nn\\
&&  - \frac{p_\perp}{\rho} \Big[\bhat\cdot\nabla p_\parallel + (p_\parallel-p_\perp)\nabla\cdot\bhat + (\nabla\cdot\boldsymbol{\Pi})\cdot\bhat\Big]
  -\frac{1}{\rho}\Big[ \nabla p_\perp + (p_\parallel-p_\perp)\bhat\cdot\nabla\bhat +\nabla\cdot\boldsymbol{\Pi}\Big] \cdot\boldsymbol{\Pi}\cdot\bhat \nn\\
&& +Q^{\textrm{ng}}_\perp =0, \label{eq:QperpGen}
\end{eqnarray}
where $Q^{\textrm{ng}}_\perp\equiv (\boldsymbol{Q}^{\textrm{ng}}:(\boldsymbol{I}-\bhat\bhat)/2)\cdot\bhat$.
The term containing $\br$ in the above equation can be rewritten to many possible forms, for example
\begin{eqnarray}
  \frac{1}{2}\Big[ \big( \trace \nabla\cdot\br \big)\cdot\bhat - \bhat\cdot(\nabla\cdot\boldsymbol{r}):\bhat\bhat \Big]
  &=& \frac{1}{2}\Big[ (\pr_l r_{liik})\hat{b}_k -(\pr_l r_{lijk})\hat{b}_i \hat{b}_j \hat{b}_k \Big]
  = \frac{1}{2}\Big[ \big( \trace \nabla\cdot\br \big) - (\nabla\cdot\boldsymbol{r}):\bhat\bhat \Big]\cdot\bhat \nn\\
  &=&  \bhat\cdot(\nabla\cdot\boldsymbol{r}):(\boldsymbol{I}-\bhat\bhat)/2. 
\end{eqnarray}
The last expression is of course consistent with an alternative way for deriving the perpendicular heat flux equation, where instead of doing Trace of the
heat flux tensor equation and subtracting the scalar parallel heat flux equation, one can directly apply operators
$:(\boldsymbol{I}-\bhat\bhat)/2$ and $\cdot\bhat$.  
\subsection{Scalar heat flux equations continued}
The parallel and perpendicular heat flux equations (\ref{eq:QparGen}), (\ref{eq:QperpGen}) contain expressions for the 4th-order moment $\boldsymbol{r}$,
and even though we will consider this moment in detail in the next section, here we want to finish the derivation of the scalar heat flux equations,
and we write down the required expressions.
Similarly to the pressure tensor and the heat flux tensor, the 4th order moment can be decomposed to its gyrotropic and non-gyrotropic part,
$\boldsymbol{r}=\boldsymbol{r}^\textrm{g}+\boldsymbol{r}^\textrm{ng}$. The gyrotropic part $\boldsymbol{r}^\textrm{g}$ is decomposed according to
(\ref{eq:RdecompFull}), and it can be shown (see later in the text), that direct calculation yields
\begin{eqnarray}
  \bhat\cdot(\nabla\cdot\boldsymbol{r}^\textrm{g}):\bhat\bhat &=& \nabla\cdot(r_{\parallel\parallel}\bhat) - 3 r_{\parallel\perp}\nabla\cdot\bhat;\label{eq:RgQverif1}\\
  \frac{1}{2}\Big[ \big( \trace \nabla\cdot\br^\textrm{g} \big)\cdot\bhat - \bhat\cdot(\nabla\cdot\boldsymbol{r}^\textrm{g}):\bhat\bhat \Big]
  &=& \nabla\cdot(r_{\parallel\perp}\bhat) + (r_{\parallel\perp}-r_{\perp\perp})\nabla\cdot\bhat, \label{eq:RgQverif2}
\end{eqnarray}
which yields the parallel heat flux equation
\begin{eqnarray}
&&  \frac{\pr q_\parallel}{\pr t}+\nabla\cdot(q_\parallel \bu) +\nabla\cdot(r_{\parallel\parallel}\bhat) - 3 r_{\parallel\perp}\nabla\cdot\bhat
  + 3 q_\parallel \bhat\cdot\nabla\bu\cdot\bhat -3\frac{p_\parallel}{\rho} \bhat\cdot\nabla p_\parallel
  -3\frac{p_\parallel}{\rho} (p_\parallel-p_\perp)\nabla\cdot\bhat \nn\\
  && \qquad  + \bhat\cdot(\nabla\cdot\boldsymbol{r}^\textrm{ng}):\bhat\bhat - 3\frac{p_\parallel}{\rho} (\nabla\cdot\boldsymbol{\Pi})\cdot\bhat
  +Q^{\textrm{ng}}_\parallel=0, \label{eq:ParHF_r}
\end{eqnarray}
and the perpendicular heat flux equation   
\begin{eqnarray}
  &&  \frac{\pr q_\perp}{\pr t} + \bu\cdot\nabla q_\perp + 2q_\perp \nabla\cdot\bu
  + \nabla\cdot(r_{\parallel\perp}\bhat) + (r_{\parallel\perp}-r_{\perp\perp})\nabla\cdot\bhat
  - \frac{p_\perp}{\rho} \Big[ \bhat\cdot \nabla p_\parallel + (p_\parallel-p_\perp)\nabla\cdot\bhat \Big] \nn\\
&&  + \frac{1}{2}\Big[ \big( \trace \nabla\cdot\br^\textrm{ng} \big)\cdot\bhat - \bhat\cdot(\nabla\cdot\boldsymbol{r}^\textrm{ng}):\bhat\bhat \Big] 
  - \frac{p_\perp}{\rho} (\nabla\cdot\boldsymbol{\Pi})\cdot\bhat
  -\frac{1}{\rho}\Big[ \nabla p_\perp + (p_\parallel-p_\perp)\bhat\cdot\nabla\bhat +\nabla\cdot\boldsymbol{\Pi}\Big] \cdot\boldsymbol{\Pi}\cdot\bhat \nn\\
&&  +Q^{\textrm{ng}}_\perp=0. \label{eq:PerpHF_r}
\end{eqnarray}
Exact nonlinear expressions for $Q^{\textrm{ng}}_\parallel$ and $Q^{\textrm{ng}}_\perp$ that represent contributions from the non-gyrotropic heat flux
$\bq^{\textrm{ng}}$ are calculated in the Appendix \ref{sec:NONGheat}, see eq. (\ref{eq:QngParX}), (\ref{eq:QngPerpX}).
The scalar heat flux equations (\ref{eq:ParHF_r}) and (\ref{eq:PerpHF_r}) are completely general at this stage, since no distribution function
was prescribed yet, and no simplification was introduced. These equations are exact. 

Nevertheless, equations (\ref{eq:ParHF_r}), (\ref{eq:PerpHF_r}) are very complicated, and at this stage, it is beneficial to simplify.
One possibility, is to cancel all the non-gyrotropic contributions $\boldsymbol{\Pi}$, $\boldsymbol{q}^{\textrm{ng}}$ and $\br^\textrm{ng}$,
and we will study such fluid models later. Another possibility, is to keep only those non-gyrotropic terms, that have some non-zero contribution at
the linear level. This eliminates the last term in the second line of (\ref{eq:PerpHF_r}) that is proportional to $[\ldots]\cdot\boldsymbol{\Pi}\cdot\bhat$.
Also, by following derivations in the Appendix \ref{sec:NONGheat}, it is easy to show that the terms $Q^{\textrm{ng}}_\parallel$, $Q^{\textrm{ng}}_\perp$ (that represent 
the non-gyrotropic heat flux $\boldsymbol{q}^{\textrm{ng}}$), do not contribute at the linear level. Therefore, the heat flux equations simplify 
\begin{eqnarray}
&&  \frac{\pr q_\parallel}{\pr t}+\nabla\cdot(q_\parallel \bu) +\nabla\cdot(r_{\parallel\parallel}\bhat) - 3 r_{\parallel\perp}\nabla\cdot\bhat
  + 3 q_\parallel \bhat\cdot\nabla\bu\cdot\bhat -3\frac{p_\parallel}{\rho} \bhat\cdot\nabla p_\parallel
  -3\frac{p_\parallel}{\rho} (p_\parallel-p_\perp)\nabla\cdot\bhat \nn\\
  && \qquad  + \bhat\cdot(\nabla\cdot\boldsymbol{r}^\textrm{ng}):\bhat\bhat - 3\frac{p_\parallel}{\rho} (\nabla\cdot\boldsymbol{\Pi})\cdot\bhat=0; \label{eq:ParHF_rX}\\
  &&  \frac{\pr q_\perp}{\pr t} + \bu\cdot\nabla q_\perp + 2q_\perp \nabla\cdot\bu
  + \nabla\cdot(r_{\parallel\perp}\bhat) + (r_{\parallel\perp}-r_{\perp\perp})\nabla\cdot\bhat
  - \frac{p_\perp}{\rho} \Big[ \bhat\cdot \nabla p_\parallel + (p_\parallel-p_\perp)\nabla\cdot\bhat \Big] \nn\\
&&  + \frac{1}{2}\Big[ \big( \trace \nabla\cdot\br^\textrm{ng} \big)\cdot\bhat - \bhat\cdot(\nabla\cdot\boldsymbol{r}^\textrm{ng}):\bhat\bhat \Big] 
  - \frac{p_\perp}{\rho} (\nabla\cdot\boldsymbol{\Pi})\cdot\bhat =0. \label{eq:PerpHF_rX}
\end{eqnarray}
Still, no specific distribution function was assumed. However, to evaluate the non-gyrotropic $\br^\textrm{ng}$, and correctly evaluate
possible cancellations with terms containing $\boldsymbol{\Pi}$, we need to use $\br^\textrm{ng}$ decomposition (\ref{eq:RngDEF}). 
Importantly, the decomposition (\ref{eq:RngDEF}) is valid only for perturbations around a bi-Maxwellian distribution function. 
By keeping only terms that have non-zero contribution at the linear level, see equations (\ref{eq:Qcontrib1}), (\ref{eq:Qcontrib2}) later in the text,  
the $\br^\textrm{ng}$ terms are evaluated as
\begin{eqnarray}
  \bhat\cdot(\nabla\cdot \br^{\textrm{ng}}):\bhat\bhat &=& \frac{3 p_\parallel}{\rho}(\nabla\cdot\boldsymbol{\Pi})\cdot\bhat; \label{eq:RngVerif1}\\
  \frac{1}{2}\Big[ \big( \trace \nabla\cdot\br^\textrm{ng} \big)\cdot\bhat - \bhat\cdot(\nabla\cdot\boldsymbol{r}^\textrm{ng}):\bhat\bhat \Big]
  &=& \frac{2p_\perp}{\rho}(\nabla\cdot\boldsymbol{\Pi})\cdot\bhat. \label{eq:RngVerif2}
\end{eqnarray}
Importantly, both non-gyrotropic contributions in the parallel heat flux equation (\ref{eq:ParHF_rX}) completely cancel out!
The cancellation demonstrates the importance of keeping the non-gyrotropic $\br^\textrm{ng}$, if one wants to keep the non-gyrotropic $\boldsymbol{\Pi}$. 
Also, there is a partial cancellation
in the perpendicular heat flux equation (\ref{eq:PerpHF_rX}). The heat flux equations therefore read
\begin{eqnarray}
&&  \frac{\pr q_\parallel}{\pr t}+\nabla\cdot(q_\parallel \bu) +\nabla\cdot(r_{\parallel\parallel}\bhat) - 3 r_{\parallel\perp}\nabla\cdot\bhat
  + 3 q_\parallel \bhat\cdot\nabla\bu\cdot\bhat -3\frac{p_\parallel}{\rho} \bhat\cdot\nabla p_\parallel
  -3\frac{p_\parallel}{\rho} (p_\parallel-p_\perp)\nabla\cdot\bhat =0; \label{eq:ParHF_Final} \\
  &&  \frac{\pr q_\perp}{\pr t} + \bu\cdot\nabla q_\perp + 2q_\perp \nabla\cdot\bu
  + \nabla\cdot(r_{\parallel\perp}\bhat) + (r_{\parallel\perp}-r_{\perp\perp})\nabla\cdot\bhat
  - \frac{p_\perp}{\rho} \Big[ \bhat\cdot \nabla p_\parallel + (p_\parallel-p_\perp)\nabla\cdot\bhat \Big] \nn\\
&& \qquad + \frac{p_\perp}{\rho} (\nabla\cdot\boldsymbol{\Pi})\cdot\bhat =0. \label{eq:PerpHF_Final}
\end{eqnarray}
The last term can be evaluated with respect to the mean magnetic field, and assuming that $\bhat_0$ is in the z-direction,
the term is equal to $\frac{p_\perp}{\rho}(\pr_x\Pi_{xz}+\pr_y\Pi_{yz})$, since $\Pi_{zz}=0$.
Essentially, the term should be written in a fully linearized form $\frac{p_\perp^{(0)}}{\rho_0}(\nabla\cdot\boldsymbol{\Pi})\cdot\bhat_0$, since other
non-gyrotropic nonlinear terms were neglected. It is important to emphasize that it is because of this one term, that
the equations (\ref{eq:ParHF_Final}), (\ref{eq:PerpHF_Final}) are valid only for perturbations around a bi-Maxwellian distribution function. 
If the term is neglected, i.e. if one neglects from the beginning the non-gyrotropic contributions in (\ref{eq:ParHF_rX}), (\ref{eq:PerpHF_rX}),  
the heat flux equations are valid for perturbations around any distribution function. We will consider bi-Kappa distribution function later. 

\subsection{Heat flux equations with ``normal'' closure}
Detailed calculations with the 4th-order moment will be presented in the next section. Here we want to finish the derivation, and
by using the bi-Maxwellian ``normal'' fluid closure
\begin{equation}
  r_{\parallel\parallel} = \frac{3p_\parallel^2}{\rho}; \quad
  r_{\parallel\perp} = \frac{p_\parallel p_\perp}{\rho}; \quad
  r_{\perp\perp} = \frac{2 p_\perp^2}{\rho},
\end{equation}
terms entering the heat flux equations directly calculate
\begin{eqnarray}
  \nabla\cdot(r_{\parallel\parallel}\bhat) - 3 r_{\parallel\perp}\nabla\cdot\bhat &=&
   3\frac{p_\parallel^2}{\rho}\nabla\cdot\bhat + 3p_\parallel\bhat\cdot\nabla\left(\frac{p_\parallel}{\rho}\right)
  + 3\frac{p_\parallel}{\rho}\bhat\cdot\nabla p_\parallel
  -3\frac{p_\parallel p_\perp}{\rho}\nabla\cdot\bhat;\\
  \nabla\cdot(r_{\parallel\perp}\bhat)+(r_{\parallel\perp}-r_{\perp\perp})\nabla\cdot\bhat &=&
  \frac{p_\perp}{\rho}\bhat\cdot\nabla p_\parallel +p_\parallel\bhat\cdot\nabla\left( \frac{p_\perp}{\rho} \right)
 +2\frac{p_\perp}{\rho}(p_\parallel-p_\perp)\nabla\cdot\bhat.
\end{eqnarray}
The use of these expressions in (\ref{eq:ParHF_Final}), (\ref{eq:PerpHF_Final}) cancels various terms, and the scalar heat flux equations read
\begin{eqnarray}
&&  \frac{\pr q_\parallel}{\pr t}+ \nabla\cdot( q_\parallel \bu )
  + 3p_\parallel \bhat\cdot\nabla\left(\frac{p_\parallel}{\rho}\right) + 3 q_\parallel \bhat\cdot\nabla\bu\cdot\bhat =0; \label{eq:ParHF_siX}\\
&&  \frac{\pr q_\perp}{\pr t} + \bu\cdot\nabla q_\perp +2q_\perp\nabla\cdot\bu +p_\parallel\bhat\cdot\nabla\left( \frac{p_\perp}{\rho} \right)
  +\frac{p_\perp}{\rho} (p_\parallel-p_\perp)\nabla\cdot\bhat + \frac{p_\perp}{\rho} (\nabla\cdot\boldsymbol{\Pi})\cdot\bhat_0 =0. \label{eq:PerpHF_siX} 
\end{eqnarray}

\newpage
\section{Fourth-order fluid moment} \label{section:4th}
The algebra of the fourth order fluid moment $\boldsymbol{r}=m\int \bc\bc\bc\bc fd^3v$ can be quite intimidating at first,
since the moment is a 4D cube (3x3x3x3). We are not going to derive the
time evolution equation for this moment step by step, nevertheless, later in the text we derive the evolution equation of the n-th order fluid moment
$\boldsymbol{X}^{(n)}$, see eq. (\ref{eq:GenTensor}), and therefore for $n=4$, the equation reads
\begin{eqnarray}
  \frac{\pr}{\pr t} \br +\nabla\cdot \big( \bX^{(5)}+\bu\br \big) +\Big[ \br\cdot\nabla\bu
    +\frac{q}{mc}\bb\times\br-\frac{1}{\rho} (\nabla\cdot\bp) \bq \Big]^S =0. \label{eq:4thTensorT}
\end{eqnarray} 
The symmetric operator ``$S$'' is here defined as
\begin{equation} \label{eq:Soper4}
A_{ijkl}^S = A_{ijkl} + A_{jkli} +A_{klij} +A_{lijk}.
\end{equation}  
Here we are not that interested in the evolution equation for $\br$, we just want to clarify its decomposition, that we need 
in the heat flux equations. The definition of gyrotropy means that the
integral has to be evaluated only over combinations of $(v_\parallel-u_\perp)\equiv c_\parallel$ and $|\bV_\perp-\bu_\perp|^2\equiv c_\perp^2$.
For the 4th moment $\br$, there are obviously only 3 possibilities: $c_\parallel^4$, $c_\parallel^2c_\perp^2$ and $(c_\perp^2)^2$, and these gyrotropic
components will be called $r_{\parallel\parallel}, r_{\parallel\perp}$ and $r_{\perp\perp}$.
We have already seen that the double contractions with $\bhat\bhat$ and $(\boldsymbol{I}-\bhat\bhat)/2$,
were very useful operators to extract the gyrotropic components for the lower order moments, $\bp$ and $\bq$. To obtain any scalar quantity from the 4th moment,
we obviously need to apply two double contractions. There are 3 possibilities: we can apply $\bhat\bhat$ twice, we can apply
$\bhat\bhat$ and $(\boldsymbol{I}-\bhat\bhat)/2$, or we can apply $(\boldsymbol{I}-\bhat\bhat)/2$ twice. Not surprisingly, these double contractions
with $\boldsymbol{r}$ indeed extract the 3 possible gyrotropic parts, as it is easy to verify
\begin{eqnarray}
 (\boldsymbol{r}:\bhat\bhat):\bhat\bhat &=& r_{ijkl} \hat{b}_i \hat{b}_j \hat{b}_k \hat{b}_l  =
  m\int c_ic_jc_kc_l f d^3v \hat{b}_i \hat{b}_j \hat{b}_k \hat{b}_l =
  m\int c_i \hat{b}_i c_j \hat{b}_j c_k \hat{b}_k c_l \hat{b}_l f d^3v \nn\\
 &=& m\int c_\parallel c_\parallel c_\parallel c_\parallel f d^3v = m\int (v_\parallel-u_\parallel)^4 f d^3v \equiv r_{\parallel\parallel}; \label{eq:rmom1}\\
  (\boldsymbol{r}:\bhat\bhat):(\boldsymbol{I}-\bhat\bhat)/2 &=& \frac{m}{2} \int c_ic_jc_kc_l \hat{b}_i \hat{b}_j (\delta_{kl}-\hat{b}_k \hat{b}_l) f d^3v
  = \frac{m}{2} \int c_\parallel^2 c_k c_l (\delta_{kl}-\hat{b}_k \hat{b}_l) f d^3v \nn\\
  &=& \frac{m}{2} \int c_\parallel^2 (|\bc|^2-c_\parallel^2) f d^3 v
  = \frac{m}{2}\int c_\parallel^2 c_\perp^2 f d^3v \nn \\
  &=& \frac{m}{2}\int (v_\parallel-u_\parallel)^2 |\bV_\perp-\bu_\perp|^2 f d^3 v \equiv r_{\parallel\perp}; \label{eq:rmom2}\\
 \Big(\boldsymbol{r}:(\boldsymbol{I}-\bhat\bhat)/2\Big):(\boldsymbol{I}-\bhat\bhat)/2 &=&
  \frac{m}{4} \int c_ic_jc_kc_l (\delta_{ij}-\hat{b}_i \hat{b}_j) (\delta_{kl}-\hat{b}_k \hat{b}_l) f d^3v = \frac{m}{4}\int c_\perp^2 c_\perp^2 f d^3v \nn\\
&&  = \frac{m}{4}\int |\bV_\perp-\bu_\perp|^4 f d^3v \equiv r_{\perp\perp}. \label{eq:rmom3}
\end{eqnarray}  
We are now ready to guess how to write the decomposition of the gyrotropic 4th order moment. Motivated with the previous decompositions,
it obviously has to be something in the form of
\begin{equation} \label{eq:Rdecomp}
  \boldsymbol{r}^\textrm{g} = r_{\parallel\parallel}\bhat\bhat\bhat\bhat + r_{\parallel\perp} \big[ \bhat\bhat (\boldsymbol{I}-\bhat\bhat) \big]^{\textrm{Sym}} +
  r_{\perp\perp} \big[ (\boldsymbol{I}-\bhat\bhat)(\boldsymbol{I}-\bhat\bhat) \big]^{\textrm{Sym}},
\end{equation}  
where we still did not determine how the symmetric operator acts here. Importantly, the symmetric operator ``Sym'' is \emph{not} equivalent to the symmetric
operator ``$S$'' that cycles all the indices around, eq. (\ref{eq:Soper4}). The fluid hierarchy obviously needs two symmetric operators, one unique ``$S$'' that is used to
derive the evolution equation of a given fluid moment $\bX^{(n)}$, and one non-unique ``Sym'' that is used for
the decomposition of that fluid moment. 
The determination of how ``Sym'' acts here is not that obvious. Nevertheless, one can consider in how many ways one can extract the
gyrotropic components from the $r_{ijkl}$. For the $r_{\parallel\parallel}$, one does two double contractions with $(\bhat\bhat)$.
The possible choices are $(\bhat\bhat)_{ij}(\bhat\bhat)_{kl}$;
$(\bhat\bhat)_{ik}(\bhat\bhat)_{jl}$ and $(\bhat\bhat)_{il}(\bhat\bhat)_{jk}$, however, all of these choices are equivalent. 
To obtain the $r_{\parallel\perp}$, we perform double contractions with $(\bhat\bhat)$ and $(\boldsymbol{I}-\bhat\bhat)/2$, where
the last operator contains a function $\delta_{ij}$. How many different delta functions we can obtain from 4 indices $i,j,k,l$?
There are $\binom{4}{2}=\frac{4!}{2!(4-2)!}=6$ different possibilities:
\begin{equation}
  \delta_{ij}; \quad \delta_{ik}; \quad \delta_{il}; \quad \delta_{jk}; \quad \delta_{jl}; \quad \delta_{kl},
\end{equation}
and all of them are non-equivalent.  The symmetric operator acting in the second term therefore has 6 components
\begin{eqnarray}
  \big[ \bhat\bhat (\boldsymbol{I}-\bhat\bhat) \big]^{\textrm{Sym}}_{ijkl} &=& (\boldsymbol{I}-\bhat\bhat)_{ij} \hat{b}_k \hat{b}_l
  + (\boldsymbol{I}-\bhat\bhat)_{ik} \hat{b}_j \hat{b}_l + (\boldsymbol{I}-\bhat\bhat)_{il} \hat{b}_j \hat{b}_k
  + (\boldsymbol{I}-\bhat\bhat)_{jk} \hat{b}_i \hat{b}_l \nn\\
  && \quad + (\boldsymbol{I}-\bhat\bhat)_{jl} \hat{b}_i \hat{b}_k + (\boldsymbol{I}-\bhat\bhat)_{kl} \hat{b}_i \hat{b}_j \nn\\
  &=& \delta_{ij} \hat{b}_k \hat{b}_l + \delta_{ik} \hat{b}_j \hat{b}_l + \delta_{il} \hat{b}_j \hat{b}_k
  + \delta_{jk} \hat{b}_i \hat{b}_l + \delta_{jl} \hat{b}_i \hat{b}_k + \delta_{kl} \hat{b}_i \hat{b}_j
  - 6 \hat{b}_i \hat{b}_j \hat{b}_k \hat{b}_l. \label{eq:Symparperp}
\end{eqnarray}  
Lastly, to obtain the $r_{\perp\perp}$ component, one applies two double contractions with $(\boldsymbol{I}-\bhat\bhat)/2$
that results in combinations of $\delta_{ij}\delta_{kl}$. How many possibilities do we have ?
Obviously, there are 6 possibilities for the first delta function,
which by pairing with the other delta function in a way that indices are not repeated, yields together 6 possibilities
$\delta_{ij}\delta_{kl}; \delta_{ik}\delta_{jl}; \delta_{il}\delta_{jk}; \delta_{jk}\delta_{il}; \delta_{jl}\delta_{ik}; \delta_{kl}\delta_{ij}$.
However, 3 possibilities have an equivalent pair, $\delta_{ij}\delta_{kl} = \delta_{kl}\delta_{ij}$;
$\delta_{ik}\delta_{jl}=\delta_{jl}\delta_{ik}$ and $\delta_{il}\delta_{jk}=\delta_{jk}\delta_{il}$ and there are only 3 non-equivalent
combinations. The symmetric operator
acting on the last term in (\ref{eq:Rdecomp}) can be determined to be
\begin{eqnarray}
  \big[ (\boldsymbol{I}-\bhat\bhat)(\boldsymbol{I}-\bhat\bhat) \big]^{\textrm{Sym}}_{ijkl} &=& \frac{1}{2}\Big[
  (\boldsymbol{I}-\bhat\bhat)_{ij}(\boldsymbol{I}-\bhat\bhat)_{kl}
  + (\boldsymbol{I}-\bhat\bhat)_{ik}(\boldsymbol{I}-\bhat\bhat)_{jl}
  + (\boldsymbol{I}-\bhat\bhat)_{il}(\boldsymbol{I}-\bhat\bhat)_{jk} \Big] \nn\\
  &=& \frac{1}{2}\Big[
    \delta_{ij}\delta_{kl} - \delta_{kl}\hat{b}_i\hat{b}_j -\delta_{ij}\hat{b}_k\hat{b}_l
    + \delta_{ik}\delta_{jl} - \delta_{jl}\hat{b}_i\hat{b}_k -\delta_{ik}\hat{b}_j\hat{b}_l
    + \delta_{il}\delta_{jk} - \delta_{jk}\hat{b}_i\hat{b}_l -\delta_{il}\hat{b}_j\hat{b}_k \nn\\
    && \qquad + 3 \hat{b}_i\hat{b}_j \hat{b}_k\hat{b}_l\Big]. \label{eq:Symperpperp}
\end{eqnarray}
The factor $1/2$ is actually not that obvious, and one needs to verify that the decomposition indeed satisfies (\ref{eq:rmom3}).   
The entire 4th-order gyrotropic moment $\boldsymbol{r}^\textrm{g}$ is decomposed in the index notation according to
\begin{eqnarray}
r_{ijkl}^\textrm{g} &=& r_{\parallel\parallel} \hat{b}_i\hat{b}_j \hat{b}_k\hat{b}_l
        + r_{\parallel\perp} \Big[ \delta_{ij} \hat{b}_k \hat{b}_l + \delta_{ik} \hat{b}_j \hat{b}_l + \delta_{il} \hat{b}_j \hat{b}_k
  + \delta_{jk} \hat{b}_i \hat{b}_l + \delta_{jl} \hat{b}_i \hat{b}_k + \delta_{kl} \hat{b}_i \hat{b}_j
  - 6 \hat{b}_i \hat{b}_j \hat{b}_k \hat{b}_l \Big] \nn\\
        &+& \frac{r_{\perp\perp}}{2} \Big[
          \delta_{ij}\delta_{kl} + \delta_{ik}\delta_{jl} + \delta_{il}\delta_{jk} -\delta_{ij}\hat{b}_k\hat{b}_l
          -\delta_{ik}\hat{b}_j\hat{b}_l-\delta_{il}\hat{b}_j\hat{b}_k - \delta_{jk}\hat{b}_i\hat{b}_l  
     - \delta_{jl}\hat{b}_i\hat{b}_k - \delta_{kl}\hat{b}_i\hat{b}_j
    + 3 \hat{b}_i\hat{b}_j \hat{b}_k\hat{b}_l\Big]. \label{eq:RdecompFull}
\end{eqnarray}
It might be tempting to rearrange this decomposition to a more compact form
\begin{eqnarray}
r_{ijkl}^\textrm{g} &=& \Big(r_{\parallel\parallel}-6r_{\parallel\perp}+\frac{3}{2}r_{\perp\perp}\Big) \hat{b}_i\hat{b}_j \hat{b}_k\hat{b}_l
        + \Big(r_{\parallel\perp}-\frac{r_{\perp\perp}}{2}\Big) \Big[ \delta_{ij} \hat{b}_k \hat{b}_l + \delta_{ik} \hat{b}_j \hat{b}_l + \delta_{il} \hat{b}_j \hat{b}_k
  + \delta_{jk} \hat{b}_i \hat{b}_l + \delta_{jl} \hat{b}_i \hat{b}_k + \delta_{kl} \hat{b}_i \hat{b}_j\Big] \nn\\
        &+& \frac{r_{\perp\perp}}{2} \Big[
          \delta_{ij}\delta_{kl} + \delta_{ik}\delta_{jl} + \delta_{il}\delta_{jk}\Big], \label{eq:RdecompFinal}
\end{eqnarray}
nevertheless, in actual calculations we find the form (\ref{eq:RdecompFull}) to be more useful. 
It is important to verity that the decomposition (\ref{eq:RdecompFull}) really works. A straightforward calculation yields
\begin{eqnarray}
  r_{ijkl}^\textrm{g} \hat{b}_i &=& r_{\parallel\parallel} \hat{b}_j \hat{b}_k\hat{b}_l + r_{\parallel\perp} \Big[ \delta_{jk}\hat{b}_l + \delta_{jl}\hat{b}_k
    +\delta_{kl}\hat{b}_j -3\hat{b}_j \hat{b}_k\hat{b}_l \Big];\\
  r_{ijkl}^\textrm{g} \hat{b}_i \hat{b}_j &=& r_{\parallel\parallel} \hat{b}_k\hat{b}_l + r_{\parallel\perp} \Big[ \delta_{kl} - \hat{b}_k \hat{b}_l \Big];\\
  r_{ijkl}^\textrm{g} \hat{b}_i \hat{b}_j \hat{b}_k &=& r_{\parallel\parallel} \hat{b}_l;\\
  r_{ijkl}^\textrm{g} \hat{b}_i \hat{b}_j \hat{b}_k \hat{b}_l &=& r_{\parallel\parallel},
\end{eqnarray}
which is consistent with (\ref{eq:rmom1}). The $r_{\parallel\perp}$ component calculates
\begin{eqnarray}
  r_{ijkl}^\textrm{g} \hat{b}_i \hat{b}_j \delta_{kl} &=& r_{\parallel\parallel} + 2r_{\parallel\perp};\\
  r_{ijkl}^\textrm{g} \hat{b}_i \hat{b}_j (\delta_{kl}-\hat{b}_k\hat{b}_l)/2 &=& r_{ijkl}^\textrm{g} \hat{b}_i \hat{b}_j \delta_{kl}/2
  - r_{ijkl}^\textrm{g} \hat{b}_i \hat{b}_j \hat{b}_k \hat{b}_l /2
  = r_{\parallel\parallel}/2 + r_{\parallel\perp} - r_{\parallel\parallel}/2 = r_{\parallel\perp},
\end{eqnarray}  
which is consistent with (\ref{eq:rmom2}). The $r_{\perp\perp}$ component calculates
\begin{eqnarray}
  r_{ijkl}^\textrm{g}\delta_{ij} &=& r_{iikl}^\textrm{g} = r_{\parallel\parallel}\hat{b}_k\hat{b}_l + r_{\parallel\perp}(\delta_{kl}+\hat{b}_k\hat{b}_l)
  + 2 r_{\perp\perp} (\delta_{kl}-\hat{b}_k\hat{b}_l);\\
  r_{ijkl}^\textrm{g}\delta_{ij}\delta_{kl} &=& r_{iikk}^\textrm{g} = r_{\parallel\parallel} + 4r_{\parallel\perp} + 4r_{\perp\perp};\\
  r_{ijkl}^\textrm{g}\delta_{ij} \hat{b}_k\hat{b}_l &=& r_{\parallel\parallel}+2r_{\parallel\perp};\\
  r_{ijkl}^\textrm{g}\delta_{ij} (\delta_{kl}-\hat{b}_k\hat{b}_l) &=& 2r_{\parallel\perp} + 4r_{\perp\perp};\\
  r_{ijkl}^\textrm{g}(\delta_{ij}-\hat{b}_i\hat{b}_j)(\delta_{kl}-\hat{b}_k\hat{b}_l)/4 &=&
  r_{ijkl}^\textrm{g}\delta_{ij}(\delta_{kl}-\hat{b}_k\hat{b}_l)/4 - r_{ijkl}^\textrm{g}\hat{b}_i\hat{b}_j(\delta_{kl}-\hat{b}_k\hat{b}_l)/4 \nn\\
  &=& r_{\parallel\perp}/2+r_{\perp\perp} - r_{\parallel\perp}/2  = r_{\perp\perp},
\end{eqnarray}  
which is consistent with (\ref{eq:rmom3}). The decomposition (\ref{eq:RdecompFull}) indeed works.

Now we need to verify expressions (\ref{eq:RgQverif1}), (\ref{eq:RgQverif2}) that were used in the scalar heat flux equations.
The first expression calculates
\begin{eqnarray}
  (\nabla\cdot\br^{\textrm{g}})_{ijk} \hat{b}_i \hat{b}_j \hat{b}_k &=& (\pr_l r^{\textrm{g}}_{ijkl}) \hat{b}_i \hat{b}_j \hat{b}_k =
  \pr_l (r^{\textrm{g}}_{ijkl} \hat{b}_i \hat{b}_j \hat{b}_k ) - r^{\textrm{g}}_{ijkl} \pr_l( \hat{b}_i \hat{b}_j \hat{b}_k) \nn\\
  &=& \pr_l (r_{\parallel\parallel} \hat{b}_l )- r^{\textrm{g}}_{ijkl} \pr_l( \hat{b}_i \hat{b}_j \hat{b}_k)
  = \nabla\cdot(r_{\parallel\parallel}\bhat) - r^{\textrm{g}}_{ijkl} \Big( (\pr_l\hat{b}_i) \hat{b}_j \hat{b}_k + \hat{b}_i (\pr_l\hat{b}_j) \hat{b}_k
  + \hat{b}_i \hat{b}_j (\pr_l \hat{b}_k) \Big); \nn \\
r^{\textrm{g}}_{ijkl} \hat{b}_j \hat{b}_k &=& r_{\parallel\parallel} \hat{b}_i\hat{b}_l + r_{\parallel\perp} \Big[ \delta_{il} - \hat{b}_i \hat{b}_l \Big];\nn\\ 
r^{\textrm{g}}_{ijkl} \hat{b}_j \hat{b}_k \pr_l \hat{b}_i &=& r_{\parallel\perp}\nabla\cdot\bhat;\nn\\
 (\nabla\cdot\br^{\textrm{g}})_{ijk} \hat{b}_i \hat{b}_j \hat{b}_k &=& \nabla\cdot(r_{\parallel\parallel}\bhat) -3r_{\parallel\perp}\nabla\cdot\bhat,
\end{eqnarray}  
and the second expression calculates similarly
\begin{eqnarray}
  (\nabla\cdot\br^{\textrm{g}})_{ijk} \delta_{ij} \hat{b}_k  &=& (\pr_l r_{iikl}^\textrm{g})\hat{b}_k = \pr_l (r_{iikl}^\textrm{g}\hat{b}_k) - r_{iikl}^\textrm{g}\pr_l \hat{b}_k; \nn\\
   r_{iikl}^\textrm{g}\hat{b}_k &=& r_{\parallel\parallel}\hat{b}_l + 2r_{\parallel\perp}\hat{b}_l;\nn\\
   r_{iikl}^\textrm{g}\pr_l\hat{b}_k &=& ( r_{\parallel\perp}  + 2r_{\perp\perp})\nabla\cdot\bhat;\nn\\
   (\nabla\cdot\br^{\textrm{g}})_{ijk} \delta_{ij} \hat{b}_k  &=&
   \nabla\cdot(r_{\parallel\parallel}\bhat) + 2\nabla\cdot(r_{\parallel\perp}\bhat) - ( r_{\parallel\perp}  + 2r_{\perp\perp})\nabla\cdot\bhat; \nn\\
  (\nabla\cdot\br^{\textrm{g}})_{ijk} (\delta_{ij}-\hat{b}_i \hat{b}_j) \hat{b}_k/2
  &=& \nabla\cdot(r_{\parallel\perp}\bhat) + (r_{\parallel\perp}-r_{\perp\perp})\nabla\cdot\bhat,
\end{eqnarray}
which verifies (\ref{eq:RgQverif1}), (\ref{eq:RgQverif2}).

Furthermore, by exploring the $\pr \br/\pr t$ equation (\ref{eq:4thTensorT}), at frequencies that are much smaller than the gyrofrequency,
the gyrotropic part of $\br$ should satisfy
\begin{equation} \label{eq:gyroR}
(\bhat\times\br^{\textrm{g}})^S = 0,
\end{equation}
in the same way that the gyrotropic parts of $\bp$ and $\bq$ satisfied this requirement. By using the gyrotropic (\ref{eq:RdecompFull}), 
it is easy to calculate for example
\begin{eqnarray}
(\bhat\times \br^{\textrm{g}})_{ijkl} = \epsilon_{irs} \hat{b}_r r^{\textrm{g}}_{sjkl} &=& r_{\parallel\perp}\Big( \epsilon_{irj}\hat{b}_r\hat{b}_k\hat{b}_l
+\epsilon_{irk}\hat{b}_r\hat{b}_j\hat{b}_l +\epsilon_{irl}\hat{b}_r\hat{b}_j\hat{b}_k \Big)\nn\\
&+& \frac{r_{\perp\perp}}{2}\Big( \epsilon_{irj}\hat{b}_r \delta_{kl} +\epsilon_{irk}\hat{b}_r \delta_{jl} +\epsilon_{irl}\hat{b}_r \delta_{jk}
-\epsilon_{irj}\hat{b}_r \hat{b}_k\hat{b}_l-\epsilon_{irk}\hat{b}_r \hat{b}_j\hat{b}_l-\epsilon_{irl}\hat{b}_r \hat{b}_j\hat{b}_k \Big),
\end{eqnarray}
and by adding together all the 4 representations of the ``$S$'' operator, one can indeed verify that all the terms cancel, yielding (\ref{eq:gyroR}).
By decomposing the entire 4th order moment to its gyrotropic and non-gyrotropic part
\begin{equation}
\br = \br^{\textrm{g}}+\br^{\textrm{ng}},
\end{equation}
the evolution equation (\ref{eq:4thTensorT}) can therefore be rewritten as
\begin{eqnarray}
  \frac{\pr}{\pr t} \br +\nabla\cdot \big( \bX^{(5)}+\bu\br \big) +\Big[ \br\cdot\nabla\bu
    +\Omega \frac{|\bb|}{B_0}\bhat\times\br^{\textrm{ng}}-\frac{1}{\rho} (\nabla\cdot\bp) \bq \Big]^S =0. \label{eq:4thTensorT2}
\end{eqnarray}

\subsection{Non-gyrotropic $\br^{\textrm{ng}}$}
Basic properties of the non-gyrotropic tensor $\boldsymbol{r}^{\textrm{ng}}$ can be easily determined with a similar procedure as those we used
for the non-gyrotropic pressure $\boldsymbol{\Pi}$ and the non-gyrotropic heat flux $\boldsymbol{q}^{\textrm{ng}}$.   
The decomposition of the entire 4th-order moment is 
\begin{equation}
  \boldsymbol{r} = r_{\parallel\parallel}\bhat\bhat\bhat\bhat + r_{\parallel\perp} \big[ \bhat\bhat (\boldsymbol{I}-\bhat\bhat) \big]^{\textrm{Sym}} +
  r_{\perp\perp} \big[ (\boldsymbol{I}-\bhat\bhat)(\boldsymbol{I}-\bhat\bhat) \big]^{\textrm{Sym}} + \boldsymbol{r}^\textrm{ng},
\end{equation}
and the meaning of the ``Sym'' operators were specified by (\ref{eq:Symparperp}), (\ref{eq:Symperpperp}). By using definitions  
(\ref{eq:rmom1}), (\ref{eq:rmom2}), (\ref{eq:rmom3}) for $r_{\parallel\parallel}$, $r_{\parallel\perp}$, $r_{\perp\perp}$ in the expression above, the full decomposition
reads
\begin{eqnarray}
  \boldsymbol{r} = && \Big[\big(\boldsymbol{r}:\bhat\bhat\big):(\bhat\bhat)\Big]\bhat\bhat\bhat\bhat +
  \Big[\big(\boldsymbol{r}:\bhat\bhat\big):(\boldsymbol{I}-\bhat\bhat)/2\Big] \big[ \bhat\bhat (\boldsymbol{I}-\bhat\bhat) \big]^{\textrm{Sym}} \nn\\
  && + \Big[\big(\boldsymbol{r}:(\boldsymbol{I}-\bhat\bhat)/2\big):(\boldsymbol{I}-\bhat\bhat)/2\Big]
  \big[ (\boldsymbol{I}-\bhat\bhat)(\boldsymbol{I}-\bhat\bhat) \big]^{\textrm{Sym}} + \boldsymbol{r}^\textrm{ng}.
\end{eqnarray}
Now, by applying $:\bhat\bhat$ twice on both sides of the equation yields
$\big(\boldsymbol{r}:\bhat\bhat\big):(\bhat\bhat) = \big(\boldsymbol{r}:\bhat\bhat\big):(\bhat\bhat) + \big(\boldsymbol{r}^\textrm{ng}:\bhat\bhat\big):(\bhat\bhat)$,
implying $\big(\boldsymbol{r}^\textrm{ng}:\bhat\bhat\big):(\bhat\bhat) =0,$ which can be also rewritten as
\begin{equation} \label{eq:labelRng1}
  \bhat\bhat:\boldsymbol{r}^\textrm{ng}:\bhat\bhat =0.
\end{equation}
Similarly, the other two properties are derived by either applying
$:\bhat$ and $:(\boldsymbol{I}-\bhat\bhat)$, or twice $:(\boldsymbol{I}-\bhat\bhat)$, yielding
\begin{equation}
 (\boldsymbol{I}-\bhat\bhat):\boldsymbol{r}^\textrm{ng}:\bhat\bhat =0; \qquad
(\boldsymbol{I}-\bhat\bhat):\boldsymbol{r}^\textrm{ng}:(\boldsymbol{I}-\bhat\bhat) =0,
\end{equation}
which by using the property (\ref{eq:labelRng1}) are further reduced to
\begin{equation}
 \boldsymbol{I}:\boldsymbol{r}^\textrm{ng}:\bhat\bhat =0; \qquad
\boldsymbol{I}:\boldsymbol{r}^\textrm{ng}:\boldsymbol{I} =0.
\end{equation}
The last two properties can be also written as $\textrm{Tr}\boldsymbol{r}^\textrm{ng}:\bhat\bhat =0$, $\textrm{Tr}\textrm{Tr}\boldsymbol{r}^\textrm{ng}=0$.
Finally, writing all 3 properties in the index notation
\begin{equation}
r^\textrm{ng}_{ijkl} \hat{b}_i\hat{b}_j\hat{b}_k\hat{b}_l=0;\qquad r^\textrm{ng}_{iikl} \hat{b}_k\hat{b}_l=0; \qquad r^\textrm{ng}_{iikk}=0.
\end{equation}
One would assume that the non-gyrotropic $\br^{\textrm{ng}}$ can be evaluated by rewriting (\ref{eq:4thTensorT2}) as 
\begin{eqnarray}
\big(\bhat\times \br^{\textrm{ng}}\big)^S = -\frac{B_0}{\Omega|\bb|} \Big[   
  \frac{\pr}{\pr t} \br +\nabla\cdot \big( \bX^{(5)}+\bu\br \big) +\Big( \br\cdot\nabla\bu -\frac{1}{\rho} (\nabla\cdot\bp) \bq \Big)^S \Big], 
\end{eqnarray}
and then expand the r.h.s. similarly to we did for the non-gyrotropic $\boldsymbol{\Pi}$
(and also the non-gyrotropic heat flux vectors in the Appendix \ref{sec:NONGheat}).  
However, the equation does not seem to yield anything useful. Instead, one needs to consider a specific example of a bi-Maxwellian distribution function.

\subsection{Bi-Maxwellian distribution}
Here we consider a special case of bi-Maxwellian distribution function
\begin{equation}
  f_0 = n \left(\frac{m}{2\pi}\right)^{3/2} \frac{1}{T_\parallel^{1/2} T_\perp}\exp
  \left[ -m \frac{(v_\parallel-u_\parallel)^2}{2T_\parallel} -m \frac{|\bV_\perp-\bu_\perp|^2}{2 T_\perp}  \right]. 
\end{equation}
It is useful to use the fluctuating velocity $\bc=\bV-\bu$, where $c^2=c_\parallel^2+c_\perp^2$ and $c_\perp^2 = c_x^2+c_y^2$.
For the brevity of calculations we introduce
\begin{equation}
  \alpha_\parallel\equiv \frac{m}{2T_\parallel}=\frac{1}{v_{\textrm{th}\parallel}^2}; \qquad \alpha_\perp \equiv \frac{m}{2 T_\perp}=\frac{1}{v_{\textrm{th}\perp}^2},
\end{equation}
where here the thermal speeds are meant to be spatially dependent, i.e. they are written with $T$ and not with $T^{(0)}$.
The bi-Maxwellian distribution therefore reads
\begin{equation}
f_0 = n \sqrt{\frac{\alpha_\parallel}{\pi}}\frac{\alpha_\perp}{\pi} e^{-\alpha_\parallel c_\parallel^2} e^{-\alpha_\perp c_\perp^2}.
\end{equation}
It is useful to remind us the following one dimensional integrals
\begin{equation}
  \int_{-\infty}^\infty e^{-\alpha x^2} dx = \sqrt{\frac{\pi}{\alpha}}; \qquad \int_{-\infty}^\infty x^2 e^{-\alpha x^2} dx = \frac{1}{2}\sqrt{\frac{\pi}{\alpha^3}};
  \qquad \int_{-\infty}^\infty x^4 e^{-\alpha x^2} dx = \frac{3}{4}\sqrt{\frac{\pi}{\alpha^5}}, \label{eq:x0_ident}
\end{equation}
and the general integral with $x^n$ where $n$ is an integer reads
\begin{eqnarray} \label{eq:xn_ident}
\int_{-\infty}^\infty x^n e^{-\alpha x^2} dx &=& \frac{(n-1)!!}{2^{n/2}}\sqrt{\frac{\pi}{\alpha^{n+1}}}; \qquad n=0,2,4\dots \quad (n=\textrm{even}); \\
&=& 0; \qquad \qquad \qquad \qquad \;\;\; n=1,3,5\dots \quad (n=\textrm{odd}).
\end{eqnarray}
The double factorial $(n-1)!!=1\cdot 3 \cdot 5 \cdot\cdot\cdot (n-1)$, with $(-1)!!=1$. 
The identity (\ref{eq:xn_ident}) is easily obtained by performing differentiation $\pr/\pr \alpha$ of the first result in (\ref{eq:x0_ident}),
and this technique is useful in the calculation of fluid moments.

To get more familiar with the bi-Maxwellian distribution, it is useful to verify if integrals over $f_0$ indeed yield the expected fluid moments.
Since the macroscopic velocity $\bu$ is independent of $\bV$, changing from variable $\bV$ to $\bc$ just yields $d^3 v=d^3 c$ (similarly to substitution
y=x+constant that yields dy=dx). We use notation $d^3c=dc_\parallel d^2c_\perp$ and $d^2 c_\perp = dc_x dc_y$. Also, because the integrals are evaluated from $-\infty$ to $\infty$,
substitution from $\bV$ to $\bc$ does not change these bounds. Integration in velocity space therefore yields
\begin{eqnarray}
  \int_{-\infty}^\infty e^{-\alpha_\parallel c_\parallel^2} dc_\parallel &=& \sqrt{\frac{\pi}{\alpha_\parallel}};\qquad\quad
  \int_{-\infty}^\infty e^{-\alpha_\perp c_\perp^2} d^2c_\perp = \int e^{-\alpha_\perp (c_x^2+c_y^2 )} dc_x dc_y = \frac{\pi}{\alpha_\perp}.
\end{eqnarray}
For simplicity we integrated in Cartesian coordinates (instead of perhaps more elegant cylindrical coordinates), which has an additional
benefit in that we can stop writing the integral bounds,
since the integrals are always from $-\infty$ to $\infty$.
 By using the bi-Maxwellian $f_0$ it is very easy to verify that 
\begin{equation}
 n = \int f_0 d^3v; \qquad n\bu = \int \bV f_0 d^3 v; \qquad p_\parallel = m \int c_\parallel^2 f_0 d^3 v; \qquad   p_\perp = \frac{m}{2}\int c_\perp^2 f_0 d^3 v.
\end{equation}  
Continuing with the heat flux components  yields
\begin{eqnarray}
q_\parallel = m \int c_\parallel^3 f_0 d^3 v &=&  m n \sqrt{\frac{\alpha_\parallel}{\pi}} \underbrace{\int c_\parallel^3 e^{-\alpha_\parallel c_\parallel^2}d c_\parallel}_{=0}
\frac{\alpha_\perp}{\pi} \int e^{-\alpha_\perp c_\perp^2} d^2 c_\perp =0; \\
q_\perp = \frac{m}{2} \int c_\parallel c_\perp^2 f_0 d^3 v &=& \frac{m}{2} n \sqrt{\frac{\alpha_\parallel}{\pi}} \underbrace{\int c_\parallel e^{-\alpha_\parallel c_\parallel^2}d c_\parallel}_{=0}
\frac{\alpha_\perp}{\pi} \int c_\perp^2 e^{-\alpha_\perp c_\perp^2} d^2 c_\perp =0. \label{eq:Heatflux_important}
\end{eqnarray}
It is important to emphasize, that the zero heat flux values were obtained by assuming some prescribed equilibrium $f_0$, here bi-Maxwellian.
A crucial principle used in constructing fluid models with higher-order moments is that the specific $f_0$ is assumed \emph{only for the last retained moment}.
Indeed, if one closes the hierarchy by prescribing $q_\parallel=0, q_\perp=0$, the CGL fluid model is obtained. However, to go higher in the fluid hierarchy, 
evolution equations for $q_\parallel$ and $q_\perp$ cannot be eliminated, even if the hierarchy will be eventually closed, for example at the
4th-order moment level, by assuming a bi-Maxwellian $f_0$. Or in another words, the zero heat flux values were obtained for a distribution function
that is strictly $f_0$, however, fluctuations/perturbations around $f_0$ are still allowed. 
Therefore, to obtain the next available model after the CGL, the heat flux equations must be retained, and a closure is performed on the 4th-order moment. Only this last moment is
calculated by assuming specific distribution function, that is strictly $f_0$. For a bi-Maxwellian, the gyrotropic 4th-order moments are calculated according to
\begin{eqnarray}
  r_{\parallel\parallel} \equiv m\int c_\parallel^4 f_0 d^3 v &=& m n \underbrace{\sqrt{\frac{\alpha_\parallel}{\pi}} \int c_\parallel^4 e^{-\alpha_\parallel c_\parallel^2}d c_\parallel}_{=3/(4\alpha_\parallel^2)}
  \underbrace{\frac{\alpha_\perp}{\pi} \int e^{-\alpha_\perp c_\perp^2} d^2 c_\perp}_{=1}  = \frac{3}{4}\frac{mn}{\alpha_\parallel^2}
  = 3 \frac{n^2 T_\parallel^2}{nm} = 3\frac{p_\parallel^2}{\rho};\\
  r_{\parallel\perp} \equiv \frac{m}{2}\int c_\parallel^2 c_\perp^2 f_0 d^3 v &=&
  \frac{m}{2} n \underbrace{\sqrt{\frac{\alpha_\parallel}{\pi}} \int c_\parallel^2 e^{-\alpha_\parallel c_\parallel^2}d c_\parallel}_{=1/(2\alpha_\parallel)}
  \underbrace{\frac{\alpha_\perp}{\pi} \int c_\perp^2 e^{-\alpha_\perp c_\perp^2} d^2 c_\perp}_{=1/\alpha_\perp} = \frac{mn}{4\alpha_\parallel \alpha_\perp}
  = \frac{n^2 T_\parallel T_\perp}{mn} = \frac{p_\parallel p_\perp}{\rho}; \\
  r_{\perp\perp} \equiv \frac{m}{4}\int c_\perp^4 f_0 d^3 v &=& \frac{m}{4} n \underbrace{\sqrt{\frac{\alpha_\parallel}{\pi}} \int e^{-\alpha_\parallel c_\parallel^2}d c_\parallel}_{=1}
  \underbrace{\frac{\alpha_\perp}{\pi} \int c_\perp^4 e^{-\alpha_\perp c_\perp^2} d^2 c_\perp}_{=2/\alpha_\perp^2}  = \frac{mn}{2\alpha_\perp^2} = 2 \frac{n^2 T_\perp^2}{m n}
  = 2\frac{p_\perp^2}{\rho}.
\end{eqnarray}
Therefore, for bi-Maxwellian, the following closure can be constructed
\begin{equation}
  r_{\parallel\parallel} = \frac{3p_\parallel^2}{\rho}; \quad
  r_{\parallel\perp} = \frac{p_\parallel p_\perp}{\rho}; \quad
  r_{\perp\perp} = \frac{2 p_\perp^2}{\rho},
\end{equation}
and this closure is known as the ``normal'' closure, a name suggested by \cite{ChustBelmont2006}.
\subsection{Bi-Maxwellian non-gyrotropic $\br^{\textrm{ng}}$ contributions}
Let's consider a specific case of a bi-Maxwellian distribution function. 
Then, an expansion procedure analogous to the one developed by \cite{Grad1949} for rarefied gases can be used, where the distribution function is expanded in a series of
Hermite polynomials. By keeping only first few terms, one can express the 4th-order moment through 2nd-order (pressure) moments.
The procedure is written down in \cite{Oraevskii1968}, and the entire 4th-order moment can be decomposed in the following way:
see equation (2.31) in \cite{Oraevskii1968} (where a small typo on the l.h.s. is present, where instead of $q_{\alpha\beta\gamma\epsilon}$,
there should be $q_{\alpha\beta\gamma\delta}$), equation (24) in \cite{Goswami2005}, equation (9) in \cite{PassotSulem2007}
\begin{equation}
\rho r_{ijkl} = p_{ij}^\textrm{g} p_{kl}^\textrm{g} + p_{ik}^\textrm{g} p_{jl}^\textrm{g} + p_{il}^\textrm{g} p_{jk}^\textrm{g} + \rho r^\textrm{ng}_{ijkl},
\end{equation}
where the non-gyrotropic contributions read
\begin{equation} \label{eq:RngDEF}
  \rho r^\textrm{ng}_{ijkl} = p_{ij}^\textrm{g} \Pi_{kl} + p_{ik}^\textrm{g} \Pi_{jl} + p_{il}^\textrm{g} \Pi_{jk}
  + \Pi_{ij} p_{kl}^\textrm{g} + \Pi_{ik} p_{jl}^\textrm{g} + \Pi_{il} p_{jk}^\textrm{g},
\end{equation}
and where terms $\Pi_{ij}\Pi_{lk}+\Pi_{ik}\Pi_{jl}+\Pi_{il}\Pi_{jk}$ were neglected.
By using $p_{ij}^\textrm{g}=p_\parallel \hat{b}_i \hat{b}_j + p_\perp (\delta_{ij}-\hat{b}_i \hat{b}_j)$, it can be indeed shown by direct straightforward
calculation that the gyrotropic part
\begin{eqnarray}
\rho r_{ijkl}^\textrm{g} &=& p_{ij}^\textrm{g} p_{kl}^\textrm{g} + p_{ik}^\textrm{g} p_{jl}^\textrm{g} + p_{il}^\textrm{g} p_{jk}^\textrm{g} =
  3p_\parallel^2 \hat{b}_i\hat{b}_j \hat{b}_k\hat{b}_l \nn\\
&&   + p_\parallel p_\perp \Big[ \delta_{ij} \hat{b}_k \hat{b}_l + \delta_{ik} \hat{b}_j \hat{b}_l + \delta_{il} \hat{b}_j \hat{b}_k
  + \delta_{jk} \hat{b}_i \hat{b}_l + \delta_{jl} \hat{b}_i \hat{b}_k + \delta_{kl} \hat{b}_i \hat{b}_j
  - 6 \hat{b}_i \hat{b}_j \hat{b}_k \hat{b}_l \Big] \nn\\
&&  + p_\perp^2 \Big[
          \delta_{ij}\delta_{kl} + \delta_{ik}\delta_{jl} + \delta_{il}\delta_{jk} -\delta_{ij}\hat{b}_k\hat{b}_l
          -\delta_{ik}\hat{b}_j\hat{b}_l-\delta_{il}\hat{b}_j\hat{b}_k - \delta_{jk}\hat{b}_i\hat{b}_l  
     - \delta_{jl}\hat{b}_i\hat{b}_k - \delta_{kl}\hat{b}_i\hat{b}_j
    + 3 \hat{b}_i\hat{b}_j \hat{b}_k\hat{b}_l\Big].
\end{eqnarray}
This agrees with the decomposition (\ref{eq:RdecompFull}) valid for a general distribution function, after one specifies that for
a bi-Maxwellian distribution $r_{\parallel\parallel} = \frac{3p_\parallel^2}{\rho}$, $r_{\parallel\perp} = \frac{p_\parallel p_\perp}{\rho}$, and 
$r_{\perp\perp} = \frac{2 p_\perp^2}{\rho}$ (last term in (\ref{eq:RdecompFull}) has there $\frac{r_{\perp\perp}}{2}$).

Now it is possible to calculate the non-gyrotropic contributions $\br^{\textrm{ng}}$, by using equation (\ref{eq:RngDEF}). It is useful to pre-calculate the trace
of (\ref{eq:RngDEF}) that reads
\begin{equation}
\rho r_{iikl} = (p_\parallel+6p_\perp) \Pi_{kl} + 2(p_\parallel-p_\perp)\hat{b}_i \big( \hat{b}_k\Pi_{il} +\hat{b}_l \Pi_{ik}\big).
\end{equation}  
The two terms that enter the parallel and perpendicular heat flux equations (\ref{eq:ParHF_r}), (\ref{eq:PerpHF_r}) calculate
\begin{eqnarray}
  \bhat\cdot(\nabla\cdot \br^{\textrm{ng}}):\bhat\bhat &=& \frac{3 p_\parallel}{\rho}(\nabla\cdot\boldsymbol{\Pi})\cdot\bhat
  +\frac{3p_\parallel}{\rho}(\bhat\cdot\nabla\boldsymbol{\Pi}):\bhat\bhat +3\Big(\nabla\frac{p_\parallel}{\rho}\Big)\cdot\boldsymbol{\Pi}\cdot\bhat;\\
  \boldsymbol{I}:(\nabla\cdot \br^{\textrm{ng}})\cdot\bhat &=& \Big( \nabla\frac{3p_\parallel+4p_\perp}{\rho}\Big)\cdot\boldsymbol{\Pi}\cdot\bhat
  +\frac{(3p_\parallel+4p_\perp)}{\rho}(\nabla\cdot\boldsymbol{\Pi})\cdot\bhat \nn\\
  && +\frac{2}{\rho}(p_\parallel-p_\perp)\Big[\boldsymbol{\Pi}:(\nabla\bhat)+(\bhat\cdot\nabla\bhat)\cdot\boldsymbol{\Pi}\cdot\bhat
    +\bhat\cdot(\nabla\boldsymbol{\Pi}):\bhat\bhat \Big].
\end{eqnarray}
At this stage, we need to further simplify, and we keep only those terms that contribute at the linear level, yielding
\begin{eqnarray}
  \bhat\cdot(\nabla\cdot \br^{\textrm{ng}}):\bhat\bhat &=& \frac{3 p_\parallel}{\rho}(\nabla\cdot\boldsymbol{\Pi})\cdot\bhat; \label{eq:Qcontrib1}\\
  \boldsymbol{I}:(\nabla\cdot \br^{\textrm{ng}})\cdot\bhat &=& \frac{(3p_\parallel+4p_\perp)}{\rho}(\nabla\cdot\boldsymbol{\Pi})\cdot\bhat,
\end{eqnarray}
so the term entering the $q_\perp$ equation (\ref{eq:PerpHF_r}) reads
\begin{equation} \label{eq:Qcontrib2}
\frac{1}{2}(\boldsymbol{I}-\bhat\bhat):(\nabla\cdot \br^{\textrm{ng}})\cdot\bhat = \frac{2p_\perp}{\rho}(\nabla\cdot\boldsymbol{\Pi})\cdot\bhat.
\end{equation}  
Note that $(\bhat\cdot\nabla\boldsymbol{\Pi}):\bhat\bhat=-\boldsymbol{\Pi}:\big(\bhat(\bhat\cdot\nabla)\bhat\big)^S$, and the term does not contribute at the linear level. 
Of course, many nonlinear terms were neglected, and technically, the only fully consistent procedure is to explicitly write those terms at the linear level, in the form
\begin{eqnarray}
  \bhat\cdot(\nabla\cdot \br^{\textrm{ng}}):\bhat\bhat &=& \frac{3 p_\parallel^{(0)}}{\rho_0}(\nabla\cdot\boldsymbol{\Pi})\cdot\bhat_0
  = \frac{3 p_\parallel^{(0)}}{\rho_0} \big(\pr_x \Pi_{xz}+\pr_y \Pi_{yz} \big); \\
  \frac{1}{2}(\boldsymbol{I}-\bhat\bhat):(\nabla\cdot \br^{\textrm{ng}})\cdot\bhat &=& \frac{2p_\perp^{(0)}}{\rho_0}(\nabla\cdot\boldsymbol{\Pi})\cdot\bhat_0
  = \frac{2p_\perp^{(0)}}{\rho_0}\big(\pr_x \Pi_{xz}+\pr_y \Pi_{yz} \big).
\end{eqnarray} 
The results show that if the $\boldsymbol{\Pi}$ contributions are kept in the heat flux equations, the $\br^{\textrm{ng}}$
contributions can not just be straightforwardly neglected, since the $\boldsymbol{\Pi}$ and $\br^{\textrm{ng}}$ contributions partially cancel out.

\clearpage
\section{Bi-Maxwellian fluid model, 2nd-order CGL (CGL2)} \label{section:CGL2}
In the previous section we have seen that for a Bi-Maxwellian distribution function, one can close the fluid hierarchy by prescribing the ``normal'' closure
\begin{equation}
  r_{\parallel\parallel} = \frac{3p_\parallel^2}{\rho}; \quad
  r_{\parallel\perp} = \frac{p_\parallel p_\perp}{\rho}; \quad
  r_{\perp\perp} = \frac{2 p_\perp^2}{\rho}.\nn
\end{equation}
It is useful to summarize the fully nonlinear model that we want to consider here. We consider only proton species, and we make the electrons massless and cold.
We simplify the pressure equations, by neglecting the FLR stress forces (which enter only nonlinearly), and we keep only those \emph{non-gyrotropic} contributions
in the pressure and heat flux equations that have a non-zero contributions. The nonlinear model reads
\begin{eqnarray}
&& \frac{\pr \rho}{\pr t} + \nabla\cdot (\rho \bu )=0; \label{eq:RhoSimpleL}\\
&& \frac{\pr \bu}{\pr t} +\bu \cdot\nabla \bu +\frac{1}{\rho}\nabla\cdot \Big[ p_\parallel\bhat\bhat + p_\perp (\boldsymbol{I}-\bhat\bhat) +\boldsymbol{\Pi} \Big]
-\frac{1}{4\pi\rho} (\nabla\times\bb)\times\bb=0; \label{eq:Mom_XXX}\\
&& \frac{\pr \bb}{\pr t} = \nabla\times(\bu\times\bb)-\frac{1}{\Omega_p}\frac{B_0}{4\pi}\nabla\times\Big[ \frac{1}{\rho}(\nabla\times\bb)\times\bb\Big];\\
&& \frac{\pr p_\parallel}{\pr t} +\bu\cdot\nabla p_\parallel + p_\parallel \nabla\cdot \bu +2p_\parallel \bhat\cdot\nabla\bu\cdot\bhat
  +\nabla\cdot(q_\parallel\bhat) - 2q_\perp \nabla\cdot\bhat +\nabla\cdot\boldsymbol{S}^\parallel_\perp = 0; \label{eq:PparSimpleL}\\
&&  \frac{\pr p_\perp}{\pr t} + \bu\cdot\nabla p_\perp +  2 p_\perp\nabla\cdot\bu -p_\perp \bhat\cdot\nabla\bu\cdot\bhat
  +\nabla\cdot(q_\perp\bhat) + q_\perp \nabla\cdot\bhat +\nabla\cdot\boldsymbol{S}^\perp_\perp=0; \label{eq:PperpSimpleL}\\
&&  \frac{\pr q_\parallel}{\pr t}+ \bu\cdot\nabla q_\parallel + q_\parallel\nabla\cdot \bu 
  + 3p_\parallel \bhat\cdot\nabla\left(\frac{p_\parallel}{\rho}\right) + 3 q_\parallel \bhat\cdot\nabla\bu\cdot\bhat =0; \label{eq:ParHF_siXXX}\\
&&  \frac{\pr q_\perp}{\pr t} + \bu\cdot\nabla q_\perp +2q_\perp\nabla\cdot\bu +p_\parallel\bhat\cdot\nabla\left( \frac{p_\perp}{\rho} \right)
  +\frac{p_\perp}{\rho} (p_\parallel-p_\perp)\nabla\cdot\bhat + \frac{p_\perp}{\rho} (\nabla\cdot\boldsymbol{\Pi})\cdot\bhat_0 =0. \label{eq:PerpHF_siXXX} 
\end{eqnarray}
To develop a vocabulary, let's first consider the above fluid model with all the non-gyrotropic fluctuations neglected, i.e. when all the terms with
$\boldsymbol{\Pi}$, $\boldsymbol{S}^\parallel_\perp$, $\boldsymbol{S}^\perp_\perp$ are neglected. Additionally, let's neglect the Hall-term in the
induction equation. Such fluid model is non-dispersive, and represents generalization of the non-dispersive CGL model. 
Even though such fluid model was not explicitly considered by Chew, Goldberger and Low, the pressure equations that include the
gyrotropic heat flux contributions $q_\parallel, q_\perp$, were indeed given by \cite{Chew1956}. As discussed previously,
the authors just used a different notation with $q_n=q_\parallel-3q_\perp$ and $q_s=q_\perp$, but their pressure equations contain the
gyrotropic heat flux. The authors did not consider evolution equations for the gyrotropic heat fluxes, eq. (\ref{eq:ParHF_siXXX}), (\ref{eq:PerpHF_siXXX}).
Nevertheless, the name CGL is so well recognized by the community, i.e. the name CGL is essentially recognized as \emph{the collisionless MHD}, 
that we suggest to call the non-dispersive fluid model that used the above gyrotropic heat flux equations, as the ``the 2nd-order CGL'', abbreviated as ``CGL2''.
The name CGL2 has a nice advantage, that the abbreviations for many models considered previously, can be easily and naturally generalized.

The CGL2 fluid model does not contain any dispersive effects and it is length-scale invariant, similarly to the CGL and MHD models. 
If the Hall term in the induction equation is considered, yields the ``Hall-CGL2'' model. Considering also the first-order FLR corrections to the pressure
tensor (FLR1) or the second-order corrections (FLR2), yields the ``Hall-CGL2-FLR1'' and the ``Hall-CGL2-FLR2'' fluid models.
Finally, considering the non-gyrotropic heat flux fluctuations $\boldsymbol{S}^\parallel_\perp$, $\boldsymbol{S}^\perp_\perp$, yields the Hall-CGL2-FLR3 model.
Fluid models with the Hall term neglected, can be easily abbreviated as the ``CGL2-FLR1'', ``CGL2-FLR2'' etc. Even though, if the dispersive effects by the FLR corrections
are considered, there is really no reason to neglect the simple Hall-term, and such fluid models are discouraged to use. 
Obviously, the name ``CGL2'' is extremely beneficial and very natural for the classification of fluid models, and we will use this name henceforth. 

It is important to correctly normalize the heat flux, and the normalization is according to $\widetilde{q}_{\parallel,\perp}=q_{\parallel,\perp} / (p_\parallel^{(0)} V_A)$,
so both the parallel and perpendicular heat flux are normalized the same way (similarly, both $p_\parallel$ and $p_\perp$ are normalized with respect to $p_\parallel^{(0)}$).
By dropping the tilde, the scalar pressure equations (\ref{eq:PparSimpleL}), (\ref{eq:PperpSimpleL}) remain unchanged, and the normalized
nonlinear heat flux equations read
\begin{eqnarray}
&&  \frac{\pr q_\parallel}{\pr t}+ \bu\cdot\nabla q_\parallel + q_\parallel\nabla\cdot \bu
  + 3\frac{\bpar}{2} p_\parallel \bhat\cdot\nabla\left(\frac{p_\parallel}{\rho}\right) + 3 q_\parallel \bhat\cdot\nabla\bu\cdot\bhat =0;  \label{eq:CGL2_rit6}\\
&&  \frac{\pr q_\perp}{\pr t} + \bu\cdot\nabla q_\perp +2q_\perp\nabla\cdot\bu +\frac{\bpar}{2}p_\parallel\bhat\cdot\nabla\left( \frac{p_\perp}{\rho} \right)
  +\frac{\bpar}{2}\frac{p_\perp}{\rho} (p_\parallel-p_\perp)\nabla\cdot\bhat+\frac{\bpar}{2}\frac{p_\perp}{\rho}(\nabla\cdot\boldsymbol{\Pi})\cdot\bhat_0=0.  \label{eq:CGL2_rit7}
\end{eqnarray}
Now we need to linearize the system, where one prescribes $q_{\parallel}^{(0)}=0$ and $q_{\perp}^{(0)}=0$. 
Once normalized, linearized, transformed to Fourier space and written in the x-z plane, the model reads (dropping tilde everywhere)
\begin{eqnarray}
&& -\omega \rho + k_\perp u_x + k_\parallel u_z =0; \label{eq:CGL2_rit10}\\
&& -\omega u_x +\frac{\bpar}{2}k_\perp p_\perp -v_{A\parallel}^2 k_\parallel B_x  +k_\perp B_z +\frac{\bpar}{2}\big( k_\perp \Pi_{xx} +k_\parallel \Pi_{xz} \big)=0;\\
&& -\omega u_y - v_{A\parallel}^2 k_\parallel B_y +\frac{\bpar}{2}\big(k_\perp \Pi_{xy}+k_\parallel \Pi_{yz}  \big)=0;\\
&& -\omega u_z +\frac{\bpar}{2}k_\parallel p_\parallel +\frac{\bpar}{2}(1-a_p)k_\perp B_x + \frac{\bpar}{2}k_\perp \Pi_{xz}=0;\\
&& -\omega B_x -k_\parallel u_x -i k_\parallel^2 B_y=0;\\
&& -\omega B_y - k_\parallel u_y -ik_\parallel k_\perp B_z +ik_\parallel^2 B_x=0;\\
&& -\omega B_z + k_\perp u_x+ik_\parallel k_\perp B_y=0;\\
&& -\omega p_\parallel + k_\perp u_x +3 k_\parallel u_z + k_\parallel q_\parallel +k_\perp S^\parallel_\perp= 0;\\
&&  -\omega p_\perp  + 2a_p k_\perp u_x +a_p k_\parallel u_z + k_\parallel q_\perp +k_\perp S^\perp_\perp=0;\label{eq:CGL2_rit11}\\
&& -\omega q_\parallel +\frac{3}{2}\bpar k_\parallel (p_\parallel-\rho) =0;\\
&& -\omega q_\perp +\frac{\bpar}{2}k_\parallel (p_\perp -a_p\rho) + \frac{\bpar}{2}a_p(1-a_p)k_\perp B_x +\frac{\bpar}{2}a_p k_\perp \Pi_{xz}=0. \label{eq:CGL2_rit12}
\end{eqnarray}
As previously, the $v_{A\parallel}^2=1+\frac{\bpar}{2}(a_p-1)$.
\subsection{CGL2 dispersion relation}
Neglecting all the dispersive effects, by prescribing $\boldsymbol{\Pi}=0$, $\boldsymbol{S}^\parallel_\perp=0$, $\boldsymbol{S}^\perp_\perp=0$, and
eliminating the Hall term in the induction equation (i.e. terms proportional to $ik^2$ in the above induction equation), yields the CGL2 model.
By exploring the system, it is obvious that the Alfv\'en mode separates from the system in the $u_y,B_y$ components.
The dispersion relation of the Alfv\'en mode in the (non-dispersive) CGL2 model is therefore the same as in the CGL model, and reads
(the normalization tildes are dropped)
\begin{equation} \label{eq:CGL2_Alf}
\omega = \pm k_\parallel v_{A\parallel} = \pm k\cos(\theta) \sqrt{1+\frac{\bpar}{2}(a_p-1)}. 
\end{equation}
The ``hard'' threshold of the oblique firehose instability is therefore the same as in the CGL model, and correctly reproduced.
The entire dispersion relation for the CGL2 fluid model can be written in the following form
\begin{eqnarray}
&&  \Big(\omega^2-v_{A\parallel}^2 k_\parallel^2\Big)\Big(\omega^8-A_6\omega^6+A_4\omega^4-A_2\omega^2+A_0\Big)=0; \label{eq:8thP}\\
  &&  A_6 = k_\parallel^2\Big( v_{A\parallel}^2 + \frac{7}{2}\bpar\Big) +k_\perp^2 (1+a_p\beta);\\
  &&  A_4 = k_\parallel^2\bpar \Big\{ k_\parallel^2\Big( \frac{7}{2}v_{A\parallel}^2 +\frac{9}{4}\bpar \Big)
  + k_\perp^2\frac{7}{2}\Big( 1+a_p\bpar-\frac{1}{7}a_p^2\bpar \Big) \Big\};\\
  &&  A_2 = k_\parallel^4\bpar^2 \Big\{ k_\parallel^2\Big( \frac{9}{4} v_{A\parallel}^2+\frac{3}{8}\bpar \Big)
  +k_\perp^2\frac{9}{4} \Big( 1+a_p\bpar-\frac{5}{9}a_p^2\bpar \Big) \Big\};\\
  &&  A_0 = \frac{3}{8}k_\parallel^6 \bpar^3 \Big\{ k_\parallel^2 v_{A\parallel}^2 + k_\perp^2 \Big( 1+a_p\bpar-a_p^2\bpar\Big) \Big\}.
\end{eqnarray}
The CGL2 fluid model therefore contains 5 forward and 5 backward propagating modes. The oblique Alfv\'en mode is separated as stated above, and
is incompressible. The remaining 4 forward and 4 backward waves
are generally compressible and coupled together through the 8th-order polynomial in $\omega$ (4th-order polynomial in $\omega^2$).
Let's ignore for a moment the distinction between the forward and backward modes, since these will always be symmetric in $\omega$.
The CGL2 model contains 5 waves/modes. The oblique Alfv\'en mode is separated, and the remaining 4 compressible modes are in general strongly coupled.
In contrast to the simpler CGL (and MHD) model, that contains only 2 compressible modes (slow and fast), we therefore
have 2 ``new'' modes, that do not have an analogy in the usual CGL \& MHD descriptions. There is no standardized vocabulary on how the modes should be
called, perhaps an expression of ``higher-order modes'' is the most appropriate.
For the highly oblique propagation, considering the mirror instability, sometimes an expression of ``mirror modes'' is used.
Nevertheless, since none of the 4 compressible modes match the slow and fast CGL dispersion relations, the distinction between modes becomes
blurry. The reader has to get used to the fact that by considering higher-order fluid moments, we perhaps
came closer to the kinetic theory, which admits an infinite number of modes that are difficult to classify, unless a
specific situation is considered. The difference between CGL2 model and kinetic description is,
that unless a firehose or mirror instability threshold is reached, the fluid modes are un-damped.   

For strictly parallel propagation $(k_\perp=0)$, in addition to the parallel Alfv\'en mode (\ref{eq:CGL2_Alf}) $\omega = \pm k_\parallel v_{A\parallel}$
that was already separated, the solution of the 8th-order polynomial contains another Alfv\'en mode $\omega = \pm k_\parallel v_{A\parallel}$, which was of
course expected and is equivalent to the CGL model, since the components $u_x,u_y,B_x,B_y$ de-couple for parallel propagation.
The remaining solutions are
\begin{eqnarray}
  \omega &=& \pm k_\parallel \sqrt{\bpar\Big(\frac{3}{2}+\sqrt{\frac{3}{2}} \Big)}=\pm 1.65 k_\parallel \sqrt{\bpar}; \label{eq:CGL2_rit1}\\
  \omega &=& \pm k_\parallel \sqrt{\bpar\Big(\frac{3}{2}-\sqrt{\frac{3}{2}} \Big)}=\pm 0.52 k_\parallel \sqrt{\bpar}; \label{eq:CGL2_rit3}\\
  \omega &=& \pm k_\parallel \sqrt{\frac{\bpar}{2}} = \pm 0.71 k_\parallel \sqrt{\bpar}. \label{eq:CGL2_rit2}
\end{eqnarray}
Solutions (\ref{eq:CGL2_rit1}) and  (\ref{eq:CGL2_rit3}) can be obtained by considering pure 1D geometry (where $p_\perp$ and $q_\perp$ disappear since
the kinetic velocity $v_\perp$ disappear) and considering only fluctuations in quantities $\rho,u_z,p_\parallel,q_\parallel$. The $q_\parallel$ heat flux fluctuations
in the CGL2 fluid model are therefore responsible for ``splitting'' of the CGL ion-acoustic mode $\omega=\pm k_\parallel \sqrt{3\bpar/2}$ to two modes
(\ref{eq:CGL2_rit1}) and  (\ref{eq:CGL2_rit3}).

Considering solutions for strictly perpendicular propagation $(k_\parallel=0)$, the only non-zero mode is the fast mode 
\begin{equation} \label{eq:CGL2_fast}
 \omega = \pm  k_\perp \sqrt{1+a_p\bpar},
\end{equation}
and the dispersion relation is unchanged from the CGL model (for perpendicular propagation the $q_\parallel,q_\perp$ contributions in the equations
for $p_\parallel,p_\perp$ naturally vanish and the CGL2 model is equivalent to the CGL model).

\subsection{Mirror instability}
A very interesting direction of propagation for this fluid model is the highly oblique limit, $k_\perp\gg k_\parallel$,
since the CGL2 fluid model contains the correct mirror instability threshold. The correct mirror threshold can be already
seen in the quantity $A_0$, where in contrast to the usual CGL model, the factor $1/6$ disappeared in the last term $\sim a_p^2\bpar$.

In the highly oblique limit, the fast mode (\ref{eq:CGL2_fast}) can be separated from the coupled 8-th degree polynomial
(actually 4-th degree polynomial in $\omega^2$) in the following way. Importantly, one can not prescribe a completely perpendicular
propagation $k_\parallel=0$, and a highly oblique limit $k_\perp\gg k_\parallel$, or $\kpar\to 0$, has to be considered. The polynomial coefficients can be approximated as
\begin{eqnarray}
  &&  A_6 = k_\perp^2 (1+a_p\beta);\\
  &&  A_4 = k_\parallel^2\bpar  k_\perp^2\frac{7}{2}\Big( 1+a_p\bpar-\frac{1}{7}a_p^2\bpar \Big) ;\\
  &&  A_2 = k_\parallel^4\bpar^2 k_\perp^2\frac{9}{4} \Big( 1+a_p\bpar-\frac{5}{9}a_p^2\bpar \Big);\\
  &&  A_0 = \frac{3}{8}k_\parallel^6 \bpar^3 k_\perp^2 \Big( 1+a_p\bpar-a_p^2\bpar\Big).
\end{eqnarray}
An alternative approximation is to use $k_\perp\to k$ in the above expressions.
The coefficient $A_6$ does not contain any $\kpar$ and is very large compared to
$A_4,A_2,A_0$ that all contain $\kpar$. The solutions will inevitably be one fast mode, and 3 very slow modes.
In this specific case, the fast mode can be quickly separated from the 8th order polynomial in (\ref{eq:8thP}) with a neat trick
\begin{eqnarray}
&& \Big(\omega^8-A_6\omega^6+A_4\omega^4-A_2\omega^2+A_0\Big)=0;\nn \\
&& \Big(\omega^2- A_6\Big)\Big(-A_6\omega^6+A_4\omega^4-A_2\omega^2+A_0\Big)=0. \label{eq:CGL2_trick}
\end{eqnarray}
The trick can be verified by multiplying both brackets term by term, dividing by $-A_6$, and observing that $(A_4,A_2,A_0)/A_6$ is negligible.  
The fast mode is always obtained correctly in this way (when the $A_6$ is much larger than all the other coefficients),
however, the dispersion relation for the 3 slow modes is correct only in this specific case when the system is non-dispersive, and all 3 slow modes have
constant phase speeds. A proper limit and expansion should be always checked. 

Nevertheless, in the highly oblique limit, the fast mode $\omega = \pm k_\perp \sqrt{1+a_p\bpar}$ is separated, and the rest of the dispersion 
is of 6th order in $\omega$ and contains 3 ``slow'' modes. By canceling out the $k_\perp^2$ that is common in all the remaining coefficients (or $k^2)$, an obvious substitution 
is offered that reads $\bar{\omega} = \omega/(\kpar\sqrt{\bpar})$, and that yields a dispersion equation of 3rd order in $\bar{\omega}^2$ in the form
\begin{equation}
 -\Big(1+a_p\bpar\Big)(\bar{\omega}^2)^3+\frac{7}{2}\Big( 1+a_p\bpar-\frac{1}{7}a_p^2\bpar \Big)(\bar{\omega}^2)^2 -\frac{9}{4} \Big( 1+a_p\bpar-\frac{5}{9}a_p^2\bpar \Big)(\bar{\omega}^2)
+\frac{3}{8}\Big( 1+a_p\bpar-a_p^2\bpar\Big)=0.
\end{equation}
Now, at exactly the mirror threshold $a_p^2\bpar = 1+a_p\bpar$, the last term disappears and we have one solution $\bar{\omega}^2 = 0$, 
and two solutions
\begin{equation} \label{eq:mirrorSolut}
  \bar{\omega}^2 = (3\pm \sqrt{5})/2,
\end{equation}
that are both positive. Slightly below or beyond the mirror threshold, we can prescribe
$a_p^2\bpar = (1+a_p\bpar)(1+\epsilon)$, where $\epsilon$ is a small parameter, meaning that
for $\epsilon<0$, the system is slightly \emph{below} the expected mirror threshold, i.e. stable.
For $\epsilon>0$, the system is slightly \emph{beyond} the expected mirror threshold, i.e. unstable.
The polynomial transforms to
\begin{equation} \label{eq:mirrorPica}
(\bar{\omega}^2)^3-(3-\frac{\epsilon}{2})(\bar{\omega}^2)^2+(1-\frac{5}{4}\epsilon)(\bar{\omega}^2)+\frac{3}{8}\epsilon=0.
\end{equation}
It is possible to show that this polynomial has a discriminant\footnote{Discriminant of a cubic polynomial
$ax^3+bx^2+cx+d=0$ is calculated according to $\triangle = 18abcd-4b^3d+b^2c^2-4ac^3-27a^2d^2.$} $\triangle>0$, implying the
existence of 3 distinct real roots. Now, when $\epsilon$ is small, the two solutions (\ref{eq:mirrorSolut}) will remain almost the same and change only
slightly, importantly, the solutions will remain positive. Solutions of a cubic polynomial $(x-x_1)(x-x_2)(x-x_3)=0$ form
the last coefficient in that polynomial, $-x_1 x_2 x_3$,   
here equal to $\frac{3}{8}\epsilon$, and here $x_1$ and $x_2$ are positive. Therefore, $\epsilon<0$ (slightly below the threshold) implies $x_3>0$;
and $\epsilon>0$ (slightly beyond the threshold) implies $x_3<0$. The negative solution (in $(\bar{\omega}^2)$) represents the mirror instability, which 
finishes the analytic proof that the (highly oblique) mirror threshold in CGL2 model is indeed $a_p^2\bpar = 1+a_p\bpar$.
The CGL2 fluid model therefore contains the same mirror instability threshold as is found in kinetic theory.

The conclusion can be double checked by numerically exploring solutions of (\ref{eq:mirrorPica})
for several values of $\epsilon$, and it is indeed possible to conclude that for $\epsilon <0$ (slightly below the threshold)
the signs of solutions are $\bar{\omega}^2=+,+,+$; and that for $\epsilon>0$ (slightly beyond the mirror threshold), the signs of solutions are $\bar{\omega}^2=+,+,-$. 
The solution with the minus sign represents the mirror instability.
It is noted that the Landau fluid description allows for the construction of a fluid model with quasi-static heat fluxes, that also correctly reproduces 
the ``hard'' mirror instability threshold \citep{SulemPassot2012}.
Considering Landau fluid models, the mirror instability was numerically investigated in detail for example by \cite{PassotSulem2007,PSH2012,SulemPassot2015}.
As a suggestion for possible future work, it would be beneficial to numerically calculate ``Hellinger's'' contours  (see subsection \ref{HellingerFigures})
with the prescribed maximum growth rate
for the mirror instability. If one uses Landau fluid models with sufficiently precise FLR corrections, the results should match the
kinetic contours very precisely.

\subsection{Hall-CGL2 dispersion relation}
Considering the Hall term, the dispersion relation of the Hall-CGL2 fluid model reads
\begin{eqnarray}
  &&  \Big(\omega^2-v_{A\parallel}^2 k_\parallel^2\Big)\Big(\omega^8-A_6\omega^6+A_4\omega^4-A_2\omega^2+A_0\Big)=\nn\\
  && \quad k^2 k_\parallel^2 \omega^2\Big[\omega^6 -\omega^4 \bpar\Big( \frac{7}{2} k_\parallel^2+a_p k_\perp^2 \Big)
    +\omega^2\bpar^2 k_\parallel^2\Big( \frac{9}{4}k_\parallel^2+3a_p k_\perp^2\Big)
    -k_\parallel^4 \bpar^3 \Big( \frac{3}{8}k_\parallel^2 + a_p k_\perp^2\Big)  \Big], \label{eq:HallCGL2_disp}
\end{eqnarray}
where the l.h.s. represents the CGL2 dispersion relation (\ref{eq:8thP}) and the r.h.s. is the Hall-term contribution.
For strictly parallel propagation, the solutions are the usual Hall-CGL whistler and ion-cyclotron waves
$\omega=\pm \kpar^2/2+\kpar\sqrt{v_{A\parallel}^2+\kpar^2/4}$, $\omega=\pm \kpar^2/2-\kpar\sqrt{v_{A\parallel}^2+\kpar^2/4}$, that are accompanied
by the CGL2 solutions (\ref{eq:CGL2_rit1}), (\ref{eq:CGL2_rit2}) and (\ref{eq:CGL2_rit3}). For strictly perpendicular propagation, the solution
is (\ref{eq:CGL2_fast}) since the Hall term vanishes. The dispersion relation of the Hall-CGL-FLR1 model was already quite large to write down, and
we do not provide the final dispersion relation for the Hall-CGL2-FLR1 model, instead we recommend to work with the full system of linearized equations
and solve it numerically or to use analytic software such as Maple.

\subsection{``Static'' closure - generalized isothermal closure} \label{sec:StaticClosure}
There exists even simpler fluid closure that recovers the correct mirror threshold, called ``static'' or
``quasi-static'' closure \citep{Constantinescu2002,ChustBelmont2006,PassotRubanSulem2006}.
By using heat flux equations (\ref{eq:ParHF_siXXX}), (\ref{eq:PerpHF_siXXX}) and considering quasi-static
regime with $\pr q_\parallel/\pr t=0$, $\pr q_\perp/\pr t=0$, $\bu=0$ yields
\begin{equation}
  \bhat\cdot\nabla\left(\frac{p_\parallel}{\rho}\right)=0; \qquad
  p_\parallel\bhat\cdot\nabla\left( \frac{p_\perp}{\rho} \right)
  +\frac{p_\perp}{\rho} (p_\parallel-p_\perp)\nabla\cdot\bhat = 0,
\end{equation}
where the FLR pressure tensor $\boldsymbol{\Pi}$ was neglected in the second equation. It is useful to define $\pr_\parallel \equiv \bhat\cdot\nabla$ which represents
gradient along the magnetic field lines, and to use the identity 
\begin{equation}
\nabla\cdot\bhat = -\frac{1}{|\bb|} \bhat\cdot\nabla |\bb|,
\end{equation}
further yielding
\begin{equation} \label{eq:Tisothermal}
\pr_\parallel T_\parallel = 0; \qquad \frac{\pr_\parallel T_\perp}{T_\perp} = \Big( 1-\frac{T_\perp}{T_\parallel} \Big) \frac{\pr_\parallel |\bb|}{|\bb|}. 
\end{equation}  
Eq. (\ref{eq:Tisothermal}) describes evolution of parallel and perpendicular temperature fluctuations along the magnetic field lines, and it can be
verified that the solution is
\begin{equation} \label{eq:TisoX}
T_\parallel = T_\parallel^{(0)}; \qquad T_\perp = T_\perp^{(0)} \frac{\frac{|\bb|}{B_0}}{1-a_p +a_p\frac{|\bb|}{B_0}},
\end{equation}
where $a_p=T_\perp^{(0)}/T_\parallel^{(0)}$. Solution (\ref{eq:TisoX}) is referred to as the ``static'' closure and it is
for example equivalent to eq. 19, 20 of \cite{PassotRubanSulem2006}, and eq. 2 of \cite{Constantinescu2002}.  
For isotropic mean temperatures $a_p=1$, eq. (\ref{eq:TisoX}) yields $T_\perp = T_\perp^{(0)}=T_\parallel^{(0)}$ and both parallel and perpendicular
temperatures are isothermal. A similar result is obtained for general $a_p$ with $|\bb|=B_0$, which yields $T_\perp = T_\perp^{(0)}$.
The closure (\ref{eq:TisoX}) can be therefore viewed as a generalization of isothermal closure in the presence of temperature
anisotropy and variations of magnetic field strength. The closure is used directly in the momentum equation (\ref{eq:Mom_XXX}), and the time-dependent pressure
equations (\ref{eq:PparSimpleL}), (\ref{eq:PperpSimpleL}) are disregarded. Therefore, even though the closure was derived by considering 
4th-order moments, the resulting fluid model is actually simpler than CGL2 and CGL models. 
As discussed for example by \cite{PassotRubanSulem2006}, the closure prescribes limitation
for the minimal value of $|\bb|/B_0>1-1/a_p$ that is required to prevent temperature singularity. Or in other words, by separating
$|\bb|=B_0+B^{(1)}$ the requirement reads $B^{(1)}>-1/a_p$, meaning that the fluctuating $B^{(1)}$ can not be too negative. Mirror instability is
often associated with nonlinear structures in the form of magnetic holes and humps, and the requirement implies that the magnetic holes
can not be too deep. To easily analyze dispersion relations, it is useful to write the closure at the linear
level, in the following form
\begin{equation} \label{eq:ClosureX}
  p_\parallel = n T_\parallel^{(0)}; \qquad p_\perp \overset{\textrm{\tiny lin}}{=} n T_\perp^{(0)} + p_\perp^{(0)} (1-a_p) \frac{B_z^{(1)}}{B_0}.
\end{equation}
The result can be obtained by linearizing (\ref{eq:TisoX}) or (\ref{eq:Tisothermal}), and at the
linear level $\pr_i |\bb|\overset{\textrm{\tiny lin}}{=}\pr_i B_z$.
The normalized closure reads (tilde are dropped) $p_\parallel=n$ and $p_\perp=a_p n+a_p(1-a_p)B_z$.
Considering the simplest non-dispersive model (by neglecting the Hall term and $\boldsymbol{\Pi}$), the dispersion relation of this
fluid model reads
\begin{eqnarray} \label{eq:IdontKnow}
&&  \Big(\omega^2-k_\parallel^2 v_{A\parallel}^2\Big)\Big(\omega^4-A_2\omega^2+A_0\Big)=0;\\
  && A_2 = k_\parallel^2 \Big( v_{A\parallel}^2+\frac{\bpar}{2}\Big) +k_\perp^2 \Big( 1+a_p\bpar-a_p^2\frac{\bpar}{2}\Big);\nn \\
  && A_0 = k_\parallel^2\frac{\bpar}{2} \Big[ k_\parallel^2 v_{A\parallel}^2 + k_\perp^2 \Big( 1+a_p\bpar-a_p^2\bpar \Big)\Big]. \nn
\end{eqnarray}
In comparison with the dispersion relation (\ref{eq:SFbigmatrix}) of the adiabatic CGL model, the ``static'' closure   
eliminates the erroneous $1/6$ factor in the $A_0$ coefficient and yields the correct mirror threshold. The erroneous $1/6$ factor in the
CGL model can be therefore interpreted as a result of inadequacy of adiabatic closures in the very slow-dynamics context, such as
the mirror instability. For completeness, 
solutions for parallel propagation ($k_\perp=0$) are $\omega=\pm v_{A\parallel}k_\parallel$ and $\omega=\pm \sqrt{\frac{\bpar}{2}}k_\parallel$.
For perpendicular propagation ($k_\parallel=0$) the solution is $\omega=\pm \sqrt{1+a_p\bpar-a_p^2\frac{\bpar}{2}} k_\perp$. Obviously, somewhere
beyond the mirror threshold, the perpendicular fast mode experiences unphysical instability, which can be interpreted as a result of inadequacy of
``static'' (isothermal) closures in the fast-dynamics context.

To clearly demonstrate that variations of magnetic field in the ``static'' closure (\ref{eq:ClosureX}) are indeed crucial in recovering
the mirror instability, let's neglect them and quickly consider strictly isothermal closure $p_\parallel = n T_\parallel^{(0)}$; $p_\perp=n T_\perp^{(0)}$, which
yields the following dispersion relation
\begin{eqnarray}
&&  \Big(\omega^2-k_\parallel^2 v_{A\parallel}^2\Big)\Big(\omega^4-A_2\omega^2+A_0\Big)=0; \label{eq:IsothX}\\
  && A_2 = k_\parallel^2 \Big( v_{A\parallel}^2+\frac{\bpar}{2}\Big) +k_\perp^2 \Big( 1+a_p\frac{\bpar}{2}\Big);\nn \\
  && A_0 = k_\parallel^2\frac{\bpar}{2} \Big[ k_\parallel^2 v_{A\parallel}^2 + k_\perp^2 \Big( 1+a_p\frac{\bpar}{2}-a_p^2\frac{\bpar}{2} \Big)\Big]. \nn
\end{eqnarray}
As can be seen from the $A_0$ coefficient, this model does not recover the mirror instability. 
For parallel propagation the solutions are the same as for the ``static'' closure, nevertheless, the perpendicular fast
mode is now always stable with $\omega=\pm \sqrt{1+a_p\frac{\bpar}{2}}k_\perp$. As a double check, after prescribing $a_p=1$, the dispersion relation 
(\ref{eq:IsothX}) is equal to (\ref{eq:IdontKnow}), since the $B_z$ contributions in the ``static'' closure disappear. 
Also, we considered models with polytropic
indices $\gamma_\parallel,\gamma_\perp$ in Section \ref{sec:EmpirModels}, and the dispersion relation (\ref{eq:IsothX}) is
consistent with (\ref{eq:GenPolypPica}), after prescribing $\gamma_\parallel=1$, $\gamma_\perp=1$.

\clearpage
\section{Bi-kappa fluid model (BiKappa)} \label{section:BiKappa}
\subsection{Bi-kappa distribution function}
Often in the solar wind and in the space plasma physics, distribution functions with elongated high-energy ``suprathermal'' tails are observed, that
cannot be modeled with the bi-Maxwellian distribution. A slightly more general distribution function is the bi-kappa distribution,
that contains one free parameter $\kappa$, see for example \cite{SummersThorne1992,PierrardLazar2010,LivadiotisMcComas2013} and references therein.
 The $\kappa$ parameter can be different in the parallel and transverse directions (see e.g. \cite{Basu2009,DosSantos2015}). Here we consider only
one free $\kappa$ parameter. The distribution has the following form
\begin{equation}
  f_0 = n \frac{\Gamma(\kappa+1)}{\pi^{3/2} \kappa^{3/2} \Gamma(\kappa-\frac{1}{2})}\frac{1}{\theta_\parallel \theta_\perp^2}
    \bigg[ 1+\frac{(v_\parallel-u_\parallel)^2}{\kappa\theta_\parallel^2} + \frac{|\bV_\perp-\bu_\perp|^2}{\kappa\theta_\perp^2} \bigg]^{-\kappa-1}, \label{eq:F_biKappa}
\end{equation}
where the generalized thermal speeds
\begin{equation} \label{eq:biKappa-th}
\theta_\parallel = \sqrt{1-\frac{3}{2\kappa}} \sqrt{\frac{2T_\parallel}{m}}; \qquad \theta_\perp = \sqrt{1-\frac{3}{2\kappa}} \sqrt{\frac{2T_\perp}{m}},
\end{equation}
and $\Gamma$ is the usual gamma function. In the limit $\kappa\rightarrow \infty$, the generalized thermal speeds are the usual thermal speeds
and the distribution function is bi-Maxwellian, since
\begin{equation}
\lim_{x \to \infty} (1+\frac{a}{x})^{-x} = e^{-a}; \qquad  \lim_{x \to \infty} (1+\frac{a}{x})^{-x-1} = e^{-a}.
\end{equation}  
Why the generalized thermal speeds are defined this way will become clear from
later calculations of moments, and the choice of the power law index $-(\kappa+1)$, instead of perhaps more logical $-\kappa$, is attributed to a purely historical reasons.
Calculating 2nd-order (pressure) integrals with this distribution requires for convergence $\kappa>3/2$, and this value is also the minimum value
for the generalized thermal speeds (\ref{eq:biKappa-th}) to have real values. Sometimes, in the definition of the isotropic kappa distribution,
a value of $\Gamma(3/2)=\sqrt{\pi}/2$ is used and $\pi^{3/2}=2\pi\sqrt{\pi}/2=2\pi\Gamma(3/2)$.
Again, for the brevity of calculations, it is useful to introduce $\alpha_\parallel=1/(\kappa\theta_\parallel^2)$,
$\alpha_\perp = 1/(\kappa\theta_\perp^2)$, and rewrite the bi-kappa distribution (\ref{eq:F_biKappa}) to the following form
\begin{equation} \label{eq:biKappa}
  f_0 = n \frac{\Gamma(\kappa+1)}{\Gamma(\kappa-\frac{1}{2})}
   \sqrt{\frac{\alpha_\parallel}{\pi}} \frac{\alpha_\perp}{\pi}  \bigg[ 1+\alpha_\parallel c_\parallel^2 + \alpha_\perp c_\perp^2 \bigg]^{-\kappa-1}.
\end{equation}
The important integral is
\begin{equation} \label{eq:biInt}
\int_{-\infty}^\infty (1+x^2)^{-a}dx = \sqrt{\pi}\frac{\Gamma(a-\frac{1}{2})}{\Gamma(a)}, \quad a>\frac{1}{2}.
\end{equation}  
For the critical value of $a=1/2$, the integral $\int 1/\sqrt{1+x^2}dx = arcsinh(x)$, and since $\lim_{x\to\pm \infty} arcsinh(x) =\pm \infty$, the
integral diverges. For an isotropic kappa distribution $f(c^2)$, the 3-dimensional integrals can be conveniently evaluated by transformation to spherical
co-ordinates
\begin{equation}
\int_{-\infty}^\infty f(c^2) d^3 c = 4\pi \int_{0}^\infty c^2 f(c^2) dc. 
\end{equation}
In this case, the important integral is
\begin{equation}
\int_{0}^\infty x^2(1+x^2)^{-a}dx = \frac{\sqrt{\pi}}{4}\frac{\Gamma(a-\frac{3}{2})}{\Gamma(a)}; \quad a>\frac{3}{2}.
\end{equation}
and by a direct substitution
\begin{eqnarray}
  \int_{0}^\infty x^2(1+\alpha x^2)^{-(\kappa+1)} dx = \frac{\sqrt{\pi}}{4\alpha^{3/2}}
  \frac{\Gamma(\kappa-\frac{1}{2})}{\Gamma(\kappa+1)}; \quad \kappa>\frac{1}{2}.
\end{eqnarray}
For an isotropic kappa distribution therefore
\begin{equation}
\int_{-\infty}^\infty (1+\alpha c^2)^{-(\kappa+1)} d^3c = 4\pi \int_{0}^\infty c^2(1+\alpha c^2)^{-(\kappa+1)} dc = \left(\frac{\pi}{\alpha}\right)^{3/2}
  \frac{\Gamma(\kappa-\frac{1}{2})}{\Gamma(\kappa+1)},
\end{equation}  
which explains the normalization factors. For an anisotropic bi-kappa distributions $f(c_\parallel^2,c_\perp^2)$, the integration is slightly more complicated
because in contrast to a bi-Maxwellian distribution, the integration cannot be fully separated to two independent integrations over
parallel and perpendicular velocity components, and have to be performed successively. It is possible to use a cylindrical coordinate system
and integrate
\begin{equation}
\int_{-\infty}^\infty f(c_\parallel^2,c_\perp^2) d^3 c = 2\pi \int_{c_\perp=0}^\infty \int_{c_\parallel=-\infty}^\infty f(c_\parallel^2,c_\perp^2) c_\perp dc_\parallel d c_\perp. 
\end{equation}
However, here we keep the old fashioned Cartesian coordinates, and at the expense of a slightly
bit more algebra (i.e. one more integration), we will integrate over $d^3 c = dc_\parallel d^2 c_\perp = dc_\parallel dc_x dc_y$. A small added benefit
is that we do not have to follow if the integrals are $\int_{-\infty}^\infty$ or $\int_0^\infty$, and we can drop writing the boundaries.   

The most important integral is a slight generalization of (\ref{eq:biInt}) in the form
\begin{equation}
\int_{-\infty}^\infty (1+b+\alpha x^2)^{-a}dx = \sqrt{\frac{\pi}{\alpha}}\frac{\Gamma(a-\frac{1}{2})}{\Gamma(a)} (1+b)^{-(a-\frac{1}{2})}; \quad a>\frac{1}{2},
\end{equation} 
where for simplicity $b,\alpha$ are assumed to be positive constants. So one can calculate
\begin{equation}
  \int_{-\infty}^\infty (1+\alpha_x x^2 + \alpha_y y^2 + \alpha_z z^2 )^{-a}dx = \sqrt{\frac{\pi}{\alpha_x}}\frac{\Gamma(a-\frac{1}{2})}{\Gamma(a)}
  (1+\alpha_y y^2 + \alpha_z z^2 )^{-(a-\frac{1}{2})},
\end{equation}
where importantly, the ``power law'' changed from $-a$ to $-(a-1/2)$. Performing a successive 3-dimensional integral therefore yields
\begin{eqnarray}
&&  \int_{-\infty}^\infty (1+\alpha_x x^2 + \alpha_y y^2 + \alpha_z z^2 )^{-a}dxdydz = \sqrt{\frac{\pi}{\alpha_x}}\frac{\Gamma(a-\frac{1}{2})}{\Gamma(a)}
  \int (1+\alpha_y y^2 + \alpha_z z^2 )^{-(a-\frac{1}{2})} dy dz \nn\\
&&  = \sqrt{\frac{\pi}{\alpha_x}}\sqrt{\frac{\pi}{\alpha_y}}\frac{\Gamma(a-\frac{1}{2})}{\Gamma(a)} \frac{\Gamma(a-1)}{\Gamma(a-\frac{1}{2})}
  \int (1+\alpha_z z^2 )^{-(a-1)} dz
  =  \sqrt{\frac{\pi}{\alpha_x}}\sqrt{\frac{\pi}{\alpha_y}}\sqrt{\frac{\pi}{\alpha_z}}
  \frac{\Gamma(a-\frac{1}{2})}{\Gamma(a)} \frac{\Gamma(a-1)}{\Gamma(a-\frac{1}{2})}
  \frac{\Gamma(a-\frac{3}{2})}{\Gamma(a-1)} \nn\\
&&  = \sqrt{\frac{\pi}{\alpha_x}}\sqrt{\frac{\pi}{\alpha_y}}\sqrt{\frac{\pi}{\alpha_z}} \frac{\Gamma(a-\frac{3}{2})}{\Gamma(a)}; \quad a>\frac{3}{2},
\end{eqnarray}
and integrating the bi-kappa distribution over the velocity space
\begin{eqnarray}
  \int_{-\infty}^\infty (1+\alpha_\parallel c_\parallel^2 + \alpha_\perp c_\perp^2)^{-(\kappa+1)} dc_\parallel d^2 c_\perp =
  \sqrt{\frac{\pi}{\alpha_\parallel}}\frac{\pi}{\alpha_\perp}\frac{\Gamma(\kappa-\frac{1}{2})}{\Gamma(\kappa+1)}; \quad\kappa>\frac{1}{2},
\end{eqnarray}
which verifies that the normalization constants in the bi-kappa definition (\ref{eq:biKappa}) are indeed correct, and that $\int f_0 d^3 c = n$. 
For later calculations, it is useful to write down two partial integrals when the integration is done over $dc_\parallel$ and $d^2c_\perp$
that read
\begin{eqnarray}
  \int_{-\infty}^\infty (1+\alpha_\parallel c_\parallel^2)^{-a} dc_\parallel &=& \sqrt{\frac{\pi}{\alpha_\parallel}} \frac{\Gamma(a-\frac{1}{2})}{\Gamma(a)};\\
  \int_{-\infty}^\infty (1+\alpha_\parallel c_\parallel^2 + \alpha_\perp c_\perp^2)^{-(\kappa+1)} dc_\parallel
  &=& \sqrt{\frac{\pi}{\alpha_\parallel}} \frac{\Gamma(\kappa+\frac{1}{2})}{\Gamma(\kappa+1)}(1+\alpha_\perp c_\perp^2)^{-(\kappa+\frac{1}{2})};\\
  \int f_0 dc_\parallel &=& n \frac{\Gamma(\kappa+\frac{1}{2})}{\Gamma(\kappa-\frac{1}{2})} \frac{\alpha_\perp}{\pi}(1+\alpha_\perp c_\perp^2)^{-(\kappa+\frac{1}{2})}, \label{eq:CparInt}
\end{eqnarray}
and
\begin{eqnarray}
  \int_{-\infty}^\infty (1+ \alpha_\perp c_\perp^2)^{-a} d^2c_\perp &=& \frac{\pi}{\alpha_\perp} \frac{\Gamma(a-1)}{\Gamma(a)};\\
  \int (1+\alpha_\parallel c_\parallel^2 + \alpha_\perp c_\perp^2)^{-(\kappa+1)} d^2c_\perp
  &=& \frac{\pi}{\alpha_\perp} \frac{\Gamma(\kappa)}{\Gamma(\kappa+1)}(1+\alpha_\parallel c_\parallel^2)^{-\kappa};\\
  \int f_0 d^2 c_\perp &=& n \frac{\Gamma(\kappa)}{\Gamma(\kappa-\frac{1}{2})}\sqrt{\frac{\alpha_\parallel}{\pi}}(1+\alpha_\parallel c_\parallel^2)^{-\kappa}. \label{eq:CperpInt}
\end{eqnarray}  
Continuing with the calculation of velocity moments,  it is easy to show that $\int \bc f_0 d^3 c = 0$, which verifies $n\bu = \int \bV f_0 d^3v$.
Continuing with the higher order moments, it is possible to calculate those from table integrals (a great help for double-checking is an analytic
software like Maple or Mathematica) or similarly to a bi-Maxwellian distribution use a trick with the differentiation with respect to $\alpha$ as
\begin{equation}
  \int_{-\infty}^\infty x^2 (1+\alpha x^2)^{-(\kappa+1)}dx = -\frac{1}{\kappa}\frac{\pr}{\pr\alpha} \int_{-\infty}^\infty (1+\alpha x^2)^{-\kappa} dx
  = -\frac{1}{\kappa} \sqrt{\pi} \frac{\Gamma(\kappa-\frac{1}{2})}{\Gamma(\kappa)}\frac{\pr}{\pr\alpha} \alpha^{-1/2}
  = \frac{\sqrt{\pi}}{2\alpha^{3/2}}\frac{\Gamma(\kappa-\frac{1}{2})}{\Gamma(\kappa+1)},
\end{equation}
where we have also used a property of Gamma function $x\Gamma(x)=\Gamma(x+1)$, and the last integral requires $\kappa>\frac{1}{2}$.
A slightly more general integral is
\begin{equation}
  \int_{-\infty}^\infty x^2 (1+b+\alpha x^2)^{-a}dx = \frac{\sqrt{\pi}}{2\alpha^{3/2}}\frac{\Gamma(a-\frac{3}{2})}{\Gamma(a)} (1+b)^{-(a-\frac{3}{2})},
\end{equation}
which can be used to evaluate the following integrals
\begin{eqnarray}
  \int_{-\infty}^\infty c_\parallel^2 (1+\alpha_\parallel c_\parallel^2)^{-a}dc_\parallel &=& \frac{\sqrt{\pi}}{2\alpha_\parallel^{3/2}}\frac{\Gamma(a-\frac{3}{2})}{\Gamma(a)};\\
  \int_{-\infty}^\infty c_x^2 (1+\alpha_\perp c_\perp^2)^{-a}d^2c_\perp &=& \int_{-\infty}^\infty c_x^2 (1+\alpha_\perp c_x^2+\alpha_\perp c_y^2)^{-a}dc_x dc_y
  = \frac{\sqrt{\pi}}{2\alpha_\perp^{3/2}} \frac{\Gamma(a-\frac{3}{2})}{\Gamma(a)}\int_{-\infty}^\infty (1+\alpha_\perp c_y^2)^{-(a-\frac{3}{2})}dc_y \nn\\
  &=& \frac{\sqrt{\pi}}{2\alpha_\perp^{3/2}} \frac{\Gamma(a-\frac{3}{2})}{\Gamma(a)} \sqrt{\frac{\pi}{\alpha_\perp}}
  \frac{\Gamma(a-2)}{\Gamma(a-\frac{3}{2})} = \frac{\pi}{2\alpha_\perp^2}\frac{\Gamma(a-2)}{\Gamma(a)};\\
   \int_{-\infty}^\infty c_\perp^2 (1+\alpha_\perp c_\perp^2)^{-a}d^2c_\perp &=& \frac{\pi}{\alpha_\perp^2}\frac{\Gamma(a-2)}{\Gamma(a)}.
\end{eqnarray}
Since we like to double check everything, the above integral can be also calculated as
\begin{eqnarray}
  \int_{-\infty}^\infty c_\perp^2 (1+\alpha_\perp c_\perp^2)^{-a}d^2c_\perp &=& -\frac{1}{a-1}\frac{\pr}{\pr \alpha_\perp} \int_{-\infty}^\infty (1+\alpha_\perp c_\perp^2)^{-(a-1)}d^2c_\perp
  = -\frac{1}{a-1}\frac{\Gamma(a-2)}{\Gamma(a-1)}\frac{\pr}{\pr \alpha_\perp} \frac{\pi}{\alpha_\perp} \nn\\
  &=& \frac{\pi}{\alpha_\perp^2} \frac{\Gamma(a-2)}{\Gamma(a)}.
\end{eqnarray}  
For the calculation of the parallel pressure, we first integrate over $d^2c_\perp$ so we can use result (\ref{eq:CperpInt}) and then over $dc_\parallel$
\begin{eqnarray}
  p_\parallel \equiv m \int c_\parallel^2 f_0 d^2 c_\perp dc_\parallel
  &=& m n \frac{\Gamma(\kappa)}{\Gamma(\kappa-\frac{1}{2})}\sqrt{\frac{\alpha_\parallel}{\pi}}\int c_\parallel^2(1+\alpha_\parallel c_\parallel^2)^{-\kappa} dc_\parallel 
  = m n \frac{\Gamma(\kappa)}{\Gamma(\kappa-\frac{1}{2})}\sqrt{\frac{\alpha_\parallel}{\pi}} \frac{\sqrt{\pi}}{2\alpha_\parallel^{3/2}}
  \frac{\Gamma(\kappa-\frac{3}{2})}{\Gamma(\kappa)}\nn\\
  &=& mn \frac{\Gamma(\kappa-\frac{3}{2})}{\Gamma(\kappa-\frac{1}{2})} \frac{1}{2\alpha_\parallel}
  = mn \frac{\Gamma(\kappa-\frac{3}{2})}{\Gamma(\kappa-\frac{1}{2})} \frac{1}{2} (\kappa-\frac{3}{2})\frac{2 T_\parallel}{m} = n T_\parallel,
\end{eqnarray}
where we have used $(\kappa-\frac{3}{2})\Gamma(\kappa-\frac{3}{2})=\Gamma(\kappa-\frac{1}{2})$.
Calculating the perpendicular pressure, we here first integrate over $dc_\parallel$ so we can use result (\ref{eq:CparInt}) and then over $d^2 c_\perp$
\begin{eqnarray}
  p_\perp \equiv \frac{m}{2} \int c_\perp^2 f_0 d c_\parallel d^2 c_\perp
  &=& \frac{m}{2} n \frac{\Gamma(\kappa+\frac{1}{2})}{\Gamma(\kappa-\frac{1}{2})} \frac{\alpha_\perp}{\pi} \int c_\perp^2 (1+\alpha_\perp c_\perp^2)^{-(\kappa+\frac{1}{2})} d^2 c_\perp
  = \frac{m}{2} n \frac{\Gamma(\kappa+\frac{1}{2})}{\Gamma(\kappa-\frac{1}{2})} \frac{\alpha_\perp}{\pi} \frac{\pi}{\alpha_\perp^2}
  \frac{\Gamma(\kappa-\frac{3}{2})}{\Gamma(\kappa+\frac{1}{2})} \nn\\
  &=& mn \frac{\Gamma(\kappa-\frac{3}{2})}{\Gamma(\kappa-\frac{1}{2})} \frac{1}{2\alpha_\perp}
  = mn \frac{\Gamma(\kappa-\frac{3}{2})}{\Gamma(\kappa-\frac{1}{2})} \frac{1}{2} (\kappa-\frac{3}{2})\frac{2 T_\perp}{m} = n T_\perp.
\end{eqnarray}
The last two calculations clarify the reasoning behind the definition of the bi-kappa generalized thermal speeds $\theta_\parallel$, $\theta_\perp$
(here rewritten with the notation $\alpha_\parallel$ and $\alpha_\perp$), where one starts with a bi-kappa distribution with unspecified
$\alpha_\parallel$ and $\alpha_\perp$, and by calculating the parallel and perpendicular pressure integrals, yields the required forms
$\alpha_\parallel^{-1}=(\kappa-\frac{3}{2})\frac{2T_\parallel}{m}$; $\alpha_\perp^{-1}=(\kappa-\frac{3}{2})\frac{2T_\perp}{m}$. 
Then the split to $\alpha_\parallel^{-1}=\kappa\theta_\parallel^2$; $\alpha_\perp^{-1}=\kappa\theta_\perp^2$ is dictated by the requirement that for
$\kappa\to\infty$ the generalized thermal speeds converge to the usual thermal speeds, and which also yields that in this limit 
the distribution is bi-Maxwellian.

The bi-kappa distribution was specified with perhaps a bit obscure power law $-(\kappa+1)$ instead of more logical $-\kappa$, which is attributed
to purely historical reasons. If one wants to define an anisotropic power-law distribution, a first guess would be
\begin{equation}
\widetilde{f}_0 \sim n \bigg[ 1+\alpha_\parallel c_\parallel^2 + \alpha_\perp c_\perp^2 \bigg]^{-\kappa},
\end{equation}
which then yields normalization constants
\begin{equation}
  \widetilde{f}_0 = n \frac{\Gamma(\kappa)}{\Gamma(\kappa-\frac{3}{2})}\sqrt{\frac{\alpha_\parallel}{\pi}}\frac{\alpha_\perp}{\pi}
  \bigg[ 1+\alpha_\parallel c_\parallel^2 + \alpha_\perp c_\perp^2 \bigg]^{-\kappa}.
\end{equation}
Then, by calculating the parallel and perpendicular pressures yields the requirement
$\alpha_{\parallel,\perp}^{-1}=(\kappa-\frac{5}{2})\frac{2T_{\parallel,\perp}}{m}$. By further writing $\alpha_{\parallel,\perp}^{-1} = \kappa \theta_{\parallel,\perp}^2$
yields a distribution function
\begin{equation}
  \widetilde{f}_0 = n \frac{\Gamma(\kappa)}{\pi^{3/2}\kappa^{3/2}\Gamma(\kappa-\frac{3}{2})} \frac{1}{\theta_\parallel \theta_\perp^2}
  \bigg[ 1+\frac{c_\parallel^2}{\kappa\theta_\parallel^2} + \frac{c_\perp^2}{\kappa\theta_\perp^2} \bigg]^{-\kappa}, \label{eq:F_biKappa_mod}
\end{equation}
with generalized thermal speeds
\begin{equation}
  \theta_{\parallel}=\sqrt{1-\frac{5}{2\kappa}}\sqrt{\frac{2T_{\parallel}}{m}}; \qquad
  \theta_{\perp}=\sqrt{1-\frac{5}{2\kappa}}\sqrt{\frac{2T_{\perp}}{m}}. 
\end{equation}
The convergence of the pressure integrals requires $\kappa>\frac{5}{2}$, which is also the limiting case for the thermal speeds to be real.
In the limit $\kappa\to\infty$ the generalized thermal speeds converge to the usual thermal speeds
and the distribution converges to the bi-Maxwellian.
To double-check that the distribution (\ref{eq:F_biKappa_mod}) is really correct, one can of course substitute $\kappa\to\kappa+1$,
which fully recovers the original bi-kappa distribution (\ref{eq:F_biKappa}), since
$\kappa \theta^2 \rightarrow (\kappa+1)(1-\frac{5}{2(\kappa+1)})\frac{2T}{m}=\kappa(1-\frac{3}{2\kappa})\frac{2T}{m}$.
From now on, lets continue calculations with the original bi-kappa distribution (\ref{eq:F_biKappa}), that uses the $-(\kappa+1)$ power-law index.

\subsection{Bi-kappa fluid closure}
Now that we are familiar with the bi-Kappa distribution, we are ready to calculate higher-order moments.
The heat fluxes are zero and easy to calculate, since both $q_\parallel,q_\perp$ are anti-symmetric in $c_\parallel$, and by a direct calculation
\begin{equation}
  \int_{-\infty}^\infty c_\parallel^3 (1+\alpha_\parallel c_\parallel^2 +\alpha_\perp c_\perp^2)^{-a} dc_\parallel d^2 c_\perp
  = -\frac{1}{a-1} \frac{\pr}{\pr \alpha_\parallel} \underbrace{\int_{-\infty}^\infty c_\parallel (1+\alpha_\parallel c_\parallel^2 +\alpha_\perp c_\perp^2)^{-(a-1)} dc_\parallel}_{=0} d^2 c_\perp =0,
\end{equation}
implying
\begin{equation}
q_\parallel \equiv m\int c_\parallel^3 f_0 d^3 c = 0; \qquad q_\perp \equiv \frac{m}{2}\int c_\parallel c_\perp^2 f_0 d^3 c = 0.
\end{equation}
As discussed previously for the bi-Maxwellian distribution, the heat fluxes must be kept in the general form, since we want to consider
closure performed at the 4-th order moment level. 
To calculate the 4-th order fluid moments, we will need the following integrals
\begin{eqnarray}
  \int_{-\infty}^\infty c_\parallel^4 (1+\alpha_\parallel c_\parallel^2)^{-a}d c_\parallel &=&
  \frac{3}{4}\frac{\sqrt{\pi}}{\alpha_\parallel^{5/2}}\frac{\Gamma(a-\frac{5}{2})}{\Gamma(a)}; \qquad a>\frac{5}{2};\\
  \int_{-\infty}^\infty c_\parallel^2 (1+\alpha_\parallel c_\parallel^2+\alpha_\perp c_\perp^2)^{-a}d c_\parallel &=&
  \frac{\sqrt{\pi}}{2\alpha_\parallel^{3/2}} \frac{\Gamma(a-\frac{3}{2})}{\Gamma(a)} (1+\alpha_\perp c_\perp^2)^{-(a-\frac{3}{2})}; \qquad a>\frac{3}{2};\\
  \int_{-\infty}^\infty c_\parallel^2 c_\perp^2 (1+\alpha_\parallel c_\parallel^2+\alpha_\perp c_\perp^2)^{-a}d c_\parallel d^2 c_\perp &=&
  \frac{\sqrt{\pi}}{2\alpha_\parallel^{3/2}}\frac{\pi}{\alpha_\perp^2}\frac{\Gamma(a-\frac{7}{2})}{\Gamma(a)}; \qquad a>\frac{7}{2};\\
\int_{-\infty}^\infty c_\perp^4 (1+\alpha_\perp c_\perp^2)^{-a}d^2 c_\perp &=& \frac{2\pi}{\alpha_\perp^3} \frac{\Gamma(a-3)}{\Gamma(a)}; \qquad a>3. 
\end{eqnarray}  
The $r_{\parallel\parallel}$ moment then calculates
\begin{eqnarray}
  r_{\parallel\parallel} &\equiv& m\int c_\parallel^4 f_0 d^2 c_\perp dc_\parallel = mn\frac{\Gamma(\kappa)}{\Gamma(\kappa-\frac{1}{2})} \sqrt{\frac{\alpha_\parallel}{\pi}}
  \int c_\parallel^4 (1+\alpha_\parallel c_\parallel^2)^{-\kappa} dc_\parallel
  = mn\frac{\Gamma(\kappa)}{\Gamma(\kappa-\frac{1}{2})} \sqrt{\frac{\alpha_\parallel}{\pi}}\frac{3}{4}\frac{\sqrt{\pi}}{\alpha_\parallel^{5/2}}\frac{\Gamma(\kappa-\frac{5}{2})}{\Gamma(\kappa)} \nn\\
  &=& mn \frac{3}{4} \frac{\Gamma(\kappa-\frac{5}{2})}{\Gamma(\kappa-\frac{1}{2})} \frac{1}{\alpha_\parallel^2}
  = mn \frac{3}{4} \frac{\Gamma(\kappa-\frac{5}{2})}{\Gamma(\kappa-\frac{1}{2})} (\kappa-\frac{3}{2})^2 \frac{4 T_\parallel^2}{m^2}
  = 3 \frac{p_\parallel^2}{\rho}\frac{(\kappa-3/2)}{(\kappa-5/2)}; \qquad \kappa>\frac{5}{2},
\end{eqnarray}
where in the last step we used that $\Gamma(x)(x+1)^2/\Gamma(x+2)=(x+1)/x$ with $x=\kappa-5/2$. 
The $r_{\parallel\perp}$ moment calculates
\begin{eqnarray}
  r_{\parallel\perp} &\equiv& \frac{m}{2}\int c_\parallel^2 c_\perp^2 f_0 dc_\parallel d^2c_\perp
  =  \frac{m}{2} n \frac{\Gamma(\kappa+1)}{\Gamma(\kappa-\frac{1}{2})}
  \sqrt{\frac{\alpha_\parallel}{\pi}} \frac{\alpha_\perp}{\pi}  \frac{\sqrt{\pi}}{2\alpha_\parallel^{3/2}}\frac{\pi}{\alpha_\perp^2}\frac{\Gamma(\kappa-\frac{5}{2})}{\Gamma(\kappa+1)}
  = \frac{1}{4}mn \frac{\Gamma(\kappa-\frac{5}{2})}{\Gamma(\kappa-\frac{1}{2})}\frac{1}{\alpha_\parallel \alpha_\perp} \nn\\
  &=& \frac{1}{4}mn \frac{\Gamma(\kappa-\frac{5}{2})}{\Gamma(\kappa-\frac{1}{2})} (\kappa-\frac{3}{2})^2 \frac{4T_\parallel T_\perp}{m^2}
  = \frac{p_\parallel p_\perp}{\rho}\frac{(\kappa-3/2)}{(\kappa-5/2)}; \qquad \kappa>\frac{5}{2},
\end{eqnarray}
and finally the $r_{\perp\perp}$ moment calculates
\begin{eqnarray}
  r_{\perp\perp} &\equiv& \frac{m}{4}\int c_\perp^4 f_0 dc_\parallel d^2c_\perp
  = \frac{m}{4} n \frac{\Gamma(\kappa+\frac{1}{2})}{\Gamma(\kappa-\frac{1}{2})} \frac{\alpha_\perp}{\pi} \int c_\perp^4 (1+\alpha_\perp c_\perp^2)^{-(\kappa+\frac{1}{2})} d^2 c_\perp \nn\\
  &=& \frac{m}{4} n \frac{\Gamma(\kappa+\frac{1}{2})}{\Gamma(\kappa-\frac{1}{2})} \frac{\alpha_\perp}{\pi} \frac{2\pi}{\alpha_\perp^3} \frac{\Gamma(\kappa-\frac{5}{2})}{\Gamma(\kappa+\frac{1}{2})}
  = \frac{1}{2}mn \frac{\Gamma(\kappa-\frac{5}{2})}{\Gamma(\kappa-\frac{1}{2})} \frac{1}{\alpha_\perp^2}
  = 2\frac{p_\perp^2}{\rho} \frac{(\kappa-3/2)}{(\kappa-5/2)}; \qquad \kappa>\frac{5}{2}.
\end{eqnarray}
To summarize, the following closure can be constructed
\begin{equation}
  r_{\parallel\parallel} = \alpha_{\kappa}\frac{3p_\parallel^2}{\rho}; \quad
  r_{\parallel\perp} = \alpha_{\kappa}\frac{p_\parallel p_\perp}{\rho}; \quad
  r_{\perp\perp} = \alpha_{\kappa}\frac{2 p_\perp^2}{\rho},\label{eq:biKappa_closure}
\end{equation}
where for the bi-kappa distribution (\ref{eq:F_biKappa}), the $\alpha_\kappa$ coefficient reads
\begin{equation}
 \alpha_{\kappa} = \frac{\kappa-\frac{3}{2}}{\kappa-\frac{5}{2}}; \qquad \kappa>\frac{5}{2}. \label{eq:biKappa_closure2}
\end{equation} 
It is noteworthy that all three 4-th order moments are just multiplied by the same constant $\alpha_\kappa$. Obviously, in the limit $\kappa\to\infty$,
the $\alpha_\kappa\to 1$, and the ``normal'' bi-Maxwellian closure is obtained.

As suggested for example by \cite{ChustBelmont2006}, a very large class of distribution
functions can be accounted for by considering closures
\begin{equation}
  r_{\parallel\parallel} = \alpha_{\parallel\parallel}\frac{3p_\parallel^2}{\rho}; \quad
  r_{\parallel\perp} = \alpha_{\parallel\perp}\frac{p_\parallel p_\perp}{\rho}; \quad
  r_{\perp\perp} = \alpha_{\perp\perp}\frac{2 p_\perp^2}{\rho},
\end{equation}  
where the 3 proportionality constants $\alpha_{\parallel\parallel}, \alpha_{\parallel\perp}, \alpha_{\perp\perp}$ have to be determined, once a
specific distribution function is prescribed. Such unspecified closure, can account for a very wide class of distribution functions.
Also, a much more complicated distribution functions can be modeled by
considering multi-fluid models. For example, one could consider plasma consisting of 3 different electron species,
that would describe the core, halo and the tail/strahl parts of the electron distribution function. Further information about the core, halo and strahl components
of the electron distribution function can be found for example in \cite{Vocks2005,Vocks2008,Pierrard2011} and references therein.

Here we consider closure  (\ref{eq:biKappa_closure}), (\ref{eq:biKappa_closure2}). Nevertheless, in the following calculations one can we keep the value of $\alpha_\kappa$
unspecified. Therefore, the results are actually valid for a much larger class of fluid models, and not just for a bi-kappa distribution.
Let's completely neglect the non-gyrotropic $\boldsymbol{\Pi}$ and $\br^\textrm{ng}$ in the heat flux equations 
(\ref{eq:ParHF_rX}) and (\ref{eq:PerpHF_rX}). We need to calculate
\begin{eqnarray}
  \nabla\cdot(r_{\parallel\parallel}\bhat) - 3 r_{\parallel\perp}\nabla\cdot\bhat &=& \alpha_{\kappa}\bigg[ 
   3\frac{p_\parallel^2}{\rho}\nabla\cdot\bhat + 3p_\parallel\bhat\cdot\nabla\left(\frac{p_\parallel}{\rho}\right)
  + 3\frac{p_\parallel}{\rho}\bhat\cdot\nabla p_\parallel
  -3\frac{p_\parallel p_\perp}{\rho}\nabla\cdot\bhat \bigg];\\
  \nabla\cdot(r_{\parallel\perp}\bhat)+(r_{\parallel\perp}-r_{\perp\perp})\nabla\cdot\bhat &=& \alpha_{\kappa}\bigg[ 
  \frac{p_\perp}{\rho}\bhat\cdot\nabla p_\parallel +p_\parallel\bhat\cdot\nabla\left( \frac{p_\perp}{\rho} \right)
 +2\frac{p_\perp}{\rho}(p_\parallel-p_\perp)\nabla\cdot\bhat \bigg],
\end{eqnarray}
and the nonlinear heat flux equations for the bi-kappa fluid model read
\begin{eqnarray}
&&  \frac{\pr q_\parallel}{\pr t}+ \bu\cdot\nabla q_\parallel + q_\parallel\nabla\cdot \bu
  + \alpha_\kappa 3p_\parallel \bhat\cdot\nabla\left(\frac{p_\parallel}{\rho}\right) + 3 q_\parallel \bhat\cdot\nabla\bu\cdot\bhat \nn\\
&&  \qquad\qquad + (\alpha_\kappa-1) \frac{3p_\parallel}{\rho}\Big[(p_\parallel-p_\perp)\nabla\cdot\bhat +\bhat\cdot\nabla p_\parallel \Big]=0; \label{eq:BiKappa_par}\\
&&  \frac{\pr q_\perp}{\pr t} + \bu\cdot\nabla q_\perp +2q_\perp\nabla\cdot\bu +\alpha_\kappa p_\parallel\bhat\cdot\nabla\left( \frac{p_\perp}{\rho} \right)
  +(2\alpha_\kappa-1)\frac{p_\perp}{\rho} (p_\parallel-p_\perp)\nabla\cdot\bhat \nn\\
&&  \qquad\qquad + (\alpha_\kappa-1) \frac{p_\perp}{\rho}\bhat\cdot\nabla p_\parallel =0. \label{eq:BiKappa_perp}
\end{eqnarray}
Obviously for $\alpha_\kappa=1$ (i.e. for $\kappa\to\infty$), the bi-Maxwellian heat flux equations are obtained. 
The model is now closed, and it is accompanied by equations (\ref{eq:RhoSimpleL})-(\ref{eq:PperpSimpleL}).
The model is nonlinear and numerical simulations for a chosen (and fixed) value of $\kappa$ can be considered. Even though the $\kappa$ coefficient is fixed,
similarly to a bi-Maxwellian nonlinear simulations, the temperature anisotropy is free to develop, and complicated plasma heating process can be studied
(we recommend to add the FLR pressure stress forces to the scalar pressure equations). 
We call the non-dispersive fluid model that uses the above heat flux equations with abbreviation ``BiKappa''. If the Hall term is
considered in the induction equation, ``Hall-BiKappa'' fluid model is obtained. Similarly, by considering FLR1 and FLR2 corrections to the pressure tensor, 
yields ``Hall-BiKappa-FLR1'' and ``Hall-BiKappa-FLR2'' fluid models.
The heat flux equations are normalized, linearized, written in the x-z plane, and Fourier transformed according to
\begin{eqnarray}
&&  -\omega q_\parallel +\alpha_\kappa \frac{3}{2}\bpar \kpar(p_\parallel-\rho) +(\alpha_\kappa-1)\frac{3}{2}\bpar
  \Big[(1-a_p)\kperp B_x +\kpar p_\parallel \Big] =0; \\
  && -\omega q_\perp +\alpha_\kappa \frac{\bpar}{2}\kpar (p_\perp-a_p\rho) +(2\alpha_\kappa-1)\frac{\bpar}{2}a_p(1-a_p)\kperp B_x
  +(\alpha_\kappa-1)\frac{\bpar}{2}a_p\kpar p_\parallel =0,
\end{eqnarray}
where we have used the same plasma beta definition as always $\bpar/2=p_\parallel^{(0)}/(V_A^2\rho_0)$.
The heat flux equations are accompanied by the linearized system (\ref{eq:CGL2_rit10})-(\ref{eq:CGL2_rit11}).

To consider FLR3 corrections, one needs to derive non-gyrotropic heat flux vectors
$\boldsymbol{S}^\parallel_\perp$, $\boldsymbol{S}^\perp_\perp$ for the bi-kappa distribution. For the first-order vectors the derivation is straightforward,
and by following calculations in the Appendix \ref{sec:NONGheat}, it is easy to show that the nonlinear
\begin{eqnarray}
  \boldsymbol{S}^{\parallel(1)}_\perp  &=& \frac{B_0}{\Omega|\bb|} \bhat\times \Big[ \alpha_\kappa p_\perp \nabla\Big(\frac{p_\parallel}{\rho}\Big)
    +(\alpha_\kappa-1)\frac{p_\parallel}{\rho}\nabla p_\perp
  +(3\alpha_\kappa-1)\frac{p_\parallel}{\rho}(p_\parallel-p_\perp) \bhat\cdot\nabla\bhat
  + 2 q_\parallel \bhat\cdot\nabla \bu
  +2q_\perp \bhat\times\boldsymbol{\omega}\Big]; \\
  \boldsymbol{S}_\perp^{\perp (1)}  &=& \frac{B_0}{\Omega|\bb|} \bhat\times \Big[
    2\alpha_\kappa p_\perp \nabla \Big( \frac{p_\perp}{\rho}\Big) +2(\alpha_\kappa-1)\frac{p_\perp}{\rho}\nabla p_\perp +
    2(\alpha_\kappa-1)\frac{p_\perp}{\rho}(p_\parallel-p_\perp)\bhat\cdot\nabla\bhat+4q_\perp \bhat\cdot\nabla\bu \Big].
\end{eqnarray}
Here we neglected perturbations $\widetilde{r}$, since we will not consider Landau fluid models for bi-kappa distribution.
For $\alpha_\kappa=1$, the results are of course equivalent to bi-Maxwellian expressions
(\ref{eq:SP2015typo}), (\ref{eq:SperpPerpF}). The second-order heat flux vectors are not addressed here, since we would have to derive
$\br^{\textrm{ng}}$ for bi-kappa distribution.

\subsection{BiKappa dispersion relation}
The BiKappa dispersion relation can be written in the following form
\begin{eqnarray}
&&  \Big(\omega^2-v_{A\parallel}^2 k_\parallel^2\Big)\Big(\omega^8-A_6\omega^6+A_4\omega^4-A_2\omega^2+A_0\Big)=0; \label{eq:BiKappa_disper}\\
  &&  A_6 = k_\parallel^2\Big( v_{A\parallel}^2 + \alpha_\kappa \frac{7}{2}\bpar\Big) +k_\perp^2 (1+a_p\beta);\\
  &&  A_4 = \alpha_\kappa k_\parallel^2\bpar \Big\{ k_\parallel^2\Big( \frac{7}{2}v_{A\parallel}^2 +\frac{(3+6\alpha_\kappa)}{4}\bpar \Big)
  + k_\perp^2\frac{7}{2}\Big( 1+a_p\bpar-\frac{1}{7}a_p^2\bpar \Big) \Big\};\\
  &&  A_2 = \alpha_\kappa k_\parallel^4\bpar^2 \Big\{ k_\parallel^2\Big( \frac{(3+6\alpha_\kappa)}{4} v_{A\parallel}^2+\alpha_\kappa \frac{3}{8}\bpar \Big)
  +k_\perp^2\frac{9}{4} \Big( \frac{(1+2\alpha_\kappa)}{3}(1+a_p\bpar)-\frac{(1+9\alpha_\kappa)}{18}a_p^2\bpar \Big) \Big\};\\
  &&  A_0 = \alpha_\kappa^2 \frac{3}{8}k_\parallel^6 \bpar^3 \Big\{ k_\parallel^2 v_{A\parallel}^2 + k_\perp^2 \Big( 1+a_p\bpar-a_p^2\bpar\Big) \Big\},\label{eq:BiKappa_disper2}
\end{eqnarray}
and for $\alpha_\kappa=1$ it is of course equivalent to the CGL2 dispersion relation (\ref{eq:8thP}). The oblique Alfv\'en mode is not effected, and
is identical to (\ref{eq:CGL2_Alf}), implying that the long-wavelength threshold of the oblique firehose instability is identical as well. 

For strictly parallel propagation $(k_\perp=0)$, the solutions of the remaining 8th-order polynomial are
\begin{eqnarray}
  \omega &=& \pm  k_\parallel v_{A\parallel};\\
  \omega &=& \pm k_\parallel \sqrt{\bpar\Big(\frac{3}{2}\alpha_\kappa +\frac{1}{2}\sqrt{3\alpha_\kappa(3\alpha_\kappa-1)} \Big)}; \label{eq:kappa_rit1}\\
  \omega &=& \pm k_\parallel \sqrt{\alpha_\kappa \frac{\bpar}{2}}; \label{eq:kappa_rit2}\\
  \omega &=& \pm k_\parallel \sqrt{\bpar\Big( \frac{3}{2}\alpha_\kappa -\frac{1}{2}\sqrt{3\alpha_\kappa(3\alpha_\kappa-1)}\Big)}. \label{eq:kappa_rit3}
\end{eqnarray}
Note that in the last solution (\ref{eq:kappa_rit3}), the expression under the square root remains always positive, regardless how large $\alpha_\kappa$ is
(i.e. when the $\kappa$-index approaches 5/2).
For strictly perpendicular propagation $(\kpar=0)$, the fast mode is not effected, and is equivalent to (\ref{eq:CGL2_fast}). 
\subsection{Mirror instability}
By considering the highly oblique limit $k_\perp\gg \kpar$, and by assuming that $\alpha_\kappa$ is not enormously large 
(i.e. that $\kappa$ is not very close to $5/2$) so that $\alpha_\kappa k_\parallel$ can be neglected with respect to $k_\perp$,
the polynomial coefficients can be approximated as 
\begin{eqnarray}
  &&  A_6 = k_\perp^2 (1+a_p\bpar);\\
  &&  A_4 = \alpha_\kappa k_\parallel^2\bpar k_\perp^2\frac{7}{2}\Big( 1+a_p\bpar-\frac{1}{7}a_p^2\bpar \Big);\\
  &&  A_2 = \alpha_\kappa k_\parallel^4\bpar^2 k_\perp^2\frac{9}{4} \Big( \frac{(1+2\alpha_\kappa)}{3}(1+a_p\bpar)-\frac{(1+9\alpha_\kappa)}{18}a_p^2\bpar \Big);\\
  &&  A_0 = \alpha_\kappa^2 \frac{3}{8}k_\parallel^6 \bpar^3 k_\perp^2 \Big( 1+a_p\bpar-a_p^2\bpar\Big).
\end{eqnarray}
The fast mode can be quickly separated with the trick (\ref{eq:CGL2_trick}), and by using substitution $\bar{\omega}=\omega/(\kpar\sqrt{\bpar})$, the resulting
polynomial reads
\begin{eqnarray}
&&  -(1+a_p\bpar) \bar{\omega}^6+\alpha_\kappa \frac{7}{2}\Big( 1+a_p\bpar-\frac{1}{7}a_p^2\bpar \Big)\bar{\omega}^4
  -\alpha_\kappa \frac{9}{4} \Big( \frac{(1+2\alpha_\kappa)}{3}(1+a_p\bpar)-\frac{(1+9\alpha_\kappa)}{18}a_p^2\bpar \Big)\bar{\omega}^2 \nn\\
&& \qquad  +\alpha_\kappa^2 \frac{3}{8} \Big( 1+a_p\bpar-a_p^2\bpar\Big) =0.
\end{eqnarray}
At exactly the (assumed) mirror threshold $a_p^2\bpar=1+a_p\bpar$, the coefficient $A_0$ disappears, so we have one solution $\bar{\omega}^2=0$;
from the remaining polynomial the $1+a_p\bpar$ can be factored-out and the solutions are
\begin{equation} \label{eq:Bikappa_tired}
\bar{\omega}^2=\frac{3}{2}\alpha_\kappa\pm\frac{1}{4}\sqrt{10\alpha_\kappa(3\alpha_\kappa-1)},
\end{equation}
both always positive.
Slightly below $(\epsilon<0)$ or beyond $(\epsilon>0)$ the mirror threshold, one can prescribe  
$a_p^2\bpar = (1+a_p\bpar)(1+\epsilon)$, where $\epsilon$ is a small parameter and the polynomial transforms to
\begin{equation}
  (\bar{\omega}^2)^3-\alpha_\kappa (3-\frac{\epsilon}{2})(\bar{\omega}^2)^2+\frac{\alpha_\kappa}{8}\Big(5+3\alpha_\kappa-\epsilon(1+9\alpha_\kappa)\Big)(\bar{\omega}^2)
  +\frac{3}{8}\alpha_\kappa^2 \epsilon=0,
\end{equation}
and which for $\alpha_\kappa=1$ is equivalent to the bi-Maxwellian polynomial (\ref{eq:mirrorPica}). As previously, for small $\epsilon$ the solutions
(\ref{eq:Bikappa_tired}) remain almost the same, and positive. Furthermore, since multiplying all 3 solutions together must yield $ -\frac{3}{8}\alpha_\kappa^2 \epsilon$,
it implies that below $(\epsilon<0)$ the threshold the 3rd solution is positive, and beyond $(\epsilon>0)$ the threshold the 3rd solution is negative,
which analytically proves that the (highly oblique) mirror instability threshold is $a_p^2\bpar=1+a_p\bpar$, regardless of $\alpha_\kappa$.
The long-wavelength limit or ``hard'' mirror threshold for the bi-kappa distribution is therefore the same as 
for the bi-Maxwellian distribution.

\subsubsection*{Conclusion}
We investigated solutions of the (non-dispersive) BiKappa fluid model. The calculations were done with closure (\ref{eq:biKappa_closure}) for a general
$\alpha_\kappa$ coefficient, i.e. without using the specific value (\ref{eq:biKappa_closure2}) for the bi-kappa distribution. The calculations are therefore more
general and valid for any distribution function that can be closed as (\ref{eq:biKappa_closure}). We conclude that all 3 long-wavelength ``hard'' thresholds
- for the parallel and oblique firehose instability and the highly oblique mirror instability - are not effected by the value of $\alpha_\kappa$.

\subsection{Hall-BiKappa dispersion relation}
The Hall-BiKappa dispersion relation can be written in the following form
\begin{eqnarray}
  &&  \Big(\omega^2-v_{A\parallel}^2 k_\parallel^2\Big)\Big(\omega^8-A_6\omega^6+A_4\omega^4-A_2\omega^2+A_0\Big)=\nn\\
  && \quad k^2 k_\parallel^2 \omega^2\Big[\omega^6 -\omega^4 \bpar\Big( \frac{7}{2}\alpha_\kappa k_\parallel^2+a_p k_\perp^2 \Big)
    +\omega^2 \alpha_\kappa \bpar^2 k_\parallel^2\Big( \frac{3}{4}(1+2\alpha_\kappa)k_\parallel^2+3a_p k_\perp^2\Big)
    -\alpha_\kappa k_\parallel^4 \bpar^3 \Big( \frac{3}{8}\alpha_\kappa k_\parallel^2 + \frac{1}{8}(5+3\alpha_\kappa) a_p k_\perp^2\Big)  \Big],\nn\\
  \label{eq:Hall-biKappaD}
\end{eqnarray}
where we kept the BiKappa dispersion relation (\ref{eq:BiKappa_disper})-(\ref{eq:BiKappa_disper2}) on the l.h.s., and the Hall term contributions 
on the r.h.s.. For the bi-Maxwellian value $\alpha_\kappa=1$, the result is equivalent to the Hall-CGL2 dispersion relation
(\ref{eq:HallCGL2_disp}).
For strictly parallel propagation, the solutions are the usual dispersive ion-cyclotron and whistler waves (that are not affected
by the $\alpha_\kappa$ coefficient) and the non-dispersive solutions (\ref{eq:kappa_rit1})-(\ref{eq:kappa_rit3}) of the BiKappa model.
For strictly perpendicular propagation, the Hall term naturally disappears and the fast mode solution is (\ref{eq:CGL2_fast}).

\clearpage
\section{Collisionless damping in fluid models - Landau fluid models} \label{section:Landau}
In the previous sections, we have calculated the fluid hierarchy up to the 4th-order fluid moment $\br$. We have seen that unless we crossed some instability threshold,
the dispersion relations yielded frequencies $\omega$ that were purely real, and there was no collisionless damping present.
 Remarkably, collisionless damping, i.e. Landau damping, will be absent even if we continue to develop the fluid hierarchy to higher-order moments.
In fact, we will see in the next section by considering parallel 1D propagation (where all the fluctuations are along $\boldsymbol{B}_0$),
that fluid models closed by a bi-Maxwellian
fluid closure at higher-order moments than $\br$, are always unstable. Such fluid models are physically ill-posed.  
To go higher in the fluid hierarchy, one needs to incorporate Landau damping phenomenon into the fluid framework, and consider Landau fluid closures. 

Nevertheless, to capture the Landau damping, it is not necessary to go beyond the 4th-order moment and closures performed at the 4th moment level describe the
Landau damping with very good accuracy. Even though the entire Part 2 of this text is dedicated to Landau fluid closures, it is beneficial to briefly show,
why the Landau damping was omitted in the previous calculations.     
Let's consider the simplest example of 1D electrostatic propagation. The collisionless Vlasov equation reads
\begin{equation} \label{eq:Vlasov1DD}
\frac{\pr f}{\pr t} +v_z \pr_z f +\frac{q}{m} E_z \frac{\pr f}{\pr v_z}=0. 
\end{equation}
Direct integration of the Vlasov equation (over $v_z$) for a general distribution function $f$, of course yields the usual density equation, $\pr n/\pr t +\pr_z (nu_z)=0$.  
The same result is obtained if one considers specific $f_0$, as for example 1D Maxwellian $f_0=n\sqrt{\frac{\alpha}{\pi}}e^{-\alpha (v_z-u_z)^2}$,
1D kappa distribution, etc.. The last term of the Vlasov equation $\sim E_z$ vanishes by the integration, since $\int^{\infty}_{-\infty} \frac{\pr f}{\pr v_z} dv_z=0$.
Now let's do the calculation differently, by expanding the distribution function $f=f_0+f^{(1)}$, where $f_0$ does not have any time or spatial dependence, so
$\pr f_0/\pr t=0,\pr_z f_0=0$. We can consider Maxwellian $f_0=n_0\sqrt{\frac{\alpha}{\pi}}e^{-\alpha v_z^2}$ where $\alpha=m/(2T^{(0)})=1/v_{\textrm{th}}^2$.
The electric field is also expanded
$E_z=E_{z}^{(0)}+E_z^{(1)}$, and one assumes that there is no large-scale electric field, so $E_{z}^{(0)}=0$. By neglecting term with two small quantities $E_z^{(1)}\pr f^{(1)}/\pr v_z$,
the Vlasov equation reads
\begin{equation}
\frac{\pr f^{(1)} }{\pr t} +v_z \pr_zf^{(1)}  +\frac{q}{m} E_z^{(1)} \frac{\pr f_0 }{\pr v_z}=0, 
\end{equation}
and transformation to Fourier space yields
\begin{equation} \label{eq:Vlasov1DDD}
  -i(\omega-k_z v_z) f^{(1)} +\frac{q}{m} E_z^{(1)} \frac{\pr f_0 }{\pr v_z}=0.   
\end{equation}
The equation (\ref{eq:Vlasov1DDD}) is actually not that much different from (\ref{eq:Vlasov1DD}), the Vlasov equation is just fully linearized and written in Fourier space.
However, in kinetic calculations, the (\ref{eq:Vlasov1DDD}) is not directly integrated in this form. The crucial technique that reveals the presence of Landau damping in the Vlasov
equation is, that the equation (\ref{eq:Vlasov1DDD}) is first divided by $(\omega-k_z v_z)$, and an expression for $f^{(1)}$ is obtained. Only then the equation is integrated
over the velocity space, yielding 
\begin{equation}
  f^{(1)} = -\frac{q}{m} iE_z^{(1)} \frac{\frac{\pr f_0 }{\pr v_z}}{\omega-k_z v_z}; \quad =>
  \quad n^{(1)} = \int_{-\infty}^{\infty} f^{(1)} dv_z = -\frac{q}{m} iE_z^{(1)}  \int_{-\infty}^{\infty} \frac{\frac{\pr f_0 }{\pr v_z}}{\omega-k_z v_z}dv_z. \label{eq:IntegralBlaa}   
\end{equation}
The integral contains a ``singularity'' in the denominator, and it is not obvious how to calculate this integral. Such singularities are called
\emph{wave-particle resonances},
and can have a more general form $\omega-k_z v_z \pm l\Omega$, where $\Omega$ is the cyclotron frequency of the considered species, and $l$ is an integer. 
The resonance for $l=0$ is called the Landau resonance, and the other resonances for $l\neq 0$ are called cyclotron resonances.
It is the presence of these resonances in integrals such as (\ref{eq:IntegralBlaa}), that yields collisionless damping mechanisms, the Landau damping and the cyclotron damping. 
In Part 2 of our guide, we will see that the integral actually
cannot be ``calculated'', i.e. the integral cannot be expressed through elementary functions, even if $f_0$ is simple Maxwellian.
The integral is just expressed through the famous \emph{plasma dispersion function} $Z(\zeta)$,
where $\zeta=\omega/(|k_z| v_{\textrm{th}})$, since the plasma dispersion function was specifically developed to describe this integral.
Skipping many technical (but very important) details such as the definition of $Z(\zeta)$, for Maxwellian $f_0$ the density integral is simply expressed as
\begin{equation}
n^{(1)} = -\frac{q}{m} \frac{iE_z^{(1)}}{k_z} \Big( 1+\zeta Z(\zeta) \Big).
\end{equation}  
One can define function $R(\zeta)=1+\zeta Z(\zeta)$, that is called the \emph{plasma response function}. Now for example considering proton-electron plasma with $T_e^{(0)}=T_p^{(0)}$
at wavelengths that are much longer than the Debye length, yields the following dispersion relation and numerical solution
\begin{eqnarray}
 R(\zeta_p) + R(\zeta_e) =0 ;\quad => \quad \zeta_p=\frac{\omega}{|k_z|v_{\textrm{th} p}} = \pm 1.457 -0.627i.
\end{eqnarray}
The negative imaginary part represents the Landau damping, and the effect comes from the Landau resonance in the integral (\ref{eq:IntegralBlaa}).
Obviously, integrating the Vlasov equation directly in the form (\ref{eq:Vlasov1DD}) or (\ref{eq:Vlasov}) neglects the wave-particle resonances in velocity space,
which clarifies why traditional fluid models do not contain collisionless damping mechanisms.
It is the $f^{(1)}$ - the fluctuations/perturbations around $f_0$ - where the wave-particle resonances are written down explicitly,
that yields the collisionless damping mechanisms in kinetic theory. 
To introduce collisionless damping into fluid models, we have no other choice, and we have to calculate the hierarchy of moments again,
this time by integrating over the ``kinetic'' $f^{(1)}$. We find it useful to introduce the following vocabulary:
\begin{itemize}
\item fluid moments = integrals over a general (unspecified) distribution function $f$, or over a specific $f_0$. Evolution equations for fluid moments are obtained by
  direct integration of the Vlasov equation written in a form that does not explicitly contain wave-particle resonances. The resulting expressions can be
  nonlinear or linear and contain only fluid variables, i.e. expressions do not contain $Z(\zeta)$.
\item kinetic moments = integrals over $f^{(1)}$ (perturbations $f^{(1)}=f-f_0$, where a specific equilibrium $f_0$ is assumed), where the wave-particle resonances
  in velocity space are written down explicitly. The resulting expressions are linear and contain $Z(\zeta)$, which is not a fluid variable.
\end{itemize}

Importantly, we want to keep the entire nonlinear fluid hierarchy discussed so far, including the nonlinear heat flux equations.    
Therefore, considering bi-Maxwellian distribution $f_0$, instead of the ``normal'' closure, the 4th-order moment is decomposed in the following form
\begin{eqnarray}
  r_{\parallel\parallel} = \frac{3p_\parallel^2}{\rho} + \widetilde{r}_{\parallel\parallel}; \qquad
  r_{\parallel\perp} = \frac{p_\parallel p_\perp}{\rho} + \widetilde{r}_{\parallel\perp}; \qquad
  r_{\perp\perp} = \frac{2 p_\perp^2}{\rho} + \widetilde{r}_{\perp\perp} \label{eq:r3}.
\end{eqnarray}
The first terms represent the ``normal'' closure, i.e. when an exact bi-Maxwellian $f_0$ is prescribed. The tilde components 
represent fluctuations/perturbations around $f_0$, and a good notation would also be $\delta r$.
The decomposition (\ref{eq:r3}) yields the following form of the nonlinear heat flux equations
\begin{eqnarray} \label{eq:HFpar_final}
&& \frac{\pr q_\parallel}{\pr t}+\nabla\cdot(q_\parallel \bu) +\nabla\cdot(\widetilde{r}_{\parallel\parallel}\bhat) - 3 \widetilde{r}_{\parallel\perp}\nabla\cdot\bhat
  + 3p_\parallel \bhat\cdot\nabla\left(\frac{p_\parallel}{\rho}\right) + 3 q_\parallel \bhat\cdot\nabla\bu\cdot\bhat =0;\\
&&  \frac{\pr q_\perp}{\pr t} + \nabla\cdot(q_\perp \bu) + q_\perp\nabla\cdot\bu + \nabla\cdot(\widetilde{r}_{\parallel\perp}\bhat) +
  (\widetilde{r}_{\parallel\perp}-\widetilde{r}_{\perp\perp})\nabla\cdot\bhat +p_\parallel\bhat\cdot\nabla\left( \frac{p_\perp}{\rho} \right) \nn\\
&&\qquad  +\frac{p_\perp}{\rho} (p_\parallel-p_\perp)\nabla\cdot\bhat+\frac{p_\perp}{\rho}(\nabla\cdot\boldsymbol{\Pi})\cdot\bhat_0=0.  
\end{eqnarray}
Similarly to decomposition (\ref{eq:r3}), ``kinetic'' corrections with wave-particle resonances can also be considered for the non-gyrotropic $\br^{\textrm{ng}}$,
which is not addressed in this guide. 
Neglecting those contributions, the heat flux equations are equivalent for example to eq. 14, 15 of \cite{PassotSulem2007}, and
eq. 9, 10 of \cite{PSH2012}.

In Part 2 of this guide, we will calculate the ``kinetic moments'' for $\widetilde{r}_{\parallel\parallel}$, $\widetilde{r}_{\parallel\perp}$, $\widetilde{r}_{\perp\perp}$,
by integrating over $f^{(1)}$. In the 3D electromagnetic geometry, we will use more sophisticated $f^{(1)}$ in the gyrotropic limit, which describes
both Landau damping and its magnetic analogue, the transit-time damping. The expressions of the entire ``kinetic'' hierarchy will contain the
$Z(\zeta)$, or actually the $R(\zeta)$, which is not a fluid variable. Nevertheless, we will see that 
by introducing the concept of Pad\'e approximation to $R(\zeta)$, that sometimes a rare possibility arises,
when the ``kinetic'' moment $\widetilde{r}$ can be expressed through lower-order ``kinetic'' moments. 
Such a closure, where the last retained kinetic moment (in this case $\widetilde{r}$) is expressed through lower-order moments,
in a way that the kinetic $Z(\zeta)$ function is completely eliminated and that the closure is valid for all $\zeta$ values,
will be called a \emph{Landau fluid closure}.  
For the impatient reader, we will reveal, that one of the possibilities are going to be (written in Fourier space)
\begin{eqnarray}
  \widetilde{r}_{\parallel\parallel} &=& \frac{32-9\pi}{2(3\pi-8)} n_0 v_{\textrm{th}\parallel}^2 T_{\parallel}
  - \frac{2\sqrt{\pi}}{3\pi-8} v_{\textrm{th}\parallel} \frac{ik_\parallel}{|k_\parallel|} q_{\parallel};\\
  \widetilde{r}_{\parallel\perp} &=& -\frac{\sqrt{\pi}}{2} v_{\textrm{th}\parallel} \frac{ik_\parallel}{|k_\parallel|} q_{\perp};\\
\widetilde{r}_{\perp \perp} &=& 0.
\end{eqnarray}
The first closure was obtained in the slab geometry by \cite{HammettPerkins1990} and the remaining closures were obtained in the 3D geometry by \cite{Snyder1997}
(note that thermal speeds in those papers are defined without the factor of 2). In Part 2 of our guide, we are going to map all the Landau fluid
closures that can be constructed at the 4th-moment level, and we will identify the most precise closures. A summary of these findings,
can be found in \cite{HunanaPRL2018}.

\clearpage
\section{Evolution equation for n-th order fluid moment} \label{section:nth}
Now we are proficient with calculating the fluid moments for general $f$, and before we start with kinetic calculations,
it is useful to derive the time-dependent evolution equation for a general n-th order fluid moment. 
Before going through this part, we highly recommend to go through the parts of the paper where the pressure tensor and heat flux tensor
equations are obtained. Then the algebra and the notation used here will look hopefully logical and natural. We define the n-th order fluid moment $\bX^{(n)}$ according to
\begin{equation} \label{eq:Xmom}
X^{(n)}_{ij\ldots n} = m \int c_i c_j\ldots c_n f d^3 v.  
\end{equation}
The symmetric operator is defined that it cycles around through all possibilities according to
\begin{equation}
\big[ X^{(n)}\big]^S_{ijk\ldots n} = X^{(n)}_{ijk\ldots n} + X^{(n)}_{jk\ldots ni} + X^{(n)}_{k\ldots nij}+ \cdots\cdots + X^{(n)}_{nijk\ldots n-1},
\end{equation}
and it is compatible with the previously used symmetric operators for a 2nd-order tensor
$[X^{(2)}]^S_{ij}=X^{(2)}_{ij}+X^{(2)}_{ji}$ and for a 3rd-order tensor $[X^{(3)}]^S_{ijk}=X^{(3)}_{ijk}+X^{(3)}_{jki}+X^{(3)}_{kij}$. We multiply the Vlasov equation by the mass m
and by the velocities $c_ic_j\ldots c_n$ and integrate over $d^3 v$. To integrate the first term of the Vlasov equation we will need
\begin{eqnarray}
  \frac{\pr}{\pr t} \big(  c_i c_j\ldots c_n    \big) &=& -\big(\frac{\pr u_i}{\pr t}c_j\ldots c_n\big) - \big(c_i \frac{\pr u_j}{\pr t}\ldots c_n\big)- \cdots\cdots
  -\big(c_i c_j\ldots \frac{\pr u_n}{\pr t}\big);
\end{eqnarray}  
and the first term of the Vlasov equation calculates
\begin{eqnarray}
\bigcirc{ }\!\!\!\!\mbox{\small 1} &=&  m \int c_i c_j\ldots c_n \frac{\pr f}{\pr t} d^3 v = \frac{\pr}{\pr t}\Big(\underbrace{m\int c_i c_j \ldots c_n f d^3 v}_{X_{ij\ldots n}^{(n)}}\Big)
- m\int f\frac{\pr}{\pr t}( c_i c_j\ldots c_n ) d^3 v = \frac{\pr}{\pr t} X_{ij\ldots n}^{(n)} \nn\\
  &&+ \frac{\pr u_i}{\pr t} \underbrace{m \int f c_j c_k \ldots c_n d^3v}_{=X_{jk\ldots n}^{(n-1)}} 
  +\frac{\pr u_j}{\pr t} \underbrace{m\int c_i c_k\ldots c_n d^3v}_{=X_{ik\ldots n}^{(n-1)}} + \cdots\cdots
  + \frac{\pr u_n}{\pr t} \underbrace{m\int f c_i c_j c_k\ldots c_{n-1}}_{=X_{ij\ldots n-1}^{(n-1)}} \nn\\
  &=& \Big\{ \frac{\pr}{\pr t} \bX^{(n)} +\Big[ \frac{\pr\bu}{\pr t}\bX^{(n-1)}\Big]^S \Big\}_{ij\ldots n}.
\end{eqnarray}
For the second term we will need
\begin{eqnarray}
  m\int c_i c_j c_k \ldots c_n v_l f d^3 v &=& m \int c_i c_j c_k \ldots c_n (\underbrace{v_l-u_l}_{=c_l} +u_l) f d^3 v
  = \underbrace{m \int c_i c_j c_k\ldots c_n c_l f d^3 v}_{=X^{(n+1)}_{ij\ldots nl}} + u_l \underbrace{m \int c_i c_j c_k\ldots c_n f d^3 v}_{=X_{ij..n}^{(n)}} \nn\\
  &=& X^{(n+1)}_{ij\ldots nl}  + X_{ij..n}^{(n)} u_l ;
\end{eqnarray}
\begin{eqnarray}
&& \pr_l \big(  c_i c_j\ldots c_n    \big) = -\big((\pr_l u_i) c_j\ldots c_n\big) - \big(c_i (\pr_l u_j) \ldots c_n\big)- \cdots\cdots
    -\big(c_i c_j\ldots (\pr_l u_n) \big); \\
&&  m \int v_l f \pr_l \big( c_i c_j c_k \ldots c_n \big) d^3 v
  = m\int v_l f\Big[  -\big((\pr_l u_i) c_j\ldots c_n\big) - \big(c_i (\pr_l u_j) \ldots c_n\big) -\cdots\cdots -\big(c_i c_j\ldots (\pr_l u_n) \big) \Big] d^3 v \nn\\
&& \quad = -(\pr_l u_i) m\int f v_l c_j\ldots c_n d^3 v - (\pr_l u_j)m \int f v_l c_i\ldots c_n d^3 v(\pr_l u_n)
  -\cdots\cdots -(\pr_l u_n) m\int f v_l c_i c_j\ldots  c_{n-1}  d^3 v \nn\\
&& \quad = -(\pr_l u_i) \Big( X^{(n)}_{jk\ldots nl} + X^{(n-1)}_{jk\ldots n} u_l \Big)- (\pr_l u_j) \Big( X^{(n)}_{ik\ldots nl} + X^{(n-1)}_{ik\ldots n} u_l \Big)
  -\cdots\cdots -(\pr_l u_n)\Big( X^{(n)}_{ij\ldots (n-1)l} + X^{(n-1)}_{ij\ldots n-1} u_l \Big)  \nn\\
&&  \quad = - \Big( X^{(n)}_{jk\ldots nl} + X^{(n-1)}_{jk\ldots n} u_l \Big)(\pr_l u_i) - \Big( X^{(n)}_{ik\ldots nl} + X^{(n-1)}_{ik\ldots n} u_l \Big) (\pr_l u_j)
 -\cdots\cdots -\Big( X^{(n)}_{ij\ldots (n-1)l} + X^{(n-1)}_{ij\ldots n-1} u_l \Big)(\pr_l u_n)  \nn\\
&& \quad = - \Big\{ \Big( \bX^{(n)} + \bX^{(n-1)}\bu \Big)\cdot \nabla \bu \Big\}^S_{ij\ldots n}.   
\end{eqnarray}  
The second term calculates
\begin{eqnarray}
\bigcirc{ }\!\!\!\!\mbox{\small 2} =  && m\int c_i c_j c_k\ldots c_n \underbrace{\bV \cdot \nabla}_{=v_l \pr_l} f d^3 v
   = \pr_l \Big( m\int c_i c_j c_k \ldots c_n v_l f d^3 v\Big) -m \int v_l f \pr_l \big( c_i c_j c_k \ldots c_n \big) d^3 v\nn\\
  && = \pr_l \Big( X^{(n+1)}_{ij\ldots nl}  + u_l  X_{ij..n}^{(n)} \Big)+ \Big\{ \Big( \bX^{(n)} + \bX^{(n-1)} \bu \Big)\cdot \nabla \bu \Big\}^S_{ij\ldots n} \nn\\
  && = \Big\{ \nabla\cdot \Big( \bX^{(n+1)}+\bu \bX^{(n)} \Big) \Big\}_{ij\ldots n} + \Big\{ \Big( \bX^{(n)} + \bX^{(n-1)}\bu \Big)\cdot \nabla \bu \Big\}^S_{ij\ldots n}.
\end{eqnarray}  
The third term calculates
\begin{eqnarray}
  \bigcirc{ }\!\!\!\!\mbox{\small 3} &=&  q\int c_i c_j \ldots c_n E_l \frac{\pr f}{\pr v_l} d^3 v
  = q E_l \underbrace{\int \frac{\pr}{\pr v_l} \big( c_i c_j \ldots c_n f \big) d^3 v}_{\to 0} - q E_l \int f \frac{\pr}{\pr v_l} \big( c_i c_j \ldots c_n \big) d^3v \nn\\
  && = -q E_l \int f \big( \delta_{il} c_j\ldots c_n + c_i \delta_{jl}\ldots c_n +\cdots\cdots + c_i c_j \ldots c_{n-1}\delta_{nl} \big) d^3 v \nn\\
  && = -q E_i \int f c_j\ldots c_n d^3 v -q E_j \int f c_i\ldots c_n d^3 v -\cdots\cdots -q E_n \int f c_i c_j\ldots c_{n-1} d^3 v \nn\\
  && = -\frac{q}{m}E_i X^{(n-1)}_{jk\ldots n} - \frac{q}{m}E_j \underbrace{X^{(n-1)}_{ik\ldots n}}_{=X^{(n-1)}_{k\ldots ni}} -\cdots\cdots -\frac{q}{m}E_n X^{(n-1)}_{ij\ldots n-1} \nn\\
  && = -\frac{q}{m}\Big[ \bE \bX^{(n-1)}\Big]^S_{ij\ldots n}.
\end{eqnarray}
For the fourth term we will need
\begin{eqnarray}
  \frac{\pr}{\pr v_l} \big[ c_i c_j \ldots c_n (\bV\times \bb)_l \big]
  &=& \delta_{il} c_j\ldots c_n (\bV\times \bb)_l + c_i \delta_{jl}\ldots c_n (\bV\times \bb)_l +\cdots\cdots + c_ic_j\ldots c_{n-1}\delta_{nl} (\bV\times \bb)_l \nn\\
  \quad && +c_i c_j \ldots c_n \underbrace{\frac{\pr}{\pr v_l}(\bV\times \bb)_l}_{=0} \nn\\
  &=&  c_j\ldots c_n (\bV\times \bb)_i + c_i\ldots c_n (\bV\times \bb)_j +\cdots\cdots + c_ic_j\ldots c_{n-1}(\bV\times \bb)_n
\end{eqnarray}  
\begin{eqnarray}
&&  \int c_j c_k \ldots c_n (\bV\times \bb)_i f d^3 v = \int c_j c_k \ldots c_n \epsilon_{irs} v_r B_s f d^3v =
  \epsilon_{irs} B_s \int c_j c_k \ldots c_n v_r f d^3 v \nn\\
  &&\quad   = \frac{1}{m}\underbrace{\epsilon_{irs}}_{=-\epsilon_{isr}} B_s \Big( X^{(n)}_{jk\ldots nr}  + X_{jk..n}^{(n-1)} u_r \Big)
  = -\frac{1}{m}\epsilon_{isr} B_s \Big( X^{(n)}_{rjk\ldots n}  + u_r X_{jk..n}^{(n-1)} \Big) \nn\\
  &&\quad = - \frac{1}{m}\Big[ \bb\times\Big( \bX^{(n)} +\bu\bX^{(n-1)} \Big) \Big]_{ijk\ldots n}.
\end{eqnarray}  
The fourth term calculates
\begin{eqnarray}
  \bigcirc{ }\!\!\!\!\mbox{\small 4} &=& \frac{q}{c}\int c_i c_j\ldots c_n (\bV\times \bb)_l \frac{\pr f}{\pr v_l} d^3 v
  = \frac{q}{c} \underbrace{\int \frac{\pr}{\pr v_l}\big[ c_i c_j\ldots c_n (\bV\times \bb)_l f\big] d^3v}_{\to 0}
  -\frac{q}{c}\int f \frac{\pr}{\pr v_l} \big[ c_i c_j\ldots c_n (\bV\times \bb)_l \big] d^3 v \nn\\
  &=& -\frac{q}{c}\int f \Big( c_j\ldots c_n (\bV\times \bb)_i + c_i\ldots c_n (\bV\times \bb)_j +\cdots\cdots + c_ic_j\ldots c_{n-1}(\bV\times \bb)_n \Big) d^3v \nn\\
  &=& +\frac{q}{mc} \Big[ \bb\times\Big( \bX^{(n)} +\bu\bX^{(n-1)} \Big) \Big]^S_{ijk\ldots n}
\end{eqnarray}
Combining all the results together
$\bigcirc{ }\!\!\!\!\mbox{\small 1}+\bigcirc{ }\!\!\!\!\mbox{\small 2}+\bigcirc{ }\!\!\!\!\mbox{\small 3} + \bigcirc{ }\!\!\!\!\mbox{\small 4}=0$,
the evolution equation for n-th order fluid moment $\bX^{(n)}_{ij\ldots n}$ reads (written in the vector notation without indices)
\begin{eqnarray}
&&  \frac{\pr}{\pr t} \bX^{(n)} + \nabla\cdot \Big( \bX^{(n+1)}+\bu \bX^{(n)} \Big) +\bigg[ \frac{\pr\bu}{\pr t}\bX^{(n-1)} 
   +  \Big( \bX^{(n)} + \bX^{(n-1)}\bu \Big)\cdot \nabla \bu  -\frac{q}{m} \bE \bX^{(n-1)} \nn\\
&&\qquad  +\frac{q}{mc} \bb\times\Big( \bX^{(n)} +\bu\bX^{(n-1)} \Big) \bigg]^S =0. \label{eq:GenTensorG}
\end{eqnarray}  
The final result should have only time derivative of $\bX^{(n)}$ and similarly to the heat flux tensor, we need to use the momentum equation to eliminate $\pr_t\bu$.
We subtract a total of n momentum equations combined with $\bX^{(n)}$ in the form
\begin{eqnarray}
&& \Big[ \Big(  \frac{\pr\bu}{\pr t} + \bu\cdot\nabla\bu + \frac{1}{\rho} \nabla\cdot\bp - \frac{q}{m}\bE
    - \frac{q}{mc}\bu\times\bb \Big)\bX^{(n-1)} \Big]^S=0.
\end{eqnarray}
Again, because of the symmetric operator, it does not matter if the $\bX^{(n)}$ is applied to the momentum equation from the left or right.
The subtraction also conveniently eliminates the electric field, however it introduces $\nabla\cdot\bp$.
We also need an identity
\begin{eqnarray}
\Big[ (\bu\times\bb)\bX^{(n-1)}\Big]_{ijk\ldots n} &=& (\bu\times\bb)_i X^{(n-1)}_{jk\ldots n} = -(\bb\times\bu)_i X^{(n-1)}_{jk\ldots n} = -\Big[ (\bb\times\bu) \bX^{(n-1)} \Big]_{ijk\ldots n} \nn\\
  &=& -\Big[ \bb\times \Big( \bu \bX^{(n-1)} \Big)\Big]_{ijk\ldots n},
\end{eqnarray}  
where in the last step we emphasized that it does not matter if the $\bb\times$ applies only to $\bu$ or to the entire tensor $\bu\bX^{(n-1)}$, since the vector product operates
only through the first component of the tensor anyway. The final evolution equation reads
\begin{equation}
\boxed{  
  \frac{\pr}{\pr t} \bX^{(n)} +\nabla\cdot \big( \bX^{(n+1)}+\bu\bX^{(n)} \big) +\Big[ \bX^{(n)}\cdot\nabla\bu
    +\frac{q}{mc}\bb\times\bX^{(n)}-\frac{1}{\rho} (\nabla\cdot\bp) \bX^{(n-1)} \Big]^S =0,} \label{eq:GenTensor}
\end{equation}  
and it is valid for $n\ge 2$. Evaluation at $m=3$ recovers the heat flux tensor equation (\ref{eq:HFgenF}) and evaluation at $m=2$ recovers the pressure tensor
equation (\ref{eq:PtensorF}). Note that $\bX^{(4)}=\br$, $\bX^{(3)}=\bq$, $\bX^{(2)}=\bp$, however, and very importantly,
according to our definition (\ref{eq:Xmom}), $\bX^{(1)}=0$ and it is \emph{not} equal to the velocity $\bu$, which is defined according
to (\ref{eq:defVel}).
Additionally, $\bX^{(0)}=\rho$. 
The momentum equation can still be recovered by evaluating (\ref{eq:GenTensorG}) at $n=1$, and by dividing by $\rho$
(where naturally the symmetric operator does not have any influence on a vector, and $u_i^S=u_i$). Similar equation to our (\ref{eq:GenTensorG})
was also obtained for example by \cite{Waelbroeck2010}, eq. 22, even though we did not verify if it is consistent with ours since $\bX^{(n)}$ in that work
is defined with $v$ instead of our $c$.

We want to always decompose the tensor $\bX^{(n)}$ to its gyrotropic and nongyrotropic part, and to have a space to write the ``g'' and ``ng'', lets move
the index $(n)$ down, so that
\begin{equation}
\bX_{(n)} = \bX_{(n)}^{\textrm{g}}+ \bX_{(n)}^{\textrm{ng}}.
\end{equation}
By generalizing the results we have seen for the gyrotropic $\bp^{\textrm{g}},\bq^{\textrm{g}},\br^{\textrm{g}}$, the same result is obtained for the
n-th moment, and at low frequencies the gyrotropic part must satisfy
\begin{equation}
\Big[ \bhat\times \bX_{(n)}^{\textrm{g}}\Big]^S = 0,
\end{equation}
which can be indeed viewed as a definition of gyrotropy. The general (\ref{eq:GenTensor}) therefore rewrites 
\begin{eqnarray}
  \frac{\pr}{\pr t} \bX_{(n)} +\nabla\cdot \big( \bX_{(n+1)}+\bu\bX_{(n)} \big) +\Big[ \bX_{(n)}\cdot\nabla\bu
    +\Omega\frac{|\bb|}{B_0}\bhat\times\bX_{(n)}^{\textrm{ng}}-\frac{1}{\rho} (\nabla\cdot\bp) \bX_{(n-1)} \Big]^S =0, \label{eq:GenTensor2}
\end{eqnarray}
yielding an implicit equation for the non-gyrotropic part in the form
\begin{eqnarray}
  \Big[\bhat\times\bX_{(n)}^{\textrm{ng}}\Big]^S = -\frac{B_0}{\Omega|\bb|} \Big\{ \frac{\pr}{\pr t} \bX_{(n)}
  +\nabla\cdot \big( \bX_{(n+1)}+\bu\bX_{(n)} \big) +\Big[ \bX_{(n)}\cdot\nabla\bu
    -\frac{1}{\rho} (\nabla\cdot\bp) \bX_{(n-1)} \Big]^S \Big\}. \label{eq:XngPica}
\end{eqnarray}
Ideally, an ``inversion procedure'' should be found for the l.h.s., so that an equation for $\bX_{(n)}^{\textrm{ng}}$ can be found, and then by expanding
the r.h.s. (for example at the leading order by considering only gyrotropic $\bX^{\textrm{g}}$), a fully nonlinear expressions
for $\bX_{(n)}^{\textrm{ng}}$ can be obtained. Nevertheless, we have seen that the inversion procedure was already quite complicated for the matrix
$(\bhat\times\boldsymbol{\Pi})^S$, see eq. (\ref{eq:PIextract}).
For a tensor of 3rd-rank, such as the $(\bhat\times\bq^{\textrm{ng}})^S$, the full inversion procedure is actually not addressed in this
guide and an advanced reader is referred to eq. 43 of \cite{Ramos2005}. The inversion procedure was obtained only for a part of the $\bq^{\textrm{ng}}$, that can
be decomposed to the non-gyrotropic heat flux vectors $\boldsymbol{S}^\parallel_\perp$, $\boldsymbol{S}^\perp_\perp$, which is done
in detail in the Appendix  \ref{sec:NONGheat}.
For the rest of the $\bq^{\textrm{ng}}$ that consists of heat flux tensor $\boldsymbol{\sigma}$, the inversion procedure cannot be avoided.
Additionally, we have seen that for 4th-order moment $\br^{\textrm{ng}}$, the eq. (\ref{eq:XngPica}) does not seem to yield anything useful even at the linear
level, and a different procedure was used, where the specific bi-Maxwellian distribution was expanded in Hermite polynomials.
Therefore, for higher-order moments, the eq. (\ref{eq:XngPica}) is immensely complicated.

It is useful to briefly consider, how many scalar gyrotropic moments has an n-th order tensor $\bX_{(n)}$.
The scalar gyrotropic moments can be constructed only from combinations of $c_\parallel$ and $c_\perp^2$. It is beneficial to write down the
following self-explanatory table
\begin{equation}
\begin{tabular}{| c | l | l | c |}
\hline
$\bX_{(n)}$  & possible combinations of $c_\parallel, c_\perp^2$          & common names                          & number \\
\hline  
  $0$      &  $1$                                                    &  $n$                                  & $1$ \\
  $1$      &  $c_\parallel$                                            &  $u_\parallel$                          & $1$ \\
  $2$      &  $c_\parallel^2;\quad c_\perp^2$                            &  $p_\parallel;\; p_\perp$                   & $2$ \\
  $3$      &  $c_\parallel^3;\quad c_\parallel c_\perp^2$                  &  $q_\parallel;\; q_\perp$                   & $2$ \\
  $4$      &  $c_\parallel^4;\quad c_\parallel^2 c_\perp^2;\quad c_\perp^4$  & $r_{\parallel\parallel};\; r_{\parallel\perp};\; r_{\perp\perp}$ & $3$ \\
  $5$      &  $c_\parallel^5;\quad c_\parallel^3 c_\perp^2;\quad c_\parallel c_\perp^4$ &                              & $3$ \\
  $6$      &  $c_\parallel^6;\quad c_\parallel^4 c_\perp^2;\quad c_\parallel^2 c_\perp^4;\quad c_\perp^6$ &              & $4$ \\
  $7$      &  $c_\parallel^7;\quad c_\parallel^5 c_\perp^2;\quad c_\parallel^3 c_\perp^4;\quad c_\parallel c_\perp^6$ &    & $4$ \\
  $8$      &  $c_\parallel^8;\quad c_\parallel^6 c_\perp^2;\quad c_\parallel^4 c_\perp^4;\quad c_\parallel^2 c_\perp^6; \quad c_\perp^8$  & & $5$ \\
  $9$      &  $c_\parallel^9;\quad c_\parallel^7 c_\perp^2;\quad c_\parallel^5 c_\perp^4;\quad c_\parallel^3 c_\perp^6; \quad c_\parallel c_\perp^8$  & & $5$ \\
\hline
\end{tabular} \nn
\end{equation}
Obviously, the number of gyrotropic moments for tensor $\bX_{(n)}$ is equal to $1+\textrm{int}[n/2]$. The function ``int'' means integer part. In mathematics,
the integer function is often abbreviated with square brackets, as $1+[n/2]$. However, this notation might be possibly confusing (we used square brackets in many other
places), and we prefer to write explicitly ``int''. Another possibility would be to use the ``floor'' function.

\subsection{Hierarchy of moments in 1D geometry (electrostatic)}
Let's see if we can use the n-th order moment equation (\ref{eq:GenTensor}) for something useful, since the equation is immensely complicated. 
Let's consider the special case of strictly parallel propagation along $B_0$ (in the z-direction), where all the moments (including the velocity) are along $B_0$,
i.e. let's consider the electrostatic case. Let's change the index of a moment from $(n)$ to $(l)$, so that the index is not confused with the number density.
Let's stop writing the $\parallel$ subscript on all the variables starting from the pressure, so $p_\parallel$, $q_\parallel$, $r_{\parallel\parallel}$ are simply
$p,q,r$ and the l-th order parallel moment is just $X^{(l)}\equiv m\int c_\parallel^l f dv$. We keep the $z$-subscript for $u_z$ to remind us we are writing 1D equations.
In 1D geometry, the nonlinear evolution equation (\ref{eq:GenTensor}) simplifies to
\begin{equation} \label{eq:Nth1Dnonlinear}
 \frac{\pr}{\pr t} X^{(l)} +\pr_z \Big( X^{(l+1)} +u_z X^{(l)} \Big) +l (\pr_z u_z) X^{(l)} - \frac{l}{\rho} (\pr_z p) X^{(l-1)} =0; \qquad l\ge 2.
\end{equation}
If one wants to be more general, instead of propagation along $B_0$, it is possible to consider propagation along magnetic field line $\bhat$, and
write the same equation with $u_\parallel$ and $\pr_\parallel$. The equation is accompanied by the density and momentum equations
\begin{eqnarray}
  && \frac{\pr}{\pr t} \rho +\pr_z \big( u_z \rho \big) = 0; \label{eq:1D_pic0}\\
  && \frac{\pr}{\pr t} u_z +u_z\pr_z u_z +\frac{1}{\rho}\pr_z p -\frac{q_r}{m} E_z  =0, \label{eq:1D_pic1}
\end{eqnarray}
where for the charge $q_r$ it is useful to keep the species index $r$, so that it is not confused with the heat flux $q$. 
The last term disappears anyway since in the case considered here $E_z=0$.
Evaluating equation (\ref{eq:Nth1Dnonlinear}) for $l=2,3,4,5,6\ldots$ yields
\begin{eqnarray}
&& \frac{\pr}{\pr t} p +\pr_z \Big( q +u_z p \Big) +2 (\pr_z u_z) p  =0; \label{eq:1D_pic2}\\
&& \frac{\pr}{\pr t} q +\pr_z \Big( r +u_z q \Big) +3 (\pr_z u_z) q - \frac{3}{\rho} (\pr_z p) p =0; \label{eq:1D_pic3}\\
&& \frac{\pr}{\pr t} r +\pr_z \Big( X^{(5)} +u_z r \Big) +4 (\pr_z u_z) r - \frac{4}{\rho} (\pr_z p) q =0; \label{eq:1D_pic4}\\
&& \frac{\pr}{\pr t} X^{(5)} +\pr_z \Big( X^{(6)} +u_z X^{(5)} \Big) +5 (\pr_z u_z)X^{(5)}  - \frac{5}{\rho} (\pr_z p) r =0;\\
&& \frac{\pr}{\pr t} X^{(6)} +\pr_z \Big( X^{(7)} +u_z X^{(6)} \Big) +6 (\pr_z u_z)X^{(6)}  - \frac{6}{\rho} (\pr_z p) X^{(5)} =0, 
\end{eqnarray}
and so on. The nonlinear equation (\ref{eq:Nth1Dnonlinear}) can be linearized, in the first step by assuming zero mean $u_{z0}$ value, yielding
\begin{equation}
 \frac{\pr}{\pr t} X^{(l)} +\pr_z X^{(l+1)} +(l+1) X^{(l)}_0 (\pr_z u_z) - \frac{l}{\rho_0} X^{(l-1)}_0 (\pr_z p) =0,
\end{equation}
and calculating it for $l=2-8$ and by prescribing $q_0=0$, $X_0^{(5)}=0$ and in general for $l$ odd $X_0^{(l)}=0$, yields
\begin{eqnarray}
&&  \frac{\pr}{\pr t} p +\pr_z q +3 p_0 (\pr_z u_z)=0;\\
&&  \frac{\pr}{\pr t} q +\pr_z r - 3\frac{p_0}{\rho_0} (\pr_z p) =0;\\
&&  \frac{\pr}{\pr t} r +\pr_z X^{(5)} +5 r_0 (\pr_z u_z) =0;\\
&&  \frac{\pr}{\pr t} X^{(5)} +\pr_z X^{(6)}  - 5 \frac{r_0}{\rho_0} (\pr_z p) =0; \label{eq:NCGLX5n}\\
&&  \frac{\pr}{\pr t} X^{(6)} +\pr_z X^{(7)} +7 X^{(6)}_0 (\pr_z u_z) =0;\label{eq:NCGLX6n}\\
&&  \frac{\pr}{\pr t} X^{(7)} +\pr_z X^{(8)}  - 7 \frac{X^{(6)}_0}{\rho_0} (\pr_z p) =0;\label{eq:NCGLX7n}\\
&&  \frac{\pr}{\pr t} X^{(8)} +\pr_z X^{(9)} +9 X^{(8)}_0 (\pr_z u_z) =0.  \label{eq:NCGLX8n}
\end{eqnarray}
Alternatively, we can split the linearized equations for even and odd $l$ according to
\begin{eqnarray}
&&  l=\textrm{even}: \qquad \frac{\pr}{\pr t} X^{(l)} +\pr_z X^{(l+1)} +(l+1) X^{(l)}_0 (\pr_z u_z) =0; \label{eq:LevenLin}\\
&&  l=\textrm{odd}: \qquad  \frac{\pr}{\pr t} X^{(l)} +\pr_z X^{(l+1)} - \frac{l}{\rho_0} X^{(l-1)}_0 (\pr_z p) =0. \label{eq:LevenLin2}
\end{eqnarray}
When closing the system, one has to be careful how to deal with $l$=even quantities, since we want to separate the ``deviations'', similarly as
we did for $r=3p^2/\rho+\widetilde{r}$. For a Maxwellian $f_0$ and $l\ge 4$ and even, we therefore write
\begin{equation}
l=\textrm{even}: \qquad X^{(l)} = \frac{l-1}{\rho} p X^{(l-2)}+\widetilde{X}^{(l)}, 
\end{equation}
or equivalently
\begin{equation}
l=\textrm{even}: \qquad X^{(l)} = (l-1)!!\frac{p^{\frac{l}{2}}}{\rho^{\frac{l}{2}-1}}+\widetilde{X}^{(l)},
\end{equation}
so for example $r_0=3p_0^2/\rho_0$, $X_0^{(6)}=15p_0^3/\rho_0^2$, $X_0^{(8)}=105p_0^4/\rho_0^3$, and linearization of the l-th order moment yields
\begin{eqnarray} \label{eq:Xnlin}
l=\textrm{even}: \qquad X^{(l)} &\overset{\textrm{\tiny lin}}{=}& X_0^{(l)} \Big[ \frac{l}{2}\frac{p}{p_0} - \Big(\frac{l}{2}-1\Big)\frac{\rho}{\rho_0} \Big] +\widetilde{X}^{(l)};\\
&\overset{\textrm{\tiny lin}}{=}& X_0^{(l)} \Big[ \frac{l}{2}\frac{T}{T_0} +\frac{\rho}{\rho_0} \Big] +\widetilde{X}^{(l)}.
\end{eqnarray}
Using (\ref{eq:Xnlin}) in the first term of $l$=even equation (\ref{eq:LevenLin}), and using the pressure and density equations,
yields linear evolution equation for deviations of even moments
\begin{equation}
l=\textrm{even}: \qquad \frac{\pr}{\pr t} \widetilde{X}^{(l)} +\pr_z X^{(l+1)} -X_0^{(l)}\frac{l}{2p_0}\pr_z q =0,
\end{equation}
and using (\ref{eq:Xnlin}) in the second term of $l$=odd equation (\ref{eq:LevenLin2}), and $T/T_0=p/p_0-\rho/\rho_0$
yields linear evolution equation for deviations of odd moments
\begin{equation}
  l=\textrm{odd}: \qquad \frac{\pr}{\pr t} X^{(l)} +\pr_z \widetilde{X}^{(l+1)} +X_0^{(l-1)}\frac{l(l-1)}{2m}\pr_z T=0.
\end{equation}
Finishing the calculation with evaluations of $X_0^{(l)}$ values, the evolution equations of deviations are
\begin{eqnarray}
  l=\textrm{even}: \qquad \frac{\pr}{\pr t} \widetilde{X}^{(l)} +\pr_z X^{(l+1)} -(l-1)!!\frac{l}{2}\Big(\frac{p_0}{\rho_0} \Big)^{\frac{l}{2}-1} \pr_z q =0;\\
  l=\textrm{odd}: \qquad \frac{\pr}{\pr t} X^{(l)} +\pr_z \widetilde{X}^{(l+1)} +l!! \frac{(l-1)}{2} n_0 \Big(\frac{p_0}{\rho_0}\Big)^{\frac{l-1}{2}}  \pr_z T=0, 
\end{eqnarray}
or by using $p_0/\rho_0=v_{\textrm{th}}^2/2$, equivalently
\begin{eqnarray}
  l=\textrm{even}: \qquad \frac{\pr}{\pr t} \widetilde{X}^{(l)} +\pr_z X^{(l+1)} -(l-1)!!\frac{l}{2^{\frac{l}{2}}} v_{\textrm{th}}^{l-2}  \pr_z q =0;\\
  l=\textrm{odd}: \qquad \frac{\pr}{\pr t} X^{(l)} +\pr_z \widetilde{X}^{(l+1)} +l!! \frac{(l-1)}{2^{\frac{l+1}{2}}} n_0 v_{\textrm{th}}^{l-1}  \pr_z T=0, 
\end{eqnarray}
Continuing to higher orders, the hierarchy of deviations of Maxwellian moments therefore evaluates
\begin{eqnarray}
  &&  \frac{\pr}{\pr t} q +\pr_z \widetilde{r} +\frac{3}{2}n_0 v_{\textrm{th}}^{2} \pr_z T=0;\\
  &&  \frac{\pr}{\pr t} \widetilde{r} +\pr_z X^{(5)} -3 v_{\textrm{th}}^{2} \pr_z q =0;\\
  &&  \frac{\pr}{\pr t} X^{(5)} +\pr_z \widetilde{X}^{(6)} +\frac{15}{2}n_0 v_{\textrm{th}}^{4} \pr_z T=0;\\
  &&  \frac{\pr}{\pr t} \widetilde{X}^{(6)} +\pr_z X^{(7)} -\frac{45}{4}v_{\textrm{th}}^{4} \pr_z q =0;\\ 
  &&  \frac{\pr}{\pr t} X^{(7)} +\pr_z \widetilde{X}^{(8)} +\frac{315}{8}n_0 v_{\textrm{th}}^{6} \pr_z T=0;\\
  &&  \frac{\pr}{\pr t} \widetilde{X}^{(8)} +\pr_z X^{(9)} -\frac{105}{2}v_{\textrm{th}}^{6} \pr_z q =0.  
\end{eqnarray}
The equations are written in physical units, and these results will be useful in Part 2, where we will calculate Landau fluid closures for deviations of various moments.  
\subsection{Impossibility to go beyond CGL2 without Landau fluid closures} \label{sec:Impossible}
One can analyze the dispersion relations easily in physical units, but since in this Part 1 we used normalized units with
$\widetilde{k}=k V_A/\Omega_p$, $\widetilde{\omega}=\omega/\Omega_p$ and $\beta=v_{\textrm{th}}^2/V_A^2$ in almost all dispersion relations,
let's rewrite the fluid hierarchy to normalized units, so that everything feels more familiar. The $X^{(5)}$ is normalized with $p_{0}V_A^3$,
the $X^{(6)}$ with $p_{0}V_A^4$, and $X^{(n)}$ with $p_{0}V_A^{n-2}$. By dropping the normalization tilde as usual
(but obviously not for even moments such as $\widetilde{r}$, that should be perhaps called $\delta r$), the equations in Fourier space read 
\begin{eqnarray}
&& -\omega \rho  + k u_z =0;\label{eq:1D-1}\\
&& -\omega u_z +\frac{\beta}{2}k p =0;\\
&& -\omega p +3 k u_z + k q = 0; \label{eq:1D-3}\\
&& -\omega q +k \widetilde{r} + \frac{3}{2}\beta k (p-\rho) =0; \label{eq:1D-4}\\
&& -\omega \widetilde{r} +k X^{(5)} -3\beta k q =0; \label{eq:1D-5}\\
&& -\omega X^{(5)} +k \widetilde{X}^{(6)} +\frac{15}{2} \beta^2 k (p-\rho)=0; \label{eq:1D-6}\\ 
&& -\omega \widetilde{X}^{(6)} +k X^{(7)} -\frac{45}{4} \beta^2 k q =0. \label{eq:1D-7}
\end{eqnarray}
Now, by using just first 3 equations (\ref{eq:1D-1})-(\ref{eq:1D-3}), and closing the system with a closure $q=0$, yields the CGL model,
 with solution
\begin{equation}
\textrm{CGL:} \qquad \omega= \pm k \sqrt{\beta \frac{3}{2}},
\end{equation}
the familiar CGL ion-acoustic mode. By going higher in the fluid hierarchy and using the first 4 equations (\ref{eq:1D-1})-(\ref{eq:1D-4}), 
and closing the system with $\widetilde{r}=0$ (which is equivalent to prescribing $r=3p^2/\rho$), yields the CGL2 model.
The dispersion relation is $\omega^4-3\beta k^2\omega^2 +\frac{3}{4}\beta^2 k^4=0$, with solutions
\begin{eqnarray}
  \textrm{CGL2:} \qquad \omega &=& \pm k \sqrt{\beta\Big(\frac{3}{2}+\sqrt{\frac{3}{2}} \Big)};\\
  \omega &=& \pm k \sqrt{\beta\Big(\frac{3}{2}-\sqrt{\frac{3}{2}} \Big)},
\end{eqnarray}
the already obtained (\ref{eq:CGL2_rit1}), (\ref{eq:CGL2_rit3}). The heat flux fluctuations in the CGL2 model therefore did ``split'' the
CGL ion-acoustic mode to two modes.  One might assume that this will always be the case when going higher and higher in the fluid hierarchy.
  With the next example we demonstrate that this does not happen. Let us consider fluctuations in $\widetilde{r}$, and use the first 5 equations
(\ref{eq:1D-1})-(\ref{eq:1D-5}) with the closure $X^{(5)}=0$. We might call this model CGL3.
The dispersion relation reads $\omega^4=\frac{15}{4}\beta^2 k^4$, and the solutions are
\begin{eqnarray}
  \textrm{CGL3:} \qquad \omega = \pm k  \sqrt{\beta} \Big(\frac{15}{4} \Big)^{1/4}; \qquad \omega = \pm i k  \sqrt{\beta} \Big(\frac{15}{4} \Big)^{1/4}.
\end{eqnarray}
Importantly, the last two modes are imaginary, and one is \emph{unstable}. Therefore, such a fluid model does not make any physical sense.
Why did this happen ? For example, expressing everything through pressure fluctuations, and putting the $\widetilde{r}$ fluctuations on the r.h.s yields
\begin{eqnarray}
  \Big( \omega^4-3\beta k^2\omega^2 +\frac{3}{4}\beta^2 k^4 \Big) p &=& k^2 \omega^2 \widetilde{r};\label{eq:Ncgl_pic1}\\
   &=& k^3 \omega X^{(5)} +\Big( -3\beta k^2 \omega^2 + \frac{9}{2}\beta^2 k^4 \Big) p.\label{eq:Ncgl_pic2}
\end{eqnarray}
Therefore, if the $\widetilde{r}=0$, so when (\ref{eq:Ncgl_pic1}) is used, the CGL2 model is obtained. However, when fluctuations in $\widetilde{r}$ are
considered, so when (\ref{eq:Ncgl_pic2}) is used, the term $-3\beta k^2\omega^2$ cancels on both sides, with the resulting dispersion
relation
\begin{equation}
 \Big( \omega^4 - \frac{15}{4}\beta^2 k^4 \Big) p = k^3\omega X^{(5)},
\end{equation}  
and by prescribing closure $X^{(5)}=0$, yields the dispersion relation of the CGL3 model.

Now it is easy to go higher in the fluid hierarchy. By considering the first 6 equations (\ref{eq:1D-1})-(\ref{eq:1D-6}), the dispersion relation
reads
\begin{equation}
 \Big( \omega^6 - \frac{45}{4}\beta^2 k^4 \omega^2 +\frac{15}{4}\beta^3 k^6 \Big) p = k^4\omega^2 \widetilde{X}^{(6)},
\end{equation}
and by prescribing  closure $\widetilde{X}^{(6)}=0$ (which is equivalent to prescribing $X^{(6)}=15p^3/\rho^2$), yields a model that we can call CGL4, with
numerical solutions
\begin{equation}
\textrm{CGL4:}\qquad \frac{\omega}{k\sqrt{\beta}}= \pm 0.58; \qquad \frac{\omega}{k\sqrt{\beta}}=\pm 1.75; \qquad \frac{\omega}{k\sqrt{\beta}}=\pm 1.87i,
\end{equation}
again, two modes being imaginary, and one unstable.

The variable $\omega/(k\sqrt{\beta})$, written here in normalized units, is easily rewritten to physical units (reintroducing tilde for clarity)
\begin{equation}
\frac{\widetilde{\omega}}{\widetilde{k}\sqrt{\beta}} = \frac{\frac{\omega}{\Omega_p}}{\frac{kV_A}{\Omega_p}\sqrt{\frac{v_{\textrm{th}}^2}{V_A^2}}} = \frac{\omega}{kv_{\textrm{th}} },
\end{equation}
which is directly related to the ``kinetic'' variable $\zeta$, that we will use in Part 2 of the text, defined as
\begin{equation}
\zeta \equiv \frac{\omega}{|k| v_{\textrm{th}}},
\end{equation}
where the absolute value does not make any difference now (since all fluid models here contain $\zeta^2$),
and will become important only when Landau fluid closures are considered.
The hierarchy of CGL dispersion relations then can be written as
\begin{eqnarray}
  \textrm{CGL:} &&\qquad \zeta^2 - \frac{3}{2}=0; \qquad \label{eq:CGLzeta}\\
  \textrm{CGL2:}&&\qquad \zeta^4 - 3\zeta^2 +\frac{3}{4}=0;\\
  \textrm{CGL3:}&&\qquad \zeta^4 - \frac{15}{4}=0;\\
  \textrm{CGL4:}&&\qquad \zeta^6 - \frac{45}{4}\zeta^2+\frac{15}{4}=0;\\
  \textrm{CGL5:}&&\qquad \zeta^6 - \frac{105}{8}=0;\\
  \textrm{CGL6:}&&\qquad \zeta^8 - \frac{105}{2}\zeta^2+\frac{315}{16}=0;\\
  \textrm{CGL7:}&&\qquad \zeta^8 - \frac{945}{16}=0. \label{eq:CGL7zeta}
\end{eqnarray}
All the models beyond CGL2 contain modes that are unstable.   
\subsubsection*{Double-check}
Let's double check our results, to verify that we calculated everything correctly. As stated previously, a closure should be performed only at the
last retained fluid moment, and this is especially true when nonlinear numerical simulations are performed.\footnote{An alternative
  approach that would describe the Landau damping very precisely, is to keep the fully nonlinear CGL2 equations,
  and supplement them with a hierarchy of time-dependent linear equations for higher-order moments, closed with an appropriate Landau fluid closure. }
The CGL and CGL2 systems were already discussed at great length. Let's check the CGL3 dispersion relation, and let's work directly in physical units.
The system of equations that we have in mind reads
\begin{eqnarray}
 && \frac{\pr}{\pr t} \rho +\rho_0 \pr_z  u_z = 0; \nn \\
 && \frac{\pr}{\pr t} u_z +\frac{1}{\rho_0}\pr_z p =0; \nn \\
 &&  \frac{\pr}{\pr t} p +\pr_z q +3 p_0 (\pr_z u_z)=0;\nn \\
 &&  \frac{\pr}{\pr t} q +\pr_z r - 3\frac{p_0}{\rho_0} (\pr_z p) =0;\nn \\
 &&  \frac{\pr}{\pr t} r +\pr_z X^{(5)} +5 r_0 (\pr_z u_z) =0, \label{eq:SystemDouble}
\end{eqnarray}
and the closure is performed by setting $X^{(5)}=0$. Importantly, the 4th-order moment $r$ is kept undetermined, and it is not separated to its ``core''
and $\widetilde{r}$, nor any specific form of a distribution function is prescribed yet (other than the mean values of odd moments $X_0^{(l)}=0$).
Calculating dispersion relation of this system (a matrix multiplied by vector $(\rho,u_z,p,q,r)$ ) yields result $\omega^4-5k^4r_0/\rho_0=0$, which for  
$r_0=3p_0^2/\rho_0$ recovers the CGL3 dispersion relation. Our calculations were therefore done right. Similarly, double checking the CGL4 dispersion,
the system (\ref{eq:SystemDouble}) is supplemented with equation (\ref{eq:NCGLX5n}),
and a closure is performed with $X^{(6)}=15p^3/\rho$. Again, the $r$ is left untouched, and dispersion relation of this system recovers the CGL4 result.
Checking the CGL5 dispersion relation, the system is supplemented with (\ref{eq:NCGLX6n}),
and a closure $X^{(7)}=0$. Again, the $X^{(6)}$ is kept untouched and generally undetermined. Calculating the dispersion relation yields
$\omega^6-7k^6X_0^{(6)}/\rho_0=0$, and prescribing Maxwellian $X_0^{(6)}=15p_0^3/\rho_0^2$ recovers the CGL5 result.
Going higher in the hierarchy, and using closure $X^{(8)}=105p^4/\rho^3$ recovers the CGL6 results. And finally, keeping all the variables up to $X^{(8)}$
undetermined and using closure $X^{(9)}=0$ yields dispersion relation $\omega^8-9k^8X_0^{(8)}/\rho_0=0$, which for Maxwellian $X_0^{(8)}=105p_0^4/\rho_0^3$
recovers the CGL7 result.

Or in general, without yet performing a closure, and going higher and higher in the hierarchy step by step, a straightforward 
calculation yields
\begin{eqnarray}
  \textrm{CGL:} &&\qquad \Big( -\omega^2 +3\frac{p_0}{\rho_0}k^2 \Big) p = -k\omega q; \label{eq:CGLnpic}\\
  \textrm{CGL2:}&& \qquad -\omega^2 p = -k^2 r;\\
  \textrm{CGL3:}&& \qquad \Big( -\omega^4 +5\frac{r_0}{\rho_0}k^4 \Big) p = -k^3 \omega X^{(5)};\\
  \textrm{CGL4:}&& \qquad -\omega^4 p = -k^4 X^{(6)};\\
  \textrm{CGL5:}&& \qquad \Big( -\omega^6 +7\frac{X_0^{(6)}}{\rho_0}k^6 \Big) p = -k^5 \omega X^{(7)};\\
  \textrm{CGL6:}&& \qquad -\omega^6 p = -k^6 X^{(8)};\\
  \textrm{CGL7:}&& \qquad \Big( -\omega^8 +9\frac{X_0^{(8)}}{\rho_0}k^8 \Big) p = -k^7 \omega X^{(9)},
\end{eqnarray}
and for higher orders therefore
\begin{eqnarray}
  \textrm{CGLl:} &&\; l=\textrm{odd;}\qquad \Big( -\omega^{l+1} +(l+2)\frac{X_0^{(l+1)}}{\rho_0}k^{l+1} \Big) p = -k^l \omega X^{(l+2)};\label{eq:CGLnpic3}\\
  \textrm{CGLl:} &&\; l=\textrm{even;}\qquad -\omega^l p = -k^l X^{(l+2)},\label{eq:CGLnpic2}
\end{eqnarray}  
where the validity can be proven by induction and using (\ref{eq:LevenLin}), (\ref{eq:LevenLin2}). CGLl model (or CGL of l-th order) is therefore defined as performing
a closure on $X^{(l+2)}$ moment. In the equations (\ref{eq:CGLnpic})-(\ref{eq:CGLnpic2}) we used the word CGL, even though no specific distribution function
(nor any closure) was prescribed yet. 

The closure in each model is performed by specifying the r.h.s. of (\ref{eq:CGLnpic3}), (\ref{eq:CGLnpic2}), which yields dispersion relations
\begin{eqnarray}
  \textrm{CGLl:} &&\; l=\textrm{odd;}\quad \textrm{closure}\; X^{(l+2)}=0; \quad => \quad
   \Big(\frac{\omega}{k}\Big)^{l+1} -(l+2)\frac{X_0^{(l+1)}}{\rho_0} = 0; \label{eq:CGLPicaaa}\\
   \textrm{CGLl:} &&\; l=\textrm{even;}\quad \textrm{closure}\; X^{(l+2)} = (l+1)!!\frac{p^{\frac{l+2}{2}}}{\rho^{\frac{l}{2}}}; \quad => \quad
   \Big(\frac{\omega}{k}\Big)^{l+2} -\Big(\frac{l}{2}+1\Big) \frac{X_0^{(l+2)}}{p_0}\Big(\frac{\omega}{k}\Big)^{2} +\frac{l}{2}\frac{X_0^{(l+2)}}{\rho_0}=0.\label{eq:CGLPicaaa2}
\end{eqnarray}
The results are written in physical units. Alternatively, by using the $\zeta$ variable, the results are 
\begin{eqnarray}
 \textrm{CGLl:} &&\; l=\textrm{odd;}\quad => \quad  \zeta^{l+1} -\frac{(l+2)!!}{2^{\frac{l+1}{2}}}= 0;\\
 \textrm{CGLl:} &&\; l=\textrm{even;} \quad => \quad \zeta^{l+2}-(l+1)!! \frac{l+2}{2^{\frac{l}{2}+1}}\zeta^2 +(l+1)!!\frac{l}{2^{\frac{l}{2}+2}}=0,
\end{eqnarray}
which for specific $l$ values recovers results (\ref{eq:CGLzeta})-(\ref{eq:CGL7zeta}). 

The results (\ref{eq:CGLPicaaa})-(\ref{eq:CGLPicaaa2}) are written for a closure performed on moment $X_0^{(l+2)}$, which is dictated by the
CGLl vocabulary, and which might be perhaps a bit confusing. Alternatively, one can reformulate the results for closures performed on
moment $X_0^{(l)}$ (corresponding name for a Maxwellian is $\textrm{CGL}(l-2)$) and that read
\begin{eqnarray}
l=\textrm{odd};l\ge 3 &&\quad \textrm{closure}\; X^{(l)}=0; \quad => \quad \Big(\frac{\omega}{k}\Big)^{l-1} -l \frac{X_0^{(l-1)}}{\rho_0}=0;\\
l=\textrm{even};l\ge 4 && \quad \textrm{closure}\; X^{(l)} = (l-1)!!\frac{p^{\frac{l}{2}}}{\rho^{\frac{l}{2}-1}}; \quad => \quad
  \Big( \frac{\omega}{k}\Big)^{l} -\frac{l}{2} \Big(\frac{\omega}{k}\Big)^{2} \frac{X_0^{(l)}}{p_0} +\Big(\frac{l}{2}-1\Big) \frac{X_0^{(l)}}{\rho_0}=0,
\end{eqnarray}
or rewritten with the $\zeta$ variable
\begin{empheq}[box=\fbox]{align}
l=\textrm{odd};l\ge 3 &\quad \textrm{closure}\; X^{(l)}=0; \quad => \quad \zeta^{l-1} -\frac{l!!}{2^{(l-1)/2}}=0;\nn\\
l=\textrm{even};l\ge 4 & \quad \textrm{closure}\; X^{(l)} = (l-1)!!\frac{p^{\frac{l}{2}}}{\rho^{\frac{l}{2}-1}}; \quad => \quad
\zeta^{l}-\frac{(l-1)!!}{2^{l/2}}\big( l\zeta^2-\frac{l}{2}+1 \big)=0, \label{eq:BiGResult}
\end{empheq}
a result reported in \cite{HunanaPRL2018}, eq. (10).
Importantly, solutions of the above dispersion relations for $l>4$, will always yield results with some complex numbers, and some solutions will be unstable.

Therefore, the last physically meaningful fluid model (without considering Landau fluid closures)
is the CGL2 model, and the last closure is the ``normal'' closure, $r=3p^2/\rho$. To go higher in the fluid hierarchy, we necessarily need to make connection to
kinetic theory, and consider Landau fluid closures. These closures are described in Part 2 of our guide, and for direct comparison with the simplest Landau fluid 
dispersion relations, see Part 2, Section 3.12 ``Parallel ion-acoustic (sound) mode, cold electrons''.

\clearpage
\section{Conclusions}
We offer a brief summary of the major results discussed throughout the text.  
\begin{itemize}  
\item Collisionless plasma is completely described by the kinetic Vlasov equation. By directly integrating the Vlasov equation over velocity space,
  yields evolution equations for an infinite hierarchy of fluid moments. In addition to the usual density and momentum equations, it is possible
  to derive an evolution equation for a general n-th order fluid moment ($n\ge 2$), see eq. (\ref{eq:GenTensor}). Then, evolution equations for the
  pressure tensor and heat flux tensor can be obtained by simply evaluating (\ref{eq:GenTensor}) at $n=2$ and $n=3$. 

\item Even though the fluid hierarchy is infinite, the reformulation of the kinetic description to a fluid formalism is not necessarily complete, 
  since the usual calculations neglect wave-particle resonances in velocity space. These wave-particle resonances that are implicitly present in
  the Vlasov equation are responsible for collisionless damping mechanisms, such as Landau damping, transit-time damping and cyclotron damping.

\item The presence (or absence) of collisionless damping mechanisms in a fluid hierarchy is determined by the type of closure
  selected to truncate that fluid hierarchy.
  We differentiate between two classes of closures: 1) ``Classical'' (non-Landau fluid) closures, that
  neglect wave-particle resonances, and 2) Landau fluid closures, that account for Landau wave-particle resonances and associated
  Landau damping \& transit-time damping, addressed in Part 2. 

\item There are currently no known closures that account for cyclotron wave-particle resonances and associated cyclotron damping.
  Nevertheless, there is a priori no reason why such fluid closures can not be found in the future, at least for the simplified case of
  electromagnetic propagation along the magnetic field (slab geometry). 
  
\item Considering ``classical'' closures, the hierarchy of fluid moments therefore does not contain collisionless damping mechanisms,
  regardless of the order to which the hierarchy is developed.
  Or in another words, collisionless damping is beyond all-orders in the classical hierarchy of fluid moments.

\item In fact, it is impossible to go beyond the 4th-order moment with classical bi-Maxwellian
  fluid closures $\boldsymbol{X}^{(n)}=\boldsymbol{X}^{(n)}_0+\widetilde{\boldsymbol{X}}^{(n)}$, where $\boldsymbol{X}^{(n)}_0$ represents bi-Maxwellian
  value of $\boldsymbol{X}^{(n)}$ and its perturbation $\widetilde{\boldsymbol{X}}^{(n)}=0$, since
  all the fluid models contain higher-order modes that are unstable. This is perhaps the most surprising result discussed in Part 1.
  For a detailed proof, see section \ref{sec:Impossible}, where 1D geometry is considered and
  a bi-Maxwellian closure is used at the n-th order fluid moment, which yields dispersion relation (\ref{eq:BiGResult}). The same dispersion relation
  will be valid in 3D geometry for the propagation along the magnetic field, where additional de-coupled modes will be present as well. For $n>4$, all the fluid models
  contain unphysical instabilities. The same result is expected for other distribution functions.   
  
\item The classical bi-Maxwellian closure at the 4th-order moment reads
  $r_{\parallel\parallel} = 3\frac{p_\parallel^2}{\rho}; r_{\parallel\perp} = \frac{p_\parallel p_\perp}{\rho}; r_{\perp\perp} = 2\frac{p_\perp^2}{\rho}$
  (or equivalently $\widetilde{\boldsymbol{r}}=0$), and is called the ``normal'' closure. Therefore, the
  ``normal'' closure is the last classical fluid closure $(\widetilde{\boldsymbol{X}}^{(n)}= 0$) and beyond the 4th-order moment, Landau fluid
  closures $(\widetilde{\boldsymbol{X}}^{(n)}\neq 0$) are required. 
  
\item From a collisionless (or weakly collisional) perspective, the long-wavelength low-frequency limit of a distribution function is not necessarily
  an isotropic Maxwellian, but a general distribution function that is gyrotropic, i.e. isotropic only in its perpendicular velocity components.
  Fluid moments are therefore typically decomposed into their
  gyrotropic and non-gyrotropic parts. For example the pressure tensor $\bp=\bp^{\textrm{g}}+\boldsymbol{\Pi}$, the heat flux tensor $\bq=\bq^{\textrm{g}}+\bq^{\textrm{ng}}$,
  the 4th-order moment $\br=\br^{\textrm{g}}+\br^{\textrm{ng}}$.
  The non-gyrotropic parts are also referred to as the Finite Larmor Radius (FLR) corrections, since they represent deviations from
  gyrotropy at small spatial scales, when the Larmor radius (gyroradius) is not infinitely small. 

\item A general n-th order fluid moment contains $1+\textrm{int}[n/2]$ scalar gyrotropic moments, where the function ``int'' means integer part. 

\item The pressure tensor is 2nd-order fluid moment and contains two gyrotropic moments, $p_\parallel$ and $p_\perp$.
  Therefore, any fluid description of collisionless (or weakly collisional) plasmas, requires two separate evolution equations for $p_\parallel$ and $p_\perp$. 
  Importantly, from a linear perspective the evolution equations remain different even when the distribution function is isotropic,
  i.e. even when the \emph{mean} pressure values are equal ($p_\parallel^{(0)}= p_\perp^{(0)}$), since pressure \emph{fluctuations}
  in the directions parallel and perpendicular to the local magnetic field remain anisotropic. 
  We often use proton pressure (temperature) anisotropy coefficient $a_p=T_\perp^{(0)}/T_\parallel^{(0)} = p_\perp^{(0)}/p_\parallel^{(0)}$.
  
\item The simplest fluid model that describes collisionless plasma in an adiabatic regime is the CGL description (model) - named after
  Chew-Goldberger-Law \citep{Chew1956}. The model is non-dispersive and obtained by a closure with zero heat flux.  
  The CGL description should be viewed as \emph{collisionless magnetohydrodynamics} (collisionless MHD),
  since the usual MHD description with isotropic scalar pressure is highly-collisional implicitly.
  
\item The CGL pressure tensor is decomposed with respect to the direction of the
  local magnetic field lines according to $\bp^{\textrm{g}}=p_\parallel\bhat\bhat+p_\perp(\boldsymbol{I}-\bhat\bhat)$. 
  It is possible to switch to the reference frame
  of the magnetic field lines, where the pressure tensor is a diagonal matrix. However, it should not be forgotten that in a such a reference frame,
  imposing external (mean) magnetic field is not straightforward. For nonlinear numerical simulations of turbulence and plasma heating,
  the laboratory reference frame is strongly recommended.  
  
\item The two pressure equations of the CGL model can be interpreted as conservation laws for the first and second adiabatic invariants,
  see Section \ref{sec:PhysMeaningCGL}.
  The first adiabatic invariant is a conservation of the magnetic moment of a particle that is periodically gyrating around a mean magnetic field. The second
  adiabatic invariant is a conservation of the average parallel momentum of a particle that is completely trapped and periodically bouncing inside of a magnetic bottle.  

\item Without performing (yet) any kind of closure and leaving the heat flux tensor $\bq$ and FLR pressure tensor $\boldsymbol{\Pi}$ unspecified, it is
  possible to derive rigorously exact evolution equations for $p_\parallel$ and $p_\perp$, see eq. (\ref{eq:Gppar})-(\ref{eq:Gpperp}).
  
\item  Rewriting the pressure equations to an alternative form, eq. (\ref{eq:VelkaPica1})-(\ref{eq:VelkaPica2}), shows that the 
  adiabatic invariants in the CGL model are broken by the inclusion of the Hall-term in the induction equation, gyrotropic heat flux,
  non-gyrotropic FLR corrections to the pressure and heat flux, as well as by coupling of various species together.
  All of these contributions therefore yield very complicated anisotropic plasma heating processes,
  that can be studied only by nonlinear numerical simulations.

\item The anisotropic plasma heating does not simplify much even in the case of periodic boundary conditions (i.e. when the system can be viewed as completely isolated)
  and when averaging over the entire spatial volume is performed and expressed as a conservation of total energy, see eq. (\ref{eq:GpparPer2})-(\ref{eq:EMAGG}).
  Only when the anisotropic heating is neglected, i.e. when only total plasma heating studied, the conservation of energy significantly simplify,
  see eq. (\ref{eq:Ekin_2})-(\ref{eq:Ekin_2x}).  
  
\item By using the concept of polytropic indices, the pressure equations in the ideal CGL model can be interpreted as having $\gamma_\parallel=3$ and $\gamma_\perp=2$.
  Consequently, the CGL dispersion relation is in general not equal to the MHD dispersion relation, even when $a_p=1$. The exception is the
  Alfv\'en mode, which for $a_p=1$ propagates with the same phase speed $\omega/k=V_A\cos\theta$ in both CGL and MHD models, for all the propagation
  directions $\theta$.

\item Polytropic indices can be further related to the number of degrees of freedom $i$ through $\gamma=(i+2)/i$,
  which for the CGL model yields $i_\parallel=1$ and $i_\perp=2$.
  The CGL pressure equations can be therefore viewed as being composed of strongly coupled 1D and 2D dynamics,
  whereas the dynamics in the MHD model with $\gamma=5/3$ and $i=3$ can be viewed as isotropicaly 3-dimensional.   

\item In the CGL model, the Alfv\'en mode
  propagates with the phase speed $\omega/k=V_A\cos\theta\sqrt{1+\frac{\bpar}{2}(a_p-1)}$, which matches the (collisionless) kinetic theory in the long-wavelength limit.
  Consequently, for $a_p=1$, the Alfv\'en mode in the MHD model matches the kinetic theory as well.
  The CGL result is very useful when numerically solving
  kinetic dispersion relations (for example with the WHAMP solver), since the Alfv\'en mode can be easily identified at long-wavelengths. 

\item  In contrast to MHD where the ordering of phase speeds is always $v_s\le v_A\le v_f$ (slow, Alfv\'en, fast), in collisionless
  plasmas the oblique slow mode can become faster than the oblique Alfv\'en mode. In general, oblique slow and fast modes in the CGL model do not
  necessarily match the kinetic theory in the long-wavelength limit. Nevertheless, the effect when $v_s>v_A$ is present in the CGL model
  and exists even for $a_p=1$, see Section \ref{sec:SAreversal}. The effect is very important for the parallel firehose instability. 
  
\item Considering strictly perpendicular propagation ($\theta=90^\circ$ or $k_\parallel=0$), the fast mode in the CGL model propagates with the
  phase speed $\omega/k=V_A\sqrt{1+a_p\bpar}$, which matches the kinetic theory in the long-wavelength limit. The expression can be rewritten
  as $(\omega/k)^2=V_A^2+v_{\textrm{th}\perp}^2=V_A^2+2\frac{p_\perp^{(0)}}{\rho_0}$, and can be contrasted with the MHD result
  $(\omega/k)^2=V_A^2+\frac{5}{3}\frac{p^{(0)}}{\rho_0}$. The factor of 2 in the CGL result comes from $\gamma_\perp=2$, which shows that the
  adiabatic assumption of the CGL model is appropriate for the perpendicular fast mode.  

\item Contributions of the Hall term completely disappear for strictly perpendicular propagation.
  
\item It is more difficult to analytically show where the adiabatic CGL value $\gamma_\parallel=3$ is appropriate,
  since for example the parallel propagating ion-acoustic (sound) mode is strongly Landau
  damped in kinetic theory (unless electrons are hot), which also modifies its real frequency.
  The ion-acoustic mode in the CGL model therefore does not match kinetic theory.
  One needs to consider Landau fluid models with proton and electron
  species, which is addressed in Part 2.
  
\item Nevertheless, when electrons are hot ($T_e\gg T_i$) and the ion-acoustic mode is Landau damped only weakly, the real phase speed obtained from
  kinetic theory in the long-wavelength limit reads $\omega/k=\sqrt{(T_e+3T_i)/m_i}$. Thus, in this example the kinetic theory is matched
  by a fluid model where protons are described adiabatically (with the CGL value $\gamma_\parallel=3$), while electrons are isothermal (with $\gamma_\parallel=1$).
  Also, a good example for $\gamma_\parallel=3$ is the real frequency of the Langmuir mode, $\omega^2=\omega_{pe}^2+3(T_{\parallel e}^{(0)}/m_e)k^2$, which is addressed in Part 2.
  
\item For anisotropic temperatures ($a_p\neq 1$), some modes can become unstable. A threshold of an instability that is obtained
  in the long-wavelength (non-dispersive) limit is called ``hard'' threshold. The CGL model contains 3 instabilities, the oblique firehose
  instability, the parallel firehose instability and the mirror instability. The mirror instability is not described correctly.

\item The oblique firehose instability is an instability of the oblique Alfv\'en mode. Since the Alfv\'en mode is described correctly by the CGL model,
  its instability threshold $1+\frac{\bpar}{2}(a_p-1)<0$ matches the kinetic theory. The instability requires  $a_p<1$ and (in the long-wavelength limit) $\bpar>2$.
  Since the CGL model is non-dispersive, here the instability growth rate only has a simple $\cos\theta$ dependence.
  Once dispersive effects are considered, the instability reaches maximum growth rate at some wavenumber and angle $\theta$ that is oblique,
  see Figure \ref{fig:OF-beta4-ap049}.

\item The parallel firehose instability is an instability of the whistler mode, which can be shown by considering the Hall-CGL model.
    However, considering usual (non-causal) analytic solutions (\ref{eq:Hall-IC1})-(\ref{eq:Hall-IC2}), one arrives at a contradiction that for $\omega_r>0$
    the whistler mode is unstable, and that for $\omega_r<0$ the whistler mode is stable.
    The contradiction arises, because at the range of wavenumbers where the firehose instability exists, the ion-cyclotron and whistler modes are
    completely degenerate and distinguishing between them looses sense. The problem is resolved by introducing a small $(\sim \epsilon)$ causal dissipation into the
    momentum equations, which is later removed by the limit $\lim_{\epsilon\to 0^+}$. The procedure yields causal dispersion relations     
    (\ref{eq:Hall-IC1corr})-(\ref{eq:Hall-IC2corr}), where the whistler mode is firehose unstable for both positive and negative $\omega_r$,
    and the ion-cyclotron mode is stable.

\item The parallel firehose instability occurs for quasi-parallel (small $\theta$) propagation directions,
  with the maximum growth rate at $\theta=0^\circ$, see Figure \ref{fig:OF-beta4-ap049}.
  In the non-dispersive CGL model it corresponds to the instability of the slow mode.
  Its threshold obtained for $\theta=0^\circ$ is equivalent to the oblique firehose instability (since $v_s=v_A$), and matches the kinetic theory.
  At first it might sound surprising that the slow mode connects to
  the whistler mode once dispersive effects are considered. However, considering $\theta=0^\circ$, $\bpar>2$, $a_p<1$
  yields that the fast mode is the ion-acoustic (sound) mode and $v_s=v_A$, see Figure \ref{fig:CGL-th}. 
  After dispersive effects are introduced, which split the Alfv\'en and slow mode to the ion-cyclotron and
  whistler mode, it is clarified that the instability is associated with  the whistler mode. For quasi-parallel propagation directions
  the situation is more obvious, since $v_s>v_A$ (see Figure \ref{fig:CGL-th}), and the slow mode clearly becomes the whistler mode.

\item In the non-dispersive CGL model, all 3 instabilities are non-propagating, i.e. purely growing with zero real frequency.
  Once dispersive effects are introduced, the parallel firehose instability becomes propagating.   
  
\item The mirror instability requires $a_p>1$ and usually develops for highly-oblique propagation angles. In the CGL model it corresponds to the instability
  of the slow mode. Its threshold obtained in the highly-oblique (but not completely perpendicular) limit   
  reads $1+a_p\bpar -\frac{1}{6}a_p^2\bpar <0$, and with respect to kinetic theory the $1/6$ factor is erroneous.
  The erroneous $1/6$ factor shows that the adiabatic CGL closure is not appropriate for the very-slow dynamics of the mirror instability.

\item The simplest fluid closure that recovers the correct threshold of the mirror instability is the ``static'' closure, see Section \ref{sec:StaticClosure}.
  This closure is derived from
  the ``normal'' closure in a simplified quasi-static approximation, and represents a generalization of the isothermal closure (in presence of temperature anisotropy
  and variations of magnetic field strength). The ``static'' closure is very useful since it clarifies, that the highly-oblique very-slow dynamics of the 
  slow mode (and of the associated mirror instability), is better described by a generalized isothermal closure than the adiabatic CGL closure.

\item However, the ``static'' closure modifies the dynamics of the perpendicular fast mode, that now does not match the kinetic theory.
  Additionally, the fast mode now experiences an uphysical instability, even though the instability is beyond the mirror threshold,
  and should perhaps not play a role in numerical simulations. The erroneous result shows that the generalized isothermal closure is not appropriate for the
  relatively fast dynamics of the perpendicular fast mode, where the adiabatic CGL closure is appropriate.
  
\item Heuristically, the correct mirror threshold can be also obtained by keeping the CGL value $\gamma_\perp=2$ and modifying the parallel
  polytropic index to $\gamma_\parallel=1/2$.
  Such a closure does not alter the dynamics of the perpendicular fast mode, since the phase speed depends only on $\gamma_\perp$ (and not
  on $\gamma_\parallel$). Thresholds of parallel and oblique firehose instability are not altered either. Nevertheless, the $\gamma_\parallel=1/2$ is
  even below the isothermal value $\gamma_\parallel=1$, the closure is heuristic, and it is mentioned only as a curiosity. 
  
\item  Obviously, to correctly capture both the slow dynamics of the highly-oblique mirror instability, as well as the fast dynamics of the perpendicular fast mode,
  neither adiabatic nor (generalized) isothermal closures are sufficient. Considering classical (rigorously derived) closures, one has no other choice but to keep the
  correct adiabatic CGL values $\gamma_\parallel=3$, $\gamma_\perp=2$ unaltered, and instead, break the (long-wavelength) adiabaticity by going higher
  in the fluid hierarchy. One needs to consider fluid models with evolution equations for the heat flux tensor $\boldsymbol{q}$,
  that has two gyrotropic components, $q_\parallel$ and $q_\perp$. The system is closed at the 4th-order moment $\boldsymbol{r}$, that have 3 gyrotropic components,
  $r_{\parallel\parallel}; r_{\parallel\perp}; r_{\perp\perp}$. To perform a closure, a specific distribution function (or a class of distribution functions) has to
  be assumed.

\item Considering a bi-Maxwellian distribution function, the closure $\widetilde{\boldsymbol{r}}=0$ is known as the ``normal'' closure (see above).
  We call this non-dispersive bi-Maxwellian fluid model the ``2nd-order CGL'', abbreviated as CGL2. 
  The abbreviation CGL2 seems beneficial, since the name CGL is  
  associated with collisionless magnetohydrodynamics, now just taken to one higher order. Additionally, by showing that all classical bi-Maxwellian models beyond
  CGL2 do not make physical sense, fluid models can be easily classified as based on CGL or CGL2 descriptions.
  For direct comparison of the two systems, see dispersion relations of the CGL model (\ref{eq:SFbigmatrix}) and the CGL2 model (\ref{eq:8thP}).
  The Hall term introduces simplest dispersive effects, see dispersion relations of the Hall-CGL model (\ref{eq:HallCGL-dispR}),
  and the Hall-CGL2 model (\ref{eq:HallCGL2_disp}).

\item  By comparing the dispersion relations, it is obvious that the gyrotropic heat flux \emph{fluctuations} in the CGL2 model
  modify the general dynamics (the mean heat flux values are zero $q_\parallel^{(0)}=q_\perp^{(0)}=0$).
  One of the exceptions is the perpendicular fast mode, which is not altered by the gyrotropic heat fluxes, and therefore described
  correctly. Other exceptions are the oblique Alfv\'en mode (the oblique firehose instability), and once dispersive effects are introduced also
  the parallel propagating ion-cyclotron and whistler modes (the parallel firehose instability).
  
\item Two additional evolution equations typically create two additional modes. For example, for the case of
  highly-oblique CGL2 propagation, one obtains the Alfv\'en mode, the fast mode, and 3 ``slow'' modes. One of these slow modes is responsible for the
  mirror instability, and is sometimes called the mirror mode. The CGL2 mirror threshold matches the kinetic theory. The CGL2 model is therefore the simplest
  classical model that correctly captures (in the long-wavelength limit) both the  mirror instability threshold, as well as the perpendicular fast mode. 
  Nevertheless, only the ``hard'' mirror threshold is correctly recovered,
    and capturing the (long-wavelength) mirror instability growth rate requires Landau damping.
  
\item One can construct closures for other distribution functions than bi-Maxwellians.
  For example, one can consider the bi-Kappa distribution function (\ref{eq:F_biKappa}) with a free parameter $\kappa>3/2$. The closure then reads
  $r_{\parallel\parallel} = 3\alpha_\kappa\frac{p_\parallel^2}{\rho}; r_{\parallel\perp} = \alpha_\kappa\frac{p_\parallel p_\perp}{\rho}; r_{\perp\perp} = 2\alpha_\kappa \frac{p_\perp^2}{\rho}$,
  where the coefficient $\alpha_{\kappa} = (\kappa-\frac{3}{2})/(\kappa-\frac{5}{2})$. The closure is restricted to $\kappa>\frac{5}{2}$, which can be
  seen from the $\alpha_\kappa$ coefficient, and the same requirement is necessary for the convergence of the velocity integrals. We call this closure
  and resulting fluid model simply as ``biKappa''.

\item If a closure for the bi-Kappa distribution is performed at a higher nth-order moment (which is presumably not possible with classical closures since it is expected
  that for $n>4$ all models contain unphysical instabilities), the closure will be restricted to $\kappa>(n+1)/2$, which increases with $n$.     
  The requirement for the minimum $\kappa$ value is there, since the power-law in the bi-Kappa distribution is assumed all the way till infinite velocities,
  and the requirement is necessary for the convergence of the velocity integrals. Therefore, by going higher and higher in the bi-Kappa fluid hierarchy,
  the minimum value of $\kappa$ increases, and in the limit $\kappa\to\infty$ the required distribution function converges towards bi-Maxwellian. 

\item The dispersion relation for the biKappa model is given by (\ref{eq:BiKappa_disper}). A general $\alpha_\kappa$ coefficient is used, so the dispersion
  relation is valid for a much larger class of distribution functions that can be closed with an analogous closure with just one $\alpha_\kappa$ coefficient. 
  The dispersion relation shows that the general dynamics is of course altered by the value of the $\alpha_\kappa$ coefficient. Nevertheless, few exceptions are:
  the oblique Alfv\'en mode (oblique firehose threshold), the perpendicular fast mode, the parallel firehose threshold, and the mirror threshold.
  When dispersive effects are considered, and figures similar to Figure  \ref{fig:OF-beta4-ap049} created, the growth rates for oblique propagation will of 
  course be affected by the $\alpha_\kappa$ value. The Hall-biKappa dispersion relation is given by (\ref{eq:Hall-biKappaD}).
  
\item  The non-gyrotropic (FLR) corrections $\boldsymbol{\Pi}$ to the pressure tensor were studied in Section \ref{section:FLR}.
  The tensor $\boldsymbol{\Pi}$ is described by the pressure tensor equation implicitly. To prevent introducing several new independent fluid variables for 
  components of the tensor $\boldsymbol{\Pi}$, each with its own evolution equation, expansion on temporal and spatial scales is required.
  Correct evaluation of the FLR tensor with respect to magnetic field lines is quite cumbersome, and at the leading order (in temporal and spatial scales)
  one obtains (\ref{eq:FLR_Nice})-(\ref{eq:FLR_Nice2}), or equivalently (\ref{eq:FLR_Ramos})-(\ref{eq:h_Ramos}).
  The FLR tensor $\boldsymbol{\Pi}$ is therefore often evaluated in the linear approximation, which is appropriate when the magnetic field lines
  are not too distorted.

\item  Even though the FLR pressure tensor $\boldsymbol{\Pi}$ has a zero trace, its contributions should not be called
  as ``off-diagonal'', since the diagonal components are non-zero. In the collisionless case considered here with the mean field in the z-direction, only $\Pi_{zz}=0$,
  and $\Pi_{xx}=-\Pi_{yy}\neq 0$. In the collisional case (which was not considered) also $\Pi_{zz}\neq 0$.
  
\item  For the purpose of comparing various contributions,
  we differentiate between FLR1, FLR2 and FLR3 corrections, even though the classification
  can be a bit blurry and there are various possibilities. Perhaps the best definition that we used later is as follows.
  The classical FLR1 tensor $\boldsymbol{\Pi}$ contains only velocity gradients, see (\ref{eq:FLR1s}).
  The FLR2 tensor also contains $\pr \boldsymbol{\Pi} /\pr t$ and the Hall-term from the induction
  equation, see (\ref{eq:FLR2_o4}). It can also contain gyrotropic heat flux contributions, see (\ref{eq:FLR-GyroX}). The FLR3 contains
  non-gyrotropic heat flux vectors $\boldsymbol{S}^{\parallel}_\perp$, $\boldsymbol{S}^{\perp}_\perp$, described in Section \ref{sec:FLR3x},
  with the detailed algebra presented in Appendix \ref{sec:NONGheat}.
  It can also contain the non-gyrotropic heat flux tensor $\boldsymbol{\sigma}$, which we neglected. For complete clarity in analyzing the dispersion
  relations, the entire Hall-CGL-FLR3 model (linearized in the x-z plane, normalized and Fourier transformed) is written down
  in Section \ref{sec:Hall-CGL-FLR3}, see eq. (\ref{eq:HallFLR3-P})-(\ref{eq:SsuperCool1}).

\item Considering the strictly perpendicular fast mode, kinetic theory yields the leading order FLR corrections (in the long-wavelength limit) to the
  phase speed in the following form $(\omega/k)^2 = V_A^2 (1-\frac{1}{8}k^2\rho_i^2) + v_{\textrm{th}\perp}^2 (1-\frac{5}{16} k^2\rho_i^2)$.
  The kinetic result is reproduced by the CGL-FLR3 model (the Hall contributions are zero).
  Importantly, both the first and second order non-gyrotropic heat flux vectors have to be retained.
  
\item  If the second order non-gyrotropic heat flux contributions are neglected in the FLR3 model, yields solution for the perpendicular fast mode (\ref{eq:PerpPIC2}).
  In such a model, the correction to the Alfv\'en speed is captured correctly, however, the correction to the thermal speed is $+\frac{1}{16} k^2\rho_i^2$ instead of
  $-\frac{5}{16} k^2\rho_i^2$, i.e., the correction has a wrong sign.
  This is surprising, since the solution of the FLR2 model, eq. (\ref{eq:DelSarto2}), has a correction to the thermal speed $-\frac{1}{16} k^2\rho_i^2$,
  i.e. at least the sign is correct. Finally, the FLR1 model yields solution (\ref{eq:DelSarto1}), and does not capture any correction to the Alfv\'en speed, additionally,
  the correction to the thermal speed has a wrong sign as well. 

\item We used the Hall-CGL-FLR3 model to investigate the parallel and oblique firehose instability, and compare it with results of the FLR2 and FLR1 models, see Figures
  \ref{fig:firehose-bump}-\ref{fig:both_firehose}. It is shown that the growth rates of the parallel and oblique firehose instability are
  strongly enhanced by the non-gyrotropic heat flux vectors of the FLR3 model, see Figure \ref{fig:OF-beta4-ap049} (see the scales of colorbars).

\item In general, when the maximum growth rates are sufficiently large (let's say $\omega_i/\Omega\sim 0.1$) and not tiny (such as $0.001$), the FLR3 model reproduces the
  kinetic results with unexpectedly good accuracy. See for example bottom panels of Figure \ref{fig:OF-beta4-ap049}. Not only is the value of the maximum growth rate
  for the oblique firehose instability (red lines) approximately the same in the FLR3 model and kinetic theory, the maximum growth rate is also
  reached approximately for the same angle of propagation and the same wavenumber.  
  
\item The oblique firehose instability can be (in general) reproduced with better accuracy than the parallel firehose instability. Nevertheless, for very high $\bpar$ values,
  the parallel firehose instability is reproduced very accurately, see Figure \ref{fig:firehose-beta100}. 

\item  Importantly, we show that the non-gyrotropic heat flux vectors in the FLR3 model,
partially reproduce the large ``bump'' in the imaginary phase speed (growth rate normalized to the wavenumber), when the plasma is close to the
long wavelength limit ``hard'' firehose threshold, see Figure \ref{fig:firehose-bump}. 
The result clearly shows, that similarly to kinetic theory, fluid models can develop firehose instability at small spatial scales, even when they
are stable in the long-wavelength limit. Or in other words, fluid models can become unstable even before the ``hard'' firehose threshold is reached. 
This unexpected result is perhaps the second most surprising result
discussed in Part 1.

\end{itemize}

\vspace{1cm}


\section{Acknowledgments}
We acknowledge support of the NSF EPSCoR RII-Track-1 Cooperative Agreement No. OIA-1655280 ``Connecting the Plasma Universe to Plasma
Technology in Alabama'', led by Gary P. Zank. This work was supported by the European Research Council in the frame
of the Consolidating Grant ERC-2017-CoG771310-PI2FA ``Partial Ionisation: Two-Fluid Approach'', led by Elena Khomenko.  
Anna Tenerani acknowledges support of the NASA Heliophysics Supporting Research Grant \#80NSSC18K1211.
PH thanks Thierry Passot, Monica Laurenza, Nikola Vitas, Petr Hellinger and S. Peter Gary for many useful discussions. 
We are also very thankful to two (out of three) anonymous referees whose comments and suggestions had a great impact on this text.
Significant effort has been made to eliminate all the misprints from the equations.
However, we will amend possible misprints, if found, in a corrigendum.

\clearpage
\appendix
\section{Fourier transformations} \label{sec:AppendixA}
If we have an equation in Real space, we decompose each quantity as a superposition of waves according to
\begin{equation} \label{eq:FourierBoth}
f(\boldsymbol{x},t) = \frac{1}{(2\pi)^4}\int_{-\infty}^\infty \int_{-\infty}^\infty \hat{f}(\boldsymbol{k},\omega) e^{i\boldsymbol{k}\cdot\boldsymbol{x}-i\omega t} d^3 k d\omega.
\end{equation}
 Most of plasma physics books and kinetic papers use $e^{i\boldsymbol{k}\cdot\boldsymbol{x}-i\omega t}$
in the definition (\ref{eq:FourierBoth}). Decomposing the frequency to real and imaginary part $\omega=\omega_r+i\omega_i$, so $e^{-i\omega t}=e^{-i\omega_r t}e^{+\omega_i t}$,
 implies that a wave with $\omega_i>0$ grows (i.e. has a positive growth rate) and a wave with $\omega_i<0$ is damped.  
Of course, an alternative definition of (\ref{eq:FourierBoth}) with $e^{-i\boldsymbol{k}\cdot\boldsymbol{x}+i\omega t}$
 is allowed, one just has to remember that a wave with  $\omega_i>0$ is damped and a wave with $\omega_i<0$ grows. 
  However, in part 2 of the text, we will calculate the Landau damping and we will see that following all the correct minus signs
   in the Landau integral can be very confusing.
  We therefore  recommend to use the first decomposition, where waves with $\omega_i<0$ are damped. This choice
  yields that Fourier transformations in the x-z plane (with $B_0$ in the z-direction) are performed according to a shortcut
\begin{equation}
\frac{\pr}{\pr t}\leftrightarrow -i\omega; \qquad \pr_z\leftrightarrow ik_\parallel; \qquad \pr_x\leftrightarrow ik_\perp. 
\end{equation}
  
The transformation (\ref{eq:FourierBoth}) is technically the inverse/backward Fourier transform $\mathcal{F}^{-1}\hat{f}(\boldsymbol{k},\omega)$. The forward Fourier transform reads
\begin{equation}
f(\boldsymbol{k},\omega) = \int_{-\infty}^\infty \int_{-\infty}^\infty \hat{f}(\boldsymbol{x},t) e^{-i\boldsymbol{k}\cdot\boldsymbol{x}+i\omega t} d^3x dt.
\end{equation}
The location of normalization constants $1/(2\pi)$ is an ad-hoc choice, one just needs to be consistent in using them, especially when calculating convolutions.
\section{Generalized vector (cross) product} \label{sec:AppendixB}
In the collisionless fluid hierarchy one encounters a vector product between vectors and tensors, as for example in the pressure tensor equation (\ref{eq:Ptensor}).
It is useful to clarify the definition of this generalized vector product. Consider matrix $\boldsymbol{A}$,
which is written as a tensor product between vectors $\boldsymbol{a},\boldsymbol{b}$, so $\boldsymbol{A}=\boldsymbol{a}\otimes\boldsymbol{b}=\boldsymbol{a}\boldsymbol{b}$,
where as everywhere in this text, we omit writing the tensor product $\otimes$. Or in the index notation $A_{ij}=a_ib_j$. Now consider vector $\boldsymbol{c}$,
and how can one define a generalized vector product $\boldsymbol{c}\times\boldsymbol{A}$, by using the usual vector product between two vectors.
The natural \emph{definition} is
$\boldsymbol{c}\times\boldsymbol{A} = \boldsymbol{c}\times(\boldsymbol{a}\boldsymbol{b}) \equiv (\boldsymbol{c}\times\boldsymbol{a}) \boldsymbol{b}$, 
or in the index notation 
$(\boldsymbol{c}\times\boldsymbol{A})_{ij} = \epsilon_{ikl} c_k a_l b_j = \epsilon_{ikl} c_k \boldsymbol{A}_{lj}.$ Now let's consider how to define the much more confusing $\boldsymbol{A}\times\boldsymbol{c}$. The natural definition is
$\boldsymbol{A}\times\boldsymbol{c} = (\boldsymbol{a}\boldsymbol{b})\times\boldsymbol{c} \equiv \boldsymbol{a}(\boldsymbol{b}\times\boldsymbol{c})$,
and in the index notation
$(\boldsymbol{A}\times\boldsymbol{c})_{ij} =  a_i \epsilon_{jkl} b_k c_l = \epsilon_{jkl} \boldsymbol{A}_{ik} c_l.$
 Now it is straightforward to show that
\begin{equation}
  \boldsymbol{A}\times\boldsymbol{c} = - (\boldsymbol{c}\times\boldsymbol{A}^T)^T.
\end{equation}  
Generalizing the vector product to higher order tensors is easy  and in the fluid hierarchy we always encounter a vector $\bhat$, and tensors $\bp,\bq,\br$,
  let's therefore write
\begin{equation}
\begin{tabular}{ l  l }
  $\big(\bhat\times\bp \big)_{ij} = \epsilon_{irs} \hat{b}_r p_{sj}$; &   $ \big( \bp\times\bhat  \big)_{ij} = \epsilon_{jrs} p_{ir} \hat{b}_s$;\\
  $\big(\bhat\times\bq \big)_{ijk} = \epsilon_{irs} \hat{b}_r q_{sjk}$; &  $ \big( \bq\times\bhat  \big)_{ijk} = \epsilon_{krs} q_{ijr} \hat{b}_s$;\\
  $\big(\bhat\times\br \big)_{ijkl} = \epsilon_{irs} \hat{b}_r r_{sjkl}$; &  $ \big( \br\times\bhat  \big)_{ijkl} = \epsilon_{lrs} r_{ijkr} \hat{b}_s$.
\end{tabular}
\end{equation}
The vector product is very useful for decomposing quantities to the directions parallel and perpendicular with respect to $\bhat$.
For example, the usual decomposition of the velocity $\bu$ reads
\begin{equation}
\bu = \underbrace{(\bu\cdot\bhat)\bhat}_{\bu_\parallel} + \underbrace{\bu\cdot(\boldsymbol{I}-\bhat\bhat)}_{\bu_\perp}.
\end{equation}
Nevertheless, it is possible to directly obtain $\bu_\perp$, by applying $\bhat\times$ twice on $\bu$, so that 
\begin{equation} \label{eq:Udecomp2}
\bu = (\bu\cdot\bhat)\bhat -\bhat\times(\bhat\times\bu),
\end{equation}
or $\bu_\perp=\bhat\times\bu\times\bhat$. Similar decompositions can be made for tensors.
For example, a general matrix $\boldsymbol{\Pi}$ can be decomposed according to
\begin{equation}
  \boldsymbol{\Pi} = (\boldsymbol{\Pi}\cdot\bhat)\bhat + \boldsymbol{\Pi}\cdot(\boldsymbol{I}-\bhat\bhat),
\end{equation}
where the perpendicular component can be written as $\boldsymbol{\Pi}\cdot(\boldsymbol{I}-\bhat\bhat)=- (\boldsymbol{\Pi}\times\bhat)\times\bhat$,
i.e., where all the $\bhat$ act on $\boldsymbol{\Pi}$ from the right. Alternatively, one can write a decomposition when all the $\bhat$ act on the $\boldsymbol{\Pi}$
from the left, $\boldsymbol{\Pi} = \bhat (\bhat\cdot\boldsymbol{\Pi}) + (\boldsymbol{I}-\bhat\bhat)\cdot\boldsymbol{\Pi}$, where the perpendicular part
$ (\boldsymbol{I}-\bhat\bhat)\cdot\boldsymbol{\Pi}=-\bhat\times(\bhat\times \boldsymbol{\Pi})$.

\section{MHD dispersion relation} \label{sec:AppendixC}
We assume that many people reading this text, will have some previous experience with MHD. It is therefore beneficial to obtain the 
MHD dispersion relation, so that it will be clear how more advanced fluid models are treated in this text. The MHD fluid model reads
\begin{eqnarray}
&& \frac{\pr \rho}{\pr t} + \nabla\cdot (\rho \bu )=0; \nn \\
&& \frac{\pr \bu}{\pr t} +\bu \cdot\nabla \bu +\frac{1}{\rho}\nabla p 
  -\frac{1}{4\pi\rho} (\nabla\times\bb)\times\bb=0; \nn \\
&& \frac{\pr \bb}{\pr t} = \nabla\times(\bu\times\bb); \nn\\  
&& \frac{\pr p}{\pr t} + \bu\cdot\nabla p +\gamma p \nabla\cdot \bu=0, 
\end{eqnarray}
where $\gamma=5/3$. As in advanced fluid models, it is beneficial to normalize the speed with respect to the Alfv\'en speed $V_A$, and normalize the length with
respect to the ion-inertial length $d_i=V_A/\Omega_p$, i.e. by using normalizations (\ref{eq:NormDef})-(\ref{eq:NormDef2}), which in Fourier space yields
$\widetilde{k}=k d_i$, $\widetilde{\omega}=\omega/\Omega_p$. Normalizing the MHD equations and \emph{dropping the tilde} yields unchanged density, induction and pressure
equations, and the momentum equation reads
\begin{equation}
 \frac{\pr \bu}{\pr t} +\bu \cdot\nabla \bu +\frac{p^{(0)}}{\rho_0 V_A^2} \frac{1}{\rho}\nabla p
- \frac{1}{\rho}(\nabla\times\bb)\times\bb=0.\\
\end{equation}
One can decide how to rewrite the $p^{(0)}/(\rho_0 V_A^2)$, either by introducing the usual MHD sound speed $C_s^2=\gamma p^{(0)}/\rho_0$, or by introducing the plasma beta
\begin{equation} \label{eq:VelkaPica}
  \beta = \frac{v_{\textrm{th}}^2}{V_A^2} = \frac{2 T^{(0)}/m_p}{V_A^2} = \frac{2 p^{(0)}/\rho_0}{V_A^2} = \frac{p^{(0)}}{B_0^2/(8\pi)}; \qquad => \qquad
  \frac{p^{(0)}}{\rho_0 V_A^2} = \frac{1}{\gamma}\frac{C_s^2}{V_A^2} = \frac{\beta}{2},
\end{equation}
where we use the Boltzmann constant $k_B=1$, see the footnote after the definition of thermal speeds (\ref{eq:vthDef}).    
By specifying the mean magnetic field to be in the z-direction, the normalized equations are linearized according to
\begin{eqnarray}
&& \frac{\pr \rho}{\pr t} + \nabla\cdot \bu =0; \nn \\
&& \frac{\pr u_x}{\pr t} +\frac{\beta}{2}\pr_x p  +\pr_x B_z-\pr_z B_x=0;\nn \\
&& \frac{\pr u_y}{\pr t} +\frac{\beta}{2}\pr_y p  +\pr_y B_z-\pr_z B_y=0; \nn \\
  && \frac{\pr u_z}{\pr t} +\frac{\beta}{2}\pr_z p =0; \nn \\
  && \frac{\pr B_x}{\pr t} = \pr_z u_x ;\qquad \frac{\pr B_y}{\pr t} = \pr_z u_y; \qquad \frac{\pr B_z}{\pr t} = -\pr_x u_x - \pr_y u_y ; \nn \\
&& \frac{\pr p}{\pr t} + \gamma \nabla\cdot\bu = 0,
\end{eqnarray}
and without a loss of generality,  we consider propagation in the x-z plane (with $\pr_y =0$). 
By considering a wave propagating with wavenumber $\boldsymbol{k}=(k\sin\theta,0,k\cos\theta)=(k_x,0,k_z)=(k_\perp,0,k_\parallel)$, i.e. a wave propagating in the direction
that makes an angle $\theta$ with respect to $\bb_0$, the  MHD system written in Fourier space reads
\begin{eqnarray}
&& -\omega \rho + (k\sin\theta) u_x +(k\cos\theta) u_z =0; \nn \\
&& -\omega u_x +\frac{\beta}{2}(k\sin\theta) p  +(k\sin\theta) B_z-(k\cos\theta) B_x =0;\nn \\
&& -\omega u_y -(k\cos\theta) B_y=0; \qquad  -\omega u_z +\frac{\beta}{2}(k\cos\theta) p =0; \nn \\
  && -\omega B_x - (k\cos\theta) u_x=0; \qquad -\omega B_y - (k\cos\theta) u_y =0;\qquad -\omega B_z + (k\sin\theta) u_x =0; \nn \\
&& -\omega p + \gamma (k\sin\theta) u_x +\gamma (k\cos\theta) u_z = 0.
\end{eqnarray}
By calculating determinant of this system yields MHD dispersion relation in normalized units (bringing back tildes for clarity) that reads 
\begin{equation}
  \Big(\widetilde{\omega}^2-\widetilde{k}^2\cos^2\theta\Big)\Big(\widetilde{\omega}^4-(1+\gamma\frac{\beta}{2})\widetilde{k}^2\widetilde{\omega}^2
  +\gamma\frac{\beta}{2} \widetilde{k}^4\cos^2\theta\Big) = 0,
\end{equation}
or alternatively by using (\ref{eq:VelkaPica}) with the sound speed
\begin{equation}
  \Big(\widetilde{\omega}^2-\widetilde{k}^2\cos^2\theta\Big)\Big(\widetilde{\omega}^4-(1+\frac{C_s^2}{V_A^2})\widetilde{k}^2\widetilde{\omega}^2
  +\frac{C_s^2}{V_A^2} \widetilde{k}^4\cos^2\theta\Big) = 0,
\end{equation}
or in physical units
\begin{equation} \label{MHDappendix}
\Big(\omega^2-V_A^2 k^2\cos^2\theta\Big)\Big(\omega^4-(V_A^2+C_s^2)k^2\omega^2+V_A^2 C_s^2 k^4\cos^2\theta\Big) = 0,
\end{equation}  
yielding the MHD dispersion relation for the Alfv\'en mode $\omega_A=\pm V_A k\cos\theta=\pm V_A \kpar$ (sometimes called the shear Alfv\'en mode),
and the slow and fast modes (\ref{eq:MHD1}).

In the limit $\beta\rightarrow 0$ that corresponds to a cold plasma ($C_s\rightarrow 0$) or
a plasma with a very strong magnetic field ($V_A\rightarrow\infty$), the slow mode $\omega_s = \pm C_sk_\parallel$,
and the fast mode $\omega_f = \pm V_A k$ (sometimes called the compressional Alfv\'en mode). In the limit
$\beta\rightarrow \infty$ that corresponds to the incompressible MHD ($C_s\rightarrow \infty$) or a plasma with vanishing magnetic
field ($V_A\rightarrow 0$), the slow mode $\omega_s=\pm V_A k_\parallel$,
and the fast mode $\omega_f = \pm C_s k$. Note the difference in $k_\parallel$ and $k$ in these two limits.

\section{Non-gyrotropic heat flux tensor $\lowercase{\boldsymbol{q}^{\textrm{ng}}}$} \label{sec:NONGheat}
The heat flux tensor equation (\ref{eq:Qngtensor}) is rewritten as
\begin{equation} \label{eq:Qngexpand}
\big(\bhat\times\bq^\textrm{ng} \big)^S = -\frac{B_0}{\Omega|\bb|} \Big[ 
  \frac{d}{d t}\bq +\nabla\cdot\br +\bq\nabla\cdot\bu   + \big(\bq\cdot\nabla\bu\big)^S  - \frac{1}{\rho}\Big(\bp(\nabla\cdot\bp)\Big)^S \Big]. 
\end{equation}
The ``inversion'' procedure for the l.h.s. of this is equation is very complicated, and will not be addressed here.
Nevertheless, the inversion procedure exist, and an interested reader can check eq. (43) of \cite{Ramos2005}. The leading-order $\bq^\textrm{ng}$ (first order in
frequency and wavenumber) can be obtained by making the quantities $\bq,\br,\bp$ on the r.h.s. gyrotropic, and we want to solve
\begin{equation}
\big(\bhat\times\bq^\textrm{ng} \big)^S = -\frac{B_0}{\Omega|\bb|} \Big[ 
  \frac{d}{d t}\bq^{\textrm{g}} +\nabla\cdot\br^{\textrm{g}} +\bq^{\textrm{g}}\nabla\cdot\bu   + \big(\bq^{\textrm{g}}\cdot\nabla\bu\big)^S
  - \frac{1}{\rho}\Big(\bp^{\textrm{g}}(\nabla\cdot\bp^{\textrm{g}})\Big)^S \Big]. \label{eq:QngEq2}
\end{equation}
The heat flux tensor decomposition is $\bq=\bq^{\textrm{g}}+\bq^{\textrm{ng}}$, or alternatively $\bq=\boldsymbol{S}+\boldsymbol{\sigma}$,
where $\boldsymbol{\sigma}:\bhat\bhat=0$, $\boldsymbol{\sigma}:\boldsymbol{I}=0$. Since
\begin{eqnarray}
  \underbrace{\bq:\bhat\bhat}_{\boldsymbol{S}^\parallel} &=& \underbrace{\bq^{\textrm{g}}:\bhat\bhat}_{q_\parallel\bhat} +\bq^{\textrm{ng}}:\bhat\bhat;\\
  \underbrace{\bq:(\boldsymbol{I}-\bhat\bhat)/2}_{\boldsymbol{S}^\perp} &=& \underbrace{\bq^{\textrm{g}}:(\boldsymbol{I}-\bhat\bhat)/2}_{q_\perp\bhat}
  +\bq^{\textrm{ng}}:(\boldsymbol{I}-\bhat\bhat)/2,
\end{eqnarray}  
one can define perpendicular components of the heat flux vectors
\begin{eqnarray}
  \boldsymbol{S}^\parallel_\perp &=& \bq^{\textrm{ng}}:\bhat\bhat;\\
  \boldsymbol{S}_\perp^\perp &=& \bq^{\textrm{ng}}:(\boldsymbol{I}-\bhat\bhat)/2.
\end{eqnarray}
Therefore, to obtain these vectors, we want to apply $:\bhat\bhat$, and $:(\boldsymbol{I}-\bhat\bhat)/2$ on the entire equation (\ref{eq:QngEq2}).
A very useful expression also is
\begin{equation}
\textrm{Tr} \bq^{\textrm{ng}} = \bq^{\textrm{ng}}:\boldsymbol{I} = \boldsymbol{S}^\parallel_\perp+2\boldsymbol{S}^\perp_\perp,
\end{equation}  
and the trace of the entire heat flux tensor reads
\begin{equation}
\textrm{Tr} \bq = q_\parallel\bhat+2q_\perp\bhat+ \boldsymbol{S}^\parallel_\perp+2\boldsymbol{S}^\perp_\perp.
\end{equation}
\subsection{Non-gyrotropic heat flux vector $\boldsymbol{S}_\perp^\parallel$}
Applying $:\bhat\bhat$ on the l.h.s. of (\ref{eq:QngEq2}) yields
\begin{eqnarray}
  \big(\bhat\times\bq^\textrm{ng} \big)^S_{ijk} \hat{b}_i \hat{b}_j &=& \Big( \epsilon_{irs} \hat{b}_r q^\textrm{ng}_{sjk} 
  +\epsilon_{jrs} \hat{b}_r q^\textrm{ng}_{ski}
  +\epsilon_{krs} \hat{b}_r q^\textrm{ng}_{sij} \Big) \hat{b}_i \hat{b}_j \nn\\
  &=& \cancel{\epsilon_{irs} \hat{b}_r \hat{b}_i \hat{b}_j q^\textrm{ng}_{sjk}} 
  +\cancel{\epsilon_{jrs} \hat{b}_r \hat{b}_i \hat{b}_j q^\textrm{ng}_{ski}}
  +\epsilon_{krs} \hat{b}_r q^\textrm{ng}_{sij} \hat{b}_i \hat{b}_j \nn\\
  &=& \epsilon_{krs} \hat{b}_r (\boldsymbol{S}^\parallel_\perp)_s = \big( \bhat\times \boldsymbol{S}^\parallel_\perp \big)_k.
\end{eqnarray}  
Terms on the r.h.s. of (\ref{eq:QngEq2}) calculate
\begin{eqnarray}
&&  \Big( \frac{d}{dt}\bq^{\textrm{g}}\Big)_{ijk} \hat{b}_i \hat{b}_j =\frac{d q_\parallel}{d t}\hat{b}_k +q_\parallel \frac{d \hat{b}_k}{dt}
  -2q_\perp \frac{d \hat{b}_k}{dt}; \\
&&  \Big( \nabla\cdot\br^{\textrm{g}} \Big)_{ijk} \hat{b}_i \hat{b}_j = \hat{b}_k \bhat\cdot\nabla r_{\parallel\parallel}
  + (r_{\parallel\parallel}-3r_{\parallel\perp}) \big( \bhat\cdot\nabla \hat{b}_k +\hat{b}_k \nabla\cdot\bhat \big) -\hat{b}_k \bhat\cdot\nabla r_{\parallel\perp}
  + \pr_k r_{\parallel\perp};\\
  && q^{\textrm{g}}_{ijk}\hat{b}_i = q_{\parallel}\hat{b}_j\hat{b}_k +q_\perp(\delta_{jk}-\hat{b}_j\hat{b}_k);\\
  && q^{\textrm{g}}_{ijk}\hat{b}_i\hat{b}_j \nabla\cdot\bu = q_\parallel \hat{b}_k \nabla\cdot\bu;\\
&&  \big( \bq^{\textrm{g}}\cdot\nabla\bu\big)^S_{ijk} \hat{b}_i \hat{b}_j = q_\parallel \bhat\cdot\nabla u_k +2(q_\parallel-q_\perp) \hat{b}_k \bhat\cdot(\nabla\bu)\cdot\bhat
  +2q_\perp (\pr_k \bu)\cdot\bhat.
\end{eqnarray}
For the last term on the r.h.s. of (\ref{eq:QngEq2})
\begin{eqnarray}
  &&  \Big(\bp(\nabla\cdot\bp)\Big)^S_{ijk} \hat{b}_i \hat{b}_j = p_\parallel (\nabla\cdot\bp)_k +2(p_\parallel \hat{b}_k +\Pi_{jk}\hat{b}_j ) (\nabla\cdot\bp)\cdot\bhat;\\
  && (\nabla\cdot \bp)_k = \hat{b}_k \bhat\cdot\nabla(p_\parallel-p_\perp) +(p_\parallel-p_\perp)\big( \bhat\cdot\nabla \hat{b}_k +\hat{b}_k \nabla\cdot\bhat \big)
  +\pr_k p_\perp +(\nabla\cdot\boldsymbol{\Pi})_k;\\
  && (\nabla\cdot\bp)\cdot\bhat = \bhat\cdot\nabla p_\parallel +(p_\parallel-p_\perp)\nabla\cdot\bhat +(\nabla\cdot\boldsymbol{\Pi})\cdot\bhat,
\end{eqnarray}
and so for the gyrotropic
\begin{eqnarray}
  \Big(\bp^{\textrm{g}}(\nabla\cdot\bp^{\textrm{g}})\Big)^S_{ijk} \hat{b}_i \hat{b}_j = p_\parallel \hat{b}_k \bhat\cdot \nabla (3p_\parallel-p_\perp)
  +p_\parallel (p_\parallel-p_\perp) \big( \bhat\cdot\nabla \hat{b}_k +3\hat{b}_k \nabla\cdot\bhat \big) +p_\parallel \pr_k p_\perp.
\end{eqnarray}
Collecting all the results together yields
\begin{eqnarray}
  \big( \bhat\times \boldsymbol{S}^\parallel_\perp \big)_k &=& -\frac{B_0}{\Omega|\bb|} \bigg[
  \frac{d q_\parallel}{d t}\hat{b}_k +q_\parallel \frac{d \hat{b}_k}{dt}
  -2q_\perp \frac{d \hat{b}_k}{dt}  + \hat{b}_k \bhat\cdot\nabla r_{\parallel\parallel}
  + (r_{\parallel\parallel}-3r_{\parallel\perp}) \big( \bhat\cdot\nabla \hat{b}_k +\hat{b}_k \nabla\cdot\bhat \big) -\hat{b}_k \bhat\cdot\nabla r_{\parallel\perp}
  + \pr_k r_{\parallel\perp}\nn\\
  &&+q_\parallel \hat{b}_k \nabla\cdot\bu + q_\parallel \bhat\cdot\nabla u_k +2(q_\parallel-q_\perp) \hat{b}_k \bhat\cdot(\nabla\bu)\cdot\bhat
  +2q_\perp (\pr_k \bu)\cdot\bhat \nn\\
  && - \frac{p_\parallel}{\rho} \hat{b}_k \bhat\cdot \nabla (3p_\parallel-p_\perp)
  -\frac{p_\parallel}{\rho} (p_\parallel-p_\perp) \big( \bhat\cdot\nabla \hat{b}_k +3\hat{b}_k \nabla\cdot\bhat \big) -\frac{p_\parallel}{\rho} \pr_k p_\perp \bigg]. \label{eq:SparperpPica}
\end{eqnarray}
The equation appears very complicated, however, it is not overly so, since it is only a vector equation. The equation
just describes perpendicular components of $\boldsymbol{S}^\parallel_\perp$.
By multiplying it with $\hat{b}_k$, the l.h.s. is zero, and on the r.h.s. one recovers the gyrotropic parallel heat flux equation (\ref{eq:ParHF_Final}).
To understand the equation better, let's simplify for a moment and evaluate it with respect to $\bhat_0=(0,0,1)$. The x-component
$( \bhat_0\times \boldsymbol{S}^\parallel_\perp )_x = - (\boldsymbol{S}^\parallel_\perp)_y$, so that
\begin{eqnarray}
  (\boldsymbol{S}^\parallel_\perp)_y &=& \frac{1}{\Omega} \Big[
  q_\parallel \frac{d \hat{b}_x}{dt}
  -2q_\perp \frac{d \hat{b}_x}{dt}
  + (r_{\parallel\parallel}-3r_{\parallel\perp}) \pr_z \hat{b}_x  + \pr_x r_{\parallel\perp}\nn\\
  && + q_\parallel \pr_z u_x  +2q_\perp \pr_x u_z 
  -\frac{p_\parallel}{\rho} (p_\parallel-p_\perp) \pr_z \hat{b}_x -\frac{p_\parallel}{\rho} \pr_x p_\perp \Big]
\end{eqnarray}
and similarly for the y-component $( \bhat_0\times \boldsymbol{S}^\parallel_\perp )_y = + (\boldsymbol{S}^\parallel_\perp)_x$, yielding
\begin{eqnarray}
  (\boldsymbol{S}^\parallel_\perp)_x &=& -\frac{1}{\Omega} \Big[
  q_\parallel \frac{d \hat{b}_y}{dt}
  -2q_\perp \frac{d \hat{b}_y}{dt}
  + (r_{\parallel\parallel}-3r_{\parallel\perp}) \pr_z \hat{b}_y  + \pr_y r_{\parallel\perp}\nn\\
  && + q_\parallel \pr_z u_y  +2q_\perp \pr_y u_z 
  -\frac{p_\parallel}{\rho} (p_\parallel-p_\perp) \pr_z \hat{b}_y -\frac{p_\parallel}{\rho} \pr_y p_\perp \Big].
\end{eqnarray}
Now, by using the leading order MHD induction equation $d\hat{b}_x/dt=\pr_z u_x$ and $d\hat{b}_y/dt=\pr_z u_y$
(which was for example used to get the FLR1 pressure tensor), and by further  
considering specific example of a bi-Maxwellian expressions for the 4th-order moments, i.e. $r_{\parallel\parallel}=3p_\parallel^2/\rho+\widetilde{r}_{\parallel\parallel}$ etc., yields 
\begin{eqnarray}
  (\boldsymbol{S}^\parallel_\perp)_y &=& \frac{1}{\Omega} \Big[ p_\perp \pr_x \Big( \frac{p_\parallel}{\rho}\Big) + 2\frac{p_\parallel}{\rho} (p_\parallel-p_\perp)\pr_z\hat{b}_x 
 +2 q_\parallel \pr_z u_x   +2q_\perp (\pr_x u_z -\pr_z u_x) \nn\\
 && \quad + \big(\widetilde{r}_{\parallel\parallel}-3\widetilde{r}_{\parallel\perp} \big) \pr_z \hat{b}_x 
  +\pr_x \widetilde{r}_{\parallel\perp}  \Big]; \label{eq:PicaRitKurva1}\\
  (\boldsymbol{S}^\parallel_\perp)_x &=& -\frac{1}{\Omega} \Big[ p_\perp \pr_y \Big( \frac{p_\parallel}{\rho}\Big) + 2\frac{p_\parallel}{\rho} (p_\parallel-p_\perp)\pr_z\hat{b}_y
 +2 q_\parallel \pr_z u_y   +2q_\perp (\pr_y u_z -\pr_z u_y) \nn\\
 && \quad + \big(\widetilde{r}_{\parallel\parallel}-3\widetilde{r}_{\parallel\perp} \big) \pr_z \hat{b}_y
  +\pr_y \widetilde{r}_{\parallel\perp}  \Big]. \label{eq:PicaRitKurva2}
\end{eqnarray}
The above equations are very useful if we ever need linearized expressions for $\boldsymbol{S}^\parallel_\perp$, which is beneficial to do right now, and the
partially linearized expressions read
\begin{eqnarray}
  (\boldsymbol{S}^\parallel_\perp)_y &\overset{\textrm{\tiny lin}}{=}& \frac{1}{\Omega} \Big[ p_\perp \pr_x \Big( \frac{p_\parallel}{\rho}\Big) + 2\frac{p_\parallel}{\rho} (p_\parallel-p_\perp)\pr_z\hat{b}_x  +\pr_x \widetilde{r}_{\parallel\perp}  \Big]; \label{eq:SparPerpPLy}\\
  (\boldsymbol{S}^\parallel_\perp)_x &\overset{\textrm{\tiny lin}}{=}& -\frac{1}{\Omega} \Big[ p_\perp \pr_y \Big( \frac{p_\parallel}{\rho}\Big) + 2\frac{p_\parallel}{\rho} (p_\parallel-p_\perp)\pr_z\hat{b}_y +\pr_y \widetilde{r}_{\parallel\perp}  \Big]. \label{eq:SparPerpPLx}
\end{eqnarray}
Note that the mean values of perturbations $\widetilde{r}^{(0)}=0$. It is also noteworthy to point out, that the contributions from the induction equation $d\bhat/dt$
are completely eliminated by the linearization process, and more elaborate forms of induction equation will not bring additional precision at the linear level. 
Coming back to the expressions (\ref{eq:PicaRitKurva1}), (\ref{eq:PicaRitKurva2}),
it is also possible to introduce vorticity $\boldsymbol{\omega}=\nabla\times\bu$, with components $\omega_x=(\pr_y u_z-\pr_z u_y)$ and $\omega_y= -(\pr_x u_z-\pr_z u_x)$.
By noting that $(\bhat_0\times\boldsymbol{\omega})_x=-\omega_y=(\pr_x u_z-\pr_z u_x)$ and $(\bhat_0\times\boldsymbol{\omega})_y=\omega_x=(\pr_y u_z-\pr_z u_y)$,
both equations can be written together as
\begin{eqnarray}
 \boldsymbol{S}^\parallel_\perp &=&  \frac{1}{\Omega} \bhat_0\times \Big[p_\perp \nabla \Big( \frac{p_\parallel}{\rho}\Big) + 2\frac{p_\parallel}{\rho} (p_\parallel-p_\perp)\pr_z\bhat 
 +2 q_\parallel \pr_z \bu   +2q_\perp \bhat_0\times\boldsymbol{\omega} \nn\\
 && \quad\qquad + \big(\widetilde{r}_{\parallel\parallel}-3\widetilde{r}_{\parallel\perp} \big) \pr_z \bhat 
  +\nabla \widetilde{r}_{\parallel\perp}  \Big].
\end{eqnarray}
Now that we understand that the equation (\ref{eq:SparperpPica}) just describes perpendicular components of $\boldsymbol{S}^\parallel_\perp$ with respect to $\bhat$,
since the parallel components are zero (directly from the decomposition of $\boldsymbol{S}^\parallel$)
\begin{equation}
\boldsymbol{S}_\perp^\parallel \cdot\bhat =0; \qquad \boldsymbol{S}_\perp^\perp \cdot\bhat =0,
\end{equation} 
we actually do not have to evaluate the eq. (\ref{eq:SparperpPica}) with respect to $\bhat_0$. For a general vector $\boldsymbol{a}$, the following identity
holds $\bhat\times(\bhat\times\boldsymbol{a})=\bhat(\bhat\cdot\boldsymbol{a})-\boldsymbol{a}$. So for a vector $\boldsymbol{a}$ which does not have
any parallel components to $\bhat$, so that $\bhat\cdot\boldsymbol{a}=0$, the identity is $\bhat\times(\bhat\times\boldsymbol{a})=-\boldsymbol{a}$, and
\begin{equation}
  \bhat\times\big(\bhat\times\boldsymbol{S}_\perp^\parallel\big) =-\boldsymbol{S}_\perp^\parallel; \qquad
  \bhat\times\big(\bhat\times\boldsymbol{S}_\perp^\perp\big) =-\boldsymbol{S}_\perp^\perp.
\end{equation}
We can apply $\bhat\times$ on the entire equation (\ref{eq:SparperpPica}), and get the fully nonlinear expression for $\boldsymbol{S}_\perp^\parallel$.
Let's apply $\bhat\times$ on each term separately, since we will need these results later, the terms calculate 
\begin{eqnarray}
&&  \bhat\times \Big[ \Big( \frac{d}{dt}\bq^{\textrm{g}}\Big): \bhat \bhat\Big] = \bhat\times\Big[ q_\parallel \frac{d \bhat}{dt} -2q_\perp \frac{d \bhat}{dt}\Big]; \label{eq:QQNQ1}\\
&&  \bhat\times \Big[ ( \nabla\cdot\br^{\textrm{g}}) : \bhat \bhat\Big] = \bhat\times\Big[ 
   (r_{\parallel\parallel}-3r_{\parallel\perp})  \bhat\cdot\nabla \bhat  + \nabla r_{\parallel\perp} \Big];\\
  && \bhat\times \Big[ \bq^{\textrm{g}} (\nabla\cdot\bu):\bhat\bhat\Big] = 0;\\
&&  \bhat\times \Big[  \big( \bq^{\textrm{g}}\cdot\nabla\bu\big)^S:\bhat\bhat\Big] = \bhat\times\Big[ q_\parallel \bhat\cdot\nabla \bu 
    +2q_\perp (\nabla \bu)\cdot\bhat \Big];\\
&& \bhat\times \Big[  \Big(\bp^{\textrm{g}}(\nabla\cdot\bp^{\textrm{g}})\Big)^S:\bhat\bhat\Big] = \bhat\times\Big[
  p_\parallel (p_\parallel-p_\perp) \bhat\cdot\nabla \bhat +p_\parallel \nabla p_\perp \Big].\label{eq:QQNQ2}
\end{eqnarray}
To summarize, by applying $:\bhat\bhat$ and $\bhat\times$ on the equation (\ref{eq:QngEq2}), yields the following nonlinear expression
\begin{eqnarray}
  \boldsymbol{S}^\parallel_\perp  &=& \frac{B_0}{\Omega|\bb|} \bhat\times \Big[
  q_\parallel \Big( \frac{d \bhat}{dt} +\bhat\cdot\nabla \bu \Big)
  +2q_\perp \Big((\nabla \bu)\cdot\bhat -\frac{d \bhat}{dt}\Big)   + (r_{\parallel\parallel}-3r_{\parallel\perp}) \bhat\cdot\nabla \bhat   \nn\\
  &&  \qquad \qquad -\frac{p_\parallel}{\rho} (p_\parallel-p_\perp) \bhat\cdot\nabla \bhat + \nabla r_{\parallel\perp} -\frac{p_\parallel}{\rho} \nabla p_\perp \Big].
\end{eqnarray}
Now the calculations proceed in a similar way as before. The usual (MHD) induction equation can be written as
\begin{equation}
  \frac{d\bhat}{dt} =\bhat\cdot\nabla\bu -\bhat \Big[ \bhat\cdot(\nabla\bu)\cdot\bhat\Big]; \quad => \quad \bhat\times \frac{d\bhat}{dt} =\bhat\times (\bhat\cdot\nabla\bu),\nn
\end{equation}
and since $(\nabla\bu)\cdot\bhat-\bhat\cdot\nabla\bu=\bhat\times(\nabla\times\bu)=\bhat\times\boldsymbol{\omega}$, further yielding
\begin{eqnarray}
  \boldsymbol{S}^\parallel_\perp  &=& \frac{B_0}{\Omega|\bb|} \bhat\times \Big[
  2 q_\parallel \bhat\cdot\nabla \bu
  +2q_\perp \bhat\times\boldsymbol{\omega}   + (r_{\parallel\parallel}-3r_{\parallel\perp}) \bhat\cdot\nabla \bhat   \nn\\
  &&  \qquad\qquad -\frac{p_\parallel}{\rho} (p_\parallel-p_\perp) \bhat\cdot\nabla \bhat + \nabla r_{\parallel\perp}
  -\frac{p_\parallel}{\rho} \nabla p_\perp \Big]. \label{eq:Macmahon1965_eq14}
\end{eqnarray}
Finally, by prescribing bi-Maxwellian perturbations for the 4th-order moment yields
\begin{eqnarray}
  \boldsymbol{S}^\parallel_\perp  &=& \frac{B_0}{\Omega|\bb|} \bhat\times \Big[ p_\perp \nabla\Big(\frac{p_\parallel}{\rho}\Big)
  +2\frac{p_\parallel}{\rho}(p_\parallel-p_\perp) \bhat\cdot\nabla\bhat
  + 2 q_\parallel \bhat\cdot\nabla \bu
  +2q_\perp \bhat\times\boldsymbol{\omega}  \nn\\
  &&\qquad\qquad + (\widetilde{r}_{\parallel\parallel}-3\widetilde{r}_{\parallel\perp}) \bhat\cdot\nabla \bhat + \nabla \widetilde{r}_{\parallel\perp} \Big]. \label{eq:SP2015typo}
\end{eqnarray}
We can report a sign typo in eq. (3.7) of \cite{SulemPassot2015}, whose term $3\widetilde{r}_{\parallel\perp}$ has an opposite sign. 

\subsection{Conversion of other notations}
In the paper of \cite{Ramos2005}, the following notation is used
\begin{eqnarray}
&&  q_{B\parallel}=\frac{1}{2}q_\parallel; \quad q_{T\parallel}=q_\perp;\quad  \boldsymbol{q}_{B\perp} = \frac{1}{2}\boldsymbol{S}^\parallel_\perp; \quad
  \boldsymbol{q}_{T\perp} = \boldsymbol{S}_\perp^\perp; \nn \\
&&  \widetilde{r}_{\parallel}^{(0)} = m\big( \frac{1}{2} \widetilde{r}_{\parallel\parallel} + \widetilde{r}_{\parallel\perp} \big); \quad
  \widetilde{r}_{B \perp}^{(0)} = \frac{m}{2} \widetilde{r}_{\parallel\perp}; \quad
  \widetilde{r}_{\perp}^{(0)} = m \big( \frac{1}{2} \widetilde{r}_{\parallel\perp} +\widetilde{r}_{\perp\perp} \big),
\end{eqnarray}
and his eq. 54, 57 are indeed equivalent to (\ref{eq:SP2015typo}).

In the paper by \cite{MacMahon1965}, the following notation is used
\begin{eqnarray}
  && \boldsymbol{q}^\parallel = \frac{1}{2}\boldsymbol{S}^\parallel;
  \quad \boldsymbol{q}^\perp = \boldsymbol{S}^\perp; \quad q_{\parallel}^\parallel = \frac{1}{2} q_\parallel; \quad q_{\parallel}^\perp = q_\perp;
  \quad \boldsymbol{q}^\parallel_\perp = \frac{1}{2}\boldsymbol{S}^\parallel_\perp;
  \quad \boldsymbol{q}^\perp_\perp = \boldsymbol{S}^\perp_\perp; \nn \\
  && R_1 = r_{\parallel\parallel}; \quad R_2 = r_{\parallel\perp}; \quad R_3 = 2 r_{\perp\perp}.
\end{eqnarray}
and his eq. 14 is equivalent to (\ref{eq:Macmahon1965_eq14}), if in \cite{MacMahon1965} we continue the calculation of
\begin{equation}
\bhat\times(\nabla\cdot\bp^{\textrm{g}}) = \bhat\times \Big[(p_\parallel-p_\perp)\bhat\cdot\nabla\bhat+\nabla p_\perp\Big].\nn
\end{equation}
\subsection{Non-gyrotropic heat flux vector $\boldsymbol{S}_\perp^\perp$}
Now we want to apply $:(\boldsymbol{I}-\bhat\bhat)/2$ on the equation (\ref{eq:QngEq2}). Additionally, we now know that  $\boldsymbol{S}_\perp^\perp$ has only
perpendicular components, and that afterward we will be also applying $\bhat\times$.
The l.h.s. calculates trivially
\begin{eqnarray} 
  \delta_{ij} \big(\bhat\times\bq^\textrm{ng} \big)^S_{ijk}  &=& \delta_{ij} \Big( \epsilon_{irs} \hat{b}_r q^\textrm{ng}_{sjk} 
  +\epsilon_{jrs} \hat{b}_r q^\textrm{ng}_{ski}
  +\epsilon_{krs} \hat{b}_r q^\textrm{ng}_{sij} \Big)\nn\\
  &=& \hat{b}_r \cancel{\epsilon_{irs} q^\textrm{ng}_{sik}} 
  +\hat{b_r} \cancel{\epsilon_{irs} q^\textrm{ng}_{ski}}
  +\epsilon_{krs} \hat{b}_r q^\textrm{ng}_{sii} = \big( \bhat\times (\bq^{\textrm{ng}}:\boldsymbol{I}) \big)_k;\\
\big(\bhat\times\bq^\textrm{ng} \big)^S:(\boldsymbol{I}-\bhat\bhat)/2 &=& \bhat\times \boldsymbol{S}_\perp^\perp  
\end{eqnarray}
and applying $\bhat\times$ therefore yields
\begin{equation}
 \bhat\times \Big[ \big(\bhat\times\bq^\textrm{ng} \big)^S:(\boldsymbol{I}-\bhat\bhat)/2 \Big]= - \boldsymbol{S}_\perp^\perp.
\end{equation}  
Let's continue with applying trace on the equation (\ref{eq:QngEq2}) term by term
\begin{eqnarray}
  && \delta_{ij} \Big( \frac{d}{dt} \bq^{\textrm{g}} \Big)_{ijk} = \hat{b}_k \frac{d}{dt} (q_\parallel+2q_\perp) + (q_\parallel+2q_\perp)\frac{d}{dt}\hat{b}_k;\\
  && \delta_{ij} (\nabla\cdot\br^{\textrm{g}})_{ijk} = \hat{b}_k \bhat\cdot\nabla \big(r_{\parallel\parallel}+r_{\parallel\perp}-2r_{\perp\perp} \big) +
  \big(r_{\parallel\parallel}+r_{\parallel\perp}-2r_{\perp\perp} \big)(\bhat\cdot\nabla \hat{b}_k +\hat{b}_k\nabla\cdot\bhat\big) +\pr_k r_{\parallel\perp}+2\pr_k r_{\perp\perp};\\
  && \delta_{ij} q_{ijk}^\textrm{g} \nabla\cdot\bu = (q_\parallel+2q_\perp) \hat{b}_k \nabla\cdot\bu;\\
  && \delta_{ij} \Big( \bq^{\textrm{g}}\cdot\nabla\bu\Big)^S_{ijk} = q_\parallel \bhat\cdot\nabla u_k +2q_\parallel \hat{b}_k \bhat\cdot(\nabla\bu)\cdot\bhat
  +2q_\perp\Big[ 2\bhat \cdot\nabla u_k +(\pr_k \bu)\cdot\bhat +\hat{b}_k \nabla\cdot\bu -3\hat{b}_k \bhat\cdot(\nabla\bu)\cdot\bhat \Big];\\
  && \delta_{ij}\Big( \bp^\textrm{g} (\nabla\cdot\bp^\textrm{g})\Big)^S_{ijk} = (3p_\parallel+2p_\perp)(p_\parallel-p_\perp)\hat{b}_k\nabla\cdot\bhat
  +(p_\parallel+4p_\perp)(p_\parallel-p_\perp)\bhat\cdot\nabla\hat{b}_k \nn\\
  && \qquad \qquad +(p_\parallel+4p_\perp)\pr_k p_\perp -(p_\parallel+4p_\perp) \hat{b}_k \bhat\cdot\nabla p_\perp
  +(3p_\parallel+2p_\perp)\hat{b}_k \bhat\cdot \nabla p_\parallel,
\end{eqnarray}
and by further applying $\bhat\times$ on these terms yields
\begin{eqnarray}
  && \bhat\times\textrm{Tr} \Big( \frac{d}{dt} \bq^{\textrm{g}} \Big) = \bhat\times\Big[ (q_\parallel+2q_\perp)\frac{d}{dt}\bhat\Big];\\
  && \bhat\times\textrm{Tr} (\nabla\cdot\br^{\textrm{g}}) = \bhat\times\Big[ \big(r_{\parallel\parallel}+r_{\parallel\perp}-2r_{\perp\perp} \big)\bhat\cdot\nabla \bhat
    +\nabla r_{\parallel\perp}+2\nabla r_{\perp\perp} \Big];\\
  && \bhat\times\textrm{Tr} \bq^\textrm{g} \nabla\cdot\bu = 0;\\
  && \bhat\times\textrm{Tr}\Big( \bq^{\textrm{g}}\cdot\nabla\bu\Big)^S = \bhat\times\Big[ q_\parallel \bhat\cdot\nabla \bu 
  +2q_\perp\Big( 2\bhat \cdot\nabla \bu +(\nabla \bu)\cdot\bhat \Big) \Big];\\
  && \bhat\times\textrm{Tr}\Big( \bp^\textrm{g} (\nabla\cdot\bp^\textrm{g})\Big)^S = \bhat\times\Big[ 
  (p_\parallel+4p_\perp)(p_\parallel-p_\perp)\bhat\cdot\nabla\bhat +(p_\parallel+4p_\perp)\nabla p_\perp\Big],
\end{eqnarray}
and direct subtraction with results (\ref{eq:QQNQ1})-(\ref{eq:QQNQ2}) gives
\begin{eqnarray}
&&  \bhat\times \Big[ \Big( \frac{d}{dt}\bq^{\textrm{g}}\Big): (\boldsymbol{I}-\bhat \bhat)/2\Big] = \bhat\times\Big[ 2q_\perp \frac{d \bhat}{dt}\Big]; \\
&&  \bhat\times \Big[ ( \nabla\cdot\br^{\textrm{g}}) : (\boldsymbol{I}-\bhat \bhat)/2\Big] = \bhat\times\Big[ 
   (2r_{\parallel\perp}-r_{\perp\perp})  \bhat\cdot\nabla \bhat  + \nabla r_{\perp\perp} \Big];\\
&&  \bhat\times \Big[  \big( \bq^{\textrm{g}}\cdot\nabla\bu\big)^S:(\boldsymbol{I}-\bhat \bhat)/2\Big] = \bhat\times\Big[ 2q_\perp \bhat\cdot\nabla\bu \Big];\\
&& \bhat\times \Big[  \Big(\bp^{\textrm{g}}(\nabla\cdot\bp^{\textrm{g}})\Big)^S:(\boldsymbol{I}-\bhat \bhat)/2\Big] = \bhat\times\Big[
  2 p_\perp (p_\parallel-p_\perp) \bhat\cdot\nabla \bhat +2p_\perp \nabla p_\perp \Big].
\end{eqnarray}
To summarize, by applying $:(\boldsymbol{I}-\bhat\bhat)/2$ and $\bhat\times$ on the equation (\ref{eq:QngEq2}) yields the following nonlinear expression 
\begin{eqnarray}
  \boldsymbol{S}_\perp^\perp  &=& \frac{B_0}{\Omega|\bb|} \bhat\times \Big[
  2q_\perp \frac{d \bhat}{dt} + (2r_{\parallel\perp}-r_{\perp\perp})  \bhat\cdot\nabla \bhat  + \nabla r_{\perp\perp}
  +2q_\perp \bhat\cdot\nabla\bu -2 \frac{p_\perp}{\rho} (p_\parallel-p_\perp) \bhat\cdot\nabla \bhat -2\frac{p_\perp}{\rho} \nabla p_\perp \Big].
\end{eqnarray}
By using the simple MHD induction equation gives
\begin{eqnarray}
  \boldsymbol{S}_\perp^\perp  &=& \frac{B_0}{\Omega|\bb|} \bhat\times \Big[
   (2r_{\parallel\perp}-r_{\perp\perp})  \bhat\cdot\nabla \bhat  + \nabla r_{\perp\perp}
  +4q_\perp \bhat\cdot\nabla\bu -2 \frac{p_\perp}{\rho} (p_\parallel-p_\perp) \bhat\cdot\nabla \bhat -2\frac{p_\perp}{\rho} \nabla p_\perp \Big], \label{eq:TypoMacMohan}
\end{eqnarray}
and finally prescribing bi-Maxwellian values for the 4-th order moments yields
\begin{eqnarray}
  \boldsymbol{S}_\perp^\perp  &=& \frac{B_0}{\Omega|\bb|} \bhat\times \Big[
    2p_\perp \nabla \Big( \frac{p_\perp}{\rho}\Big)   +4q_\perp \bhat\cdot\nabla\bu + (2\widetilde{r}_{\parallel\perp}-\widetilde{r}_{\perp\perp})  \bhat\cdot\nabla \bhat
    +\nabla \widetilde{r}_{\perp\perp} \Big]. \label{eq:SperpPerpF}
\end{eqnarray}
Additionally, evaluation from linear kinetic theory in the gyrotropic limit (at long wavelengths) yields $\widetilde{r}_{\perp\perp}=0$, as we will see in Part 2.
With this requirement, the equation (\ref{eq:SperpPerpF}) is equivalent to eq. 3.6 of \cite{SulemPassot2015}. The result (\ref{eq:SperpPerpF}) is also equivalent to
expressions 55, 58 of \cite{Ramos2005}. However, eq. 15 of \cite{MacMahon1965} appears to have a typo, where in comparison to our (\ref{eq:TypoMacMohan}),
his definition of $\boldsymbol{T}$ appears to be missing a factor of 2 in front of the $\nabla\bu$ term. \cite{Ramos2005} however states that his results
are equivalent to those of \cite{MacMahon1965}, so we might not correctly understand the notation in that paper. 

Partial linearization of result (\ref{eq:SperpPerpF}) yields that by components
\begin{eqnarray}
  (\boldsymbol{S}^\perp_\perp)_y &\overset{\textrm{\tiny lin}}{=}& \frac{1}{\Omega} \Big[ 2p_\perp \pr_x \Big( \frac{p_\perp}{\rho}\Big)  +\pr_x \widetilde{r}_{\perp\perp}  \Big];
  \label{eq:SperpPerpPLy}\\
  (\boldsymbol{S}^\perp_\perp)_x &\overset{\textrm{\tiny lin}}{=}& -\frac{1}{\Omega} \Big[ 2p_\perp \pr_y \Big( \frac{p_\perp}{\rho}\Big) +\pr_y \widetilde{r}_{\perp\perp}  \Big],
  \label{eq:SperpPerpPLx}
\end{eqnarray}
where we have kept the $\widetilde{r}_{\perp\perp}$ contributions, just in case we need them in the future. 
Note that similarly to the linearization of the vector $\boldsymbol{S}^\parallel_\perp$, the induction equation is completely eliminated at the linear level, and more
elaborate forms of induction equation will not have additional contributions.

\subsection{2nd-order heat flux vectors}
In a previous section we have seen that by applying $:\bhat\bhat$, $:(\boldsymbol{I}-\bhat\bhat)/2$ and $\bhat\times$ on the equation
(\ref{eq:Qngexpand}), the fully nonlinear (but implicit) expressions for the non-gyrotropic heat flux vectors read
\begin{eqnarray}
\boldsymbol{S}^\parallel_\perp &=& \frac{B_0}{\Omega|\bb|}\bhat\times \Big\{\bhat\bhat: \Big[ 
  \frac{d}{d t}\bq +\nabla\cdot\br +\bq\nabla\cdot\bu   + \big(\bq\cdot\nabla\bu\big)^S  - \frac{1}{\rho}\Big(\bp(\nabla\cdot\bp)\Big)^S \Big] \Big\};\\
\boldsymbol{S}^\perp_\perp &=& \frac{B_0}{\Omega|\bb|}\bhat\times \Big\{ (\boldsymbol{I}-\bhat\bhat)/2 : \Big[ 
  \frac{d}{d t}\bq +\nabla\cdot\br +\bq\nabla\cdot\bu   + \big(\bq\cdot\nabla\bu\big)^S  - \frac{1}{\rho}\Big(\bp(\nabla\cdot\bp)\Big)^S \Big] \Big\}.
\end{eqnarray}
Here we are interested only in the linear level contributions, which we want to use in the FLR pressure tensor. Therefore, since we know that here
we will linearize everything at the end, we can immediately get rid of the two heat flux terms $\bq\nabla\cdot\bu$ and $(\bq\cdot\nabla\bu)^S$, and
also replace the $d/dt$ by $\pr/\pr t$, so partial linearization yields
\begin{eqnarray}
\boldsymbol{S}^\parallel_\perp &\overset{\textrm{\tiny lin}}{=}& \frac{1}{\Omega} \bhat \times \Big\{ \bhat\bhat:\Big[ 
  \frac{\pr}{\pr t}\bq +\nabla\cdot\br - \frac{1}{\rho}\Big(\bp(\nabla\cdot\bp)\Big)^S \Big] \Big\}; \\
\boldsymbol{S}^\perp_\perp &\overset{\textrm{\tiny lin}}{=}& \frac{1}{\Omega} \bhat \times \Big\{ (\boldsymbol{I}-\bhat\bhat)/2:\Big[ 
  \frac{\pr}{\pr t}\bq +\nabla\cdot\br - \frac{1}{\rho}\Big(\bp(\nabla\cdot\bp)\Big)^S \Big] \Big\}.
\end{eqnarray}
Additionally, the $\pr \bq /\pr t$ term can be easily precalculated as
\begin{eqnarray}
  \Big( \frac{\pr}{\pr t}\bq \Big):\bhat\bhat &=& \frac{\pr}{\pr t} (q_\parallel\bhat+\boldsymbol{S}^\parallel_\perp) -\bq :\frac{\pr}{\pr t}\Big( \bhat\bhat\Big)
\overset{\textrm{\tiny lin}}{=} \frac{\pr q_\parallel}{\pr t} \bhat+\frac{\pr}{\pr t} \boldsymbol{S}^\parallel_\perp; \\
  \Big( \frac{\pr}{\pr t}\bq \Big):(\boldsymbol{I}-\bhat\bhat)/2 
  &=& \frac{\pr}{\pr t} (q_\perp\bhat +\boldsymbol{S}^\perp_\perp) +\bq :\frac{\pr}{\pr t}\Big(\bhat\bhat/2\Big)
\overset{\textrm{\tiny lin}}{=}\frac{\pr q_\perp}{\pr t} \bhat+ \frac{\pr}{\pr t} \boldsymbol{S}^\perp_\perp,
\end{eqnarray}
and by applying $\bhat\times$, terms proportional to $\bhat$ disappear, yielding
\begin{eqnarray}
  \boldsymbol{S}^\parallel_\perp &\overset{\textrm{\tiny lin}}{=}& \frac{1}{\Omega} \bhat \times \Big\{ \frac{\pr}{\pr t} \boldsymbol{S}^\parallel_\perp
  + \bhat\bhat:\Big[ 
   \nabla\cdot\br - \frac{1}{\rho}\Big(\bp(\nabla\cdot\bp)\Big)^S \Big] \Big\}; \\
\boldsymbol{S}^\perp_\perp &\overset{\textrm{\tiny lin}}{=}& \frac{1}{\Omega} \bhat \times \Big\{ \frac{\pr}{\pr t} \boldsymbol{S}^\perp_\perp
+ (\boldsymbol{I}-\bhat\bhat)/2:\Big[ 
  \nabla\cdot\br - \frac{1}{\rho}\Big(\bp(\nabla\cdot\bp)\Big)^S \Big] \Big\}.
\end{eqnarray}
The first-order vectors were obtained by keeping only gyrotropic quantities on the right hand side. Linearizing results (\ref{eq:SP2015typo}), (\ref{eq:SperpPerpF})
indeed yields that at the linear level
\begin{eqnarray}
  \boldsymbol{S}^{\parallel (1)}_\perp &\overset{\textrm{\tiny lin}}{=}&  \frac{1}{\Omega} \bhat_0\times \Big[ p_\perp^{(0)} \nabla\Big(\frac{p_\parallel}{\rho}\Big)
  +2\frac{p_\parallel^{(0)}}{\rho_0}(p_\parallel^{(0)}-p_\perp^{(0)}) \pr_z \bhat \Big];\\
  \boldsymbol{S}_\perp^{\perp(1)}  &\overset{\textrm{\tiny lin}}{=}& \frac{1}{\Omega} \bhat_0\times \Big[
    2p_\perp^{(0)} \nabla \Big( \frac{p_\perp}{\rho}\Big) \Big],
\end{eqnarray}
where the gradients are meant to be further linearized, and where we also neglected perturbations $\widetilde{r}$,
since right now we do not want to calculate these quantities from linear kinetic theory.
To obtain the second-order heat flux vectors at the linear level, we need to calculate
 \begin{eqnarray}
  \boldsymbol{S}^{\parallel (2)}_\perp &\overset{\textrm{\tiny lin}}{=}& \frac{1}{\Omega} \bhat \times \Big\{ \frac{\pr}{\pr t} \boldsymbol{S}^{\parallel (1)}_\perp
  + \bhat\bhat:\Big[ 
   \nabla\cdot\br^{\textrm{ng}} - \frac{1}{\rho}\Big(\bp^{\textrm{g}}(\nabla\cdot\boldsymbol{\Pi})\Big)^S \Big] \Big\}; \\
 \boldsymbol{S}^{\perp (2)}_\perp &\overset{\textrm{\tiny lin}}{=}& \frac{1}{\Omega} \bhat \times \Big\{ \frac{\pr}{\pr t} \boldsymbol{S}^{\perp (1)}_\perp
+ (\boldsymbol{I}-\bhat\bhat)/2:\Big[ 
  \nabla\cdot\br^{\textrm{ng}} - \frac{1}{\rho}\Big(\bp^{\textrm{g}}(\nabla\cdot\boldsymbol{\Pi})\Big)^S \Big] \Big\}.
\end{eqnarray}
To calculate the $\br^{\textrm{ng}}$ contributions it is enough to work with partially linearized (valid only for bi-Maxwellian)
\begin{eqnarray}
  \pr_l r_{ijkl}^{\textrm{ng}} \overset{\textrm{\tiny lin}}{=} \frac{1}{\rho_0} \Big[ p^{\textrm{g}}_{ij}\pr_l \Pi_{kl} +p^{\textrm{g}}_{ik}\pr_l \Pi_{jl}
    +p^{\textrm{g}}_{il}\pr_l \Pi_{jk} +p^{\textrm{g}}_{kl}\pr_l \Pi_{ij} +p^{\textrm{g}}_{jl}\pr_l \Pi_{ik}+p^{\textrm{g}}_{jk}\pr_l \Pi_{il} \Big],
\end{eqnarray}
which further yields
\begin{eqnarray}
  \big( \pr_l r_{ijkl}^{\textrm{ng}} \big) \hat{b}_i \hat{b}_j &\overset{\textrm{\tiny lin}}{=}& \frac{1}{\rho_0} \Big[
    p_\parallel \pr_l \Pi_{kl} +2p_\parallel \hat{b}_k (\pr_l \Pi_{il})\hat{b}_i +2p_\parallel \hat{b}_l (\pr_l \Pi_{ik})\hat{b}_i
    + p_{kl}^\textrm{g} (\pr_l\Pi_{ij}) \hat{b}_i\hat{b}_j\Big]; \\
  \big( \pr_l r_{ijkl}^{\textrm{ng}} \big) \delta_{ij} &\overset{\textrm{\tiny lin}}{=}& \frac{1}{\rho_0} \Big[
    (p_\parallel+6p_\perp)\pr_l \Pi_{kl} +2(p_\parallel-p_\perp)\hat{b}_k (\pr_l \Pi_{il}) \hat{b}_i +2(p_\parallel-p_\perp)\hat{b}_l (\pr_l \Pi_{ik}) \hat{b}_i \Big],
\end{eqnarray}
and further evaluation with $\bhat_0$
(e.g. $\hat{b}_l\pr_l\rightarrow \pr_z$, or $\Pi_{ik}\hat{b}_i\rightarrow \Pi_{zk}$, but keeping the $\hat{b}_k$ intact for now) yields
\begin{eqnarray}
  \Big[ \bhat\bhat : (\nabla\cdot \br^{\textrm{ng}}) \Big]_k &\overset{\textrm{\tiny lin}}{=}&
  \frac{1}{\rho_0} \Big[ p_\parallel \pr_l \Pi_{kl} +2p_\parallel \hat{b}_k \pr_l \Pi_{zl} +2p_\parallel \pr_z \Pi_{zk} \Big]; \\
  \Big[ \boldsymbol{I} : (\nabla\cdot \br^{\textrm{ng}}) \Big]_k &\overset{\textrm{\tiny lin}}{=}&
  \frac{1}{\rho_0} \Big[ (p_\parallel+6p_\perp) \pr_l \Pi_{kl} +2(p_\parallel-p_\perp)\hat{b}_k \pr_l \Pi_{zl}+2(p_\parallel-p_\perp) \pr_z \Pi_{zk} \Big]; \\
  \Big[ (\boldsymbol{I}-\bhat\bhat)/2 : (\nabla\cdot \br^{\textrm{ng}})\Big]_k &\overset{\textrm{\tiny lin}}{=}&
  \frac{1}{\rho_0} \Big[ 3p_\perp \pr_l \Pi_{kl} -p_\perp \hat{b}_k\pr_l\Pi_{zl}-p_\perp \pr_z \Pi_{zk} \Big].
\end{eqnarray}
We purposely kept the $\hat{b}_k$ for now, since multiplying the results by $\hat{b}_k$ and further linearizing yields
\begin{eqnarray}
  \bhat\cdot \Big[ \bhat\bhat : (\nabla\cdot \br^{\textrm{ng}}) \Big] &\overset{\textrm{\tiny lin}}{=}&  3\frac{p_\parallel^{(0)}}{\rho_0} \pr_l \Pi_{zl};\\
  \bhat\cdot \Big[ (\boldsymbol{I}-\bhat\bhat)/2 : (\nabla\cdot \br^{\textrm{ng}})\Big] &\overset{\textrm{\tiny lin}}{=}& 2\frac{p_\perp^{(0)}}{\rho_0} \pr_l \Pi_{zl},
\end{eqnarray}
which verifies that the contributions (\ref{eq:RngVerif1}), (\ref{eq:RngVerif2}) to the gyrotropic heat flux equations were indeed calculated correctly. 
In contrast, applying $\bhat\times$ eliminates terms proportional to $\hat{b}_k$, implying 
\begin{eqnarray}
  \bhat\times \Big[ \bhat\bhat : (\nabla\cdot \br^{\textrm{ng}}) \Big]_k &\overset{\textrm{\tiny lin}}{=}&
  \bhat_0\times \frac{p_\parallel^{(0)}}{\rho_0} \Big[ \pr_l \Pi_{kl} +2 \pr_z \Pi_{kz} \Big]; \\
  \bhat\times \Big[ (\boldsymbol{I}-\bhat\bhat)/2 : (\nabla\cdot \br^{\textrm{ng}}) \Big]_k &\overset{\textrm{\tiny lin}}{=}&
  \bhat_0\times \frac{p_\perp^{(0)}}{\rho_0} \Big[ 3 \pr_l \Pi_{kl} - \pr_z \Pi_{kz} \Big],
\end{eqnarray}
where we have used perhaps a bit strange mixture of vector and index notations, but which should be otherwise clear. 
For any vector $\boldsymbol{a}$, the notation just means $\bhat_0\times\boldsymbol{a}=(-a_y,a_x,0)$.
Similarly straightforward calculation yields
\begin{eqnarray}
 \bhat\times \Big[ \bhat\bhat : (\bp^{\textrm{g}} \nabla\cdot \boldsymbol{\Pi})^S \Big) \Big]_k &\overset{\textrm{\tiny lin}}{=}&
 \bhat_0\times p_\parallel^{(0)} \pr_l \Pi_{lk};\\
 \bhat\times \Big[ \boldsymbol{I} : (\bp^{\textrm{g}} \nabla\cdot \boldsymbol{\Pi})^S \Big) \Big]_k &\overset{\textrm{\tiny lin}}{=}&
 \bhat_0\times (p_\parallel^{(0)}+4p_\perp^{(0)}) \pr_l \Pi_{lk};\\
 \bhat\times \Big[ (\boldsymbol{I}-\bhat\bhat)/2 : (\bp^{\textrm{g}} \nabla\cdot \boldsymbol{\Pi})^S \Big) \Big]_k &\overset{\textrm{\tiny lin}}{=}&
 \bhat_0\times 2p_\perp^{(0)} \pr_l \Pi_{lk}.
\end{eqnarray}
The final form of the second-order non-gyrotropic heat flux vectors reads
 \begin{eqnarray}
  (\boldsymbol{S}^{\parallel}_{\perp})^{(2)}_k &\overset{\textrm{\tiny lin}}{=}& \frac{1}{\Omega} \bhat_0 \times \Big[ \frac{\pr}{\pr t} (\boldsymbol{S}^{\parallel}_{\perp})^{(1)}_k
  + 2\frac{p_\parallel^{(0)}}{\rho_0} \pr_z \Pi_{kz} \Big]; \label{eq:Spar_2nd}\\
 (\boldsymbol{S}^{\perp}_{\perp})^{(2)}_k &\overset{\textrm{\tiny lin}}{=}& \frac{1}{\Omega} \bhat_0 \times \Big[ \frac{\pr}{\pr t} (\boldsymbol{S}^{\perp}_{\perp})^{(1)}_k
+ \frac{p_\perp^{(0)}}{\rho_0}\Big( \pr_l \Pi_{lk} -\pr_z \Pi_{kz}  \Big) \Big]. \label{eq:Sperp_2nd}
 \end{eqnarray}
 Alternatively, one can write $\pr_l \Pi_{lk} -\pr_z \Pi_{kz} = \pr_x \Pi_{xk}+\pr_y \Pi_{yk}$. The second order heat flux vectors contribute to scalar (gyrotropic)
 pressure equations, and also to the FLR pressure tensor $\boldsymbol{\Pi}$. The non-gyrotropic pressure tensor $\boldsymbol{\Pi}$, and the non-gyrotropic
 heat flux vectors $\boldsymbol{S}^{\parallel}_{\perp}$, $\boldsymbol{S}^{\perp}_{\perp}$ are therefore generally coupled. To make the system easily solvable,
 terms with $\boldsymbol{\Pi}$ on the r.h.s. of (\ref{eq:Spar_2nd}), (\ref{eq:Sperp_2nd}) are approximated by $\boldsymbol{\Pi}^{(1)}$. 
 The equations (\ref{eq:Spar_2nd}), (\ref{eq:Sperp_2nd}) can be written in a more elegant vector form, see eq. (\ref{eq:Spar_2nd3}), (\ref{eq:Sperp_2nd3}),
 and the result is equivalent to equations 53, 54 of \cite{PSH2012}. 

 Instead of splitting of $\boldsymbol{\Pi}$, $\boldsymbol{S}^{\parallel}_{\perp}$ and $\boldsymbol{S}^{\perp}_{\perp}$ to the first and second order components,
 one might be interested in the future to check the behaviour of the coupled system. I.e., one might be interested in solving the pressure tensor
 equation, that is coupled to the heat flux vectors, by making the associated variables as independent quantities. 
 For example, one might be interested to double check the solutions for the parallel firehose instability. The procedure how to obtain time dependent equations
 for $\boldsymbol{S}^\parallel_\perp$ and $\boldsymbol{S}^\perp_\perp$ can be perhaps a bit blurry from the steps outlined above, so let's write it one more time, only
 for $\boldsymbol{S}^\parallel_\perp$. One can work directly with the heat flux tensor equation (\ref{eq:Qngtensor}), and
keeping only terms that contribute at the linear level, the partially linearized heat flux tensor equation reads
\begin{eqnarray}
  \frac{\pr}{\pr t}\bq 
  + \Omega\big(\bhat\times\bq^\textrm{ng})^S +\nabla\cdot\br - \frac{1}{\rho_0}\Big(\bp(\nabla\cdot\bp)\Big)^S = 0. \nn
\end{eqnarray}
By applying $:\bhat\bhat$ and $\bhat\times$, yields that at the linear level
\begin{eqnarray}
  \bhat_0\times \frac{\pr}{\pr t} \boldsymbol{S}^{\parallel}_\perp -\Omega \boldsymbol{S}^{\parallel}_\perp +\bhat_0\times \Big[ p_\perp^{(0)}\nabla\big( \frac{p_\parallel}{\rho} \big)
   +2\frac{p_\parallel^{(0)}}{\rho_0}(p_\parallel^{(0)}-p_\perp^{(0)})\pr_z \bhat \Big] +\bhat_0\times 2\frac{p_\parallel^{(0)}}{\rho_0} \pr_z \vec{\boldsymbol{\Pi}}_z =0, 
\end{eqnarray}
or in the index notation
\begin{eqnarray}
 \frac{\pr}{\pr t} ({S}^{\parallel}_\perp)_x -\Omega ({S}^{\parallel}_\perp)_y + p_\perp^{(0)} \pr_x \big( \frac{p_\parallel}{\rho} \big)
   +2\frac{p_\parallel^{(0)}}{\rho_0}(p_\parallel^{(0)}-p_\perp^{(0)})\pr_z \hat{b}_x  + 2\frac{p_\parallel^{(0)}}{\rho_0} \pr_z \Pi_{xz} =0;\\
  \frac{\pr}{\pr t} ({S}^{\parallel}_\perp)_y +\Omega ({S}^{\parallel}_\perp)_x + p_\perp^{(0)} \pr_y \big( \frac{p_\parallel}{\rho} \big)
   +2\frac{p_\parallel^{(0)}}{\rho_0}(p_\parallel^{(0)}-p_\perp^{(0)})\pr_z \hat{b}_y  + 2\frac{p_\parallel^{(0)}}{\rho_0} \pr_z \Pi_{yz} =0.
\end{eqnarray}
Importantly, bi-Maxwellian distribution function was assumed. Normalizing the equations, writing them in the x-z plane and Fourier transforming yields
\begin{eqnarray}
&& -i\omega({S}^{\parallel}_\perp)_x - ({S}^{\parallel}_\perp)_y +\frac{\bpar}{2}a_p i k_\perp (p_\parallel-\rho) 
   +\bpar(1-a_p)ik_\parallel B_x  + \bpar ik_\parallel \Pi_{xz} =0;\\
&&  -i\omega ({S}^{\parallel}_\perp)_y + ({S}^{\parallel}_\perp)_x
   +\bpar(1-a_p)ik_\parallel B_y  + \bpar ik_\parallel \Pi_{yz} =0.
\end{eqnarray}
Considering parallel propagation ($k_\perp=0$), these equations are coupled with equations for $\Pi_{xz},\Pi_{yz}$, eq. (\ref{eq:PixzRit}), (\ref{eq:PiyzRit}),
and also with equations for $u_x, u_y, B_x, B_y$, system (\ref{eq:ParallelFirehoseKurva}). Now it should be easy to derive the dispersion relation, and
for example check, if the solutions for the parallel firehose instability are improved. 

\subsection{Contributions of $\boldsymbol{q}^{\textrm{ng}}$ to various equations}
Here we calculate contributions of $\boldsymbol{q}^{\textrm{ng}}$ to the FLR pressure tensor, scalar heat flux equations, and scalar
pressure equations. Contributions to the FLR pressure tensor are calculated at the linear level by neglecting $\boldsymbol{\sigma}$.
Contributions to scalar heat flux equations and scalar pressure equations are calculated nonlinearly. In these calculations, no specific form of a distribution
function is assumed.
\subsubsection*{Contributions to FLR pressure tensor}
For contributions to the FLR tensor $\boldsymbol{\Pi}$, we need to calculate $\nabla\cdot\bq^{\textrm{ng}}$. To calculate it, we will neglect the heat flux
tensor $\boldsymbol{\sigma}$, and use the decomposition of the heat flux tensor $\boldsymbol{S}$, that can be figured out to be 
\begin{equation}
  \boldsymbol{S} = q_\parallel \bhat\bhat\bhat + q_\perp \Big[ \bhat(\boldsymbol{I}-\bhat\bhat)\Big]^S
  +\Big[ \boldsymbol{S}^\parallel_\perp \bhat\bhat  \Big]^S + \frac{1}{2}\Big[ \boldsymbol{S}^\perp_\perp (\boldsymbol{I}-\bhat\bhat)\Big]^S,
\end{equation}
or in the index notation
\begin{eqnarray}
  S_{ijk} &=&q_\parallel \hat{b}_i\hat{b}_j\hat{b}_k + q_\perp \Big[ \hat{b}_i\delta_{jk} +\hat{b}_j \delta_{ik}+\hat{b}_k \delta_{ij} -3\hat{b}_i \hat{b}_j \hat{b}_k \Big] \nn\\
  && +\big(S^\parallel_\perp\big)_i \hat{b}_j \hat{b}_k + \big(S^\parallel_\perp\big)_j \hat{b}_i \hat{b}_k
  + \big(S^\parallel_\perp\big)_k \hat{b}_i \hat{b}_j \nn\\
  &&+ \frac{1}{2}\Big[ \big( S^\perp_\perp)_i (\delta_{jk}-\hat{b}_j\hat{b}_k) +\big( S^\perp_\perp)_j (\delta_{ik}-\hat{b}_i\hat{b}_k)
    +\big( S^\perp_\perp)_k (\delta_{ij}-\hat{b}_i\hat{b}_j) \Big].
\end{eqnarray}
Here we want to calculate only contributions at the linear level, and linearizing the $\nabla\cdot\boldsymbol{S}$ yields
\begin{eqnarray} 
  \pr_k S_{ijk} &\overset{\textrm{\tiny lin}}{=}& \hat{b}_i \hat{b}_j \pr_z q_\parallel +\hat{b}_i \pr_j q_\perp +\hat{b}_j \pr_i q_\perp +\delta_{ij}\pr_z q_\perp
  -3\hat{b}_i \hat{b}_j \pr_z q_\perp\nn\\
  &+& \hat{b}_j\pr_z (S^\parallel_\perp)_i  + \hat{b}_i \pr_z (S^\parallel_\perp)_j + \hat{b}_i\hat{b}_j\pr_k (S^\parallel_\perp)_k \nn\\
  &+& \frac{1}{2}\Big[ \pr_j (S^\perp_\perp)_i -\hat{b}_j \pr_z (S^\perp_\perp)_i  +\pr_i (S^\perp_\perp)_j -\hat{b}_i \pr_z (S^\perp_\perp)_j
     +\delta_{ij}\pr_k (S^\perp_\perp)_k -\hat{b}_i\hat{b}_j \pr_k (S^\perp_\perp)_k \Big],
\end{eqnarray}
and straightforward evaluation gives
\begin{eqnarray}
  \big( \nabla\cdot\boldsymbol{S} \big)_{xx} &=& \pr_z q_\perp +\frac{1}{2} \big[ 3\pr_x (S^\perp_\perp)_x +\pr_y (S^\perp_\perp)_y \big];\\
  \big( \nabla\cdot\boldsymbol{S} \big)_{xy} &=& \frac{1}{2} \big[ \pr_y (S^\perp_\perp)_x +\pr_x (S^\perp_\perp)_y\big];\\
  \big( \nabla\cdot\boldsymbol{S} \big)_{xz} &=& \pr_x q_\perp +\pr_z (S^\parallel_\perp)_x;\\
  \big( \nabla\cdot\boldsymbol{S} \big)_{yy} &=& \pr_z q_\perp +\frac{1}{2}\big[ \pr_x (S^\perp_\perp)_x +3\pr_y (S^\perp_\perp)_y \big];\\
  \big( \nabla\cdot\boldsymbol{S} \big)_{yz} &=& \pr_y q_\perp +\pr_z (S^\parallel_\perp)_y;\\
  \big( \nabla\cdot\boldsymbol{S} \big)_{zz} &=& \pr_z q_\parallel + \pr_x (S^\parallel_\perp)_x +  \pr_y (S^\parallel_\perp)_y.
\end{eqnarray}
\subsubsection*{Contributions to scalar heat flux equations}
When we calculated the scalar heat flux equations, we separated the non-gyrotropic heat flux $\bq^\textrm{ng}$ by defining a quantity
\begin{equation}
\boldsymbol{Q}^{\textrm{ng}} =   \frac{d}{d t}\bq^\textrm{ng} +\bq^\textrm{ng}\nabla\cdot\bu
  + \Big[\bq^\textrm{ng} \cdot\nabla\bu + \Omega\frac{|\bb|}{B_0}\bhat\times\bq^\textrm{ng} \Big]^S,\nn
\end{equation} 
and we need to calculate 
\begin{equation}
Q^{\textrm{ng}}_\parallel\equiv (\boldsymbol{Q}^{\textrm{ng}}:\bhat\bhat)\cdot\bhat; \qquad  
Q^{\textrm{ng}}_\perp\equiv (\boldsymbol{Q}^{\textrm{ng}}:(\boldsymbol{I}-\bhat\bhat)/2)\cdot\bhat.
\end{equation}  
Here we want to obtain exact nonlinear expressions. Direct calculations yields
\begin{eqnarray}
  \Big( \frac{d}{d t} q^{\textrm{ng}}_{ijk} \Big) \hat{b}_i\hat{b}_j\hat{b}_k &=& \frac{d}{d t}\big( \underbrace{q^{\textrm{ng}}_{ijk} \hat{b}_i\hat{b}_j\hat{b}_k}_{=0} \big)
  -q^{\textrm{ng}}_{ijk} \frac{d}{d t}\big( \hat{b}_i\hat{b}_j\hat{b}_k \big) =-3\boldsymbol{S}^\parallel_\perp \cdot\frac{d\bhat}{dt};\\
  (\bq^\textrm{ng} \cdot\nabla\bu)^S_{ijk} \hat{b}_i\hat{b}_j\hat{b}_k &=& 3 \boldsymbol{S}^\parallel_\perp \cdot (\nabla\bu)\cdot\bhat; \\
  (\bhat\times\bq^\textrm{ng})^S_{ijk} \hat{b}_i\hat{b}_j\hat{b}_k &=& 0, 
\end{eqnarray}
and the final fully nonlinear result reads
\begin{equation} \label{eq:QngParX}
Q^{\textrm{ng}}_\parallel = -3\boldsymbol{S}^\parallel_\perp \cdot\frac{d\bhat}{dt} + 3 \boldsymbol{S}^\parallel_\perp \cdot (\nabla\bu)\cdot\bhat.
\end{equation}
Similar calculations yield nonlinear expression for the $Q^{\textrm{ng}}_\perp$, which calculates
\begin{eqnarray}
  \Big( \frac{d}{d t} q^{\textrm{ng}}_{ijk} \Big) \delta_{ij}\hat{b}_k &=& \frac{d}{d t}\big( \underbrace{q^{\textrm{ng}}_{ijk} \delta_{ij}\hat{b}_k}_{=0} \big)
  -q^{\textrm{ng}}_{ijk} \delta_{ij}\frac{d}{d t}\hat{b}_k = -(\bq^{\textrm{ng}}:\boldsymbol{I})\cdot \frac{d\bhat}{dt} =
  -(\boldsymbol{S}^\parallel_\perp+2\boldsymbol{S}^\perp_\perp)\cdot\frac{d\bhat}{dt};\\
  \Big( \frac{d}{d t} q^{\textrm{ng}}_{ijk} \Big) (\delta_{ij}-\hat{b}_i\hat{b}_j)\hat{b}_k/2 &=& (\boldsymbol{S}^\parallel_\perp-\boldsymbol{S}^\perp_\perp)\cdot\frac{d\bhat}{dt};\\
  (\bhat\times\bq^\textrm{ng})^S_{ijk} \delta_{ij}\hat{b}_k &=& 0;\\
  (\bq^\textrm{ng} \cdot\nabla\bu)^S_{ijk} \delta_{ij}\hat{b}_k &=& \big(q^{\textrm{ng}}_{ijl}\pr_l u_k+q^{\textrm{ng}}_{jkl}\pr_l u_i+q^{\textrm{ng}}_{kil}\pr_l u_j\big)\delta_{ij}\hat{b}_k
  = (\bq^{\textrm{ng}}:\boldsymbol{I})_l (\pr_l u_k)\hat{b}_k + 2  q^{\textrm{ng}}_{ilk} \hat{b}_k \pr_l u_i\nn\\
  &=& (\boldsymbol{S}^\parallel_\perp+2\boldsymbol{S}^\perp_\perp)\cdot(\nabla\bu)\cdot\bhat + 2 (\bq^{\textrm{ng}}\cdot\bhat):\nabla\bu.
\end{eqnarray}
To continue the calculation, we need to use (an exact expression)
\begin{equation}
\bq^{\textrm{ng}} = \Big[ \boldsymbol{S}^\parallel_\perp \bhat\bhat  \Big]^S + \frac{1}{2}\Big[ \boldsymbol{S}^\perp_\perp (\boldsymbol{I}-\bhat\bhat)\Big]^S +\boldsymbol{\sigma},
\end{equation}  
which allows us to calculate
\begin{eqnarray}
  q^{\textrm{ng}}_{ijk}\hat{b}_k &=& (S^{\parallel}_\perp)_i \hat{b}_j + (S^{\parallel}_\perp)_j \hat{b}_i +\sigma_{ijk}\hat{b}_k; \\
  q^{\textrm{ng}}_{ilk}\hat{b}_k \pr_l u_i &=&  (S^{\parallel}_\perp)_i \hat{b}_l\pr_l u_i + (S^{\parallel}_\perp)_l (\pr_l u_i)\hat{b}_i +\sigma_{ilk}\hat{b}_k\pr_l u_i\nn\\
  &=& \boldsymbol{S}^\parallel_\perp \cdot (\bhat\cdot\nabla\bu) + \boldsymbol{S}^\parallel_\perp \cdot (\nabla\bu)\cdot\bhat + (\boldsymbol{\sigma}\cdot\bhat):\nabla\bu,
\end{eqnarray}  
and therefore
\begin{eqnarray}
  (\bq^\textrm{ng} \cdot\nabla\bu)^S_{ijk} \delta_{ij}\hat{b}_k &=&
  (3\boldsymbol{S}^\parallel_\perp+2\boldsymbol{S}^\perp_\perp)\cdot(\nabla\bu)\cdot\bhat + 2\boldsymbol{S}^\parallel_\perp \cdot (\bhat\cdot\nabla\bu)
  +2 (\boldsymbol{\sigma}\cdot\bhat):\nabla\bu;\\
  (\bq^\textrm{ng} \cdot\nabla\bu)^S_{ijk} (\delta_{ij}-\hat{b}_i\hat{b}_j)\hat{b}_k/2 &=& \boldsymbol{S}^\perp_\perp\cdot(\nabla\bu)\cdot\bhat
  + \boldsymbol{S}^\parallel_\perp \cdot (\bhat\cdot\nabla\bu) +\bhat\cdot\boldsymbol{\sigma}:\nabla\bu.
\end{eqnarray}
The final fully nonlinear result reads
\begin{equation}
Q^{\textrm{ng}}_\perp = (\boldsymbol{S}^\parallel_\perp-\boldsymbol{S}^\perp_\perp)\cdot\frac{d\bhat}{dt} + \boldsymbol{S}^\perp_\perp\cdot(\nabla\bu)\cdot\bhat
  + (\bhat\cdot\nabla\bu)\cdot\boldsymbol{S}^\parallel_\perp +\bhat\cdot\boldsymbol{\sigma}:\nabla\bu. \label{eq:QngPerpX}
\end{equation}  
However, it is easy to see that at the linear level both $Q^{\textrm{ng}}_\parallel\overset{\textrm{\tiny lin}}{=} 0$ and $Q^{\textrm{ng}}_\perp\overset{\textrm{\tiny lin}}{=} 0$.
\subsubsection*{Contributions to scalar pressure equations}
The non-gyrotropic heat flux term entering the parallel pressure equation calculate
\begin{eqnarray}
  (\nabla\cdot\bq^{\textrm{ng}}):\bhat\bhat &=& (\pr_k q^{\textrm{ng}}_{kij}) \hat{b}_i\hat{b}_j = \pr_k \big(  q^{\textrm{ng}}_{kij} \hat{b}_i\hat{b}_j \big)
  - q^{\textrm{ng}}_{kij} \pr_k (\hat{b}_i\hat{b}_j ) \nn\\
  &=& \pr_k \big( S^{\parallel}_\perp \big)_k - 2q^{\textrm{ng}}_{kij} \hat{b}_i \pr_k \hat{b}_j,
\end{eqnarray}
and since
\begin{eqnarray}
  q^{\textrm{ng}}_{ijk}\hat{b}_i &=& (S^{\parallel}_\perp)_j \hat{b}_k + (S^{\parallel}_\perp)_k \hat{b}_j +\sigma_{ijk}\hat{b}_i; \\
  q^{\textrm{ng}}_{ijk}\hat{b}_i\pr_k \hat{b}_j &=& (S^{\parallel}_\perp)_j \hat{b}_k \pr_k \hat{b}_j +  (S^{\parallel}_\perp)_k \underbrace{\hat{b}_j\pr_k \hat{b}_j}_{=0}
  + \sigma_{ijk}\hat{b}_i \pr_k \hat{b}_j  = (\bhat\cdot\nabla\bhat)\cdot \boldsymbol{S}^\parallel_\perp +\bhat\cdot\boldsymbol{\sigma}:\nabla\bhat,
\end{eqnarray}
which yields the final nonlinear result
\begin{eqnarray}
  (\nabla\cdot\bq^{\textrm{ng}}):\bhat\bhat &=& \nabla\cdot \boldsymbol{S}^\parallel_\perp
  -2(\bhat\cdot\nabla\bhat)\cdot \boldsymbol{S}^\parallel_\perp -2\bhat\cdot\boldsymbol{\sigma}:\nabla\bhat. \label{eq:QngPressure1}
\end{eqnarray}
Similarly, the term entering the perpendicular pressure equation calculates
\begin{eqnarray}
  (\nabla\cdot\bq^{\textrm{ng}}):\boldsymbol{I} &=& (\pr_k q^{\textrm{ng}}_{kij}) \delta_{ij} = \nabla\cdot (\bq^{\textrm{ng}}:\boldsymbol{I}) =
  \nabla\cdot( \boldsymbol{S}^\parallel_\perp + 2\boldsymbol{S}^\perp_\perp ); \\
  (\nabla\cdot\bq^{\textrm{ng}}):(\boldsymbol{I}-\bhat\bhat)/2 &=& \nabla\cdot\boldsymbol{S}^\perp_\perp+(\bhat\cdot\nabla\bhat)\cdot \boldsymbol{S}^\parallel_\perp
  +\bhat\cdot\boldsymbol{\sigma}:\nabla\bhat. \label{eq:QngPressure2}
\end{eqnarray}  
The nonlinear results (\ref{eq:QngPressure1}), (\ref{eq:QngPressure2}) are exact, and here we do not address how to further decompose the tensor $\boldsymbol{\sigma}$.
Finally, considering contributions at the linear level (assuming that the mean values of the entire heat flux $\bq^{(0)}=0$), it is easy to see that
\begin{eqnarray}
  (\nabla\cdot\bq^{\textrm{ng}}):\bhat\bhat &\overset{\textrm{\tiny lin}}{=}&
  \nabla\cdot \boldsymbol{S}^{\parallel}_\perp  = \pr_x \big( S^{\parallel}_\perp \big)_x + \pr_y \big( S^{\parallel}_\perp \big)_y; \label{eq:ScontribP1}\\
  (\nabla\cdot\bq^{\textrm{ng}}):(\boldsymbol{I}-\bhat\bhat)/2 &\overset{\textrm{\tiny lin}}{=}&
  \nabla\cdot\boldsymbol{S}^{\perp}_\perp = \pr_x \big( S^{\perp}_\perp \big)_x + \pr_y \big( S^{\perp}_\perp \big)_y. \label{eq:ScontribP2}
\end{eqnarray} 
The heat flux tensor $\boldsymbol{\sigma}$ therefore does not contribute to the gyrotropic pressure equations at the linear level.

\bibliographystyle{jpp}
\bibliography{hunana_mhd}

\end{document}